%% file: etdrtemplate.tex
\documentclass[final,letterpaper,12pt,oneside]{class_diss}

\usepackage{graphicx} % Extended graphics package.
\usepackage{amsmath} % American Mathematics Society standards
\usepackage{amsxtra} % Additional math symbols
\usepackage{amssymb} % Additional math symbols
\usepackage{amsthm} % Additional math symbols
\usepackage{latexsym} % Additional math symbols
\usepackage{setspace} % Controls line spacing
\usepackage{nomencl} % Enables nomenclature list addition 
\usepackage[margin=1in]{geometry} % Sets page margins to 1 inch on all sides
\usepackage[titles]{tocloft} % Adds leader dots to all entries in the table of contents
\usepackage{comment}

\usepackage{amsmath, nccmath, amssymb}
\usepackage{graphicx}
\usepackage{multicol}
\usepackage{threeparttable}
\usepackage{upgreek}
\usepackage{graphics}
\usepackage{bm}
\usepackage{longtable}
\usepackage{appendix}

\usepackage{subfig}
\usepackage{comment}
\usepackage{longtable}
\usepackage{caption}
\usepackage{threeparttable}
\usepackage{booktabs}
\usepackage[figuresright]{rotating}
\newcommand{\be}{\begin{equation}}
\newcommand{\ee}{\end{equation}}
\newcommand{\bea}{\begin{eqnarray}}
\newcommand{\eea}{\end{eqnarray}}
\newcommand{\hunit}{$\rm{km \ s^{-1} \ Mpc^{-1}}$}
\usepackage{array}
\makeatletter
\newcommand{\thickhline}{%
    \noalign {\ifnum 0=`}\fi \hrule height 1pt
    \futurelet \reserved@a \@xhline
}
\newcolumntype{"}{@{\hskip\tabcolsep\vrule width 1pt\hskip\tabcolsep}}
\makeatother

\newcommand{\hiig}{H\,\textsc{ii}G}
\newcommand{\hii}{H\,\textsc{ii}}

\usepackage{array}
\usepackage{subfig}
\usepackage{comment}
\usepackage{longtable}
\usepackage{caption}
\usepackage{placeins}
\usepackage{threeparttable}

\newcommand{\lcdm}{$\Lambda$CDM}
\newcommand{\pcdm}{$\phi$CDM}
\newcommand{\om}{\Omega_{m0}}
\newcommand{\ol}{\Omega_{\Lambda}}
\newcommand{\ok}{\Omega_{k0}}
\newcommand{\FT}[1]{}

\usepackage[authoryear,sort&compress]{natbib}
\setcitestyle{authoryear}
\usepackage[pdftex, plainpages=false, pdfpagelabels]{hyperref}

\hypersetup{
    linktocpage=true,
    colorlinks=true,
    bookmarks=true,
    citecolor=blue,
    urlcolor=blue,
    linkcolor=blue,
    citebordercolor={1 0 0},
    urlbordercolor={1 0 0},
    linkbordercolor={.7 .8 .8},
    breaklinks=true,
    pdfpagelabels=true,
    }

\begin{document}

% +--------------------------------------------------------------------+
% | ******Masters Students -- You Need to Make Some Changes Here******
% |
% | The Abstract Title page and Abstract page following the Abstract
% | Title page are required only for doctoral dissertations.  For
% | masters theses or reports, comment out or delete the lines:
% |
% | \input{abstracttitle.tex} through \end{abstract}.
% |
% | You will also need to uncomment the two lines following the
% | \begin{abstract} command:
% |    %\setcounter{page}{-1}
% |    %\pdfbookmark[0]{Abstract}{PDFAbstractPage}
% |
% | Don't uncomment the lines above.  Scroll down several lines until
% | you see the section "For masters theses or reports, uncomment
% | the commands..." and uncomment the lines in that section.
% +--------------------- ----------------------------------------------+

\input{abstracttitle.tex}  % Masters students - comment or delete this line

\begin{abstract} % Masters students - comment or delete this line
   \setcounter{page}{-1} % Masters students - comment or delete this line
   \pdfbookmark[0]{Abstract}{PDFAbstractPage} % Masters students - comment or delete this line
   \input{abstract.tex} % Masters students - comment or delete this line
   \vfill % Masters students - comment or delete this line
\end{abstract} % Masters students - comment or delete this line

% +--------------------------------------------------------------------+
% | Title Page -- Required for both Doctoral and Masters Students
% +--------------------------------------------------------------------+

\input{title.tex}

% +--------------------------------------------------------------------+
% | Copyright Page -- Required for both Doctoral and Masters Students
% +--------------------------------------------------------------------+

\input{copyright.tex}

% +--------------------------------------------------------------------+
% |  Abstract -- Required for both Doctoral and Masters Students
% +--------------------------------------------------------------------+

\begin{abstract}

% +--------------------------------------------------------------------+
% | For masters theses or reports, uncomment the commands on the next
% | two lines (\setcounter and \pdfbookmark)
% +--------------------------------------------------------------------+

   %\setcounter{page}{-1}
   %\pdfbookmark[0]{Abstract}{PDFAbstractPage}

\input{abstract.tex}
\vfill
\end{abstract}

% +--------------------------------------------------------------------+
% | The following commands start a new page and set the page numbering
% | to lowercase roman numerals.
% +--------------------------------------------------------------------+

\newpage
\pagenumbering{roman}

% +--------------------------------------------------------------------+
% |
% | *********************** IMPORTANT ******************************
% |
% | In the \setcounter command below, set the number to represent the
% | page number of the table of contents page.  For example, if the
% | table of contents page is the 6th page of your document, enter 6
% | in the brackets.  This number may vary, depending on the length of
% | your abstract.
% |
% | Numbers do not appear on the title and abstract pages, but they
% | are included in the page count.  The table of contents page is the
% | first page on which page numbers are displayed.
% +--------------------------------------------------------------------+

\setcounter{page}{6}

% +--------------------------------------------------------------------+
% | The following command creates a bookmark for the table of contents
% | in the final PDF document.
% +--------------------------------------------------------------------+

\pdfbookmark[0]{\contentsname}{contents}

% +--------------------------------------------------------=-----------+
% | The following command adds dot leaders for all entries in the
% | table of contents.
% +--------------------------------------------------------------------+

\renewcommand{\cftchapleader}{\cftdotfill{\cftdotsep}}

% +--------------------------------------------------------------------+
% | The following commands makes all entries and page numbers in the
% | table of contents appear in normal weight font (not bold).
% +--------------------------------------------------------------------+

\renewcommand{\cftchapfont}{\mdseries}
\renewcommand{\cftchappagefont}{\mdseries}

% +--------------------------------------------------------------------+
% | These commands add the table of contents, list of figures, and
% | list of tables.
% +--------------------------------------------------------------------+

\makenomenclature
\nomenclature{symbol}{definition}
%  Some fields use the nomenclature for a List of Symbols or Terms.  
%  This list is sometimes emplaced within chapters.  In other cases, it is introduced early and follows the List of Figures and List of Tables.  
%  In the location where you want your nomenclature to appear, use the following command <\printnomenclature>
%  If you put the nomenclature in the Table of Contents, make sure that you are referring to that below  

\tableofcontents
\listoffigures
\listoftables

% +--------------------------------------------------------------------+
% | Nomenclature Page
% |
% | If you choose not to have a Nomenclature page, comment out
% | or delete the following 3 lines.
% +--------------------------------------------------------------------+

%\newpage
%\phantomsection
%\addcontentsline{toc}{chapter}{List of Nomenclature}
%\input{nomenclature.tex}

% +--------------------------------------------------------------------+
% | Acknowledgements Page
% |
% | If you choose not to have an Acknowledgements page, comment out
% | or delete the following 3 lines.
% +--------------------------------------------------------------------+

\newpage
\phantomsection
\addcontentsline{toc}{chapter}{Acknowledgements}
\input{acknowledge.tex}

% +--------------------------------------------------------------------+
% | Dedication Page
% |
% | If you choose not to have a Dedication page, comment out
% | or delete the following 3 lines.
% +--------------------------------------------------------------------+
\begin{comment}
\newpage
\phantomsection
\addcontentsline{toc}{chapter}{Dedication}
\input{dedication.tex}
\end{comment}

% +--------------------------------------------------------------------+
% | Preface Page
% |
% | If you choose not to have a Dedication page, comment out
% | or delete the following 4 lines.
% +--------------------------------------------------------------------+
\begin{comment}
\newpage
\phantomsection
\addcontentsline{toc}{chapter}{Preface}
\input{preface.tex}
\end{comment}

% +--------------------------------------------------------------------+
% | This is where the chapter content of your ETDR begins.
% +--------------------------------------------------------------------+

%\phantomsection
\newpage
\pagenumbering{arabic}
\setcounter{page}{1}

% +--------------------------------------------------------------------+
% | Individual chapters of your ETDR are added using the \input
% | command.
% +--------------------------------------------------------------------+

\input{chapter1.tex}

\input{chapter2.tex}
\input{chapter3.tex}
\input{chapter4.tex}
\input{chapter5.tex}
\input{chapter7.tex}
\input{chapter8.tex}

\input{chapter9.tex}
\input{chapter6.tex}
\input{chapter10.tex}
\input{conclusion.tex}

% +--------------------------------------------------------------------+
% | Uncomment the lines below to add additional chapters.
% +--------------------------------------------------------------------+

%\input{chapter4.tex}
%\input{chapter5.tex}

% +--------------------------------------------------------------------+
% | References
% +--------------------------------------------------------------------+

% +--------------------------------------------------------------------+
% | Included for Gather Purpose only.  Do NOT uncomment the next line.
%input "references.bib"
% | In order for the WinEDT editor to index references correctly, it
% | has to know where the "references.bib" file resides.  This
% | command will be ignored completely by LaTeX
% |
% | WinEDT can read file path names with either "\" or "/". LaTeX,
% | however,doesn't like "\", so it's easier to store a path name
% | using forward slashes "/".
% +--------------------------------------------------------------------+

\cleardoublepage
\phantomsection

% +--------------------------------------------------------------------+
% | This template uses the BibTeX program to format references.  The
% | lines below create a separate Bibliography section and add
% | an entry for "Bibliography" to the Table of Contents.  The actual
% | data for your references (author, title, journal, date, etc.) are
% | entered in the references.bib file.  See "Citations and Bibliography"
% | for details on to creating citations and formatting references.
% +--------------------------------------------------------------------+

\addcontentsline{toc}{chapter}{Bibliography}
\bibdata{references}
\bibliography{references}

% +--------------------------------------------------------------------+
% | The following commands add the appendices  To add or delete
% | appendices, add or remove the line
% |
% |     \input{appendixX.tex}
% |
% | where "X" is the letter designation of the appendix (A, B, C,
% | etc.) You should have one \input{appendixX.tex} line and a
% | corresponding file appendixX.tex for each appendix.
% |
% |If you do not have any appendices, comment out or delete the three
% |lines below.
% +--------------------------------------------------------------------+

\appendix
\input{appendixA.tex}
\input{appendixC.tex}

\end{document}

%% file: abstracttitle.tex
% +--------------------------------------------------------------------+
% | Abstract Title Page
% |
% |This page is required only for doctoral dissertations.
% +--------------------------------------------------------------------+

% +--------------------------------------------------------------------+
% | This page should not contain a page number.  We use the
% | \thispagestyle[empty] command below to suppress page numbers
% | and other style elements.
% +--------------------------------------------------------------------+

\thispagestyle{empty}

% +--------------------------------------------------------------------+
% | The Abstract Title page begins here
% +--------------------------------------------------------------------+

\pdfbookmark[0]{Title Page}{PDFTitlePage}

\begin{center}

   \vspace{1cm}

% +--------------------------------------------------------------------+
% | Enter the title of your ETDR below.  For 2017 and on, use "Sentence case" (not ALL CAPS).
% | For details, see:  k-state.edu/grad/etdr/create/sentencecase.html 
% +--------------------------------------------------------------------+

   \large Observational constraints on the cosmological expansion rate and spatial curvature\\

   \vspace{0.5cm}

   by\\

   \vspace{0.5cm}

% +--------------------------------------------------------------------+
% | Enter your name below in standard name format (not ALL CAPS).
% +--------------------------------------------------------------------+

   \large Joseph Ryan\\

   \vspace{0.5cm}

% +--------------------------------------------------------------------+
% | On the line below, replace "Enter Your Previous Degrees"
% | with your previous degrees in mixed case. Include the abbreviation
% | for the degree, the name of the university, and the year separated
% | by commas. For example:
% |
% |    B.A., University of Illinois, 2000
% |
% | If desired, it is acceptable to include a city or country with
% | the university name. For example:
% |
% |    B.S., Jillian University, China, 2002
% |
% | Each degree should appear on a separate line.  Use the \\
% | command to create a line break.
% +--------------------------------------------------------------------+

   B.S., Wichita State University, 2013\\

   \vspace{0.55cm}
   \rule{2in}{0.5pt}\\
   \vspace{0.75cm}

   {\large AN ABSTRACT OF A DISSERTATION}\\

   \vspace{0.5cm}
   \begin{singlespace}
   submitted in partial fulfillment of the\\
   requirements for the degree\\
   \end{singlespace}

   \vspace{0.5cm}

% +--------------------------------------------------------------------+
% | On the line below, enter the name of your earned degree in ALL
% | CAPITAL LETTERS.  For example: DOCTOR OF PHILOSOPHY
% +--------------------------------------------------------------------+

   {\large DOCTOR OF PHILOSOPHY}\\
   \vspace{0.5cm}

% +--------------------------------------------------------------------+
% | On the two lines below, enter the name of your department and the
% | name of the college in mixed case.  For example:
% |
% |     Biochemistry Department
% |     College of Arts and Sciences
% +--------------------------------------------------------------------+

   \begin{singlespace}
   Department of Physics\\
   College of Arts and Sciences\\
   \end{singlespace}

   \vspace{0.5cm}

   \begin{singlespace}
   {\Large KANSAS STATE UNIVERSITY}\\
   Manhattan, Kansas\\
   \end{singlespace}

% +--------------------------------------------------------------------+
% | On the line below, replace "Graduation Year" with the four-digit year
% | of your graduation. For example:
% |
% |     2017
% +--------------------------------------------------------------------+

   2021\\
   \vspace{1cm}

\end{center}

%% file: abstract.tex
% +--------------------------------------------------------------------+
% | Abstract Page
% +--------------------------------------------------------------------+

\pagestyle{empty}
%\vspace{1cm}
\setlength{\baselineskip}{0.8cm}

%\indent

% +--------------------------------------------------------------------+
% | Enter the text of your abstract below.  The abstract should does not have an upper word limit.  
% +--------------------------------------------------------------------+

Observations conducted over the last few decades show that the expansion of the Universe is accelerating. In the standard model of cosmology, this accelerated expansion is attributed to a dark energy in the form of a cosmological constant. It is conceivable, however, for the dark energy to exhibit mild dynamics (so that its energy density changes with time rather than having a constant value), or for the accelerated expansion of the Universe to be caused by some mechanism other than dark energy. In this work I will investigate both of these possibilities by using observational data to place constraints on the parameters of simple models of dynamical dark energy as well as cosmological models without dark energy. I find that these data favor the standard model while leaving some room for dynamical dark energy.

The standard model also holds that the Universe is flat on large spatial scales. The same observational data used to test dark energy dynamics can be used to constrain the large-scale curvature of the Universe, and these data generally favor spatial flatness, with some mild preference for spatial curvature in some data combinations.

%% file: title.tex
% +--------------------------------------------------------------------+
% | Title Page
% +--------------------------------------------------------------------+

\newpage

% +--------------------------------------------------------------------+
% | This page should not contain a page number.  We use the
% | \thispagestyle[empty] command below to suppress page numbers
% | and other style elements.
% +--------------------------------------------------------------------+

\thispagestyle{empty}

% +--------------------------------------------------------------------+
% | The Title page begins here.
% +--------------------------------------------------------------------+

\begin{center}

   \vspace{1cm}

% +--------------------------------------------------------------------+
% | Enter the title of your ETDR below. For works from 2017 and on, use "Sentence case" (not ALL CAPS).
% | For details, see: k-state.edu/grad/etdr/create/sentencecase.html 
% +--------------------------------------------------------------------+

\large Observational constraints on the cosmological expansion rate and spatial curvature\\

\vspace{0.5cm}

by\\

\vspace{0.5cm}

% +--------------------------------------------------------------------+
% | Enter your name below in standard name format (not ALL CAPS).
% +--------------------------------------------------------------------+

\large Joseph Ryan\\

 \vspace{0.3cm}

% +--------------------------------------------------------------------+
% | On the line below, replace "Enter Your Previous Degrees"
% | with your previous degrees in mixed case. Include the abbreviation
% | for the degree, the name of the university, and the year separated
% | by commas. For example:
% |
% |    B.A., University of Illinois, 2000
% |
% | If desired, it is acceptable to include a city or country with
% | the university name. For example:
% |
% |    B.S., Jillian University, China, 2002
% |
% | Each degree should appear on a separate line.  Use the \\
% | command to create a line break.
% +--------------------------------------------------------------------+

   B.S., Wichita State University, 2013\\

   \vspace{0.35cm}
   \rule{2in}{0.5pt}\\
   \vspace{0.65cm}

   {\large A DISSERTATION}\\

   \vspace{0.3cm}
   \begin{singlespace}
   submitted in partial fulfillment of the\\
   requirements for the degree\\
   \end{singlespace}

   \vspace{0.3cm}

% +--------------------------------------------------------------------+
% | On the line below, replace "ENTER YOUR DEGREE NAME" with the name
% | of your earned degree in ALL CAPITAL LETTERS.
% +--------------------------------------------------------------------+

   {\large DOCTOR OF PHILOSOPHY}\\
   \vspace{0.3cm}

% +--------------------------------------------------------------------+
% | On the two lines below, replace "Enter Your Department Name" and
% | "Enter Your College Name" with the name of your department and the
% | name of the college in mixed case.  For example:
% |
% |     Biochemistry Department
% |     College of Arts and Sciences
% +--------------------------------------------------------------------+

   \begin{singlespace}
   Department of Physics\\
   College of Arts and Sciences\\
   \end{singlespace}

   \vspace{0.3cm}

   \begin{singlespace}
   {\large KANSAS STATE UNIVERSITY}\\
   Manhattan, Kansas\\
   \end{singlespace}

% +--------------------------------------------------------------------+
% | On the line below, replace "Graduation Year" with the four-digit
% | year of your graduation.  For example:
% |
% |     2016
% +--------------------------------------------------------------------+

   2021\\
   \vspace{0.3cm}

    \end{center}

    \begin{flushright}
    Approved by:\\
    \vspace{0.3cm}
    \begin{singlespace}
    Major Professor

% +--------------------------------------------------------------------+
% | On the line below, replace "Enter Your Major Professor's Name"
% | with  the name of your major professor in mixed case.  Use the
% | format Firstname Lastname.  For example:
% |
% |     Lori Goetsch
% |
% +--------------------------------------------------------------------+

    Bharat Ratra\\
    \end{singlespace}
    \end{flushright}

% +--------------------------------------------------------------------+
% | If you have co-major professors, comment out the lines above from
% | \begin{flushright} through \end{flushright} and uncomment the
% | lines below.  Enter your co-major professors' names where indicated.
% +--------------------------------------------------------------------+

%\begin{flushright}
%   Approved by:\\
%  \vspace{ 0.3cm}
%   \begin{singlespace}
%   Co-Major Professor\\
%   Enter Your Co-Major Professor's Name\\
%   \vspace{.25cm}
%   Co-Major Professor\\
%   Enter Your Co-Major Professor's Name\\
%   \end{singlespace}
%\end{flushright}

%% file: copyright.tex
% +--------------------------------------------------------------------+
% | Copyright Page
% +--------------------------------------------------------------------+

\newpage

\thispagestyle{empty}

\vspace*{0.9cm}

\begin{center}

{\bf \Huge Copyright}

\vspace{1cm}

% +--------------------------------------------------------------------+
% | The Graduate School began using a 1-line copyright format in 2017 with 
% | the symbol, author name, graduation year, and a period at the end.
% | On the line below, replace "Enter Your Name" with your name.
% | Use the same form of your name as it appears on your title page, and 
% | use Mixed Case (no more ALL CAPS). 
% | Replace "YYYY" with the four-digit year of your graduation. 
% | Be sure to include the period after the year. 
% | EXAMPLE: © Abigail J. Adams 2019.
% +--------------------------------------------------------------------+

\Large\copyright\ Joseph Ryan 2021.\\

\vspace{0.5cm}

\end{center}

%% file: acknowledge.tex
% +--------------------------------------------------------------------+
% | Acknowledgements Page (Optional)
% +--------------------------------------------------------------------+

\newpage
\vspace*{0.9cm}
\begin{center}
{\bf \Huge Acknowledgments}
\end{center}

\setlength{\baselineskip}{0.8cm}

%\pdfbookmark[0]{Acknowledgements}{PDF_Acknowledgements}

% +--------------------------------------------------------------------+
% | Enter text for your acknowledgements in the space below this box.
% |                                                                    
% +--------------------------------------------------------------------+

Thanks first of all are due to my major professor, Dr. Bharat Ratra, for his unfailing support, patience, and advice. Thanks also to my other coauthors Sanket Doshi, Yun Chen, Shulei Cao, and Narayan Khadka. Our collaboration thus far has been a fruitful one (as evidenced by the length of this document), and it would be a pleasure to work with any of you again in the future.

Thanks to Dr. Lado Samushia and Dr. Larry Weaver for helpful discussions, and for always being willing to answer my questions on various physics topics.

Thanks to the members of the Beocat support staff, particularly Dr. Dave Turner and Adam Tygart, for helping me work out the kinks in my codes. Much of the computing for this work was performed on the Beocat Research Cluster at Kansas
State University, which is funded in part by NSF grants
CNS-1006860, EPS-1006860, EPS-0919443, ACI-1440548,
CHE-1726332, and NIH P20GM113109. Additionally, this work was partially funded by Department of Energy grant DE-SC0011840.

Finally, thanks to my parents, William and Debra Ryan, to whom this work is dedicated. To say it all would require another dissertation in its own right, so I'll be brief here: you made this long journey possible, and I couldn't have done it without you behind me at every step of the way.

%% file: chapter1.tex
% +--------------------------------------------------------------------+
% | Sample Chapter 1
% |
% | This file provides examples of how to
% | - insert a figure with a caption
% | - construct a table with a caption
% | - create subsections within the chapter
% | - insert a reference to a Figure or Table
% | - make a citation
% +--------------------------------------------------------------------+

\cleardoublepage

% +--------------------------------------------------------------------+
% | Replace "Chapter Title" below with the title of your chapter.
% | LaTeX will automatically number the chapters.
% +--------------------------------------------------------------------+

\chapter{Fundamentals of theoretical cosmology}

\label{Chapter1}

%%
%Section: History
%%

\emph{Physical} cosmology is the scientific study of the origin, evolution, and fate of the Universe. More specifically, physical cosmologists use the laws of physics to understand how the Universe evolves across the largest conceivable length and time scales. Physical cosmologists concern themselves with such questions as: ``Did the Universe have a beginning? If so, how did it begin?", ``How old is the Universe?", ``What is the composition of the matter in the Universe, and how is it distributed?", ``What is the overall geometry of the Universe?", ``Will the Universe ever come to an end?". It is a testament to the remarkable scientific progress made in the last several hundred years that these kinds of questions are now beginning to be answered quantitatively and precisely. Here I will briefly review the key elements of the theoretical side of physical cosmology, from the basics of general relativity up to the derivation of the Friedmann equations that govern the background evolution of the Universe on large scales. I will not attempt to be comprehensive in this chapter, as some results will be worked out carefully and in detail while others will be stated without proof. I intend merely to give an overview of the material that I take to be most important for an understanding of the later chapters; more details can be found elsewhere (e.g. \citealp{Peebles_1993, Dodelson, Mukhanov, weinberg, zee, MCP}).

\begin{comment}
%%
%Section: Newtonian cosmology
%%

\section{Newtonian Cosmology [Newtonian physics?]}

\noindent\enquote{\itshape Nature and Nature's laws lay hid in night; God said `Let Newton be' and all was light.}\bigbreak

\hfill Alexander Pope, 1727

\vspace{5mm}

Isaac Newton laid the foundations of physical cosmology in [...] when he published his theory of universal gravity. In Newtonian physics, the magnitude ($F_{\rm G}$) of the force of gravity acting on a particle that has a mass of $m$, is equal to
\begin{equation}
\label{eq:F_G}
    F_{\rm G} = \frac{GMm}{r^2}
\end{equation}
where $M$ is the mass of the body that exerts the gravitational force, and $r$ is the distance between said body and the particle.
\end{comment}

%%
%Section: Einsteinian cosmology
%%

\section{Einsteinian Cosmology}

Physical cosmology is based on the general theory of relativity, and Einstein's great insight into the nature of gravity begins with the simplest of observations: that the inertial mass $m_{\rm I}$, in Newton's Second Law
\begin{equation}
    \vec{F}_{\rm net} = m_{\rm I}\vec{a}
\end{equation}
is equal to the gravitational mass $m_{\rm G}$ in Newton's law of gravity:
\begin{equation}
    \vec{F}_{\rm G} = \frac{GMm_{\rm G}}{r^3}\vec{r}.
\end{equation}
Near the surface of the Earth, the magnitude of the force of gravity acting on a particle reduces to
\begin{equation}
    F_{\rm G} = m_{\rm G}g
\end{equation}
where
\begin{equation}
\label{eq:g=GM/R^2}
    g := \frac{GM}{R^2_{\rm E}},
\end{equation}
and $R_{\rm E}$ is the radius of the Earth. Because $m_{\rm I} = m_{\rm G}$, it follows that, if the force of gravity is the only force acting on a particle,
\begin{equation}
    a = g.
\end{equation}
This is the \textbf{Equivalence Principle} which, stated informally, says that a falling particle ``does not feel its own weight", owing to the equivalence of gravitational and inertial mass. Before Einstein, the fact that inertial mass is equivalent to gravitational mass was merely a curious, unexplained coincidence. General relativity however, takes this principle as its very foundation, and even provides a framework for explaining it, as we will see in the next secion.

%Subsection 1
\subsection{The geometry of spacetime}
\label{subsec:ch1_geometry_of_spacetime}

In the special theory of relativity, space and time are unified into spacetime, as expressed by the infinitesimal line element (in Cartesian coordinates and with $c = 1$)
\begin{equation}
\label{eq:ds^2}
    ds^2 = -dt^2 + dx^2 + dy^2 + dz^2.
\end{equation}
This quantity is invariant under Lorentz transformations, i.e., transformations between inertial frames that satisfy Einstein's two postulates of special relativity.\footnote{Namely: 1.) the laws of physics are the same in all inertial reference frames, and 2.) all inertial observers measure the same value for the speed of light in vacuum \citep{Thornton_Rex}} Lore\-ntz transformations are linear, so to generalize the special theory of relativity we can consider non-linear transformations between coordinates. Under a general, non-linear transformation, the line element will be invariant if it takes the form
\begin{equation}
\label{eq:ds^2=g}
    ds^2 = g_{\mu \nu}(x)dx^{\mu}dx^{\nu},
\end{equation}
where $g_{\mu \nu}(x)$ is a tensor known as the \textbf{metric tensor}, which depends, in general, on the spacetime coordinates $x$. In eq. (\ref{eq:ds^2=g}) I have employed the Einstein summation convention whereby matching upper and lower indices are summed.\footnote{In this work, Greek indices run from 0 to 4.} To show that eq. (\ref{eq:ds^2=g}) is invariant under general coordinate transformations, we must recall that an arbitrary two-index tensor field $T_{\mu\nu}\left(x\right)$ transforms like
\begin{equation}
    \bar{T}_{\mu\nu}\left(\bar{x}\right) = T_{\alpha\beta}\left(x\right) \frac{\partial x^{\alpha}}{\partial \bar{x}^{\mu}}\frac{\partial x^{\beta}}{\partial \bar{x}^{\nu}}
\end{equation}
whenever its coordinates are subjected to a general transformation (\citealp{zee}). Similarly, the coordinate differentials transform like
\begin{equation}
    d\bar{x}^{\alpha} = \frac{\partial \bar{x}^{\alpha}}{\partial x^{\mu}}dx^{\mu},
\end{equation}
so
\begin{equation}
    \begin{aligned}
    ds^2 & = g_{\mu \nu}(x)dx^{\mu}dx^{\nu}\\
    & = \bar{g}_{\alpha\beta}\left(\bar{x}\right)\frac{\partial x^{\alpha}}{\partial \bar{x}^{\mu}}\frac{\partial x^{\beta}}{\partial \bar{x}^{\nu}}dx^{\mu}dx^{\nu}\\
    & = \bar{g}_{\alpha\beta}\left(\bar{x}\right)d\bar{x}^{\alpha}d\bar{x}^{\beta}.\\
    \end{aligned}
\end{equation}
This agrees nicely with our intuition, because a line is a geometric object, whose length in spacetime should not depend on the coordinate system the observer has chosen.

In making the line element dependent on both position and time, we have opened up the possibility that spacetime can be curved. The line element of special relativity (eq. \ref{eq:ds^2}), describes a flat spacetime, or one in which the angles of a triangle always add up to $180^{\rm o}$, and parallel lines always remain parallel. In contrast, in a curved spacetime (described in general by the line element of eq. \ref{eq:ds^2=g}), the angles of a triangle may add up to more or less than $180^{\rm o}$, and parallel lines may either converge or diverge. This is due to the fact that the coefficients of the coordinate differentials in the general line element (namely, the components of the metric tensor) depend on $x$, so depending on one's location in spacetime, the relative scale of one coordinate may be greater or less than the relative scale of another coordinate.
\begin{comment}
Integrating the quantity
\begin{equation}
ds = \sqrt{g_{\mu \nu}(x)dx^{\mu}dx^{\nu}}
\end{equation}
therefore produces, in general, a curved line [be careful with statements like this; a flat spacetime doesn't necessarily have to look flat].
The introduction of spacetime curvature has two profound consequences: (1.) it explains the Equivalence Principle, and (2.) it endows spacetime with dynamics. Regarding (1.), consider again the uniform gravitational field of eq. (\ref{eq:g=GM/R^2}). 
\end{comment}

Spacetime curvature is what explains the equivalence principle. While it is possible to say the equivalence of inertial and gravitational mass explains the fact that objects fall at the same rate in a vacuum, this is somewhat unsatisfying because the equivalence itself remains unexplained. General relativity, on the other hand, turns this explanation around: it says that particles fall at the same rate in vacuum, independently of the materials of which they're composed, because they're following the same (curved) paths in spacetime. More precisely, two particles freely falling in a vacuum on the surface of the earth will fall toward the earth's center at the same rate because they are both following the ``straightest-possible'' paths toward the center.
\begin{comment}
by transforming from an inertial frame in a uniform gravitational field $g$ to a non-inertial frame that has a constant acceleration equal to $g$, the gravitational field can be ``transformed away'' locally (that is, in a region sufficiently small compared to the scale over which the gravitational field varies). This means that, to a particle undergoing free fall, a frame co-moving with it is an inertial frame. Within this frame, the particle follows a straight path. Other observers, not undergoing free fall, see that the particle follows a curved path, and before Einstein came along these observers would have attributed the particle's motion to a force called gravity. General relativity tells us, however, that the particle moves the way it does simply because it's following a curved path in spacetime (albeit one that, from the particle's point of view, looks like a straight line). The fact that inertial mass is equivalent to gravitational mass is therefore due to the fact that freely-falling particles follow the straightest-possible paths in spacetime, independently of the materials of which they're composed; objects having different masses fall at the same rate because they follow the same paths in spacetime. What do I mean by ``straightest-possible path''? 
%\end{comment}
In the co-moving frame (which is described by locally-flat coordinates $y^{\nu}$), the particle in question travels along a straight line:
\end{comment}
Specifically, if we follow one of the particles on its way down, then we (who occupy a frame co-moving with the particle, having coordinates $y^{\nu}$) will describe the particle's motion by
\begin{equation}
\label{eq:ch1_ddoty=0}
    \frac{d^2y^{\nu}}{d\tau^2} = 0,
\end{equation}
where $\tau = \int\sqrt{-ds}$ is the particle's \textbf{proper time}, that is the time measured by a co-moving observer. Eq. (\ref{eq:ch1_ddoty=0}) says that the particle follows a straight-line path, with no acceleration. Observers resting on the surface of the earth, however, will use coordinates $x^{\mu}$ to describe the particle's motion. To translate eq. (\ref{eq:ch1_ddoty=0}) from the $y^{\nu}$ coordinate system to the $x^{\mu}$ coordinate system, we must simply express $y^{\nu}$ in terms of $x^{\mu}$, and then differentiate this quantity:
\begin{equation}
\label{eq:dy/ds}
    \frac{d}{d\tau}y^{\nu}(x^{\mu}) = \frac{\partial y^{\nu}}{\partial x^{\mu}}\frac{dx^{\mu}}{d\tau}.
\end{equation}
Taking the second derivative of this expression yields
\begin{equation}
\label{eq:d2y/ds2}
    \frac{d^2y^{\nu}}{d\tau^2} = \frac{d}{d\tau}\left(\frac{\partial y^{\nu}}{\partial x^{\mu}}\right)\frac{dx^{\mu}}{d\tau} + \frac{\partial y^{\nu}}{\partial x^{\mu}}\frac{d}{d\tau}\left(\frac{dx^{\mu}}{d\tau}\right) = 0
\end{equation}
From eq. (\ref{eq:dy/ds}), we have
\begin{equation}
    \frac{d}{d\tau}\left(\frac{\partial y^{\nu}}{\partial x^{\mu}}\right)\frac{dx^{\mu}}{d\tau} = \frac{\partial^2 y^{\nu}}{\partial x^{\mu} \partial x^{\rho}}\frac{dx^{\rho}}{d\tau}\frac{dx^{\mu}}{d\tau}.
\end{equation}
Plugging this back into eq. (\ref{eq:d2y/ds2}), we get
\begin{equation}
    \frac{\partial^2 y^{\nu}}{\partial x^{\mu} \partial x^{\rho}}\frac{dx^{\rho}}{d\tau}\frac{dx^{\mu}}{d\tau} + \frac{\partial y^{\nu}}{\partial x^{\lambda}}\frac{d^2 x^{\lambda}}{d\tau^2} = 0,
\end{equation}
where I have renamed a dummy index. After a little rearrangement, this becomes
\begin{equation}
    \frac{d^2 x^{\lambda}}{d\tau^2} + \frac{\partial x^{\lambda}}{\partial y^{\nu}}\frac{\partial^2 y^{\nu}}{\partial x^{\mu} \partial x^{\rho}}\frac{dx^{\rho}}{d\tau}\frac{dx^{\mu}}{d\tau} = 0.
\end{equation}
where I have used the identity
\begin{equation}
    \frac{\partial y^{\tau}}{\partial y^{\lambda}}\frac{\partial y^{\lambda}}{\partial y^{\epsilon}} = \delta^{\tau}_{\epsilon}.
\end{equation}
Upon making the identification
\begin{equation}
    \Gamma^{\lambda}_{\rho \mu} := \frac{\partial^2 y^{\nu}}{\partial x^{\rho} \partial x^{\mu}},
\end{equation}
the geodesic equation takes the standard form:
\begin{equation}
\label{eq:ch1_geodesic}
    \boxed{\frac{d^2 x^{\lambda}}{d\tau^2} + \Gamma^{\lambda}_{\rho \mu}\frac{dx^{\rho}}{d\tau}\frac{dx^{\mu}}{d\tau} = 0}.
\end{equation}
$\Gamma^{\lambda}_{\rho\mu}$ are known as the \textbf{Christoffel symbols}, and they account for the acceleration produced by the change in the coordinates (similar to the centrifugal and Coriolis accelerations that arise in a rotating frame of reference). The effect of gravity on a particle is encapsulated by the Christoffel symbols, so that a particle moving through empty spacetime, subject only to gravitational interactions, will follow a trajectory governed by eq. (\ref{eq:ch1_geodesic}), called a \textbf{geodesic}. Geodesic paths are the ``straightest-possible'' paths alluded to earlier; they are the analogues, in curved spacetime, of straight lines in flat spacetime. It is possible to show that the Christoffel symbols can be written, in terms of the metric tensor components,
\begin{equation}
\label{eq:ch1_Gamma_def}
    \Gamma^{\lambda}_{\rho\mu} = \frac{1}{2}g^{\lambda\tau}\left(\partial_{\rho}g_{\mu\tau} + \partial_{\mu}g_{\rho\tau} - \partial_{\tau}g_{\rho\mu}\right)
\end{equation}
\citep{zee}. This form will be very useful later. It can also be used to justify the assertion that the Christoffel symbols account for the action of gravity on the particle. As we have seen, if no non-gravitational forces act on the particle, then the particle follows a trajectory whose coordinates $x^{\lambda}\left(\tau\right)$ satisfy eq. (\ref{eq:ch1_geodesic}).
\begin{comment}
such that the net non-gravitational force per unit mass is given by $f^{\lambda}$, then
\begin{equation}
\label{eq:ch1_geodesic_f}
    \frac{d^2 x^{\lambda}}{d\tau^2} + \Gamma^{\lambda}_{\rho \mu}\frac{dx^{\rho}}{d\tau}\frac{dx^{\mu}}{d\tau} = f^{\lambda}.
\end{equation}
\end{comment}
\begin{comment}
This is Einstein's version of Newton's second law $\vec{a} = \frac{\vec{F}_{\rm net}}{m}$, applicable to general spacetimes which may or may not be curved.
\end{comment}
This should reduce to $\vec{a} = \vec{g}$ in the non-relativistic limit. If the particle moves slowly, then $\frac{dt}{d\tau} \gg \frac{dx^{i}}{d\tau}$.
\begin{comment}
In the non-relativistic limit $v \ll c$, the proper time of the particle is almost equal to its coordinate time because:
\begin{equation}
    d\tau^2 = dt^2\left(1 - \frac{d\vec{x}\cdot d\vec{x}}{c^2 dt^2}\right) \approx dt^2.
\end{equation}
\end{comment}
This implies
\begin{equation}
    \frac{d^2 x^{\lambda}}{d\tau^2} + \Gamma^{\lambda}_{00}\frac{dx^{0}}{d\tau}\frac{dx^{0}}{d\tau} + \Gamma^{\lambda}_{ij}\frac{dx^{i}}{d\tau}\frac{dx^{j}}{d\tau} \approx \frac{d^2 x^{\lambda}}{d\tau^2} + \Gamma^{\lambda}_{00}\frac{dx^{0}}{d\tau}\frac{dx^{0}}{d\tau} = 0
\end{equation}
If we further stipulate that the gravitational field is weak, such that $g_{\rho \mu} \approx \eta_{\rho \mu} + h_{\rho \mu}$ where $h_{\rho \mu}$ is a small perturbation, and that $h_{\rho \mu}$ does not depend on time, then
\begin{equation}
    \Gamma^{i}_{00} = -\frac{1}{2}g^{ij}\partial_j h_{00} \approx -\frac{1}{2}\eta^{ij}\partial_{j}h_{00} = -\frac{1}{2}\partial_{i}h_{00},
\end{equation}
and
\begin{equation}
    \Gamma^0_{00} = -\frac{1}{2}\partial_0 h_{00} = 0.
\end{equation} 
Then
\begin{equation}
\label{eq:ch1_f^0}
    \frac{d^2 x^0}{d\tau^2} = 0,
\end{equation}
and
\begin{equation}
    \frac{d^2 x^{i}}{d\tau^2} = \frac{1}{2}\partial_{i}h_{00}.
\end{equation}
Eq. (\ref{eq:ch1_f^0}) says that $\frac{dx^0}{d\tau} = \frac{dt}{d\tau}$ is a constant. We can argue that this constant is approximately unity because, in the non-relativistic limit $v \ll c$,
\begin{equation}
    d\tau^2 = dt^2\left(1 - \frac{d\vec{x}\cdot d\vec{x}}{c^2 dt^2}\right) = dt^2\left(1 - \frac{v^2}{c^2}\right) \approx dt^2.
\end{equation}
\begin{comment}
The 4-force per unit mass is
\begin{equation}
    f^{\lambda} = \frac{1}{m}\frac{dp^{\lambda}}{d\tau} = \frac{1}{m}\left(\frac{dE}{d\tau}, \frac{d\vec{p}}{d\tau}\right),
\end{equation}
where $E$ is the particle's total energy, and $\vec{p}$ is its 3-momentum. Eq. (\ref{eq:ch1_f^0}) then implies
\begin{equation}
    \frac{dt}{d\tau} = \frac{E}{m}.
\end{equation}
If the particle moves slowly, then $E \approx m$, so
\begin{equation}
    \frac{dt}{d\tau} \approx 1.
\end{equation}
\end{comment}
Consequently
\begin{equation}
    \frac{d^2 x^{i}}{d\tau^2} \approx \frac{d^2 x^{i}}{dt^2} = \frac{1}{2}\partial_{i}h_{00}
\end{equation}
If we define $h_{00} := -2\Phi$, where $\Phi$ is the Newtonian gravitational potential, then
\begin{equation}
\label{eq:ch1_Newt_geo}
    \frac{d^2\vec{x}}{dt^2} = -\vec{\nabla}\Phi,
\end{equation}
or $\vec{a} = \vec{g}$, as expected.

%%
%Section: Covariant derivatives
%%
\section{Covariant derivatives}
To do physics, we need to be able to take derivatives of functions. For example, to compute the components of the electric field $\vec{E}$ created by the charge density $\rho$, we could integrate Gauss's Law
\begin{equation}
    \vec{\nabla} \cdot \vec{E} = \frac{\rho}{\epsilon_0},
\end{equation}
where
\begin{equation}
    \vec{\nabla}\cdot \vec{E} = \partial_{x}E_{x} + \partial_{y}E_{y} + \partial_{z}E_{z}
\end{equation}
in 3-dimensional Cartesian coordinates. 
\begin{comment}
The point is that we assume we can compute these derivatives uniquely anywhere in space. In Euclidean space this is true, but it is not true in general. This is because 
\end{comment}
In Euclidean space it is easy to compute the difference of two vectors; simply slide one vector to the location of the other vector until their tails lie on the same point, and subtract them component-by-component. In flat space the subtraction of two vectors gives the intrinsic change of the vector, independent of the locations of the vectors in space or of one's choice of coordinates. In curved space (or curved spacetime), the basis vectors associated with the coordinate system have different values at different points, so to compute the change of a vector in a curved space (or curved spacetime), we also need to account for the change in the basis vectors when we ``slide'' one vector over to the location of the other. This is most easily seen by noting that a 4-vector transforms like
\begin{equation}
    A^{\bar{\mu}} = \frac{\partial \bar{x}^{\mu}}{\partial x^{\nu}}A^{\nu}
\end{equation}
under a general coordinate transformation $x^{\nu} \rightarrow \bar{x}^{\mu}(x^{\nu})$ (\citealp{zee}). If we differentiate $A^{\nu}$, then transform the coordinates of the resulting two-index object, we obtain:
\begin{equation}
\label{eq:ch1_wrong_deriv}
    \partial_{\bar{\rho}}A^{\bar{\mu}} = \frac{\partial x^{\lambda}}{\partial \bar{x}^{\rho}}\frac{\partial^2 \bar{x}^{\mu}}{\partial x^{\lambda}\partial x^{\nu}}A^{\nu} + \frac{\partial x^{\lambda}}{\partial \bar{x}^{\rho}}\frac{\partial \bar{x}^{\mu}}{\partial x^{\nu}}\partial_{\lambda}A^{\nu}.
\end{equation}
Recall that an arbitrary tensor $T_{\lambda}^{\nu}$ transforms like:
\begin{equation}
    T_{\bar{\rho}}^{\bar{\mu}} = \frac{\partial x^{\lambda}}{\partial \bar{x}^{\rho}}\frac{\partial \bar{x}^{\mu}}{\partial x^{\nu}}T_{\lambda}^{\nu}
\end{equation}
\citep{zee}. It is clear that eq. (\ref{eq:ch1_wrong_deriv}) does not transform like a two-index tensor, because we have not accounted for the change in the basis vectors of the coordinates (the first term ruins the transformation). If eq. (\ref{eq:ch1_wrong_deriv}) did transform correctly, we could write an equation of the form
\begin{equation}
    \partial_{\lambda}A^{\nu} = T^{\nu}_{\lambda},
\end{equation}
where $T^{\nu}_{\lambda}$ is again an arbitrary tensor, and the equality would be valid in all coordinate systems (because the $\frac{\partial x^{\lambda}}{\partial \bar{x}^{\rho}}\frac{\partial \bar{x}^{\mu}}{\partial x^{\nu}}$ terms would cancel on both sides of the equation). As physicists, we seek equations that are valid irrespective of the observer's arbitrary choice of coordinates; only these deserve to be called ``laws of physics''. Therefore, if we want to formulate the laws of physics in curved spacetime, we need a derivative that transforms like a tensor. If turns out that if we define the \textbf{covariant derivative} according to
\begin{equation}
    D_{\lambda}A^{\nu} := \partial_{\lambda}A^{\nu} + \Gamma_{\lambda\tau}^{\nu} A^{\tau}
\end{equation}
(\citealp{zee}) then $D_{\lambda}A^{\nu}$ will be a tensor. Similarly, the covariant derivative of a vector with a lower index is
\begin{equation}
    D_{\lambda}A_{\nu} = \partial_{\lambda}A_{\nu} - \Gamma^{\tau}_{\lambda\nu}A_{\tau}.
\end{equation}
Covariant differentiation of tensors works in much the same way as the covariant differentiation of vectors. For a tensor with two upper indices,
\begin{equation}
    D_{\lambda}T^{\mu\nu} = \partial_{\lambda}T^{\mu\nu} + \Gamma^{\mu}_{\rho\lambda}T^{\rho \nu} + \Gamma^{\nu}_{\rho\lambda}T^{\mu\rho},
\end{equation}
whereas a tensor with two lower indices has
\begin{equation}
\label{eq:ch1_Dg_lower}
    D_{\lambda}T_{\mu\nu} = \partial_{\lambda}T_{\mu\nu} - \Gamma^{\rho}_{\mu\lambda}T_{\rho \nu} - \Gamma^{\rho}_{\nu\lambda}T_{\mu\rho}
\end{equation}
\citep{zee}. From these equations, we can see that the sign of the Christoffel symbols in the covariant derivatives of vectors and tensors with lower indices is negative. This generalizes easily to mixed tensors (that is, tensors with some upper indices and some lower indices) having any number of total indices. Using eqs. (\ref{eq:ch1_Dg_lower}) and (\ref{eq:ch1_Gamma_def}), we can derive an important identity, namely
\begin{equation}
\label{eq:ch1_Dg=0}
    D_{\lambda}g_{\mu\nu} = 0.
\end{equation}
Writing this out explicitly gives
\begin{equation}
    D_{\lambda}g_{\mu\nu} = \partial_{\lambda}g_{\mu\nu} - \Gamma^{\rho}_{\mu\lambda}g_{\rho \nu} - \Gamma^{\rho}_{\nu\lambda}g_{\mu\rho}
\end{equation}
where
\begin{equation}
    \Gamma^{\rho}_{\mu\lambda}g_{\rho \nu} = \frac{1}{2}\left(\partial_{\mu}g_{\lambda\nu} + \partial_{\lambda}g_{\mu\nu}-\partial_{\nu}g_{\mu\lambda}\right),
\end{equation}
and
\begin{equation}
    \Gamma^{\rho}_{\nu\lambda}g_{\mu\rho} = \frac{1}{2}\left(\partial_{\nu}g_{\lambda\mu} + \partial_{\lambda}g_{\mu\nu}-\partial_{\mu}g_{\nu\lambda}\right).
\end{equation}
From these equations, it's clear that eq. (\ref{eq:ch1_Dg=0}) is satisfied since $g_{\mu\nu} = g_{\nu\mu}$.

%%
%Section: Einstein equations
%%
\section{Einstein's gravitational field equations}
\label{sec:EFE}
\begin{comment}
[We have seen] that the non-meshing of inertial frames results from spacetime curvature, similar to how the non-meshing of coordinate frames on the surface of the earth is caused by the earth's curvature (\citealp{MCP}). 
\end{comment}
In Sec. \ref{subsec:ch1_geometry_of_spacetime} we saw that gravity can be attributed to the curvature of spacetime, through the Christoffel symbol term in the geodesic equation. This tells us how matter moves under the influence of a given spacetime geometry, but it tells us nothing about how to determine the geometry in the first place. In Newtonian physics, the gravitational fields are generated by matter distributions, according to
\begin{equation}
\label{eq:ch1_Gauss_law_for_gravity}
    \vec{\nabla} \cdot \vec{g} = -4\pi G\rho,
\end{equation}
where $G$ is the gravitational constant and $\rho$ is a matter density. Given a matter density $\rho$, we can solve eq. (\ref{eq:ch1_Gauss_law_for_gravity}) for the gravitational field $\vec{g}$. The goal, then, is to figure out how to solve for the relativistic analogue of $\vec{g}$. Because gravity can be attributed to spacetime curvature, determining the field equations for gravity in four-dimensional spacetime should be equivalent to working out a set of field equations for the curvature of four-dimensional spacetime. One way to do this is to proceed by analogy with the way we derive the relativistic form of the electromagnetic field equations: we postulate an action which depends on some scalar function of the coordinates $x$
\begin{equation}
    S \propto \int F\left(x\right)\sqrt{-g}d^4 x,
\end{equation}
and demand that this action be stationary\footnote{Not extremal! See \cite{Gray_Taylor_2007}.} under arbitrary small variations $\delta g_{\mu \nu}$ of the metric (the dynamical variable being $g_{\mu \nu}$ in this case).

One of the consequences of the equivalence principle is that, in locally-flat coordinates, the Christoffel symbols vanish: $\Gamma^{\alpha}_{\beta\gamma} = 0$. The derivatives of the Christoffel symbols, however, do not vanish in locally-flat coordinates. 
\begin{comment}
It is the derivatives of the Christoffel symbols (i.e. the second derivatives of the metric) that describe curvature, which cannot be transformed away by going to locally-flat coordinates. 
\end{comment}
Presumably, then, $F\left(x\right)$ must somehow be made from a tensor containing two derivatives of the metric (otherwise $F(x)$, being a scalar function, would vanish in all frames). The tensor that has the necessary form is the \textbf{Riemann curvature tensor}
\begin{equation}
    \label{eq:ch1_def_Riemann}
    R^{\alpha}_{\beta \mu \nu} = \left(\partial_{\mu}\Gamma^{\alpha}_{\nu\beta} + \Gamma^{\alpha}_{\mu\lambda}\Gamma^{\lambda}_{\nu\beta}\right) - \left(\partial_{\nu}\Gamma^{\alpha}_{\mu\beta} + \Gamma^{\alpha}_{\nu\lambda}\Gamma^{\lambda}_{\mu\beta}\right).
\end{equation}
If we contract the first and third indices of this tensor, we obtain the \textbf{Ricci tensor}
\begin{equation}
    \label{eq:ch1_def_Ricci}
    R^{\mu}_{\beta\mu\nu} = R_{\beta \nu} = \left(\partial_{\mu}\Gamma^{\mu}_{\nu\beta} + \Gamma^{\mu}_{\mu\lambda}\Gamma^{\lambda}_{\nu\beta}\right) - \left(\partial_{\nu}\Gamma^{\mu}_{\mu\beta} + \Gamma^{\mu}_{\nu\lambda}\Gamma^{\lambda}_{\mu\beta}\right),
\end{equation}
which can then be contracted to form the \textbf{scalar curvature}
\begin{equation}
    \label{eq:ch1_def_R}
    R^{\nu}_{\nu} = R = g^{\beta\nu}\left[\left(\partial_{\mu}\Gamma^{\mu}_{\nu\beta} + \Gamma^{\mu}_{\mu\lambda}\Gamma^{\lambda}_{\nu\beta}\right) - \left(\partial_{\nu}\Gamma^{\mu}_{\mu\beta} + \Gamma^{\mu}_{\nu\lambda}\Gamma^{\lambda}_{\mu\beta}\right)\right].
\end{equation}
Being a scalar, $R$ is left invariant by general coordinate transformations. This, combined with the fact that $R$ contains two derivatives of the metric, means that $R$ is the function $F\left(x\right)$ we seek. Therefore, the action for the gravitational field is
\begin{equation}
    S \propto \int R\sqrt{-g}d^4 x.
\end{equation}
(\citealp{zee}). It turns out (and we will justify this later) that the proportionality constant in this action has the value $\frac{1}{16 \pi G}$ in units where $c = 1$. This is known as the \textbf{Einstein-Hilbert action},
\begin{equation}
\label{eq:ch1_S_EH}
    S_{\rm EH} = \frac{1}{16\pi G}\int R\sqrt{-g}d^4x,
\end{equation}
after its discoverers. Given eq. (\ref{eq:ch1_S_EH}), we can derive Einstein's equations governing the curvature of spacetime by demanding that the variation of $S_{\rm EH}$ with respect to small changes $\delta g_{\mu\nu}$ in the metric vanish:
\begin{equation}
\delta S_{\rm EH} = 0.
\end{equation}
The variation of eq. (\ref{eq:ch1_S_EH}) is:
\begin{equation}
\label{eq:ch1_delta_SEH_integrand}
\delta S_{\rm EH} = \frac{1}{16 \pi G}\int \delta\left(R\sqrt{-g}\right)d^4x.
\end{equation}
The first thing to do here is to write $R$ in terms of the metric and the Ricci tensor,
\begin{equation}
R = g^{\alpha \beta}R_{\alpha \beta},
\end{equation}
so that eq. (\ref{eq:ch1_delta_SEH_integrand}) becomes
\begin{equation}
\label{eq:S_EH}
\delta S_{\rm EH} = \frac{1}{16\pi G}\int \delta\left(g^{\alpha \beta} R_{\alpha \beta}\sqrt{-g}\right) d^4x.
\end{equation}
Expand the integrand via the product rule
\begin{equation}
\label{eq:EH_integrand}
\delta\left(g^{\alpha \beta}R_{\alpha \beta}\sqrt{-g}\right) = \delta g^{\alpha \beta} R_{\alpha \beta} \sqrt{-g} + g^{\alpha \beta}\delta R_{\alpha \beta} \sqrt{-g} + R \delta \sqrt{-g},
\end{equation}
Then use the fact that
\begin{equation}
\delta g^{\alpha \beta} = -g^{\alpha \mu}\delta g_{\mu \nu} g^{\nu \beta}
\end{equation}
to re-write the first term on the RHS of eq. (\ref{eq:EH_integrand}):
\begin{equation}
 \delta g^{\alpha \beta} R_{\alpha \beta} \sqrt{-g} =\sqrt{-g} R^{\mu \nu}\delta g_{\mu \nu}.
\end{equation}
This already has the form a function multiplied by $g_{\mu \nu}$, so nothing more needs to be done. The second term on the RHS of eq. (\ref{eq:EH_integrand}) will require a little more work, because it depends on the variation $\delta R_{\alpha \beta}$ of the Ricci tensor. This variation is equal to
\begin{equation}
\begin{aligned}
    \delta R_{\alpha \beta} & = \delta \left(\partial_{\sigma}\Gamma^{\sigma}_{\alpha \beta} + \Gamma^{\lambda}_{\tau \lambda}\Gamma^{\tau}_{\alpha \beta}\right) - \delta \left(\partial_{\beta}\Gamma^{\sigma}_{\alpha \sigma} + \Gamma^{\lambda}_{\tau \beta}\Gamma^{\tau}_{\alpha \lambda}\right)\\
    & = \partial_{\sigma}\delta \Gamma^{\sigma}_{\alpha \beta} + \delta \Gamma^{\lambda}_{\tau \lambda}\Gamma^{\tau}_{\alpha \beta} + \Gamma^{\lambda}_{\tau \lambda}\delta \Gamma^{\tau}_{\alpha \beta} - \partial_{\beta}\delta \Gamma^{\sigma}_{\alpha \sigma} - \delta\Gamma^{\lambda}_{\tau \beta}\Gamma^{\tau}_{\alpha \lambda} - \Gamma^{\lambda}_{\tau \beta}\delta \Gamma^{\tau}_{\alpha \lambda}.\\
\end{aligned}
\end{equation}
Now, we can simplify this expression by using locally-flat coordinates (in which $\Gamma^{\alpha}_{\beta \gamma} = 0$). Doing this produces
\begin{equation}
\begin{aligned}
    \delta R_{\alpha \beta} & = \partial_{\sigma}\delta \Gamma^{\sigma}_{\alpha \beta} - \partial_{\beta}\delta \Gamma^{\sigma}_{\alpha \sigma},\\
    & = D_{\sigma}\delta \Gamma^{\sigma}_{\alpha \beta} - D_{\beta}\delta \Gamma^{\sigma}_{\alpha \sigma},
\end{aligned}
\end{equation}
where the second line follows because covariant derivatives are equivalent to partial derivatives in locally-flat coordinates. After integrating by parts, the second term on the RHS of eq. (\ref{eq:EH_integrand}) becomes
\begin{equation}
\begin{aligned}
    \int g^{\alpha \beta}\left(D_{\sigma}\delta \Gamma^{\sigma}_{\alpha \beta} - D_{\beta}\delta \Gamma^{\sigma}_{\alpha \sigma}\right)\sqrt{-g}d^4x & = \int\left(-D_{\sigma}g^{\alpha\beta}\delta \Gamma^{\sigma}_{\alpha \beta} + D_{\beta}g^{\alpha\beta}\delta \Gamma^{\sigma}_{\alpha \sigma}\right)\sqrt{-g}d^4x\\
    & = 0,
\end{aligned}
\end{equation}
because $D_{\alpha}g^{\beta\gamma} = 0$. It follows that the second term on the RHS of eq. (\ref{eq:EH_integrand}) does not contribute to the integral on the RHS of eq. (\ref{eq:S_EH}). To bring the third term on the RHS of eq. (\ref{eq:EH_integrand}) to the required form, use the fact that
\begin{equation}
\label{eq:ch1_delta_sqrt(-g)}
\delta\sqrt{-g} = \frac{1}{2}\sqrt{-g}g^{\mu\nu}\delta g_{\mu \nu}
\end{equation}
to re-write the third term on the RHS of eq. (\ref{eq:EH_integrand}) as
\begin{equation}
R \delta \sqrt{-g} = \frac{1}{2}\sqrt{-g}g^{\mu\nu}R\delta g_{\mu \nu}
\end{equation}
so that the variation of $S_{\rm EH}$ becomes
\begin{equation}
\delta S_{\rm EH} = \frac{1}{16\pi G}\int \left(-R^{\mu\nu} + \frac{1}{2}Rg^{\mu\nu}\right)\delta g_{\mu\nu}\sqrt{-g}d^4x.
\end{equation}
We want to make the action stationary, so we require
\begin{equation}
0 = \int\left(-R^{\mu\nu} + \frac{1}{2}Rg^{\mu\nu}\right)\delta g_{\mu\nu}\sqrt{-g}d^4x,
\end{equation}
which implies, by the fundamental lemma of variational calculus,
\begin{equation}
R^{\mu\nu} - \frac{1}{2}Rg^{\mu\nu} = 0,
\end{equation}
or
\begin{equation}
\boxed{R_{\mu\nu} - \frac{1}{2}Rg_{\mu\nu} = 0}.
\end{equation}
These are \textbf{Einstein's equations} for gravitation in empty spacetime. More precisely, these equations describe \textit{the curvature of spacetime itself} in the absence of sources (i.e. non-gravitational forms of energy).

There is one simple way to generalize the Einstein-Hilbert action, and that is by adding a constant $\lambda$ to $R$ in the integrand:
\begin{equation}
\tilde{S}_{\rm EH} =  \frac{1}{16\pi G}\int (R + \lambda)\sqrt{-g}d^4x.
\end{equation}
Now, in Newtonian mechanics (i.e. mechanics with a fixed geometry), adding a constant to the action wouldn't change the dynamics because
\begin{equation}
\delta\int \lambda d^3x = \int\delta\lambda d^3x = 0.
\end{equation}
This is why, in Newtonian mechanics, energy differences (as opposed to absolute energies) determine dynamics. Einsteinian mechanics, however, is different. Because $\sqrt{-g}$ is sensitive to changes in the metric (after all, it \textit{is} a function of the metric), the geometry knows about ``arbitrary'' constants. Therefore, when we vary $\tilde{S}_{\rm EH}$, we get
\begin{equation}
\delta \tilde{S}_{\rm EH} = \frac{1}{16\pi G}\int\left[\delta\left(R\sqrt{-g}\right) + \lambda \delta \sqrt{-g}\right]d^4x.
\end{equation}
We already know what the first term in the integrand looks like, and we can use eq. (\ref{eq:ch1_delta_sqrt(-g)}) again to rewrite the second term, so that with almost no extra work, we obtain the most general form of Einstein's equations in the absence of sources:
\begin{equation}
R_{\mu \nu} -\frac{1}{2}\left(R + \lambda\right)g_{\mu \nu} = 0.
\end{equation}
In the literature (and most textbooks) this is usually written
\begin{equation}
\boxed{R_{\mu \nu} -\frac{1}{2}Rg_{\mu \nu} + \Lambda g_{\mu \nu} = 0},
\end{equation}
where $\Lambda$ is the \textbf{cosmological constant}. Getting our equations into this form is just a matter of redefining the constant we started out with:
\begin{equation}
\lambda := -2\Lambda,
\end{equation}
so that the modified Einstein-Hilbert action is
\begin{equation}
\tilde{S}_{\rm EH} = \frac{1}{16\pi G}\int \left(R -2\Lambda\right)\sqrt{-g}d^4x.
\end{equation}

Neglecting the cosmological constant for a moment, what if our spacetime isn't empty, but is rather filled with stuff? After all, stuff gravitates, so there must be a way to describe how it interacts with geometry. It turns out that this is pretty easy to do. First, write the total action (action of gravity plus matter) as
\begin{equation}
S = S_{\rm EH} + S_{\rm M}.
\end{equation}
Here I am using the word ``matter'', construed in its broadest possible sense, to refer to any non-gravitational parts of the action. If we define the energy-momentum tensor
\begin{equation}
T^{\mu \nu} := \frac{2}{\sqrt{-g}}\frac{\delta S_{\rm M}}{\delta g_{\mu\nu}}
\end{equation}
\citep{zee}, then vary the action $S$ with respect to $g_{\mu \nu}$, we obtain
\begin{equation}
\delta S = \frac{1}{16\pi G}\int\left(-R^{\mu \nu} + \frac{1}{2}Rg^{\mu \nu}\right)\delta g_{\mu \nu}d^4x + \int \frac{\delta S_{\rm M}}{\delta g_{\mu\nu}}\delta g_{\mu \nu}d^4x.
\end{equation}
Now use eq. (23) to re-write the matter term:
\begin{equation}
\delta S = \frac{1}{16\pi G}\int\left(-R^{\mu \nu} + \frac{1}{2}Rg^{\mu \nu}\right)\delta g_{\mu \nu}\sqrt{-g}d^4x + \frac{1}{2}\int T^{\mu \nu}\delta g_{\mu \nu}\sqrt{-g}d^4x.
\end{equation}
Just like before, if we demand that the variation vanishes,
\begin{equation}
0 = \int\left(-R^{\mu \nu} + \frac{1}{2}Rg^{\mu \nu} + 8\pi G T^{\mu \nu}\right)\delta g_{\mu \nu}\sqrt{-g}d^4x
\end{equation}
then we find, after invoking the fundamental lemma again,
\begin{equation}
\label{eq:Einstein_0}
\boxed{R_{\mu \nu} - \frac{1}{2}Rg_{\mu \nu} = 8\pi G T_{\mu \nu}}.
\end{equation}
If we take the trace of this equation, we find
\begin{equation}
\label{eq:ch1_R_trace}
    \begin{aligned}
        R^{\nu}_{\nu} - \frac{1}{2}Rg^{\nu}_{\nu} & = 8 \pi G T^{\nu}_{\nu}\\
        R - \frac{4}{2}R & = 8 \pi G T\\
        -R & = 8 \pi G T,\\
    \end{aligned}
\end{equation}
so that
\begin{equation}
    R_{\mu \nu} + \frac{1}{2}\left(8\pi GT\right)g_{\mu \nu} = 8\pi GT_{\mu \nu},
\end{equation}
or
\begin{equation}
\label{eq:Einstein}
    \boxed{R_{\mu \nu} = 8\pi G\left(T_{\mu \nu} - \frac{1}{2}Tg_{\mu \nu}\right)}.
\end{equation}
This form will be convenient later. For now, however, let us examine eq. (\ref{eq:Einstein_0}) again. This equation has the important property that the covariant divergence of both sides vanishes:
\begin{equation}
    D_{\mu}\left(R^{\mu \nu} - \frac{1}{2}Rg^{\mu \nu}\right) = D_{\mu}T^{\mu \nu} = 0
\end{equation}
This is important because the equality on the RHS is the generalization of the energy and momentum conservation laws we know from Newtonian physics; it is the \textbf{law of conservation of energy-momentum} (``energy-momentum'' here referring to the relativistic unification of energy and momentum, similar to the relativistic unification of space and time). To show that this conservation law holds, take the covariant derivate of the Riemann tensor:
\begin{equation}
    D_{\lambda}R^{\alpha}_{\beta\gamma\delta} = D_{\lambda}\left[\left(\partial_{\gamma}\Gamma^{\alpha}_{\beta\delta} + \Gamma^{\alpha}_{\sigma\gamma}\Gamma^{\sigma}_{\beta\delta}\right) - \left(\partial_{\delta}\Gamma^{\alpha}_{\beta\gamma} + \Gamma^{\alpha}_{\sigma\delta}\Gamma^{\sigma}_{\beta\gamma}\right)\right].
\end{equation}
Because this is a tensor equation, it will also hold in locally-flat coordinates. In such a coordinate system the connection coefficients vanish (but their gradients do not vanish) so that
\begin{equation}
\label{eq:DRiemann}
    D_{\lambda}R^{\alpha}_{\beta\gamma\delta} = \partial_{\lambda}\partial_{\gamma}\Gamma^{\alpha}_{\beta\delta} - \partial_{\lambda}\partial_{\delta}\Gamma^{\alpha}_{\beta\gamma}.
\end{equation}
If we add $D_{\gamma}R^{\alpha}_{\beta\delta\lambda}$ to this equation, we find:
\begin{equation}
\label{eq:2DRiemann}
\begin{aligned}
    D_{\lambda}R^{\alpha}_{\beta\gamma\delta} + D_{\gamma}R^{\alpha}_{\beta\delta\lambda} & = \partial_{\lambda}\partial_{\gamma}\Gamma^{\alpha}_{\beta\delta} - \partial_{\lambda}\partial_{\delta}\Gamma^{\alpha}_{\beta\gamma} + \partial_{\gamma}\partial_{\delta}\Gamma^{\alpha}_{\beta\lambda} - \partial_{\gamma}\partial_{\lambda}\Gamma^{\alpha}_{\beta\delta}\\
    & = \partial_{\gamma}\partial_{\delta}\Gamma^{\alpha}_{\beta\lambda} - \partial_{\lambda}\partial_{\delta}\Gamma^{\alpha}_{\beta\gamma}\\
    & = -D_{\delta}R^{\alpha}_{\beta\lambda\gamma}.\\
\end{aligned}
\end{equation}
Therefore
\begin{equation}
    \label{eq:Bianchi}
    D_{\lambda}R^{\alpha}_{\beta\gamma\delta} + D_{\gamma}R^{\alpha}_{\beta\delta\lambda} + D_{\delta}R^{\alpha}_{\beta\lambda\gamma} = 0.
\end{equation}
Contracting both sides of this equation with $g^{\gamma}_{\alpha}$ produces
\begin{equation}
    D_{\lambda}R^{\gamma}_{\beta\gamma\delta} + D_{\gamma}R^{\gamma}_{\beta\delta\lambda} + D_{\delta}R^{\gamma}_{\beta\lambda\gamma} = 0,
\end{equation}
or
\begin{equation}
    D_{\lambda}R_{\beta\delta} + D_{\gamma}R^{\gamma}_{\beta\delta\lambda} - D_{\delta}R_{\beta\lambda} = 0.
\end{equation}
Contracting with $g^{\beta\delta}$ produces
\begin{equation}
    D_{\lambda}R - D^{\gamma}R_{\gamma\lambda} - D^{\beta}R_{\beta\lambda} = 0.
\end{equation}
or
\begin{equation}
    D_{\beta}R^{\beta}_{\lambda} - \frac{1}{2}D_{\lambda}R = 0.
\end{equation}
Contracting one last time with $g^{\lambda\alpha}$, and noting that $D^{\alpha} = g^{\alpha\beta}D_{\beta}$, we obtain
\begin{equation}
\label{eq:GR_consv}
    \boxed{D_{\beta}\left(R^{\beta\alpha} - \frac{1}{2}Rg^{\beta\alpha}\right) = 0}.
\end{equation}
This identity also gives us another way to introduce the cosmological constant, $\Lambda$. Because $D_{\beta}g_{\mu \nu} = 0$, it follows that we can add a term $\Lambda g_{\mu \nu}$ to Einstein's equations without altering the conservation law of eq. (\ref{eq:GR_consv}). If we happen to live in a universe with a non-vanishing cosmological constant, then
\begin{equation}
\label{eq:ch1_full_EFE}
\boxed{R_{\mu \nu} - \frac{1}{2}Rg_{\mu \nu} + \Lambda g_{\mu \nu} = 8\pi G T_{\mu \nu}},
\end{equation}
or
\begin{equation}
G_{\mu\nu} + \Lambda g_{\mu \nu} = 8\pi G T_{\mu \nu},
\end{equation}
where
\begin{equation}
G_{\mu \nu} := R_{\mu \nu} - \frac{1}{2}Rg_{\mu \nu},
\end{equation}
which is usually called the \textbf{Einstein tensor}. It is also common to carry $\Lambda$ over to the RHS of eq. (\ref{eq:ch1_full_EFE}), so that
\begin{equation}
G_{\mu \nu} = 8\pi G T_{\mu \nu}
\end{equation}
where, in this case,
\begin{equation}
T_{\mu \nu} = T^{\rm M}_{\mu \nu} + T^{\Lambda}_{\mu \nu},
\end{equation}
with
\begin{equation}
\label{eq:T_Lambda}
T^{\Lambda}_{\mu \nu} := -\frac{\Lambda}{8 \pi G}g_{\mu \nu}.
\end{equation}

Finally, we can show that the numerical factor $\frac{1}{16\pi G}$ in the Einstein-Hilbert action is correct by examining the Newtonian limit of Einstein's field equations. We know that the Ricci tensor is
\begin{equation}
    R_{\mu \nu} = \left(\partial_{\tau}\Gamma^{\tau}_{\mu \nu} + \Gamma^{\lambda}_{\rho\lambda}\Gamma^{\rho}_{\mu\nu}\right) - \left(\partial_{\nu}\Gamma^{\tau}_{\mu \tau} + \Gamma^{\tau}_{\lambda \nu}\Gamma^{\lambda}_{\mu \tau}\right).
\end{equation}
In the Newtonian limit, which is the weak field limit $g_{\mu \nu} = \eta_{\mu \nu} + h_{\mu \nu}$ where $h_{\mu \nu}$ is a small perturbation to the Minkowski metric, only the derivatives of the Christoffel symbols survive when we expand $R_{\mu \nu}$ to first order:
\begin{equation}
    R_{\mu \nu} \approx \partial_{\tau}\Gamma^{\tau}_{\mu \nu} - \partial_{\nu}\Gamma^{\tau}_{\mu \tau}.
\end{equation}
The Christoffel symbols, to first order, are
\begin{equation}
    \Gamma^{\tau}_{\mu \nu} = \frac{1}{2}\eta^{\tau \rho}\left(\partial_{\mu}h_{\nu\rho} + \partial_{\nu}h_{\mu\rho} - \partial_{\rho}h_{\mu\nu}\right),
\end{equation}
and
\begin{equation}
    \Gamma^{\tau}_{\mu \tau} = \frac{1}{2}\eta^{\tau \rho}\left(\partial_{\mu}h_{\tau\rho} + \partial_{\tau}h_{\mu\rho} - \partial_{\rho}h_{\mu\tau}\right).
\end{equation}
In the Newtonian limit, $h_{\mu \nu}$ has no time dependence (no gravitational waves), and $h_{00} \gg h_{ij}$ (so that the $h_{ij}$ components can be neglected).\footnote{This can be justified on the grounds that, for a non-relativistic particle, $\frac{dx}{dt} \ll c$ or $c dt \gg dx$. Measured in meters, the ``distance'' that the particle moves along the time axis is much greater than the distance it moves along any of the spatial axes. Therefore the particle's worldline samples a much greater portion of the time-time component of the metric compared to the space-space components, so $h_{ij}$ must be negligible (I adapted this argument from \citealp{Price_2016}).} We can also neglect the diagonal components $h_{i0} = h_{0i}$. These conditions result in
\begin{equation}
    \Gamma^{i}_{00} = -\frac{1}{2}\eta^{i j}\partial_{j}h_{00}
\end{equation}
being the only non-vanishing Christoffel symbols. If we identify $h_{00} = -2\Phi$ (as we did in the text immediately above eq. \ref{eq:ch1_Newt_geo}) then
\begin{equation}
    R_{\mu \nu} = R_{00} = -\frac{1}{2}\partial_{i} \partial^{i} h_{00} = \nabla^2 \Phi,
\end{equation}
to first order in $h_{\mu \nu}$. The second term on the LHS of eq. (\ref{eq:Einstein_0}) must be
\begin{equation}
    \frac{1}{2}Rg_{\mu\nu} = \frac{1}{2}R\eta_{\mu\nu}
\end{equation}
because $R_{\mu\nu}$ is first order in $h_{\mu\nu}$; since $R$ is the trace of $R_{\mu\nu}$, it must also be of first order in $h_{\mu\nu}$, so only the lowest order term in $g_{\mu\nu}$ survives. From eq. (\ref{eq:ch1_R_trace}), we know that $R = -8\pi G T$, where $T$ is the trace of $T_{\mu\nu}$. In the Newtonian limit, matter moves non-relativistically (such that $v \ll c$), so the energy density $T_{00} = \rho$ of the matter distribution must be much greater than the other components of the energy-momentum tensor. This implies $T = -\rho$ and $R = 8\pi G\rho$. Therefore
\begin{equation}
\begin{aligned}
    R_{00} - \frac{1}{2}R\eta_{00} & = 8\pi G T_{00}\\
    \nabla^2 \Phi - \frac{1}{2}\left(8\pi G\rho\right)\left(-1\right) & = 8\pi G\rho\\
    \nabla^2 \Phi & = 4\pi G\rho\\
    \vec{\nabla}\cdot\vec{g} & = -4\pi G\rho,\\
\end{aligned}
\end{equation}
as expected.

%%
%Section: Friedmann equations
%%
\section{The Friedmann equations}
\subsection{The Friedmann-Lema\^{i}tre-Robertson-Walker (FLRW) metric}
\label{sec:FLRW_metric}
The metric of a homogeneous, isotropic, expanding spacetime has the form
\begin{equation}
\label{eq:FLRW}
ds^2 = -dt^2 + a^2(t)\gamma_{ij}dx^idx^j
\end{equation}
where $i,j = 1,2,3$ and 
\begin{equation}
\label{eq:ch1_lower_gamma}
\gamma_{ij} := \frac{dr^2}{1-kr^2} + r^2d\Omega^2.
\end{equation}
This form of the metric can be justified on the following grounds (\citealp{Peebles_1993}, though my sign conventions are different from his). First, consider a general line element
\begin{equation}
    ds^2 = -g_{00}dt^2 + 2g_{0i}dtdx^{i} + g_{ij}dx^{i}dx^{j}
\end{equation}
where $g_{0i}$ is assumed to be equal to $g_{i0}$. If $\tau$ is the proper time of an observer co-moving with the local motion of matter in his or her vicinity, then the square of the proper time interval he or she measures on his or her clock will be
\begin{equation}
    d\tau^2 = dt^2 = g_{00}dt^2
\end{equation}
if he or she assigns coordinates $\left(t, x^{i}\right)$ to the events within his or her reference frame. Because the Universe is assumed to be isotropic, $g_{0i}$ must vanish (being a vector). This plus the assumption of homogeneity further imply (see \citealp{Peebles_1993} for more details), that all co-moving observers can synchronize their clocks. Therefore all co-moving observers use the time coordinate $t$ and the same spatial coordinates $x^{i}$ (they must occupy the same constant-time hypersurfaces). Therefore the line element of a homogeneous, isotropic universe has the form
\begin{equation}
    ds^2 = -dt^2 + g_{ij}dx^{i}dx^{j}.
\end{equation}
We can place further restrictions on the form of this line element by noting that a curved four-dimensional spacetime can be embedded within a flat Minkowski spacetime of one higher dimension such that
\begin{equation}
\label{eq:ch1_embedding}
    g_{\mu\nu}dx^{\mu}dx^{\nu} = -dt^2 + dx^2 + dy^2 + dz^2 + dw^2
\end{equation}
where $\left(x, y, z, w\right)$ are Cartesian coordinates in the five-dimensional spacetime. A curved three-dimensional space is then equivalent to a surface on a four-dimensional spatial hypersurface of the larger five-dimensional spacetime. To show this, we note that the coordinates of the three-dimensional surface must obey the following constraint:
\begin{equation}
    L^2 = x^2 + y^2 + z^2 + w^2
\end{equation}
where $L$ is the surface's radius of curvature. If we define $r^2 := x^2 + y^2 + z^2$, then $w$ must satisfy
\begin{equation}
    w^2 = L^2 - r^2,
\end{equation}
so that
\begin{equation}
    wdw = -rdr.
\end{equation}
This implies
\begin{equation}
\label{eq:ch1_dw^2}
    dw^2 = \frac{r^2 dr^2}{w^2} = \frac{r^2 dr^2}{L^2 - r^2}.
\end{equation}
Since the Universe is assumed to be isotropic, it is natural to rewrite the line element in spherical coordinates, so that
\begin{equation}
\label{eq:ch1_dx^2+dy^2+dz^2}
    dx^2 + dy^2 + dz^2 = dr^2 + r^2 d\Omega^2,
\end{equation}
where $d\Omega^2 := d\theta^2 + {\rm sin}^2\theta d\phi^2$. Inserting eqs. (\ref{eq:ch1_dw^2}) and (\ref{eq:ch1_dx^2+dy^2+dz^2}) into eq. (\ref{eq:ch1_embedding}) then produces
\begin{equation}
\begin{aligned}
    g_{ij}dx^{i}dx^{j} & = \left(1 + \frac{r^2}{L^2 - r^2}\right)dr^2 + r^2 d\Omega^2\\
    & = \frac{dr^2}{1 - \frac{r^2}{L^2}} + r^2 d\Omega^2.\\
\end{aligned}
\end{equation}
In general the sign of $L^2$ can be positive, negative, or zero, corresponding to spatially closed, spatially open, or spatially flat hypersurfaces (\citealp{zee, weinberg}).\footnote{Here, and in all following portions of this work, the word ``hypersurface'' refers to three-dimensional hypersurfaces rather than four-dimensional hypersurfaces.} It is convenient to define
\begin{equation}
    k := \frac{|L^2|}{L^2}
\end{equation}
so that the line element becomes
\begin{equation}
    ds^2 = -dt^2 + \frac{dr^2}{1 - k\frac{r^2}{|L^2|}} + r^2 d\Omega^2
\end{equation}
with $k = 1, -1, 0$, corresponding to the cases listed above. If the Universe can expand or contract homogeneously, then the physical distance $x_{\rm ph}$ between any two points will be related to the coordinate separation $x$ of those two points by
\begin{equation}
    x_{\rm ph}\left(t\right) = a\left(t\right)x,
\end{equation}
where $a(t)$ is known as the \textbf{scale factor} of the Universe. Therefore the line element of a homogeneously expanding or contracting universe must be
\begin{equation}
    ds^2 = -dt^2 + a\left(t\right)\left[\frac{dr^2}{1 - k\frac{r^2}{|L^2|}} + r^2 d\Omega^2\right].
\end{equation}
It is also common to redefine
\begin{equation}
    \frac{r}{|L|} \rightarrow r.
\end{equation}
If we then absorb $|L|$ into the scale factor,
\begin{equation}
    a\left(t\right)|L| \rightarrow a\left(t\right),
\end{equation}
the line element takes the form
\begin{equation}
    ds^2 = -dt^2 + a\left(t\right)\left[\frac{dr^2}{1 - kr^2} + r^2 d\Omega^2\right],
\end{equation}
in agreement with eqs. (\ref{eq:FLRW}) and (\ref{eq:ch1_lower_gamma}).

%Subsection: deriving Friedmann
\subsection{Deriving the Friedmann equations}
\label{sec:Friedmann_derivation}
Given Einstein's equations,
\begin{equation}
\label{eq:ch1_R_T-T/2}
R_{\mu \nu} = 8\pi G\left(T_{\mu \nu} -\frac{1}{2} T g_{\mu \nu}\right),
\end{equation}
we can derive equations that determine how $a(t)$, the scale factor, evolves with time. These are known as the \textbf{Friedmann equations}.

First, assume that the universe consists only of an ideal fluid with energy density given by $\rho$ and pressure given by $p$. This fluid has an energy-momentum tensor given by
\begin{equation}
\label{eq:T ideal fluid}
T_{\mu \nu} = (\rho + p)u_{\mu}u_{\nu} + pg_{\mu \nu}.
\end{equation}
I wrote Einstein's equations in terms of the trace of the energy-momentum tensor in eq. (\ref{eq:ch1_R_T-T/2}) because the trace of eq. (\ref{eq:T ideal fluid}) is easy to calculate:
\begin{equation}
T \equiv T_{\nu}^{\nu} = 3p - \rho.
\end{equation}
Remember that $u_{\nu}^{\nu} = -1$ and $g_{\nu}^{\nu} = 4$.

The final assumption we require is that the observer is in a co-moving reference frame with respect to the fluid (he or she is at rest with respect to the local movement of fluid in his or her vicinity). This means
\begin{equation}
u_{\mu}^{\rm T} = \left(1, 0, 0, 0\right).
\end{equation}
Now we can write down Einstein's equations in terms of $\rho$ and $p$:
\begin{equation}
R_{00} = 4\pi G\left(\rho + 3p\right),
\end{equation}
\begin{equation}
R_{0i} = 0,
\end{equation}
\begin{equation}
R_{ij} = 4\pi G\left(\rho - p\right)a^2 \gamma_{ij}.
\end{equation}
$R_{00}$ is easier to calculate than $R_{ij}$, so I'll start with that one. In terms of the Christoffel symbols, the time-time component of the Ricci tensor is given by
\begin{equation}
R_{00} = \partial_{\tau}\Gamma^{\tau}_{00} + \Gamma^{\tau}_{\tau \rho}\Gamma^{\rho}_{00} - \partial_0\Gamma^{\tau}_{\tau 0} -  \Gamma^{\tau}_{0 \rho}\Gamma^{\rho}_{\tau 0}.
\end{equation}
Remember that the Christoffel symbols are given in terms of the metric by
\begin{equation}
\Gamma^{\sigma}_{\beta \alpha} = \frac{1}{2}g^{\sigma \lambda}\left(\partial_{\beta}g_{\alpha \lambda} + \partial_{\alpha}g_{\beta \lambda} - \partial_{\lambda}g_{\beta \alpha}\right),
\end{equation}
so, in this case,
\begin{equation}
\Gamma^{\tau}_{00} = \frac{1}{2}g^{\tau\lambda}\left(\partial_{0}g_{0 \lambda} + \partial_{0}g_{0 \lambda} - \partial_{\lambda}g_{00}\right) = 0,
\end{equation}
\begin{equation}
\label{eq:Gamma^rho_tau0}
\Gamma^{\rho}_{\tau 0} = \frac{1}{2}g^{\rho\lambda}\left(\partial_{\tau}g_{0 \lambda} + \partial_{0}g_{\tau \lambda} - \partial_{\lambda}g_{\tau 0}\right) = \frac{\dot{a}}{a}\gamma^{i}_{j}.
\end{equation}
eq. (\ref{eq:Gamma^rho_tau0}) implies
\begin{equation}
\Gamma^{\tau}_{\rho 0}\Gamma^{\rho}_{\tau 0} = \left(\frac{\dot{a}}{a}\right)^2\gamma^{i}_{i} = 3\left(\frac{\dot{a}}{a}\right)^2,
\end{equation}
so that
\begin{equation}
R_{00} = -3\frac{\ddot{a}}{a}.
\end{equation}
Now we know how the expansion of the universe accelerates with time:
\begin{equation}
\boxed{\frac{\ddot{a}}{a} = -\frac{4 \pi G}{3}\left(\rho + 3p\right)}.
\end{equation}

The space-space component of the Ricci tensor requires a little more work. First, divide
\begin{equation}
R_{ij} = \partial_{\tau}\Gamma^{\tau}_{ji} + \Gamma^{\tau}_{\tau \rho}\Gamma^{\rho}_{ji} - \partial_j\Gamma^{\tau}_{\tau i} -  \Gamma^{\tau}_{j \rho}\Gamma^{\rho}_{\tau i}
\end{equation}
 into
\begin{equation}
R_{ij} = R^{\left({\rm time}\right)}_{ij} + R^{\left({\rm space}\right)}_{ij},
\end{equation}
such that
\begin{equation}
\label{eq:ch1_R^time}
R^{\left({\rm time}\right)}_{ij} = \partial_{0}\Gamma^{0}_{ji} + \Gamma^{i}_{i0}\Gamma^{0}_{ji} - \partial_j\Gamma^{0}_{0 i} -  \Gamma^{0}_{j 0}\Gamma^{0}_{0 i},
\end{equation}
and
\begin{equation}
R^{\left({\rm space}\right)}_{ij} = \partial_{k}\Gamma^{k}_{ji} + \Gamma^{k}_{k n}\Gamma^{n}_{ji} - \partial_j\Gamma^{k}_{k i} -  \Gamma^{k}_{j n}\Gamma^{n}_{k i}.
\end{equation}
We already know
\begin{equation}
\Gamma^{\rho}_{0 \tau} = \frac{\dot{a}}{a}\gamma^{i}_{j},
\end{equation}
from eq. (\ref{eq:Gamma^rho_tau0}). We also need
\begin{equation}
\Gamma^{0}_{ij} = a\dot{a}\gamma_{ij},
\end{equation}
Substituting these into eq. (\ref{eq:ch1_R^time}) yields
\begin{equation}
R^{\left({\rm time}\right)}_{ij} = \left(\dot{a}^2 + a\ddot{a}\right)\gamma_{ij} + \dot{a}^2\gamma_{ij}.
\end{equation}
That was the easy part. The hard part is calculating $R^{\rm (space)}_{ij}$; we'll have to go component-by-component. The non-zero Christoffel symbols that we need are:
\begin{equation}
\Gamma^{r}_{rr} = \frac{kr}{1-kr^2},
\end{equation}
\begin{equation}
\Gamma^{r}_{\theta \theta} = r\left(kr^2 - 1\right),
\end{equation}
\begin{equation}
\Gamma^{r}_{\phi \phi} = r\left(kr^2 - 1\right){\rm {\rm sin}}^2\theta,
\end{equation}
\begin{equation}
\Gamma^{\theta}_{r \theta} = \Gamma^{\phi}_{r \phi} = \frac{1}{r},
\end{equation}
\begin{equation}
\Gamma^{\theta}_{\phi \phi} = -{\rm sin}\theta {\rm cos}\theta,
\end{equation}
\begin{equation}
\Gamma^{\phi}_{\theta \phi} = {\rm cot} \theta.
\end{equation}
Suppose $ij = rr$. Then
\begin{equation}
R^{\left({\rm space}\right)}_{rr} = \partial_{r}\Gamma^{r}_{rr} + 2\Gamma^{\theta}_{\theta r}\Gamma^{r}_{rr} - \partial_r\Gamma^{r}_{r r} +  2\partial_r\Gamma^{\theta}_{\theta r} -  \Gamma^{r}_{r r}\Gamma^{r}_{k r} - 2\Gamma^{\theta}_{r \theta}\Gamma^{\theta}_{\theta r},
\end{equation}
which is
\begin{equation}
R^{\left({\rm space}\right)}_{rr} = 2k\gamma_{rr}.
\end{equation}
If, on the other hand, $ij = \theta \theta$, then
\begin{equation}
R^{\left({\rm space}\right)}_{\theta \theta} = \partial_{r}\Gamma^{r}_{\theta \theta} + 2\Gamma^{\theta}_{\theta r}\Gamma^{r}_{\theta \theta} - \partial_{\theta}\Gamma^{r}_{r \theta} -  2\Gamma^{\theta}_{\theta r}\Gamma^{r}_{\theta \theta}.
\end{equation}
The $\theta \theta$ component of the Ricci tensor is, therefore,
\begin{equation}
R^{\left({\rm space}\right)}_{\theta \theta} = 2k\gamma_{\theta \theta}.
\end{equation}
Finally, suppose $ij = \phi \phi$. Then
\begin{equation}
R^{\left({\rm space}\right)}_{\phi \phi} = \partial_{r}\Gamma^{r}_{\phi \phi} + \partial_{\theta}\Gamma^{\theta}_{\phi \phi} +  \Gamma^{r}_{rr}\Gamma^{r}_{\phi \phi} + 2\Gamma^{\phi}_{\phi r}\Gamma^{r}_{\phi \phi} -  \Gamma^{\phi}_{\phi r}\Gamma^{r}_{\phi \phi} + \Gamma^{\theta}_{\phi \phi}\Gamma^{\phi}_{\theta \phi}
\end{equation}
or
\begin{equation}
R^{\left({\rm space}\right)}_{\phi \phi} = 2k\gamma_{\phi \phi}.
\end{equation}
Putting all three components together, we get
\begin{equation}
\label{eq:ch1_R^space}
R^{\left({\rm space}\right)}_{ij} = 2k\gamma_{ij}.
\end{equation}
When eq. (\ref{eq:ch1_R^space}) is combined with $R^{\left({\rm time}\right)}_{ij}$, we find that the space-space components of the Ricci tensor are
\begin{equation}
R_{ij} =\left(2\dot{a}^2 + a\ddot{a} + 2k\right)\gamma_{ij}
\end{equation}
so the space-space components of Einstein's equations are
\begin{equation}
\left(2\dot{a}^2 + a\ddot{a} + 2k\right)\gamma_{ij} = 4\pi G\left(\rho - p\right)a^2\gamma_{ij},
\end{equation}
which implies
\begin{equation}
2\left(\frac{\dot{a}}{a}\right)^2 + \frac{\ddot{a}}{a} + \frac{2k}{a^2} = 4\pi G\left(\rho - p\right).
\end{equation}
Insert eq. (16) into eq. (38) to eliminate $\frac{\ddot{a}}{a}$:
\begin{equation}
2\left(\frac{\dot{a}}{a}\right)^2 + \frac{2k}{a^2} = 4\pi G\left(\rho - p\right) + \frac{4\pi G}{3}\left(\rho + 3p\right)
\end{equation}
and simplify to get
\begin{equation}
\label{eq:Friedmann_1}
\boxed{\left(\frac{\dot{a}}{a}\right)^2 = \frac{8 \pi G}{3}\rho - \frac{k}{a^2}}.
\end{equation}
This equation, which relates the expansion rate of the universe to its energy density, is commonly known as the \textbf{first Friedmann equation}. The equation for the acceleration of the expansion rate,
\begin{equation}
\label{eq:Friedmann_2}
\frac{\ddot{a}}{a} = -\frac{4 \pi G}{3}\left(\rho + 3p\right),
\end{equation}
is often referred to as the \textbf{second Friedmann equation}. 

%Subsection 1
\subsection{Matter dynamics}
\label{subsec:matter}
To solve either eq. (\ref{eq:Friedmann_1}) or (\ref{eq:Friedmann_2}), we need to know how $\rho$ depends on $a$. In cosmology, we typically assume that the various types of matter in the Universe behave, on large scales, like ideal fluids that are separately conserved. Under this assumption, the energy-momentum tensor on the RHS of eq. (\ref{eq:Einstein}) takes the form given in eq. (\ref{eq:T ideal fluid}). If we make the further assumption that the energy density $\rho$ and pressure $p$ of these ideal fluids can each be described by an equation of state $w$ such that
\begin{equation}
\label{eq:w_def}
    w := \frac{p}{\rho}
\end{equation}
then their respective energy-momentum tensors have the form
\begin{equation}
    \label{eq:T_with w}
    \begin{aligned}
    T^{\mu \nu} & = \left(\rho + p\right)u^{\mu}u^{\nu} + pg^{\mu \nu}\\
    & = \rho\left(1 + w\right)u^{\mu}u^{\nu} + \rho w g^{\mu \nu}.\\
    \end{aligned}
\end{equation}
Because the energy-momentum tensor is conserved according to eq. (\ref{eq:GR_consv}):
\begin{equation}
    D_{\mu}T^{\mu \nu} = 0,
\end{equation}
it follows that
\begin{equation}
\label{eq:DT_X = 0}
    \begin{aligned}
    0 & = D_{\mu}T^{\mu \nu}\\
    & = D_{\mu}\left[\left(\rho + p\right)u^{\mu}\right]u^{\nu} + \left(\rho + p\right)u^{\mu}\left(D_{\mu}u^{\nu}\right) + \left(D_{\mu}p\right)g^{\mu \nu} + p\left(D_{\mu}g^{\mu \nu}\right)\\
    & = \left[\partial_{\mu}\left(\rho + p\right)\right]u^{\mu}u^{\nu} + \left(\rho + p\right)\left(D_{\mu}u^{\mu}\right)u^{\nu} + \left(\rho + p\right)u^{\mu}\left(D_{\mu}u^{\nu}\right) + \left(\partial_{\mu}p\right)g^{\mu \nu},\\
    \end{aligned}
\end{equation}
where, in going from the second line to the third line, I used the fact that 
\begin{equation}
    D_{\mu}f = \partial_{\mu}f
\end{equation}
for any scalar function $f$, and
\begin{equation}
    D_{\mu}g^{\mu \nu} = 0.
\end{equation}
Now, eq. (\ref{eq:DT_X = 0}) is a relation between tensors, so it will hold in any frame. In the rest frame of the X-fluid, the spatial components of the X-fluid's four-velocity vanish, and the time component of its four-velocity equals unity. Therefore $\mu = \nu = 0$, and eq. (\ref{eq:DT_X = 0}) becomes
\begin{equation}
    \begin{aligned}
    0 & = \left[\partial_{0}\left(\rho + p\right)\right]u^{0}u^{0} + \left(\rho + p\right)\left(D_{\mu}u^{\mu}\right)u^{0} + \left(\rho + p\right)u^{0}\left(\partial_{0}u^{0}\right) + \left(\partial_{0}p\right)g^{0 0}\\
    & = \dot{\rho} + w\dot{\rho} + \left(\rho + p\right)\left(\partial_{0}u^{0} + \Gamma^{\lambda}_{\lambda 0}u^0\right) - w\dot{\rho}\\
    & = \dot{\rho} + \rho\left(1 + w\right)\left(\frac{3\dot{a}}{a}\right).\\
    \end{aligned}
\end{equation}
where, in going from the first line to the second line, I used eqs. (\ref{eq:w_def}), (\ref{eq:FLRW}), and (\ref{eq:Gamma^rho_tau0}). After a bit of rearrangement, this becomes
\begin{equation}
    \frac{d\rho}{\rho} = -3\left(1 + w\right)\frac{da}{a},
\end{equation}
which is solved by
\begin{equation}
\label{eq:rho}
    \rho = Ca^{-3\left(1 + w\right)},
\end{equation}
where $C$ is an arbitrary constant. In the present, $\rho = \rho_{0}$ and $a = a_{0}$, by definition, so
\begin{equation}
\label{eq:ch1_rho_3(1+w)}
    \rho_{0} = Ca_{0}^{-3\left(1 + w\right)}.
\end{equation}
With this equation, we can eliminate the constant $C$ from eq. (\ref{eq:rho}), which then takes the form
\begin{equation}
    \boxed{\rho = \rho_{0}\left(\frac{a_0}{a}\right)^{3\left(1 + w\right)}}.
\end{equation}
For non-relativistic matter (i.e. cold dust), $w = 0$ because $p << \rho$. For relativistic matter (i.e. radiation), $w = \frac{1}{3}$ (see e.g. \citealp{Lemons_Thermo}). Therefore the energy density of non-relativistic matter, as a function of $a$, is
\begin{equation}
    \label{eq:rho_m = 1/a^3}
    \rho_{m} = \rho_{m0}\left(\frac{a_0}{a}\right)^3
\end{equation}
and the energy density of relativistic matter, as a function of $a$, is
\begin{equation}
    \label{eq:rho_g = 1/a^4}
    \rho_{\gamma} = \rho_{\gamma0}\left(\frac{a_0}{a}\right)^4.
\end{equation}
In these equations, $\rho_{m0}$, $\rho_{\gamma 0}$, and $a_0$ are the current values of $\rho_{m}$, $\rho_{\gamma}$, and $a$, respectively. To incorporate the cosmological constant $\Lambda$, consider eqs. (\ref{eq:T_Lambda}) and (\ref{eq:T ideal fluid}). The energy-momentum tensor of the cosmological constant is proportional to the metric tensor, so the proportionality constant must equal
\begin{equation}
    p_{\Lambda} = -\frac{\Lambda}{8 \pi G}
\end{equation}
Also, because $T^{\Lambda}_{\mu \nu}$ does not depend on the four-velocity $u^{\nu}$, it must also be the case that
\begin{equation}
\label{eq:rho_Lambda}
    \rho_{\Lambda} = -p_{\Lambda} = \frac{\Lambda}{8 \pi G},
\end{equation}
so the energy density of the cosmological constant does not depend on $a$ (hence the name ``cosmological constant''). Inserting eq. (\ref{eq:rho_Lambda}) into the first Friedmann equation produces
\begin{equation}
\begin{aligned}
    \left(\frac{\dot{a}}{a}\right)^2 & = \frac{8 \pi G}{3}\left(\rho_{m} + \rho_{\gamma} + \rho_{\Lambda}\right) - \frac{k}{a^2}\\
    & = \frac{8 \pi G}{3}\left(\rho_{m} + \rho_{\gamma}\right) + \frac{8 \pi G}{3}\frac{\Lambda}{8 \pi G} - \frac{k}{a^2}\\
    & = \frac{8 \pi G}{3}\left(\rho_{m} + \rho_{\gamma}\right) - \frac{k}{a^2} + \frac{\Lambda}{3}.\\
\end{aligned}
\end{equation}

%%
%Section: Scalar field dynamics
%%

\section{Scalar field dynamics}
\label{sec:scalar_field_dynamics}
The standard model posits that the Universe underwent a period of accelerated expansion, known as inflation, early in its history (this accelerated expansion creating the initial conditions for later structure formation). We will not consider the details of inflation in this work; for our purposes it is sufficient to know that the simplest version of the inflation paradigm is built on the hypothesis that a scalar field $\phi$ powered the early period of accelerated expansion. It is natural, then, to consider the possibility that the current era of accelerated expansion might also be driven by one or more slowly evolving scalar fields, rather than a cosmological constant $\Lambda$. I will describe a well-studied model in this class in Sec. (\ref{sec:phiCDM}), while here I will derive a few results pertaining to scalar field dynamics in general.

The relativistic action for a scalar field $\phi$ is
\begin{equation}
S_{\phi} = \int\left[-\frac{1}{2}g^{\mu\nu}\partial_{\mu}\phi\partial_{\nu}\phi - V\left(\phi\right)\right]\sqrt{-g}d^4x.
\end{equation}
where $V\left(\phi\right)$ is an arbitrary potential energy density associated with the field (see e.g. \citealp{weinberg}, but note that my sign conventions are different from his). If we vary this action with respect to $\phi$, we get
\begin{equation}
\delta S_{\phi} = \int \left[-\frac{1}{2}g^{\mu \nu}\partial_{\mu}\delta\left(\partial_{\nu}{\phi}\right)  -\frac{1}{2}g^{\mu\nu}\partial_{\nu}\delta\left(\partial_{\mu}{\phi}\right) - V'(\phi) \right]\sqrt{-g} d^4 x.
\end{equation}
Where the prime denotes differentiation with respect to $\phi$. After renaming dummy indices,
\begin{equation}
\delta S_{\phi} = \int \left[-g^{\mu \nu}\partial_{\mu}\phi \delta\left(\partial_{\nu}{\phi}\right)- V'(\phi) \right]\sqrt{-g} d^4 x.
\end{equation}
Because
\begin{equation}
\delta\left(\partial_{\nu} \phi\right) = \partial_{\nu}\delta \phi,
\end{equation}
It follows that
\begin{equation}
\delta S_{\phi} = \int \left[-g^{\mu \nu}\partial_{\mu} \phi \partial_{\nu}\delta \phi- V'(\phi) \right]\sqrt{-g} d^4 x,
\end{equation}
and we can integrate the first term in the integral by parts:
\begin{equation}
\int -g^{\mu \nu}\partial_{\mu} \phi \partial_{\nu}\delta \phi \sqrt{-g} d^4 x = \left[-\sqrt{-g}g^{\mu \nu}\partial_{\mu} \phi \delta \phi \right]^{\infty}_{-\infty} + \int \partial_{\nu}\left(\sqrt{-g} g^{\mu \nu}\partial_{\mu}\phi\right)\delta \phi d^4x.
\end{equation}
The boundary term vanishes because $\delta \phi$ is assumed to vanish on the surface at infinity. That leaves
\begin{equation}
\delta S_{\phi} = \int \left[\partial_{\nu}\left(\sqrt{-g} g^{\mu \nu}\partial_{\mu}\phi\right) - \sqrt{-g}V'(\phi)\right] \delta \phi d^4 x.
\end{equation}
If we demand that $\delta S_{\phi} = 0$, then by the fundamental lemma of variational calculus,
\begin{equation}
\boxed{\partial_{\nu}\left(\sqrt{-g} g^{\mu \nu}\partial_{\mu}\phi\right) = \sqrt{-g}V'(\phi)}.
\end{equation}
This can be written in manifestly covariant form by dividing both sides by $\sqrt{-g}$:
\begin{equation}
\label{eq:ch1_phi_EOM_covariant}
\boxed{D_{\nu}\left(g^{\mu \nu} \partial_{\mu}\phi\right) = V'(\phi)},
\end{equation}
where $D_{\nu}$ represents the covariant derivative.

	If we define the energy-momentum tensor
\begin{equation}
T_{\phi}^{\mu \nu} := \frac{2}{\sqrt{-g}}\frac{\delta S_{\phi}}{\delta g_{\mu \nu}},
\end{equation}
(see e.g. \citealp{zee}) and vary the scalar field action with respect to the metric, we find
\begin{equation}
\label{eq:ch1_delta_Sphi}
\delta S_{\phi} = -\int\frac{1}{2}\delta g^{\alpha \beta}\partial_{\alpha} \phi \partial_{\beta} \phi\sqrt{-g}d^4x -  \int\left[\frac{1}{2}g^{\alpha \beta}\partial_{\alpha} \phi \partial_{\beta} \phi + V\left(\phi\right)\right]\delta \sqrt{-g}d^4x.
\end{equation}
Use
\begin{equation}
\delta g^{\alpha \beta} = -g^{\alpha \mu} \delta g_{\mu \nu} g^{\nu \beta},
\end{equation}
and
\begin{equation}
\delta \sqrt{-g} = \frac{1}{2}\sqrt{-g}g^{\mu \nu} \delta g_{\mu \nu},
\end{equation}
to re-write eq. (\ref{eq:ch1_delta_Sphi}):
\begin{equation}
\delta S_{\phi} = \int \frac{1}{2}\left[g^{\alpha \mu} g^{\nu \beta} \partial_{\alpha} \phi \partial_{\beta} \phi - \left(\frac{1}{2}g^{\alpha \beta}\partial_{\alpha} \phi \partial_{\beta} \phi + V\left(\phi\right)\right)g^{\mu \nu}\right] \delta g_{\mu \nu} \sqrt{-g} d^4x.
\end{equation}
Then use the definition of the functional derivative
\begin{equation}
\delta S_{\phi} = \int \frac{\delta S_{\phi}}{\delta g_{\mu \nu}}\delta g_{\mu \nu} d^4x,
\end{equation}
(\citealp{zee}) in conjunction with eq. (11), to write down the energy-momentum tensor for the scalar field:
\begin{equation}
\label{eq:ch1_T_sc}
T_{\phi}^{\mu \nu} = g^{\alpha \mu} g^{\nu \beta} \partial_{\alpha} \phi \partial_{\beta} \phi - \left(\frac{1}{2}g^{\alpha \beta}\partial_{\alpha} \phi \partial_{\beta} \phi + V\left(\phi\right)\right)g^{\mu \nu}.
\end{equation}
A scalar field can be considered to be an ``ideal fluid", so if we compare eq. (\ref{eq:ch1_T_sc}) to the energy-momentum tensor of an ideal fluid,
\begin{equation}
T^{\mu \nu} = \left(\rho + p\right)u^{\mu} u^{\nu} + pg^{\mu \nu}
\end{equation}
then we can identify the pressure of the scalar field
\begin{equation}
\label{eq:p_phi}
p_{\phi} = -\frac{1}{2}g^{\alpha \beta}\partial_{\alpha} \phi \partial_{\beta} \phi - V\left(\phi\right).
\end{equation}
Further, if we set
\begin{equation}
u^{\mu} = -\frac{g^{\mu\lambda}\partial_{\lambda}\phi}{\sqrt{-g^{\alpha \beta}\partial_{\alpha}\phi \partial_{\beta}\phi}},
\end{equation}
(\citealp{weinberg}) then the energy density of the field can be identified as
\begin{equation}
\label{eq:rho_phi}
\rho_{\phi} = -\frac{1}{2}g^{\alpha \beta}\partial_{\alpha} \phi \partial_{\beta} \phi + V\left(\phi\right).
\end{equation}

%% file: chapter2.tex
\cleardoublepage

\chapter{Fundamentals of observational cosmology}

\label{Chapter2}
%%
%Section 0
%%
\section{Homogeneity and Isotropy}
\label{sec:ch2_homogeneity_and_isotropy}

The FLRW metric of the Universe was derived under the assumption that the Universe is homogeneous and isotropic (to the lowest order of approximation) on large scales. There is a large body of evidence in favor of these (not independent) hypotheses.\footnote{And some contrary findings, as well. See, e.g., \cite{Park_et_al_2017, Meszaros_2019, Secrest_et_al_2021}.} In particular, recent evidence of a homogeneity scale consistent with the prediction of \lcdm\ was found by \cite{Goncalves_et_al_2018, Goncalves_et_al_2020} (for earlier results, see e.g. \citealp{Yadav_et_al_2005}). The current strongest evidence of universal isotropy comes from observations of the CMB.\footnote{``CMB'' = ``cosmic microwave background radiation''. This background radiation, in the form of microwaves, permeates the Universe. It is a relic of the Universe's early life, and its discovery in the 1960s provided strong evidence for the Big Bang model.} Over the last three decades, measurements of the fluctuations in the temperature of the CMB have consistently shown strong agreement with the hypothesis of isotropy, as predicted by the $\Lambda$CDM model (for recent measurements, see \citealp{Planck_2018_Isotropy}; for earlier results, see e.g. \cite{Wu_Lahav_Rees_1999}). \cite{Deng_Wei_2018}, have also recently found that the Pantheon sample of SNe Ia measurements (containing 1048 data points) is consistent with isotropy. These findings are significant because evidence of universal isotropy (at least, to the lowest order of approximation) is also evidence of homogeneity \citep{weinberg}. If we make the reasonable assumption that we are not located at a special place in the Universe (this is just the extrapolation of the Copernican Principle to cosmic scales), then the fact that we see an isotropic Universe implies that observers located elsewhere (these locations presumably also not being special) will do the same. If this is true, however, then the Universe must be homogeneous because any large-scale homogeneity would be seen, by an observer close enough to it to see it, as an anisotropic feature on that observer's sky.

To be sure, the Universe is not completely homogeneous, as there are large, gravitationally collapsed structures (otherwise we would not exist). The CMB is also not completely isotropic, as there are small anisotropies in the temperature distribution. The existence of these anisotropies (relics of fluctuations in the matter density of the early Universe) is one of the key predictions of the \lcdm\ model, and their discovery in the early 1990's was one of the factors that led to the adoption of the \lcdm\ model as the ``standard model'' of cosmology (see any of the books cited at the beginning of Chapter \ref{Chapter1} and \citealp{Smoot_et_al_1992}). We will not concern ourselves here with the details of structure formation, but will instead focus on constraining models of the background evolution (that is, models of the evolution of the average distributions of matter and energy), as described in Chapter \ref{Chapter3}.

%%
%Section 1
%%
\section{Cosmic expansion}
\label{sec:Cosmic_expansion}

Investigations conducted by Vesto Slipher and Edwin Hubble in the early part of the last century showed that distant galaxies are moving away from us, the recession rate increasing with their distance from us according to
\begin{equation}
    \label{eq:ch2_Hubble_law}
    cz = H_0d
\end{equation}
where $c$ is the speed of light, $d$ is the distance to the galaxy, $H_0$ is a proportionality constant known as the ``Hubble constant'', and $z$ is the redshift of the light emitted by the galaxy \citep{Ryden_Peterson}. Redshift is defined as the fractional difference between the wavelength we observe and the wavelength emitted in the galaxy's rest frame:
\begin{equation}
    \label{eq:ch2_z_def}
    z := \frac{\lambda_{\rm o} - \lambda_{\rm e}}{\lambda_{\rm e}}.
\end{equation}
So a positive redshift in the light received from a distant galaxy indicates that we observe it to have a longer wavelength than the wavelength observed in the galaxy's rest frame, a result of the galaxy moving away from us. A negative redshift (or ``blueshift'') means that the galaxy is moving toward us. By observing the spectra of distant galaxies, we can compute their redshifts from the shift in the wavelengths of their spectral lines. With a knowledge of the distance $d$ to the galaxy (see below for a discussion of distance measurements), we can compute $H_0$, which sets the characteristic scale of the expansion rate of the Universe. Eq. (\ref{eq:ch2_Hubble_law}) only has approximate validity, however; it works at small redshifts ($z \ll 1$), but for larger redshifts it must be replaced by a relativistic description. This can be seen if we identify $cz$ with the recessional velocity $v_{\rm rec}$ of the galaxy, so that
\begin{equation}
    v_{\rm rec} = cz = H_0d.
\end{equation}
This implies that galaxies separated from us by distances greater than $d = H_0/c$, such that their redshifts are greater than 1, are moving away from us with recessional velocities greater than $c$! This conclusion, however, is merely a result of the extrapolation of Eq. (\ref{eq:ch2_Hubble_law}) to distances over which it is not valid, and we need to be careful in how we interpret what we mean when we say that distant galaxies are ``moving''. For large redshifts, the curvature scale of the Universe must be taken into account \citep{Mukhanov}, and this is where the aforementioned relativistic effects come into play. At small redshifts, nearby galaxies can be incorporated into our inertial reference frame (or our inertial frame can be meshed with the inertial frame of the galaxy to form a larger inertial frame). When the curvature scale of the Universe (which is set by $H_0$) becomes important, however, this is no longer possible. At such distances, our inertial reference frame cannot be meshed with the inertial frame of the galaxy, as a result of spacetime curvature \citep{MCP}. The fundamental speed limit set by relativity pertains to speeds measured relative to a given inertial reference frame, and so can only apply to objects that can be said to be within the same inertial frame (or, more precisely, objects whose separation is very small compared to the curvature scale of the Universe). To these objects we can assign coordinate velocities $v = \frac{dx}{dt}$ measured using the clocks ($dt$) and rulers ($dx$) of the given inertial frame, and it is these velocities which cannot exceed $c$. Since objects whose separation is comparable to the curvature scale cannot be said to be within the same inertial frame, the fundamental speed limit does not apply to their relative velocity \citep{Mukhanov}. To put it another way, because the large-scale distribution of matter in the Universe defines the FLRW coordinate system we use to locate events within the Universe, distant galaxies cannot be said to be moving relative to this coordinate system; rather, it is the coordinate system itself which is changing (via $a(t)$), and the ``velocity'' associated with this change has nothing to do with the speed limit set by relativity (so it can exceed $c$).

At any rate, even though the linear Hubble law of eq. (\ref{eq:ch2_Hubble_law}) does not apply at all redshifts, the Universe is observed to expand at all redshifts. The concept of redshift is also well-defined at large separation distances, although the computation of these distances must take the curvature scale (set by $H_0$) into account. We will see how this works in more detail below.

%%
%Section 2
%%
\section{Redshift and the scale factor}

We have already seen how the general theory of relativity can be used to derive a set of equations (the Friedmann equations) that govern the rate of expansion (or contraction) of a homogeneous, isotropic Universe. For a quantitative science like physical cosmology, however, theory alone is not enough. We need to make contact with observations so that we can test our theories. Whenever we observe the Universe, we're observing it as it was in the past; it is often said that, when one looks out to a great distance in space, one is also looking far back in time, owing to the finite speed of light. To learn about the state of the Universe at some time in past, therefore, we can use observations of very distant objects. As I described in Section \ref{sec:Cosmic_expansion}, distant objects appear redder and dimmer, the reddening being represented by the redshift $z$. It turns out that the scale factor and the redshift are related by the simple equation
\begin{equation}
\label{eq:a = 1/1+z}
    \boxed{\frac{a}{a_0} = \frac{1}{1 + z}}
\end{equation}
That this is so can be seen by considering eq. (\ref{eq:rho_g = 1/a^4}) in conjunction with the definition of redshift given by eq. (\ref{eq:ch2_z_def}). From eq. (\ref{eq:rho_g = 1/a^4}), the energy density of a gas of (monochromatic) photons goes like
\begin{equation}
\label{eq:ch2_photon_energy_density}
    \rho_{\gamma}(t) = \frac{Nhf}{V(t)} \propto \frac{f}{a^3(t)} \propto \frac{1}{a^4(t)}.
\end{equation}
where $N$ is the number of photons within the (time-dependent) volume $V(t) \propto a^3(t)$, $h$ is Planck's constant, and $f$ is the frequency of the photons. Because $f = c/\lambda$, it follows that $\lambda \propto a(t)$. eq. (\ref{eq:ch2_z_def}) then becomes
\begin{equation}
    z = \frac{\lambda(t_{\rm o}) - \lambda(t_{\rm e})}{\lambda(t_{\rm e})} = \frac{a(t_{\rm o}) - a(t_{\rm e})}{a(t_{\rm e})} = \frac{a(t_{\rm o})}{a(t_{\rm e})} - 1.
\end{equation}
If we say that the time of observation $t_{\rm o}$ is equal to the present time $t_0$ and define $a_0 := a(t_0)$ along with $a := a(t_{\rm e})$, then we obtain eq. (\ref{eq:a = 1/1+z}).

With eq. (\ref{eq:a = 1/1+z}) in hand, we can write the first Friedmann equation as a function of $z$ thereby completing the bridge between theory and observations. First, we define
\begin{equation}
    \label{eq:Hubble = adot/a}
    H := \frac{\dot{a}}{a},
\end{equation}
then we insert eqs. (\ref{eq:rho_m = 1/a^3}), (\ref{eq:rho_g = 1/a^4}), and (\ref{eq:rho_Lambda}) into the first Friedmann equation, to obtain:
\begin{equation}
\begin{aligned}
\label{eq:ch2_H^2}
    H^2 & = \frac{8 \pi G}{3}\left(\rho_{m0}\left(1 + z\right)^3 + \rho_{\gamma 0}\left(1 + z\right)^4\right) - \frac{k}{a_0^2}\left(1 + z\right)^2 + \frac{\Lambda}{3}.
\end{aligned}
\end{equation}
We can write this in a more convenient form by defining
\begin{equation}
\label{eq:ch2_Om}
    \Omega_{m0} := \frac{\rho_{m0}}{\rho_c},
\end{equation}
\begin{equation}
    \Omega_{\gamma 0} := \frac{\rho_{\gamma 0}}{\rho_c},
\end{equation}
\begin{equation}
    \Omega_{k0} := -\frac{k}{\left(a_0H_0\right)^2},
\end{equation}
\begin{equation}
\label{eq:ch2_OL}
    \Omega_{\Lambda} := \frac{\rho_{\Lambda}}{\rho_c},
\end{equation}
where $\rho_c := \frac{8\pi G}{3 H_0^2}$ and $\rho_{\Lambda}$ is given by Eq. (\ref{eq:rho_Lambda}). $\rho_c$ is known as the ``critical density'' because a universe having an energy density equal to $\rho_c$ such that
\begin{equation}
    \rho_m + \rho_{\gamma} + \rho_{\Lambda} = \rho_c,
\end{equation}
is exactly spatially flat ($k = 0$). A universe with an energy density greater than $\rho_c$ is spatially closed (or positively curved such that $k = 1$ and $\Omega_{k0} < 0$), and a universe having an energy density less than $\rho_c$ is spatially open (or negatively curved such that $k = -1$ and $\Omega_{k0} > 0$). Dividing both sides of eq. (\ref{eq:ch2_H^2}) by $H_0^2$ and using the definition of the critical density, we find
\begin{equation}
    \label{eq:H(z)}
    \boxed{H(z) = H_0\sqrt{\Omega_{\gamma 0}\left(1 + z\right)^4 + \Omega_{m0}\left(1 + z\right)^3 + \Omega_{k0}\left(1 + z\right)^2 + \Omega_{\Lambda}}},
\end{equation}
where $H_0$, the Hubble constant, is equal to the current value of $H(z)$ ($z = 0$ in the present, by definition). It is also common to see this written in the form
\begin{equation}
    \label{eq:ch2_E(z)_def}
    E(z) = \sqrt{\Omega_{\gamma 0}\left(1 + z\right)^4 + \Omega_{m0}\left(1 + z\right)^3 + \Omega_{k0}\left(1 + z\right)^2 + \Omega_{\Lambda}}
\end{equation}
where $E(z) = H(z)/H_0$ is sometimes called the ``expansion factor''.

%%
%Section 3
%%
\section{Distances in an expanding universe}
Astronomical observations are made by collecting the light that distant sources emit. We can use this light to infer, among other things, the distances to these sources. Along the geodesic traveled by a photon during its journey from the site of its emission to us, the line element is
\begin{equation}
    0 = c^2dt^2 - \frac{a^2}{1 - kr^2}dr^2.
\end{equation}
In the special case that the Universe is flat ($k = 0$), the coordinate distance between us and the source is
\begin{equation}
\label{eq:ch2_r_distance}
    r = \int_{t_{\rm e}}^{t_{\rm o}} \frac{dt}{a} = \int_{a_{\rm e}}^{a_{\rm o}} \frac{da}{a^2}\frac{a}{\dot{a}} = -\int^0_{z_{\rm e}} \frac{dz}{H(z)} = \frac{c}{H_0}\int_0^{z_{\rm e}} \frac{dz}{E(z)},
\end{equation}
where, in the last equality, I have restored the factor of $c$. If the Universe is not flat, so that $k \neq 0$, then we must integrate
\begin{equation}
    \int \frac{dr}{\sqrt{1 - kr^2}} = \int_{t_{\rm e}}^{t_{\rm o}} \frac{dt}{a}.
\end{equation}
The RHS of this equation is the same as the RHS of eq. (\ref{eq:ch2_r_distance}). The LHS is
\begin{equation}
    \int \frac{dr}{\sqrt{1 - kr^2}} = \frac{-1}{\sqrt{k}}{\rm sin}^{-1}\left(-\sqrt{k}r\right)
\end{equation}
if $k > 0$, and
\begin{equation}
    \int \frac{dr}{\sqrt{1 - kr^2}} = \frac{1}{\sqrt{|k|}}{\rm sinh}^{-1}\left(\sqrt{|k|}r\right)
\end{equation}
if $k < 0$ \citep{G&R}. For all three cases, we then have
\begin{equation}
    r = 
    \begin{cases}
    \frac{c}{H_0}\int_0^{z_{\rm e}} \frac{dz}{E(z)} & \text{if}\ k = 0,\\
    \vspace{1mm}
    \frac{1}{\sqrt{k}}{\rm sin}\left(\sqrt{k}\frac{c}{H_0}\int_0^{z_{\rm e}} \frac{dz}{E(z)}\right) & \text{if}\ k = 1, \\
    \vspace{1mm}
    \frac{1}{\sqrt{|k|}}{\rm sinh}\left(\sqrt{|k|}\frac{c}{H_0}\int_0^{z_{\rm e}} \frac{dz}{E(z)}\right) & \text{if}\ k = -1,
    \end{cases}
\end{equation}
Following \cite{Hogg}, this distance function can also be written in the form
\begin{equation}
\label{eq:D_M}
    D_{\rm M}(z) = 
    \begin{cases}
    D_{\rm C} & \text{if}\ \Omega_{k0} = 0,\\
    \vspace{1mm}
    \frac{c}{H_0\sqrt{\Omega_{k0}}}{\rm sinh}\left[\sqrt{\Omega_{k0}}\frac{D_C H_0}{c}\right] & \text{if}\ \Omega_{k0} > 0, \\
    \vspace{1mm}
    \frac{c}{H_0\sqrt{|\Omega_{k0}|}}{\rm sin}\left[\sqrt{|\Omega_{k0}|}\frac{D_C H_0}{c}\right] & \text{if}\ \Omega_{k0} < 0,
    \end{cases}
\end{equation}
which Hogg calls the ``transverse co-moving distance'', with
\begin{equation}
\label{eq:D_H}
    D_{\rm H} := \frac{c}{H_0},
\end{equation}
and
\begin{equation}
\label{eq:D_C}
    D_{\rm C} := \frac{c}{H_0}\int^z_0 \frac{dz'}{E(z')},
\end{equation}
In what follows, I will follow Hogg's notational conventions. With the transverse co-moving distance in hand, we can define a number of other useful distances scales. The luminosity distance is one such distance scale, defined as
\begin{equation}
\label{eq:D_L}
    D_{\rm L}(z) = (1 + z)D_{\rm M}(z)
\end{equation}
\citep{Hogg}. To see that this is correct, recall that for a source that emits a given luminosity $L$, we can define a flux ($F$) received through
\begin{equation}
\label{eq:ch2_flux}
    F = \frac{L}{4\pi r^2},
\end{equation}
where $r$ is the coordinate distance between the receiver and the source. In a universe that is not expanding, $r$ would also be called the ``luminosity distance''. In an expanding universe, two corrections must be made to obtain the luminosity distance \citep{weinberg}. First, we must correct for the decrease of the energy of the photons as they traverse the distance from the source to the receiver (see the discussion just below eq. \ref{eq:ch2_photon_energy_density}). This requires us to divide eq. (\ref{eq:ch2_flux}) by $1 + z$. Second, the rate at which photons are received by the source will also decrease by a factor of $1 + z$. These two effects then result in
\begin{equation}
    F = \frac{L}{4\pi r^2 (1 + z)^2}
\end{equation}
If make the identification $r(z) = D_{\rm M}(z)$, then we can define the luminosity distance as in eq. (\ref{eq:D_L}). It is also possible to define an ``angular diameter distance'' $D_{\rm A}(z)$, through
\begin{equation}
\label{eq:ch2_DA}
    D_{\rm A}(z) = \frac{D_{\rm M}(z)}{1 + z}.
\end{equation}
When observing a distant object that subtends an angle $\theta$ on the sky, the transverse size of the object will be
\begin{equation}
    s = a(t)r\theta,
\end{equation}
where $a(t)r$ is the physical distance from the observer to the object. If we define
\begin{equation}
    D_{\rm A}(z) := a(t)r = \frac{r}{1 + z}
\end{equation}
such that
\begin{equation}
    s = D_{\rm A}(z)\theta,
\end{equation}
then eq. (\ref{eq:ch2_DA}) follows after identifying $r$ with $D_{\rm M}(z)$. Finally, one can define a ``volume-averaged angular diameter distance"
\begin{equation}
    \label{eq:D_V}
    D_{\rm V}(z) = \left[\frac{cz}{H_0}\frac{D_{\rm M}^2(z)}{E(z)}\right]^{1/3},
\end{equation}
(\cite{5}) which is useful in studies of cosmological constraints from baryon acoustic oscillation data (about which, see below).

%%
%Section 5
%%
\section{Accelerated expansion}
\label{sec:ch5_accel_expansion}
Stars can be classified by their flux, and astronomers do this by comparing their \textbf{apparent magnitudes} \citep{Ryden_Peterson}. Two stars (or other light sources, in general), having apparent magnitudes $m_i$ and $m_j$ that differ by unity have fluxes that differ by a factor of 100, by definition. In general,
\begin{equation}
    m_i - m_j = 2.5{\rm log}\left(\frac{F_i}{F_j}\right).
\end{equation}
We have already seen that the flux received from a source depends on the distance between the source and the receiver. In a static universe, this distance is the coordinate distance $r$; in an expanding (or contracting) universe, this distance is the luminosity distance $D_{\rm L}(z)$. Thus given a knowledge of the apparent magnitudes of two light sources, and a knowledge of their luminosities (and redshifts), we can compute the distance between them. To compute the distance $d$ to a single source, astronomers use the \textbf{distance modulus}
\begin{equation}
    m - M = 5{\rm log}\left(\frac{d}{10\hspace{1mm} {\rm pc}}\right)
\end{equation}
where the \textbf{absolute magnitude} $M$ is defined as the source's apparent magnitude at a distance of 10 pc (and $1 \hspace{1mm} {\rm pc} = 3.086\times 10^{16}\hspace{1mm}{\rm m}$). Given a knowledge of the source's luminosity, it is possible to compute the apparent and absolute magnitudes, and thereby obtain the distance to the source. Such sources are known as \textbf{standard candles}, and they can be used to study the dynamics of the Universe's expansion. As an example, consider Type Ia supernovae. Supernovae come in three flavors \citep{Ryden_Peterson}: Type Ia, Type Ib, and Type II. The differences between them have to do with the absorption lines they show in their spectra. Type Ia supernovae have no hydrogen or helium lines, Type Ib supernovae have helium lines but no hydrogen lines, and Type II supernovae have hydrogen lines. A Type Ia supernovae is the end result of a process by which a white dwarf in a binary system accretes mass from its partner \citep{Ryden_Peterson, weinberg}. Over time, the mass of the white dwarf increases until it reaches the Chandrasekhar mass
\begin{equation}
    M_{\rm C} := 1.4M_{\rm sun}
\end{equation}
where $M_{\rm sun} = 1.989 \times 10^{30}$ kg is the mass of our sun. Once the mass of the white dwarf exceeds $M_{\rm C}$, electron degeneracy pressure can no longer prevent the gravitational collapse of the dwarf; the collapse initiates rapid nuclear fusion, creating an explosion with a luminosity that does depend strongly on where or when in the Universe the explosion happens (because the masses of the white dwarfs are always close to $M_{\rm C}$). Therefore, if an analysis of a supernova's spectrum shows it to be of Type Ia, then its luminosity can be determined from the decay time of its light curve (along with empirical corrections for the metallicity of the white dwarf; \citealp{weinberg}), and that supernova can be used as a standard candle.
\begin{comment}
Type Ia supernovae can be used as standard candles because they are powered by (relatively) simple, well-known physics \citep{Ryden_Peterson, weinberg}.
\end{comment}
In an expanding universe, with source fluxes given by
\begin{equation}
    F = \frac{L}{4\pi D^2_{\rm L}(z)},
\end{equation}
the luminosity distance $D_{\rm L}(z)$ depends on the cosmological model through eqs. (\ref{eq:D_L}), (\ref{eq:D_C}), (\ref{eq:D_M}) and (\ref{eq:H(z)}). After making the appropriate unit conversions ($D_{\rm M}(z)$ and all distance scales derived from it are typically measured in Mpc), we can write the distance modulus in the form
\begin{equation}
    m - M = 5 {\rm log}D_{\rm L}(z) + 25.
\end{equation}
By examining the apparent magnitudes of Type Ia supernovae (or any other kind of standard candle) as a function of $z$, it is possible to place constraints on the dimensionless energy density parameters of eqs. (\ref{eq:ch2_Om})-(\ref{eq:ch2_OL}), thereby revealing the dynamics of the background expansion. Two independent teams \citep{Riess_et_al_1998, Perlmutter_et_al_1999} did just this at the turn of the last century, finding persuasive evidence that the expansion of the Universe is accelerating, in a fashion consistent with that of a Universe whose energy budget is dominated by \textbf{dark energy} (in the form of a cosmological constant $\Lambda$). Since then, measurements of the small temperature anisotropies in the cosmic microwave background (CMB; see e.g. \citealp{planck2018_overview}) $z \sim 1100$ have corroborated this paradigm. In what follows, we will be concerned with the constraints that can be placed on simple models of dark energy from measurements having relatively low redshifts ($z \lesssim 8$).

%% file: chapter3.tex
\cleardoublepage

\chapter{Models of cosmic acceleration}

\label{Chapter3}
Most cosmologists today attribute the currently-accelerated phase of cosmic expansion to a ``dark energy'', an additional component of the Universe's energy budget (beyond matter and radiation) which has yet to be satisfactorily explained in terms of fundamental theory. It is typical to study the behavior of dark energy phenomenologically, by constructing models for its dynamics and using observational data to place constraints on the parameters of these models. Many models of this kind can be found in the literature. The simplest, the cosmological constant model, is a key part of the standard model of cosmology known as $\Lambda$CDM.\footnote{See the textbooks cited at the beginning of Chapter \ref{Chapter1}, in addition to, e.g., \cite{Peebles_Ratra, Ratra_Vogeley, planck2018_overview, planck2018}.} Others include the XCDM parametrization \citep{5}, the CPL parametrization \citep{Chevallier_Polarski_2001, Linder_2003}, holographic dark energy \citep{Wang_Wang_Li_2017}, the Ratra-Peebles quintessence model \citep{6, ratpeeb88, pavlov13}, K-essence \citep{Armendariz-Picon_Mukhanov_Steinhardt_2000, Aremendariz-Picon_Mukhanov_Steinhardt_2001}, the Chaplyin gas model \citep{Kamenshchik_Moschella_Pasquier_2001, Bento_Bertolami_Sen_2002}, and many others (see e.g. \citealp{Copeland_Sami_Tsujikawa_2006, Yoo_Watanabe_2012}). In this work, I will examine three models of dark energy, namely the cosmological constant model, the XCDM parametrization, and the Ratra-Peebles quintessence model (also known as $\phi$CDM), as well as two models of cosmic acceleration which, properly speaking, do not belong to the ``dark energy'' category, but which nevertheless make testable predictions that can be compared to the predictions of dark energy models. These are a model of emergent cosmic acceleration due to Gregory Ryskin (which I'll call ``Ryskin's model''; \citealp{Ryskin}), and the power law model (in which the scale factor is proportional to the cosmic time raised to a constant exponent; see Chapter \ref{Chapter10} for references to the literature). Different models of dark energy can be characterized by the values that they assign to the equation of state parameter
\begin{equation}
    w := \frac{p}{\rho}.
\end{equation}
For our purposes, however, it will be more useful to characterize dark energy models through their Hubble parameter functions $H(z)$ (defined in Chap. \ref{Chapter2}). This approach also has the virtue of being applicable to models of cosmic acceleration that do not incorporate dark energy, as we will see in Secs. \ref{sec:ch3_Ryskin_model} and \ref{sec:ch3_power_law_model} below.

%%
%Section: Dark Energy Models
%%
\section{Dark Energy Models}
%%
%Section: LCDM
%%
\subsection{$\Lambda$CDM}
\label{sec:LCDM}
As we have seen, the energy density of the cosmological constant is equal to the negative of its pressure, so the equation of state in the $\Lambda$CDM model is
\begin{equation}
w_{\Lambda} = \frac{p_{\Lambda}}{\rho_{\Lambda}} = -1.
\end{equation}
In terms of the present values of the dimensionless energy density parameters, the Hubble rate function in the $\Lambda$CDM model is
\begin{equation}
\label{eq:E(z)_LCDM}
H(z) = H_0\sqrt{\Omega_{m0}\left(1 + z\right)^3 + \Omega_{k0}\left(1 + z\right)^2 + \Omega_{\Lambda}}.
\end{equation}
Note that I have set $\Omega_{\gamma 0} = 0$, which I will do in all following analyses. This is justified given the relatively low redshifts of the data I use. The energy density parameters are constrained by
\begin{equation}
    \Omega_{m0} + \Omega_{k0} + \Omega_{\Lambda} = 1,
\end{equation}
so the free parameters in the $\Lambda$CDM model are $p = \left(H_0, \Omega_{m0}, \Omega_{k0}\right)$. In the special case that the Universe is taken to be exactly spatially flat, the free parameters are $p = \left(H_0, \Omega_{m0}\right)$.

%%
%Section 2
%%
\subsection{XCDM}
\label{sec:XCDM}
In the previous section, we saw that the cosmological constant has an equation of state parameter equal to $-1$. It is possible, however, that the equation of state of dark energy has a value that is different from $-1$; this is the simplest way to generalize the $\Lambda$CDM model, and it is known as the XCDM model (or parametrization) of dark energy. If we allow the equation of state parameter (call it $w_{\rm X}$) to take values different from $-1$, then by eqs. (\ref{eq:ch1_rho_3(1+w)}) and (\ref{eq:a = 1/1+z}) the dimensionless energy density of the X-fluid must have the form
\begin{equation}
    \Omega_{X} = \Omega_{X0}\left(1 + z\right)^{3\left(1 + w_{\rm X}\right)},
\end{equation}
where $\Omega_{X0}$ is the present value of the X-fluid's dimensionless energy density. Therefore, the Hubble parameter function in the XCDM parametrization takes the form
\begin{equation}
\label{eq:E(z)_XCDM}
H(z) = H_0\sqrt{\Omega_{m0}\left(1 + z\right)^3 + \Omega_{k0}\left(1 + z\right)^2 + \Omega_{X0}\left(1 + z\right)^{3\left(1 + w_{\rm X}\right)}}.
\end{equation}
Because the energy density parameters are constrained by
\begin{equation}
    \Omega_{m0} + \Omega_{k0} + \Omega_{{\rm X}0} = 1
\end{equation}
the XCDM parametrization has, in general, four free parameters: $p = \left(H_0, \Omega_{m0}, \Omega_{k0}, w_{\rm X}\right)$. In the special case that the Universe is taken to be exactly spatially flat, the free parameters are $p = \left(H_0, \Omega_{m0}, w_{\rm X}\right)$. If $w_{\rm X} = -1$, then eq. (\ref{eq:E(z)_XCDM}) reduces to eq. (\ref{eq:E(z)_LCDM}).

%%
%Subsection 3
%%
\subsection{$\phi$CDM}
\label{sec:phiCDM}
In Sec. \ref{sec:scalar_field_dynamics} I derived the equation of motion of a general, cosmological scalar field in a universe that is homogeneous, isotropic, and expanding with time. Here I specialize those results to a scalar field model that has been extensively studied in the literature: the Ratra-Peebles ``quintessence'' model, which provides a simple, physically consistent description of dynamical dark energy \citep{6, ratpeeb88, pavlov13}. In this model, the scalar field has a potential energy density of the form
\begin{equation}
\label{eq:V(phi)}
    V = \frac{1}{2}\kappa m_{\rm p}^2 \phi^{-\alpha}
\end{equation}
where $\alpha > 0$, $m_{\rm p}^2 := G^{-1}$ is the Planck mass in units where $\hbar = 1$ and $c = 1$, and
\begin{equation}
\label{eq:ch3_kappa_def}
    \kappa = \frac{8}{3}\left(\frac{\alpha + 4}{\alpha + 2}\right)\left[\frac{2}{3}\alpha\left(\alpha + 2\right)\right]^{\alpha/2}
\end{equation}
\citep{5}. The homogeneous part of the scalar field is a function only of time, so its equation of motion is
\begin{equation}
D_{\nu}\left(g^{0 \nu} \partial_{0}\phi\right) = V'(\phi)
\end{equation}
or
\begin{equation}
\label{eq:ch1_1.168}
\partial_{\nu}\left(g^{0\nu}\partial_{0}\phi\right) + \Gamma^{\nu}_{\nu \sigma}g^{\sigma 0} \partial_{0}\phi = V'(\phi).
\end{equation}
Because the metric of a homogeneous, isotropic spacetime has the form
\begin{equation}
ds^2 = -dt^2 + a^2\left(t\right)\gamma_{ij}dx^{i}dx^{j},
\end{equation}
where $\gamma_{ij}$ is given by eq. (\ref{eq:ch1_lower_gamma}), eq. (\ref{eq:ch1_1.168}) reduces to
\begin{equation}
\label{eq:scalar_field_EOM}
\begin{aligned}
-\partial_0^2\phi - \Gamma^{\nu}_{\nu 0}\partial_0 \phi - V'\left(\phi\right) & = \ddot{\phi} + \frac{\dot{a}}{a}\gamma^i_i \dot{\phi} + V'\left(\phi\right)\\
 & = \ddot{\phi} + 3H\dot{\phi} + V'\left(\phi\right),
\end{aligned}
\end{equation}
where $H \equiv \dot{a}/a$ is the Hubble rate and I have used eq. (\ref{eq:Gamma^rho_tau0}). After inserting eq. (\ref{eq:V(phi)}), the scalar field's equation of motion takes the form
\begin{equation}
\label{eq:ch3_phiCDM_EOM}
    \boxed{\ddot{\phi} + 3\frac{\dot{a}}{a}\dot{\phi} - \frac{1}{2}\kappa \alpha m_{\rm p}^2 \phi^{-\alpha - 1} = 0.}
\end{equation}

The energy density and pressure of the scalar field are simple to derive from eqs. (\ref{eq:p_phi}) and (\ref{eq:rho_phi}). They are:
\begin{equation}
\label{eq:ch3_rho_phi}
    \rho_{\phi} = \frac{1}{2}\dot{\phi}^2 + V,
\end{equation}
and
\begin{equation}
    p_{\phi} = \frac{1}{2}\dot{\phi}^2 - V,
\end{equation}
respectively. It follows that the equation of state parameter of $\phi$CDM is
\begin{equation}
w_{\phi}  = \frac{p_{\phi}}{\rho_{\phi}} = \frac{\frac{1}{2}\dot{\phi}^2 - V(\phi)}{\frac{1}{2}\dot{\phi}^2 + V(\phi)},
\end{equation}
which, unlike in the $\Lambda$CDM model and the XCDM parametrization, changes with time. Given eq. (\ref{eq:ch3_rho_phi}), it is also easy to show that the Hubble rate function $H(z)$ in the $\phi$CDM model is
\begin{equation}
H(z) = H_0\sqrt{\Omega_{m0}\left(1 + z\right)^3 + \Omega_{k0}\left(1 + z\right)^2 + \Omega_{\phi}(z,\alpha)},
\end{equation}
where
\begin{equation}
\Omega_{\phi}(z, \alpha) := \frac{8 \pi G\rho_{\phi}}{3 H_0^2}.
\end{equation}
\begin{comment}
There is a long history, in the literature, of using time-dependent scalar fields to represent a time-varying cosmological ``constant" (i.e. a quantity that only appears to be constant on short time scales) [cite]. The dynamics of a scalar field in an expanding universe is easy to derive given the Friedmann equations and the scalar field energy-momentum tensor. Let us assume that the Universe is homogeneous and isotropic on large scales, so that the metric has the form given in eq. (\ref{eq:FLRW}), and that the scalar field depends only on time. 
\end{comment}
In contrast to $\Omega_{X}$, $\Omega_{\phi}$ is not an explicit function of a power of $\left(1 + z\right)$; it must be determined by solving the dynamics of the scalar field numerically. This can be done by solving eq. (\ref{eq:ch3_phiCDM_EOM}) together with the first Friedmann equation
\begin{equation}
\label{eq:Friedmann_scalar_field}
    \left(\frac{\dot{a}}{a}\right)^2 = \frac{8 \pi G}{3}\left(\rho_{\rm m} + \rho_{\phi}\right) - \frac{k}{a^2},
\end{equation}
after which the  parameters $p = \left(H_0, \Omega_{m0}, \Omega_{k0}, \alpha\right)$ of the $\phi$CDM model can be fitted to the available data. In the special case that the Universe is taken to be exactly spatially flat, the free parameters are $p = \left(H_0, \Omega_{m0}, \alpha\right)$. If $\alpha = 0$, then the $\phi$CDM model reduces to the $\Lambda$CDM model. 

An evaluation of the relative quality of $\Lambda$CDM, XCDM, and $\phi$CDM through an analysis of their respective fits to observational data forms the main part of this work, and is the subject of Chapters \ref{Chapter4}, \ref{Chapter5}, \ref{Chapter7}, \ref{Chapter8}, and \ref{Chapter9}.

%%
%Section: Cosmic acceleration without Dark Energy
%%
\section{Cosmic acceleration without Dark Energy}
This section is based on \cite{Ryan_2020, Ryan_power_law}
%%
%Section: Ryskin model
%%
\subsection{Ryskin's model}
\label{sec:ch3_Ryskin_model}

As we have seen in Sec. \ref{sec:LCDM}, the $\Lambda$CDM model holds that the accelerated expansion of the Universe is powered by a spatially homogeneous energy density in the form of a cosmological constant, $\Lambda$. Although this model has successfully explained many observations to date (see e.g. \cite{planck2018_overview, planck2018}), a number of theoretical problems remain unsolved (see e.g. \cite{Weinberg_1, Weinberg_2, Straumann, Martin}). Many researchers have therefore attempted to construct models of cosmic acceleration that do not incorporate the cosmological constant, or any other form of dark energy (for a review of which, see e.g. \cite{modgrav}). One such model (not covered in \cite{modgrav}) is Gregory Ryskin's model of emergent cosmic acceleration (presented in \cite{Ryskin}). In this model, the observed acceleration of the universe is argued to emerge naturally as a consequence of applying a mean-field treatment to Einstein's gravitational field equations on cosmic scales. In this way, Ryskin claims to have arrived at an explanation of cosmic acceleration that does not require any fundamentally new physics.

The main idea behind Ryskin's paper is that the standard gravitational field equations of General Relativity,
\begin{equation}
    R_{\mu \nu} - \frac{1}{2}Rg_{\mu \nu} = \kappa T_{\mu \nu},
\end{equation}
where $\kappa := 8 \pi G$, which are well-tested on the scale of the solar system, must be modified when applied to cosmological scales. Ryskin contends, in \cite{Ryskin}, that moving from sub-cosmological scales (in which matter is distributed inhomogeneously) to cosmological scales (in which matter is distributed homogeneously) introduces emergent properties to the description of the Universe (such properties being ``emergent" in the sense that they are not apparent on sub-cosmological scales) and that the observed large-scale acceleration of the Universe may be one such emergent property. According to Ryskin's model, the emergence of cosmic acceleration is therefore analogous to the emergence of properties like temperature and pressure that result from averaging over the microscopic degrees of freedom of an ideal fluid, thereby moving from a length-scale regime in which kinetic theory is valid to a length-scale regime in which the fluid must be described with continuum hydrodynamics; in the same way, Ryskin contends, cosmic acceleration ``emerges" from Einstein's field equations when these are applied to cosmological scales.  Ryskin's model, therefore, purports to offer an explanation of the origin of cosmic acceleration that does not require the introduction of dark energy. In his paper, Ryskin introduces a mean-field tensor with components $\kappa\Phi_{\mu \nu}$ into the right-hand side of eq. (\ref{eq:Einstein_0}), so that the (large-scale) field equations read
\begin{equation}
\label{eq:Ryskin}
    R_{\mu \nu} - \frac{1}{2}Rg_{\mu \nu} = \kappa \left(T_{\mu \nu} + \Phi_{\mu \nu}\right).
\end{equation}
It can be shown (see \cite{Ryskin} for details) that when the standard gravitational field equations are modified in this way, and that if the Universe has flat spatial hypersurfaces and is dominated by non-relativistic matter, then the total rest energy density and pressure of the averaged, large-scale cosmic fluid are
\begin{equation}
\label{eq:ch6_rho}
    \rho = 4\rho_m,
\end{equation}
and
\begin{equation}
\label{eq:p}
    p = -3\rho_m,
\end{equation}
respectively, where $\rho_m$ and $p_m$ are the rest energy density and pressure of the non-relativistic matter. Energy conservation then implies that $\rho_m \propto a^{-3/4}$, from which $a(t) \propto t^{8/3}$ follows. Rykin's model therefore predicts that the Hubble parameter takes the simple form
\begin{equation}
\label{eq:RyskinH}
    H(z) = H_0\left(1 + z\right)^{3/8},
\end{equation}
where the definitions of the Hubble parameter $H(t) := \frac{\dot{a}\left(t\right)}{a\left(t\right)}$ and redshift $1 + z := \frac{a\left(0\right)}{a\left(t\right)}$ have been used. Ryskin's model therefore has only one free parameter: $H_0$. We will investigate the quality of this model's fit to cosmological data in Chapter \ref{Chapter6}.

%%
%Section: The power law model
%%
\subsection{The power law model}
\label{sec:ch3_power_law_model}
Another alternative to the dark energy paradigm comes in the form of the power law model (of which Ryskin's model is a special case), in which the scale factor takes the form of a power law $a(t) \propto t^{\beta}$ with a constant exponent $\beta$. One of the virtues of this model is its simplicity: it only depends on the single parameter $\beta$, and the functional form $t^{\beta}$ is easy to integrate analytically when it appears in the form $\int \frac{dt}{a(t)}$ (as in the computation of the co-moving distance scale). Additionally, the power law model with $\beta \geq 1$ does not suffer from the horizon or flatness problems,\footnote{The horizon problem is based on the observation that distantly separated portions of the CMB sky could not have had time to reach thermal equilibrium when the CMB itself was emitted (that is, these regions are outside of each other's respective horizons, and so could not have come into causal contact). The flatness problem stems from the apparent fine-tuning required to make the Universe spatially flat to the degree we observe today. In the standard model, inflation solves both of these problems by positing a phase of accelerated expansion during the earliest moments of the Universe's life which: 1.) rapidly expands the sizes of causally connected regions, explaining the observed thermal equilibrium of the CMB and 2.) redshifting away the spatial curvature component of the early Universe's energy budget, so that the Universe is flat (or very nearly flat) on large scales. By saying that the power law model does not suffer from the horizon or flatness problems, I mean that it is a model which is purported to describe the evolution of the Universe on large scales without requiring an ``accessory'' like inflation, which was not a part of the original Big Bang model.} and produces a universe whose age is compatible with the ages of the oldest known objects in the Universe  (these being globular clusters and high-reshift galaxies; \citealp{Dev_Jain_Lohiya_2008, Sethi_Dev_Jain_2005}). Power law expansion is also a predicted feature of some alternative gravity theories that are designed to solve the cosmological constant problem dynamically \citep{Dev_Jain_Lohiya_2008, Sethi_Dev_Jain_2005}.

In the power law model, the scale factor $a(t)$ takes the form
\begin{equation} \label{eq:a}
    a(t) = Ct^{\beta},
\end{equation}
where $C$ and $\beta$ are constants. From the definition of redshift, $\frac{a_0}{a(t)} := 1 + z$ ($a_0$ is the current value of the scale factor and $z$ is the redshift) and eq. (\ref{eq:a}), we can write
\begin{equation}
    \frac{a_0}{Ct^{\beta}} = 1 + z,
\end{equation}
from which it follows that
\begin{equation} \label{eq:1/t}
    \frac{1}{t} = \left[\frac{C}{a_0}\left(1 + z\right)\right]^{1/\beta}.
\end{equation}
The definition of the Hubble parameter, $H(t) := \frac{\dot{a(t)}}{a(t)}$, with the overdot denoting the time derivative, implies $H(t) = \frac{\beta}{t}$. Therefore eq. (\ref{eq:1/t}) can be written in the form
\begin{equation}
\label{eq:ch3_Hz_PL}
    H(z) = H_0\left(1 + z\right)^{1/\beta},
\end{equation}
where I have defined the present value of the Hubble constant to be $H_0 := \beta\left(\frac{C}{a_0}\right)^{1/\beta}$. The power law model therefore has two free parameters: $H_0$ and $\beta$. The comparison of the power law model to cosmological data will be taken up in Chapter \ref{Chapter10}.

%% file: chapter4.tex
\cleardoublepage

\chapter{Constraints on dark energy dynamics and spatial curvature from Hubble parameter and baryon acoustic oscillation data}\chaptermark{Hubble parameter and BAO data constraints}

\newcommand{\hublow}{$\bar{H}_0 \pm \sigma_{H_0} = 68 \pm 2.8$ km s$^{-1}$ Mpc$^{-1}$}
\newcommand{\hubhigh}{$\bar{H}_0 \pm \sigma_{H_0} = 73.24 \pm 1.74$ km s$^{-1}$ Mpc$^{-1}$}

\label{Chapter4}

This chapter is based on \cite{Ryan_Doshi_Ratra_2018}.

%%
%Section: Intro
%%
\section{Introduction}
\label{sec. 1}
\begin{comment}
It is widely accepted that the universe is undergoing accelerated expansion today. The consensus cosmological model, $\Lambda$CDM, posits that this acceleration is driven by the spatially homogeneous, constant dark energy density $\rho_{\Lambda}$ of the cosmological constant $\Lambda$ \citep{peeb84}. For reviews of the accelerated cosmological expansion and of the $\Lambda$CDM model, see \cite{Ratra_Vogeley}, \cite{Martin}, \cite{64}, and \cite{62}. In this model, cold dark matter (CDM) is the second largest contributor to the current energy budget and, with non-relativistic baryonic matter, powered the decelerating cosmological expansion at earlier times. 
\end{comment}
The consensus $\Lambda$CDM model assumes flat spatial hypersurfaces, but observations don't rule out mildly curved spatial hypersurfaces; observations also do not rule out the possibility that the dark energy density varies slowly with time. In this chapter we will examine, in addition to the general (not necessarily spatially flat) $\Lambda$CDM model, the XCDM parametrization of dynamical dark energy, and the $\phi$CDM model in which a scalar field $\phi$ is the dynamical dark energy (see Chapter \ref{Chapter3} for details).\footnote{While cosmic microwave background (CMB) anisotropy data provide the most restrictive constraints on cosmological parameters, many other measurements have been used to constrain the XCDM parametrization and the $\phi$CDM model \citetext{see, e.g., \citealp{24}, \citealp{yashar_et_al_2009}, \citealp{samushia_ratra_2010}, \citealp{chen_ratra_2011b}, \citealp{20}, \citealp{23}, \citealp{Avsajanishvili_2015}, \citealp{26}, \citealp{Sola_perez_gomez_2018, Sola_etal_2017, 30, 28}, \citealp{18}, \citealp{22}, \citealp{32}, \citealp{91}, \citealp{sangwan_tripathi_jassal_2018}}.} In the XCDM and $\phi$CDM cases we allow for both vanishing and non-vanishing spatial curvature.

\begin{comment}
and more information can be found in \cite{5}.
\end{comment}

\cite{Ooba_Ratra_Sugiyama_2018_FpCDM} have recently shown that, in the spatially flat case, the Planck 2015 CMB anisotropy data from \cite{planck_2016} (and some baryon acoustic oscillation distance measurements) weakly favor the XCDM parametrization and the $\phi$CDM model of dynamical dark energy over the $\Lambda$CDM consensus model. The XCDM case results have been confirmed by \cite{Park_Ratra_2018_FXCDM_NFXCDM} for a much bigger compilation of cosmological data, including most available Type Ia supernova apparent magnitude observations, BAO distance measurements, growth factor data, and Hubble parameter observations.\footnote{For earlier indications favoring dynamical dark energy over the $\Lambda$CDM consensus model, based on smaller compilations of data, see \cite{39}, \cite{34}, \cite{40}, \cite{43}, \cite{28}, \cite{26}, \cite{Sola_etal_2017}, \cite{42}, \cite{Sola_perez_gomez_2018}, \cite{41}, \cite{30}, \cite{22}, \cite{33}, and \cite{35}. However, more recent analyses, based on bigger compilations of data, do not support the significant evidence for dynamical dark energy indicated in some of the earlier analyses \citep{Ooba_Ratra_Sugiyama_2018_FpCDM, Park_Ratra_2018_FXCDM_NFXCDM}.} Also, spatially flat XCDM and $\phi$CDM both reduce the tension between CMB temperature anisotropy and weak gravitational lensing estimates of $\sigma_8$, the rms fractional energy density inhomogeneity averaged over $8\hspace{1mm}h^{-1}$Mpc radius spheres, where $h$ is the Hubble constant in units of 100 km s$^{-1}$ Mpc$^{-1}$ \citep{Ooba_Ratra_Sugiyama_2018_FpCDM, Park_Ratra_2018_FXCDM_NFXCDM}.

In non-flat models nonzero spatial curvature provides an additional length scale which invalidates usage of the power-law power spectrum for energy density inhomogeneities in the non-flat case (as was assumed in the analysis of non-flat models in \cite{planck_2016}). Non-flat inflation models \citep{gott_1982, hawking_1984, ratra_1985} provide the only known physically-consistent mechanism for generating energy density inhomogeneities in the non-flat case; the resulting open and closed model power spectra are not power laws \citep{57, ratra_peebles_1995, ratra_2017}. Using these power spectra, \cite{Ooba_Ratra_Sugiyama_2017_NFLCDM} have found that the Planck 2015 CMB anisotropy data in combination with a few BAO distance measurements no longer rule out the non-flat $\Lambda$CDM case (unlike the earlier \cite{planck_2016} analyses based on the incorrect assumption of a power-law power spectrum in the non-flat model).\footnote{Currently available non-CMB measurements do not significantly constrain spatial curvature \citep{Farooq_Mania_Ratra_2015, Chen_et_al_2016, yu_wang_2016, 48, Farooq_Ranjeet_Crandall_Ratra_2017, wei_wu_2017, rana_jain_mahajan_mukherjee_2017, 60, mitra_choudhury_ratra_2018}.} \cite{Park_Ratra_2018_FLCDM} confirmed these results for a bigger compilation of cosmological data, and similar conclusions hold in the non-flat dynamical dark energy XCDM and $\phi$CDM cases \citep{Ooba_Ratra_Sugiyama_2017_NFXCDM, Ooba_Ratra_Sugiyama_2017_NFpCDM, Park_Ratra_2018_FXCDM_NFXCDM}.

Additionally, the non-flat models provide a better fit to the observed low multipole CMB temperature anisotropy power spectrum, and do better at reconciling the CMB anisotropy and weak lensing constraints on $\sigma_8$, but do a worse job at fitting the observed large multipole CMB anisotropy temperature power spectrum \citep{Ooba_Ratra_Sugiyama_2017_NFXCDM, Ooba_Ratra_Sugiyama_2017_NFXCDM, Ooba_Ratra_Sugiyama_2017_NFpCDM, Park_Ratra_2018_FXCDM_NFXCDM, Park_Ratra_2018_FLCDM}. Given the non-standard normalization of the Planck 2015 CMB anisotropy likelihood and that the flat and non-flat $\Lambda$CDM models are not nested, it is not possible to compute the relative goodness of fit between the flat and non-flat $\Lambda$CDM models quantitatively, although qualitatively the flat $\Lambda$CDM model provides a better fit to the current data \citep{Ooba_Ratra_Sugiyama_2017_NFXCDM, Ooba_Ratra_Sugiyama_2017_NFXCDM, Ooba_Ratra_Sugiyama_2017_NFpCDM, Park_Ratra_2018_FXCDM_NFXCDM, Park_Ratra_2018_FLCDM}.

In the analyses discussed above, the Planck 2015 CMB anisotropy data played the major role. Those authors found consistency between cosmological constraints derived using the CMB anisotropy data in combination with various non-CMB data sets. CMB anisotropy data are sensitive to the behavior of cosmological spatial inhomogeneities.

Here we derive constraints on similar models from a combination of all available Hubble parameter data as well as all available radial and transverse BAO data.\footnote{The $H(z)$ and radial BAO data provide a unique measure of the cosmological expansion rate over a wide redshift range, up to almost $z = 2.4$, well past the cosmological deceleration-acceleration transition redshift. These data show evidence for this transition and can be used to measure the redshift of the transition \citep{Farooq_Ratra_2013, farooq_crandall_ratra_2013, 88, 68, Farooq_Ranjeet_Crandall_Ratra_2017, 60, jesus_holanda_pereira_2018, haridasu_etal_2018}.} Unlike the CMB anisotropy data, the $H(z)$ and these BAO data are not sensitive to the behavior of cosmological spatial inhomogeneities.

The models that we study here are the flat and nonflat $\Lambda$CDM model, the flat and nonflat XCDM parametrization, and the flat and nonflat $\phi$CDM model. See Chapter \ref{Chapter3} for a description of these models and of their free parameters.\footnote{In this chapter we do not treat $H_0$ as a free parameter, treating it instead as a prior to be marginalized over. See below for a description of these priors, and of the marginalization procedure we employ.} These models, and the methods we use to analyze our data, are the same as those presented in \cite{Farooq_Ranjeet_Crandall_Ratra_2017, Farooq_Mania_Ratra_2015}. Some of the measurements we use are the same as the measurements used in those papers, although our data set is more up to date.

The constraints we derive in this chapter are consistent with, but weaker than, those of the papers cited above; this provides a necessary and useful consistency test of those results. In particular, we find that the consensus flat $\Lambda$CDM model is a reasonable fit, in most cases, to the BAO and $H(z)$ data we study here. However, depending somewhat on the Hubble constant prior we use, consensus flat $\Lambda$CDM can be 1$\sigma$ away from the best-fit parameter values in some cases, which can favor mild dark energy dynamics or non-flat spatial hypersurfaces.
\begin{comment}
In chapter \ref{Chapter3} we provide a short summary of the models we studied. Sec. \ref{sec:ch4_Data} presents the data that we used, and in Sec. \ref{sec:ch4_Methods} we describe the methods by which we analyzed these data. Sec. \ref{sec:ch4_Results} describes the results of our analyses, and our conclusions are given in Sec. \ref{sec:ch4_Conclusion}.
\end{comment}

%%
%Section: Data
%%
\section{Data}
\label{sec:ch4_Data}
\begin{comment}
%Sub-section 1
\subsection{Hubble parameter data}

Table \ref{tab:H(z)_data} lists 31 $H(z)$ measurements determined using the ``cosmic chronometer" technique, which are the same as the $H(z)$ data used in \cite{60} \citetext{see e.g. \citealp{70} [cite more recent papers?] for a discussion of cosmic chronometers}. With this method, the Hubble rate as a function of redshift is determined by using
\begin{equation}
    H(z) = -\frac{1}{\left(1 + z\right)}\frac{dz}{dt}.
\end{equation}
Although this determination of $H(z)$ does not depend on a cosmological model, it does depend on the quality of the measurement of $dz/dt$, which requires an accurate determination of the age-redshift relation for a given chronometer. See \cite{70} and \cite{72} for discussions of the strengths and weaknesses of this method [cite more recent papers?]. While their approach requires accurate knowledge of the star formation history and metallicity of massive, passively evolving early galaxies, and although the two different techniques they use give slightly different values, they also point out that the measurement of $H(z)$ from this method is relatively insensitive to changes in the chosen stellar population synthesis model [cite more recent studies of systematic errors?].

%Sub-section 2
\subsection{Baryon acoustic oscillation data}
\end{comment}
Baryon acoustic oscillations provide observers with a ``standard ruler" which can be used to measure cosmological distances (see \citealp{75} for a review). These distances can be computed in a given cosmological model, so measurements of them can be used to constrain the parameters of the model in question. The BAO distance measurements are listed in Table \ref{tab:ch4_BAO_data}. See Chapter \ref{Chapter2} for the definitions of the various distance functions listed in Table \ref{tab:ch4_BAO_data}.

\begin{table}
\centering
\caption[BAO data.]{BAO data. $D_M \left(r_{s,{\rm fid}}/r_s\right)$ and $D_V \left(r_{s,{\rm fid}}/r_s\right)$ have units of Mpc, while $H(z)\left(r_s/r_{s,{\rm fid}}\right)$ has units of ${\rm km}\hspace{1mm}{\rm s}^{-1}{\rm Mpc}^{-1}$ and $r_s$ has units of Mpc.}
\label{tab:ch4_BAO_data}
\begin{tabular}{ccccc}
\hline
$z$ & Measurement & Value & $\sigma$ & Ref.\\
\hline
$0.38$ & $D_M\left(r_{s,{\rm fid}}/r_s\right)$ & 1518 & 22 & \cite{Alam_et_al_2017}\\
\hline
$0.51$ & $D_M\left(r_{s,{\rm fid}}/r_s\right)$ & 1977 & 27 & \cite{Alam_et_al_2017}\\
\hline
$0.61$ & $D_M\left(r_{s,{\rm fid}}/r_s\right)$ & 2283 & 32 & \cite{Alam_et_al_2017}\\
\hline
$0.38$ & $H(z)\left(r_s/r_{s,{\rm fid}}\right)$ & 81.5 & 1.9 & \cite{Alam_et_al_2017}\\
\hline
$0.51$ & $H(z)\left(r_s/r_{s,{\rm fid}}\right)$ & 90.4 & 1.9 & \cite{Alam_et_al_2017}\\
\hline
$0.61$ & $H(z)\left(r_s/r_{s,{\rm fid}}\right)$ & 97.3 & 2.1 & \cite{Alam_et_al_2017}\\
\hline
$0.106$ & $r_s/D_V$ & 0.336 & 0.015 & \cite{10}\\
\hline
$0.15$ & $D_V\left(r_{s,{\rm fid}}/r_s\right)$ & $664$ & $25$ & \cite{2}\\
\hline
$1.52$ & $D_V\left(r_{s,{\rm fid}}/r_s\right)$ & $3855$ & $170$ & \cite{3}\\
\hline
$2.33$ & $\frac{\left(D_H\right)^{0.7} \left(D_{M}\right)^{0.3}}{r_s}$ & 13.94 & 0.35 & \cite{9}\\
\hline
$2.36$ & $c/\left(r_s H(z)\right)$ & 9.0 & 0.3 & \cite{11}\\
\hline
\end{tabular}
\end{table}

All of the measurements in Table \ref{tab:ch4_BAO_data} are scaled by the size of the sound horizon at the drag epoch ($r_{\rm s}$). This quantity is \citetext{see \citealp{8} for a derivation}:
\begin{equation}
    r_{\rm s} = \frac{2}{3k_{\rm eq}}\sqrt{\frac{6}{R_{\rm eq}}}{\rm ln}\left[\frac{\sqrt{1 + R_d} + \sqrt{R_d + R_{\rm eq}}}{1 + \sqrt{R_{\rm eq}}}\right]
\end{equation}
where $R_d \equiv R(z_d)$ and $R_{\rm eq} \equiv R(z_{\rm eq})$ are the values of $R$, the ratio of the baryon to photon momentum density,
\begin{equation}
    R = \frac{3\rho_b}{4\rho_{\gamma}}
\end{equation}
at the drag epoch and matter-radiation equality epoch, respectively. Here $k_{\rm eq}$ is the scale of the particle horizon at the matter-radiation equality epoch, and $\rho_b$ and $\rho_{\gamma}$ are the baryon and photon mass densities. In this analyses, where appropriate, the original data listed in Table \ref{tab:ch4_BAO_data} have been rescaled to a fiducial sound horizon $r_{\rm s, fid} = 147.60$ Mpc (from Table 4, column 3, of \cite{planck_2016}). This fiducial sound horizon was determined by using the $\Lambda$CDM model, so its value is model dependent, though not to a significant degree (as can be seen by comparing the computed $r_s$ of the \cite{planck_2016} baseline model to that measured using the spatially open \lcdm\ and flat XCDM parametrization of \cite{planck_2016}).

Table \ref{tab:H(z)_data} lists 31 $H(z)$ measurements determined using the cosmic chronometric technique, which are the same as the cosmic chronometric $H(z)$ data used in \cite{60} \citetext{see e.g. \citealp{70} for a discussion of cosmic chronometers}. With this method, the Hubble rate as a function of redshift is determined by using
\begin{equation}
    H(z) = -\frac{1}{\left(1 + z\right)}\frac{dz}{dt}.
\end{equation}
Although this determination of $H(z)$ does not depend on a cosmological model, it does depend on the quality of the measurement of $dz/dt$, which requires an accurate determination of the age-redshift relation for a given chronometer. See \cite{70} and \cite{72} for discussions of the strengths and weaknesses of this method. While their approach requires accurate knowledge of the star formation history and metallicity of massive, passively evolving early galaxies, and although the two different techniques they use give slightly different values, they also point out that the measurement of $H(z)$ from this method is relatively insensitive to changes in the chosen stellar population synthesis model.

\section{Methods}
\label{sec:ch4_Methods}

	To determine the values of the best-fit parameters, we minimized
\begin{equation}
\chi^2(p) \equiv -2 {\rm ln} \mathcal{L}(p),
\end{equation}
where $\mathcal{L}$ is the likelihood function and $p$ is the set of parameters of the model under consideration. If the likelihood function $\mathcal{L}(p, \nu)$ depends on an uninteresting nuisance parameter $\nu$ with a probability distribution $\pi(\nu)$, we marginalize the likelihood function by integrating $\mathcal{L}(p, \nu)$ over $\nu$
\begin{equation}
\mathcal{L}(p) = \int \mathcal{L}(p, \nu) \pi(\nu) d\nu.
\end{equation}
In our $H(z)$ analyses $H_0$ is a nuisance parameter. We assumed a Gaussian distribution for $H_0$
\begin{equation}
    \pi\left(H_0\right) = \frac{1}{\sqrt{2\pi\sigma^2_{H_0}}}{\rm exp}\left[\frac{-\left(H_0 - \bar{H}_0\right)^2}{2\sigma^2_{H_0}}\right]
\end{equation}
and marginalized over it. We considered two cases: $\bar{H}_0 \pm \sigma_{H_0} = 68 \pm 2.8$ km s$^{-1}$ Mpc$^{-1}$ and $\bar{H}_0 \pm \sigma_{H_0} = 73.24 \pm 1.74$ km s$^{-1}$ Mpc$^{-1}$.\footnote{The lower value, $68 \pm 2.8$ km s$^{-1}$ Mpc$^{-1}$ is the most recent median statistics estimate of the Hubble constant \citep{chenratmed}. It is consistent with earlier median statistics estimates \citep{gott_etal_2001, 76}. It is also consistent with many other recent measurements of $H_0$ \citetext{\citealp{planck_2016}; \citealp{48}; \citealp{chen_etal_2017}; \citealp{wang_xu_zhao_2017}; \citealp{Linishak}; \citealp{81}; \citealp{Gomez-ValentAmendola2018}; \citealp{60}; \citealp{Park_Ratra_2018_FXCDM_NFXCDM}; \citealp{haridasu_etal_2018}}. The higher value, $73.24 \pm 1.74$ km s$^{-1}$ Mpc$^{-1}$, comes from a local expansion rate estimate \citep{riess2016}. Other local expansion rate estimates find slightly lower $H_0$'s with larger error bars \citep{rigault_etal_2015, 86, Dhawan, FernandezArenas}.}

Most of the data we analyzed are uncorrelated, however six of the data points \citetext{those from \citealp{Alam_et_al_2017}}, are correlated. For uncorrelated data points,
\begin{equation} \label{eq. 23}
\chi^2(p) = \sum^{N}_{i = 1} \frac{[A_{{\rm th}}(p; z_i) - A_{{\rm obs}}(z_i)]^2}{\sigma_i^2},
\end{equation}
where $A_{{\rm th}}(p; z_i)$ are the model predictions at redshifts $z$, and $A_{{\rm obs}}(z_i)$ and $\sigma_i$ are the central values and error bars of the measurements listed in Table \ref{tab:H(z)_data} and the last five lines of Table \ref{tab:ch4_BAO_data}. The correlated data (the first six entries in Table \ref{tab:ch4_BAO_data}) require
\begin{equation} \label{eq. 24}
\chi^2(p) = \left[\vec{A}_{{\rm th}}(p) - \vec{A}_{{\rm obs}}\right]^{T} \mathcal{C}^{-1} \left[\vec{A}_{{\rm th}}(p) - \vec{A}_{{\rm obs}}\right]
\end{equation}
where $\mathcal{C}^{-1}$ is the inverse of the covariance matrix
\begin{equation}
\label{eq:ch4_covmat}
    \mathcal{C} = 
\begin{bmatrix}
    484.0       & 9.530 & 295.2 & 4.669 & 140.2 & 2.402 \\
    9.530       & 3.610 & 7.880 & 1.759 & 5.983 & 0.9205 \\
    295.2 & 7.880 & 729.0 & 11.93 & 442.4 & 6.866 \\
    4.669 & 1.759 & 11.93 & 3.610 & 9.552 & 2.174 \\
    140.2 & 5.983 & 442.4 & 9.552 & 1024 & 16.18 \\
    2.402 & 0.9205 & 6.866 & 2.174 & 16.18 & 4.410 \\
\end{bmatrix}
\end{equation}
\citep{Alam_et_al_2017}. $\vec{A}_{\rm obs}$ (in eq. \ref{eq. 24}) are the measurements in the first six lines of Table \ref{tab:ch4_BAO_data}.
	
In addition to $\chi^2$, we also used the Bayes Information Criterion
\begin{equation}
{\rm BIC} \equiv \chi^2_{{\rm min}} + k {\rm ln} N
\end{equation}
and the Akaike Information Criterion
\begin{equation}
{\rm AIC} \equiv \chi^2_{{\rm min}} + 2k
\end{equation}
\citep{Liddle_2007}. In these equations $\chi^2_{\rm min}$ is the minimum value of $\chi^2$, $k$ is the number of parameters of the given model, and $N$ is the number of data points. BIC and AIC provide means to compare models with different numbers of parameters; they penalize models with a higher $k$ in favor of those with a lower $k$, in effect enforcing Occam's Razor in the model selection process.

To determine the confidence intervals $r_n$ on the 1d best-fit parameters, we computed one-sided limits $r^{\pm}_n$ by using
\begin{equation} \label{eq. 28}
    \frac{\int^{r^{\pm}_n}_{\bar{p}} \mathcal{L}(p)dp}{\int^{\pm \infty}_{\bar{p}} \mathcal{L}(p)dp} = \sigma_n,
\end{equation}
where $\bar{p}$ is the point at which $\mathcal{L}(p)$ has its maximimum value, such that $n = 1, 2$ and $\sigma_1 = 0.6827$, $\sigma_2 = 0.9545$. Because the one-dimensional likelihood function is not guaranteed to be symmetric about $\bar{p}$, we compute the upper and lower confidence intervals separately. In the $\Lambda$CDM model, for example, the 1-sigma confidence intervals on $\Omega_{m0}$ are computed by first integrating the likelihood function $\mathcal{L}(\Omega_{m0}, \Omega_{\Lambda})$ over $\Omega_{\Lambda}$ to obtain a marginalized likelihood function that only depends on $\Omega_{m0}$,
\begin{equation}
    \int^1_0 \mathcal{L}(\Omega_{m0}, \Omega_{\Lambda})d\Omega_{\Lambda} = \mathcal{L}(\Omega_{m0}),
\end{equation}
and then inserting this marginalized likelihood function into eq. (\ref{eq. 28}).

The ranges over which we marginalized the parameters of the $\Lambda$CDM model were $0 \leq \Omega_{\Lambda} \leq 1$ and $0.01 \leq \Omega_{m0} \leq 1$. For the spatially flat XCDM parametrization, we used $-2 \leq w_X \leq 0$ and $0.01 \leq \Omega_{m0} \leq 1$, and for the spatially flat $\phi$CDM model we used $0.01 \leq \alpha \leq 5$ and $0.01 \leq \Omega_{m0} \leq 1$. For 3-parameter XCDM, we used $-0.7 \leq \Omega_{k0} \leq 0.7$, $0.01 \leq \Omega_{m0} \leq 1$, and $-2.00 \leq w_X \leq 0$. For the 3-parameter $\phi$CDM model we considered $-0.5 \leq \Omega_{k0} \leq 0.5$, $0.01 \leq \Omega_{m0} \leq 1$, and $0.01 \leq \alpha \leq 5$. \footnote{$\Omega_{m0}, \alpha = 0.01$ were excluded because our codes ran into difficulties at those points.} 

We analyzed the data with two independent Python codes, written by Sanket Doshi and Joseph Ryan, that produced almost identical results for all 2-parameter cases as well as the 3-parameter XCDM parametrization, and results that agreed to within 1\% in the 3-parameter $\phi$CDM case.

%%
%Section: Results
%%
\section{Results}
\label{sec:ch4_Results}

 \begin{table*}
  \caption[Best-fit values for 2-parameter models.]{Best-fit values for 2-parameter models. $\Delta \chi^2$ is evaluated relative to $\chi^2$ of $\Lambda$CDM for each $H_0$ prior.}
  \label{table 3}
  \centering
  \resizebox{\columnwidth}{!}{%
  \begin{tabular}{ccccccccccc}
    \hline
    $H_0$ prior $\left({\rm km}\hspace{1mm}{\rm s}^{-1}{\rm Mpc}^{-1}\right)$ & Model & $\Omega_{m0}$ & $\Omega_{\Lambda}$ & $w_X$ & $\alpha$ & $\chi^2$ & $\Delta \chi^2$ & AIC & BIC\\
    \hline
    $68 \pm 2.8$ & $\Lambda$CDM & 0.29 & 0.68 & - & - & 25.35 & 0.00 & 29.35 & 32.83\\
     & flat XCDM & 0.29 & - & -0.94 & - & 25.04 & -0.31 & 29.04 & 32.52\\
     & flat $\phi$CDM & 0.29 & - & - & 0.16 & 25.05 & -0.30 & 29.05 & 32.53\\
    \hline
    $73.24 \pm 1.74$ & $\Lambda$CDM & 0.30 & 0.77 & - & - & 26.92 & 0.00 & 30.92 & 34.40\\
     & flat XCDM & 0.29 & - & -1.13 & - & 28.26 & 1.34 & 32.26 & 35.74\\
     & flat $\phi$CDM & 0.30 & - & - & 0.01 & 32.62 & 5.70 & 36.62 & 40.10\\
    \hline
  \end{tabular}%
  }
 \end{table*}

\begin{figure*}
\begin{multicols}{3}
    \includegraphics[width=\linewidth]{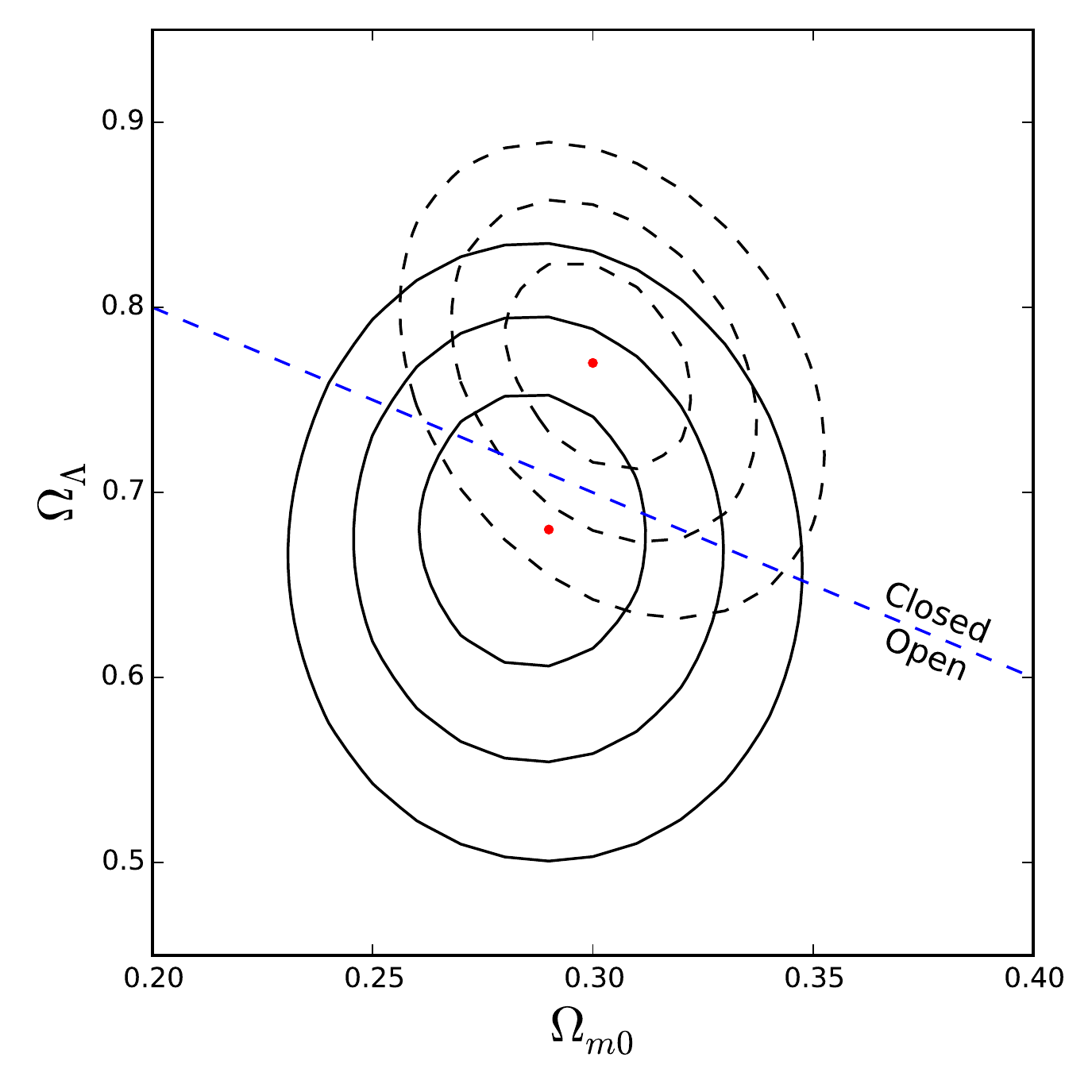}\par 
    \includegraphics[width=\linewidth]{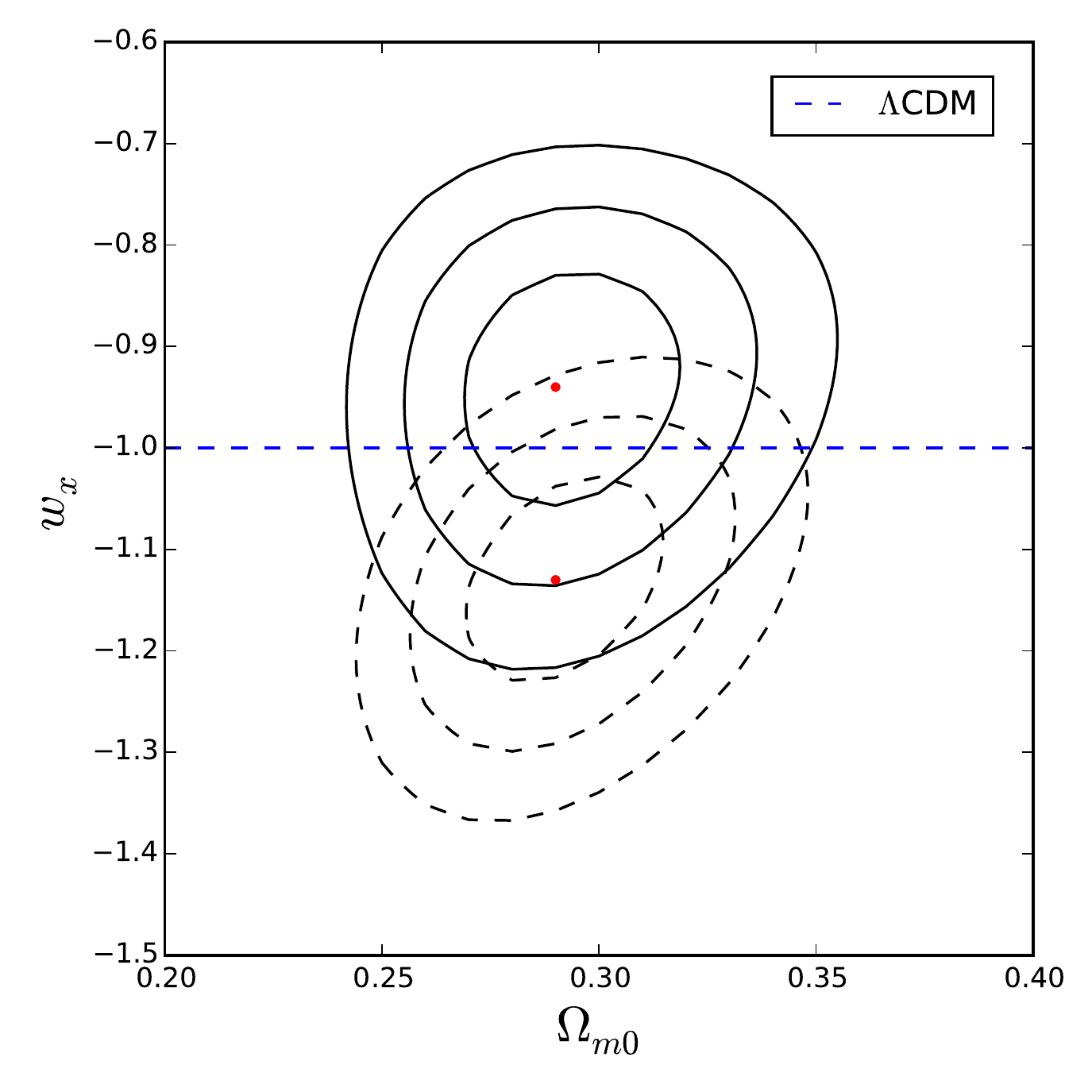}\par
    \includegraphics[width=\linewidth]{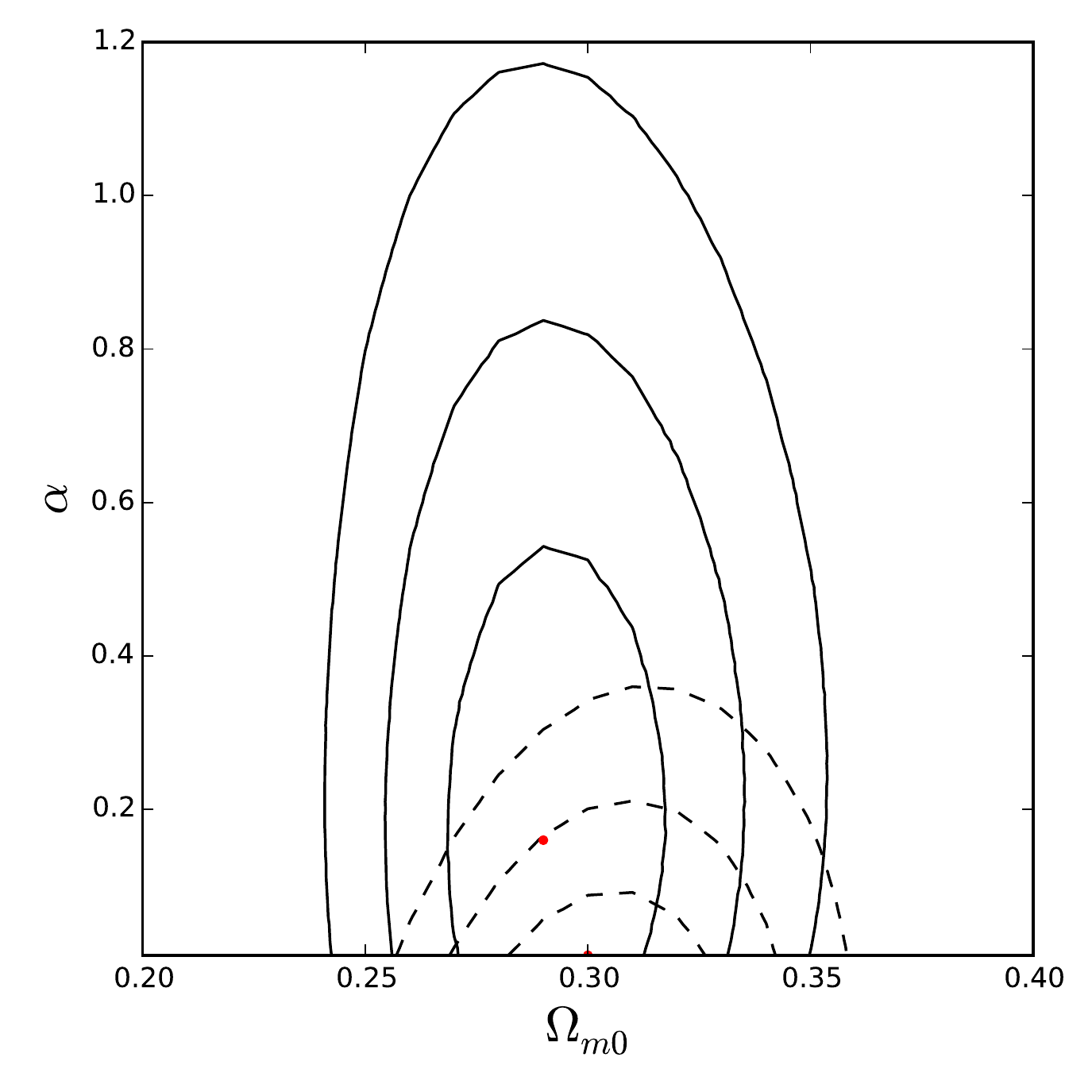}\par
\end{multicols}
\caption[Confidence contours for 2-parameter models.]{Confidence contours for 2-parameter models. Solid (dashed) 1, 2, and 3$\sigma$ contours correspond to $\bar{H}_0 \pm \sigma_{H_0} = 68 \pm 2.8 \hspace{1mm}(73.24 \pm 1.74)$ km s$^{-1}$ Mpc$^{-1}$ prior, and the red dots indicate the location of the best-fit point in each prior case. Left: $\Lambda$CDM. The blue dashed line indicates the spatially flat $\Lambda$CDM model; points above (below) the line correspond to models with closed (open) spatial hypersurfaces. Center: flat XCDM. The blue dashed line (for which $w_{\rm X} = -1$) demarcates the flat $\Lambda$CDM case. Right: flat $\phi$CDM. The horizontal $\alpha = 0$ axis corresponds to the flat $\Lambda$CDM model.}
\label{fig. 1}
\end{figure*}

\begin{figure*}
\begin{multicols}{3}
    \includegraphics[width=\linewidth]{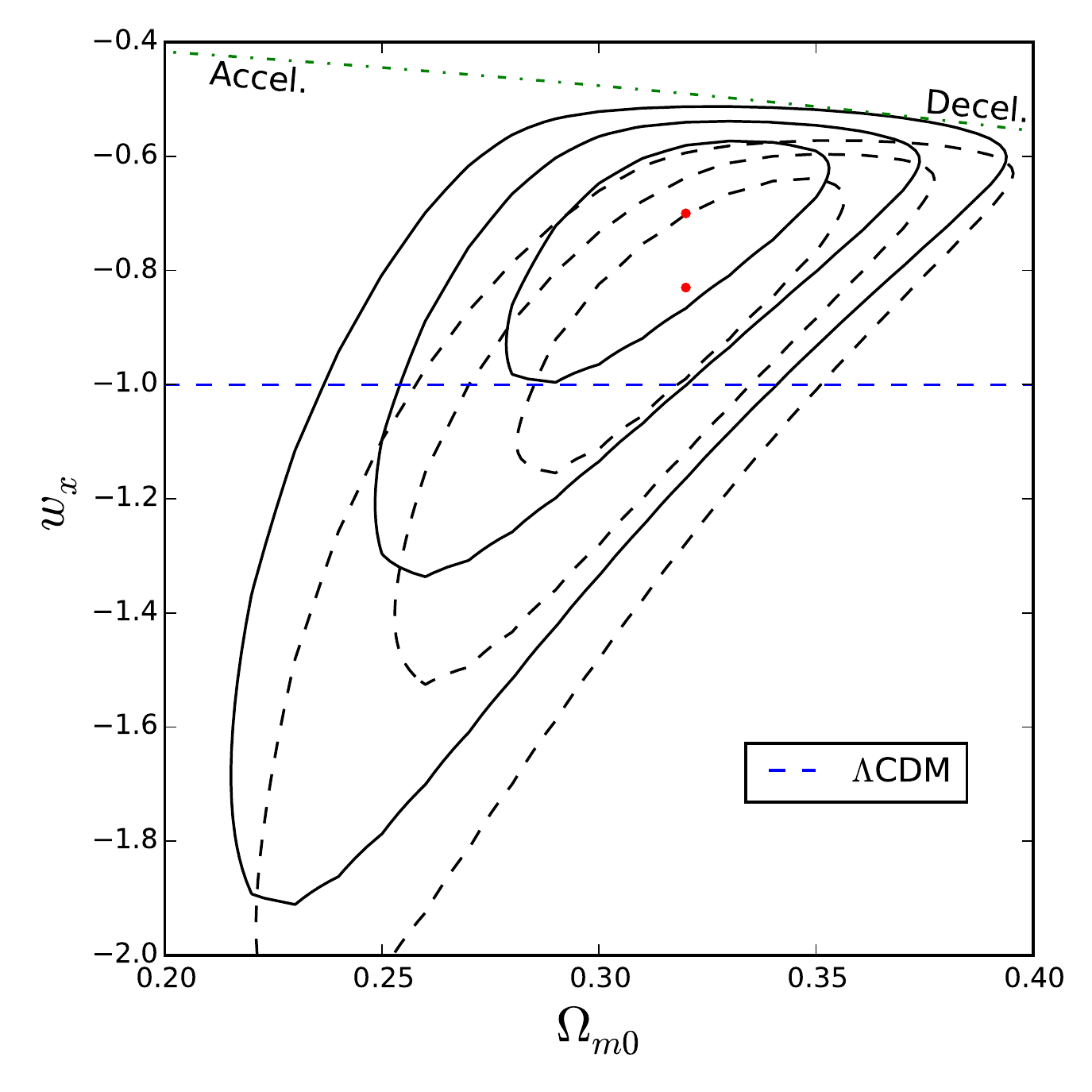}\par 
    \includegraphics[width=\linewidth]{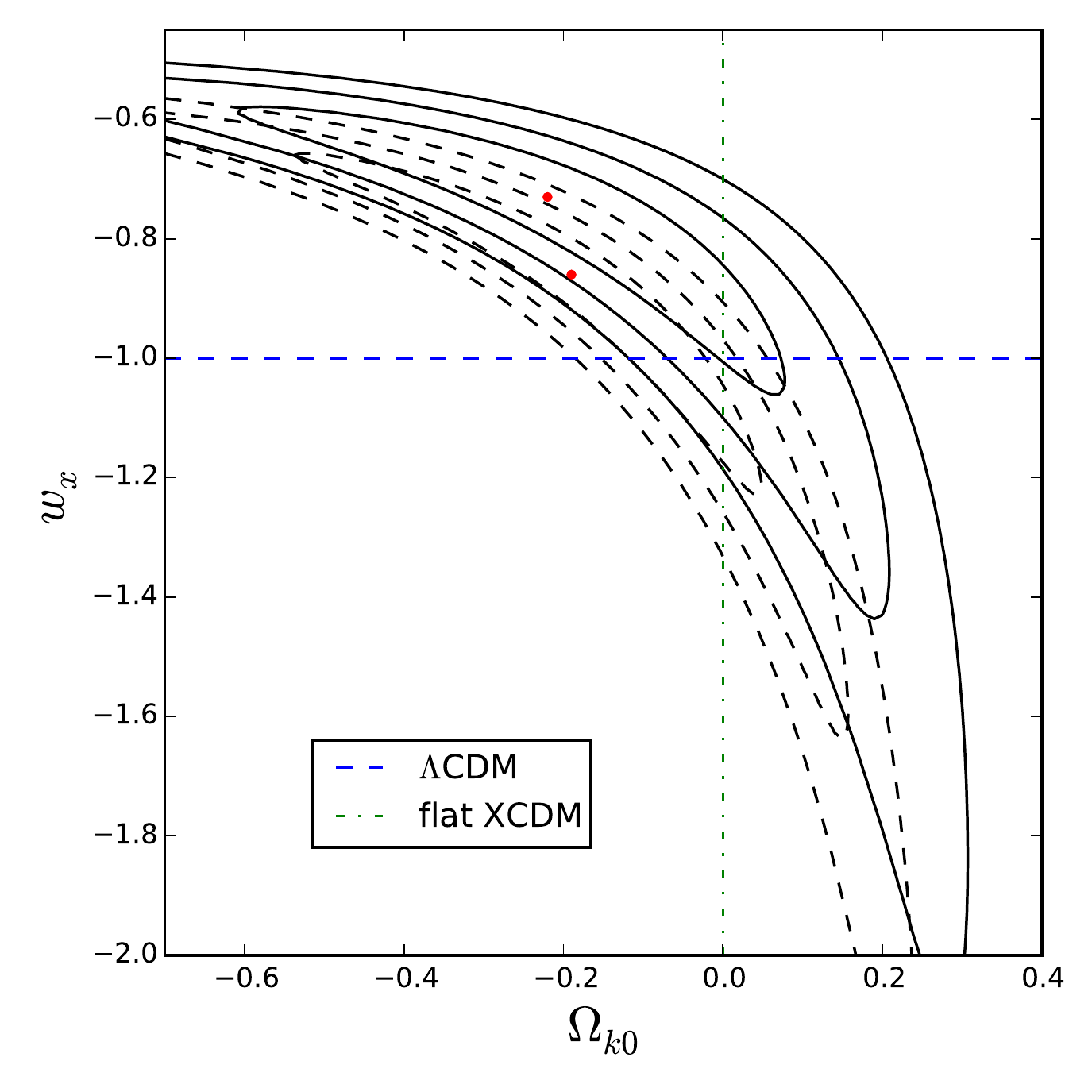}\par
    \includegraphics[width=\linewidth]{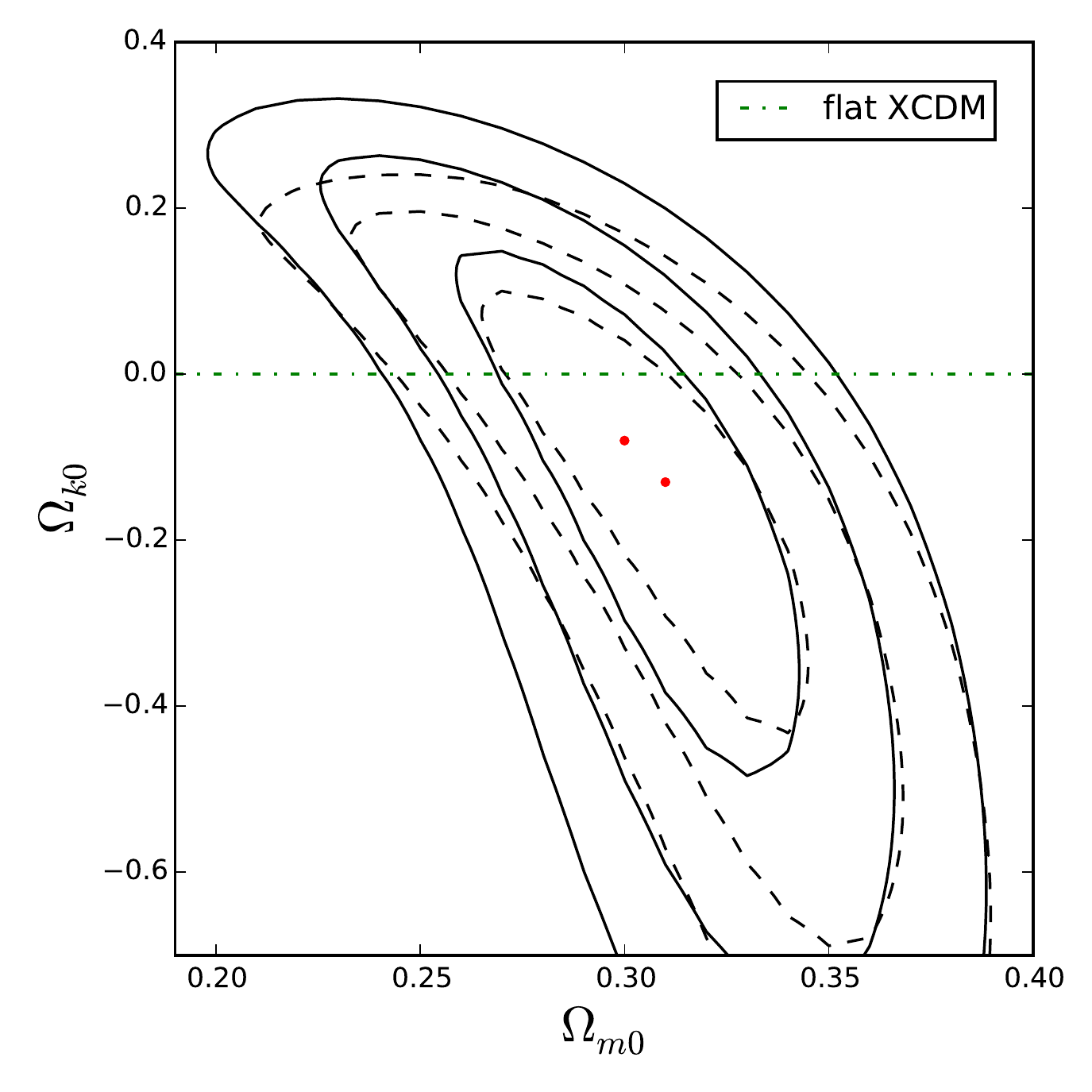}\par
\end{multicols}
\caption[Confidence contours for 3-parameter XCDM.]{Confidence contours for 3-parameter XCDM. Solid (dashed) 1, 2, and 3$\sigma$ contours correspond to $\bar{H}_0 \pm \sigma_{H_0} = 68 \pm 2.8 \hspace{1mm}(73.24 \pm 1.74)$ km s$^{-1}$ Mpc$^{-1}$ prior, and the red dots indicate the location of the best-fit point in each prior case. Left: $\Omega_{k0}$ marginalized. The blue dashed line indicates the $\Lambda$CDM model. Points above (below) the green dot-dashed curve near the top of the panel correspond to models with late-time (decelerating) accelerating expansion. Center: $\Omega_{m0}$ marginalized. The horizontal blue dashed line (for which $w_{\rm X} = -1$) demarcates the $\Lambda$CDM case, and the vertical green dot-dashed line demarcates the spatially flat XCDM case. Right: $w_X$ marginalized. The horizontal green dot-dashed line indicates the spatially flat XCDM case.}
\label{fig. 2}
\end{figure*}

\begin{figure*}
\begin{multicols}{3}
    \includegraphics[width=\linewidth]{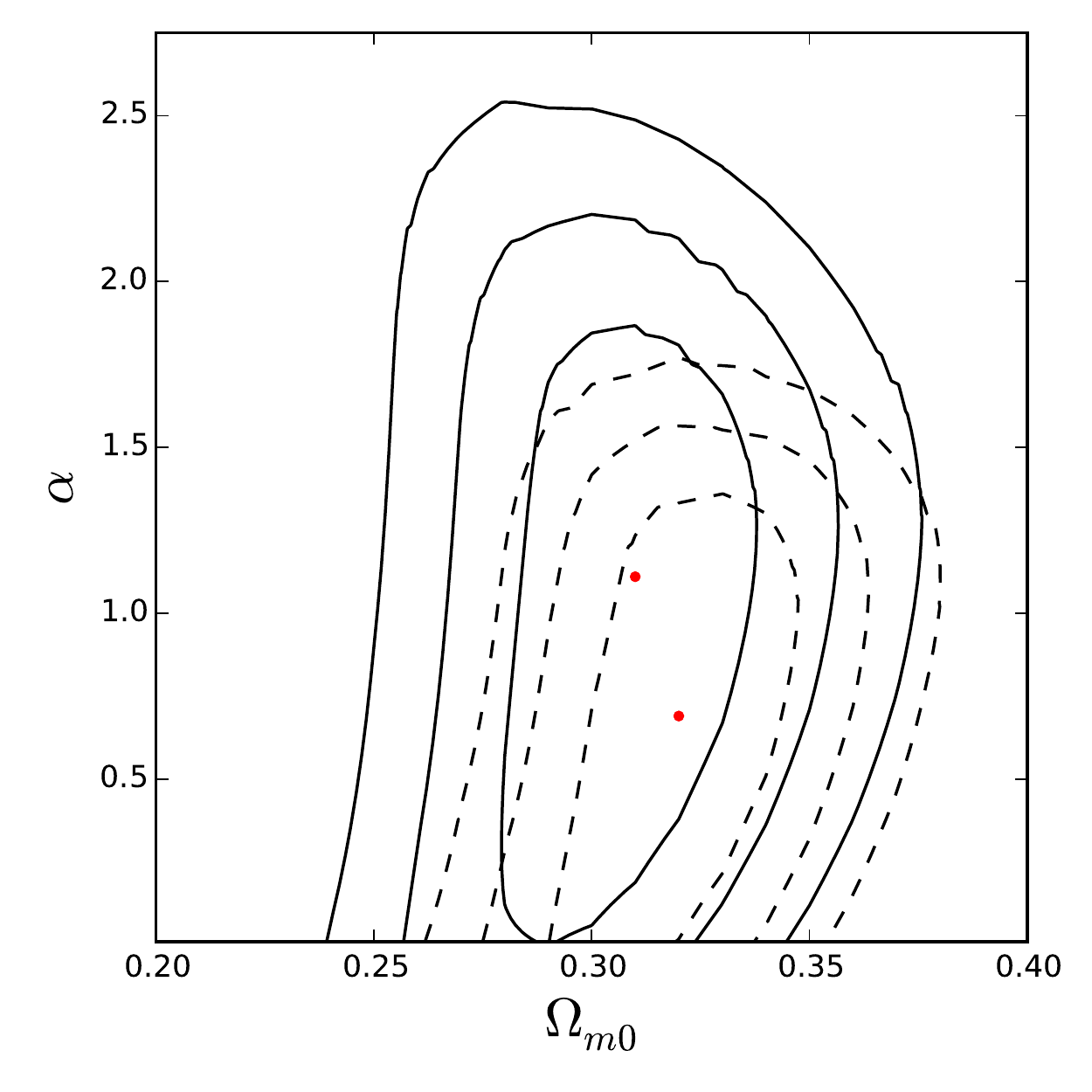}\par 
    \includegraphics[width=\linewidth]{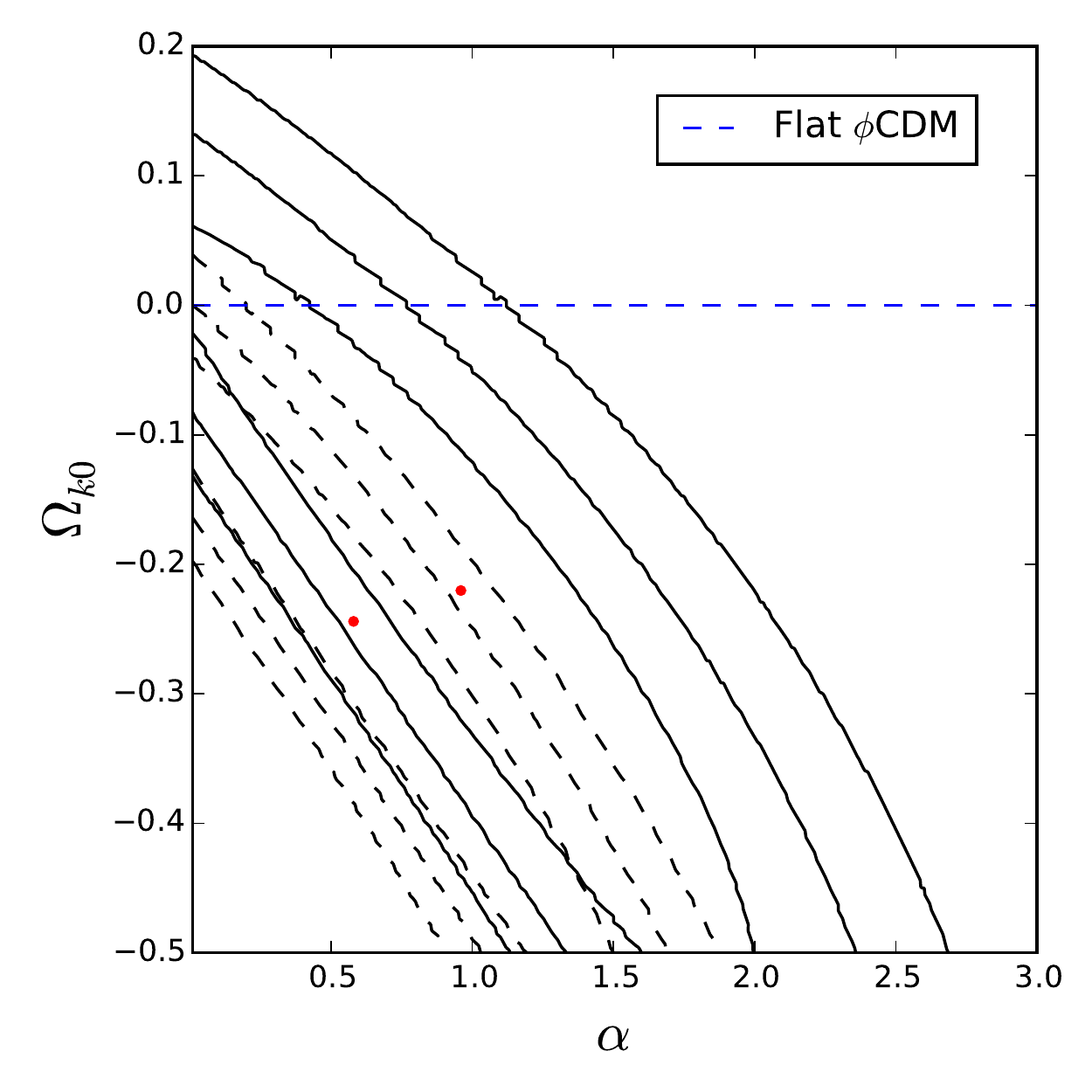}\par
    \includegraphics[width=\linewidth]{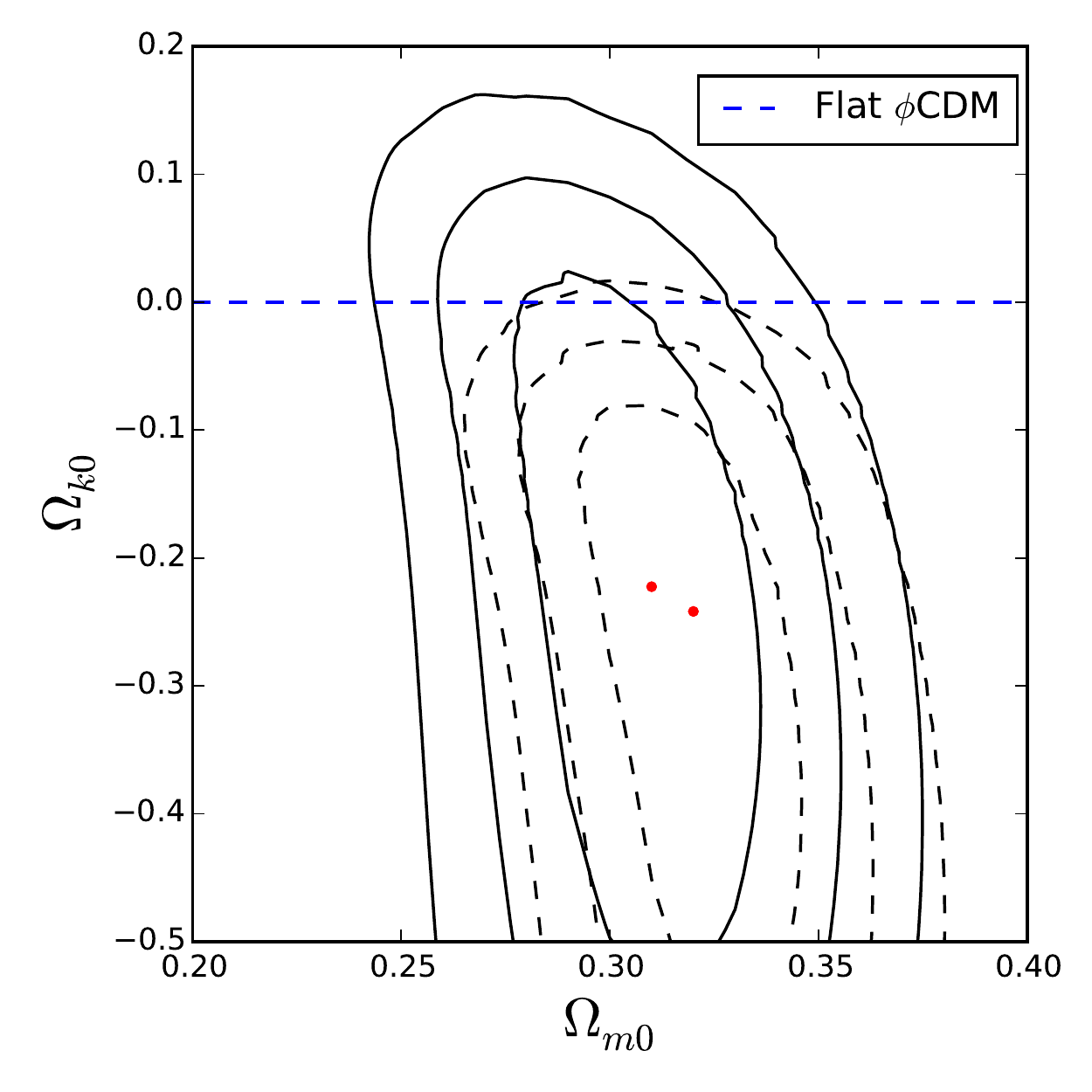}\par
\end{multicols}
\caption[Confidence contours for 3-parameter $\phi$CDM.]{Confidence contours for 3-parameter $\phi$CDM. Solid (dashed) 1, 2, and 3$\sigma$ contours correspond to $\bar{H}_0 \pm \sigma_{H_0} = 68 \pm 2.8 \hspace{1mm}(73.24 \pm 1.74)$ km s$^{-1}$ Mpc$^{-1}$ prior, and the red dots indicate the location of the best-fit point in each prior case. Left: $\Omega_{k0}$ marginalized. The horizontal $\alpha = 0$ axis corresponds to the $\Lambda$CDM model. Center: $\Omega_{m0}$ marginalized. The vertical $\alpha = 0$ axis corresponds to the $\Lambda$CDM model and the horizontal blue dashed line here and in the next panel correspond to the spatially flat $\phi$CDM case. Right: $\alpha$ marginalized.}
\label{fig. 3}
\end{figure*}

The confidence contours for the models we considered are shown in Figs. \ref{fig. 1}, \ref{fig. 2}, and \ref{fig. 3}. The solid black contours indicate the $\bar{H}_0 = 68 \pm 2.8$ km s$^{-1}$ Mpc$^{-1}$ prior constraints, the dashed black contours indicate the $H_0 = 73.24 \pm 1.74$ km s$^{-1}$ Mpc$^{-1}$ prior constraints, and the red dots indicate the best-fit point in each prior case. Our results for the parameter values of the unmarginalized and marginalized cases are collected in Tables \ref{table 3}-\ref{table 6}, along with their $\chi^2$, AIC, and BIC values. Wherever $\Delta \chi^2$, $\Delta$AIC, and $\Delta$BIC are given, these are computed relative to the $\chi^2$, AIC, and BIC of the corresponding $\Lambda$CDM model of each prior case.

 \begin{table}
  \caption[Best-fit values for 3-parameter models.]{Best-fit values for 3-parameter models. $\Delta \chi^2$, $\Delta$AIC, and $\Delta$BIC are evaluated relative to $\chi^2$, AIC, and BIC of $\Lambda$CDM for each $H_0$ prior.}
  \label{table 4}
  \centering
  \resizebox{\columnwidth}{!}{%
  \begin{tabular}{cccccccccccc}
    \hline
    $H_0$ prior $\left({\rm km}\hspace{1mm}{\rm s}^{-1}{\rm Mpc}^{-1}\right)$ & Model & $\Omega_{m0}$ & $\Omega_{k0}$ & $w_X$ & $\alpha$ & $\chi^2$ & $\Delta \chi^2$ & AIC & $\Delta$AIC & BIC & $\Delta$BIC\\
    \hline
    $68 \pm 2.8$ & XCDM & 0.31 & -0.18 & -0.76 & - & 23.65 & -1.70 & 29.65 & 0.30 & 34.86 & 2.03\\
     & $\phi$CDM & 0.31 & -0.22 & - & 0.96 & 23.82 & -1.53 & 29.82 & 0.47 & 35.03 & 2.20\\
    \hline
    $73.24 \pm 1.74$ & XCDM & 0.32 & -0.21 & -0.84 & - & 26.48 & -0.44 & 32.48 & 1.56 & 37.69 & 3.29\\
     & $\phi$CDM & 0.32 & -0.26 & - & 0.62 & 26.30 & 0.95 & 32.30 & 1.38 & 37.51 & 3.11\\
    \hline
  \end{tabular}%
  }
 \end{table}

In the 2-parameter case, the spatially flat XCDM parametrization has the lowest value of $\chi^2$ if the prior on $H_0$ is chosen to be $\bar{H}_0 = 68 \pm 2.8$ km s$^{-1}$ Mpc$^{-1}$. If, on the other hand, the $H_0$ prior is chosen to be $\bar{H}_0 = 73.24 \pm 1.74$ km s$^{-1}$ Mpc$^{-1}$ then the spatially flat $\Lambda$CDM model has the lowest value of $\chi^2$. These models also have lower AIC and BIC values than the 3-parameter XCDM parametrization and the 3-parameter $\phi$CDM model (see Tables \ref{table 3} and \ref{table 4}). On the other hand, the 3-parameter models typically have a lower $\chi^2$ than the 2-parameter $\Lambda$CDM case. These differences, however, are not statistically significant. Focusing on the $\bar{H}_0 = 68 \pm 2.8$ km s$^{-1}$ Mpc$^{-1}$ prior case, the $\chi^2$ differences indicate that the non-flat $\phi$CDM model and non-flat XCDM parametrization provide a 1.2$\sigma$ and 1.3$\sigma$ better fit to the data, respectively, while from $\Delta$AIC we find that these two models are 79\% and 86\% as probable as the 2-parameter $\Lambda$CDM model, respectively.

\begin{table*}
  \caption{1$\sigma$ and 2$\sigma$ parameter intervals for 2-parameter models.}
  \label{table 5}
  \centering
  \resizebox{\columnwidth}{!}{%
  \begin{tabular}{cccccc}
    \hline 
    $H_0$ prior $\left({\rm km}\hspace{1mm}{\rm s}^{-1}{\rm Mpc}^{-1}\right)$ & Model & Marginalization range & Best-fit & $1\sigma$ & $2\sigma$\\
    \hline \vspace{2pt}
     $68\pm 2.8$ & $\Lambda$CDM & $0 \leq \Omega_{\Lambda0} \leq 1$ & $\Omega_{m0} = 0.29$ & $0.27 \leq \Omega_{m0} \leq 0.31$ & $0.26 \leq \Omega_{m0} \leq 0.32$\\
     & & $0.01 \leq \Omega_{m0} \leq 1$ & $\Omega_{\Lambda0} = 0.68$ & $0.63 \leq \Omega_{\Lambda0} \leq 0.73$ & $0.58 \leq \Omega_{\Lambda0} \leq 0.77$\\
     \hline \vspace{2pt}
      & flat XCDM & $-2 \leq w_X \leq 0$ & $\Omega_{m0} = 0.29$ & $0.28 \leq \Omega_{m0} \leq 0.31$ & $0.26 \leq \Omega_{m0} \leq 0.33$\\
     &  & $0.01 \leq \Omega_{m0} \leq 1$ & $w_X = -0.94$ & $-1.02 \leq w_X \leq -0.87$ & $-1.10 \leq w_X \leq -0.80$\\ 
     \hline \vspace{2pt}
      & flat $\phi$CDM & $0.01 \leq \alpha \leq 5$ & $\Omega_{m0} = 0.29$ & $0.28 \leq \Omega_{m0} \leq 0.31$ & $0.26 \leq \Omega_{m0} \leq 0.33$\\
     &  & $0.01 \leq \Omega_{m0} \leq 1$ & $\alpha = 0.16$ & $0.06 \leq \alpha \leq 0.43$ & $0.02 \leq \alpha \leq 0.72$\\ 
    \hline \hline \vspace{2pt}
     $73.24 \pm 1.74$ & $\Lambda$CDM & $0 \leq \Omega_{\Lambda0} \leq 1$ & $\Omega_{m0} = 0.30$ & $0.29 \leq \Omega_{m0} \leq 0.32$ & $0.27 \leq \Omega_{m0} \leq 0.33$\\
     &  & $0.01 \leq \Omega_{m0} \leq 1$ & $\Omega_{\Lambda0} = 0.77$ & $0.73 \leq \Omega_{\Lambda0} \leq 0.81$ & $0.69 \leq \Omega_{\Lambda0} \leq 0.84$\\ 
     \hline \vspace{2pt}
     & flat XCDM & $-2 \leq w_X \leq 0$ & $\Omega_{m0} = 0.29$ & $0.28 \leq \Omega_{m0} \leq 0.31$ & $0.26 \leq \Omega_{m0} \leq 0.32$\\
     &  & $0.01 \leq \Omega_{m0} \leq 1$ & $w_X = -1.13$ & $-1.20 \leq w_X \leq -1.06$ & $-1.27 \leq w_X \leq -1.00$\\ 
     \hline \vspace{2pt}
     & flat $\phi$CDM & $0.01 \leq \alpha \leq 5$ & $\Omega_{m0} = 0.31$ & $0.29 \leq \Omega_{m0} \leq 0.32$ & $0.28 \leq \Omega_{m0} \leq 0.34$\\
     &  & $0.01 \leq \Omega_{m0} \leq 1$ & $\alpha = 0.01$ & $0.01 \leq \alpha \leq 0.09$ & $0.01 \leq \alpha \leq 0.20$\\ 
     \hline \vspace{2pt}
  \end{tabular}%
  }
 \end{table*}

In Table \ref{table 5} (\ref{table 6}), we list the 1$\sigma$ and 2$\sigma$ confidence intervals on the parameters of each of the 2-parameter (3-parameter) models. We obtained these by marginalizing the 2-parameter (3-parameter) likelihood function as described in Sec. \ref{sec:ch4_Methods}. The best-fit points in these tables correspond to the maximum value of the relevant one-dimensional marginalized likelihood function. Table \ref{table 3} (\ref{table 4}) lists the corresponding two-dimensional (three-dimensional) best-fit points.

\begin{table*}
  \caption{1$\sigma$ and 2$\sigma$ parameter intervals for 3-parameter models.}
  \label{table 6}
  \centering
  \resizebox{\columnwidth}{!}{%
  \begin{tabular}{cccccc}
    \hline 
    $H_0$ prior $\left({\rm km}\hspace{1mm}{\rm s}^{-1}{\rm Mpc}^{-1}\right)$ & Model & Marginalization range & Best-fit & $1\sigma$ & $2\sigma$\\
    \hline \vspace{2pt}
     $68\pm 2.8$ & XCDM & $-0.7 \leq \Omega_{k0} \leq 0.7$ & $\Omega_{m0} = 0.31$ & $0.28 \leq \Omega_{m0} \leq 0.33$ & $0.25 \leq \Omega_{m0} \leq 0.36$\\
     & & & $w_X = -0.70$ & $-0.93 \leq w_X \leq -0.62$ & $-1.27 \leq w_X \leq -0.57$\\
     \hline \vspace{2pt}
      &  & $0.01 \leq \Omega_{m0} \leq 1$ & $\Omega_{k0} = -0.11$ & $-0.36 \leq \Omega_{k0} \leq 0.06$ & $-0.59 \leq \Omega_{k0} \leq 0.19$\\
     & & & $w_X = -0.70$ & $-0.93 \leq w_X \leq -0.62$ & $-1.27 \leq w_X \leq -0.57$\\
     \hline \vspace{2pt}
     &  & $-2 \leq w_X \leq 0$ & $\Omega_{m0} = 0.31$ & $0.28\leq \Omega_{m0} \leq 0.33$ & $0.25 \leq \Omega_{m0} \leq 0.36$\\
     & & & $\Omega_{k0} = -0.11$ & $-0.36 \leq \Omega_{k0} \leq 0.06$ & $-0.59 \leq \Omega_{k0} \leq 0.19$\\
     \hline \vspace{2pt}
      & $\phi$CDM & $-0.5 \leq \Omega_{k0} \leq 0.5$ & $\Omega_{m0} = 0.31$ & $0.29 \leq \Omega_{m0} \leq 0.33$ & $0.27 \leq \Omega_{m0} \leq 0.35$\\
     &  & & $\alpha = 1.12$ & $0.51 \leq \alpha \leq 1.59$ & $0.11 \leq \alpha \leq 1.97$\\ 
     \hline \vspace{2pt}
    & & $0.01 \leq \Omega_{m0} \leq 1$ & $\Omega_{k0} = -0.22$ & $-0.38 \leq \Omega_{k0} \leq -0.07$ & $-0.48 \leq \Omega_{k0} \leq 0.03$\\
     &  & & $\alpha = 1.16$ & $0.53 \leq \alpha \leq 1.61$ & $0.12 \leq \alpha \leq 2.01$\\ 
    \hline \vspace{2pt}
    & & $0.01 \leq \alpha \leq 5$ & $\Omega_{m0} = 0.31$ & $0.29 \leq \Omega_{m0} \leq 0.33$ & $0.27 \leq \Omega_{m0} \leq 0.35$\\
     &  & & $\Omega_{k0} = -0.21$ & $-0.38 \leq \Omega_{k0} \leq -0.07$ & $-0.48 \leq \Omega_{k0} \leq 0.04$\\ 
     \hline \hline \vspace{2pt}
     $73.24\pm 1.74$ & XCDM & $-0.7 \leq \Omega_{k0} \leq 0.7$ & $\Omega_{m0} = 0.32$ & $0.29 \leq \Omega_{m0} \leq 0.34$ & $0.26 \leq \Omega_{m0} \leq 0.36$\\
     & & & $w_X = -0.82$ & $-1.08 \leq w_X \leq -0.71$ & $-1.44 \leq w_X \leq -0.63$\\
     \hline \vspace{2pt}
      &  & $0.01 \leq \Omega_{m0} \leq 1$ & $\Omega_{k0} = -0.11$ & $-0.33 \leq \Omega_{k0} \leq 0.03$ & $-0.55 \leq \Omega_{k0} \leq 0.14$\\
     & & & $w_X = -0.82$ & $-1.08 \leq w_X \leq -0.71$ & $-1.44 \leq w_X \leq -0.63$\\
     \hline \vspace{2pt}
     &  & $-2 \leq w_X \leq 0$ & $\Omega_{m0} = 0.32$ & $0.29 \leq \Omega_{m0} \leq 0.34$ & $0.26 \leq \Omega_{m0} \leq 0.36$\\
     & & & $\Omega_{k0} = -0.11$ & $-0.33 \leq \Omega_{k0} \leq 0.03$ & $-0.55 \leq \Omega_{k0} \leq 0.14$\\
     \hline \vspace{2pt}
      & $\phi$CDM & $-0.5 \leq \Omega_{k0} \leq 0.5$ & $\Omega_{m0} = 0.32$ & $0.30 \leq \Omega_{m0} \leq 0.34$ & $0.29 \leq \Omega_{m0} \leq 0.35$\\
     &  & & $\alpha = 0.76$ & $0.31 \leq \alpha \leq 1.14$ & $0.06 \leq \alpha \leq 1.41$\\ 
     \hline \vspace{2pt}
    & & $0.01 \leq \Omega_{m0} \leq 1$ & $\Omega_{k0} = -0.25$ & $-0.40 \leq \Omega_{k0} \leq -0.14$ & $-0.48 \leq \Omega_{k0} \leq -0.07$\\
     & & & $\alpha = 0.79$ & $0.32 \leq \alpha \leq 1.16$ & $0.06 \leq \alpha \leq 1.43$\\ 
    \hline \vspace{2pt}
    & & $0.01 \leq \alpha \leq 5$ & $\Omega_{m0} = 0.32$ & $0.30 \leq \Omega_{m0} \leq 0.34$ & $0.29 \leq \Omega_{m0} \leq 0.35$\\
     &  & & $\Omega_{k0} = -0.24$ & $-0.40 \leq \Omega_{k0} \leq -0.14$ & $-0.48 \leq \Omega_{k0} \leq -0.07$\\ 
     \hline \vspace{2pt}
  \end{tabular}%
  }
 \end{table*}

From the figures and tables, we see that the spatially flat $\Lambda$CDM model is a reasonable fit to the $H(z)$ and BAO data we use (although the flat XCDM parametrization and flat $\phi$CDM model provide slightly better fits in the \hublow case). In particular, from the figures, for the \hublow prior, flat $\Lambda$CDM is always within about 1$\sigma$ of the best-fit value. However, the \hubhigh case favors some larger deviations from flat \lcdm. For example in the middle panel of Fig. \ref{fig. 1} for the flat XCDM parametrization it favors a phantom model over flat \lcdm\ at a little more than 1$\sigma$, while in the center and right panels of Fig. \ref{fig. 3} for the non-flat \pcdm\ case it also favors a closed model at a little more than 2$\sigma$. Similar conclusions may be drawn from the parameter limits listed in Tables \ref{table 5} and \ref{table 6}.

When both dynamical dark energy and spatial curvature are present (as opposed to cases with only dynamical dark energy or only spatial curvature) it is not as easy to constrain both parameters simultaneously. This can be seen by comparing the center and right panels of Fig. \ref{fig. 1} to the left panels of Figs. \ref{fig. 2} and \ref{fig. 3}, respectively. When spatial curvature is allowed to vary, the confidence contours in the 3-parameter XCDM parametrization and the $\phi$CDM model expand along the $w_X$ and $\alpha$ axes (these are the parameters that govern the dynamics of the dark energy).

The consensus model, spatially flat $\Lambda$CDM, is consistent with current $H(z)$ + BAO data, but these data allow some nonzero spatial curvature. In particular, we find that the best-fit values of the parameters in the $\Lambda$CDM model imply a curvature energy density parameter of $\Omega_{k0} = 0.03$ for the $\bar{H}_0 \pm \sigma_{H_0} = 68 \pm 2.8$ km s$^{-1}$ Mpc$^{-1}$ prior case, and $\Omega_{k0} = -0.07$ for the $\bar{H}_0 \pm \sigma_{H_0} = 73.24 \pm 1.74$ km s$^{-1}$ Mpc$^{-1}$ prior case. More precisely, using the $\Omega_{m0}$ and $\Omega_{\Lambda}$ best-fit values and error bars for flat \lcdm\ from Table \ref{table 5}, and combining the errors in quadrature, an approximate estimate is $\Omega_{k0} = 0.03(1 \pm 1.8)$ and $\Omega_{k0} = -0.07(1 \pm 0.59)$ for the \hublow and \hubhigh priors, with the data favoring a closed model at a little over 1$\sigma$ in the second case. The 3-parameter models, in both prior cases, favor closed spatial hypersurfaces, but the error bars are so large that these results only stand out in the \hubhigh prior case of the \pcdm\ model (see the center and right panels of \ref{fig. 3}). While not very statistically significant, we note that these results are not inconsistent with those of \cite{Ooba_Ratra_Sugiyama_2017_NFXCDM, Ooba_Ratra_Sugiyama_2017_NFXCDM, Ooba_Ratra_Sugiyama_2017_NFpCDM} and \cite{Park_Ratra_2018_FXCDM_NFXCDM, Park_Ratra_2018_FLCDM}, who found that CMB anisotropy data, in conjunction with other cosmological data, were not inconsistent with mildly closed spatial hypersurfaces.

The current data are also not inconsistent with some mild dark energy dynamics, although the size of the effect varies depending on the choice of $H_0$ prior and whether or not $\Omega_{k0}$ is allowed to vary as a free parameter. In the flat \pcdm\ model, for instance, $\alpha$ can be different from zero only in the $\bar{H}_0 \pm \sigma_{H_0} = 68 \pm 2.8$ km s$^{-1}$ Mpc$^{-1}$ prior case, whereas $\alpha$ can be different from zero in both prior cases if $\Omega_{k0}$ is allowed to vary (see the right panel of \ref{fig. 1} and the left panel of \ref{fig. 3}).

\section{Conclusions}
\label{sec:ch4_Conclusion}
We analyzed a total of 42 measurements, 31 of which consisted of uncorrelated $H(z)$ data points, with the remainder coming from BAO observations (some correlated, some not), to constrain dark energy dynamics and spatial curvature, by determining how well these measurements can be described by three common models of dark energy: $\Lambda$CDM, the XCDM parametrization, and $\phi$CDM. 

The consensus flat \lcdm\ model is in reasonable agreement with these data, but depending on the model analyzed and the $H_0$ prior used, it can be a little more than 1$\sigma$ away from the best-fit model. These data are consistent with mild dark energy dynamics as well as non-flat spatial hypersurfaces. While these results are interesting and encouraging, more and better data are needed before we can make definitive statements about the spatial curvature of the universe and about dark energy dynamics.

%% file: chapter5.tex
\cleardoublepage

\chapter{Baryon acoustic oscillation, Hubble parameter, and angular size measurement constraints on the Hubble constant, dark energy dynamics, and spatial curvature}\chaptermark{Constraints from QSO, $H(z)$, and BAO data}

\label{Chapter5}

This chapter is based on \cite{Ryan_Chen_Ratra_2019}.

\section{Introduction}
\label{sec:Introduction}
\begin{comment}
[The universe is currently undergoing accelerated cosmological expansion. The simplest cosmological model compatible with this acceleration is the standard \lcdm\ model \citep{peeb84}, in which the acceleration is powered by a spatially-homogeneous energy density that is constant in time (a cosmological constant $\Lambda$). The standard \lcdm\ model is consistent with many observational constraints \citep{Alam_et_al_2017, Farooq_Ranjeet_Crandall_Ratra_2017, scolnic_et_al_2018, planck2018} if $\Lambda$ contributes about 70\% of the current energy density budget with cold dark matter (CDM) being the next largest contributor, at a little more than 25\%.
\end{comment}

As discussed in the last chapter, the standard \lcdm\ model assumes flat spatial hypersurfaces. It has been argued that cosmic microwave background (CMB) anisotropy measurements show that spatial hypersurfaces are very close to being flat, but the recent \cite{planck_2016} and \cite{planck2018} CMB anisotropy data analyses in the non-flat case are based on a somewhat arbitrary primordial power spectrum for spatial inhomogeneities. A physically consistent primordial inhomogeneity energy density power spectrum can be generated by inflation, and non-flat inflation models exist which can be used to compute such a power spectrum (for the models, see \citealp{gott_1982}, \citealp{hawking_1984}, and \citealp{ratra_1985}; for the power spectra, see \citealp{ratra_peebles_1995} and \citealp{ratra_2017}).\footnote{These non-flat inflation models are slow-roll models, so quantum mechanical fluctuations during inflation in these models result in an untilted primordial power spectrum. It is possible that these power spectra are too simple, but they are physically consistent; it is not known if the power spectrum used in the Planck non-flat CMB analyses are physically consistent.}

When these power spectra \citep{ratra_peebles_1995, ratra_2017} are used in a non-flat \lcdm\ model analysis of CMB anisotropy data (\citealp{planck_2016}) and a large compilation of non-CMB data \citep{Ooba_Ratra_Sugiyama_2017_NFLCDM, Park_Ratra_2018_FLCDM}, a mildly closed \lcdm\ model with $\sim$1\% spatial curvature contribution to the current cosmological energy budget is favored at over 5$\sigma$. A similar spatial curvature contribution is favored in dynamical dark energy XCDM and \pcdm\ models (in which dark energy is modelled as an X-fluid and scalar field, respectively; see \citealp{Ooba_Ratra_Sugiyama_2017_NFpCDM, Ooba_Ratra_Sugiyama_2017_NFXCDM}, \citealp{Park_Ratra_2018_FXCDM_NFXCDM, Park_Ratra_2018_FpCDM_NFpCDM}). These closed models provide better fits to the low multipole CMB anisotropy data, but the flat models are in better agreement with the higher multipole CMB anisotropy data. The non-flat models are in better agreement with weak lensing measurements, but do a worse job fitting higher redshift cosmic reionization data \citep{mitra_choudhury_ratra_2018, mitra_park_choudhury_ratra_2019} and deuterium abundance measurements \citep{penton_peyton_zahoor_ratra_2018}.\footnote{Overall the standard tilted flat \lcdm\ model has a lower total $\chi^2$ than the non-flat models, lower by $\Delta \chi^2 \sim 10$-20, depending on the data compilation and non-flat model used. However, the tilted flat \lcdm\ model is not nested inside any of the three untilted non-flat models, so it is not possible to convert these $\Delta \chi^2$ values to relative goodness-of-fit probabilities (\citealp{Ooba_Ratra_Sugiyama_2017_NFLCDM, Ooba_Ratra_Sugiyama_2017_NFpCDM, Ooba_Ratra_Sugiyama_2017_NFXCDM}; \citealp{Park_Ratra_2018_FXCDM_NFXCDM, Park_Ratra_2018_FLCDM, Park_Ratra_2018_FpCDM_NFpCDM}).}

It has also been found that spatially-flat dynamical dark energy XCDM and \pcdm\ models provide slightly better overall fits (lower total $\chi^2$) to the current data than does flat \lcdm\ (in the best-fit versions of these models the dark energy density has only very mild time dependence; see \citealp{Ooba_Ratra_Sugiyama_2018_FpCDM}, \citealp{Park_Ratra_2018_FXCDM_NFXCDM, Park_Ratra_2018_FpCDM_NFpCDM}, and \citealp{sola_gomez_perez_2019}).\footnote{For studies of other spatially-flat dynamical dark energy models that fit the data better than does flat \lcdm, see \cite{41}, \cite{wang_pogosian_zhao_zucca_2018}, and \cite{zhang_lee_geng_2018}.}

The constraints on spatial curvature and dark energy dynamics discussed above make use of CMB anisotropy data, which requires the assumption of a primordial spatial inhomogeneity power spectrum. As mentioned above, the only currently known physically motivated power spectra in non-flat models are untilted power spectra generated by slow-roll inflation. Such power spectra might not be general enough, so the CMB anisotropy data constraints on spatial curvature derived using these power spectra could be misleading. It is therefore of great importance to constrain spatial curvature and dark energy dynamics using non-CMB data that does not require the assumption of a primordial spatial inhomogeneity power spectrum. For recent studies along these lines, see Chapter \ref{Chapter4}, as well as \cite{Farooq_Mania_Ratra_2015}, \cite{Chen_et_al_2016}, \cite{yu_wang_2016}, \cite{Farooq_Ranjeet_Crandall_Ratra_2017}, \cite{wei_wu_2017}, \cite{rana_jain_mahajan_mukherjee_2017}, \cite{60}, \cite{Qi_et_al2018}, \cite{park_ratra_2019b}, \cite{mukherjee_paul_jassal_2019}, \cite{DES_2019}, \cite{Zheng_2019}, and \cite{ruan_etal_2019}.\footnote{For possible constraints on spatial curvature from future data, see \cite{witzemann_et_al_2018} and \cite{wei_2018}.}

In Chapter \ref{Chapter4} we used Hubble parameter and baryon acoustic oscillation (BAO) measurements to constrain spatial curvature and dark energy dynamics.\footnote{Hubble parameter data span a large enough redshift range to be able to detect and study the transition from early matter dominated cosmological deceleration to the current dark energy dominated accelerated expansion (see, e.g., \citealp{Farooq_Ratra_2013, farooq_crandall_ratra_2013, 68, Farooq_Ranjeet_Crandall_Ratra_2017, jesus_holanda_pereira_2018, gomez_valent_2018}). For other uses of Hubble parameter data, see \cite{chen_ratra_2011b}, \cite{chen_geng_cao_huang_zhu_2015}, \cite{anagnostopoulos_2018}, \cite{mamon_bamba_2018}, \cite{geng_et_al_2018}, and \cite{liu_et_al_2018}.} Here we improve upon and extend the analyses of that chapter. To do this, we:
\begin{itemize}
  \item Consider a sixth cosmological model, flat \lcdm.
  \item Update our BAO measurements.
  \item More accurately compute the size of the sound horizon at the drag epoch for the BAO constraints.
  \item Treat the Hubble constant $H_0$ as an adjustable parameter to be determined by the data we use.
  \item Use milliarcsecond quasar angular size versus redshift data \citep{Cao_et_al2017b}, alone and in combination with $H(z)$ and BAO data, to constrain cosmological parameters.
\end{itemize}
We note that, in our analyses here, we make use of the baryon density determined from the Planck 2015 TT + lowP + lensing CMB anisotropy data \citep{planck_2016}, as computed in each of the six cosmological models we consider by \cite{Park_Ratra_2018_FXCDM_NFXCDM, Park_Ratra_2018_FLCDM, Park_Ratra_2018_FpCDM_NFpCDM}, in order to calibrate the scale of the BAO sound horizon $r_{\rm s}$ (which scale is necessary for the computation of distances from BAO data; see below). This means that the constraints we obtain from the BAO data are not completely independent of the Planck 2015 CMB anisotropy data. That said, the baryon density determined from the CMB anisotropy data in the spatially flat models is very consistent with the baryon density determined from deuterium abundance measurements, although it is a little less consistent with these measurements in the non-flat models \citep{penton_peyton_zahoor_ratra_2018}.

The new data set that we incorporate in this chapter consists of measurements of quasar angular size from \cite{Cao_et_al2017b}.\footnote{For other angular size versus redshift data compilations and constraints, see \cite{daly_guerra_2002}, \cite{podariu_daly_mory_ratra_2003}, \cite{bonamente_2006}, and \cite{Chen_Ratra_2012}.} Measurements of the milliarcsecond-scale angular size of distant radio sources, from data compiled in \cite{gurvits_kellermann_frey_1999}, have been used in the past to constrain cosmological parameters; see \cite{vishwakarma_2001}, \cite{lima_alcaniz_2002}, \cite{zhu_fujimoto_2002}, and \cite{Chen_Ratra_2003}. There is, however, reason to doubt some of these earlier findings. Angular size measurements are only useful if radio sources are standard rulers, as accurate knowledge of the characteristic linear size $l_m$ of the ruler is necessary to convert measurements of the angular size distance into measurements of the angular size, and the estimates of $l_m$ used by \cite{vishwakarma_2001}, \cite{lima_alcaniz_2002}, and \cite{zhu_fujimoto_2002} were inaccurate. To account for the uncertainty in the characteristic linear size $l_m$, \cite{Chen_Ratra_2003} marginalized over $l_m$, finding only weak constraints on the cosmological parameters they studied from the angular size data. More recent studies, such as \cite{Cao_et_al2017a} and \cite{Cao_et_al2017b}, based on a sample of 120 intermediate-luminosity quasars recently compiled by \cite{Cao_et_al2017b}, have more precisely calibrated $l_m$, and these angular size versus redshift data have been used to constrain cosmological parameters \citep{Cao_et_al2017b, Li_et_al2017, Qi_et_al2017, Xu_et_al2018}. Here we use these data, in conjunction with $H(z)$ measurements and BAO distance measurements, to constrain cosmological parameters. We find that when the QSO angular size versus redshift data are used in conjunction with the $H(z)$ + BAO data combination, cosmological parameter constraints tighten a bit. We also confirm, as described below, that the QSO data have a large reduced $\chi^2 \sim 3$.

From the full data set, we measure a Hubble constant $H_0$ that is very consistent with the $H_0 = 68 \pm 2.8$ km s$^{-1}$ Mpc$^{-1}$ median statistics estimate \citep{chenratmed} but is a model-dependent 1.9$\sigma$ to 2.5$\sigma$ (from the quadrature sum of the error bars) lower than the local expansion rate measurement of $H_0 = 73.48 \pm 1.66$ km s$^{-1}$ Mpc$^{-1}$ \citep{riess2018}. In the non-flat \lcdm\ model these data are consistent with flat spatial hypersurfaces, while they favor closed geometry at 1.2$\sigma$ and 1.7$\sigma$ in the non-flat XCDM parametrization and non-flat \pcdm\ model, respectively. In some of dynamical dark energy models, both flat and non-flat, these data favor dark energy dynamics over a $\Lambda$ (up to a little more 2$\sigma$).
\begin{comment}
Another difference is that MCMC analysis uses a different algorithm to find the maximum of the likelihood distribution than what we used; we simply wrote Python [get the formatting right] scripts that calculated the likelihood distribution for each of the models we studied, and found the maximum by a direct search of the array in which the likelihood distribution was stored. This produces slightly different constraints [Is this true? Yun said something like this several months ago, but I'm not sure I'm entirely convinced; why should different algorithms yield different results, all else being equal?] [Also we kept $\Omega_b h^2$ fixed, while \cite{park_ratra_2019b} varied it, so our sound horizons are probably different]
\end{comment}
\begin{comment}
As in \cite{Ryan_Doshi_Ratra_2018}, we study spatially flat and non-flat $\Lambda$CDM, XCDM, and $\phi$CDM; see chapter \ref{Chapter3} for descriptions of these models. Sec. \ref{sec:ch5_data} summarizes the data we use, Sec. \ref{sec:methods} describes our analysis methods, Sec. \ref{sec:ch5_Results} describes our results, and we conclude in Sec. \ref{sec:ch5_conclusion}.
\end{comment}

\begin{table}
\centering
    \caption[Baryon densities.]{Baryon densities for the models we studied.}
    \begin{tabular}{c|c|c}
        \hline
        Model & $\Omega_b h^2$ & Ref.\\
        \hline
        Flat \lcdm & 0.02225 & \cite{Park_Ratra_2018_FLCDM}\\
        Nonflat \lcdm & 0.02305 & \cite{Park_Ratra_2018_FLCDM}\\
        Flat XCDM & 0.02229 & \cite{Park_Ratra_2018_FXCDM_NFXCDM}\\
        Nonflat XCDM & 0.02305 & \cite{Park_Ratra_2018_FXCDM_NFXCDM}\\
        Flat \pcdm & 0.02221 & \cite{Park_Ratra_2018_FpCDM_NFpCDM}\\
        Nonflat \pcdm & 0.02303 & \cite{Park_Ratra_2018_FpCDM_NFpCDM}\\
        \hline
    \end{tabular}
    \label{tab:Baryons}
\end{table}

\section{Data}
\label{sec:ch5_data}
We use a combination of 120 quasar angular size measurements ("QSO"), 31 expansion rate measurements ("$H(z)$"), and 11 baryon acoustic oscillation measurements ("BAO") to constrain our models. The $H(z)$ data, compiled in Table \ref{tab:H(z)_data}, are identical to the data used in Chapter \ref{Chapter4} (see that chapter for a discussion). The BAO data (see Table \ref{tab:ch5_BAO_data}) have been updated from Chapter \ref{Chapter4}; in that chapter we used the preprint value of the measurement from \cite{3}, while here we use the published version. Also, we have taken the first six measurements of Table \ref{tab:ch5_BAO_data} (and the covariance matrix of these measurements) directly from the SDSS website;\footnote{\url{https://sdss3.org/science/boss_publications.php}} in Chapter \ref{Chapter4} we did not use the full precision measurements. Additionally, our analysis of the BAO data in this paper differs from that of Chapter \ref{Chapter4}, as discussed below. The BAO measurements collected in Table \ref{tab:ch5_BAO_data} are expressed in terms of the transverse co-moving distance $D_{\rm M}(z)$ (eq. \ref{eq:D_M}), the Hubble distance $D_{\rm H}(z)$ (eq. \ref{eq:D_H}), the volume-averaged angular diameter distance $D_{\rm V}(z)$ (eq. \ref{eq:D_V}), and the line-of-sight co-moving distance $D_{\rm C}(z)$ (eq. \ref{eq:D_C}).

All measurements listed in Table \ref{tab:ch5_BAO_data} are scaled by the size of the sound horizon at the drag epoch $r_{\rm s}$. This quantity is \citetext{see \citealp{8} for a derivation}
\begin{equation}
    r_{\rm s} = \frac{2}{3k_{\rm eq}}\sqrt{\frac{6}{R_{\rm eq}}}{\rm ln}\left[\frac{\sqrt{1 + R_{\rm d}} + \sqrt{R_{\rm d} + R_{\rm eq}}}{1 + \sqrt{R_{\rm eq}}}\right],
\end{equation}
where $R_{\rm d} \equiv R(z_{\rm d})$ and $R_{\rm eq} \equiv R(z_{\rm eq})$ are the values of $R$, the ratio of the baryon to photon momentum density
\begin{equation}
    R = \frac{3\rho_b}{4\rho_{\gamma}},
\end{equation}
at the drag and matter-radiation equality redshifts $z_{\rm d}$ and $z_{\rm eq}$, respectively, and $k_{\rm eq}$ is the particle horizon wavenumber at the matter-radiation equality epoch. 

To compute $r_s$ as a function of our model parameters, we use the fitting formula presented in \cite{8}. This calculation also requires $\Omega_{b} h^2$ as input, and in Chapter \ref{Chapter4} we used $\Omega_b h^2 = 0.02227$ for all models considered. It is more accurate, however, to use the different values of $\Omega_b h^2$ computed by \cite{Park_Ratra_2018_FLCDM, Park_Ratra_2018_FXCDM_NFXCDM, Park_Ratra_2018_FpCDM_NFpCDM} for each model from the Planck 2015 TT + lowP + lensing CMB anisotropy data \citep{planck_2016}, because the values of $\Omega_{b}h^2$ estimated from CMB anisotropy data are model dependent, and vary significantly between the spatially-flat and non-flat inflation models \citep{Park_Ratra_2018_FLCDM}. The values of $\Omega_{b} h^2$ that we use are collected in Table \ref{tab:Baryons}, without their associated small uncertainties, which we do not account for in our analyses. Here we remind the reader that, because the values of $\Omega_b h^2$ in Table \ref{tab:Baryons} are computed from CMB data, our analysis of the BAO measurements in Table \ref{tab:ch5_BAO_data} (each of which requires a computation of the sound horizon, which in turn depends on $\Omega_b h^2$) is not completely independent of the CMB data. This should be borne in mind when comparing our $H_0$ measurements to local $H_0$ measurements (see \citealp{riess2018}, for example).

\begin{table*}
\caption[BAO data.]{BAO data. $D_M \left(r_{s,{\rm fid}}/r_s\right)$ and $D_V \left(r_{s,{\rm fid}}/r_s\right)$ have units of Mpc, while $H(z)\left(r_s/r_{s,{\rm fid}}\right)$ has units of ${\rm km}\hspace{1mm}{\rm s}^{-1}{\rm Mpc}^{-1}$ and $r_s$ and $r_{s, {\rm fid}}$ have units of Mpc. The uncertainty on the first six measurements is accounted for by the covariance matrix of eq. (\ref{covmat}).}
\centering
%\begin{threeparttable}
\begin{tabular}{ccccc}
\hline
$z$ & Measurement & Value & $\sigma$ & Ref.\\
\hline
$0.38$ & $D_M\left(r_{s,{\rm fid}}/r_s\right)$ & 1512.39 & - & \cite{Alam_et_al_2017}\\
\hline
$0.38$ & $H(z)\left(r_s/r_{s,{\rm fid}}\right)$ & 81.2087 & - & \cite{Alam_et_al_2017}\\
\hline
$0.51$ & $D_M\left(r_{s,{\rm fid}}/r_s\right)$ & 1975.22 & - & \cite{Alam_et_al_2017}\\
\hline
$0.51$ & $H(z)\left(r_s/r_{s,{\rm fid}}\right)$ & 90.9029 & - & \cite{Alam_et_al_2017}\\
\hline
$0.61$ & $D_M\left(r_{s,{\rm fid}}/r_s\right)$ & 2306.68 & - & \cite{Alam_et_al_2017}\\
\hline
$0.61$ & $H(z)\left(r_s/r_{s,{\rm fid}}\right)$ & 98.9647 & - & \cite{Alam_et_al_2017}\\
\hline
$0.106$ & $r_s/D_V$ & 0.336 & 0.015 & \cite{10}\\
\hline
$0.15$ & $D_V\left(r_{s,{\rm fid}}/r_s\right)$ & $664$ & $25$ & \cite{2}\\
\hline
$1.52$ & $D_V\left(r_{s,{\rm fid}}/r_s\right)$ & $3843$ & $147$ & \cite{3}\\
\hline
$2.33$ & $\frac{\left(D_H\right)^{0.7} \left(D_{M}\right)^{0.3}}{r_s}$ & 13.94 & 0.35 & \cite{9}\\
\hline
$2.36$ & $c/\left(r_s H(z)\right)$ & 9.0 & 0.3 & \cite{11}\\
\hline
\end{tabular}
%\begin{tablenotes}
%\item[a]$D_M \left(r_{s,{\rm fid}}/r_s\right)$ and $D_V \left(r_{s,{\rm fid}}/r_s\right)$ have units of Mpc, while $H(z)\left(r_s/r_{s,{\rm fid}}\right)$ has units of ${\rm km}\hspace{1mm}{\rm s}^{-1}{\rm Mpc}^{-1}$ and $r_s$ and $r_{s, {\rm fid}}$ have units of Mpc.
%\end{tablenotes}
%\end{threeparttable}
\label{tab:ch5_BAO_data}
\end{table*}

Our method of scaling the sound horizon here differs from the method used in Chapter \ref{Chapter4}. For studies that scale their measurements by $r_{s, \rm fid}/r_s$, we use the fitting formula of \cite{8} to compute both $r_s$ and $r_{s, \rm fid}$. For $r_{s, \rm fid}$ we use the parameters $(\Omega_{m0}, H_0, \Omega_b h^2)$ of the fiducial cosmology adopted in the paper in which the measurement is reported. For measurements scaled only by $r_s$, we again use the fitting formula of \cite{8}, but we modify it with a multiplicative scaling factor $147.60 {\rm \hspace{1mm} Mpc}/r_{s, \rm Planck}$, where 147.60 Mpc is the value of the sound horizon from Table 4, column 3 of \cite{planck_2016}, and $r_{s, \rm Planck}$ is the output of the sound horizon fitting formula from \cite{8} when it takes the best-fitting values of $(\Omega_{m0}, H_0, \Omega_b h^2)$ from \cite{planck_2016} as input.\footnote{We thank C.-G. Park for suggesting this.} We do this because the output of the fitting formula in \cite{8} deviates by a few per cent from CAMB's output; the scaling factor ensures that $r_s = 147.60 \hspace{1mm} \rm Mpc$ when $(\Omega_{m0}, H_0, \Omega_b h^2)$ take their best-fitting values found by \cite{planck_2016}. We believe that these modifications to the output of the fitting formula result in more accurate determinations of the size of the sound horizon than the scaling employed in Chapter \ref{Chapter4}.

Recently, \cite{Cao_et_al2017a} found that compact structures in intermediate-luminosity radio quasars could serve as standard cosmological rulers. Our QSO data come from a newly compiled sample of these standard rulers from observations of 120 intermediate-luminosity quasars taken over a redshift range of 0.46 < z < 2.76, with angular sizes $\theta_{\textrm{obs}}(z)$ and redshifts $z$ listed in Table 1 of \cite{Cao_et_al2017b}.
The corresponding theoretical predictions for the angular sizes can be obtained via
\begin{equation}
\label{eq:AngularSize}
\theta_{\textrm{th}}(z) = \frac{l_m}{D_A(z)},
\end{equation}
where $l_m = 11.03\pm0.25$ pc is the intrinsic linear size of the ruler (see \citealp{Cao_et_al2017b}), and
\begin{equation}
\label{eq:D_A(z)}
    D_A(z) = \frac{D_M(z)}{1 + z}
\end{equation}
is the angular diameter distance  at redshift $z$ (see \citealp{Hogg}).

\section{Data Analysis Methods}
\label{sec:methods}
We use the $\chi^2$ statistic to find the best-fitting parameter values and limits for a given model. Most of the data points we use are uncorrelated, so
\begin{equation}
\label{eq:chi2}
    \chi^2(p) = \sum^{N}_{i = 1} \frac{[A_{{\rm th}}(p; z_i) - A_{{\rm obs}}(z_i)]^2}{\sigma_i^2}.
\end{equation}
Here $p$ is the set of model parameters, for example $p = (H_0, \Omega_{m0})$ in the flat \lcdm\ model, $z_i$ is the redshift at which the measured value is $A_{\rm obs}(z_i)$ with one standard deviation uncertainty $\sigma_{i}$, and $A_{\rm th}(p; z_i)$ is the predicted value computed in the model under consideration. The $\chi^2$ expression in eq. (\ref{eq:chi2}) holds for the $H(z)$ measurements listed in Table \ref{tab:H(z)_data} and the BAO measurements listed in the last five lines of Table \ref{tab:ch5_BAO_data} here.

The measurements in the first six lines of Table \ref{tab:ch5_BAO_data} are correlated, in which case $\chi^2$ is given by
\begin{equation}
\label{eq:correlated chi2}
    \chi^2(p) = \left[\vec{A}_{{\rm th}}(p) - \vec{A}_{{\rm obs}}\right]^{T} C^{-1} \left[\vec{A}_{{\rm th}}(p) - \vec{A}_{{\rm obs}}\right],
\end{equation}
where $C^{-1}$ is the inverse of the covariance matrix $C = $
\begin{equation}
\label{covmat}
\begin{bmatrix}
    624.707 & 23.729 & 325.332 & 8.34963 & 157.386 & 3.57778 \\
    23.729 & 5.60873 & 11.6429 & 2.33996 & 6.39263 & 0.968056 \\
    325.332 & 11.6429 & 905.777 & 29.3392 & 515.271 & 14.1013 \\
    8.34963 & 2.33996 & 29.3392 & 5.42327 & 16.1422 & 2.85334 \\
    157.386 & 6.39263 & 515.271 & 16.1422 & 1375.12 & 40.4327 \\
    3.57778 & 0.968056 & 14.1013 & 2.85334 & 40.4327 & 6.25936 \\
\end{bmatrix}
\end{equation}
(from the SDSS website).

For the QSO data (\citealp{Cao_et_al2017b}) we use
\begin{equation}
\label{eq:chi2 QSO}
    \chi^2(p) = \sum^{N}_{i = 1} \left[\frac{\theta_{{\rm th}}(p; z_i) - \theta_{{\rm obs}}(z_i)}{\sigma_i + 0.1\theta_{{\rm obs}}(z_i)}\right]^2,
\end{equation}
where $\theta_{{\rm th}}(p; z_i)$ is the model-predicted value of the angular size, $\theta_{{\rm obs}}(z_i)$ is the measured angular size at redshift $z_i$, and $\sigma_i$ is the uncertainty on the measurement made at redshift $z_i$. The term proportional to $\theta_{{\rm obs}}(z_i)$ in the denominator is added to $\sigma_i$ in order to account for systematic uncertainties in the angular size measurements (see the discussion of this point in the first paragraph of Sec. 3 of \citealp{Cao_et_al2017b}).

%%Note A
To determine constraints on the parameters of a given model, we use the likelihood
\begin{equation}
\label{eq:GaussLike}
    \mathcal{L}(p) = e^{-\chi(p)^2/2}.
\end{equation}
We are interested in presenting two-dimensional confidence contour plots and one-dimensional likelihoods. To do this, for the models with more than two parameters, we marginalize over the parameters in turn to get one- and two-dimensional likelihoods. In general, we marginalize our likelihood functions by computing integrals of the form
\begin{equation}
    \mathcal{L}(p_x) = \int \mathcal{L}(p_x, p_y)\uppi(p_y)dp_y,
\end{equation}
where $p_x$ refers to the set of parameters not marginalized over, $p_y$ refers to the parameter to be marginalized, and $\uppi(p_y)$ is a flat, top-hat prior of the form
\begin{equation}
  \uppi(p_y) =
    \begin{cases}
      1 & \text{if $p_{y, \rm min} < p_y < p_{y, \rm max}$}\\
      0 & \text{otherwise}
    \end{cases}       
\end{equation}
(see Table \ref{tab:1d intervals} for the parameter ranges).\footnote{We compute the full (not marginalized) likelihoods on a grid. In all flat models and the non-flat \lcdm\ model each parameter has an associated step size of $\Delta p = 0.01$. In the non-flat XCDM parametrization and the non-flat \pcdm\ model, we use $\Delta H_0 = 0.1$ km s$^{-1}$ Mpc$^{-1}$ to reduce computation time (with all other parameters having $\Delta p = 0.01$).} For example, in the non-flat \lcdm\ model one of the two-dimensional likelihoods we compute is
\begin{equation}
    \mathcal{L}(\om, \ol) = \int^{85}_{50} \mathcal{L}(\om, H_0, \ol)dH_0,
\end{equation}
where we integrate the Hubble constant from 50 km s$^{-1}$ Mpc$^{-1}$ to 85 km s$^{-1}$ Mpc$^{-1}$.\footnote{In Chapter \ref{Chapter4} we used two $H_0$ priors, gaussian with central values and error bars of $\bar{H}_0 \pm \sigma_{H_0} = 68 \pm 2.8$ km s$^{-1}$ Mpc$^{-1}$ \citep{chenratmed} and $\bar{H}_0 \pm \sigma_{H_0} = 73.24 \pm 1.74$ km s$^{-1}$ Mpc$^{-1}$ \citep{riess2016}. Here we are instead treating $H_0$ as an adjustable parameter to be determined from the data we use.} We then plot the isocontours of $\chi^2(\om, \ol) = -2 {\rm ln} \mathcal{L}(\om, \ol)$ in the $\om$-$\ol$ subspace of the total parameter space (see Fig. \ref{fig:Non-flat LCDM 2D QSO+Hz+BAO}).
\begin{comment}
To obtain the one-dimensional likelihood as a function of a given parameter within a model, we marginalize the likelihood over all the other parameters of the model under consideration. The marginalization ranges that we use for the parameters of each model are listed in Table \ref{tab:1d intervals}; for all parameters we assume a flat prior distribution, and we use a step size of $\Delta p = 0.01$. The only exception to this is the step size of $H_0$ in non-flat XCDM and non-flat \pcdm, for which we use $\Delta H_0 = 0.1$ km s$^{-1}$ Mpc$^{-1}$ (to reduce computation time).
\end{comment}

In addition to plotting the one-dimensional likelihoods for each parameter of each model we consider, we also compute one-sided confidence limits on the best-fitting values of these parameters. The best-fitting value of a parameter $p$ within a given model, after marginalization over the other parameters of the model, is that value $\bar{p}$ which maximizes the one-dimensional likelihood $\mathcal{L}(p)$. To determine the confidence limits $r_{n}^{\pm}$ on either side of $\bar{p}$, we compute
\begin{equation}
    \label{eq:1d limits}
    \frac{\int^{r_{n}^{\pm}}_{\bar{p}}\mathcal{L}(p)dp}{\int^{\pm \infty}_{\bar{p}}\mathcal{L}(p)dp} = \sigma_{n}
\end{equation}
where $\sigma_{1,2} = 0.6827$, $0.9545$ and $r_{n}^{+}$, $r_{n}^{-}$ are the upper and lower confidence limits out to $\sigma_{n}$, respectively.

\begin{table*}
	\centering
	\caption{Best-fitting parameters of all models.}
	\label{tab:BFP}
	\centering
    \resizebox{\columnwidth}{!}{%
	\begin{threeparttable}
	\begin{tabular}{lccccccccccc} % four columns, alignment for each
		\hline
		Model & Data set & $\om$ & $\ol$ & $\ok$ & $w_{X}$ & $\alpha$ & $H_0$\tnote{a} & $\nu$ & $\chi^2$ & AIC & BIC\\
		\hline
		Flat \lcdm\ &  $H(z)$ + BAO & 0.30 & 0.70 & 0 & - & - & 67.99 & 39 & 23.63 & 27.63 & 31.11\\
		 & QSO & 0.32 & 0.68 & 0 & - & - & 68.49 & 117 & 352.05 & 356.05 & 361.63\\
		 & QSO + $H(z)$ + BAO & 0.31 & 0.69 & 0 & - & - & 68.43 & 159 & 376.44 & 380.44 & 386.62\\
		\hline
		Non-flat \lcdm\ & $H(z)$ + BAO & 0.30 & 0.70 & 0 & - & - & 68.46 & 38 & 23.2 & 29.2 & 34.41\\
		 & QSO & 0.27 & 1 & $-0.27$ & - & - & 74.62 & 116 & 351.3 & 357.30 & 365.66\\
		 & QSO + $H(z)$ + BAO & 0.30 & 0.73 & $-0.03$ & - & - & 69.51 & 158 & 375.38 & 381.38 & 390.64\\
		\hline
		Flat XCDM & $H(z)$ + BAO & 0.30 & - & 0 & $-0.94$ & - & 66.73 & 38 & 23.29 & 29.29 & 34.50\\
		 & QSO & 0.27 & - & 0 & $-1.97$ & - & 81.22 & 116 & 351.84 & 357.84 & 366.20\\
		 & QSO + $H(z)$ + BAO & 0.32 & - & 0 & $-0.97$ & - & 67.90 & 158 & 376.27 & 382.27 & 391.53\\
		 \hline
		Flat \pcdm\ & $H(z)$ + BAO & 0.30 & - & 0 & - & 0.14 & 66.89 & 38 & 23.41 & 29.41 & 34.62\\
		 & QSO & 0.32 & - & 0 & - & 0.01 & 68.44 & 116 & 352.05 & 358.05 & 366.41\\
		 & QSO + $H(z)$ + BAO & 0.31 & - & 0 & - & 0.07 & 67.94 & 158 & 376.39 & 382.39 & 391.65\\
		 \hline
		Non-flat XCDM & $H(z)$ + BAO & 0.32 & - & $-0.23$ & $-0.73$ & - & 66.9 & 37 & 20.94 & 28.94 & 35.89\\
		 & QSO & 0.10 & - & $-0.55$ & $-0.67$ & - & 73.9 & 115 & 350.11 & 358.11 & 369.26\\
		 & QSO + $H(z)$ + BAO & 0.31 & - & $-0.15$ & $-0.78$ & - & 66.7 & 157 & 372.95 & 380.95 & 393.30\\
		 \hline
		Non-flat $\phi$CDM & $H(z)$ + BAO & 0.31 & - & $-0.18$ & - & 0.79 & 67.5 & 37 & 21.36 & 29.36 & 36.31\\
		 & QSO & 0.10 & - & $-0.43$ & - & 2.95 & 72.3 & 115 & 351 & 359 & 370.15\\
		 & QSO + $H(z)$ + BAO & 0.31 & - & $-0.14$ & - & 0.68 & 67.3 & 157 & 373.49 & 381.49 & 393.84\\
		 \hline
	\end{tabular}
	\begin{tablenotes}
    \item[a]${\rm km}\hspace{1mm}{\rm s}^{-1}{\rm Mpc}^{-1}$.
    \end{tablenotes}
    \end{threeparttable}%
    }
\end{table*}

In addition to the $\chi^2$ statistic, we also use the Akaike Information Criterion
\begin{equation}
    AIC \equiv \chi^2_{\rm min} + 2k
    \label{eq:AIC}
\end{equation}
and the Bayes Information Criterion
\begin{equation}
    BIC \equiv \chi^2_{\rm min} + k{\rm ln}N
\end{equation}
\citep{Liddle_2007}, where $\chi^2_{\rm min}$ is the minimum value of $\chi^2$ in the given model, $k$ is the number of parameters in the model, and $N$ is the number of data points. The $AIC$ and $BIC$ penalize models with a greater number of parameters compared to those with fewer parameters, and as such they can be used to compare the effectiveness of the fits of models with different numbers of parameters.

Although we use Bayesian statistics to analyze our data, this analysis is not complete because we do not compute the Bayesian evidence (a computation which would be prohibitively expensive given that we calculate our likelihoods on a grid rather than using MCMC). Instead we approximate the full Bayesian evidence via $\chi^2$, $AIC$, and $BIC$, which we use to compare our models.

\section{Results}
\label{sec:ch5_Results}
\subsection{$H(z)$ + BAO constraints}
\label{sec:Hz+BAO}

\begin{figure*}
\begin{multicols}{3}
    \includegraphics[width=\linewidth]{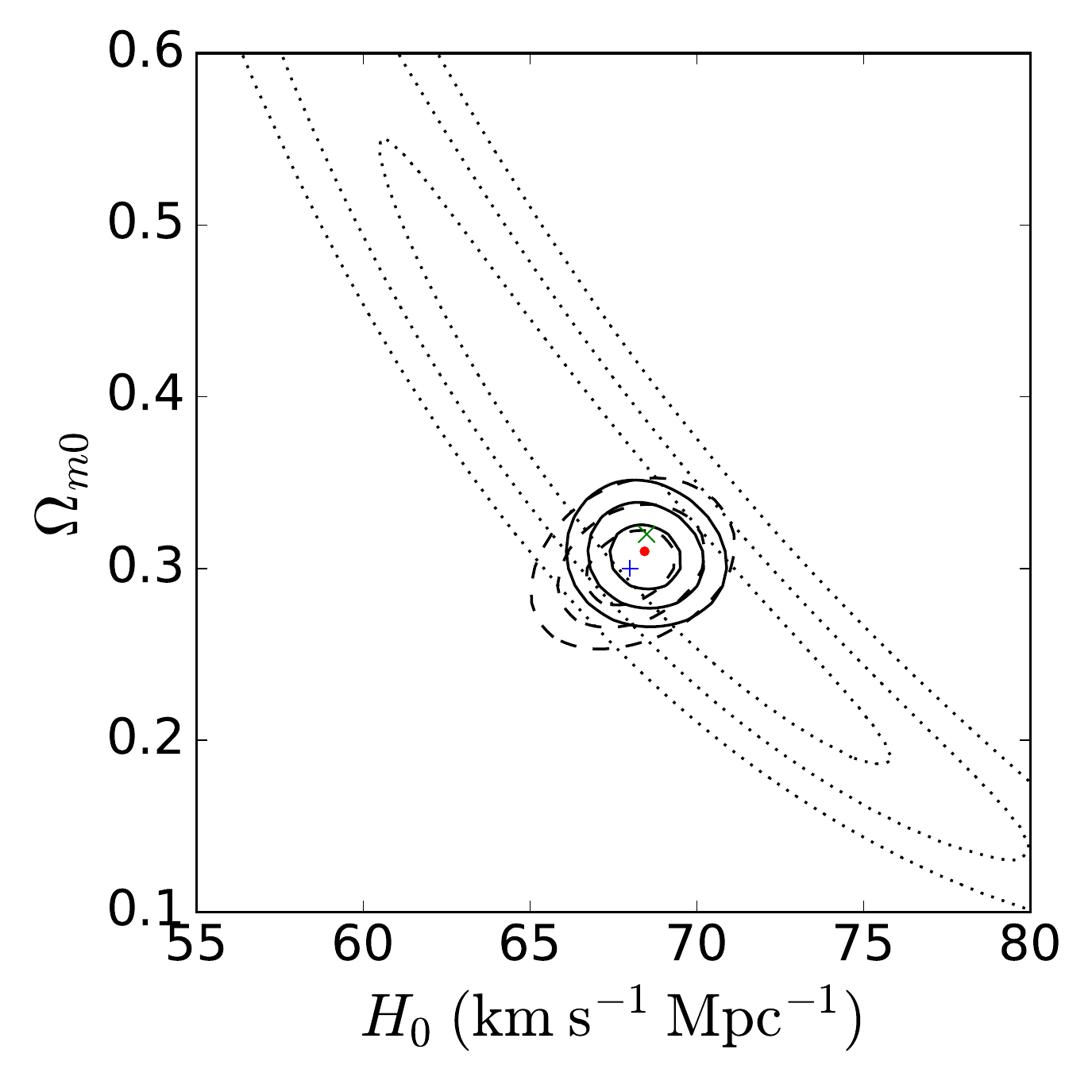}\par
    \includegraphics[width=\linewidth]{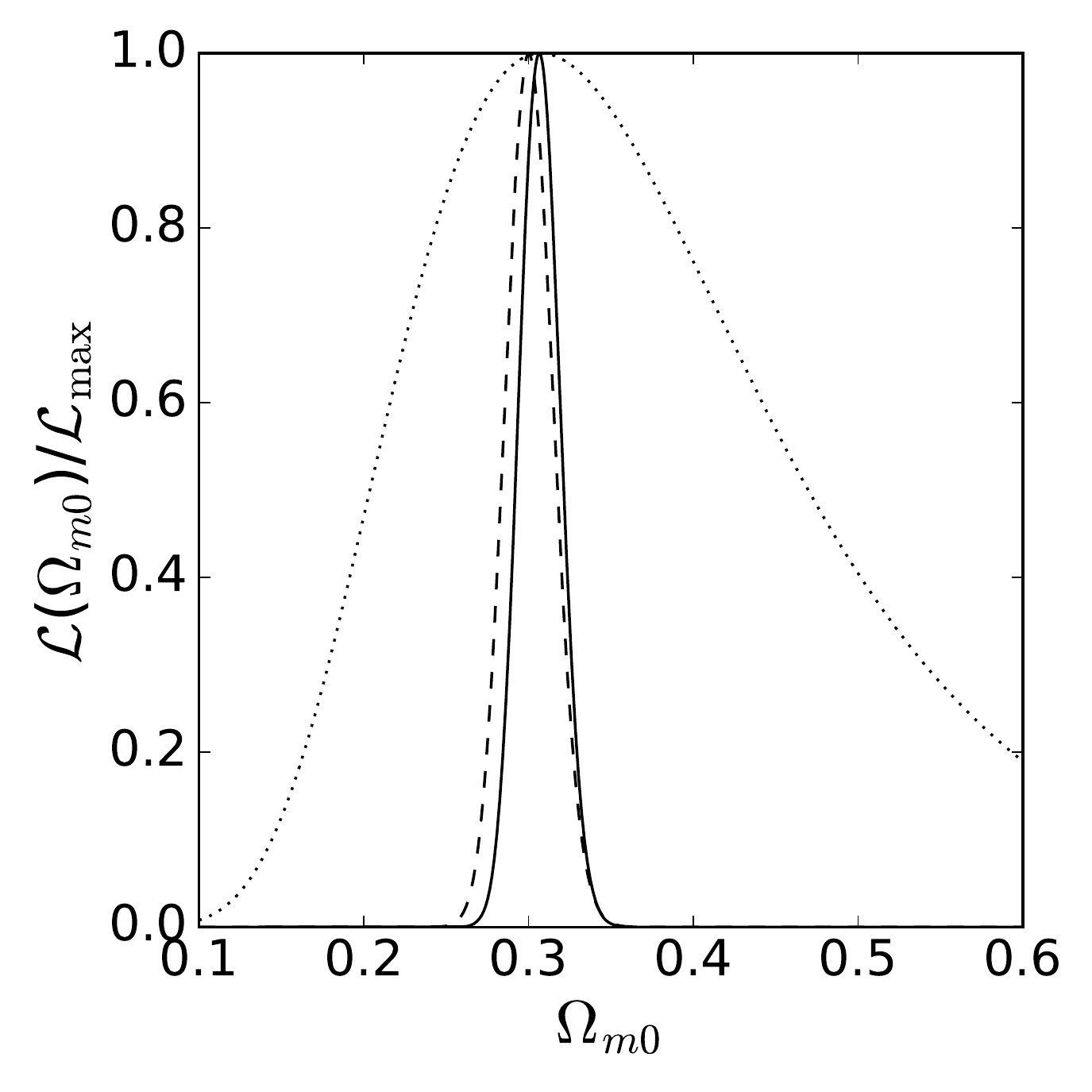}\par 
    \includegraphics[width=\linewidth]{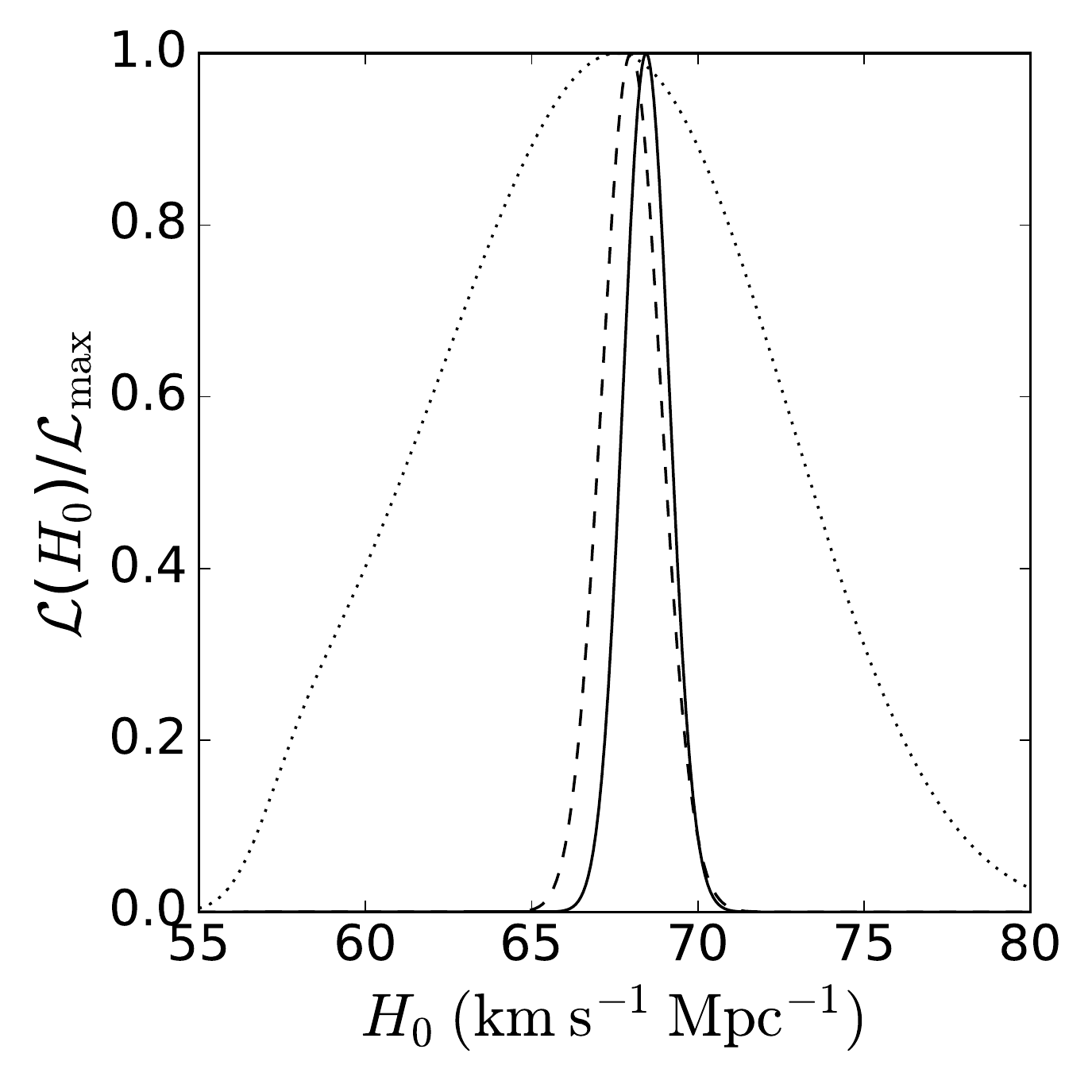}\par
\end{multicols}
\caption[Flat \lcdm\ model with QSO, $H(z)$, and BAO data.]{Flat \lcdm\ model with QSO, $H(z)$, and BAO data. Left panel: 1, 2, and 3$\sigma$ confidence contours and best-fitting points. Center and right panels: one-dimensional likelihoods for $\om$ and $H_0$. See text for description and discussion.}
\label{fig:flat LCDM 2D QSO}
\end{figure*}

\begin{figure*}
\begin{multicols}{3}
    \includegraphics[width=\linewidth]{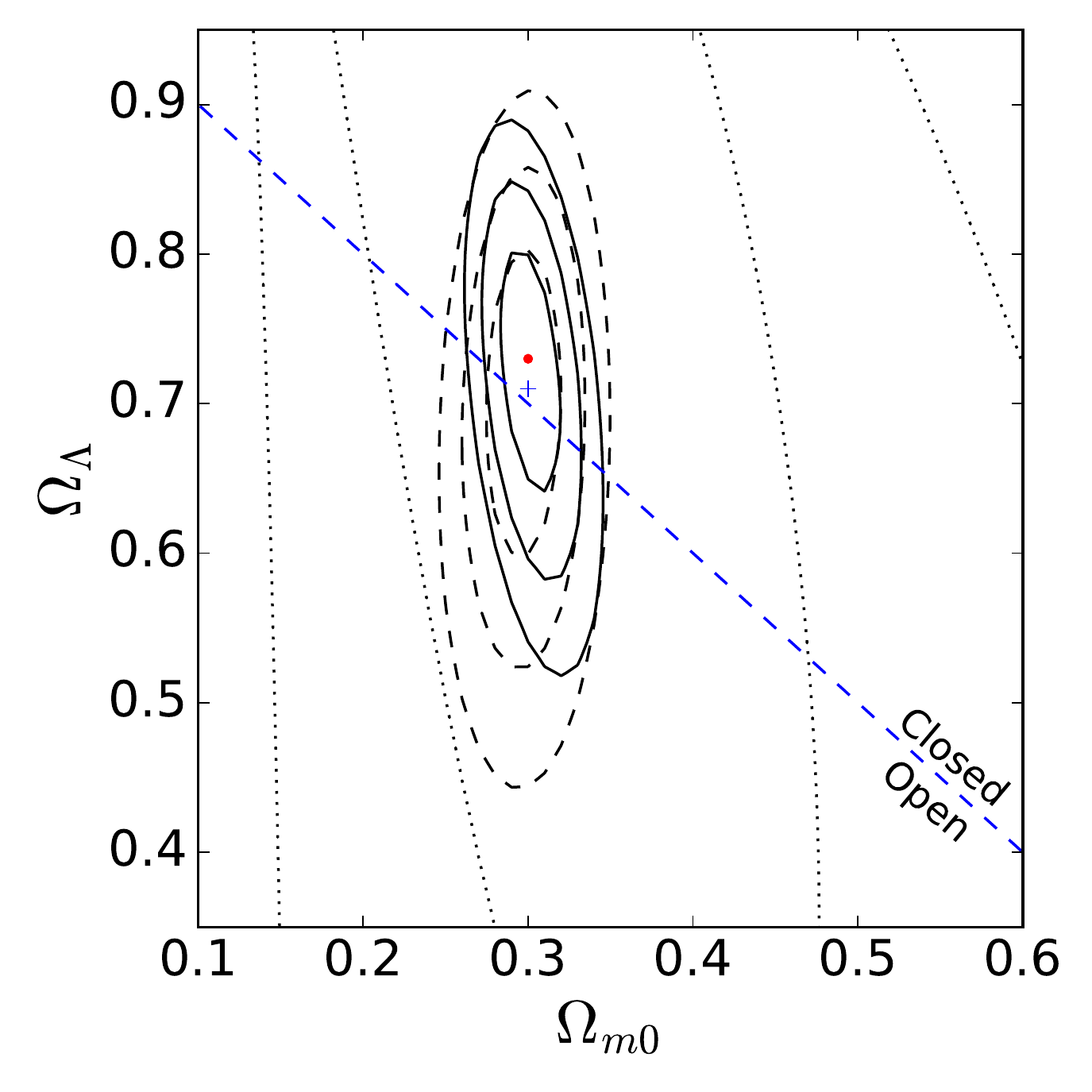}\par 
    \includegraphics[width=\linewidth]{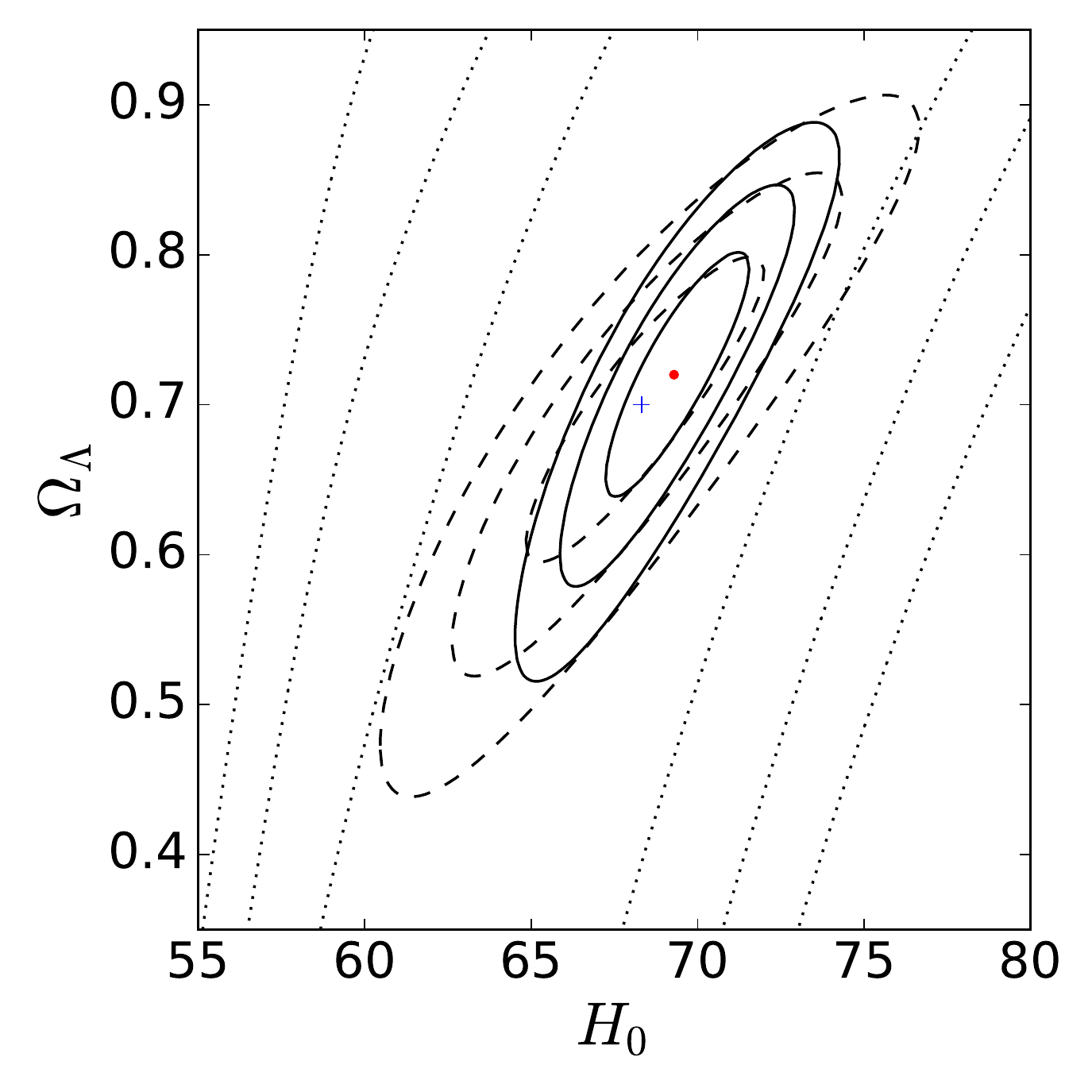}\par
    \includegraphics[width=\linewidth]{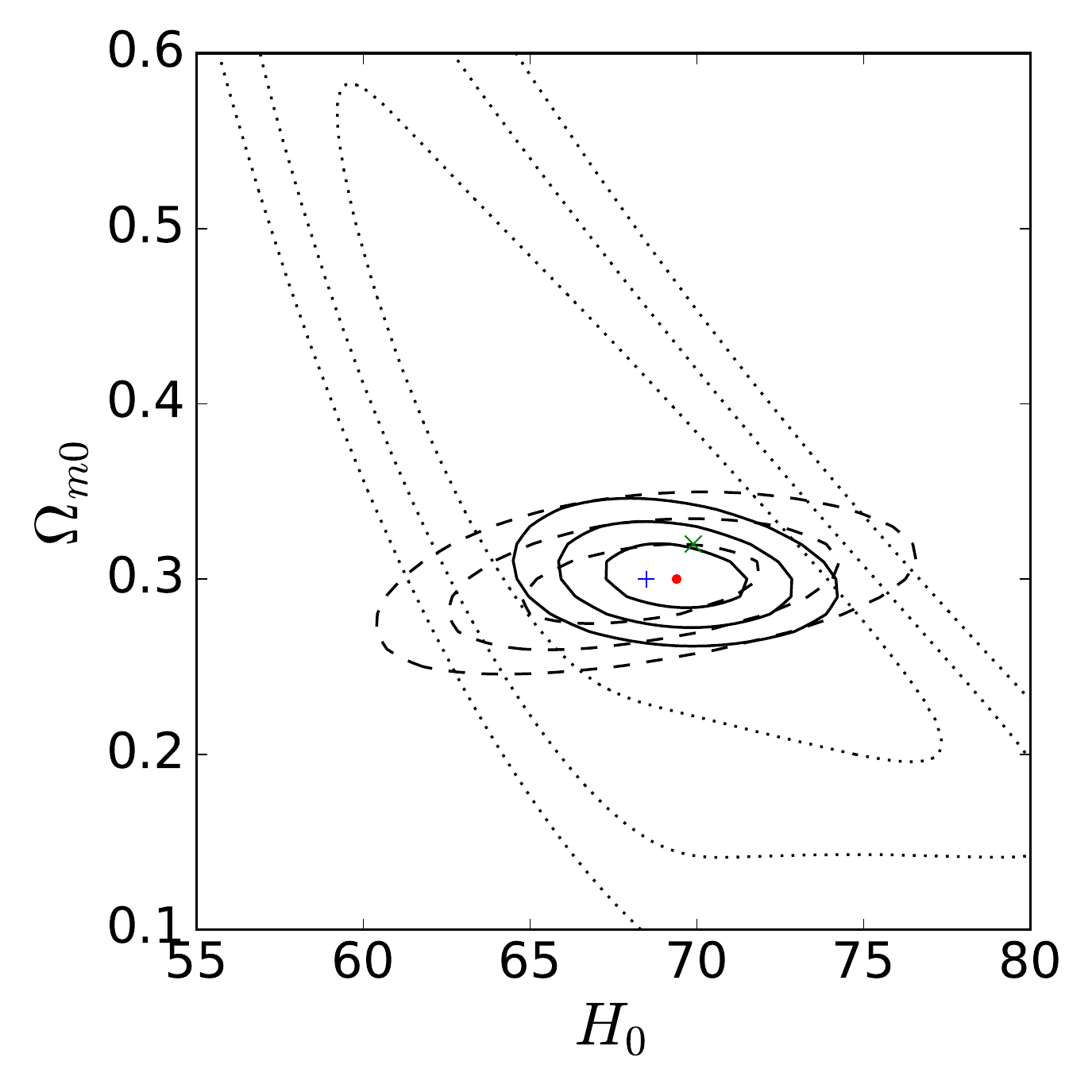}\par
\end{multicols}
\begin{multicols}{3}
    \includegraphics[width=\linewidth]{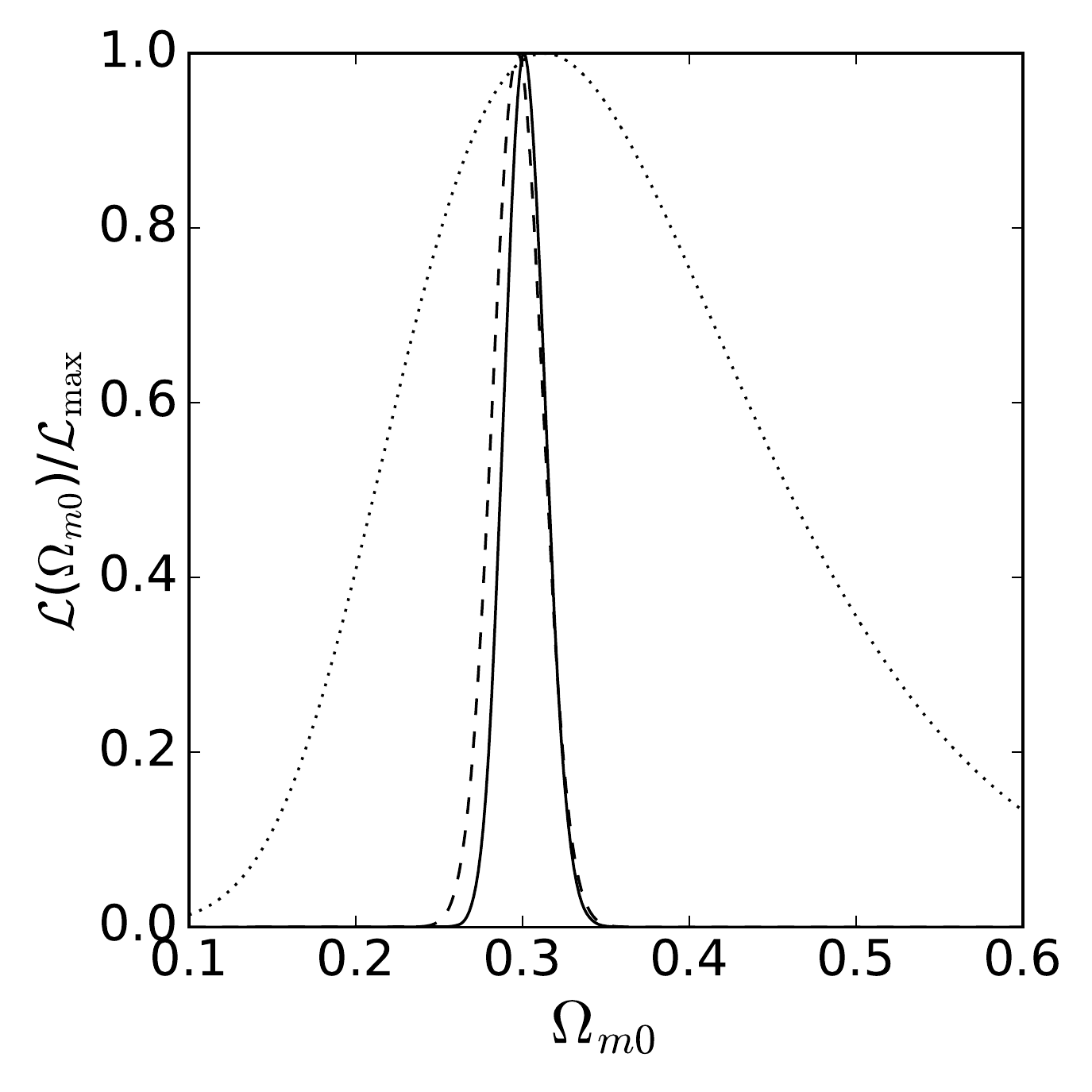}\par 
    \includegraphics[width=\linewidth]{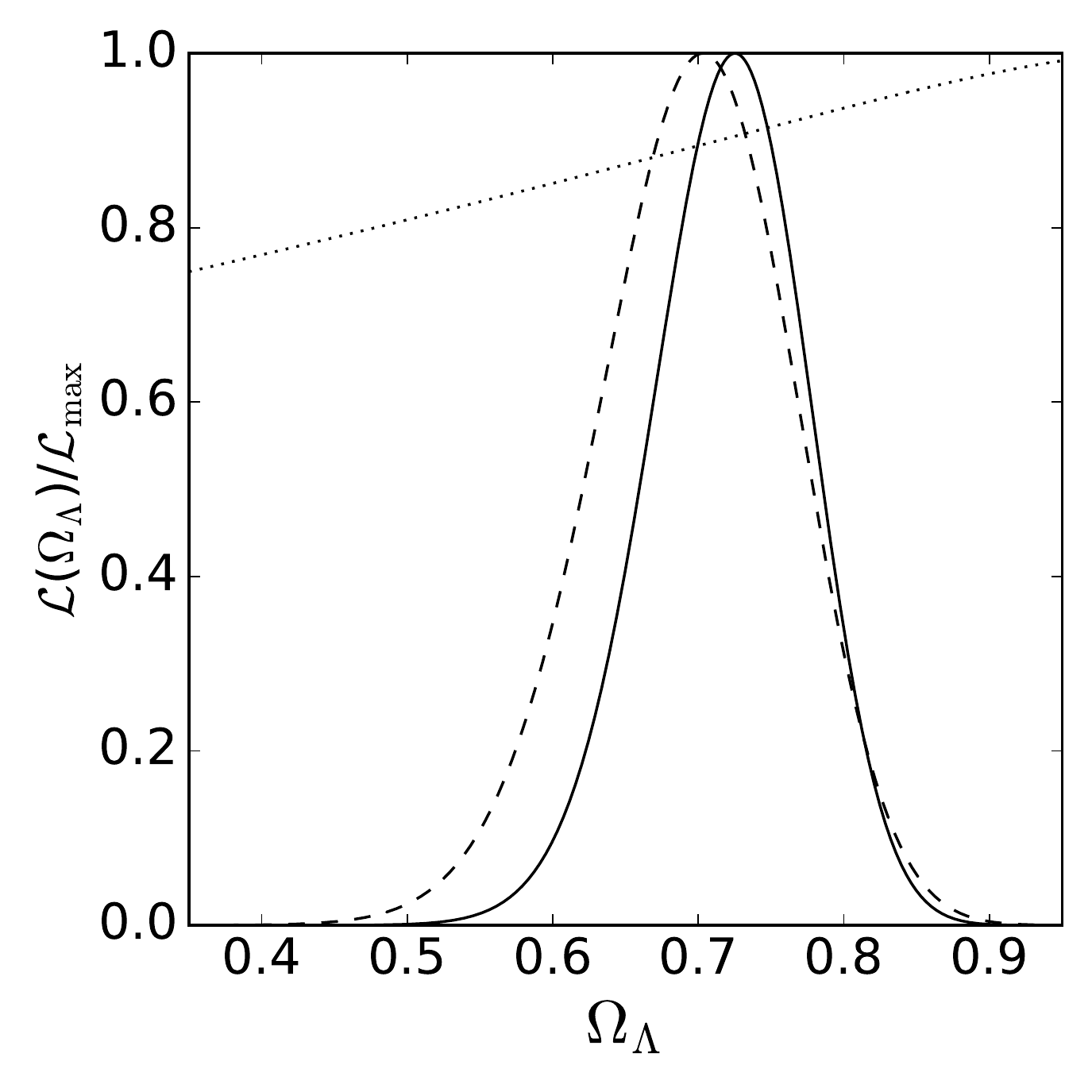}\par
    \includegraphics[width=\linewidth]{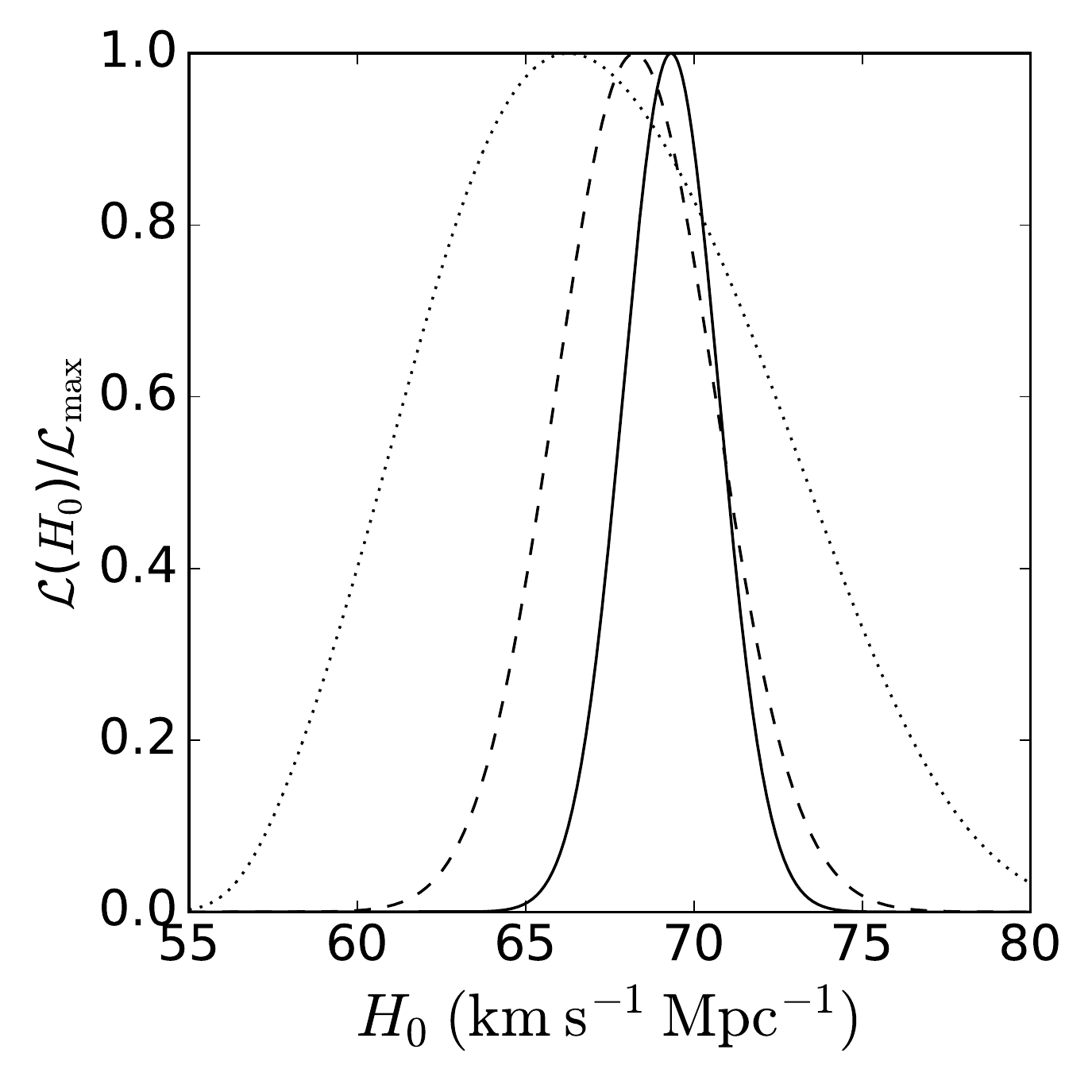}\par
\end{multicols}
\caption[Non-flat \lcdm\ model with QSO, $H(z)$, and BAO data.]{Non-flat \lcdm\ model with QSO, $H(z)$, and BAO data. In the top left panel, the blue dashed line demarcates regions of the $\om$-$\ol$ parameter space that correspond to spatially open ($\ok > 0$) and spatially closed ($\ok < 0$) models. Points on the line correspond to spatially flat models, with $\ok = 0$. Bottom panels: one-dimensional likelihoods for $\om$, $\ol$, and $H_0$. See text for description and discussion.}
\label{fig:Non-flat LCDM 2D QSO+Hz+BAO}
\end{figure*}

\begin{figure*}
\begin{multicols}{3}
    \includegraphics[width=\linewidth]{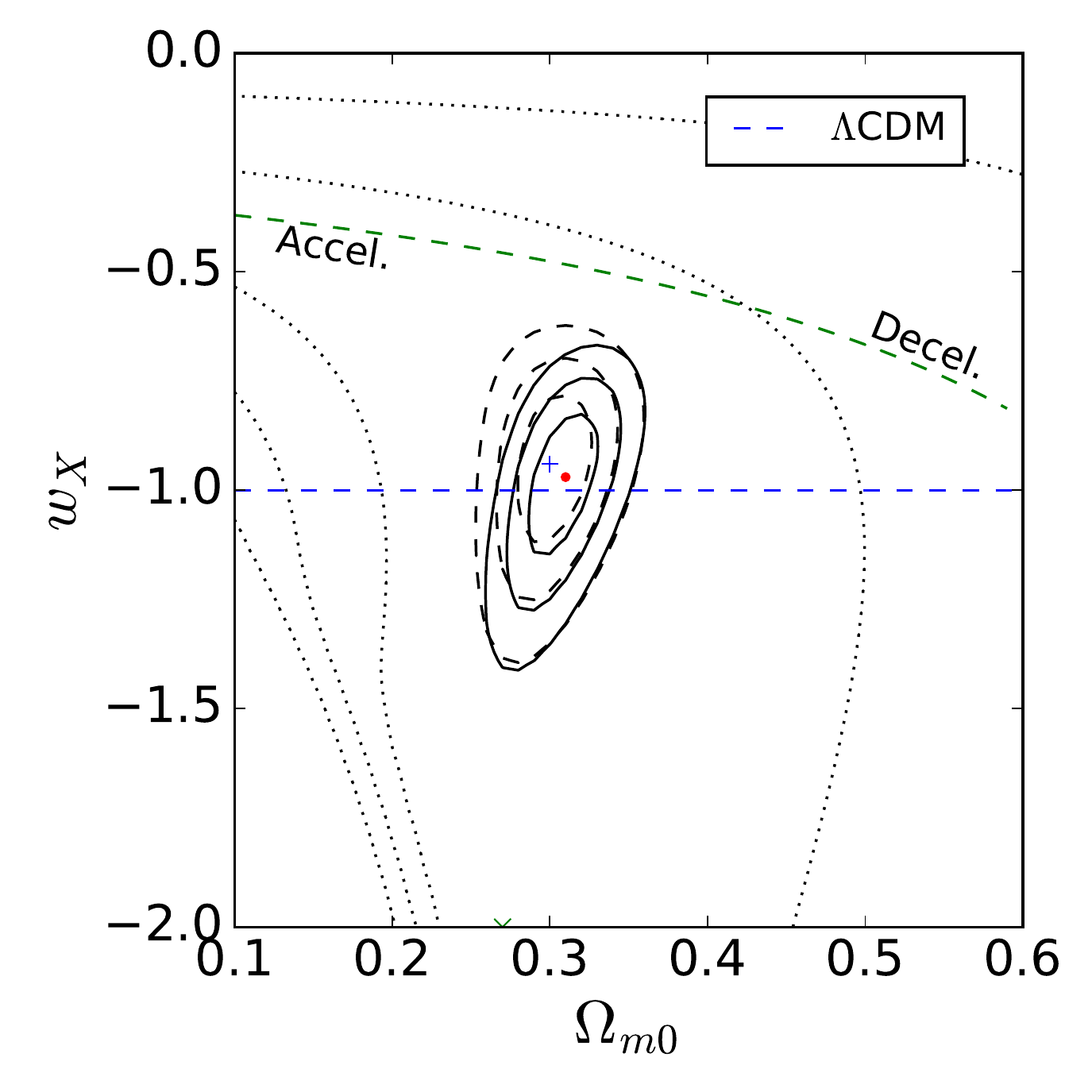}\par
    \includegraphics[width=\linewidth]{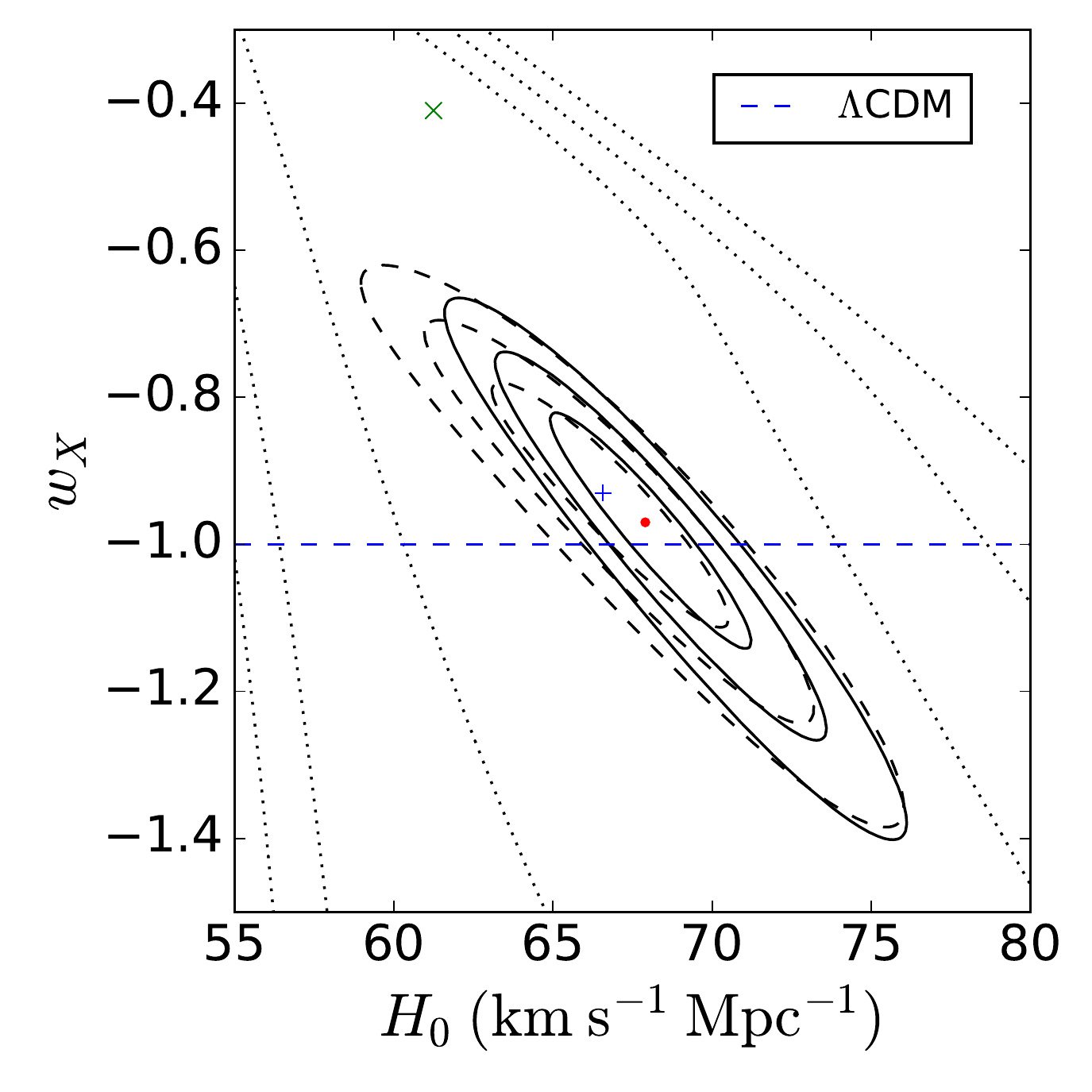}\par
    \includegraphics[width=\linewidth]{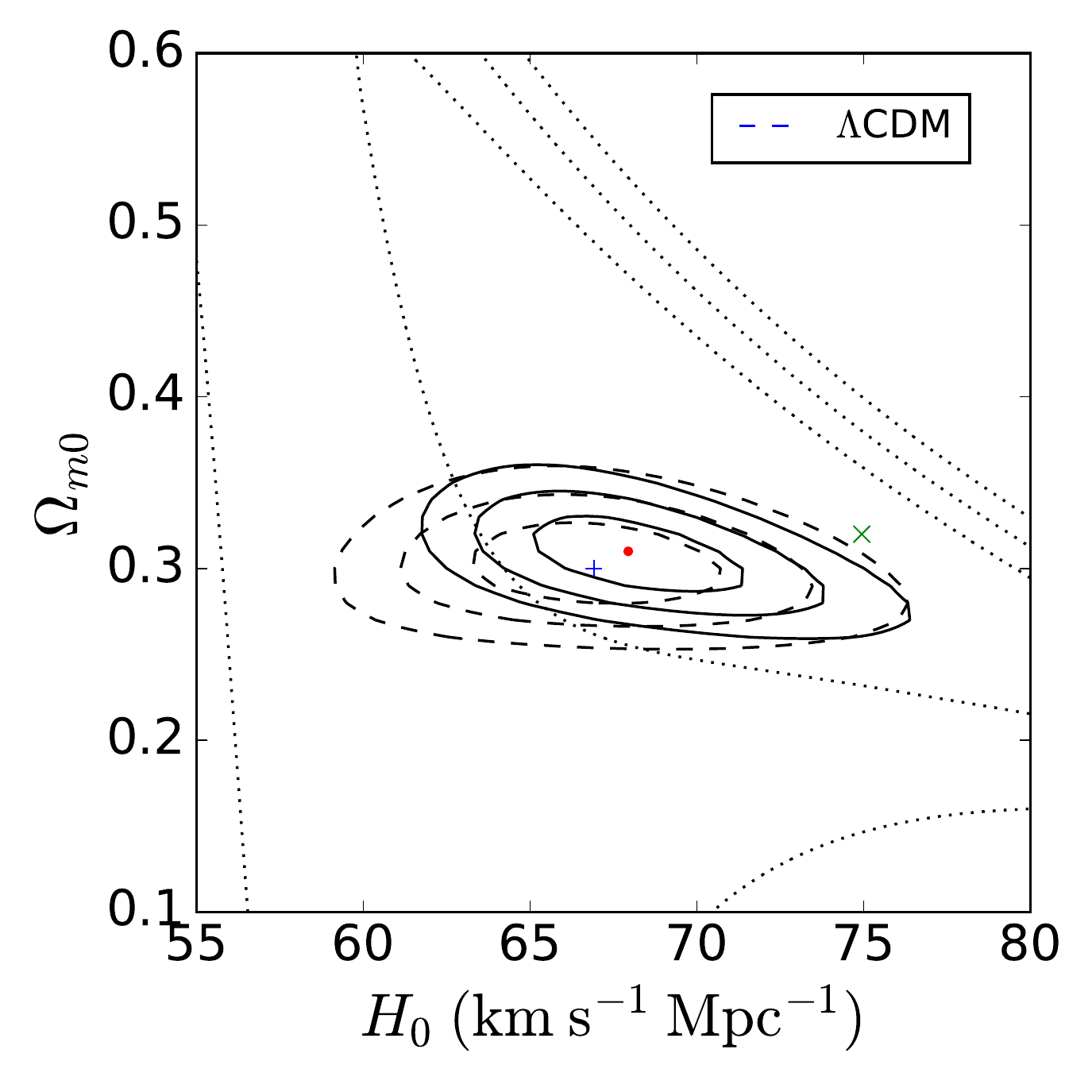}\par
\end{multicols}
\begin{multicols}{3}
    \includegraphics[width=\linewidth]{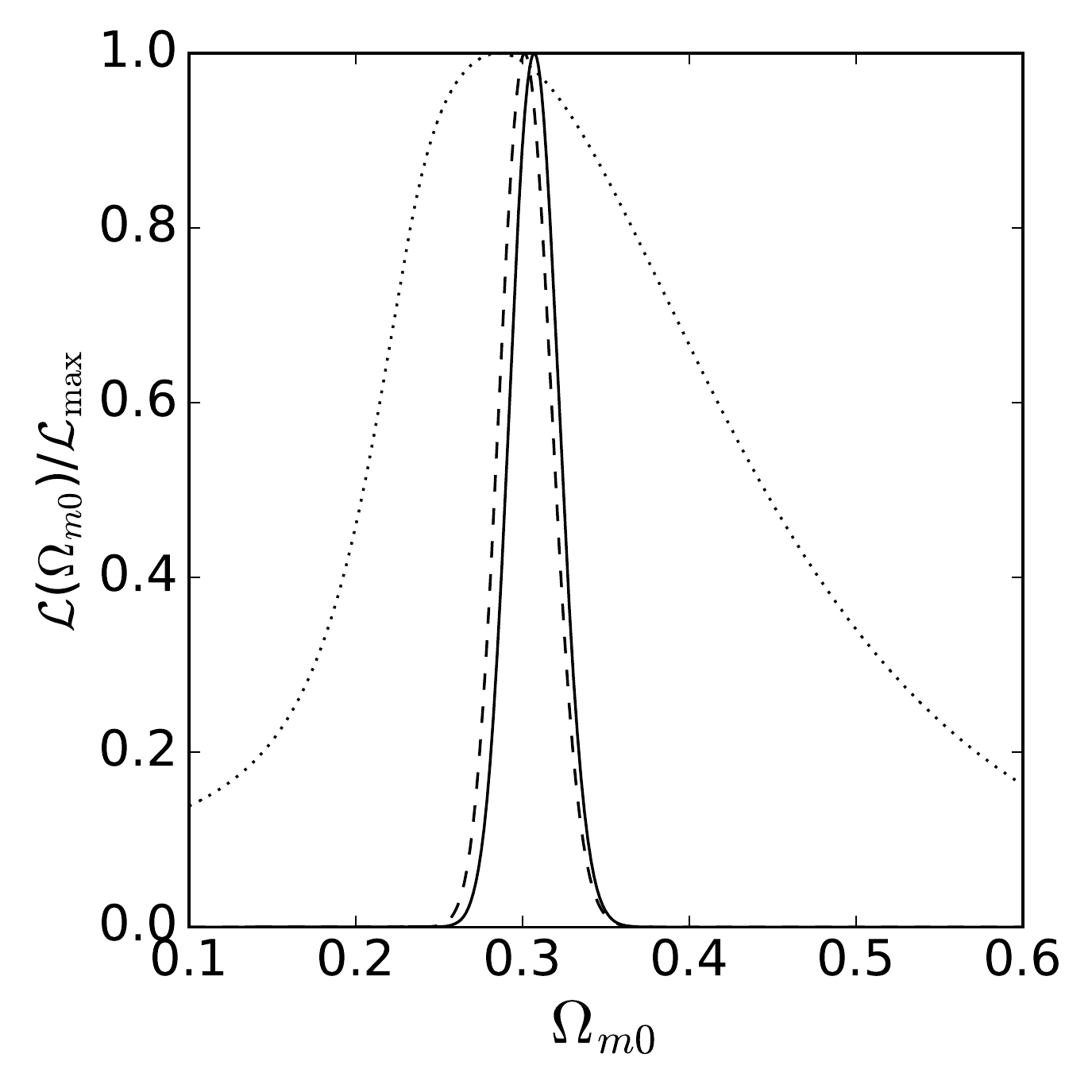}\par 
    \includegraphics[width=\linewidth]{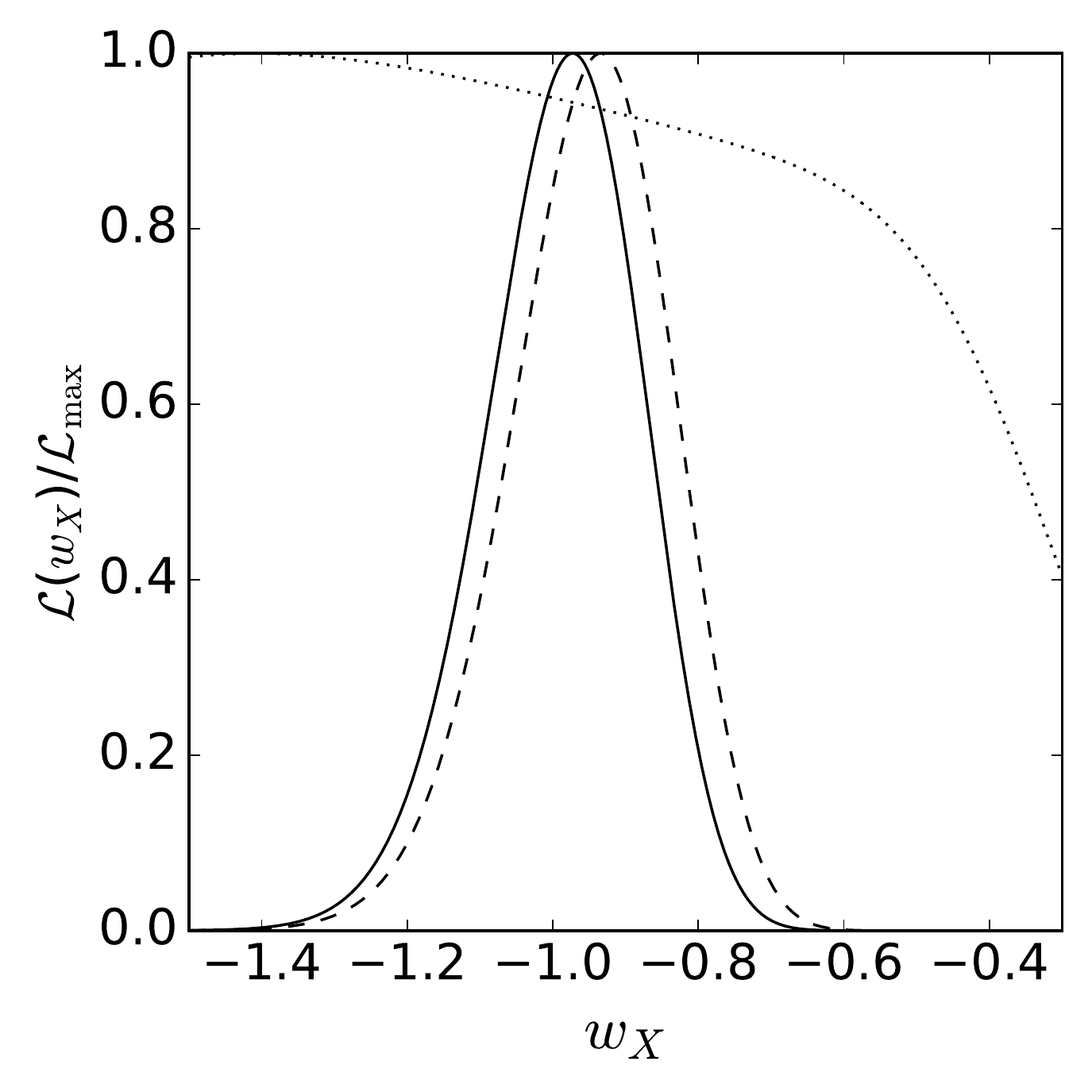}\par
    \includegraphics[width=\linewidth]{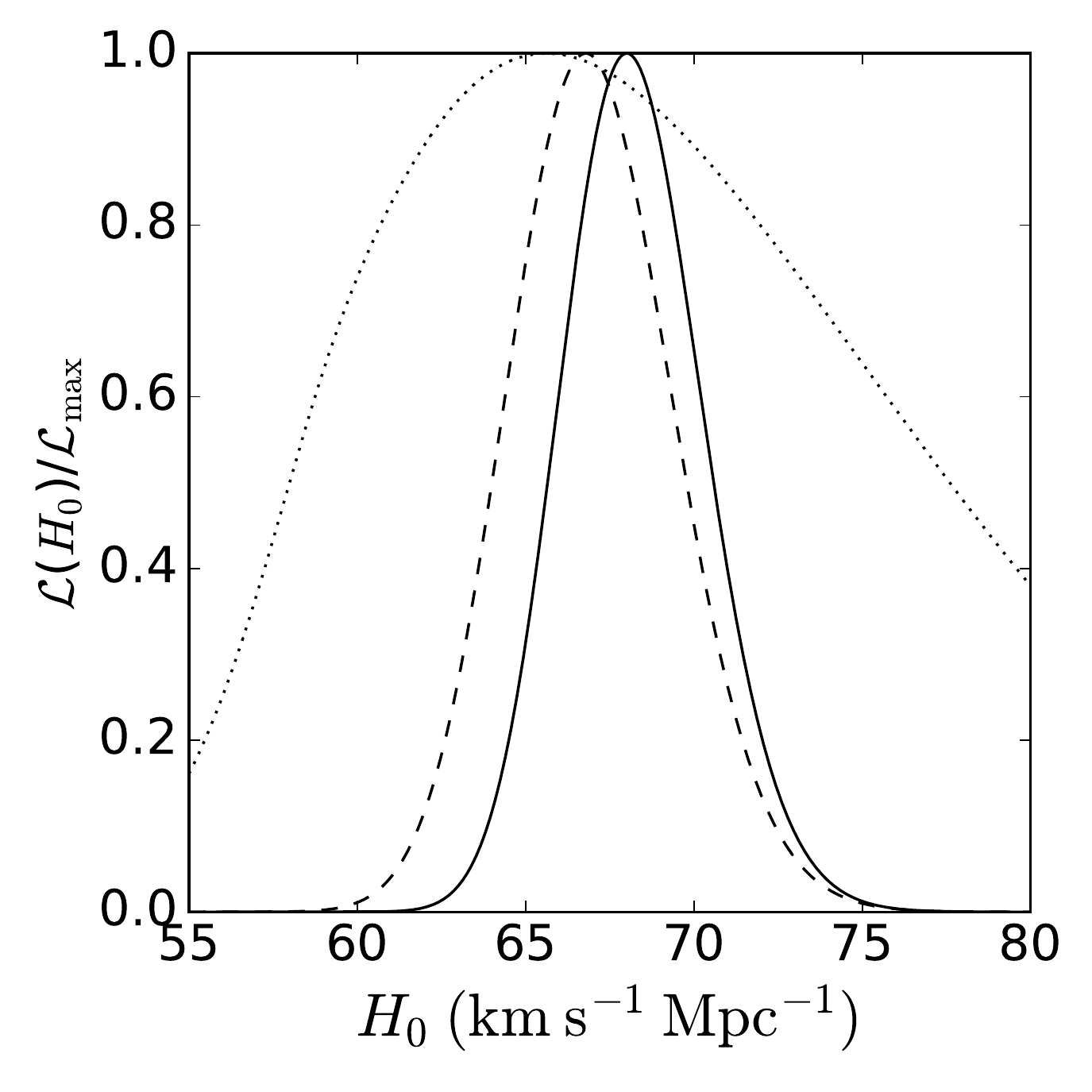}\par
\end{multicols}
\caption[Flat XCDM parametrization with QSO, $H(z)$, and BAO data.]{Flat XCDM parametrization with QSO, $H(z)$, and BAO data. Top panels: 1, 2, and 3$\sigma$ confidence contours and best-fitting points. In the top left and top center panels the horizontal blue dashed line separates quintessence-type parametrizations of dark energy (for which $w_{X} > -1$) from phantom-type parametrizations of dark energy (for which $w_{X} < -1$). Points on the blue line (for which $w_{X} = -1$) correspond to the flat $\Lambda$CDM model. The green dashed curve in the left panel separates models that undergo accelerated expansion now from models that undergo decelerated expansion now. Bottom panels: one-dimensional likelihoods for $\om$, $w_{X}$, and $H_0$. See text for description and discussion.}
\label{fig:flat XCDM 2D combined}
\end{figure*}

\begin{figure*}
\begin{multicols}{3}
    \includegraphics[width=\linewidth]{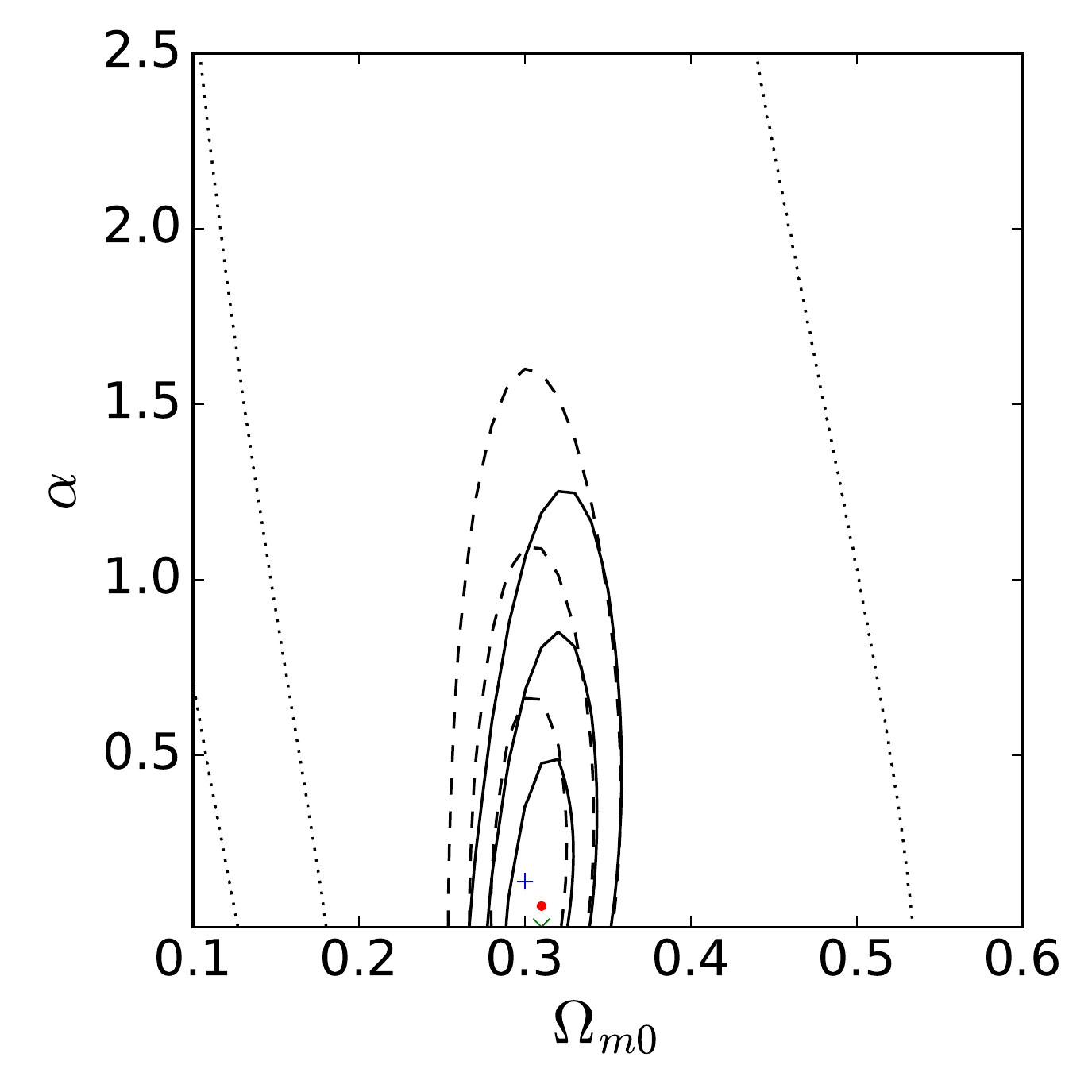}\par 
    \includegraphics[width=\linewidth]{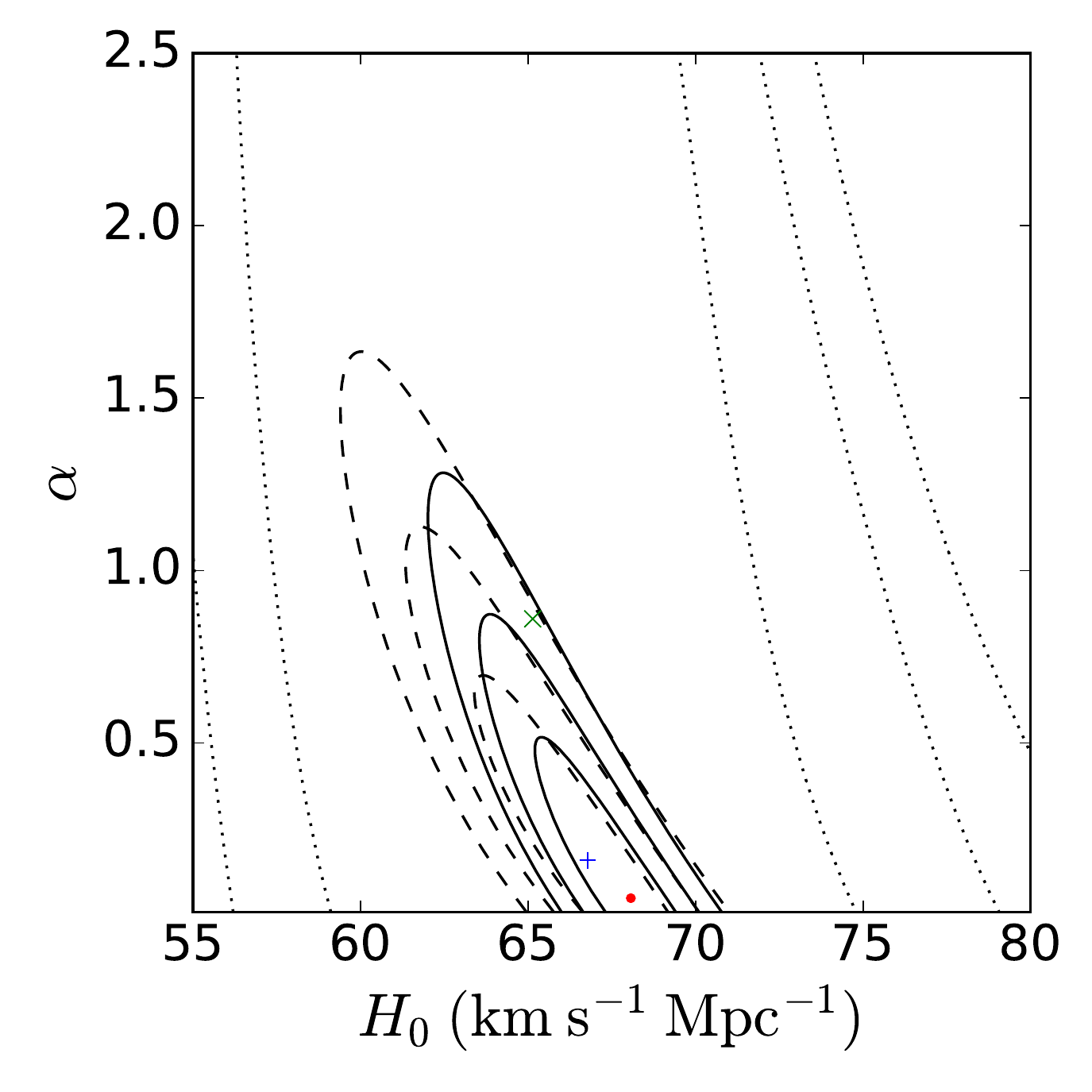}\par
    \includegraphics[width=\linewidth]{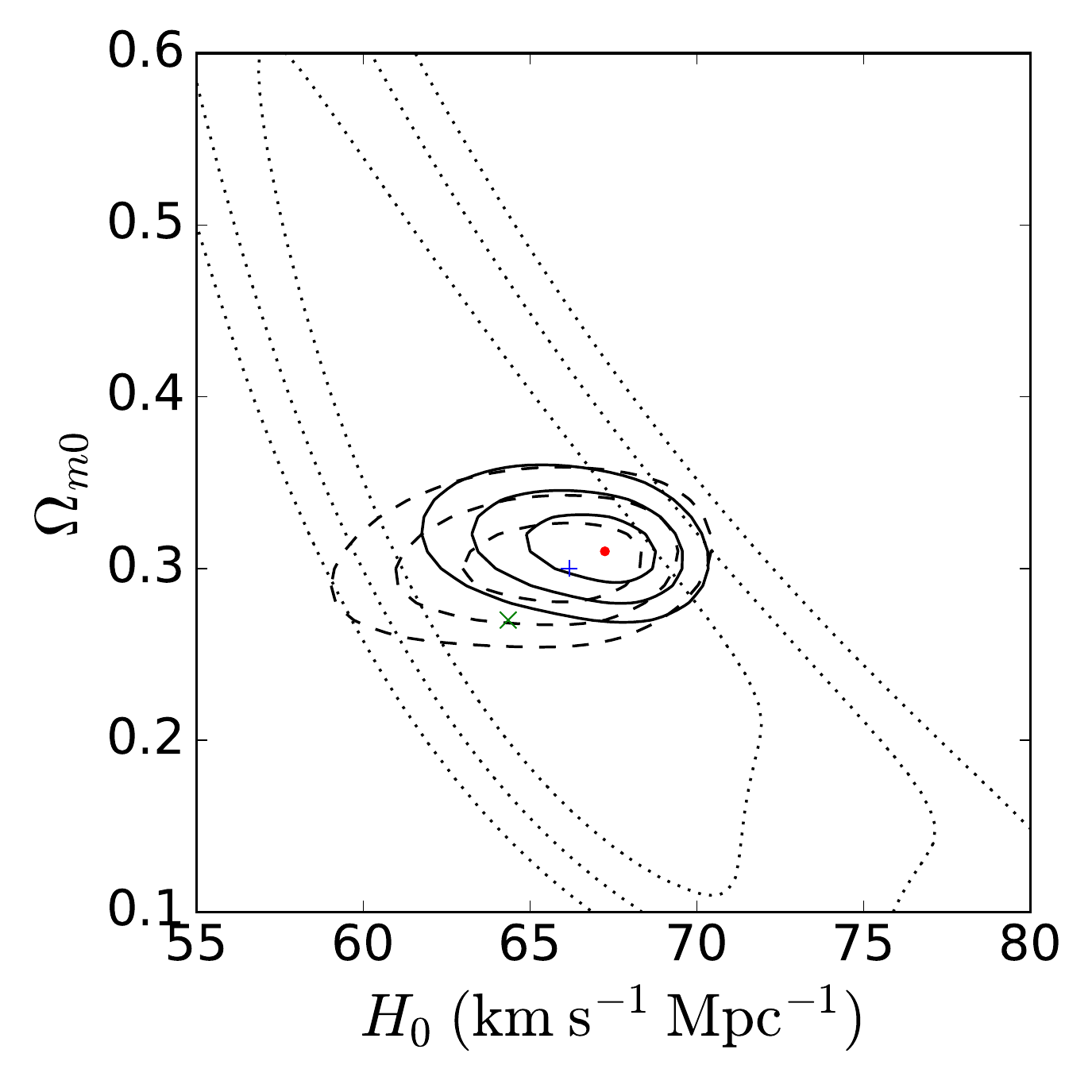}\par
\end{multicols}
\begin{multicols}{3}
    \includegraphics[width=\linewidth]{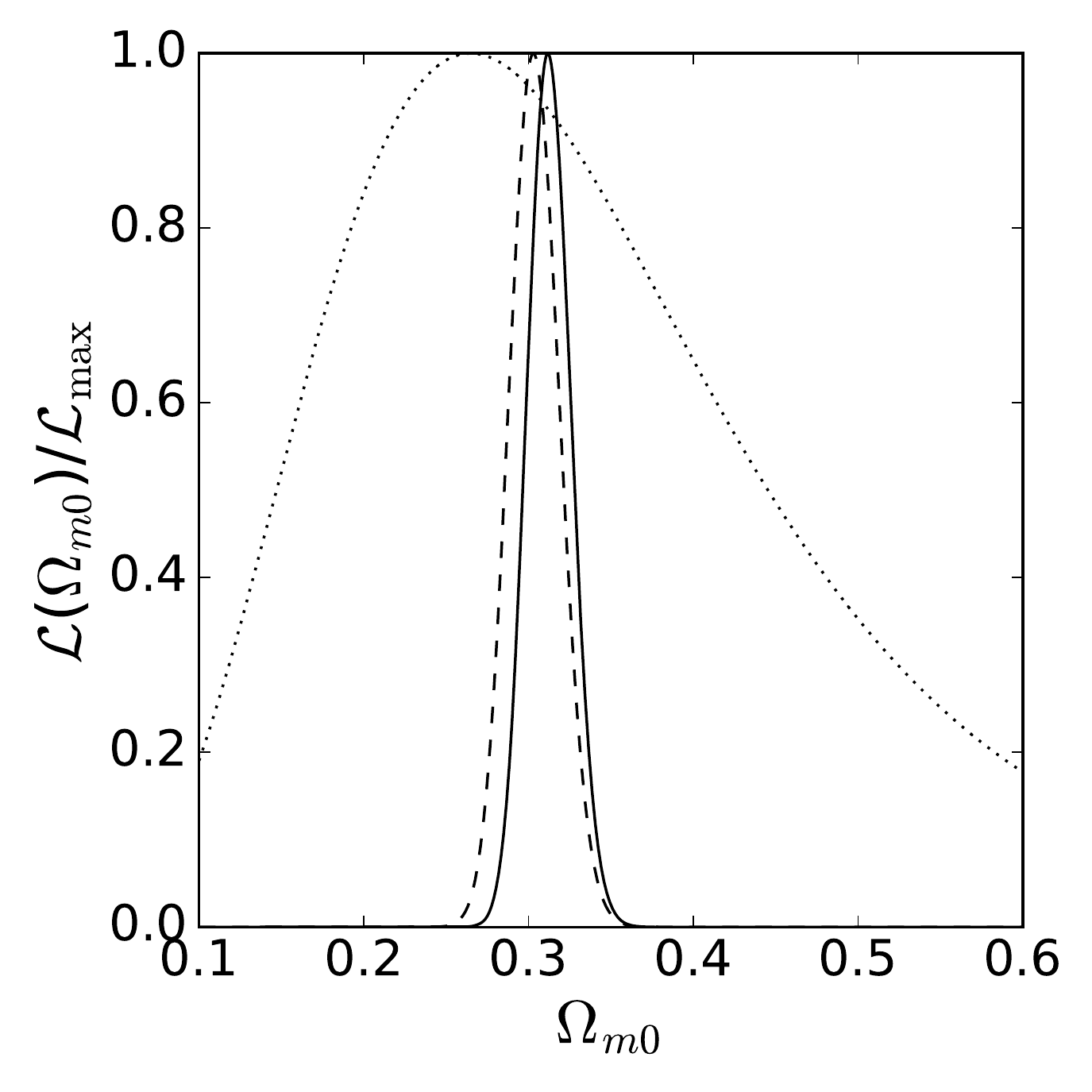}\par 
    \includegraphics[width=\linewidth]{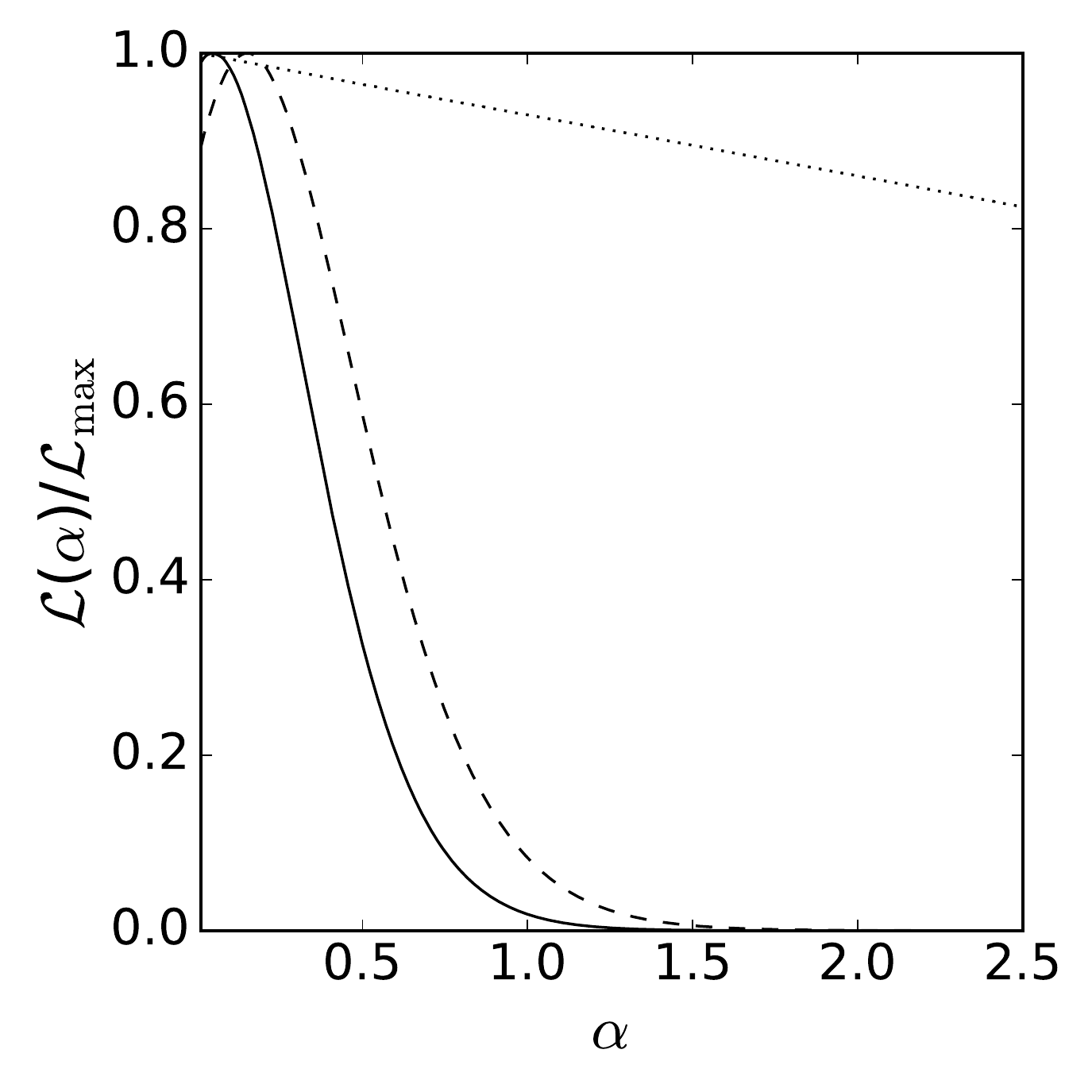}\par
    \includegraphics[width=\linewidth]{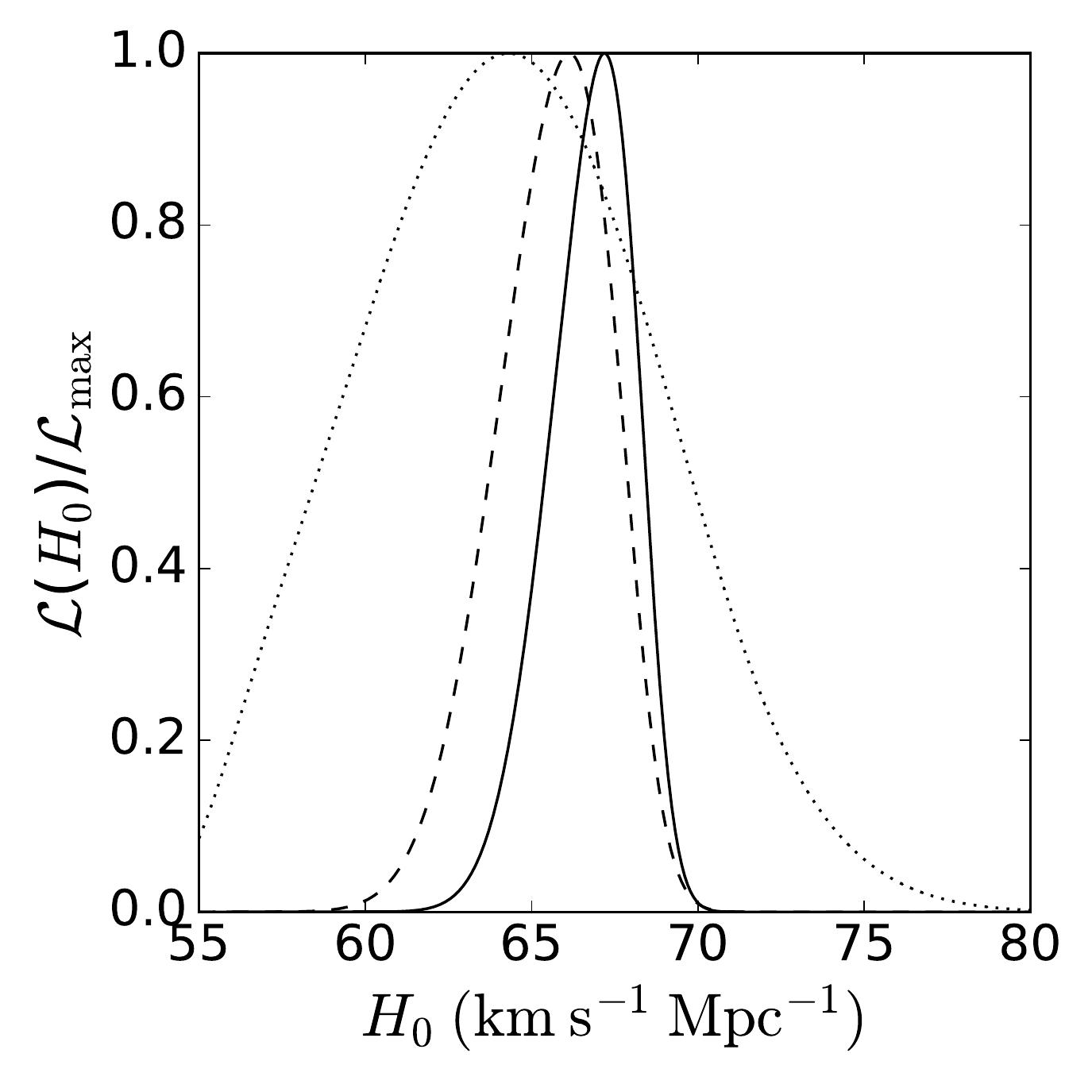}\par
\end{multicols}
\caption[Flat \pcdm\ model with QSO, $H(z)$, and BAO data.]{Flat \pcdm\ model with QSO, $H(z)$, and BAO data. Top panels: 1, 2, and 3$\sigma$ confidence contours and best-fitting points. Points on the $\alpha = 0$ line in the top left and top center panels correspond to the flat \lcdm\ model. Bottom panels: one-dimensional likelihoods of $\om$, $\alpha$, and $H_0$. See text for description and discussion.}
\label{fig:flat pCDM 2D (QSO+Hz+BAO)}
\end{figure*}

\begin{figure*}
\begin{multicols}{3}
    \includegraphics[width=\linewidth]{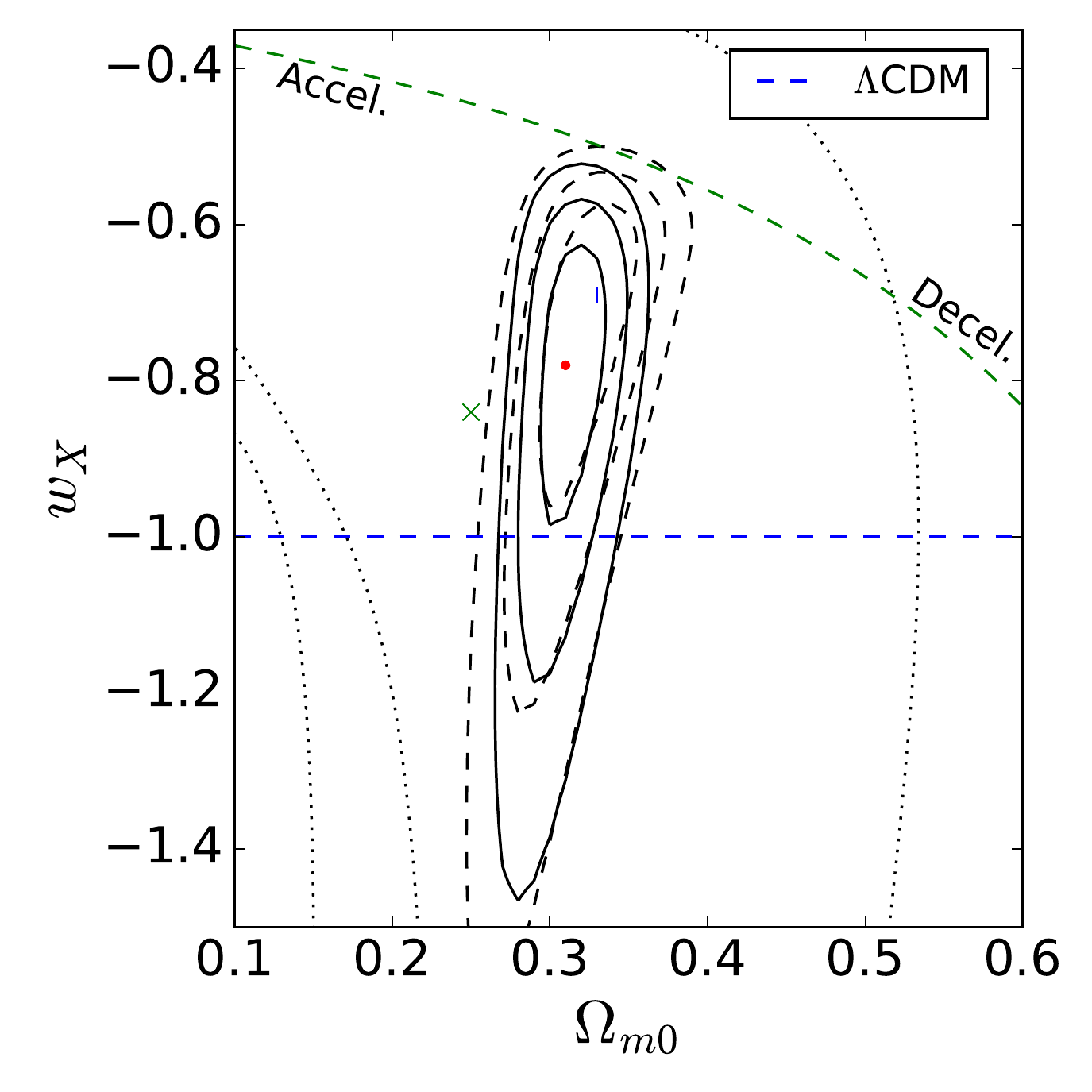}\par
    \includegraphics[width=\linewidth]{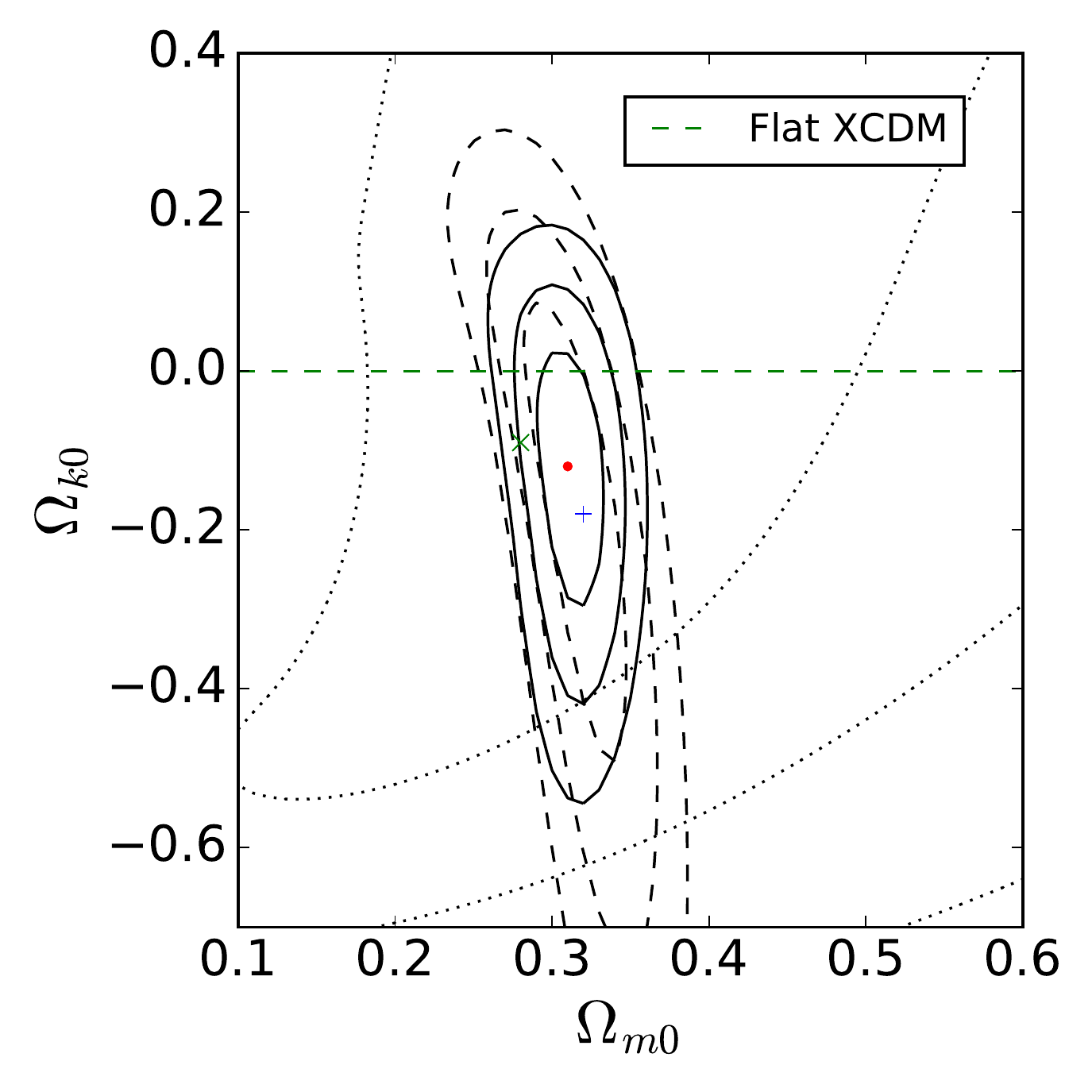}\par
    \includegraphics[width=\linewidth]{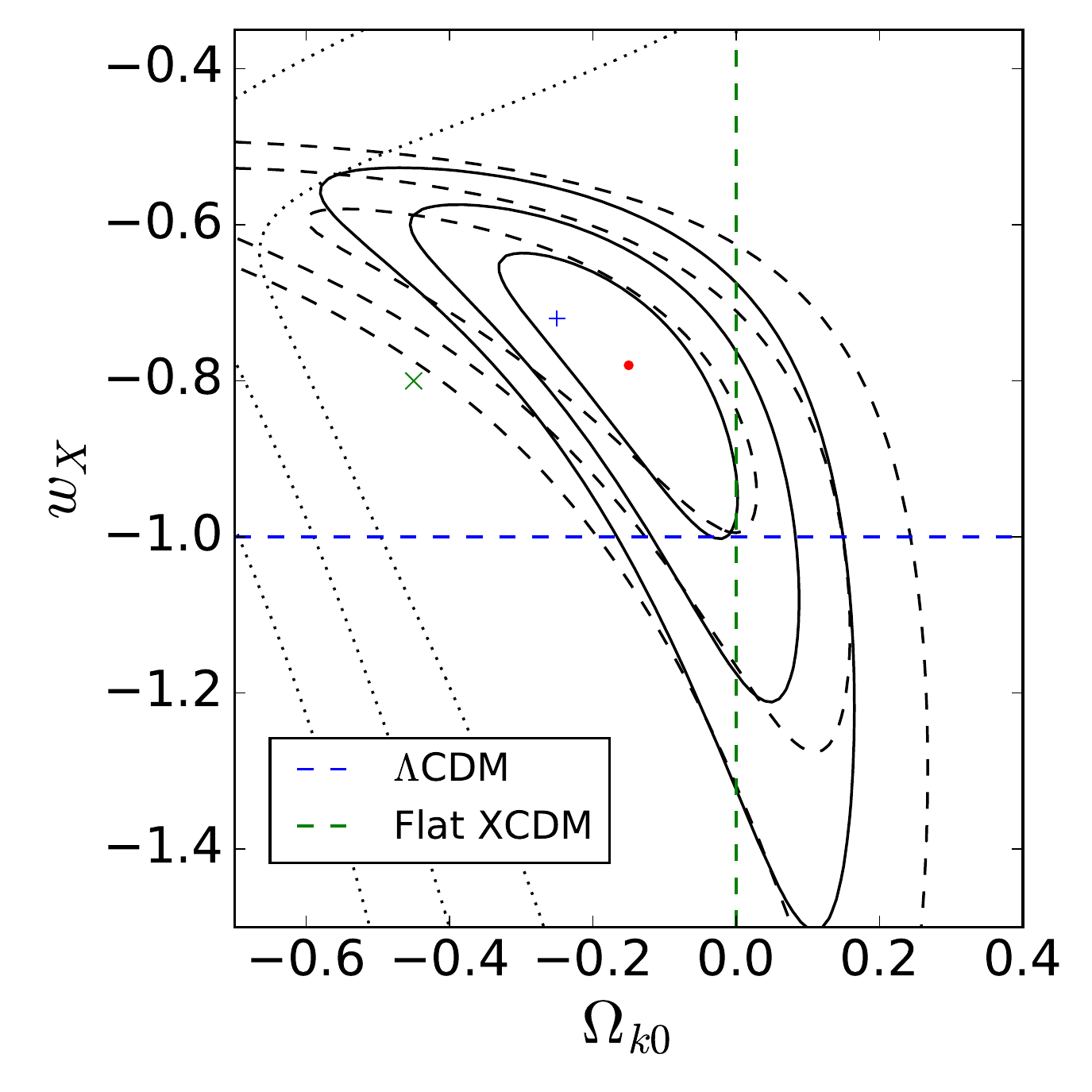}\par
    \includegraphics[width=\linewidth]{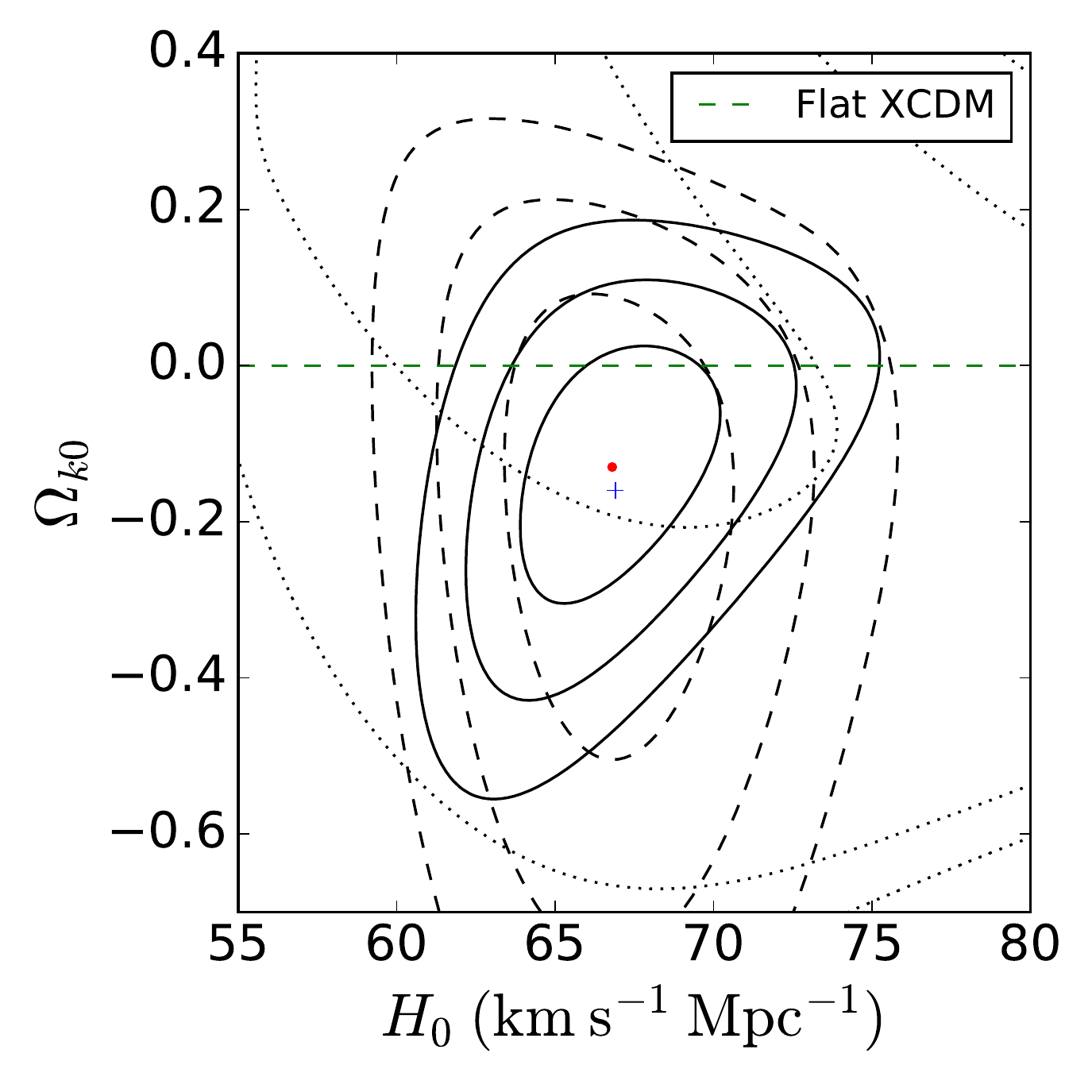}\par
    \includegraphics[width=\linewidth]{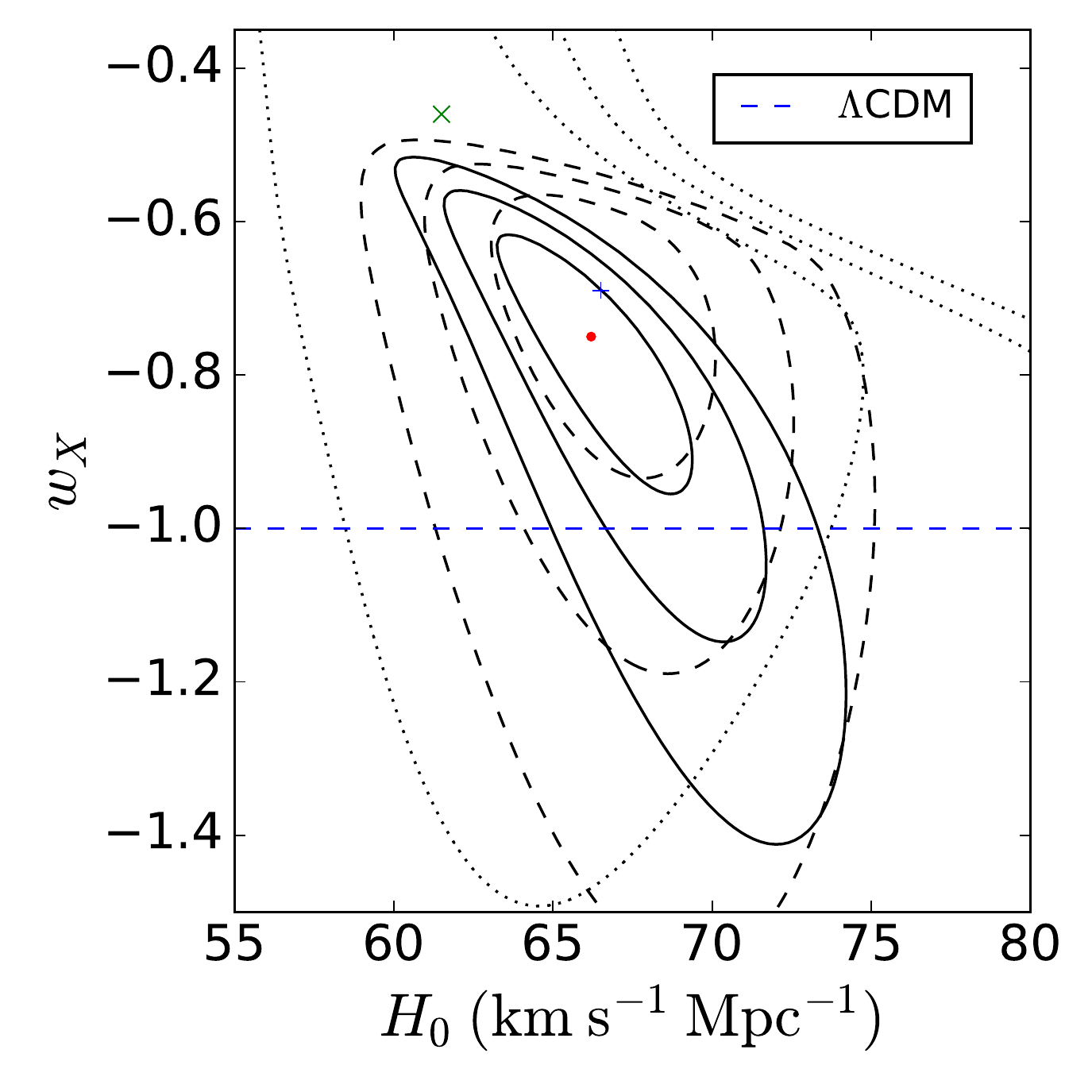}\par
    \includegraphics[width=\linewidth]{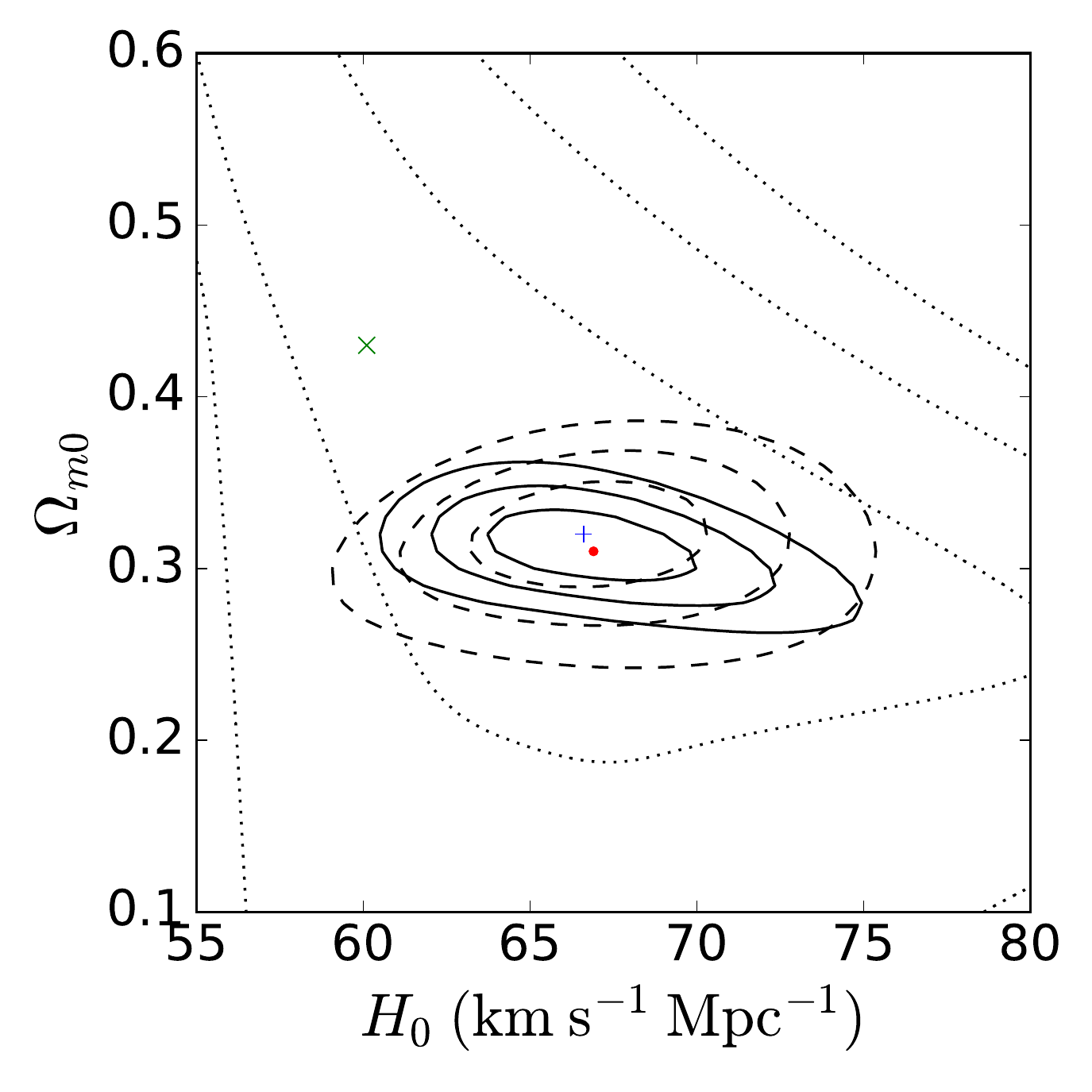}\par
\end{multicols}
\begin{multicols}{4}
    \includegraphics[width=\linewidth]{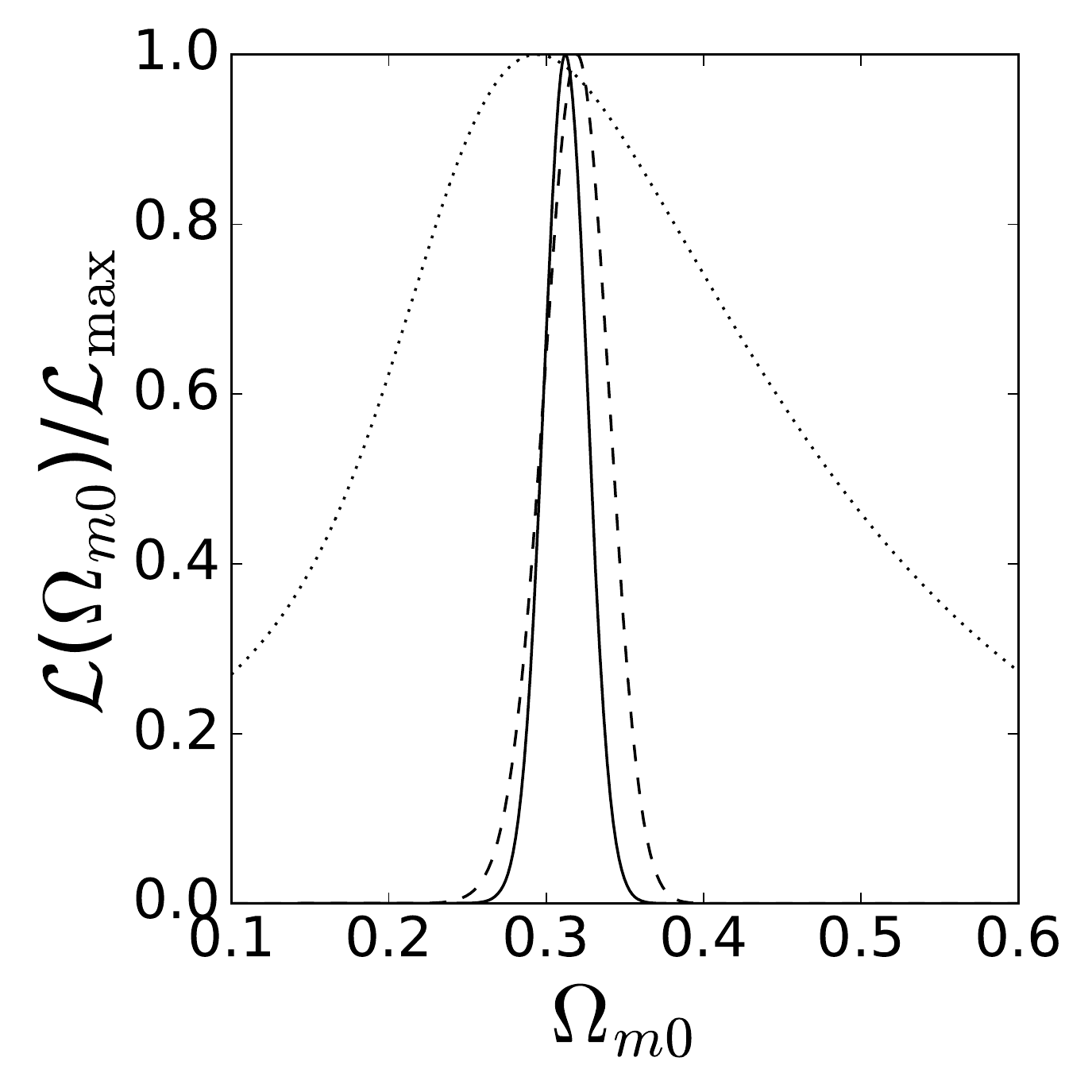}\par 
    \includegraphics[width=\linewidth]{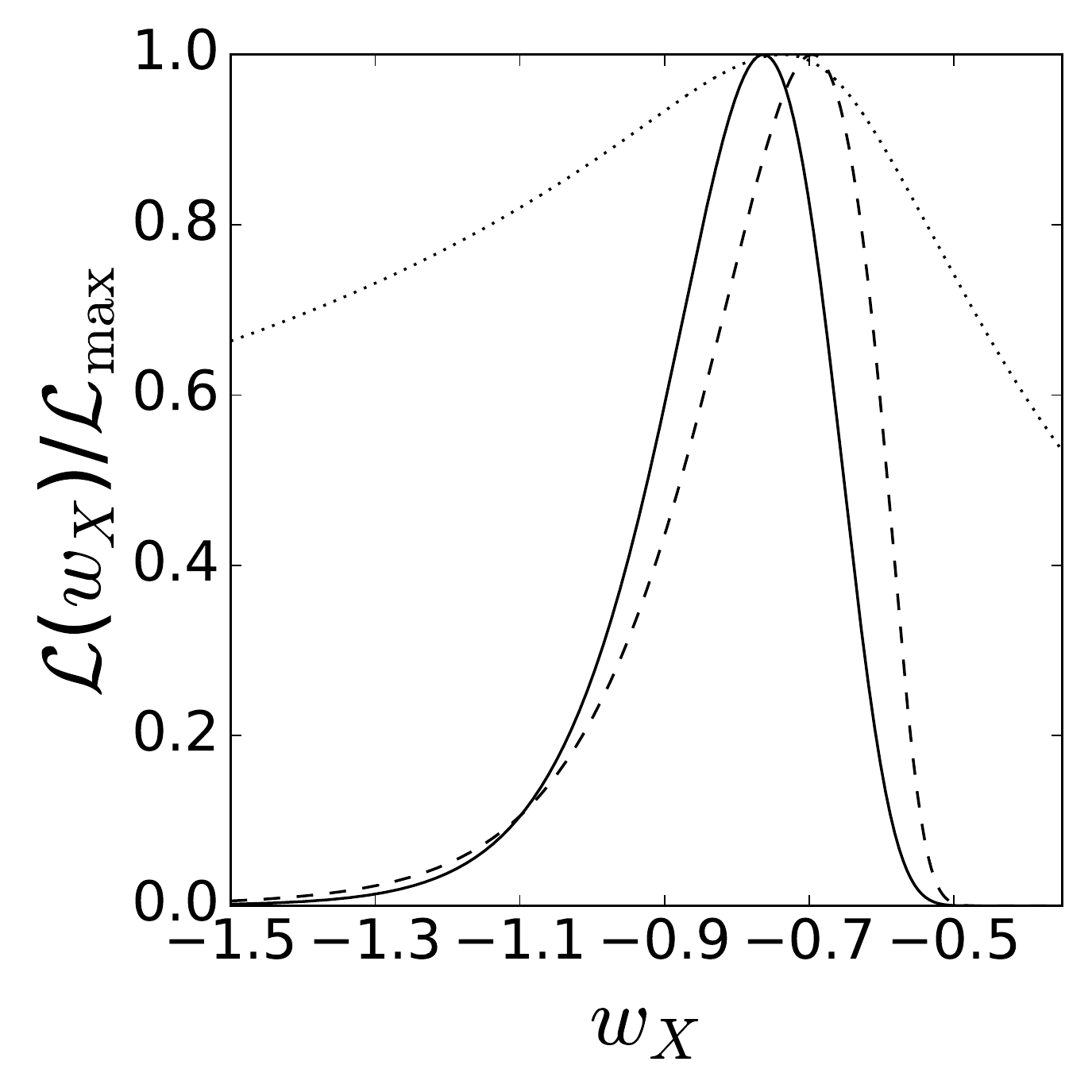}\par
    \includegraphics[width=\linewidth]{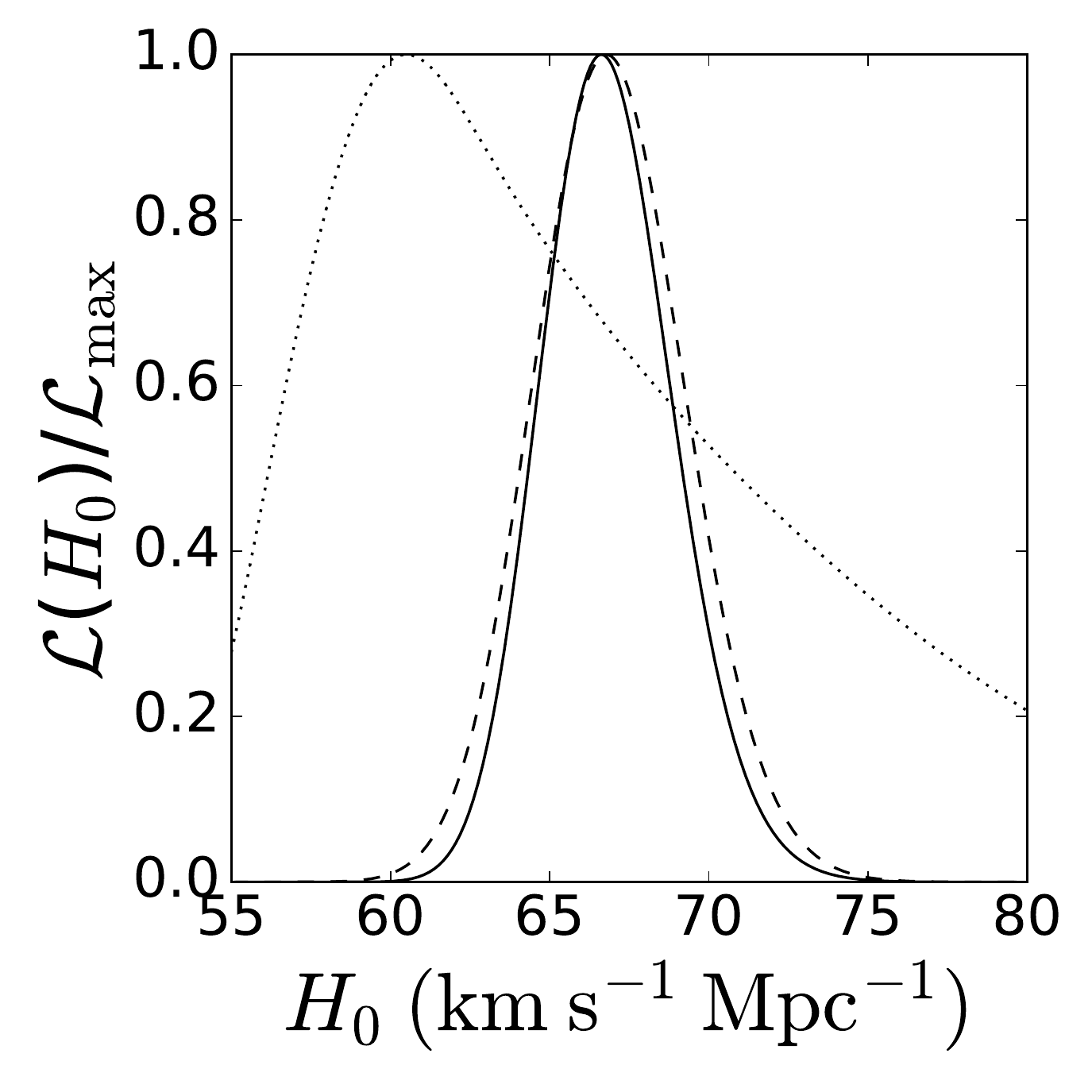}\par
    \includegraphics[width=\linewidth]{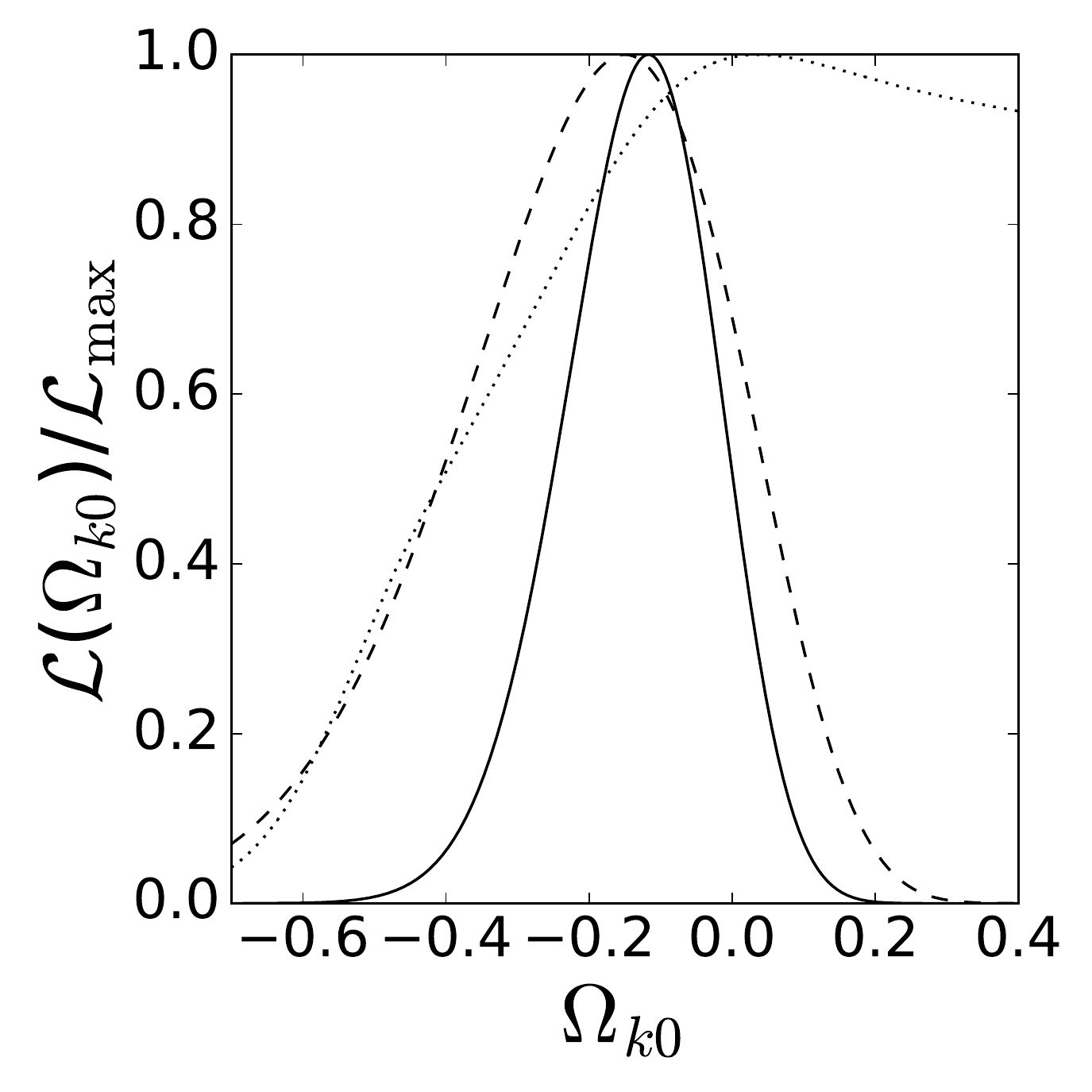}\par
\end{multicols}
\caption[Non-flat XCDM parametrization with QSO, $H(z)$ and BAO data.]{Non-flat XCDM parametrization with QSO, $H(z)$ and BAO data. Top and middle row: 1, 2, and 3$\sigma$ confidence contours and best-fitting points. In the top panels, the horizontal blue dashed line separates quintessence-type parametrizations of dark energy (for which $w_{X} > -1$) from phantom-type parametrization of dark energy (for which $w_{X} < -1$). Points on the blue line (for which $w_{X} = -1$) correspond to the non-flat $\Lambda$CDM model. The green dashed curve in the top left panel separates models that undergo accelerated expansion now from models that undergo decelerated expansion now. The vertical green dashed line in the top center panel, and the horizontal green dashed lines in the left and center panels of the middle row, separate spatially closed models (for which $\ok < 0$) from spatially open models (for which $\ok > 0$). Bottom panels: one-dimensional likelihoods for $\om$, $w_{X}$, $H_0$, and $\ok$. See text for description and discussion.}
\label{fig:Nonflat XCDM 2D (QSO+Hz+BAO)}
\end{figure*}

\begin{figure*}
\begin{multicols}{3}
    \includegraphics[width=\linewidth]{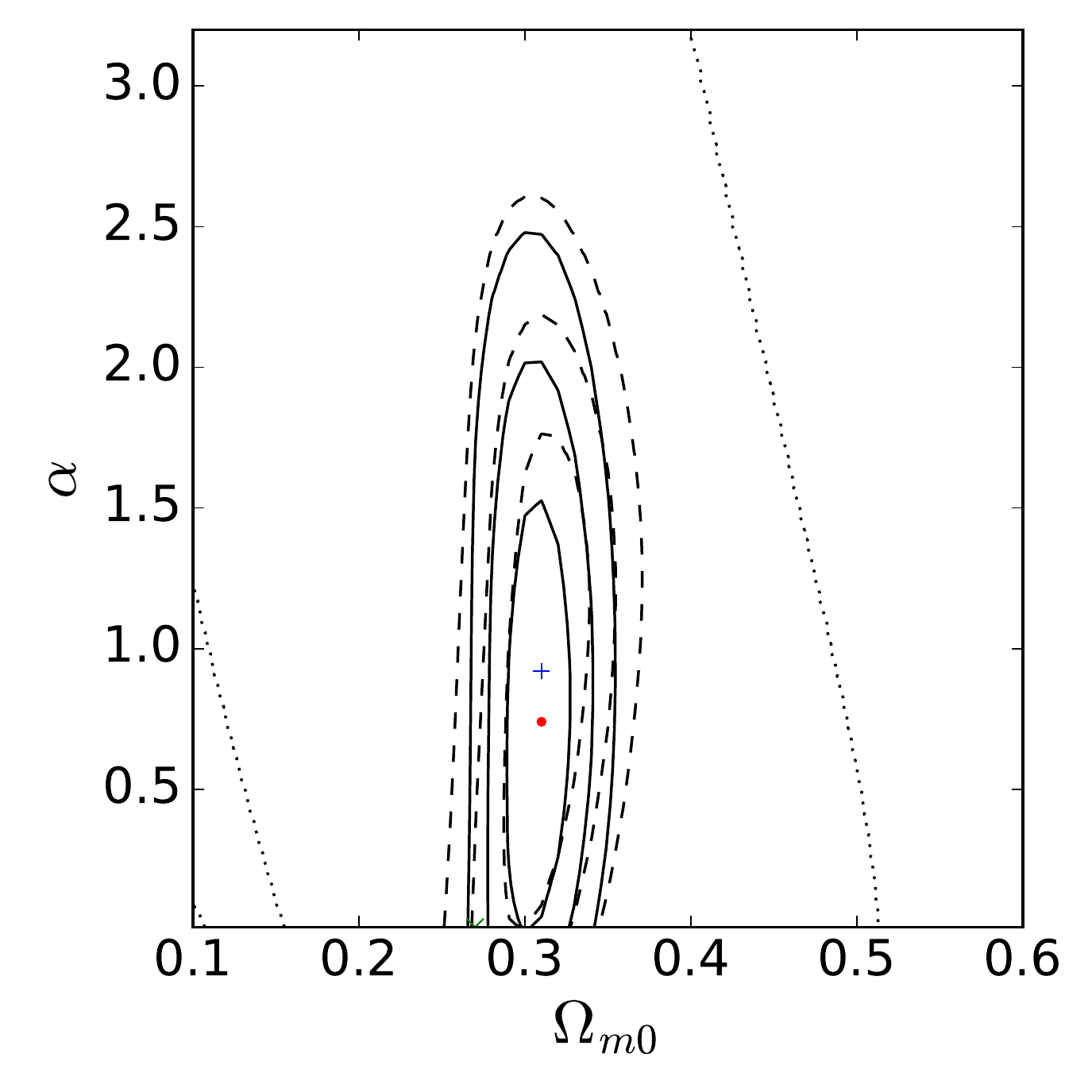}\par
    \includegraphics[width=\linewidth]{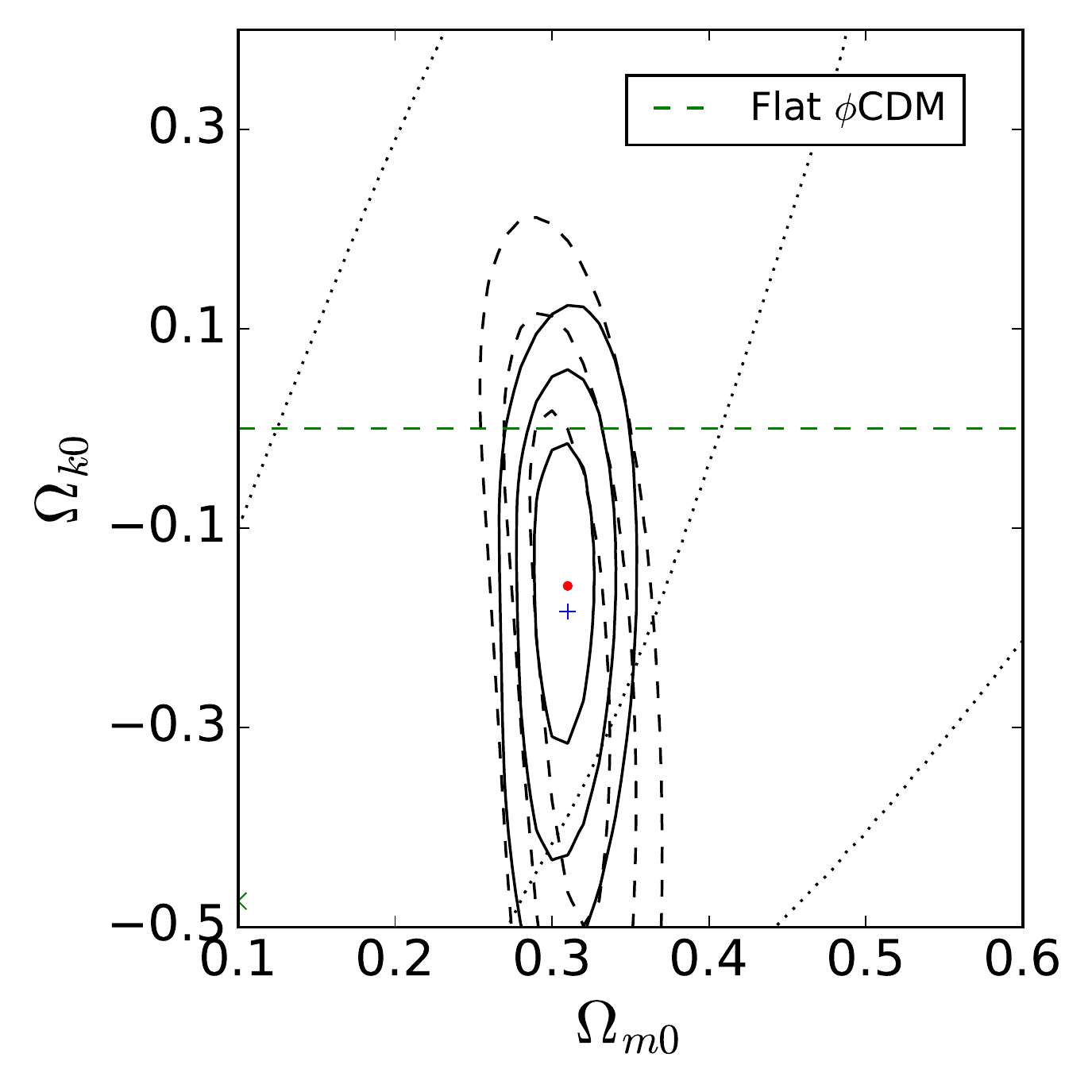}\par
    \includegraphics[width=\linewidth]{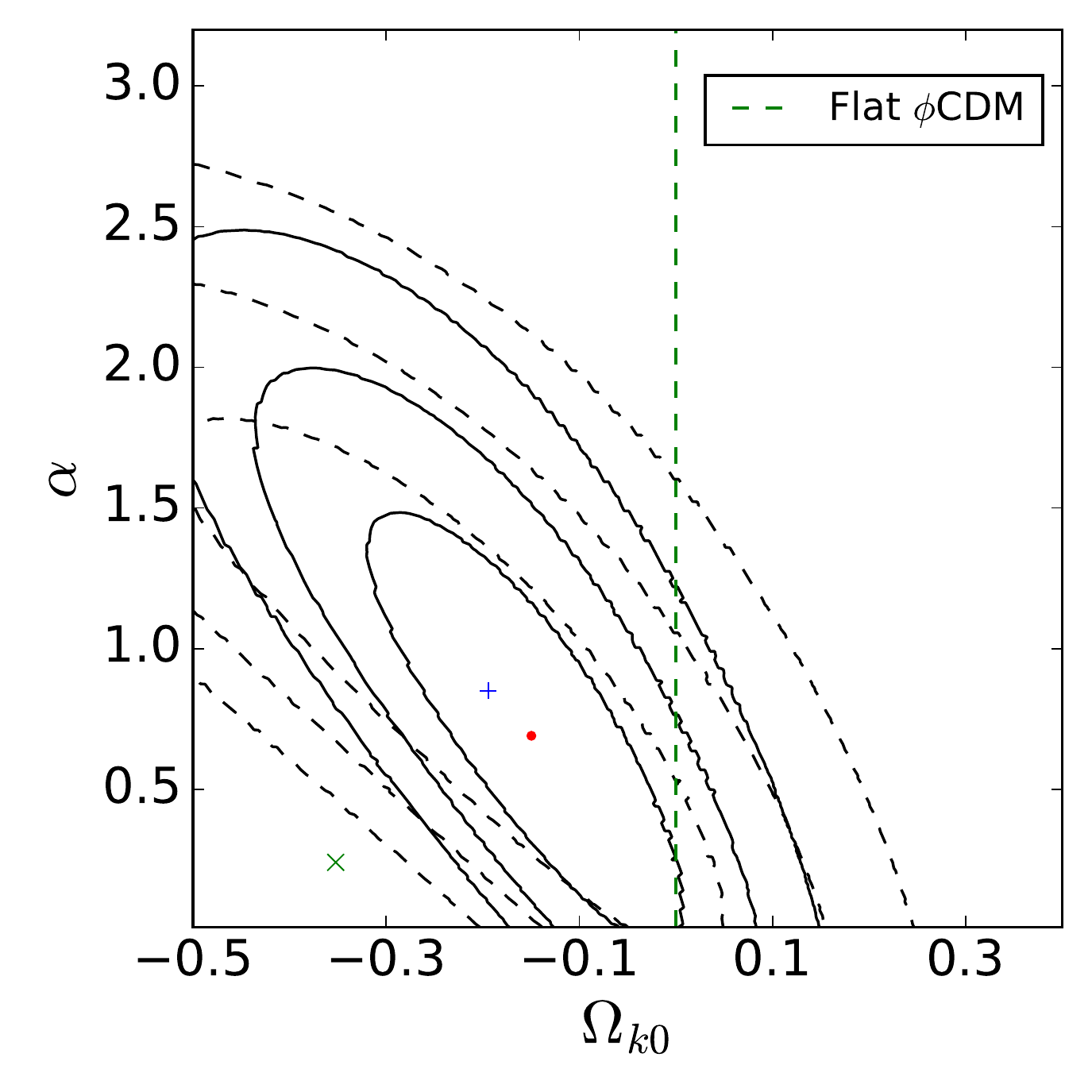}\par
    \includegraphics[width=\linewidth]{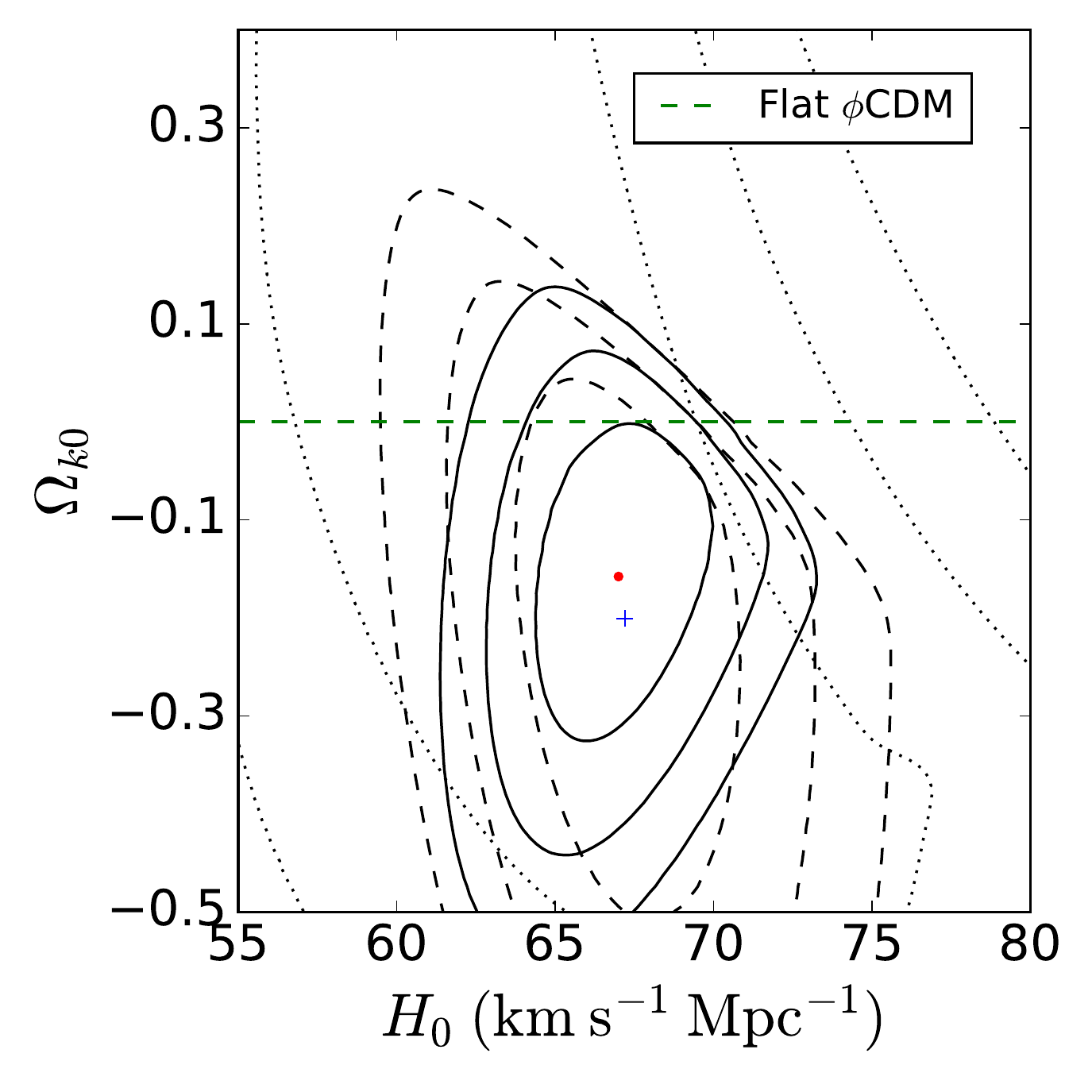}\par
    \includegraphics[width=\linewidth]{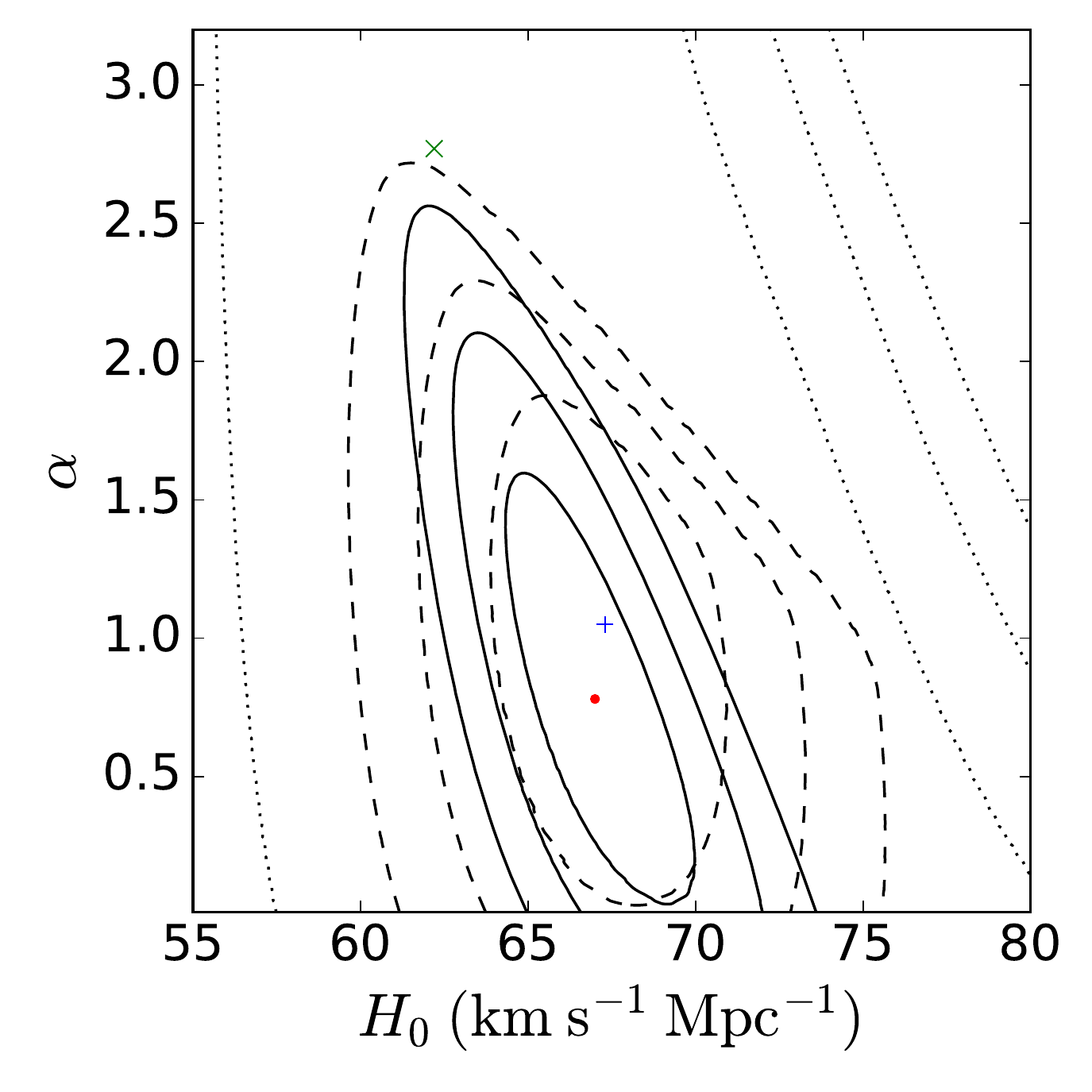}\par
    \includegraphics[width=\linewidth]{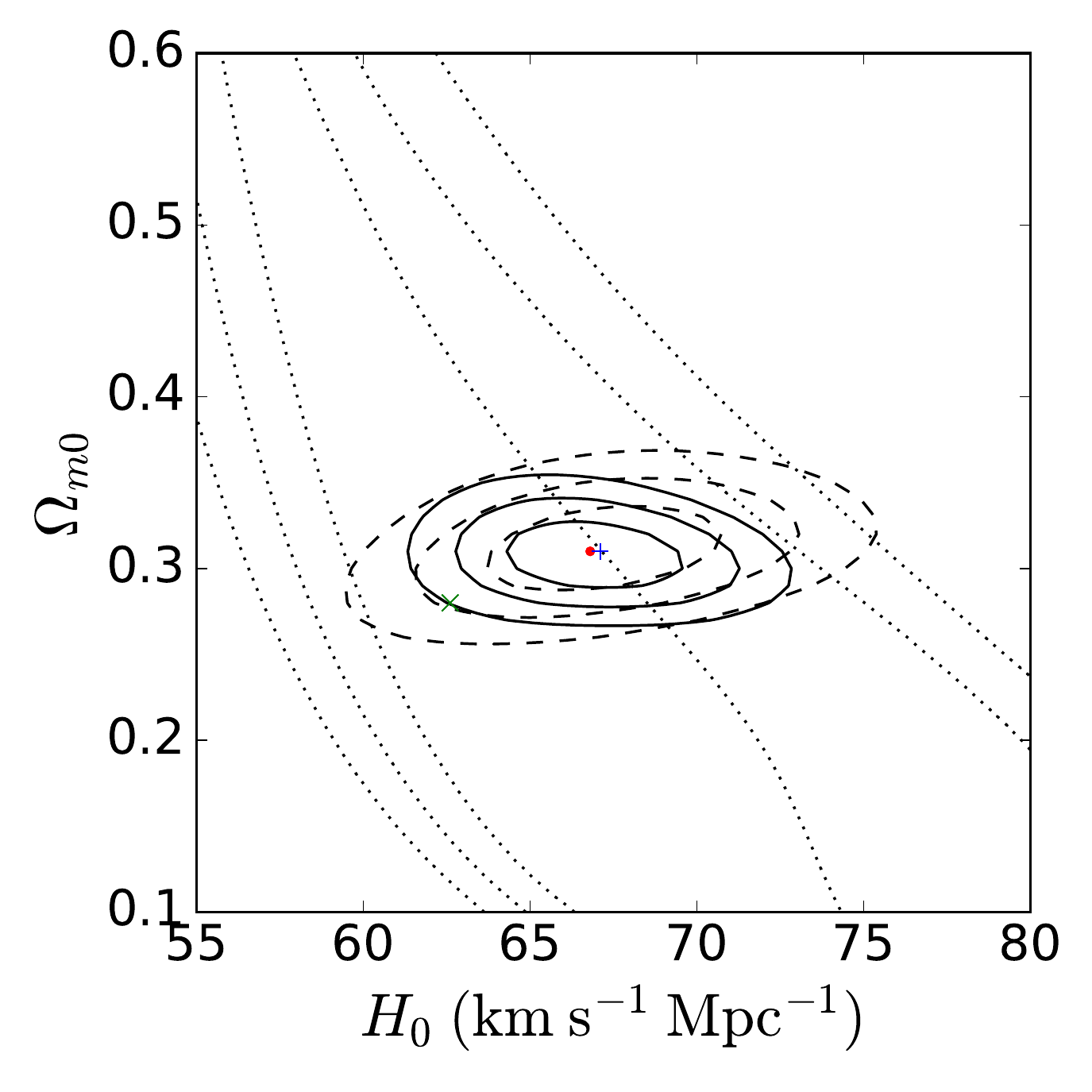}\par
\end{multicols}
\begin{multicols}{4}
    \includegraphics[width=\linewidth]{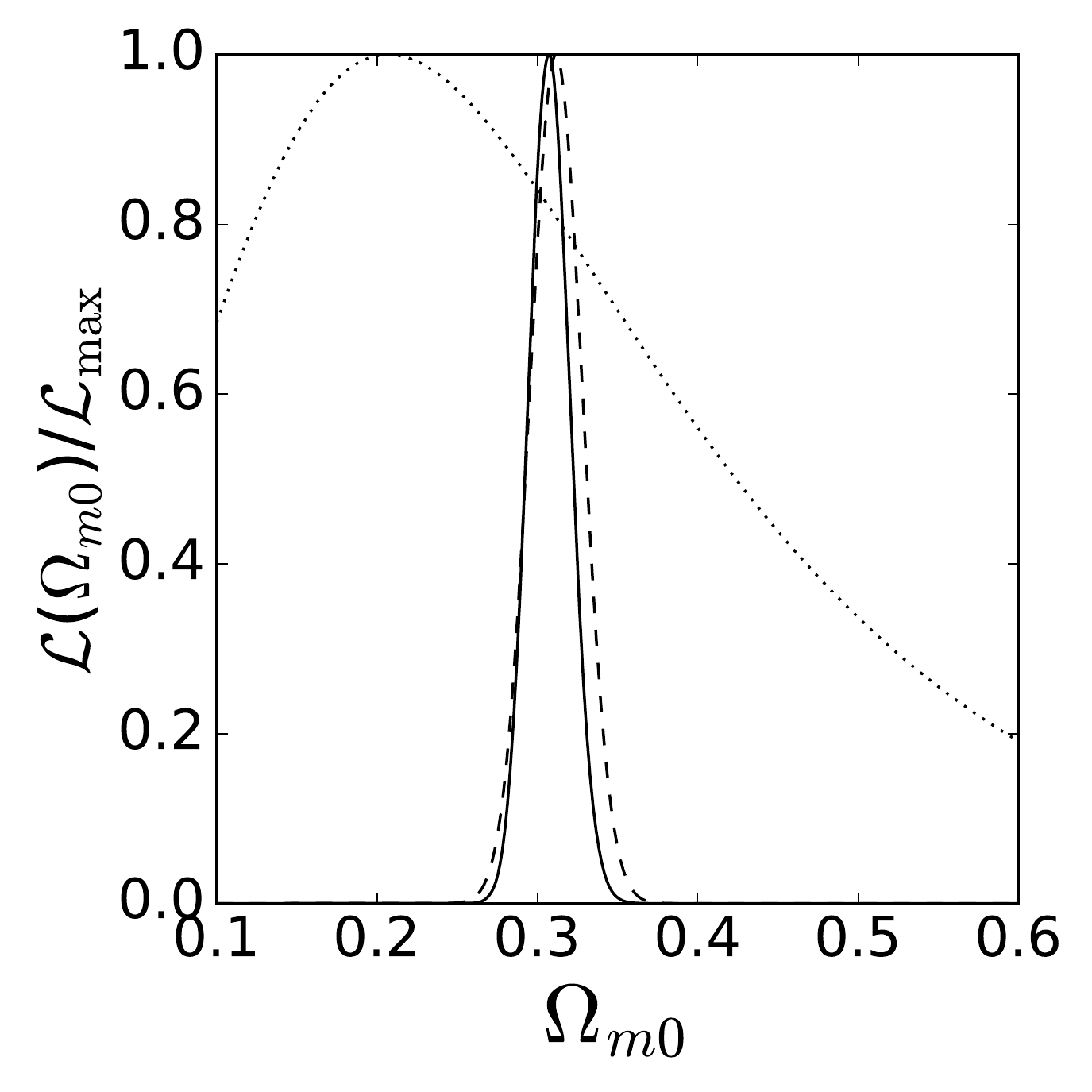}\par 
    \includegraphics[width=\linewidth]{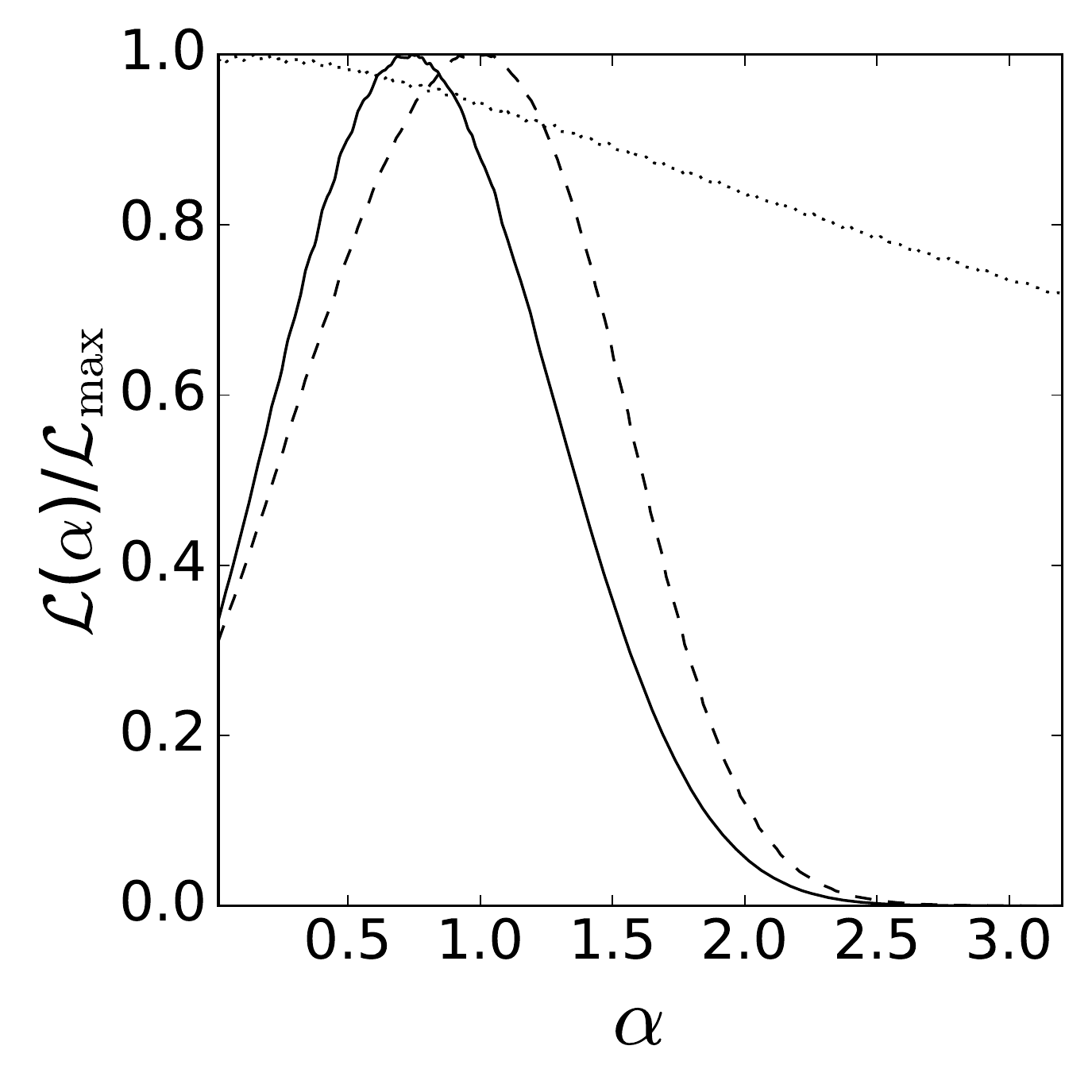}\par
    \includegraphics[width=\linewidth]{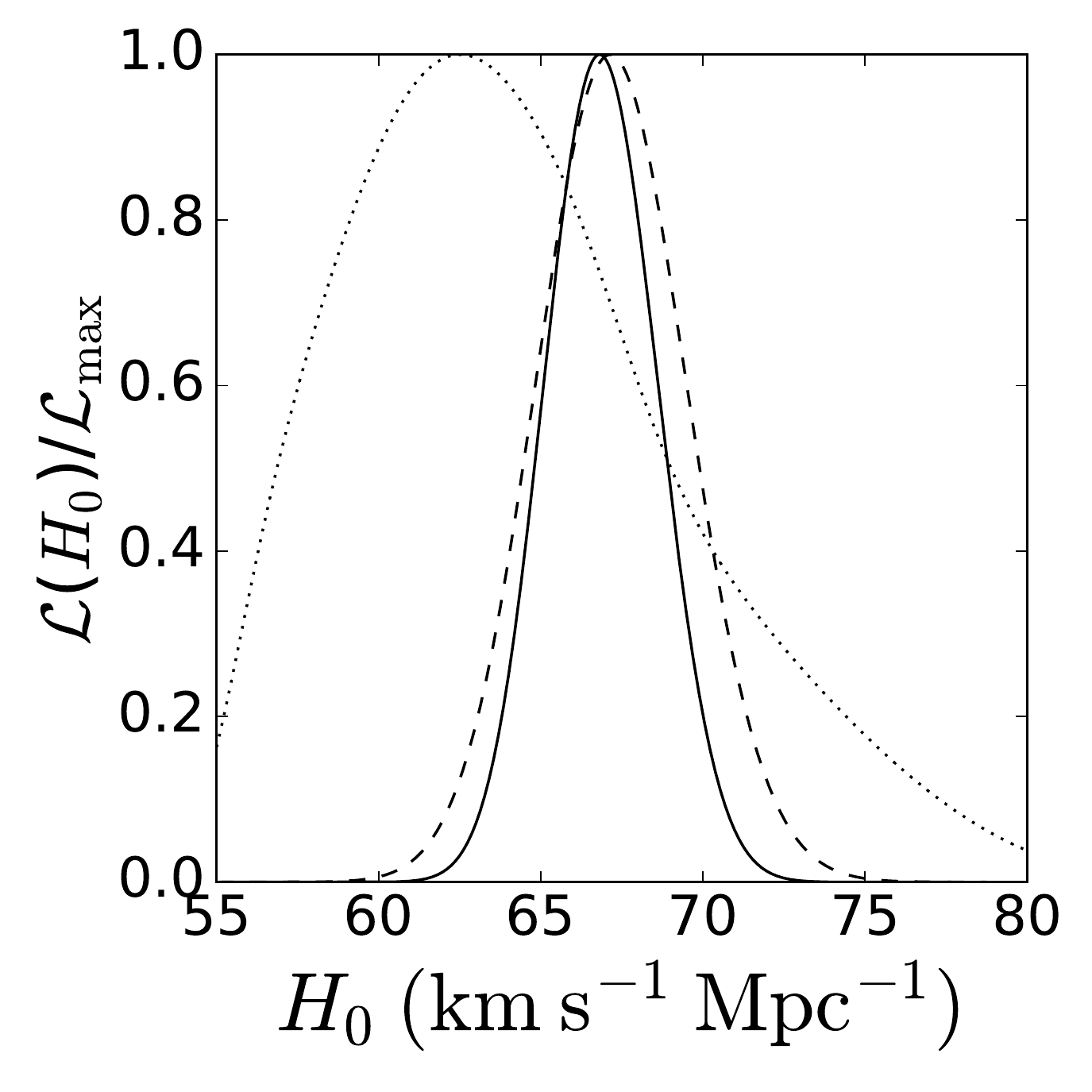}\par
    \includegraphics[width=\linewidth]{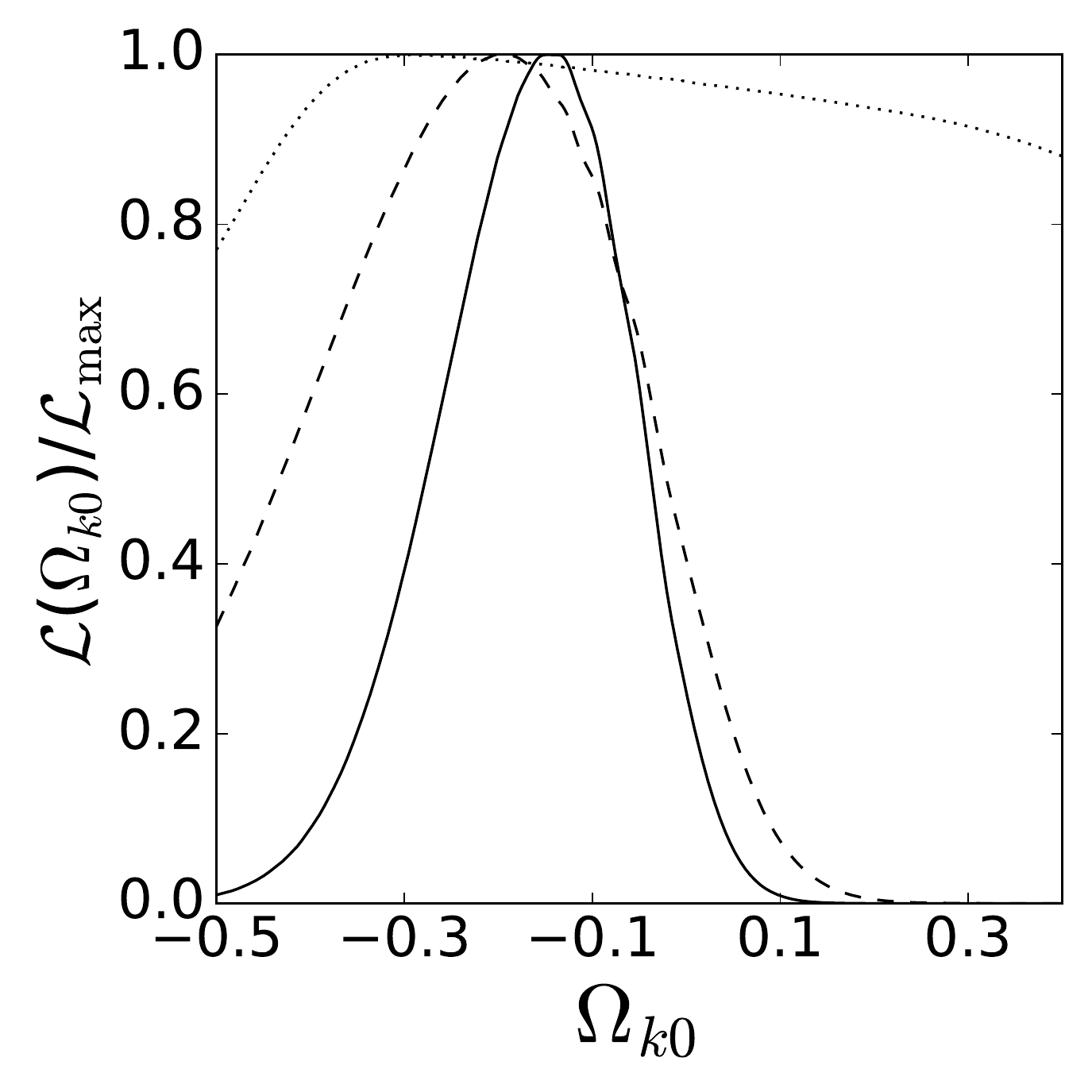}\par
\end{multicols}
\caption[Non-flat \pcdm\ model with QSO, $H(z)$, and BAO data.]{Non-flat \pcdm\ model with QSO, $H(z)$, and BAO data. Top and middle rows: 1, 2, and 3$\sigma$ confidence contours and best-fitting points. The vertical green dashed line in the top center panel, and the horizontal green dashed lines in the middle left and middle center panels, separate spatially closed models (with $\ok < 0$) from spatially open models (with $\ok > 0$). Points on the $\alpha = 0$ line in the top panels correspond to the non-flat \lcdm\ model. Bottom row: one-dimensional likelihoods for $\om$, $\alpha$, $\ok$, and $H_0$. See text for description and discussion.}
\label{fig:Nonflat phiCDM 2D QSO}
\end{figure*}

We discuss our results for the $H(z)$ + BAO data combination first (i.e. without the QSO data). The flat \lcdm\ model (with two free parameters, $H_0$ and $\Omega_{m0}$) two-dimensional $\chi^2$ confidence contours and one-dimensional normalized likelihood curves are plotted in Fig. \ref{fig:flat LCDM 2D QSO}. In Figs. \ref{fig:Non-flat LCDM 2D QSO+Hz+BAO}-\ref{fig:Nonflat phiCDM 2D QSO} we present our results for the non-flat $\Lambda$CDM model, the flat and non-flat XCDM parametrizations, and the flat and non-flat $\phi$CDM models. These results appear in Figs. \ref{fig:flat LCDM 2D QSO}-\ref{fig:Nonflat phiCDM 2D QSO} as two-dimensional dashed black likelihood contours and one-dimensional dashed black likelihood curves.

The best-fitting values of the parameters of our models, from their unmarginalized two-, three-, or four-dimensional likelihoods, are listed in Table \ref{tab:BFP}. This table also lists the number of degrees of freedom, $\nu$, and the values of $\chi^2$, $AIC$, and $BIC$ that correspond to the best-fitting parameters. The marginalized, one-dimensional best-fitting values of our model parameters, along with their 1$\sigma$ and 2$\sigma$ ranges, are listed in Table \ref{tab:1d intervals}.

\begin{table*}
	\caption{Best-fitting parameters and 1$\sigma$ and 2$\sigma$ confidence intervals for all models.}
	\label{tab:1d intervals}
	\centering
    \resizebox{\columnwidth}{!}{%
	\begin{threeparttable}
	\begin{tabular}{lccccc} % four columns, alignment for each
		\hline
		Model & Data set & Marginalization range\tnote{a} & Best-fitting & 1$\sigma$ & 2$\sigma$\\
		\hline
		Flat \lcdm\ & $H(z)$ + BAO & 0.1 $\leq \Omega_{m0} \leq$ 0.7 & $\Omega_{m0} = 0.30$ & 0.29 $\leq \Omega_{m0} \leq$ 0.32 & 0.27 $\leq \Omega_{m0} \leq$ 0.33\\
		 & & 50 $\leq H_0 \leq$ 85 & $H_0 = 67.99$ & $67.11 \leq H_0 \leq 68.90$ & $66.25 \leq H_0 \leq 68.91$\\
		\hline
		 & QSO + $H(z)$ + BAO & 0.1 $\leq \Omega_{m0} \leq$ 0.7 & $\Omega_{m0} = 0.31$ & 0.30 $\leq \Omega_{m0} \leq$ 0.32 & 0.28 $\leq \Omega_{m0} \leq$ 0.33\\
		 & & 50 $\leq H_0 \leq$ 85 & $H_0 = 68.44$ & 67.75 $\leq H_0 \leq $ 69.14 & 67.06 $\leq H_0 \leq $ 69.85\\
		\hline
		\hline
		Non-flat \lcdm\ & $H(z)$ + BAO & 0.1 $\leq \Omega_{m0} \leq$ 0.7 & $\Omega_{m0} = 0.30$ & 0.28 $\leq \Omega_{m0} \leq$ 0.31 & 0.27 $\leq \Omega_{m0} \leq$ 0.33\\
		 & & 50 $\leq H_0 \leq$ 85 & $H_0 = 68.24$ & $65.91 \leq H_0 \leq 70.63$ & $63.60 \leq H_0 \leq 73.03$\\
		 & & $0.2 \leq \ol \leq 1$ & $\ol = 0.70$ & $0.63 \leq \ol \leq 0.76$ & $0.55 \leq \ol \leq 0.82$\\
		\hline
		 & QSO + $H(z)$ + BAO & 0.1 $\leq \Omega_{m0} \leq$ 0.7 & $\Omega_{m0} = 0.30$ & 0.29 $\leq \Omega_{m0} \leq$ 0.31 & 0.28 $\leq \Omega_{m0} \leq$ 0.33\\
		 & & 50 $\leq H_0 \leq$ 85 & $H_0 = 69.32$ & $67.90 \leq H_0 \leq 70.74$ & $66.48 \leq H_0 \leq 72.16$\\
		 & & $0.2 \leq \ol \leq 1$ & $\ol = 0.73$ & $0.67 \leq \ol \leq 0.78$ & $0.61 \leq \ol \leq 0.82$\\
		\hline
		\hline
		Flat XCDM & $H(z)$ + BAO & $0.1 \leq \om \leq 0.7$ & $\om = 0.30$ & $0.29 \leq \om \leq 0.32$ & $0.27 \leq \om \leq 0.33$\\
		 & & $50 \leq H_0 \leq 85$ & $H_0 = 66.79$ & $64.47 \leq H_0 \leq 69.39$ & $62.23 \leq H_0 \leq 72.67$\\
		 & & $-2 \leq w_{X} \leq 0$ & $w_{X} = -0.93$ & $-1.05 \leq w_{X} \leq -0.83$ & $-1.19 \leq w_{X} \leq -0.74$\\
		\hline
		 & QSO + $H(z)$ + BAO & $0.1 \leq \om \leq 0.7$ & $\om = 0.31$ & $0.29 \leq \om \leq 0.32$ & $0.28 \leq \om \leq 0.34$\\
		 & & $50 \leq H_0 \leq 85$ & $H_0 = 68.00$ & $66.06 \leq H_0 \leq 70.27$ & $64.22 \leq H_0 \leq 72.67$\\
		 & & $-2 \leq w_{X} \leq 0$ & $w_{X} = -0.97$ & $-1.09 \leq w_{X} \leq -0.87$ & $-1.22 \leq w_{X} \leq -0.79$\\
		\hline
		\hline
		Flat \pcdm\ & $H(z)$ + BAO & $0.1 \leq \om \leq 0.7$ & $\om = 0.30$ & $0.29 \leq \om \leq 0.32$ & $0.27 \leq \om \leq 0.33$\\
		& & $50 \leq H_0 \leq 85$ & $H_0 = 66.13$ & $64.04 \leq H_0 \leq 67.51$ & $61.95 \leq H_0 \leq 68.73$\\
		& & $0.01 \leq \alpha \leq 3$ & $\alpha = 0.15$ & $0.06 \leq \alpha \leq 0.52$ & $0.02 \leq \alpha \leq 0.95$\\
		\hline
		& QSO + $H(z)$ +  BAO & $0.1 \leq \om \leq 0.7$ & $\om = 0.31$ & $0.30 \leq \om \leq 0.32$ & $0.29 \leq \om \leq 0.34$\\
		& & $50 \leq H_0 \leq 85$ & $H_0 = 67.19$ & $65.59 \leq H_0 \leq 68.19$ & $63.96 \leq H_0 \leq 69.09$\\
		& & $0.01 \leq \alpha \leq 3$ & $\alpha = 0.05$ & $0.02 \leq \alpha \leq 0.36$ & $0.01 \leq \alpha \leq 0.72$\\
		\hline
		\hline
		 Non-flat XCDM & $H(z)$ + BAO & $0.1 \leq \om \leq 0.7$ & $\om = 0.32$ & $0.30 \leq \om \leq 0.34$ & $0.27 \leq \om \leq 0.36$\\
		 & & $50 \leq H_0 \leq 85$ & $H_0 = 66.8$ & $64.5 \leq H_0 \leq 69.3$ & $62.3 \leq H_0 \leq 71.8$\\
		 & & $-2 \leq w_{X} \leq 0$ & $w_{X} = -0.70$ & $-0.89 \leq w_{X} \leq -0.62$ & $-1.1 \leq w_{X} \leq -0.56$\\
		 & & $-0.7 \leq \ok \leq 0.7$ & $\ok = -0.15$ & $-0.38 \leq \ok \leq 0.01$ & $-0.59 \leq \ok \leq 0.14$\\
		\hline
		 & QSO + $H(z)$ + BAO & $0.1 \leq \om \leq 0.7$ & $\om = 0.31$ & $0.30 \leq \om \leq 0.33$ & $0.28 \leq \om \leq 0.34$\\
		 & & $50 \leq H_0 \leq 85$ & $H_0 = 66.6$ & $64.7 \leq H_0 \leq 68.8$ & $62.9 \leq H_0 \leq 71.2$\\
		 & & $-2 \leq w_{X} \leq 0$ & $w_{X} = -0.76$ & $-0.92 \leq w_{X} \leq -0.68$ & $-1.1 \leq w_{X} \leq -0.61$\\
		 & & $-0.7 \leq \ok \leq 0.7$ & $\ok = -0.12$ & $-0.24 \leq \ok \leq -0.02$ & $-0.36 \leq \ok \leq 0.07$\\
		 \hline
		 \hline
		Non-flat \pcdm & $H(z)$ + BAO & $0.1 \leq \om \leq 0.7$ & $\om = 0.31$ & $0.29 \leq \om \leq 0.33$ & $0.28 \leq \om \leq 0.34$\\
		& & $50 \leq H_0 \leq 85$ & $H_0 = 67.1$ & $64.8 \leq H_0 \leq 69.5$ & $62.6 \leq H_0 \leq 71.9$\\
		& & $0.01 \leq \alpha \leq 5$ & $\alpha = 0.97$ & $0.44 \leq \alpha \leq 1.48$ & $0.01 \leq \alpha \leq 1.95$\\
		& & $-0.5 \leq \ok \leq 0.5$ & $\ok = -0.2$ & $-0.36 \leq \ok \leq -0.06$ & $-0.47 \leq \ok \leq 0.05$\\
		\hline
		& QSO + $H(z)$ + BAO & $0.1 \leq \om \leq 0.7$ & $\om = 0.31$ & $0.30 \leq \om \leq 0.32$ & $0.28 \leq \om \leq 0.34$\\
		& & $50 \leq H_0 \leq 85$ & $H_0 = 66.8$ & $65.1 \leq H_0 \leq 68.6$ & $63.5 \leq H_0 \leq 70.3$\\
		& & $0.01 \leq \alpha \leq 5$ & $\alpha = 0.74$ & $0.33 \leq \alpha \leq 1.27$ & $0.08 \leq \alpha \leq 1.79$\\
		& & $-0.5 \leq \ok \leq 0.5$ & $\ok = -0.15$ & $-0.26 \leq \ok \leq -0.06$ & $-0.38 \leq \ok \leq 0.02$\\
		\hline
	\end{tabular}
	\begin{tablenotes}
    \item[a]$H_0$ has units of ${\rm km}\hspace{1mm}{\rm s}^{-1}{\rm Mpc}^{-1}$.
    \end{tablenotes}
    \end{threeparttable}%
    }
\end{table*}

When it is measured using the $H(z)$ + BAO data combination, $\om$ has consistent best-fitting values, and tight confidence limits, across the models we studied (see Tables \ref{tab:1d intervals}). For the flat and non-flat \lcdm\ models, $\om = 0.30^{+0.02}_{-0.01}$ and $\om = 0.30^{+0.01}_{-0.02}$, respectively. For flat XCDM and flat \pcdm\ we find $\om = 0.30^{+0.02}_{-0.01}$, while non-flat XCDM and non-flat \pcdm\ favor the slightly larger values $\om = 0.32^{+0.02}_{-0.02}$ and $\om = 0.31^{+0.02}_{-0.02}$, respectively. Because our $\om$ step size is 0.01, the 1$\sigma$ error bars on $\om$ that we list here are probably somewhat inaccurate. The data, however, do determine $\om$ fairly precisely, with the error bars increasing a bit as the number of model parameters increase, as expected. These $\om$ estimates are in reasonable agreement with those made by \cite{park_ratra_2019b} from a similar compilation of $H(z)$ and BAO data.

The measurements of $H_0$ vary a bit less across the models we studied. For flat (non-flat) \lcdm\ we measure $H_0 = 67.99^{+0.91}_{-0.88} \left(68.24^{+2.39}_{-2.33}\right)$ km s$^{-1}$ Mpc$^{-1}$, while for flat (non-flat) XCDM $H_0 = 66.79^{+2.60}_{-2.32} \left(66.8^{+2.5}_{-2.3}\right)$ km s$^{-1}$ Mpc$^{-1}$, and for flat (non-flat) \pcdm\ we find $H_0 = 66.13^{+1.38}_{-2.09} \left(67.1^{+2.4}_{-2.3}\right)$ km s$^{-1}$ Mpc$^{-1}$, all with 1$\sigma$ error bars. Our step size is $\Delta H_0 = 0.01$ km s$^{-1}$ Mpc$^{-1}$ for the flat models and non-flat \lcdm, which we increased to $\Delta H_0 = 0.1$ km s$^{-1}$ Mpc$^{-1}$ for the non-flat XCDM and \pcdm\ cases, so the $H_0$ error bars are more accurate than those of the $\om$ measurements. These six measured $H_0$ values are mutually quite consistent. Aside from the flat \lcdm\ case, the $H_0$ central values and limits are very consistent with those found from a similar $H(z)$ + BAO data compilation in \cite{park_ratra_2019b}. Unlike here where we fix $\Omega_b h^2$ to the values obtained by \cite{Park_Ratra_2018_FLCDM, Park_Ratra_2018_FXCDM_NFXCDM, Park_Ratra_2018_FpCDM_NFpCDM}, \cite{park_ratra_2019b} allow the baryonic matter density parameter to vary, so the \cite{park_ratra_2019b} models have an additional free parameter compared to our models; this will have a bigger effect in the flat \lcdm\ case which has the fewest parameters. These $H_0$ measurements are more consistent with the recent median statistics estimate of $H_0 = 68 \pm 2.8$ km s$^{-1}$ Mpc$^{-1}$ \citep{chenratmed}, and with earlier median statistics estimates (\citealp{gott_etal_2001}, \citealp{76})\footnote{These $H_0$ measurements are also consistent with many other recent $H_0$ measurements (\citealp{chen_etal_2017}, \citealp{wang_xu_zhao_2017}, \citealp{Linishak}, \citealp{DES_2018}, \citealp{daSilva}, \citealp{Gomez-ValentAmendola2018}, \cite{planck2018}, \citealp{60}, \citealp{zhang_2018}, \citealp{zhanghuang}, \citealp{ruan_etal_2019}, \citealp{zhanghuangli}).} than with the recent measurement of $H_0 = 73.48 \pm 1.66$ km s$^{-1}$ Mpc$^{-1}$ determined from the local expansion rate \citep{riess2018}.\footnote{We note that other local expansion rate $H_0$ values are slightly lower, with slightly larger error bars. See, e.g., \cite{86}, \cite{Dhawan}, and \cite{FernandezArenas}.} As a comparison, both our highest and lowest $H_0$ measurements (those of non-flat \lcdm\ and flat \pcdm, respectively) are within 1$\sigma$ of the measurement made in \cite{chenratmed}, relative to the error bars of that measurement, but they are 1.8$\sigma$ (non-flat \lcdm) and 3.4$\sigma$ (flat \pcdm) lower than the \cite{riess2018} measurement (here $\sigma$ is the quadrature sum of the two measurement error bars, and these two cases span the range of differences).

As for spatial curvature, we find some evidence in favor of non-flat spatial hypersurfaces, although this evidence is fairly weak. For non-flat \lcdm, we measure $\ok = 0^{+0.06}_{-0.07}$, with 1$\sigma$ error bars, consistent with spatial flatness (see Table \ref{tab:1d intervals}). For the non-flat XCDM parametrization and non-flat \pcdm\ model, we measure $\ok = -0.15^{+0.16}_{-0.23}$, and $\ok = -0.20^{+0.14}_{-0.16}$, respectively (1$\sigma$ error bars). From these results we can see that non-flat XCDM is consistent with spatial flatness, but non-flat \pcdm\ favors closed spatial hypersurfaces at a little more than 1.4$\sigma$. For these three cases, using a similar $H(z)$ and BAO data compilation, \cite{park_ratra_2019b} find $\ok = -0.086 \pm 0.078$, $\ok = -0.32 \pm 0.11$, and $\ok = -0.24 \pm 0.15$, respectively, favoring closed geometry at 1.1$\sigma$, 2.9$\sigma$, and 1.6$\sigma$, respectively. Our measurements of spatial curvature are also consistent with the results obtained by other groups, particularly with the model-independent constraints obtained by \cite{yu_wang_2016}, \cite{rana_jain_mahajan_mukherjee_2017}, \cite{wei_wu_2017}, \cite{60}, and \cite{ruan_etal_2019}. We find some disagreement in the non-flat \pcdm\ case with the model-independent studies conducted by \cite{morescoOK} and \cite{Zheng_2019}; when compared to the measurements of $\ok$ made by these groups, we find that our non-flat \pcdm\ measurement of $\ok$ is not consistent with their measurements to 1$\sigma$, although it is consistent to 2$\sigma$, owing to the much larger error bars on our measurements.

Our results also show some evidence for dark energy dynamics, although like the evidence for $|\ok| \neq 0$ it is also weak. For example, our measurements in the flat (non-flat) XCDM cases are $w_{X} = -0.93^{+0.10}_{-0.12}\hspace{1mm} \text{at 1}\sigma \left(w_{X} = -0.70^{+0.08+0.14}_{-0.19-0.40} \hspace{1mm}\text{at 1 and 2}\sigma\right)$, which both favor quintessence-type dark energy, for which $w_{X} > -1$, over a $\Lambda$, though to different degrees of statistical significance. The best-fitting value of $w_{X}$ in the flat XCDM parametrization is within 0.7$\sigma$ of $w_{X} = -1$ (which corresponds to flat \lcdm), while the best-fitting value of $w_{X}$ in the non-flat XCDM parametrization is a little less than 1.6$\sigma$ away from $w_{X} = -1$ (non-flat \lcdm\ in this case). \cite{park_ratra_2019b} find $w_{X} = -0.72 \pm 0.16$ $(-0.604 \pm 0.099)$ for these two cases, from their $H(z)$ + BAO compilation, which favors quintessence-type dark energy over a $\Lambda$ at $1.8\sigma$ $(4\sigma)$. We find marginal evidence for dark energy dynamics in both the flat and non-flat \pcdm\ models, in which we measure $\alpha = 0.15^{+0.37+0.80}_{-0.09-0.13}$ and $\alpha = 0.97^{+0.51+0.98}_{-0.53-0.96}$, respectively (1$\sigma$ and 2$\sigma$ error bars). In both of these cases the measured values of $\alpha$ are a little more than 2$\sigma$ away from $\alpha = 0$ (corresponding to \lcdm), but this is due to the fact that, in both the flat and non-flat cases, the marginalized likelihood function for $\alpha$ terminates at $\alpha = 0$, the lower limit of our prior range on $\alpha$. The computation of a confidence limit on the low side of the marginalized likelihood function is therefore less meaningful than the computation of a confidence limit on the high side. Our results for $\alpha$ are in less precise agreement with \cite{park_ratra_2019b} than the results for our other parameters. \cite{park_ratra_2019b} find for flat \pcdm\ $\alpha = 2.5 \pm 1.6$ at 1$\sigma$ and $\alpha < 6.0$ at 2$\sigma$, while for non-flat \pcdm\ they find $\alpha = 3.1 \pm 1.5$ at 1$\sigma$.

Our results here cannot be directly compared to those of the previous chapter, because here $H_0$ is an adjustable parameter to be constrained by the data, while in Chapter \ref{Chapter4} we marginalized over $H_0$, assuming two different gaussian $H_0$ priors. However, we find that the results we have obtained, after marginalizing over $H_0$ with a flat prior, are qualitatively consistent with the results found in Chapter \ref{Chapter4}. 
%This is paragraph C
Further, although we have compared our parameter measurements to those of \cite{park_ratra_2019b}, a direct comparison of our best-fitting $\chi^2$ values to the best-fitting $\chi^2$ values of that paper is not possible because of the different numbers of parameters and data points those authors used,\footnote{\cite{park_ratra_2019b} use a BAO measurement that we do not; instead of the one gaussian approximation constraint at $z = 2.36$ from \cite{11} in Table \ref{tab:ch5_BAO_data} here, \cite{park_ratra_2019b} use the probability distribution that describes the shift of the BAO peak position in both the perpendicular and parallel directions to the line of sight.} but we agree qualitatively with their result that there are only small differences between the $\chi^2$ of the six models; for each data combination, the six models have relatively similar $\chi^2$, $AIC$, and $BIC$ values (see Table \ref{tab:BFP}).

\subsection{QSO + $H(z)$ + BAO constraints}
\label{sec:QSO+Hz+BAO}

Our results for the full data set, consisting of QSO data combined with $H(z)$ and BAO data, are presented in Tables \ref{tab:BFP}-\ref{tab:1d intervals} and in Figs. \ref{fig:flat LCDM 2D QSO}-\ref{fig:Nonflat phiCDM 2D QSO}. The two-dimensional dotted black likelihood contours and one-dimensional dotted black likelihood curves in Figs. \ref{fig:flat LCDM 2D QSO}-\ref{fig:Nonflat phiCDM 2D QSO} correspond to the QSO data alone. The two-dimensional solid black likelihood contours and one-dimensional solid black likelihood curves in Figs. \ref{fig:flat LCDM 2D QSO}-\ref{fig:Nonflat phiCDM 2D QSO} correspond to the full data set, namely QSO + $H(z)$ + BAO. By examining the two-dimensional likelihood contours and one-dimensional likelihood curves shown in Figs. \ref{fig:flat LCDM 2D QSO}-\ref{fig:Nonflat phiCDM 2D QSO}, we see that even though the QSO data by themselves are not able to tightly constrain cosmological parameters, they do contribute to a tightening of the constraints on these parameters when used in combination with $H(z)$ + BAO data.\footnote{We confirm the high reduced $\chi^2$ values for the QSO angular size data (see Tables \ref{tab:BFP}-\ref{tab:1d intervals}) found earlier by \cite{zheng_biesiada_cao_qi_zhu_2017} (see Table 2 of that paper), \cite{Qi_et_al2017} (see Table 5 of that paper), and \cite{Xu_et_al2018} (see Table 2 of that paper). What causes this is apparently not yet understood.}

\begin{comment}We retain the $H(z)$ + BAO category here so that we can see the extent to which the addition of the QSO data tightens the constraints from $H(z)$ + BAO alone, which we describe in detail below.\end{comment}

As with the $H(z)$ + BAO data combination, $\om$ has consistent central value and error bars when it is measured with the full data set (see Table \ref{tab:1d intervals}). For the flat and non-flat \lcdm\ models, $\om = 0.31^{+0.01}_{-0.01}$ and $\om = 0.30^{+0.01}_{-0.01}$, respectively. For flat XCDM and flat \pcdm\ we find $\om = 0.31^{+0.01}_{-0.02}$ and $\om = 0.31^{+0.01}_{-0.01}$. Non-flat XCDM and non-flat \pcdm\ have $\om = 0.31^{+0.02}_{-0.01}$ and $\om = 0.31^{+0.01}_{-0.01}$, respectively. These measurements have very similar central values and error bars to the measurements made using the $H(z)$ + BAO data combination, as shown by the one-dimensional likelihoods in Figs. \ref{fig:flat LCDM 2D QSO}-\ref{fig:Nonflat phiCDM 2D QSO}. 

\begin{comment}
Our results for the flat $\Lambda$CDM model are presented in fig. \ref{fig:flat LCDM 2D QSO}, with the two-dimensional confidence contours in the left panel, and the one-dimensional normalized likelihood distributions for $\om$ and $H_0$ in the center and right panels, respectively.
\end{comment}

For flat (non-flat) \lcdm\ we measure $H_0 = 68.44^{+0.70}_{-0.69} \left(69.32^{+1.42}_{-1.42}\right)$ km s$^{-1}$ Mpc$^{-1}$, while for flat (non-flat) XCDM $H_0 = 68.00^{+2.27}_{-1.94} \left(66.6^{+2.2}_{-1.9}\right)$ km s$^{-1}$ Mpc$^{-1}$, and for flat (non-flat) \pcdm\ we find $H_0 = 67.19^{+1.00}_{-1.60} \left(66.8^{+1.8}_{-1.7}\right)$ km s$^{-1}$ Mpc$^{-1}$, all 1$\sigma$ error bars. Compared to the cases without the QSO data in Sec. \ref{sec:Hz+BAO}, the central $H_0$ values here are a little larger (except in the non-flat XCDM and \pcdm\ cases) and the error bars are a little smaller. These $H_0$ estimates are still in very good agreement with that from median statistics \citep{chenratmed} but differ from that measured from the local expansion rate \citep{riess2018}, being between 1.9$\sigma$ (non-flat \lcdm) and 2.5$\sigma$ (non-flat XCDM) lower (as before, $\sigma$ refers to the quadrature sum of the error bars on the two measurements).

When we measure the curvature energy density parameter using the full data set, we find in the non-flat \lcdm\ model that $\ok = -0.03^{+0.05}_{-0.06}$, which is again consistent with flat spatial hypersurfaces, but with slightly tighter error bars. The same pattern holds when we measure $\ok$ in the non-flat XCDM parametrization and the non-flat \pcdm\ model, in which $\ok = -0.12^{+0.10}_{-0.12}$ and $\ok = -0.15^{+0.09}_{-0.11}$, respectively. Both of these measurements are slightly more consistent with closed spatial hypersurfaces than the corresponding measurements made using only the $H(z)$ + BAO data combination, being 1.2$\sigma$ (non-flat XCDM) and 1.7$\sigma$ (non-flat \pcdm) away from spatial flatness.

The parameters that govern dark energy dynamics move closer to \lcdm\ when we measure them with the full data set. In the flat (non-flat) XCDM parametrization, $w_{X} = -0.97^{+0.10}_{-0.12} \left(w_{X} = -0.76^{+0.08}_{-0.16}\right)$, with 1$\sigma$ error bars. In both cases we find that the addition of QSO data to the $H(z)$ + BAO data drives the value of $w_{X}$ closer to $w_{X} = -1$, the value that it takes in the flat and non-flat \lcdm\ models (although $w_{X}$ is still over 1$\sigma$ larger than $-1$ in the non-flat case). Something similar happens to $\alpha$; in the flat (non-flat) \pcdm\ model we measure $\alpha = 0.05^{+0.31+0.68}_{-0.03-0.04} \left(\alpha = 0.74^{+0.53+1.05}_{-0.41-0.66}\right)$, with 1 and 2$\sigma$ error bars that are tighter in the flat case than they are when $\alpha$ is measured using only $H(z)$ + BAO data. As in XCDM, the parameter that controls the dark energy dynamics, $\alpha$, is driven closer to $\alpha = 0$, the value that it takes when \pcdm\ reduces to \lcdm\ (though $\alpha$ is still measured to be about $2\sigma$ away from zero in the non-flat case).
\section{Conclusion}
\label{sec:ch5_conclusion}
We analyzed a total of 162 observations, 120 of which were measurements of the QSO angular sizes from \cite{Cao_et_al2017b}, with the remaining 42 measurements being a combination of $H(z)$ data and distance measurements from baryon acoustic oscillations (listed in Table \ref{tab:ch4_BAO_data}).

Our methods and models were largely the same in this chapter as in Chapter \ref{Chapter4}, with a few key differences. First, we treated $H_0$ as a free parameter, so as to obtain constraints on its value within the models we studied. We also presented results for each of our data sets separately and in combination (in Chapter \ref{Chapter4} we only presented results for the $H(z)$ + BAO data combination), and treated the sound horizon and $D_{M}$-$H(z)$ covariance matrix more accurately (see Sec. \ref{sec:methods}). After accounting for these differences, we find that our results for the energy density parameters $\om$ and $\ol$, the dark energy equation of state $w_{X}$, and the \pcdm\ potential energy density parameter $\alpha$ are largely consistent with those of Chapter \ref{Chapter4}.
%Any other differences?

Adding QSO data to $H(z)$ and BAO data tightens parameter constraints in some of the models we studied. In particular, using the full data set, we find that there is some evidence for closed spatial hypersurfaces in dynamical dark energy models, but that this evidence is only marginally significant (being between 1.2$\sigma$ and 1.7$\sigma$, depending on the model considered). We also find that there is marginal evidence for dark energy dynamics in both flat and non-flat models, ranging from around 0.7$\sigma$ to a little more than 2$\sigma$, depending on the model. A little more significant is the evidence we find in favor of a lower value of the Hubble constant. Our $H_0$ results are more consistent with the results of \cite{chenratmed} and \cite{planck2018} than that of \cite{riess2018}, being between 1.9$\sigma$ lower than the measurement made by \cite{riess2018} in the non-flat \lcdm\ model and 2.5$\sigma$ lower than said measurement in the non-flat XCDM parametrization (although these error bars on $H_0$ are not as wide as the error bars $H_0$ when $H_0$ is measured using only the $H(z)$ + BAO data combination).

%% file: chapter7.tex
\cleardoublepage

\chapter{Cosmological constraints from HII starburst galaxy apparent magnitude and other cosmological measurements}
\chaptermark{Constraints from HIIG and other data}

\label{Chapter7}

This chapter is based on \cite{Cao_Ryan_Ratra_2020}. Figures and tables by Shulei Cao, from analyses conducted indepedently by Shulei Cao and Joseph Ryan.

%%
%Section: Intro
%%
\section{Introduction} 
\label{sec:ch7_intro}

The major goal of this chapter is to use measurements of the redshift, apparent luminosity, and gas velocity dispersion of HII starburst galaxies to constrain cosmological parameters.\footnote{For early attempts see \cite{Siegel_2005}, \cite{Plionis_2009,Plionis_2010,Plionis_2011} and \cite{Mania_2012}. For more recent discussions see \cite{Chavez_2016}, \cite{Wei_2016}, \cite{Yennapureddy_2017}, \cite{Zheng_2019}, \cite{ruan_etal_2019}, \cite{G-M_2019}, \cite{Wan_2019}, and \cite{Wu_2020}.}
An HII starburst galaxy (hereinafter ``\hiig'') is one that contains a large HII region, an emission nebula sourced by the UV radiation from an O- or B-type star. There is a correlation between the measured luminosity ($L$) and the inferred velocity dispersion ($\sigma$) of the ionized gases within these \hiig, referred to as the $L$-$\sigma$ relation (see Section \ref{sec:ch7_Data}) which has been shown to be a useful cosmological tracer (see \citealp{Melnick_2000,Siegel_2005,Plionis_2011,Chavez_2012,Chavez_2014,Chavez_2016,Terlevich_2015,G-M_2019}, and references therein). This relation has been used to constrain the Hubble constant $H_0$ (\citealp{Chavez_2012,FernandezArenas}), and it can also be used to put constraints on the dark energy equation of state parameter $w$ (\citealp{Terlevich_2015,Chavez_2016,G-M_2019}).

\hiig\ data reach to redshift $z\sim2.4$, a little beyond that of the highest redshift baryon acoustic oscillation (BAO) data which reach to $z\sim2.3$. \hiig\ data are among a handful of cosmological observations that probe the largely unexplored part of redshift space from $z\sim2$ to $z\sim1100$. Other data that probe this region include quasar angular size measurements that reach to $z\sim2.7$ (see Chapter \ref{Chapter5}, as well as \citealp{gurvits_kellermann_frey_1999,Chen_Ratra_2003,Cao_et_al2017b} and references therein), quasar flux measurements that reach to  $z\sim5$ (\citealp{RisalitiandLusso_2015,RisalitiandLusso_2019,Yang_2019,KhadkaandRatra_2020,Khadka_2020a,Zheng_2020}, and references therein), and gamma ray burst data that reach to $z\sim8$ (\citealp{Lamb_2000,samushia_ratra_2010,Demianski_2019}, and references therein). In this paper we also use quasar angular size measurements (hereinafter ``QSO'') to constrain cosmological model parameters.

While \hiig\ and QSO data probe the largely unexplored $z\sim2.3$--2.7 part of the universe, current \hiig\ and QSO measurements provide relatively weaker constraints on cosmological parameters than those provided by more widely used measurements, such as BAO peak length scale observations or Hubble parameter (hereinafter ``$H(z)$'') observations (with these latter data being at lower redshift but of better quality than \hiig\ or QSO data). However, we find that the \hiig\ and QSO constraints are consistent with those that follow from BAO and $H(z)$ data, and so we use all four sets of data together to constrain cosmological parameters. We find that the \hiig\ and QSO data tighten parameter constraints relative to the $H(z)$ + BAO only case.

Using six different cosmological models\footnote{As in Chapters \ref{Chapter4} and \ref{Chapter5} we constrain the flat and nonflat $\Lambda$CDM model, the flat and nonflat XCDM parametrization, and the flat and nonflat $\phi$CDM model.} to constrain cosmological parameters allows us to determine which of our results are less model-dependent. In all models, the \hiig\ data favor those parts of cosmological parameter space for which the current cosmological expansion is accelerating.\footnote{This result could weaken, however, as the \hiig\ data constraint contours could broaden when \hiig\ data systematic uncertainties are taken into account. We do not incorporate any \hiig\ systematic uncertainties into our analysis; see below.} The joint analysis of the \hiig\, QSO, BAO and $H(z)$ data results in relatively model-independent and fairly tight determination of the Hubble constant $H_0$ and the current non-relativistic matter density parameter $\Omega_{m0}$.\footnote{The BAO and $H(z)$ data play a more significant role than do the \hiig\ and QSO data in setting these and other limits, but the \hiig\ and QSO data tighten the BAO + $H(z)$ constraints. We note, however, that the $H(z)$ and QSO data, by themselves, give lower central values of $H_0$ but with larger error bars. Also, because we calibrate the distance scale of the BAO measurements listed in Table \ref{tab:BAO} via the sound horizon scale at the drag epoch ($r_s$, about which see below), a quantity that depends on early-Universe physics, we would expect these measurements to push the best-fitting values $H_0$ lower when they are combined with late-Universe measurements like \hiig\ (whose distance scale is not set by the physics of the early Universe).} Depending on the model, $\Omega_{m0}$ ranges from a low of $0.309^{+0.015}_{-0.014}$ to a high of $0.319 \pm 0.013$, being consistent with most other estimates of this parameter (unless indicated otherwise, uncertainties given in this paper are $\pm 1\sigma$). The best-fitting values of $H_0$, ranging from $68.18^{+0.97}_{-0.75}$ \hunit to $69.90 \pm 1.48$ \hunit, are, from the quadrature sum of the error bars, 2.01$\sigma$ to 3.40$\sigma$ lower than the local $H_0 = 74.03 \pm 1.42$ \hunit measurement of \cite{riess_etal_2019} and only 0.06$\sigma$ to 0.60$\sigma$ higher than the median statistics $H_0 = 68 \pm 2.8$ km s$^{-1}$ Mpc$^{-1}$ estimate of \cite{chenratmed}. These combined measurements are consistent with the spatially-flat \lcdm\ model, but also do not strongly disallow some mild dark energy dynamics, as well as a little non-zero spatial curvature energy density.
\begin{comment}
This paper is organized as follows. In Section \ref{sec:ch7_Data} we introduce the data we use. Chapter \ref{Chapter3} describes the models we analyze, with a description of our analysis method in Section \ref{sec:ch7_Methods}. Our results are in Section \ref{sec:ch7_Results}, and we provide our conclusions in Section \ref{sec:ch7_conclusion}. 
\end{comment}
%%
%Section: Data
%%
\section{Data}
\label{sec:ch7_Data}
\begin{table}
\centering
\begin{threeparttable}
\caption{BAO data.}\label{tab:BAO}
\setlength{\tabcolsep}{1.5mm}{
\begin{tabular}{lccc}
\hline
$z$ & Measurement\tnote{a} & Value & Ref.\\
\hline
$0.38$ & $D_M\left(r_{s,{\rm fid}}/r_s\right)$ & 1512.39 & \cite{Alam_et_al_2017}\tnote{b}\\
$0.38$ & $H(z)\left(r_s/r_{s,{\rm fid}}\right)$ & 81.2087 & \cite{Alam_et_al_2017}\tnote{b}\\
$0.51$ & $D_M\left(r_{s,{\rm fid}}/r_s\right)$ & 1975.22 & \cite{Alam_et_al_2017}\tnote{b}\\
$0.51$ & $H(z)\left(r_s/r_{s,{\rm fid}}\right)$ & 90.9029 & \cite{Alam_et_al_2017}\tnote{b}\\
$0.61$ & $D_M\left(r_{s,{\rm fid}}/r_s\right)$ & 2306.68 & \cite{Alam_et_al_2017}\tnote{b}\\
$0.61$ & $H(z)\left(r_s/r_{s,{\rm fid}}\right)$ & 98.9647 & \cite{Alam_et_al_2017}\tnote{b}\\
$0.122$ & $D_V\left(r_{s,{\rm fid}}/r_s\right)$ & $539\pm17$ & \cite{Carter_2018}\\
$0.81$ & $D_A/r_s$ & $10.75\pm0.43$ & \cite{DES_2019b}\\
$1.52$ & $D_V\left(r_{s,{\rm fid}}/r_s\right)$ & $3843\pm147$ & \cite{3}\\
$2.34$ & $D_H/r_s$ & 8.86 & \cite{Agathe}\tnote{c}\\
$2.34$ & $D_M/r_s$ & 37.41 & \cite{Agathe}\tnote{c}\\
\hline
\end{tabular}}
\begin{tablenotes}
\item[a] $D_M \left(r_{s,{\rm fid}}/r_s\right)$, $D_V \left(r_{s,{\rm fid}}/r_s\right)$, $r_s$, and $r_{s, {\rm fid}}$ have units of Mpc, while $H(z)\left(r_s/r_{s,{\rm fid}}\right)$ has units of \hunit, and $D_A/r_s$ is dimensionless.
\item[b] The six measurements from \cite{Alam_et_al_2017} are correlated; see eq. (\ref{covmat}) for their correlation matrix.
\item[c] The two measurements from \cite{Agathe} are correlated; see eq. \eqref{CovM} below for their correlation matrix.
\end{tablenotes}
\end{threeparttable}
\end{table}

We use a combination of $H(z)$, BAO, QSO, and \hiig\ data to obtain constraints on our cosmological models. The $H(z)$ data, spanning the redshift range $0.070 \leq z \leq 1.965$, are identical to the $H(z)$ data used in Chapters \ref{Chapter4} and \ref{Chapter5} and compiled in Table \ref{tab:H(z)_data}. The QSO data compiled by \cite{Cao_et_al2017b} (listed in Table 1 of that paper) and spanning the redshift range $0.462 \leq z \leq 2.73$, are identical to that used in Chapter \ref{Chapter5}. Our BAO data (see Table \ref{tab:BAO}) have been updated relative to Chapter \ref{Chapter5} and span the redshift range $0.38 \leq z \leq 2.34$. Our \hiig\ data are new, comprising 107 low redshift ($0.0088 \leq z \leq 0.16417$) \hiig\ measurements, used in \cite{Chavez_2014}, and 46 high redshift ($0.636427 \leq z \leq 2.42935$) \hiig\ measurements, used in \cite{G-M_2019}.\footnote{15 from \cite{G-M_2019}, 25 from \cite{Erb_2006}, \cite{Masters_2014}, and \cite{Maseda_2014}, and 6 from \cite{Terlevich_2015}.} These extinction-corrected measurements (see below for a discussion of extinction correction) were very kindly provided to us by Ana Luisa Gonz\'{a}lez-Mor\'{a}n (private communications, 2019 and 2020).

In order to use BAO measurements to constrain cosmological model parameters, knowledge of the sound horizon scale at the drag epoch ($r_s$) is required. We compute this scale more accurately than in Chapter \ref{Chapter5} by using the approximate formula \citep{PhysRevD.92.123516}
\be
\label{eq:sh}
    r_s=\frac{55.154\exp{[-72.3(\Omega_{\nu 0}h^2+0.0006)^2]}}{(\Omega_{\rm b0}h^2)^{0.12807}(\Omega_{\rm cb0}h^2)^{0.25351}} \hspace{1mm}{\rm Mpc}.
\ee
Here $\Omega_{\rm cb0} = \Omega_{\rm c0} + \Omega_{\rm b0} = \Omega_{\rm m0} - \Omega_{\nu_0}$ with $\Omega_{\rm cb0}$, $\Omega_{\rm c0}$, $\Omega_{\rm b0}$, and $\Omega_{\nu_0} = 0.0014$ (following \citealp{Carter_2018}) being the current values of the CDM + baryonic matter, CDM, baryonic matter, and neutrino energy density parameters, respectively, and the Hubble constant $H_0 = 100\ h$ \hunit. Here and in what follows, a subscript of `0' on a given quantity denotes the current value of that quantity. Additionally, $\Omega_{b0}h^2$ is slightly model-dependent; the values of this parameter that we use in this paper are the same as those originally computed in \cite{Park_Ratra_2018_FpCDM_NFpCDM, Park_Ratra_2018_FXCDM_NFXCDM, Park_Ratra_2018_FLCDM} and listed in Table \ref{tab:ch5_BAO_data}.

As mentioned in Section \ref{sec:ch7_intro}, \hiig\ can be used as cosmological probes because they exhibit a tight correlation between the observed luminosity ($L$) of their Balmer emission lines and the velocity dispersion ($\sigma$) of their ionized gas (as measured from the widths of the emission lines). That correlation can be expressed in the form
\begin{equation}
\label{eq:logL}
    \log L = \beta \log \sigma + \gamma,
\end{equation}
where $\gamma$ and $\beta$ are the intercept and slope, respectively, and $\log = \log_{10}$ here and in what follows. In order to determine the values of $\beta$ and $\gamma$, it is necessary to establish the extent to which light from an \hiig\ is extinguished as it propagates through space. A correction must be made to the observed flux so as to account for the effect of this extinction. As mentioned above, the data we received from Ana Luisa Gonz\'{a}lez-Mor\'{a}n have been corrected for extinction. In \cite{G-M_2019}, the authors used the \cite{Gordon_2003} extinction law, and in so doing found
\begin{equation}
    \label{eq:Gordon_beta}
    \beta = 5.022 \pm 0.058,
\end{equation}
and
\begin{equation}
    \label{eq:Gordon_gamma}
    \gamma = 33.268 \pm 0.083,
\end{equation}
respectively. These are the values of $\beta$ and $\gamma$ that we use in the $L$-$\sigma$ relation, eq. \eqref{eq:D_L}.

Once the luminosity of an \hiig\ has been established through eq. \eqref{eq:logL}, this luminosity can be used, in conjunction with a measurement of the flux ($f$) emitted by the \hiig, to determine the distance modulus of the \hiig\ via
\begin{equation}
    \mu_{\rm obs} = 2.5\log L - 2.5\log f - 100.2
\end{equation}
(see e.g. \citealp{Terlevich_2015}, \citealp{G-M_2019}, and references therein).\footnote{For each \hiig\ in our sample we have the measured values and uncertainties of $\log f$\!, $\log \sigma$\!, and $z$.} This quantity can then be compared to the value of the distance modulus predicted within a given cosmological model
\be
\label{eq:mu_th}
    \mu_{\rm th}\left(p, z\right) = 5\log D_{\rm L}\left(p, z\right) + 25,
\ee
where $D_L(p, z)$ is the luminosity distance (see eq. \ref{eq:D_L}).

As the precision of cosmological observations has grown over the last few years, a tension between measurements of the Hubble constant made with early-Universe probes and measurements made with late-Universe probes has revealed itself (for a review, see \citealp{riess_2019}). Whether a given cosmological observation supports a lower value of $H_0$ (i.e. one that is closer to the early-Universe \textit{Planck} measurement) or a higher value of $H_0$ (i.e. one that is closer to the late-Universe value measured by \citealp{riess_etal_2019}) may depend on whether the distance scale associated with this observation has been set by early- or late-Universe physics. It is therefore important to know what distance scale cosmological observations have been calibrated to, so that the extent to which measurements of $H_0$ are pushed higher or lower by these different distance calibrations can be clearly identified.

The $H_0$ values we measure from the combined $H(z)$, BAO, QSO, and \hiig\ data are based on a combination of both early- and late-Universe distance calibrations. As mentioned above, the distance scale of our BAO measurements is set by the size of the sound horizon at the drag epoch $r_s$. The sound horizon, in turn, depends on $\Omega_{b0}h^2$, which was computed by \cite{Park_Ratra_2018_FpCDM_NFpCDM, Park_Ratra_2018_FXCDM_NFXCDM, Park_Ratra_2018_FLCDM} using early-Universe data. Our \hiig\ measurements, on the other hand, have been calibrated using cosmological model independent distance ladder measurements of the distances to nearby giant HII regions (see \citealp{G-M_2019} and references therein), so these data qualify as late-Universe probes. The distance scale of our QSO measurements is set by the intrinsic linear size ($l_m$) of the QSOs themselves, which is a late-Universe measurement (see \citealp{Cao_et_al2017b}). Finally, our $H(z)$ data depend on late-Universe astrophysics through the modeling of the star formation histories of the galaxies whose ages are measured to obtain the Hubble parameter (although the differences between different models are not thought to have a significant effect on measurements of $H(z)$ from these galaxies; see \citealp{Setting_stage_1, Setting_stage_2}).

%%
%Methods
%%
\section{Data Analysis Methodology}
\label{sec:ch7_Methods}

We perform a Markov chain Monte Carlo (MCMC) analysis with the \textsc{Python} module \textsc{emcee} \citep{Foreman-Mackey_Hogg_Lang_Goodman_2013} and maximize the likelihood function, $\mathcal{L}$, 
to determine the best-fitting values of the parameters $p$ of the models. At the most general level, a Monte Carlo analysis is one in which a given problem is solved by randomly generating a large number of trial solutions to the problem, then evaluating those solutions with a quality function. Solving the problem then amounts to locating that trial solution which maximizes the quality function \citep{Cahill_2013}. A Markov chain is a list of numbers generated by a Markov process, where a Markov process is a random process in which the probability of obtaining a given value at some step in the process depends only the value obtained in previous step of the process \citep{von_Toussaint_2011, Foreman-Mackey_Hogg_Lang_Goodman_2013, MCP}. By identifying the likelihood $\mathcal{L}$ with the quality function, we can use the MCMC method to compute the parameters that maximize $\mathcal{L}$.

To implement the MCMC method in practice, it is necessary to employ an algorithm for generating new points in the Markov chain. The most common algorithm is the Metropolis-Hastings algorithm, in which new points in the chain are accepted with the probability
\begin{equation}
\label{eq:ch7_MH_accept}
    P_{\rm accept}\left(\vec{X}_1 \rightarrow \vec{X}_2\right) = {\rm min}\left(1, \frac{P\left(\vec{X}_2|D\right)}{P\left(\vec{X}_1|D\right)}\frac{\mathcal{P}\left(\vec{X}_1, \vec{X}_2\right)}{\mathcal{P}\left(\vec{X}_2, \vec{X}_1\right)}\right),
\end{equation}
where $\mathcal{P}\left(\vec{X}_j, \vec{X}_i\right)$ is the probability that, if the chain currently occupies the point $\vec{X}_i$, the candidate point $\vec{X}_j$ will be proposed, and $P\left(\vec{X}_i|D\right)$ is the posterior probability (which is proportional to the likelihood) of the point $\vec{X}_i$ given the data $D$. In words, the chain will be attracted to regions of the parameter space having a high posterior probability, and will be repelled from regions having a low posterior probability. That is, $P_{\rm accept} = 1$ if the region that the chain wanders into has a higher probability than the region it came from, and $P_{\rm accept} < 1$ if the chain wanders into a region having a lower probability than the region it came from. This means that the distribution of points sampled from the Markov chain eventually approaches the target distribution (that is, the distribution of the quantities under study), and it is sometimes said that the MCMC distribution is \textbf{invariant} at this point \citep{von_Toussaint_2011}. The MCMC distribution becomes invariant when the condition of detailed balance,
\begin{equation}
\label{eq:ch7_detailed_balance}
    T\left(\vec{X}_2, \vec{X}_1\right)\mathcal{P}\left(\vec{X}_2, \vec{X}_1\right)P\left(\vec{X}_1|D\right) = T\left(\vec{X}_1, \vec{X}_2\right)\mathcal{P}\left(\vec{X}_1, \vec{X}_2\right)P\left(\vec{X}_2|D\right)
\end{equation}
is satisfied \citep{von_Toussaint_2011}. This simply means that the net ``flow'' of probability has stopped, and the distribution has reached its ``equilibrium'' (i.e. invariant form), the target distribution. We can show that the acceptance probability of eq. (\ref{eq:ch7_MH_accept}) is consistent with the condition of detailed balance by identifying the transition probability $T\left(\vec{X}_j, \vec{X}_i\right)$ with the acceptance probability $P_{\rm accept}\left(\vec{X}_i \rightarrow \vec{X}_j\right)$ and rewriting eq. (\ref{eq:ch7_detailed_balance}):
\begin{equation}
    \begin{aligned}
    T\left(\vec{X}_2, \vec{X}_1\right)\mathcal{P}\left(\vec{X}_2, \vec{X}_1\right)P\left(\vec{X}_1|D\right) & = {\rm min}\left(1, \frac{P\left(\vec{X}_2|D\right)}{P\left(\vec{X}_1|D\right)}\frac{\mathcal{P}\left(\vec{X}_1, \vec{X}_2\right)}{\mathcal{P}\left(\vec{X}_2, \vec{X}_1\right)}\right)\mathcal{P}\left(\vec{X}_2, \vec{X}_1\right)P\left(\vec{X}_1|D\right)\\
    & = {\rm min}\left(\mathcal{P}\left(\vec{X}_2, \vec{X}_1\right)P\left(\vec{X}_1|D\right), P\left(\vec{X}_2|D\right)\mathcal{P}\left(\vec{X}_1, \vec{X}_2\right)\right)\\
    & = {\rm min}\left(\mathcal{P}\left(\vec{X}_1, \vec{X}_2\right)P\left(\vec{X}_2|D\right), P\left(\vec{X}_1|D\right)\mathcal{P}\left(\vec{X}_2, \vec{X}_1\right)\right)\\
    & = {\rm min}\left(1, \frac{P\left(\vec{X}_1|D\right)}{P\left(\vec{X}_2|D\right)}\frac{\mathcal{P}\left(\vec{X}_2, \vec{X}_1\right)}{\mathcal{P}\left(\vec{X}_1, \vec{X}_2\right)}\right)\mathcal{P}\left(\vec{X}_1, \vec{X}_2\right)P\left(\vec{X}_2|D\right)\\
    & = T\left(\vec{X}_1, \vec{X}_2\right)\mathcal{P}\left(\vec{X}_1, \vec{X}_2\right)P\left(\vec{X}_2|D\right),\\
    \end{aligned}
\end{equation}
\citep{von_Toussaint_2011}. These considerations also illustrate the advantage of the MCMC method over the methods we employed in Chapters \ref{Chapter4} and \ref{Chapter5}: it doesn't waste time scanning low probability regions of the parameter space (and the MCMC advantage is especially pronounced when it is employed to search high-dimensional parameter spaces). The MCMC method is, in a sense, ``smarter'' than the grid scan method, because it seeks out high-probability regions of the parameter space rather than blindly searching everywhere. 

The \textsc{Python} module \textsc{emcee} \citep{Foreman-Mackey_Hogg_Lang_Goodman_2013} is a software package that can be used to quickly produce samples from a probability distribution, via an MCMC approach, although the algorithm it is based on is different from the Metropolis-Hastings algorithm. The algorithm employed by \textsc{emcee}, called the ``stretch move'' algorithm (for reasons that will become clear below) works by evolving an ensemble of ``walkers'' $\{\vec{X}_i\}$. New moves are proposed according to the rule
\begin{equation}
    \vec{X}_i \rightarrow \vec{Y} = \vec{X}_j + Z\left(\vec{X}_i - \vec{X}_j\right),
\end{equation}
where $\vec{X}_j$ is a randomly-drawn walker from the ensemble that is different from $\vec{X}_i$, and $Z$ is a random number (hence the name ``stretch move'': the proposal stretches along the line joining $\vec{X}_i$ and $\vec{X}_j$). The algorithm draws $Z$ from the distribution
\begin{equation}
g\left(Z\right) \propto
    \begin{cases}
    \frac{1}{\sqrt{Z}} & \text{if}\ Z \in \left[\frac{1}{a}, a\right],\\
    \vspace{1mm}
    0 & {\rm otherwise} \\
    \end{cases}
\end{equation}
where the authors of \textsc{emcee} set $a = 2$ (though it could take other values, in general). According to \cite{Goodman_Weare_2010}, the creators of this algorithm, the conditional probability to propose the move from $\vec{X}_i$ to $\vec{Y}$ is proportional to
\begin{equation}
    ||\vec{Y} - \vec{X}_j||^{n-1}P\left(\vec{Y}|D\right),
\end{equation}
where $n$ is the number of dimensions of the parameter space. Detailed balance is maintained if the step $\vec{X}_i \rightarrow \vec{Y}$ is accepted with probability
\begin{equation}
\label{eq:ch7_emcee_accept_prob}
\begin{aligned}
    P_{\rm accept}\left(\vec{X}_i \rightarrow \vec{Y}\right) & = {\rm min}\left(1 - \frac{||\vec{Y} - \vec{X}_j||^{n-1}P\left(\vec{Y}|D\right)}{||\vec{X}_i - \vec{X}_j||^{n-1}P\left(\vec{X_i}|D\right)}\right)\\
    & = {\rm min}\left(1, Z^{N-1}\frac{P\left(\vec{Y}|D\right)}{P\left(\vec{X}_i|D\right)}\right).\\
\end{aligned}
\end{equation}
One of the advantages of the stretch move algorithm is that it is \textbf{affine invariant}. This means that it can sample equally well any probability distributions whose independent variables are related by a linear transformation \citep{Goodman_Weare_2010}. In practical terms, affine invariant algorithms can sample highly skewed or anisotropic distributions more quickly and efficiently than traditional MCMC algorithms like the Metropolis-Hastings algorithm \citep{Goodman_Weare_2010}. The other main advantage of the stretch move algorithm is that it can be parallelized. By splitting the ensemble of walkers into two subsets of equal size, it is possible to advance the positions of the walkers in one subset by proposing moves based on the positions of the walkers in the other subset \citep{Foreman-Mackey_Hogg_Lang_Goodman_2013}. Each advancement can be computed independently, and in this way the entire algorithm can be run in parallel, greatly speeding up the process of sampling from the likelihood $\mathcal{L}$.

We use flat priors for all parameters $p$. For all models, the priors on $\Omega_{m0}$ and $h$ are non-zero over the ranges $0.1 \leq \Omega_{m0} \leq 0.7$ and $0.50 \leq h \leq 0.85$. In the non-flat \lcdm\ model the $\Omega_{\Lambda}$ prior is non-zero over $0.2 \leq \Omega_{\Lambda} \leq 1$. In the flat and non-flat XCDM parametrizations the prior range on $w_{\rm X}$ is $-2 \leq w_{\rm X} \leq 0$, and the prior range on $\Omega_{k0}$ in the non-flat XCDM parametrization is $-0.7 \leq \Omega_{k0} \leq 0.7$. In the flat and non-flat \pcdm\ models the prior range on $\alpha$ is $0.01 \leq \alpha \leq 3$ and $0.01 \leq \alpha \leq 5$, respectively, and the prior range on $\Omega_{k0}$ is also $-0.7 \leq \Omega_{k0} \leq 0.7$.

For \hiig, the likelihood function is
\be
\label{eq:LH1}
    \mathcal{L}_{\rm \hiig}= e^{-\chi^2_{\rm \hiig}/2},
\ee
where
\be
\label{eq:chi2_HIIG}
    \chi^2_{\rm \hiig}(p) = \sum^{153}_{i = 1} \frac{[\mu_{\rm th}(p, z_i) - \mu_{\rm obs}(z_i)]^2}{\epsilon_i^2},
\ee
and $\epsilon_i$ is the uncertainty of the $i_{\rm th}$ measurement. Following \cite{G-M_2019}, $\epsilon$ has the form
\be
\label{eq:err_HIIG}
    \epsilon=\sqrt{\epsilon^2_{\rm stat}+\epsilon^2_{\rm sys}},
\ee
where the statistical uncertainties are
\be
\label{eq:stat_err_HIIG}
    \epsilon^2_{\rm stat}=6.25\left[\epsilon^2_{\log f}+\beta^2\epsilon^2_{\log\sigma}+\epsilon^2_{\beta}(\log\sigma)^2+\epsilon^2_{\gamma}\right]+\left(\frac{\partial{\mu_{\rm th}}}{\partial{z}}\right)^2\epsilon^2_{z}.
\ee
Following \cite{G-M_2019} we do not account for systematic uncertainties in our analysis, so the uncertainty on the \hiig\ measurements consists entirely of the statistical uncertainty (so that $\epsilon = \epsilon_{\rm stat}$).\footnote{A systematic error budget for \hiig\ data is available in the literature, however; see \cite{Chavez_2016}.} The reader should also note here that although the theoretical statistical uncertainty depends our cosmological model parameters (through the theoretical distance modulus $\mu_{\rm th} = \mu_{\rm th}\left(p, z\right)$), the effect of this model-dependence on the parameter constraints is negligible for the current data.\footnote{In contrast to our definition of $\chi^2$ in eq. \eqref{eq:chi2_HIIG}, \cite{G-M_2019} defined an $H_0$-independent $\chi^2$ function in their eq. (27) and weighted this $\chi^2$ function by $1/\epsilon^2_{\rm stat}$ (where $\epsilon^2_{\rm stat}$ is given by their eq. (15)) which we do not do. This procedure is discussed in the literature \citep{Melnick_2017,FernandezArenas}, and when we use it we find that it leads to a reduced $\chi^2$ identical to that given in \cite{G-M_2019} (being less than 2 but greater than 1) without having a noticeable effect on the shapes or peak locations of our posterior likelihoods (hence providing very similar best-fitting values and error bars of the cosmological model parameters). As discussed below, with our $\chi^2$ definition we find reduced $\chi^2$ values $\sim2.75$. \cite{G-M_2019} note that an accounting of systematic uncertainties could decrease the reduced $\chi^2$ values towards unity.\label{fn5}}

For $H(z)$, the likelihood function is
\be
\label{eq:LH2}
    \mathcal{L}_{\rm H}= e^{-\chi^2_{\rm H}/2},
\ee
where
\be
\label{eq:chi2_Hz}
    \chi^2_{\rm H}(p) = \sum^{31}_{i = 1} \frac{[H_{\rm th}(p, z_i) - H_{\rm obs}(z_i)]^2}{\epsilon_i^2},
\ee
and $\epsilon_i$ is the uncertainty of $H_{\rm obs}(z_i)$.

For the BAO data, the likelihood function is
\be
\label{eq:LH3}
    \mathcal{L}_{\rm BAO}= e^{-\chi^2_{\rm BAO}/2},
\ee
and for the uncorrelated BAO data (lines 7-9 in Table \ref{tab:BAO}) the $\chi^2$ function takes the form
\be
\label{eq:chi2_BAO1}
    \chi^2_{\rm BAO}(p) = \sum^{3}_{i = 1} \frac{[A_{\rm th}(p, z_i) - A_{\rm obs}(z_i)]^2}{\epsilon_i^2},
\ee
where $A_{\rm th}$ and $A_{\rm obs}$ are, respectively, the theoretical and observational quantities as listed in Table \ref{tab:BAO}, and $\epsilon_{i}$ corresponds to the uncertainty of $A_{\rm obs}(z_i)$. For the correlated BAO data, the $\chi^2$ function takes the form
\be
\label{eq:chi2_BAO2}
    \chi^2_{\rm BAO}(p) = [\vec{A}_{\rm th}(p, z_i) - \vec{A}_{\rm obs}(z_i)]^T C^{-1}[\vec{A}_{\rm th}(p, z_i) - \vec{A}_{\rm obs}(z_i)],
\ee
where superscripts $T$ and $-1$ denote the transpose and inverse of the matrices, respectively. The covariance matrix $C$ for the BAO data, taken from \cite{Alam_et_al_2017}, is given in eq. (\ref{covmat}), while for the BAO data from \cite{Agathe},
\be
\label{CovM}
    C =
    \begin{bmatrix}
    0.0841 & -0.183396 \\
    -0.183396 & 3.4596
    \end{bmatrix}.
\ee

For QSO, the likelihood function is
\be
\label{eq:LH4}
    \mathcal{L}_{\rm QSO}= e^{-\chi^2_{\rm QSO}/2},
\ee
and the $\chi^2$ function takes the form
\be
\label{eq:chi2_QSO}
    \chi^2_{\rm QSO}(p) = \sum^{120}_{i = 1} \left[\frac{\theta_{\rm th}(p, z_i) - \theta_{\rm obs}(z_i)}{\epsilon_i+0.1\theta_{\rm obs}(z_i)}\right]^2,
\ee
where $\theta_{\rm th}(p, z_i)$ and $\theta_{\rm obs}(z_i)$ are theoretical and observed values of the angular size at redshift $z_i$, respectively, and $\epsilon_i$ is the uncertainty of $\theta_{\rm obs}(z_i)$ (see Chapter \ref{Chapter5} for more details).

For the joint analysis of these data, the total likelihood function is obtained by multiplying the individual likelihood functions (that is, eqs. \eqref{eq:LH1}, \eqref{eq:LH2}, \eqref{eq:LH3}, and \eqref{eq:LH4}) together in various combinations. For example, for $H(z)$, BAO, and QSO data, we have
\be
\label{LH}
    \mathcal{L}=\mathcal{L}_{\rm H}\mathcal{L}_{\rm BAO}\mathcal{L}_{\rm QSO}.
\ee

We also use the Akaike Information
Criterion ($AIC$) and Bayesian Information Criterion ($BIC$) to compare the goodness of fit of models with different numbers of parameters, where
\be
\label{AIC}
    AIC=-2\ln \mathcal{L}_{\rm max} + 2n\equiv\chi^2_{\rm min}+2n,
\ee
and
\be
\label{BIC}
    BIC=-2\ln \mathcal{L}_{\rm max} + n\ln N\equiv\chi^2_{\rm min}+n\ln N.
\ee
In these two equations, $\mathcal{L}_{\rm max}$ refers to the maximum value of the given likelihood function, $\chi^2_{\rm min}$ refers to the corresponding minimum $\chi^2$ value, $n$ is the number of parameters of the given model, and $N$ is the number of data points (for example for \hiig\ we have $N=153$, etc.).

%%
%Section: Results
%%

\section{Results}
\label{sec:ch7_Results}

We present the posterior one-dimensional (1D) probability distributions and two-dimensional (2D) confidence regions of the cosmological parameters for the six flat and non-flat models in Figs. \ref{fig01}--\ref{fig06}, in gray. The unmarginalized best-fitting parameter values are listed in Table \ref{tab:ch7_BFP}, along with the corresponding $\chi^2$, $AIC$, $BIC$, and degrees of freedom $\nu$ (where $\nu \equiv N - n$). The marginalized best-fitting parameter values and uncertainties ($\pm 1\sigma$ error bars or $2\sigma$ limits) are given in Table \ref{tab:1d_BFP}.\footnote{We plot these figures by using the Python package GetDist \citep{Lewis_2019}, which we also use to compute the central values (sample means) and uncertainties of the cosmological parameters listed in Table \ref{tab:1d_BFP}.}

From the fit to the \hiig\ data, we see that most of the probability lies in the part of the parameter space corresponding to currently accelerating cosmological expansion (see the gray contours in Figs. \ref{fig01}--\ref{fig06}). This means that the \hiig\ data favor currently accelerating cosmological expansion,\footnote{Although a full accounting of the systematic uncertainties in the \hiig\ data could weaken this conclusion.} in agreement with supernova Type Ia, BAO, $H(z)$, and other cosmological data. We also find that the constraints on the non-relativistic matter density parameter $\Omega_{m0}$ are consistent with other estimates, ranging between a high of $0.300^{+0.106}_{-0.083}$ (flat XCDM) and a low of $\Omega_{m0} = 0.210^{+0.043}_{-0.092}$ (flat \pcdm).

The \hiig\ data constraints on $H_0$ in Table \ref{tab:1d_BFP} are consistent with the estimate of $H_0 = 71.0 \pm 2.8 ({\rm stat.}) \pm 2.1 ({\rm sys.})$ \hunit made by \cite{FernandezArenas} based on a compilation of \hiig\ measurements that differs from what we have used here. The \hiig\ $H_0$ constraints listed in Table \ref{tab:1d_BFP} are also consistent with other recent measurements of $H_0$, being between $0.90\sigma$ (flat XCDM) and $1.56\sigma$ (non-flat \pcdm) lower than the recent local expansion rate measurement of $H_0 = 74.03 \pm 1.42$ \hunit \citep{riess_etal_2019},\footnote{Note that other local expansion rate measurements are slightly lower with slightly larger error bars \citep{rigault_etal_2015,86,Dhawan,FernandezArenas,freedman_etal_2019,freedman_etal_2020,rameez_sarkar_2019}.} and between $0.78\sigma$ (non-flat \pcdm) and $1.13\sigma$ (flat XCDM) higher than the median statistics estimate of $H_0=68 \pm 2.8$ \hunit \citep{chenratmed},\footnote{This is consistent with earlier median statistics estimates \citep{gott_etal_2001,76} and also with a number of recent $H_0$ measurements \citep{chen_etal_2017,DES_2018,Gomez-ValentAmendola2018,haridasu_etal_2018,planck2018,zhang_2018,dominguez_etal_2019,martinelli_tutusaus_2019,Cuceu_2019,zeng_yan_2019,schoneberg_etal_2019,lin_ishak_2019,zhang_huang_2019}.} with our measurements ranging from a low of $H_0=70.60^{+1.68}_{-1.84}$ \hunit (non-flat \pcdm) to a high of $H_0=71.85 \pm 1.96$ \hunit (flat XCDM).

As for spatial curvature, from the marginalized 1D likelihoods in Table \ref{tab:1d_BFP}, for non-flat \lcdm, non-flat XCDM, and non-flat \pcdm, we measure $\Omega_{k0}=0.094^{+0.237}_{-0.363}$,\footnote{Since $\Omega_{k0}=1-\Omega_{m0}-\Omega_{\Lambda}$, in the non-flat \lcdm\ model analysis we replace $\Omega_{\Lambda}$ with $\Omega_{k0}$ in the MCMC chains of $\{H_0, \Omega_{m0}, \Omega_{\Lambda}\}$ to obtain new chains of $\{H_0, \Omega_{m0}, \Omega_{k0}\}$ and so measure $\Omega_{k0}$ central values and uncertainties. A similar procedure, based on $\Omega_{\Lambda}=1-\Omega_{m0}$, is used to measure $\Omega_{\Lambda}$ in the flat \lcdm\ model.} $\Omega_{k0}=0.011^{+0.457}_{-0.460}$, and $\Omega_{k0}=0.291^{+0.348}_{-0.113}$, respectively. From the marginalized likelihoods, we see that non-flat \lcdm\ and XCDM models are consistent with all three spatial geometries, while non-flat \pcdm\ favors the open case at 2.58$\sigma$. However, this seems to be a little odd, especially for non-flat \pcdm, considering their unmarginalized best-fitting $\Omega_{k0}$\!'s are all negative (see Table \ref{tab:ch7_BFP}).

The fits to the \hiig\ data are consistent with dark energy being a cosmological constant but don't rule out dark energy dynamics. For flat (non-flat) XCDM, $w_{\rm X}=-1.180^{+0.560}_{-0.330}$ ($w_{\rm X}=-1.125^{+0.537}_{-0.321}$), which are both within 1$\sigma$ of $w_{\rm X}=-1$. For flat (non-flat) \pcdm, $2\sigma$ upper limits of $\alpha$ are $\alpha<2.784$ ($\alpha<4.590$), with the 1D likelihood functions, in both cases, peaking at $\alpha=0$.

Current \hiig\ data do not provide very restrictive constraints on cosmological model parameters, but when used in conjunction with other cosmological data they can help tighten the constraints.

\subsection{$H(z)$, BAO, and \hiig\ (ZBH) constraints}
\label{subsec:ZBH}

The \hiig\ constraints discussed in the previous subsection are consistent with constraints from most other cosmological data, so it is appropriate to use the \hiig\ data in conjunction with other data to constrain parameters. In this subsection we perform a full analysis of $H(z)$, BAO, and \hiig\ (ZBH) data and derive tighter constraints on cosmological parameters.

\begin{figure*}
\centering
    \includegraphics[width=3.5in,height=3.5in]{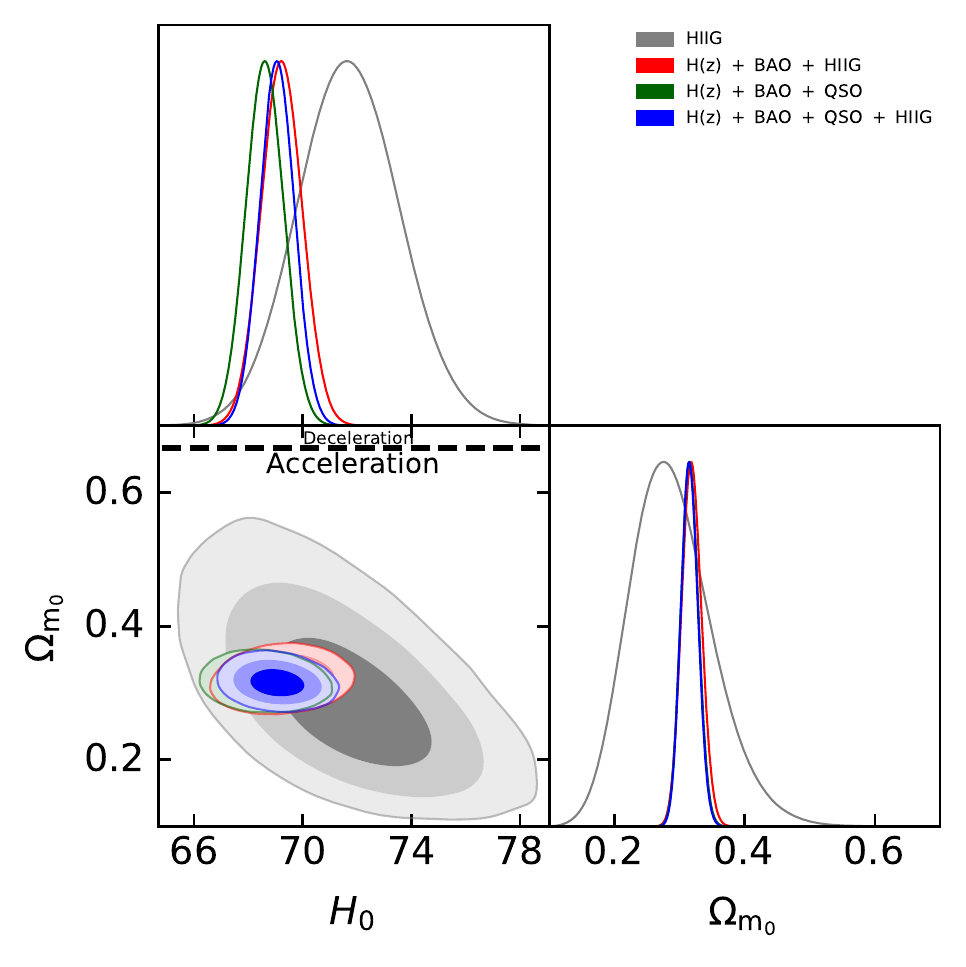}
    \includegraphics[width=3.5in,height=3.5in]{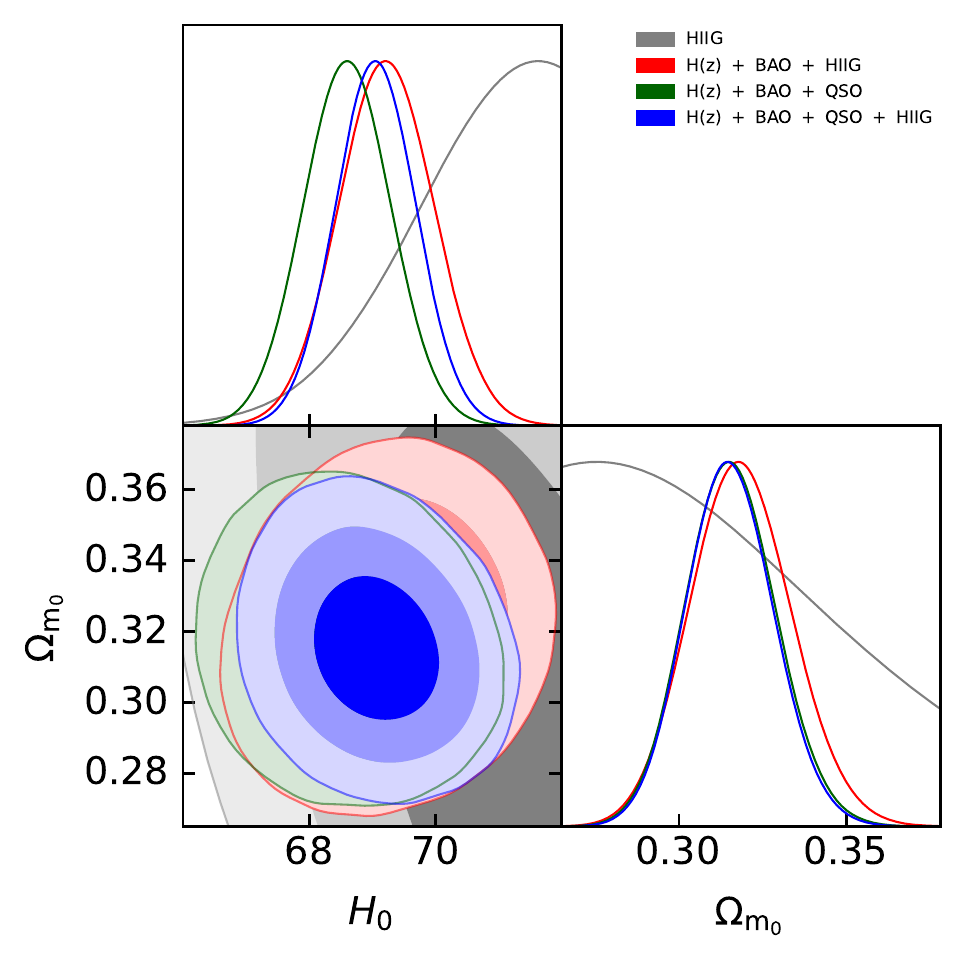}\\
\caption[1$\sigma$, 2$\sigma$, and 3$\sigma$ confidence contours for flat \lcdm.]{1$\sigma$, 2$\sigma$, and 3$\sigma$ confidence contours for flat \lcdm, where the right panel is the comparison zoomed in. The black dotted line is the zero-acceleration line, which divides the parameter space into regions associated with currently accelerated (below) and currently decelerated (above) cosmological expansion.}
\label{fig01}
\end{figure*}

\begin{figure*}
\centering
    \includegraphics[width=3.5in,height=3.5in]{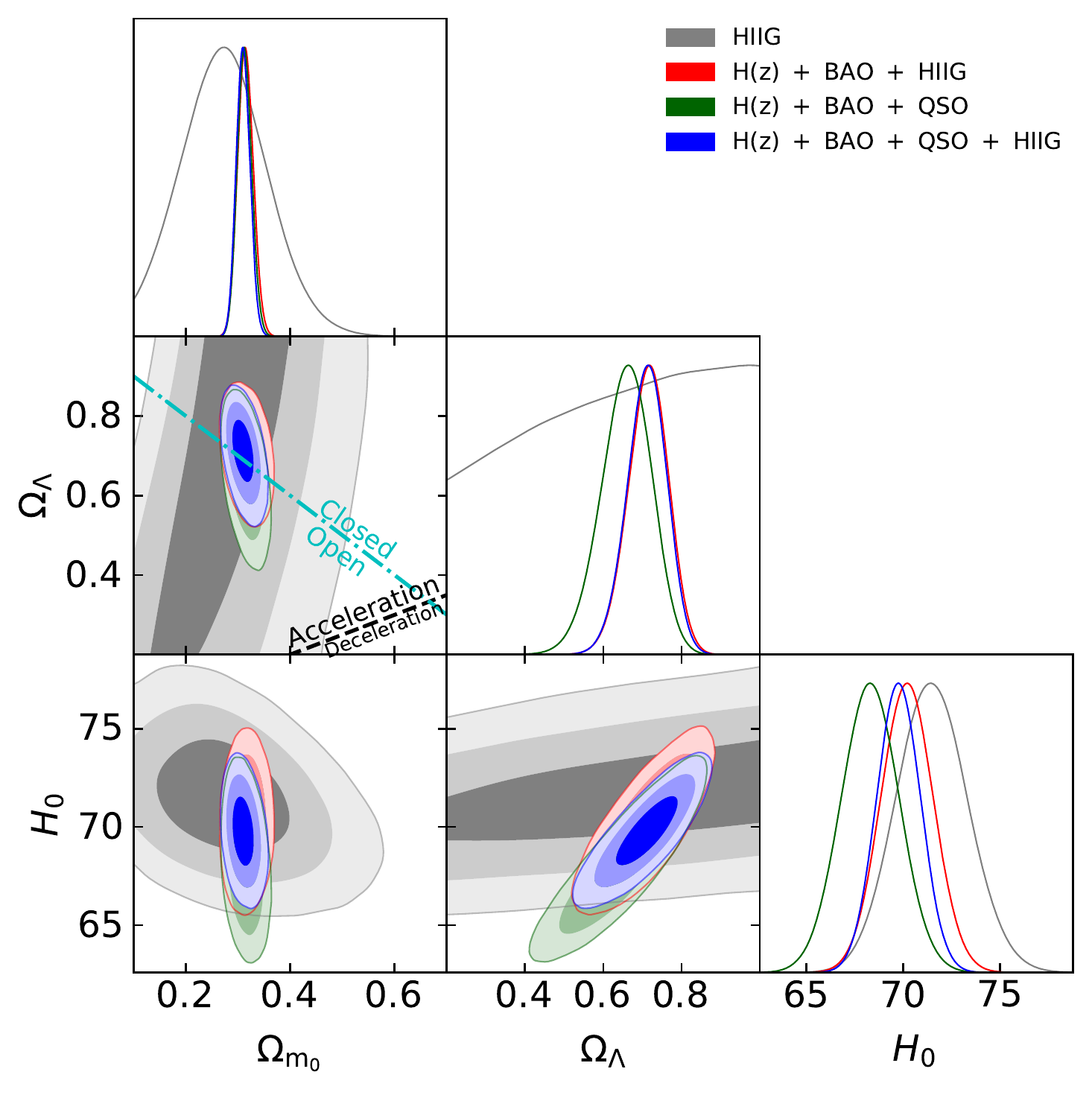}
    \includegraphics[width=3.5in,height=3.5in]{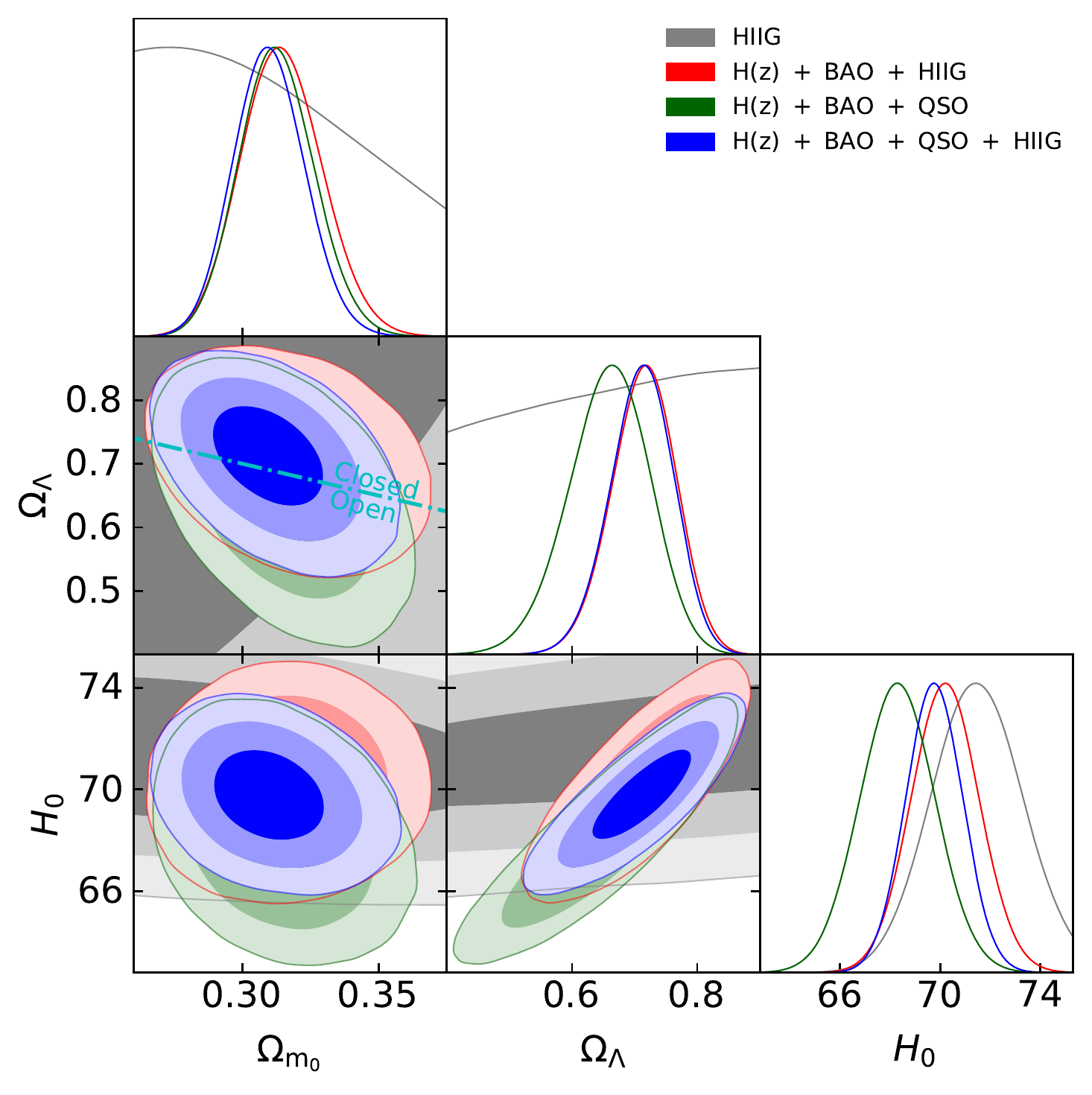}\\
\caption[1$\sigma$, 2$\sigma$, and 3$\sigma$ confidence contours for non-flat \lcdm.]{Same as Fig. \ref{fig01} but for non-flat \lcdm. The cyan dash-dot line represents the flat case, with closed spatial hypersurfaces to the upper right. The black dotted line is the zero-acceleration line, which divides the parameter space into regions associated with currently accelerated (above left) and currently decelerated (below right) cosmological expansion.}
\label{fig02}
\end{figure*}

\begin{figure*}
\centering
    \includegraphics[width=3.5in,height=3.5in]{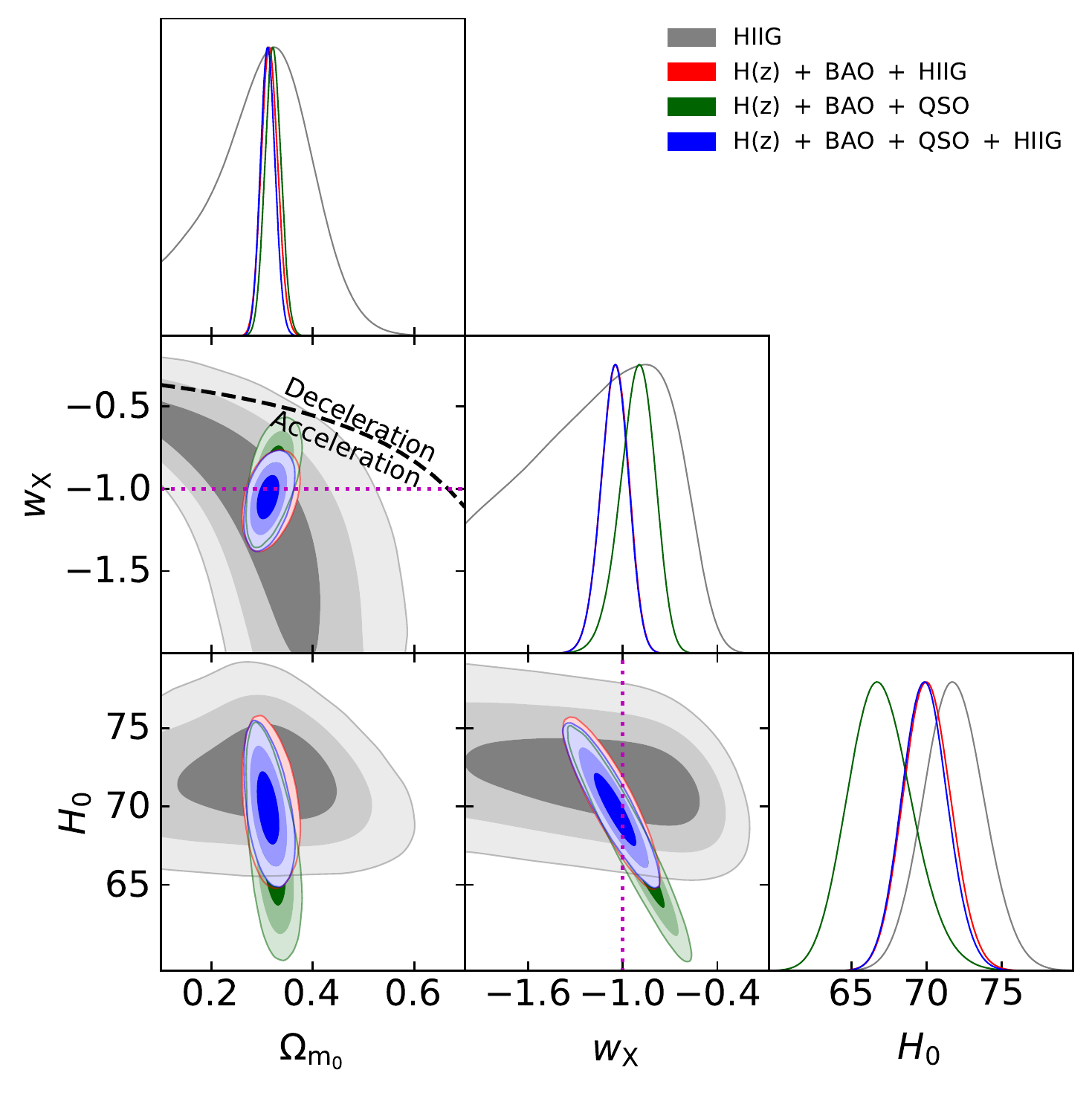}
    \includegraphics[width=3.5in,height=3.5in]{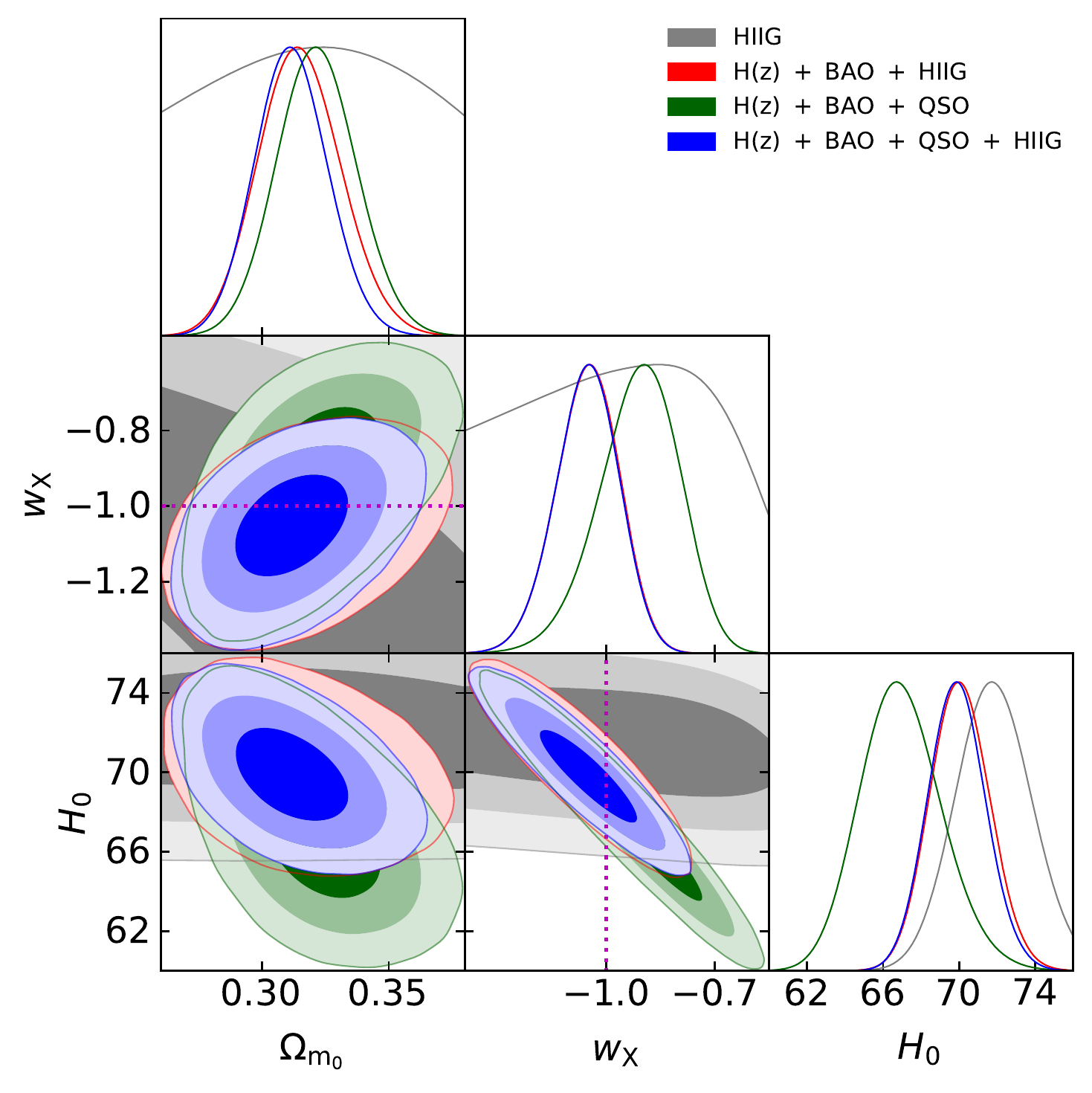}\\
\caption[1$\sigma$, 2$\sigma$, and 3$\sigma$ confidence contours for flat XCDM.]{1$\sigma$, 2$\sigma$, and 3$\sigma$ confidence contours for flat XCDM. The black dotted line is the zero-acceleration line, which divides the parameter space into regions associated with currently accelerated (below left) and currently decelerated (above right) cosmological expansion. The magenta lines denote $w_{\rm X}=-1$, i.e. the flat \lcdm\ model.}
\label{fig03}
\end{figure*}

\begin{figure*}
\centering
    \includegraphics[width=3.5in,height=3.5in]{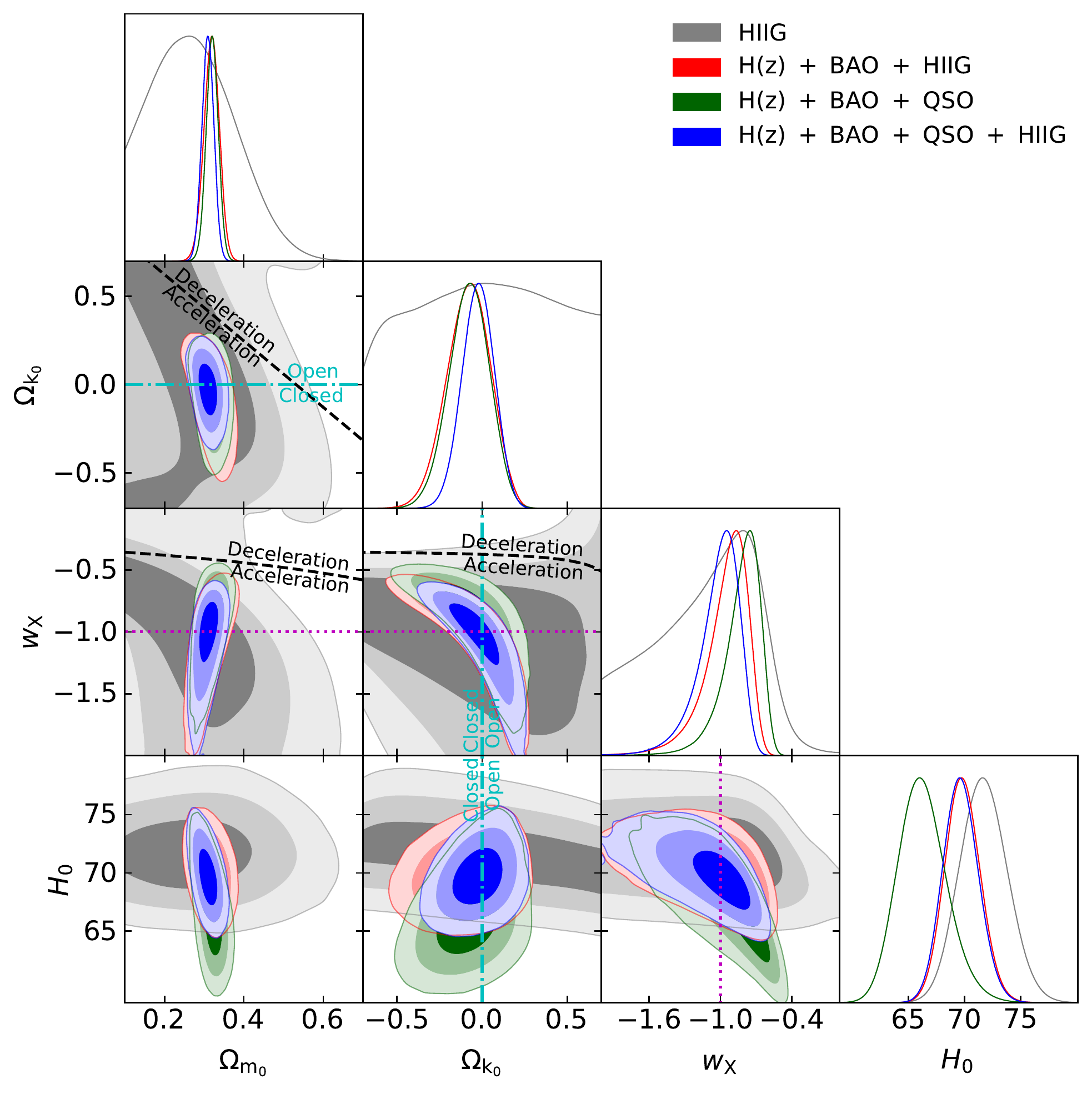}
    \includegraphics[width=3.5in,height=3.5in]{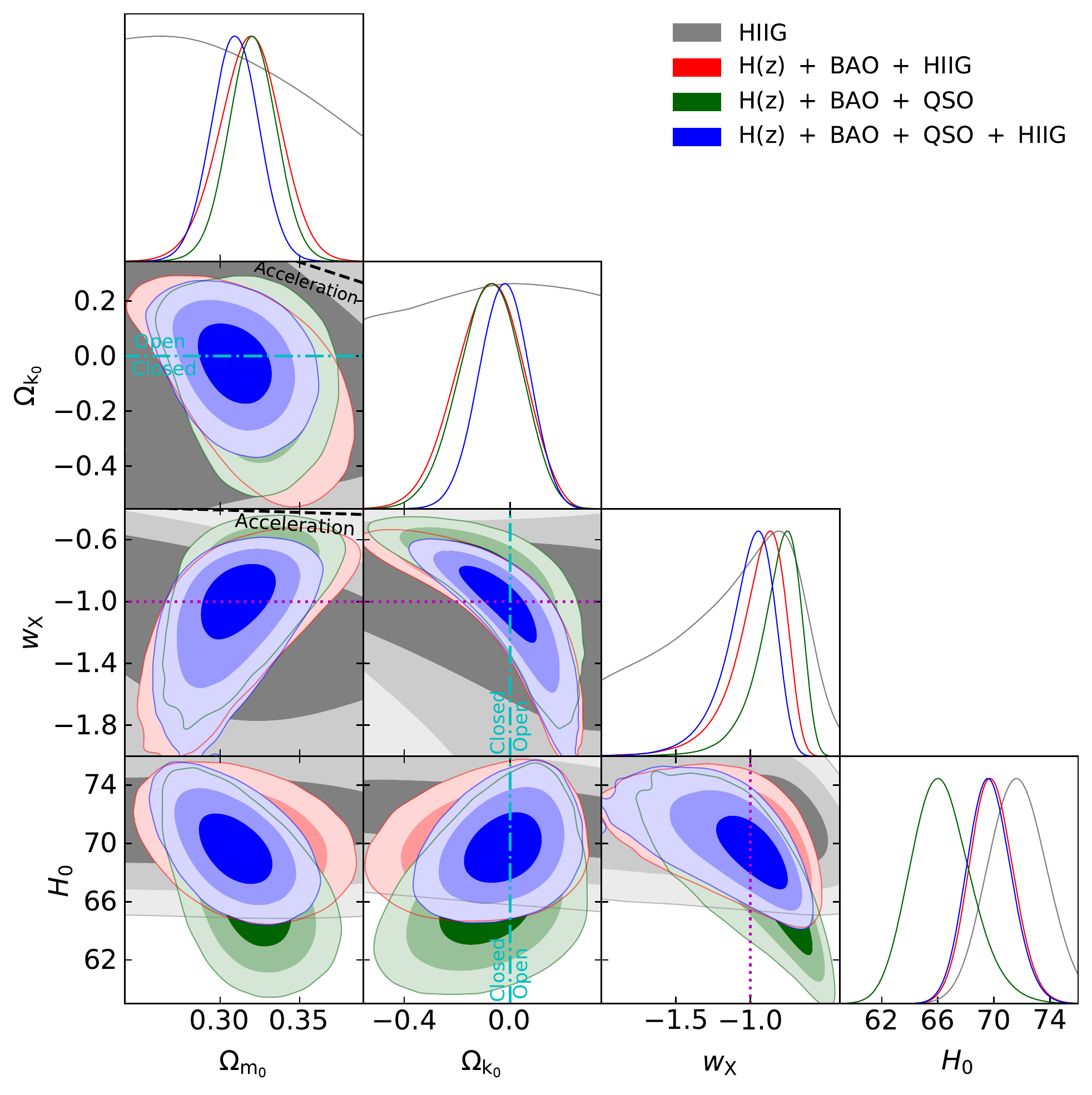}\\
\caption[1$\sigma$, 2$\sigma$, and 3$\sigma$ confidence contours for non-flat XCDM.]{Same as Fig. \ref{fig03} but for non-flat XCDM, where the zero acceleration lines in each of the three subpanels are computed for the third cosmological parameter set to the \hiig\ data only best-fitting values listed in Table \ref{tab:ch7_BFP}. Currently-accelerated cosmological expansion occurs below these lines. The cyan dash-dot lines represent the flat case, with closed spatial hypersurfaces either below or to the left. The magenta lines indicate $w_{\rm X} = -1$, i.e. the non-flat \lcdm\ model.}
\label{fig04}
\end{figure*}

\begin{figure*}
\centering
    \includegraphics[width=3.5in,height=3.5in]{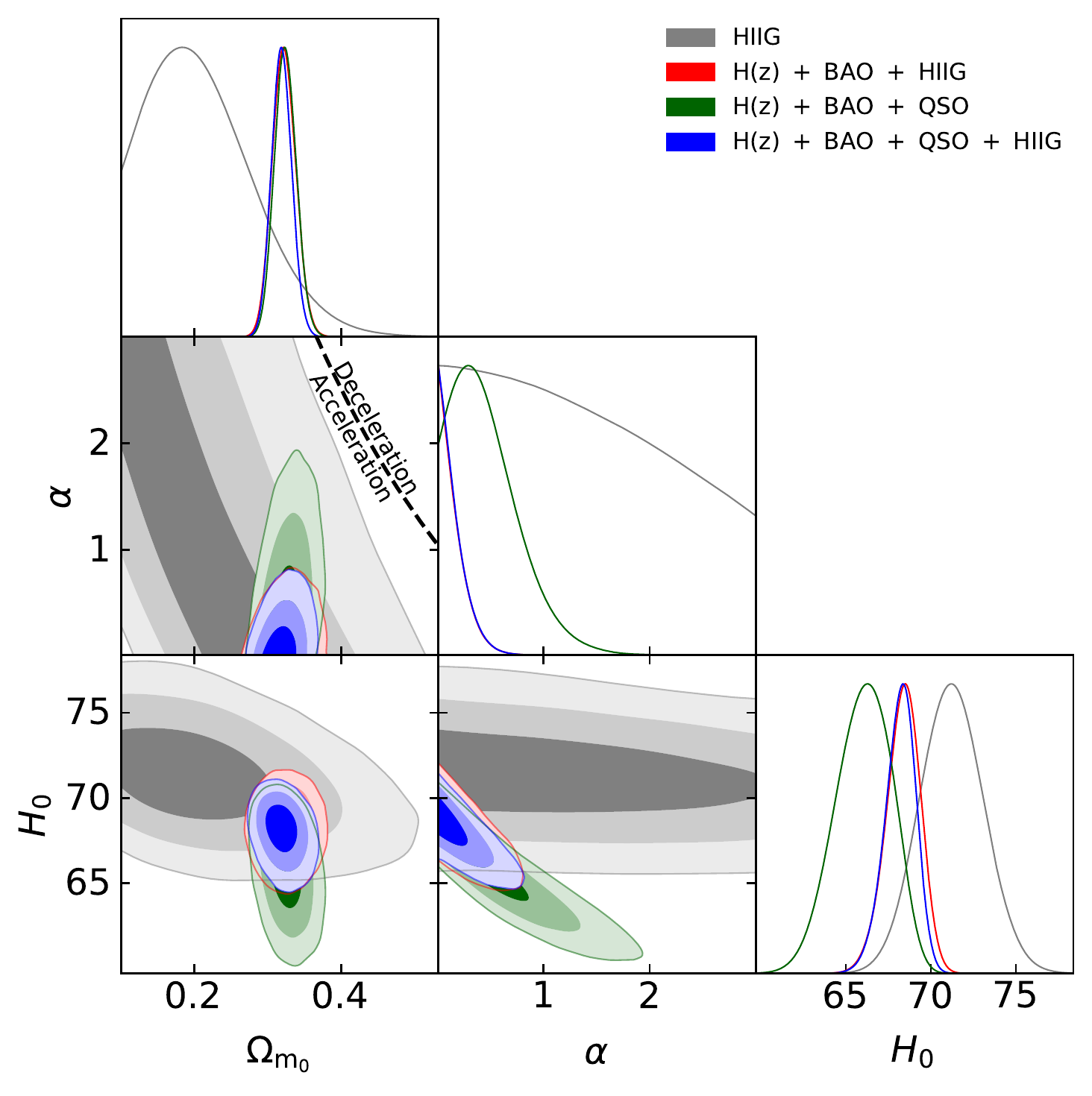}
    \includegraphics[width=3.5in,height=3.5in]{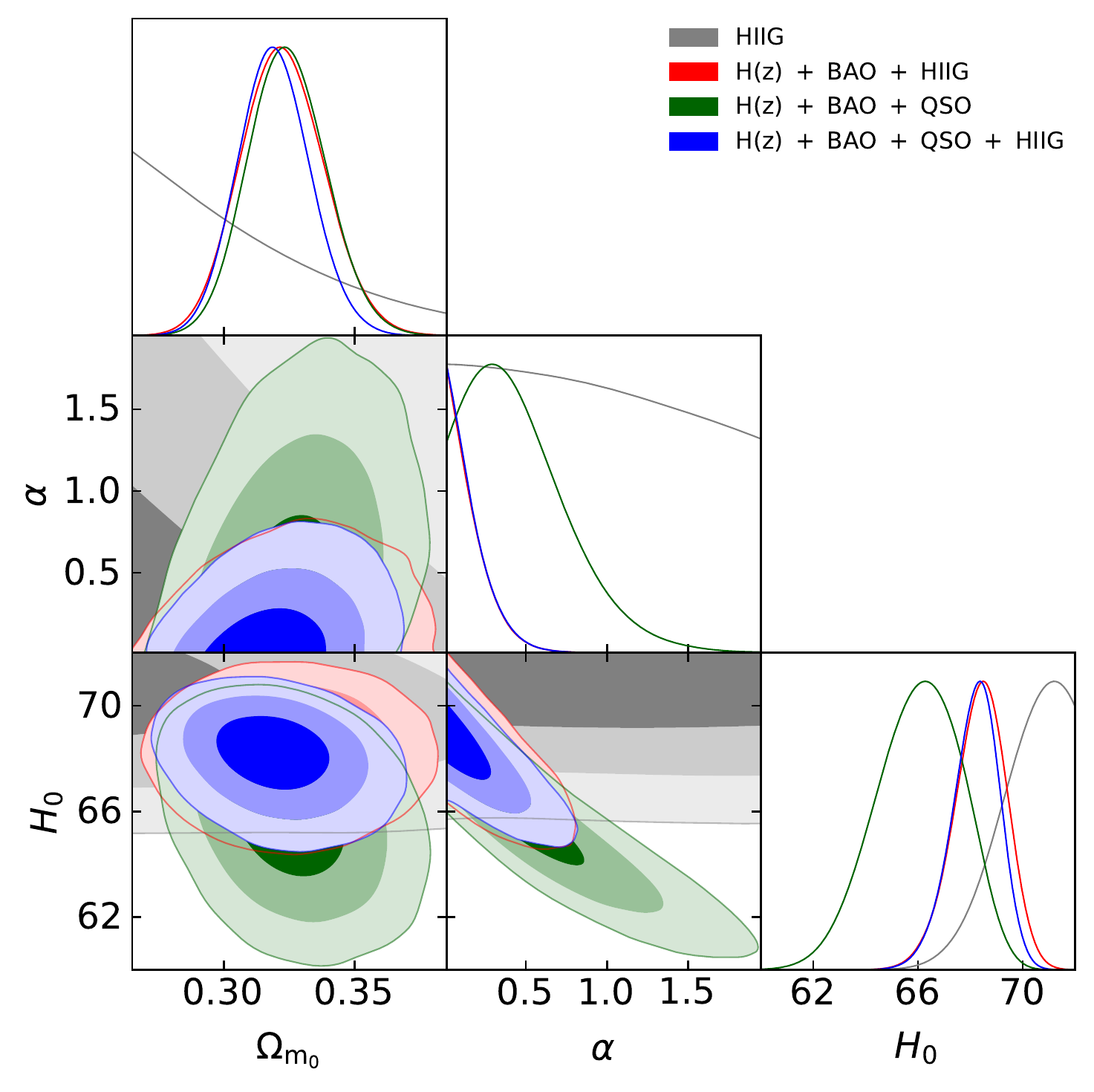}\\
\caption[1$\sigma$, 2$\sigma$, and 3$\sigma$ confidence contours for flat $\phi$CDM]{1$\sigma$, 2$\sigma$, and 3$\sigma$ confidence contours for flat $\phi$CDM. The black dotted zero-acceleration line splits the parameter space into regions of currently accelerated (below left) and currently decelerated (above right) cosmological expansion. The $\alpha = 0$ axis is the flat \lcdm\ model.}
\label{fig05}
\end{figure*}

\begin{figure*}
\centering
    \includegraphics[width=3.5in,height=3.5in]{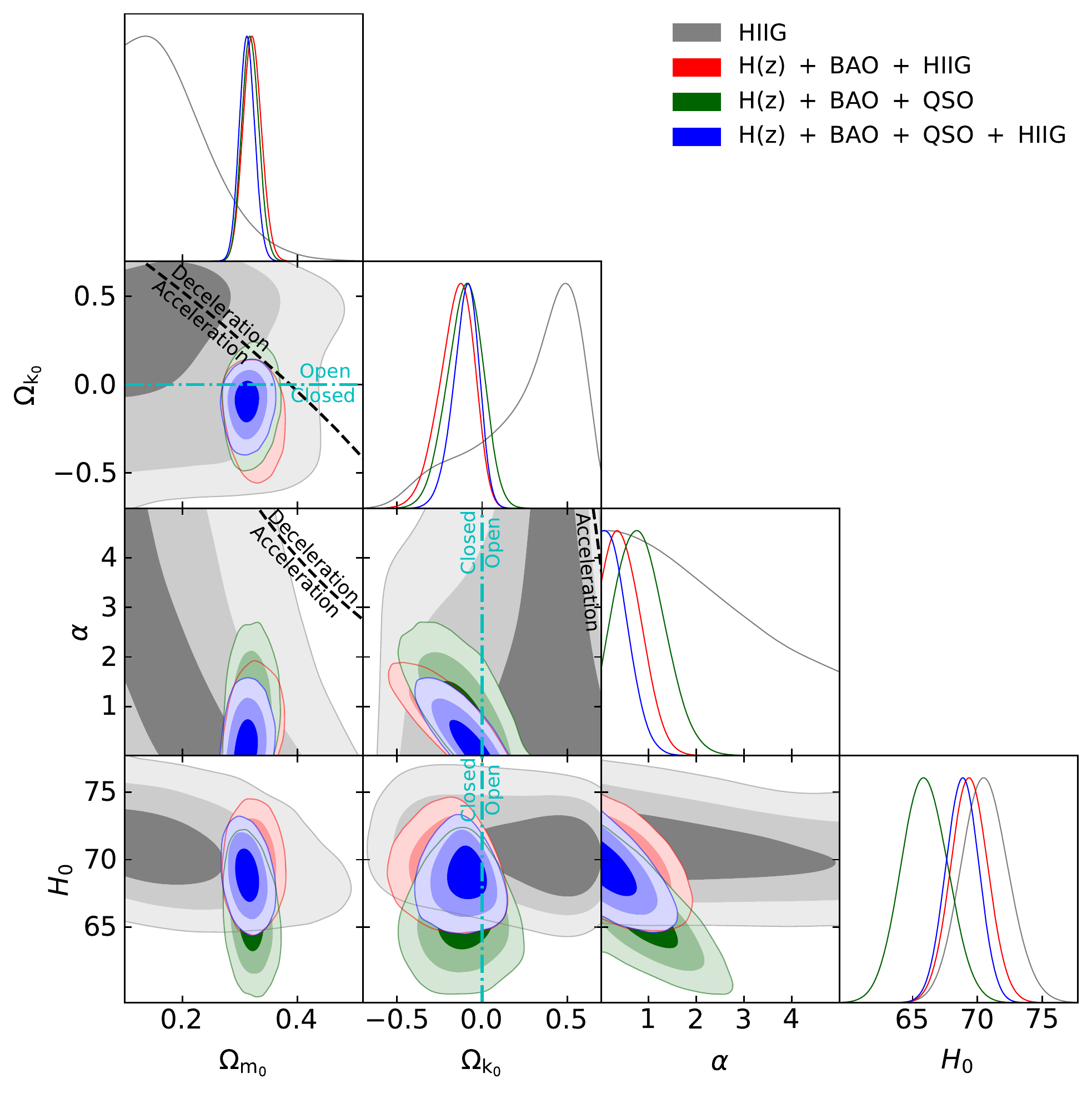}
    \includegraphics[width=3.5in,height=3.5in]{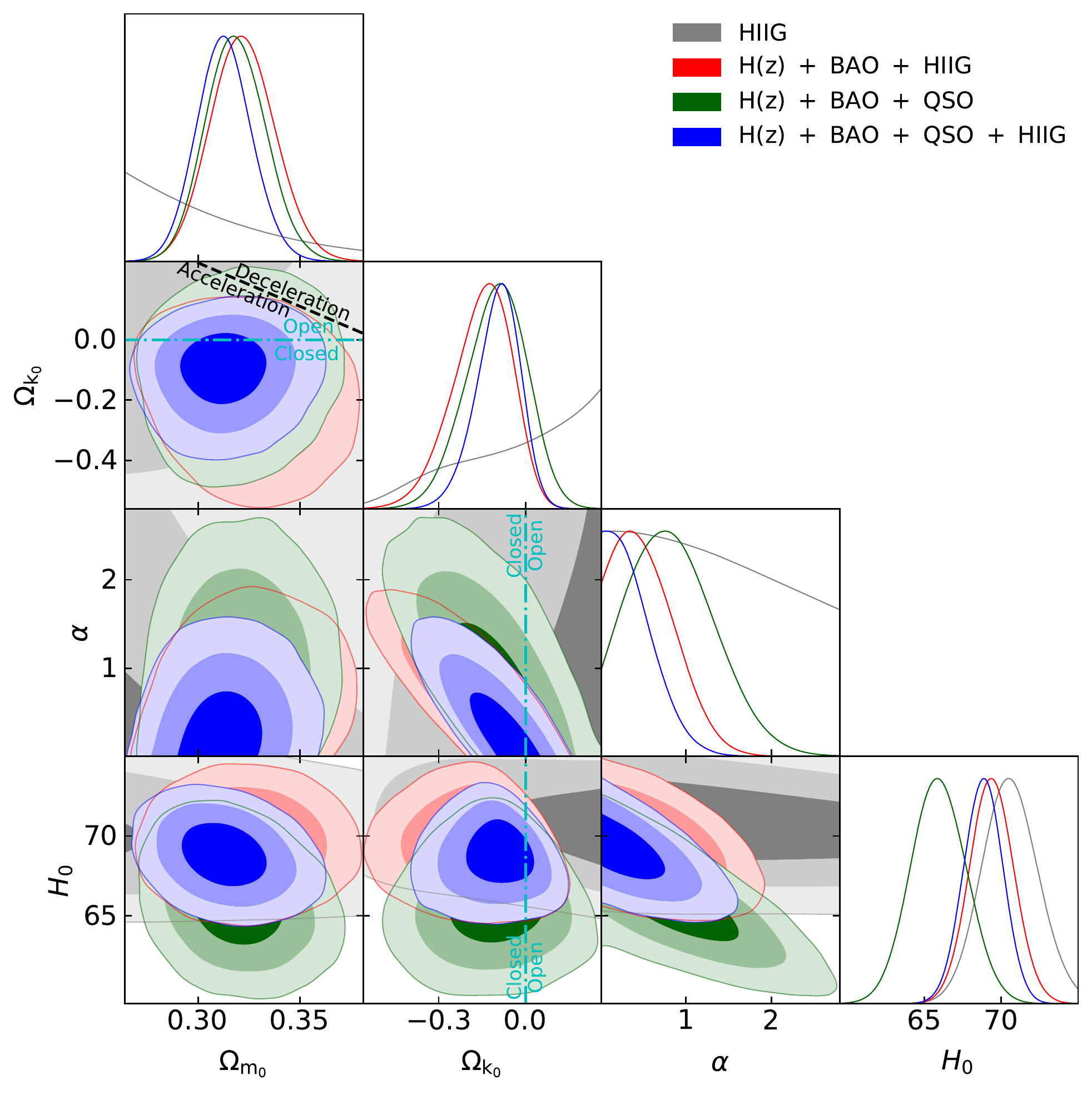}\\
\caption[1$\sigma$, 2$\sigma$, and 3$\sigma$ confidence contours for non-flat $\phi$CDM]{Same as Fig. \ref{fig05} but for non-flat \pcdm, where the zero-acceleration lines in each of the subpanels are computed for the third cosmological parameter set to the \hiig\ data only best-fitting values listed in Table \ref{tab:ch7_BFP}. currently accelerating cosmological expansion occurs below these lines. The cyan dash-dot lines represent the flat case, with closed spatial geometry either below or to the left. The $\alpha = 0$ axis is the non-flat \lcdm\ model.}
\label{fig06}
\end{figure*}

The 1D probability distributions and 2D confidence regions of the cosmological parameters for the six flat and non-flat models are shown in Figs. \ref{fig01}--\ref{fig06}, in red. The best-fitting results and uncertainties are listed in Tables \ref{tab:ch7_BFP} and \ref{tab:1d_BFP}.

When we fit our cosmological models to the ZBH data we find that the measured values of the matter density parameter $\Omega_{m0}$ fall within a narrower range in comparison to the \hiig\ only case, being between $0.314 \pm 0.015$ (non-flat \lcdm) and $0.323^{+0.014}_{-0.016}$ (flat \pcdm).

\begin{table*}
\centering
\resizebox{\columnwidth}{!}{%
\begin{threeparttable}
\caption{Unmarginalized best-fitting parameter values for all models from various combinations of data.}\label{tab:ch7_BFP}
\setlength{\tabcolsep}{2.0mm}{
\begin{tabular}{lccccccccccc}
\hline
 Model & Data set & $\Omega_{m0}$ & $\Omega_{\Lambda}$ & $\Omega_{k0}$ & $w_{\mathrm X}$ & $\alpha$ & $H_0$\tnote{a} & $\chi^2$ & $AIC$ & $BIC$ & $\nu$\\
\hline
Flat \lcdm & \hiig\ & 0.276 & 0.724 & --- & --- & --- & 71.81 & 410.75 & 414.75 & 420.81 & 151\\
 & $H(z)$ + BAO + \hiig\ & 0.318 & 0.682 & --- & --- & --- & 69.22 & 434.29 & 438.29 & 444.84 & 193 \\
 & $H(z)$ + BAO + QSO & 0.315 & 0.685 & --- & --- & --- & 68.61 & 372.88 & 376.88 & 383.06 & 160\\
 & $H(z)$ + BAO + QSO + \hiig\ & 0.315 & 0.685 & --- & --- & --- & 69.06 & 786.50 & 790.50 & 798.01 & 313\\
\\
Non-flat \lcdm & \hiig\ & 0.312 & 0.998 & $-0.310$ & --- & --- & 72.35 & 410.44 & 416.44 & 425.53 & 150\\
 & $H(z)$ + BAO + \hiig\ & 0.313 & 0.718 & $-0.031$ & --- & --- & 70.24 & 433.38 & 439.38 & 449.19 & 192\\
 & $H(z)$ + BAO + QSO & 0.311 & 0.665 & 0.024 & --- & --- & 68.37 & 372.82 & 378.82 & 388.08 & 159\\
 & $H(z)$ + BAO + QSO + \hiig\ & 0.309 & 0.716 & $-0.025$ & --- & --- & 69.82 & 785.79 & 791.79 & 803.05 & 312\\
\\
Flat XCDM & \hiig\ & 0.249 & --- & --- & $-0.892$ & --- & 71.65 & 410.72 & 416.72 & 425.82 & 150\\
 & $H(z)$ + BAO + \hiig\ & 0.314 & --- & --- & $-1.044$ & --- & 69.94 & 433.99 & 439.99 & 449.81 & 192\\
 & $H(z)$ + BAO + QSO & 0.322 & --- & --- & $-0.890$ & --- & 66.62 & 371.95 & 377.95 & 387.21 & 159\\
 & $H(z)$ + BAO + QSO + \hiig\ & 0.311 & --- & --- & $-1.045$ & --- & 69.80 & 786.19 & 792.19 & 803.45 & 312\\
\\
Non-flat XCDM & \hiig\ & 0.104 & --- & $-0.646$ & $-0.712$ & --- & 72.61 & 407.69 & 415.69 & 427.81 & 149\\
 & $H(z)$ + BAO + \hiig\ & 0.322 & --- & $-0.117$ & $-0.878$ & --- & 66.67 & 432.85 & 440.85 & 453.94 & 191\\
 & $H(z)$ + BAO + QSO & 0.322 & --- & $-0.112$ & $-0.759$ & --- & 65.80 & 370.68 & 378.68 & 391.03 & 158\\
 & $H(z)$ + BAO + QSO + \hiig\ & 0.310 & --- & $-0.048$ & $-0.957$ & --- & 69.53 & 785.70 & 793.70 & 808.71 & 311\\
\\
Flat $\phi$CDM & \hiig\ & 0.255 & --- & --- & --- & 0.261 & 71.70 & 410.70 & 416.70 & 425.80 & 150\\
 & $H(z)$ + BAO + \hiig\ & 0.318 & --- & --- & --- & 0.011 & 69.09 & 434.36 & 440.36 & 450.18 & 192\\
 & $H(z)$ + BAO + QSO & 0.321 & --- & --- & --- & 0.281 & 66.82 & 372.05 & 378.05 & 387.31 & 159\\
 & $H(z)$ + BAO + QSO + \hiig\ & 0.315 & --- & --- & --- & 0.012 & 68.95 & 786.58 & 792.58 & 803.84 & 312\\
\\
Non-flat $\phi$CDM & \hiig\ & 0.114 & --- & $-0.437$ & --- & 2.680 & 72.14 & 409.91 & 417.91 & 430.03 & 149\\
 & $H(z)$ + BAO + \hiig\ & 0.321 & --- & $-0.132$ & --- & 0.412 & 69.69 & 432.75 & 440.75 & 453.84 & 191\\
 & $H(z)$ + BAO + QSO & 0.317 & --- & $-0.106$ & --- & 0.778 & 66.27 & 370.83 & 378.83 & 391.18 & 158\\
 & $H(z)$ + BAO + QSO + \hiig\ & 0.310 & --- & $-0.054$ & --- & 0.150 & 69.40 & 785.65 & 793.65 & 808.66 & 311\\
\hline
\end{tabular}}
\begin{tablenotes}
\item [a] \hunit.
\end{tablenotes}
\end{threeparttable}%
}
\end{table*}

\begin{table*}
\centering
\resizebox{\columnwidth}{!}{%
\begin{threeparttable}
\caption{One-dimensional marginalized best-fitting parameter values and uncertainties ($\pm 1\sigma$ error bars or $2\sigma$ limits) for all models from various combinations of data.}\label{tab:1d_BFP}
\setlength{\tabcolsep}{1.8mm}{
\begin{tabular}{lccccccc}
\hline
 Model & Data set & $\Omega_{m0}$ & $\Omega_{\Lambda}$ & $\Omega_{k0}$ & $w_{\mathrm X}$ & $\alpha$ & $H_0$\tnote{a} \\
\hline
Flat \lcdm & \hiig\ & $0.289^{+0.053}_{-0.071}$ & --- & --- & --- & --- & $71.70\pm1.83$ \\
 & $H(z)$ + BAO + \hiig\ & $0.319^{+0.014}_{-0.015}$ & --- & --- & --- & --- & $69.23\pm0.74$ \\
 & $H(z)$ + BAO + QSO & $0.316^{+0.013}_{-0.014}$ & --- & --- & --- & --- & $68.60\pm0.68$ \\
 & $H(z)$ + BAO + QSO + \hiig\ & $0.315^{+0.013}_{-0.012}$ & --- & --- & --- & --- & $69.06^{+0.63}_{-0.62}$ \\
\\
Non-flat \lcdm & \hiig\ & $0.275^{+0.081}_{-0.078}$ & $>0.501$\tnote{b} & $0.094^{+0.237}_{-0.363}$ & --- & --- & $71.50^{+1.80}_{-1.81}$ \\
 & $H(z)$ + BAO + \hiig\ & $0.314\pm0.015$ & $0.714^{+0.054}_{-0.049}$ & $-0.029^{+0.049}_{-0.048}$ & --- & --- & $70.21\pm1.33$ \\
 & $H(z)$ + BAO + QSO & $0.313^{+0.013}_{-0.015}$ & $0.658^{+0.069}_{-0.060}$ & $0.029^{+0.056}_{-0.063}$ & --- & --- & $68.29\pm1.47$ \\
 & $H(z)$ + BAO + QSO + \hiig\ & $0.310\pm0.013$ & $0.711^{+0.053}_{-0.048}$ & $-0.021^{+0.044}_{-0.048}$ & --- & --- & $69.76^{+1.12}_{-1.11}$ \\
\\
Flat XCDM & \hiig\ & $0.300^{+0.106}_{-0.083}$ & --- & --- & $-1.180^{+0.560}_{-0.330}$ & --- & $71.85\pm1.96$ \\
 & $H(z)$ + BAO + \hiig\ & $0.315^{+0.016}_{-0.017}$ & --- & --- & $-1.052^{+0.092}_{-0.082}$ & --- & $70.05\pm1.54$ \\
 & $H(z)$ + BAO + QSO & $0.322^{+0.015}_{-0.016}$ & --- & --- & $-0.911^{+0.122}_{-0.098}$ & --- & $66.98^{+1.95}_{-2.30}$ \\
 & $H(z)$ + BAO + QSO + \hiig\ & $0.312\pm0.014$ & --- & --- & $-1.053^{+0.091}_{-0.082}$ & --- & $69.90\pm1.48$ \\
\\
Non-flat XCDM & \hiig\ & $0.275^{+0.084}_{-0.125}$ & --- & $0.011^{+0.457}_{-0.460}$ & $-1.125^{+0.537}_{-0.321}$ & --- & $71.71^{+2.07}_{-2.08}$ \\
 & $H(z)$ + BAO + \hiig\ & $0.318\pm0.019$ & --- & $-0.082^{+0.135}_{-0.119}$ & $-0.958^{+0.219}_{-0.098}$ & --- & $69.83^{+1.50}_{-1.62}$ \\
 & $H(z)$ + BAO + QSO & $0.320\pm0.015$ & --- & $-0.078^{+0.124}_{-0.112}$ & $-0.826^{+0.185}_{-0.088}$ & --- & $66.29^{+1.90}_{-2.35}$ \\
 & $H(z)$ + BAO + QSO + \hiig\ & $0.309^{+0.015}_{-0.014}$ & --- & $-0.025\pm0.092$ & $-1.022^{+0.208}_{-0.104}$ & --- & $69.68^{+1.49}_{-1.64}$ \\
\\
Flat $\phi$CDM & \hiig\ & $0.210^{+0.043}_{-0.092}$ & --- & --- & --- & $<2.784$ & $71.23^{+1.79}_{-1.80}$ \\
 & $H(z)$ + BAO + \hiig\ & $0.323^{+0.014}_{-0.016}$ & --- & --- & --- & $<0.411$ & $68.36^{+1.05}_{-0.86}$ \\
 & $H(z)$ + BAO + QSO & $0.324^{+0.014}_{-0.015}$ & --- & --- & --- & $0.460^{+0.116}_{-0.440}$ & $66.03^{+1.79}_{-1.42}$ \\
 & $H(z)$ + BAO + QSO + \hiig\ & $0.319\pm0.013$ & --- & --- & --- & $<0.411$ & $68.18^{+0.97}_{-0.75}$\\
\\
Non-flat $\phi$CDM & \hiig\ & $<0.321$ & --- & $0.291^{+0.348}_{-0.113}$ & --- & $<4.590$ & $70.60^{+1.68}_{-1.84}$ \\
 & $H(z)$ + BAO + \hiig\ & $0.322^{+0.015}_{-0.016}$ & --- & $-0.153^{+0.114}_{-0.079}$ & --- & $0.538^{+0.151}_{-0.519}$ & $69.39\pm1.37$ \\
 & $H(z)$ + BAO + QSO & $0.319^{+0.013}_{-0.015}$ & --- & $-0.103^{+0.111}_{-0.091}$ & --- & $0.854^{+0.379}_{-0.594}$ & $65.94^{+1.75}_{-1.73}$ \\
 & $H(z)$ + BAO + QSO + \hiig\ & $0.313^{+0.012}_{-0.014}$ & --- & $-0.098^{+0.082}_{-0.061}$ & --- & $<0.926$ & $68.83\pm1.23$ \\
\hline
\end{tabular}}
\begin{tablenotes}
\item [a] \hunit.
\item [b] This is the 1$\sigma$ lower limit. The $2\sigma$ lower limit is set by the prior, and is not shown here.
\end{tablenotes}
\end{threeparttable}%
}
\end{table*}

Similarly, the measured values of $H_0$ also fall within a narrower range when our models are fit to the ZBH data combination (and are in better agreement with the median statistics estimate of $H_0$ from \citealp{chenratmed} than with the local measurement carried out by \citealp{riess_etal_2019}; this is because the $H(z)$ and BAO data favor a lower $H_0$ value) being between $H_0=68.36^{+1.05}_{-0.86}$ \hunit (flat \pcdm) and $70.21 \pm 1.33$ \hunit (non-flat \lcdm). We assume that the tension between early- and late-Universe measurements of $H_0$ is not a major issue here, because the 2D and 1D contours in Fig. \ref{fig01} overlap, and so we compute a combined $H_0$ value (but if one is concerned about the early- vs late-Universe $H_0$ tension then one should not compare our combined-data $H_0$\!'s here, and in Secs. \ref{subsec:ZBQ} and \ref{subsec:ZBQH}, directly to the measurements of \citealp{riess_etal_2019} or of \citealp{planck2018}).

In contrast to the \hiig\ only cases, when fit to the ZBH data combination the non-flat models mildly favor closed spatial hypersurfaces. This is because the $H(z)$ and BAO data mildly favor closed spatial hypersurfaces; see, e.g. \cite{park_ratra_2019b} and Chapter \ref{Chapter5}. For non-flat \lcdm, non-flat XCDM, and non-flat \pcdm, we find $\Omega_{k0}=-0.029^{+0.049}_{-0.048}$, $\Omega_{k0}=-0.082^{+0.135}_{-0.119}$, and $\Omega_{k0}=-0.153^{+0.114}_{-0.079}$, respectively, with the non-flat \pcdm\ model favoring closed spatial hypersurfaces at 1.34$\sigma$.

The fit to the ZBH data combination produces weaker evidence for dark energy dynamics (in comparison to the \hiig\ only case) with tighter error bars on the measured values of $w_{\rm X}$ and $\alpha$. For flat (non-flat) XCDM, $w_{\rm X}=-1.052^{+0.092}_{-0.082}$ ($w_{\rm X}=-0.958^{+0.219}_{-0.098}$), with $w_{\rm X}=-1$ still being within the 1$\sigma$ range. For flat (non-flat) \pcdm, $\alpha<0.411$ ($\alpha=0.538^{+0.151}_{-0.519}$), where the former is peaked at $\alpha=0$ but for the latter, $\alpha=0$ is just out of the 1$\sigma$ range.

\subsection{$H(z)$, BAO, and QSO (ZBQ) constraints}
\label{subsec:ZBQ}

The $H(z)$, BAO, and QSO (ZBQ) data combination was studied in Chapter \ref{Chapter5}. Relative to that analysis, we use an updated BAO data compilation, a more accurate formula for $r_s$, and the MCMC formalism (instead of the grid-based $\chi^2$ approach); consequently the parameter constraints derived here slightly differ from those of Chapter \ref{Chapter5}.

The 1D probability distributions and 2D confidence regions of the cosmological parameters for all models are presented in Figs. \ref{fig01}--\ref{fig06}, in green. The corresponding best-fitting results and uncertainties are listed in Tables \ref{tab:ch7_BFP} and \ref{tab:1d_BFP}.

The measured values of $\Omega_{m0}$ here fall within a similar range to the range quoted in the last subsection, being between $0.313^{+0.013}_{-0.015}$ (non-flat \lcdm) and $0.324^{+0.014}_{-0.015}$ (flat \pcdm). This range is larger than, but still consistent with, the range of $\Omega_{m0}$ reported in Chapter \ref{Chapter5}, where the same models are fit to the ZBQ data combination.

The $H_0$ measurements in this case fall within a broader range than in the ZBH case, being between $65.94^{+1.75}_{-1.73}$ \hunit (non-flat \pcdm) and $68.60 \pm 0.68$ km s$^{-1}$ Mpc$^{-1}$ (flat \lcdm). In addition, they are lower than the corresponding measurements in the ZBH cases, and are in better agreement with the median statistics \citep{chenratmed} estimate of $H_0$ than with what is measured from the local expansion rate \citep{riess_etal_2019}. Compared with Chapter \ref{Chapter5}, the central values are lower except for the non-flat XCDM model.

For non-flat \lcdm, non-flat XCDM, and non-flat \pcdm, we measure $\Omega_{k0}=0.029^{+0.056}_{-0.063}$, $\Omega_{k0}=-0.078^{+0.124}_{-0.112}$, and $\Omega_{k0}=-0.103^{+0.111}_{-0.091}$, respectively. These results are consistent with their unmarginalized best-fittings (see Table \ref{tab:ch7_BFP}), where the best-fitting to the non-flat \lcdm\ model favors open spatial hypersurfaces, and the best-fittings to the non-flat XCDM parametrization and the non-flat \pcdm\ model both favor closed spatial hypersurfaces. Note that the central values are larger than those of Chapter \ref{Chapter5}, especially for non-flat \lcdm\ (positive instead of negative). In all three models the constraints are consistent with flat spatial hyperfurfaces.

The fit to the ZBQ data combination provides slightly stronger evidence for dark energy dynamics than does the fit to the ZBH data combination. For flat (non-flat) XCDM, $w_{\rm X}=-0.911^{+0.122}_{-0.098}$ ($w_{\rm X}=-0.826^{+0.185}_{-0.088}$), with the former barely within 1$\sigma$ of $w_{\rm X}=-1$ and the latter almost 2$\sigma$ away from $w_{\rm X}=-1$. For flat (non-flat) \pcdm, $\alpha=0.460^{+0.116}_{-0.440}$ ($\alpha=0.854^{+0.379}_{-0.594}$), with the former 1.05$\sigma$ and the latter 1.44$\sigma$ away from the $\alpha=0$ cosmological constant. In comparison with Chapter \ref{Chapter5}, the central values of $w_{\rm X}$ are larger and smaller for flat and non-flat XCDM models, respectively, and that of $\alpha$ are larger for both flat and non-flat \pcdm\ models.

\subsection{$H(z)$, BAO, QSO, and \hiig\ (ZBQH) constraints}
\label{subsec:ZBQH}

Comparing the results of the previous two subsections, we see that when used in conjunction with $H(z)$ and BAO data, the QSO data result in tighter constraints on $\Omega_{m0}$, $\Omega_{k0}$ (in non-flat XCDM), $w_{\rm X}$ (in non-flat XCDM), and $H_0$ (in flat \lcdm), while the \hiig\ data result in tighter constraints on $H_0$ (except for flat \lcdm), $\Omega_{\Lambda}$, $\Omega_{k0}$(in non-flat \lcdm\ and \pcdm), $w_{\rm X}$ (in flat XCDM), and $\alpha$. Consequently, it is useful to derive constraints from an analysis of the combined $H(z)$, BAO, QSO, and \hiig\ (ZBQH) data. We present the results of such an analysis in this subsection.

In Figs. \ref{fig01}--\ref{fig06}, we present the 1D probability distributions and 2D confidence constraints for the ZBQH cases in blue. Tables \ref{tab:ch7_BFP} and \ref{tab:1d_BFP} list the best-fitting results and uncertainties.

It is interesting that the best-fitting values of $\Omega_{m0}$ in this case are lower compared with both the ZBQ and the ZBH results, being between $0.309^{+0.015}_{-0.014}$ (non-flat XCDM) and $0.319 \pm 0.013$ (flat \pcdm). The best-fitting values of $H_0$ are higher than the ZBQ cases and have central values that are closer to those of the ZBH cases, but are still in better agreement with the lower median statistics estimate of $H_0$ \citep{chenratmed} than the higher local expansion rate measurement of $H_0$ \citep{riess_etal_2019}, being between $68.18^{+0.97}_{-0.75}$ \hunit (flat \pcdm) and $69.90 \pm 1.48$ \hunit (flat XCDM). 

For non-flat \lcdm, non-flat XCDM, and non-flat \pcdm, we measure $\Omega_{k0}=-0.021^{+0.044}_{-0.048}$, $\Omega_{k0}=-0.025 \pm 0.092$, and $\Omega_{k0}=-0.098^{+0.082}_{-0.061}$, respectively. For non-flat \lcdm\ and XCDM, the measured values of the curvature energy density parameter are within 0.48$\sigma$ and 0.27$\sigma$ of $\Omega_{k0} = 0$, respectively, while the non-flat \pcdm\ model favors a closed geometry with an $\Omega_{k0}$ that is 1.20$\sigma$ away from zero.

There is not much evidence in support of dark energy dynamics in the ZBQH case, with $\Lambda$ consistent with this data combination. For flat (non-flat) XCDM, $w_{\rm X}=-1.053^{+0.091}_{-0.082}$ ($w_{\rm X}=-1.022^{+0.208}_{-0.104}$). For flat (non-flat) \pcdm, the $2\sigma$ upper limits are $\alpha<0.411$ ($\alpha<0.926$), which indicates that $\alpha = 0$ or $\Lambda$ is consistent with these data.

\subsection{Model comparison}
\label{sec:comparison}

From Table \ref{tab:cab}, we see that the reduced $\chi^2$ for all models is relatively large (being between 2.25 and 2.75). This could probably be attributed to underestimated systematic uncertainties in the \hiig\ data.\footnote{Underestimated systematic uncertainties might also explain the large reduced $\chi^2$ of QSO data Chapter \ref{Chapter5}.} This is suggested by \cite{G-M_2019}, who also found relatively large values of $\chi^2/\nu$ in their cosmological model fits to the \hiig\ data (though not as large as ours, because they compute a different $\chi^2$, as explained in footnote \ref{fn5} in Section \ref{sec:ch7_Methods}). They note that an additional systematic uncertainty of $\sim0.22$ could bring their $\chi^2/\nu$ down to $\sim1$. As mentioned previously, we do not account for \hiig\ systematic uncertainties in our analysis.

\begin{table*}
\centering
\resizebox{\columnwidth}{!}{%
\begin{threeparttable}

\caption{$\Delta \chi^2$, $\Delta AIC$, $\Delta BIC$, and $\chi^2_{\mathrm{min}}/\nu$ values.}
\label{tab:cab}

\begin{tabular}{lccccccc}
\hline
 Quantity & Data set & Flat \lcdm & Non-flat \lcdm & Flat XCDM & Non-flat XCDM & Flat \pcdm & Non-flat \pcdm\\
\hline
 & \hiig\ & 3.06 & 2.75 & 3.03 & 0.00 & 3.01 & 2.22\\
$\Delta \chi^2$ & $H(z)$ + BAO + \hiig\ & 1.54 & 0.63 & 1.24 & 0.10 & 1.61 & 0.00 \\
 & $H(z)$ + BAO + QSO & 2.20 & 2.14 & 1.27 & 0.00 & 1.37 & 0.15\\
 & $H(z)$ + BAO + QSO + \hiig\ & 0.85 & 0.14 & 0.54 & 0.05 & 0.93 & 0.00\\
 \\
 & \hiig\ & 0.00 & 1.69 & 1.97 & 0.94 & 1.95 & 3.16\\
$\Delta AIC$ & $H(z)$ + BAO + \hiig\ & 0.00 & 1.09 & 1.70 & 2.56 & 2.07 & 2.46\\
 & $H(z)$ + BAO + QSO & 0.00 & 1.94 & 1.07 & 1.80 & 1.17 & 1.95\\
 & $H(z)$ + BAO + QSO + \hiig\ & 0.00 & 1.29 & 1.69 & 3.20 & 2.08 & 3.15\\
 \\
 & \hiig\ & 0.00 & 4.72 & 5.01 & 7.00 & 4.99 & 9.22\\
$\Delta BIC$ & $H(z)$ + BAO + \hiig\ & 0.00 & 4.35 & 4.97 & 9.10 & 5.34 & 9.00\\
 & $H(z)$ + BAO + QSO & 0.00 & 5.02 & 4.15 & 7.97 & 4.25 & 8.12\\
 & $H(z)$ + BAO + QSO + \hiig\ & 0.00 & 5.04 & 5.44 & 10.70 & 5.83 & 10.65\\
 \\
 & \hiig\ & 2.72 & 2.74 & 2.74 & 2.74 & 2.74 & 2.75\\
$\chi^2_{\mathrm{min}}/\nu$ & $H(z)$ + BAO + \hiig\ & 2.25 & 2.26 & 2.26 & 2.27 & 2.26 & 2.27\\
 & $H(z)$ + BAO + QSO & 2.33 & 2.34 & 2.34 & 2.35 & 2.34 & 2.35\\
 & $H(z)$ + BAO + QSO + \hiig\ & 2.51 & 2.52 & 2.52 & 2.53 & 2.52 & 2.53\\
\hline
\end{tabular}
\end{threeparttable}%
}
\end{table*}

One thing that is clear, regardless of the absolute size of \hiig\ or QSO systematics (and ignoring the large values of $\chi^2/\nu$), is that the flat \lcdm\ model remains the most favored model among the six models we studied, based on the $AIC$ and $BIC$ criteria (see Table \ref{tab:cab}).\footnote{Note that based on the $\Delta \chi^2$ results of Table \ref{tab:cab} non-flat XCDM has the minimum $\chi^2$ in the \hiig\ and ZBQ cases, whereas non-flat \pcdm\ has the minimum $\chi^2$ for the ZBH and ZBQH cases. The $\Delta \chi^2$ values do not, however, penalize a model for having more parameters.} In Table \ref{tab:cab} we define $\Delta \chi^2$, $\Delta AIC$, and $\Delta BIC$, respectively, as the differences between the values of the $\chi^2$, $AIC$, and $BIC$ associated with a given model and their corresponding minimum values among all models.

From the \hiig\ results for $\Delta AIC$ and $\Delta BIC$ listed in Table \ref{tab:cab}, we see that the evidence against non-flat \lcdm, flat XCDM, and flat \pcdm\ is weak (according to $\Delta AIC$) and positive (according to $\Delta BIC$) where, among these three models, the flat XCDM model is the least favored. The evidence against the non-flat XCDM model is weak regarding $\Delta AIC$ but strong based on $\Delta BIC$, while the evidence against non-flat \pcdm\ in this case is positive ($\Delta AIC$) and strong ($\Delta BIC$), respectively, with it being the least favored model overall.

Largely similar conclusions result from $\Delta AIC$ and $\Delta BIC$ values for the \hiig\ and ZBQ data. The exception is that the ZBQ $\Delta AIC$ value gives only weak evidence against non-flat \pcdm, instead of the positive evidence against it from the \hiig\ $\Delta AIC$ value.

The ZBH and ZBQH values of $\Delta AIC$ and $\Delta BIC$ result in the following conclusions:

1) the evidence against both non-flat \lcdm\ and flat XCDM is weak (ZBH) and positive (ZBQH) for $\Delta AIC$ and $\Delta BIC$;

2) the evidence against flat \pcdm\ is positive; 

3) non-flat XCDM is the least favored model with non-flat \pcdm\ doing almost as badly. $\Delta AIC$ gives positive evidence against non-flat XCDM and non-flat \pcdm, while $\Delta BIC$ strongly disfavors (ZBH) and very strongly disfavors (ZBQH) both of these non-flat models.

\section{Conclusions}
\label{sec:ch7_conclusion}
In this paper, we have constrained cosmological parameters in six flat and non-flat cosmological models by analyzing a total of 315 observations, comprising 31 $H(z)$, 11 BAO, 120 QSO, and 153 \hiig\ measurements. The QSO angular size and \hiig\ apparent magnitude measurements are particularly noteworthy, as they reach to $z\sim2.7$ and $z\sim2.4$ respectively (somewhat beyond the highest $z\sim2.3$ reached by BAO data) and into a much less studied area of redshift space. While the current \hiig\ and QSO data do not provide very restrictive constraints, they do tighten the limits when they are used in conjunction with BAO + $H(z)$ data.

By measuring cosmological parameters in a variety of cosmological models, we are able to draw some relatively model-independent conclusions (i.e. conclusions that do not differ significantly between the different models). Specifically, for the full data set (i.e the ZBQH data), we find quite restrictive constraints on $\Omega_{m0}$, a reasonable summary perhaps being $\Omega_{m0}=0.310 \pm 0.013$, in good agreement with many other recent estimates. $H_0$ is also fairly tightly constrained, with a reasonable summary perhaps being $H_0=69.5 \pm 1.2$ \hunit, which is in better agreement with the results of \cite{chenratmed} and \cite{planck2018} than that of \cite{riess_etal_2019}. The ZBQH measurements are consistent with the standard spatially-flat \lcdm\ model, but do not strongly rule out mild dark energy dynamics or a little spatial curvature energy density. More and better-quality \hiig, QSO, and other data at $z \sim 2$--4 will significantly help to test these extensions.

%% file: chapter8.tex
\cleardoublepage

\chapter{Cosmological constraints from higher-redshift gamma-ray burst, HII starburst galaxy, and quasar (and other) data}
\chaptermark{Cosmological constraints from observations}

\label{Chapter8}

This chapter is based on \cite{Cao_Ryan_Khadka_Ratra}. Figures and tables by Shulei Cao, from analyses conducted independently by Shulei Cao, Joseph Ryan, and Narayan Khadka.

%%
%Section: Intro
%%
\section{Introduction} \label{sec:ch8_intro}
\begin{comment}
There is a large body of evidence indicating that the Universe recently transitioned from a decelerated to an accelerated phase of expansion (at redshift $z \sim 3/4$; see e.g. \citealp{Farooq_Ranjeet_Crandall_Ratra_2017}) and has been undergoing accelerated expansion ever since (for reviews, see e.g. \citealp{Ratra_Vogeley,Martin,Coley_Ellis}). In the standard model of cosmology, called the \lcdm\ model \citep{peeb84}, the accelerated expansion is powered by a constant dark energy density (the cosmological constant, $\Lambda$). This model also assumes that spatial hypersurfaces are flat on cosmological scales, and that the majority of non-relativistic matter in the Universe consists of cold dark matter (CDM). 
\end{comment}
Out of all the models that have been devised to explain the observed accelerated expansion of the Universe, the \lcdm\ model is currently the most highly favored in terms of both observational data and theoretical parsimony (see e.g. \citealp{Farooq_Ranjeet_Crandall_Ratra_2017,scolnic_et_al_2018,planck2018,eBOSS_2020}). In spite of these virtues, however, there are some indications that the \lcdm\ model may not tell the whole story. On the observational side, some workers have found evidence of discrepancies between the \lcdm\ model and cosmological observations (\citealp{riess_2019, martinelli_tutusaus_2019}) and on the theoretical side, the origin of $\Lambda$ has yet to be explained in fundamental terms (e.g., \citealp{Martin}). One way to pin down the nature of dark energy is by studying its dynamics phenomenologically, as we have seen in the last three chapters. It is possible that the dark energy density may evolve in time (\citealp{6}), and many dark energy models exhibiting this behavior have been proposed.

Cosmological models have largely been tested in the redshift range $0 \lesssim z \lesssim 2.3$, with baryon acoustic oscillation (BAO\footnote{In our BAO data analyses in this paper the sound horizon computation assumes a value for the current baryonic matter physical density parameter $\Omega_{b0} h^2$, appropriate for the model under study, computed from Planck CMB anisotropy data.}) measurements probing the upper end of this range, and at $z\sim1100$, using cosmic microwave background (CMB) anisotropy data. To determine the accuracy of our cosmological models, we also need to test them in the redshift range $2.3 \lesssim z \lesssim 1100$. Quasar angular size (QSO-AS), HII starburst galaxy (HIIG), quasar X-ray and UV flux (QSO-Flux), and gamma-ray burst (GRB) measurements are some of the handful of data available in this range. The main goal of this paper is, therefore, to examine the effect that QSO-AS, HIIG, and GRB data have on cosmological model parameter constraints, in combination with each other, and in combination with more well-known probes.\footnote{We relegate the analysis of QSO-Flux data to an appendix (Sec. \ref{sec:appendix}), the reasons for which are discussed there.} 

Gamma-ray bursts are promising cosmological probes for two reasons. First, it is believed that they can be used as standardizable candles \citep{Lamb_2000, Lamb2001, Amati2002, Amati2008, Amati_2009, Ghirlanda2004, Demianski2011, Fyan2015}. Second, they cover a redshift range that is wider than most other commonly-used cosmological probes, having been observed up to $z \sim 8.2$ \citep{Amati2008, Amati_2009, Amati2019, samushia_ratra_2010, Demianski2011, Wang_2016, Demianski_2017a, Demianski_2019, Dirirsa_2019, Khadka_Ratra_2020}. In particular, the $z\sim 2.7$--8.2 part of the Universe is primarily accessed by GRBs,\footnote{Though QSO-Flux measurements can reach up to $z \sim 5.1$.} so if GRBs can be standardized, they could provide useful information about a large, mostly unexplored, part of the Universe.

QSO-AS data currently reach to $z\sim 2.7$. These data, consisting of measurements of the angular size of astrophysical radio sources, furnish a standard ruler that is independent of that provided by the BAO sound horizon scale. The intrinsic linear size $l_m$ of intermediate luminosity QSOs has recently been accurately determined by \cite{Cao_et_al2017b}, opening the way for QSOs to, like GRBs, test cosmological models in a little-explored region of redshift space.\footnote{The use of QSO-AS measurements to constrain cosmological models dates back to near the turn of the century (e.g. \citealp{gurvits_kellermann_frey_1999, vishwakarma_2001, lima_alcaniz_2002, zhu_fujimoto_2002, Chen_Ratra_2003}), but, as discussed in Chapter \ref{Chapter5}, these earlier results are suspect, because they are based on an inaccurate determination of $l_m$.}

HIIG data reach to $z\sim 2.4$, just beyond the range of current BAO data. Measurements of the luminosities of the Balmer lines in HII galaxies can be correlated with the velocity dispersion of the radiating gas, making HII galaxies a standard candle that can complement both GRBs and lower-redshift standard candles like supernovae (\citealp{Siegel_2005,Plionis_2009,Mania_2012,Chavez_2014,G-M_2019}).

Current QSO-Flux measurements reach to $z\sim 5.1$, but they favor a higher value of the current (denoted by the subscript ``0'') non-relativistic matter density parameter ($\Omega_{m0}$) than what is currently thought to be reasonable. The $\Omega_{m0}$ values obtained using QSO-Flux data, in a number of cosmological models, are in nearly 2$\sigma$ tension with the values obtained by using other well-established cosmological probes like CMB, BAO, and Type Ia supernovae (\citealp{RisalitiandLusso_2019, Yang_2019, Wei_Melia_2020, KhadkaandRatra_2020}). Techniques for standardizing QSO-Flux measurements are still under development, so it might be too early to draw strong conclusions about the cosmological constraints obtained from QSO-Flux measurements. Therefore, in this chapter, we use QSO-Flux data alone and in combination with other data to constrain cosmological parameters in four different models, and record these results in Sec. \ref{sec:appendix}.

We find that the GRB, HIIG, and QSO-AS constraints are largely mutually consistent, and that their joint constraints are consistent with those from more widely used, and more restrictive, BAO and Hubble parameter ($H(z)$) data. When used jointly with the $H(z)$ + BAO data, these higher-$z$ data tighten the $H(z)$ + BAO constraints.

\begin{comment}
This paper is organized as follows. In Section \ref{sec:data} we introduce the data we use. Section \ref{sec:model} describes the models we analyze, with a description of our analysis method in Section \ref{sec:ch8_analysis}. Our results are in Section \ref{sec:results}, and we provide our conclusions in Section \ref{sec:conclusion}. Additionally, we discuss our results for QSO-Flux measurements in Appendix \ref{sec:appendix}.
\end{comment}
%%
%Section: Data
%%
\section{Data}
\label{sec:ch8_Data}

%Subsection: GRB data
\subsection{GRB data}
\label{subsec:GRB_data}
\begin{comment}
[Gamma-ray bursts are promising cosmological probes for two reasons. First, it is believed that they can be used as standardizable candles \citep{Lamb_2000, Lamb2001, Amati2002, Amati2008, Amati_2009, Ghirlanda2004, Demianski2011, Fyan2015}. Second, they cover a redshift range that is wider than most other commonly-used cosmological probes, having been observed up to $z \sim 8.2$ \citep{Amati2008, Amati_2009, Amati2019, samushia_ratra_2010, Demianski2011, Wang_2016, Demianski_2017a, Demianski_2019, Dirirsa_2019, Khadka_Ratra_2020}. In particular, the $z\sim 2.7$--8.2 part of the Universe is primarily accessed by GRBs,\footnote{Though QSO-Flux measurements can reach up to $z \sim 5.1$.} so if GRBs can be standardized, they could provide useful information about a large, mostly unexplored, part of the Universe.]
\end{comment}
We use QSO-AS, \hiig, QSO-Flux, and GRB data to obtain constraints on the cosmological models we study. The QSO-AS data, comprising 120 measurements compiled by \cite{Cao_et_al2017b} (listed in Table 1 of that paper) and spanning the redshift range $0.462 \leq z \leq 2.73$, are also used in Chapter \ref{Chapter5}. The \hiig\ data, comprising 107 low redshift ($0.0088 \leq z \leq 0.16417$) \hiig\ measurements, used in \cite{Chavez_2014} (recalibrated by \citealp{G-M_2019}), and 46 high redshift ($0.636427 \leq z \leq 2.42935$) \hiig\ measurements, used in \cite{G-M_2019}, are also used in Chapter \ref{Chapter7}. The GRB data, spanning the redshift range $0.48 \leq z \leq 8.2$, are collected from \cite{Dirirsa_2019} (25 from Table 2 of that paper (F10), and the remaining 94 from Table 5 of the same, which are a subset of those compiled by \citealp{Wang_2016}) and also used in \cite{Khadka_Ratra_2020}. We also add 1598 QSO-Flux measurements spanning the redshift range $0.036 \leq z \leq 5.1003$, from \cite{RisalitiandLusso_2019}. These data are used in \cite{Khadka_2020b}; see that paper for details. Results related to these QSO-Flux data are discussed in Sec. \ref{sec:appendix}.

In order to be useful as cosmological probes, GRBs need to be standardized, and many phenomenological relations have been proposed for this purpose (\citealp{Amati2002}, \citealp{Ghirlanda2004}, \citealp{Liang2005}, \citealp{Muccino_2020}, and references therein). As in \cite{Khadka_Ratra_2020}, we use the Amati relation (\citealp{Amati2002}), which is an observed correlation between the peak photon energy $E_{\rm p}$ and the isotropic-equivalent radiated energy $E_{\rm iso}$ of long-duration GRBs, to standardize GRB measurements. There have been many attempts to standardize GRBs using the Amati relation. Some analyses assume a fixed value of $\Omega_{\rm m_0}$ to calibrate the Amati relation, so they favor a relatively reasonable value of $\Omega_{\rm m_0}$. Others use supernovae data to calibrate the Amati relation, while some use $H(z)$ data to calibrate it. This means that most previous GRB analyses are affected by some non-GRB external factors. In some cases this leads to a circularity problem, in which the models to be constrained by using the Amati relation are also used to calibrate the Amati relation itself (\citealp{Liu_Wei_2015, Demianski_2017a, Demianski_2019, Dirirsa_2019}). In other cases, the data used in the calibration process dominate the analysis results. To overcome these problems, we fit the parameters of the Amati relation simultaneously with the parameters of the cosmological models we study (as done in \citealp{Khadka_Ratra_2020}; also see \citealp{Wang_2016}).

The isotropic radiated energy $E_{\rm iso}$ of a source in its rest frame at a luminosity distance $D_L$ is
\begin{equation}
\label{Eiso}
    E_{\rm iso}=\frac{4\pi D_L^2}{1+z}S_{\rm bolo},
\end{equation}
where $S_{\rm bolo}$ is the bolometric fluence, and $D_L$ (defined in Chapter \ref{Chapter2}) depends on $z$ and on the parameters of our cosmological models. $E_{\rm iso}$ is connected to the source's peak energy output $E_{\rm p}$ via the Amati relation \citep{Amati2008, Amati_2009}
\begin{equation}
    \label{eq:Amati}
    \log E_{\rm iso} = a  + b\log E_{\rm p},
\end{equation}
where $a$ and $b$ are free parameters that we vary in our model fits.\footnote{$\log=\log_{10}$ is implied hereinafter.} Note here that the peak energy $E_{\rm p} = (1+z)E_{\rm p, obs}$ where $E_{\rm p, obs}$ is the observed peak energy.

See Chapters \ref{Chapter5} and \ref{Chapter7} for discussions of the QSO-AS data and the HIIG data, respectively.

%%
%Section: Methods
%%
\section{Data Analysis Methodology}
\label{sec:ch8_analysis}

By using the \textsc{python} module \textsc{emcee} \citep{Foreman-Mackey_Hogg_Lang_Goodman_2013}, we perform a Markov chain Monte Carlo (MCMC) analysis to maximize the likelihood function, $\mathcal{L}$, and thereby determine the best-fitting values of the free parameters (see Chapter \ref{Chapter7} for the details of this method). The flat cosmological parameter priors are the same as those used in Chapter \ref{Chapter7} and the flat priors of the parameters of the Amati relation are non-zero over $0\leq\sigma_{\rm ext}\leq10$ (described below), $40\leq a\leq60$, and $0\leq b\leq5$. 

The likelihood functions associated with $H(z)$, BAO, HIIG, and QSO-AS data are described in Chapter \ref{Chapter7}. For GRB data, the natural log of its likelihood function \citep{D'Agostini_2005} is
\begin{equation}
\label{eq:LH_GRB}
    \ln\mathcal{L}_{\rm GRB}= -\frac{1}{2}\Bigg[\chi^2_{\rm GRB}+\sum^{119}_{i=1}\ln\left(2\pi(\sigma_{\rm ext}^2+\sigma_{{y_i}}^2+b^2\sigma_{{x_i}}^2)\right)\Bigg],
\end{equation}
where
\begin{equation}
\label{eq:chi2_GRB}
    \chi^2_{\rm GRB} = \sum^{119}_{i=1}\bigg[\frac{(y_i-b x_i-a)^2}{(\sigma_{\rm ext}^2+\sigma_{{y_i}}^2+b^2\sigma_{{x_i}}^2)}\bigg],
\end{equation}
$x=\log\frac{E_{\rm p}}{\rm keV}$, $\sigma_{x}=\frac{\sigma_{E_{\rm p}}}{E_{\rm p}\ln 10}$, $y=\log\frac{E_{\rm iso}}{\rm erg}$, and $\sigma_{\rm ext}$ is the extrinsic scatter parameter, which contains the unknown systematic uncertainty. For the GRB with $\sigma_z$ uncertainty in $z$,
\begin{equation}
\label{eq:err_y2}
    \sigma^2_{y}=\left(\frac{\sigma_{S_{\rm bolo}}}{S_{\rm bolo}\ln 10}\right)^2+\left(\frac{2(1+z)\frac{\partial D_M}{\partial z}+D_M}{(1+z)D_M\ln 10}\sigma_z\right)^2,
\end{equation}
and for those without $z$ uncertainties $\sigma_z=0$ (the non-zero $\sigma_z$ has a negligible effect on our results).

The Akaike Information Criterion ($AIC$) and the Bayesian Information Criterion ($BIC$) are used to compare the goodness of fit of models with different numbers of parameters, where
\begin{equation}
    AIC=-2\ln \mathcal{L}_{\rm max} + 2n,
\end{equation}
and
\begin{equation}
    BIC=-2\ln \mathcal{L}_{\rm max} + n\ln N.
\end{equation}
In these equations, $\mathcal{L}_{\rm max}$ is the maximum value of the relevant likelihood function, $n$ is the number of free parameters of the model under consideration, and $N$ is the number of data points (e.g., for GRB $N=119$).

%%
%Section: Results
%%
\section{Results}
\label{sec:chi8_results}

\subsection{HIIG, QSO-AS, and GRB constraints, individually}
\label{subsec:GRB}

\begin{figure*}
\centering
    \includegraphics[width=3.5in,height=3.5in]{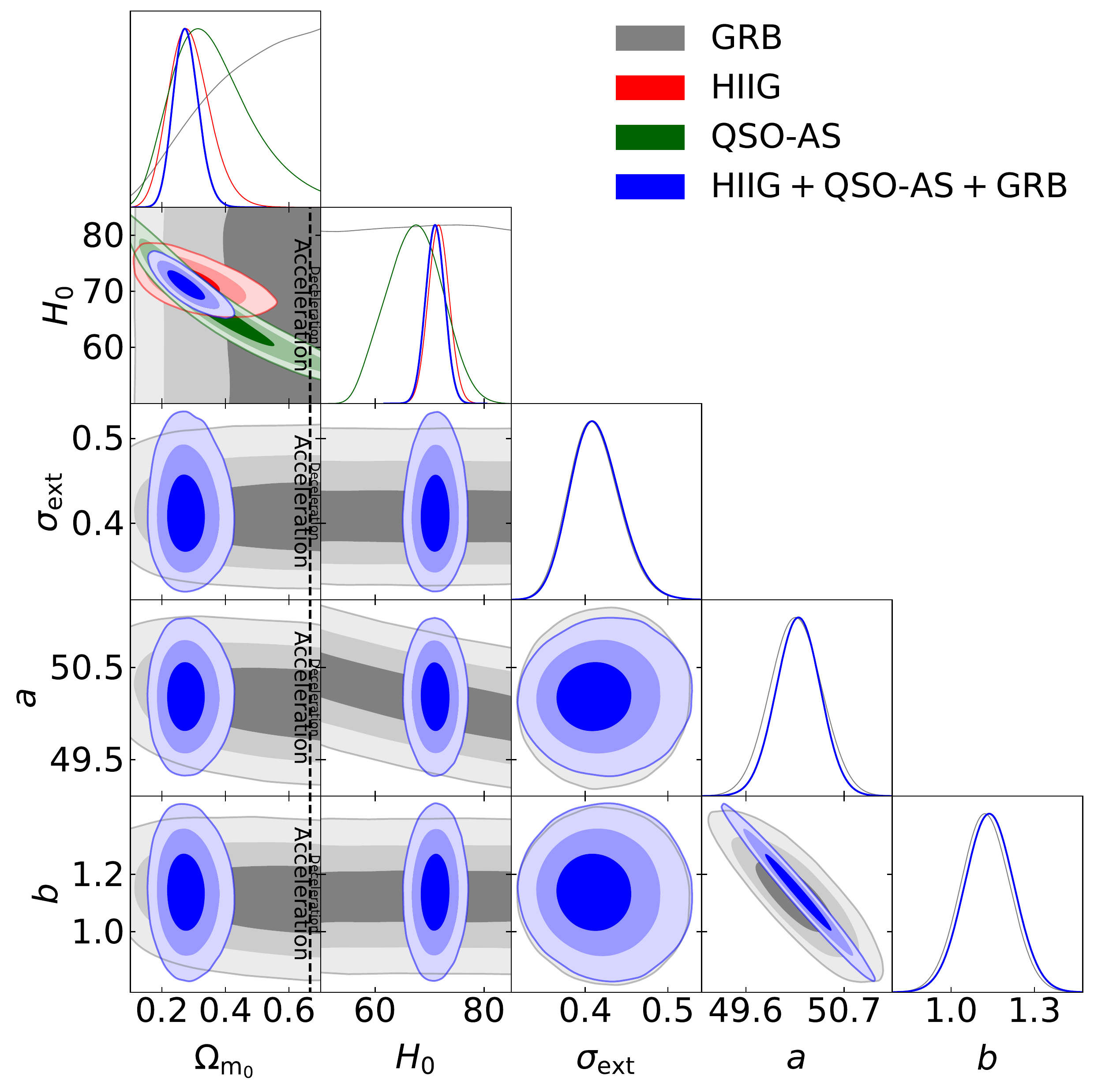}
    \includegraphics[width=3.5in,height=3.5in]{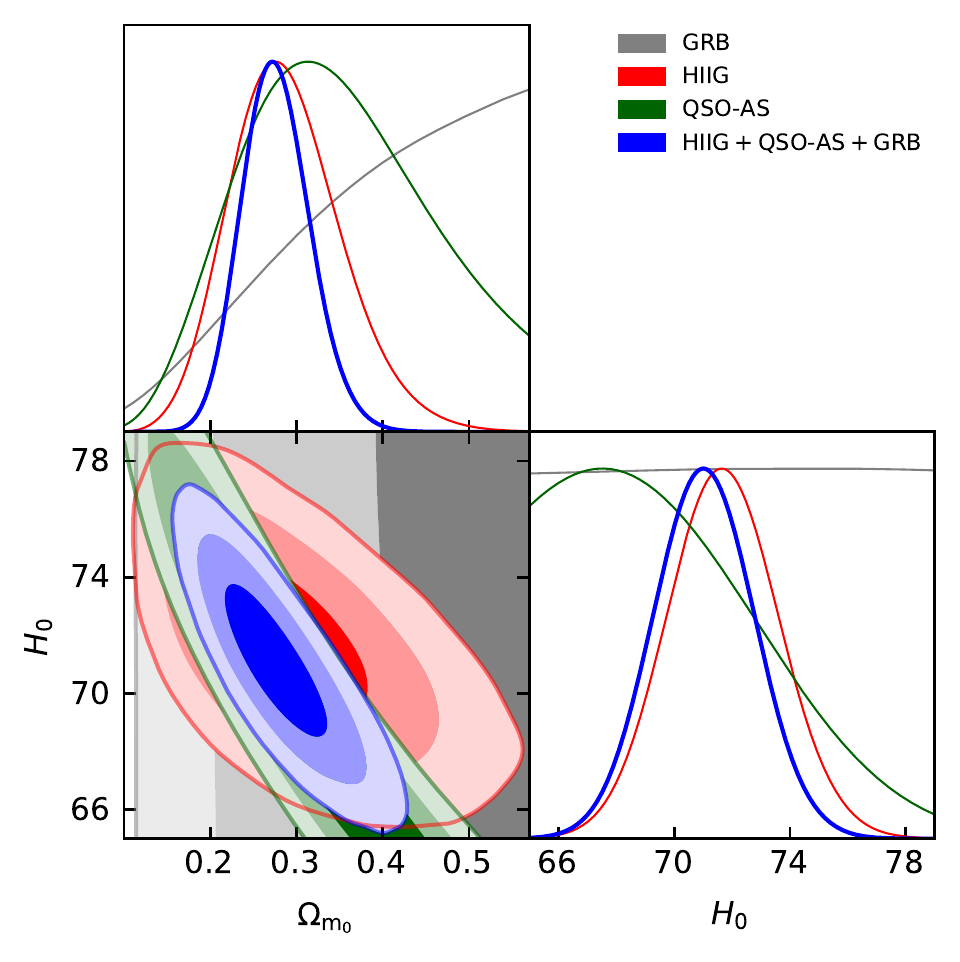}\\
\caption[1$\sigma$, 2$\sigma$, and 3$\sigma$ confidence contours for flat \lcdm.]{1$\sigma$, 2$\sigma$, and 3$\sigma$ confidence contours for flat \lcdm, where the right panel is the cosmological parameters comparison zoomed in. The black dotted lines in the left sub-panels of the left panel are the zero-acceleration lines, which divide the parameter space into regions associated with currently-accelerating (left) and currently-decelerating (right) cosmological expansion.}
\label{fig1}
\end{figure*}

\begin{figure*}
\centering
    \includegraphics[width=3.5in,height=3.5in]{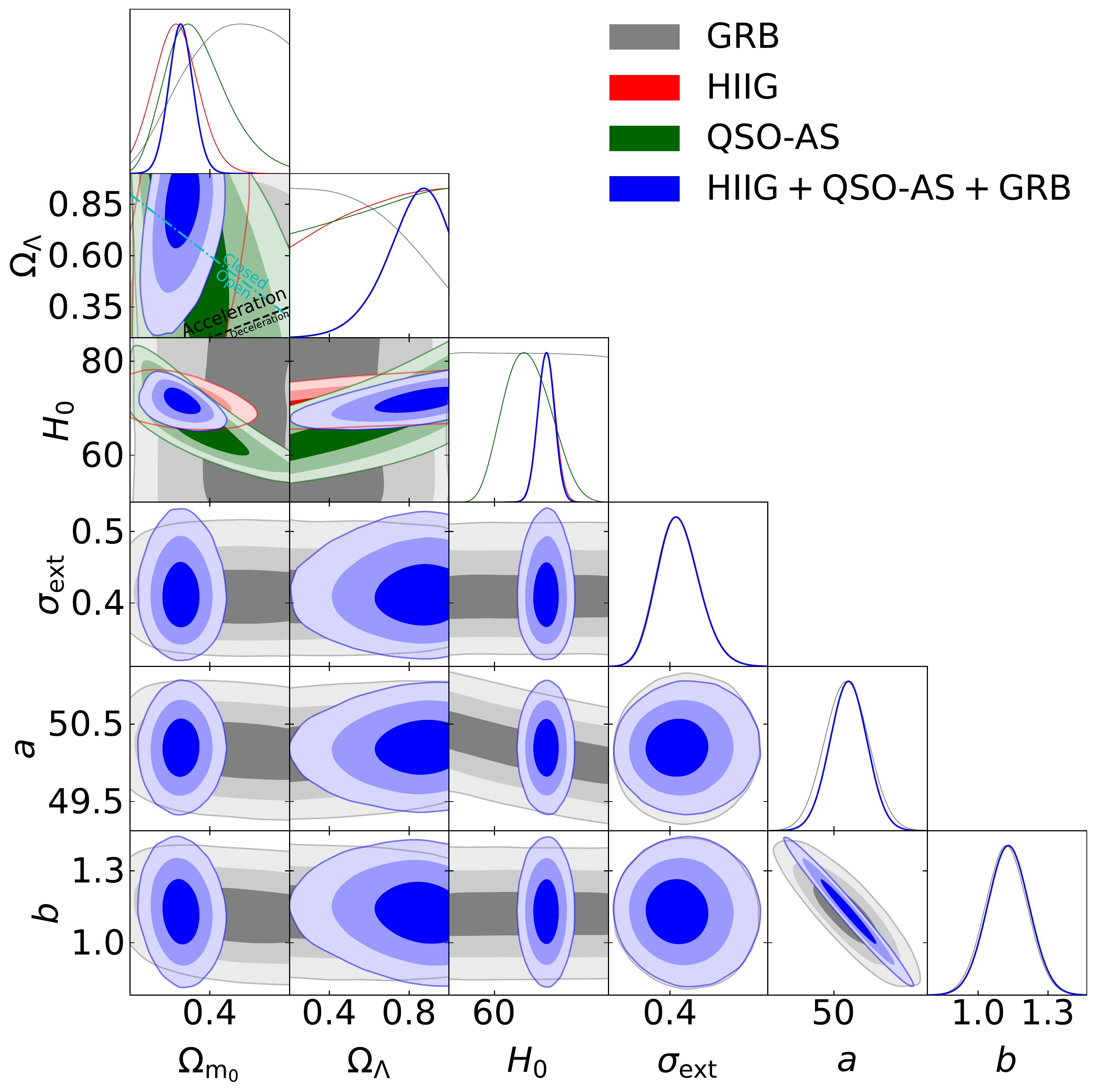}
    \includegraphics[width=3.5in,height=3.5in]{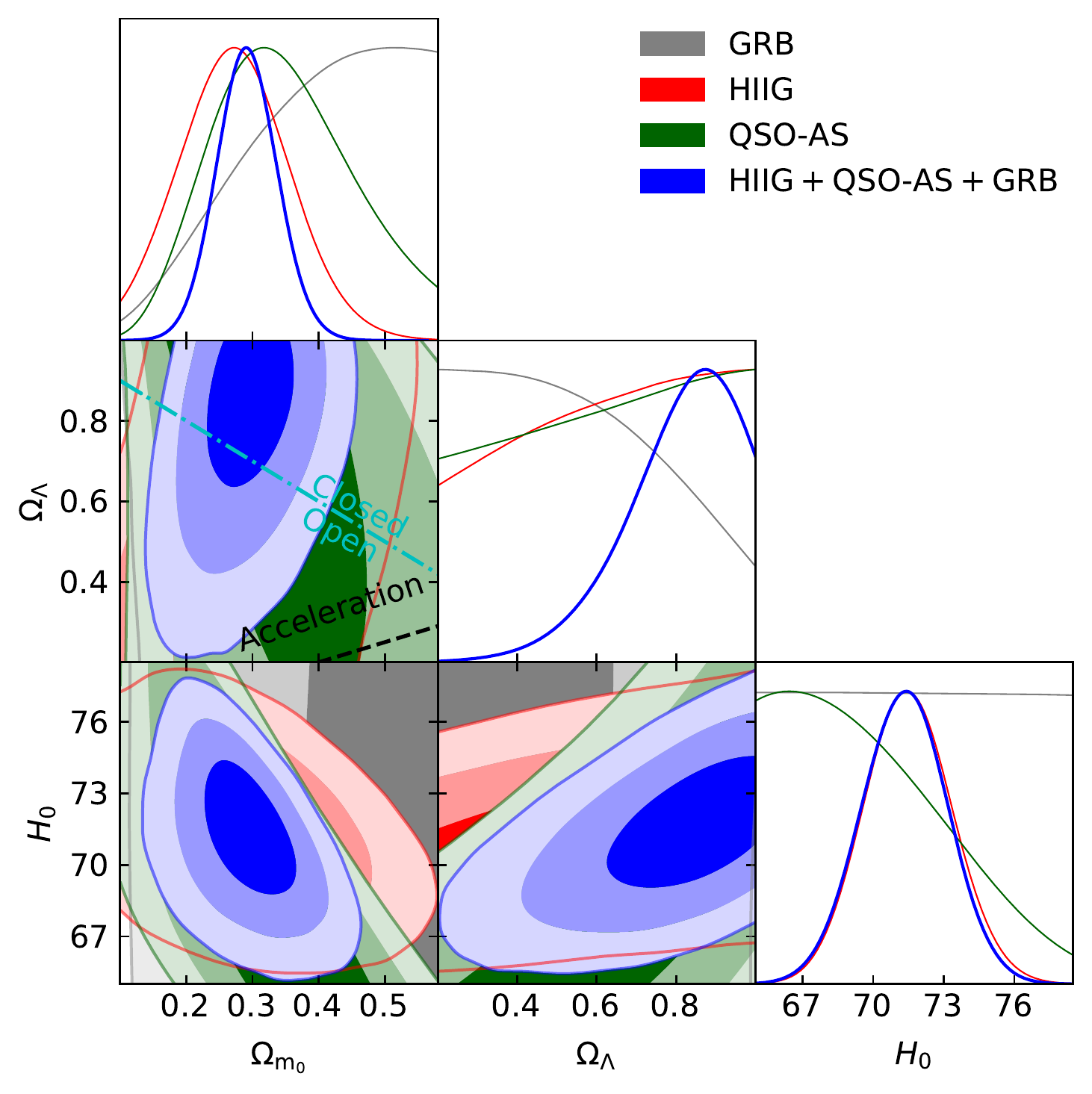}\\
\caption[1$\sigma$, 2$\sigma$, and 3$\sigma$ confidence contours for non-flat \lcdm.]{Same as Fig. \ref{fig1} but for non-flat \lcdm. The cyan dash-dot line represents the flat \lcdm\ case, with closed spatial hypersurfaces to the upper right. The black dotted line is the zero-acceleration line, which divides the parameter space into regions associated with currently-accelerating (above left) and currently-decelerating (below right) cosmological expansion.}
\label{fig2}
\end{figure*}

\begin{figure*}
\centering
    \includegraphics[width=3.5in,height=3.5in]{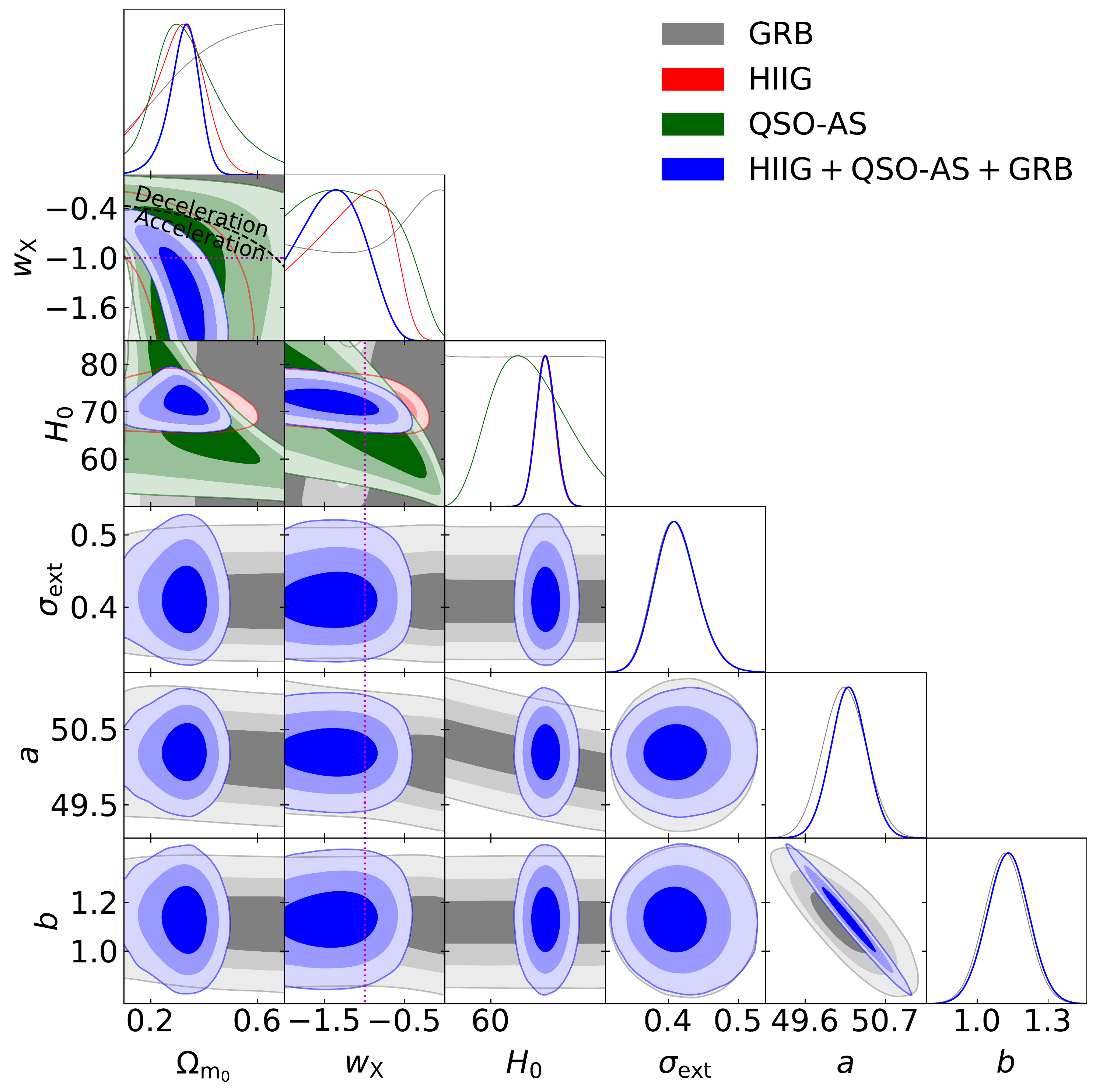}
    \includegraphics[width=3.5in,height=3.5in]{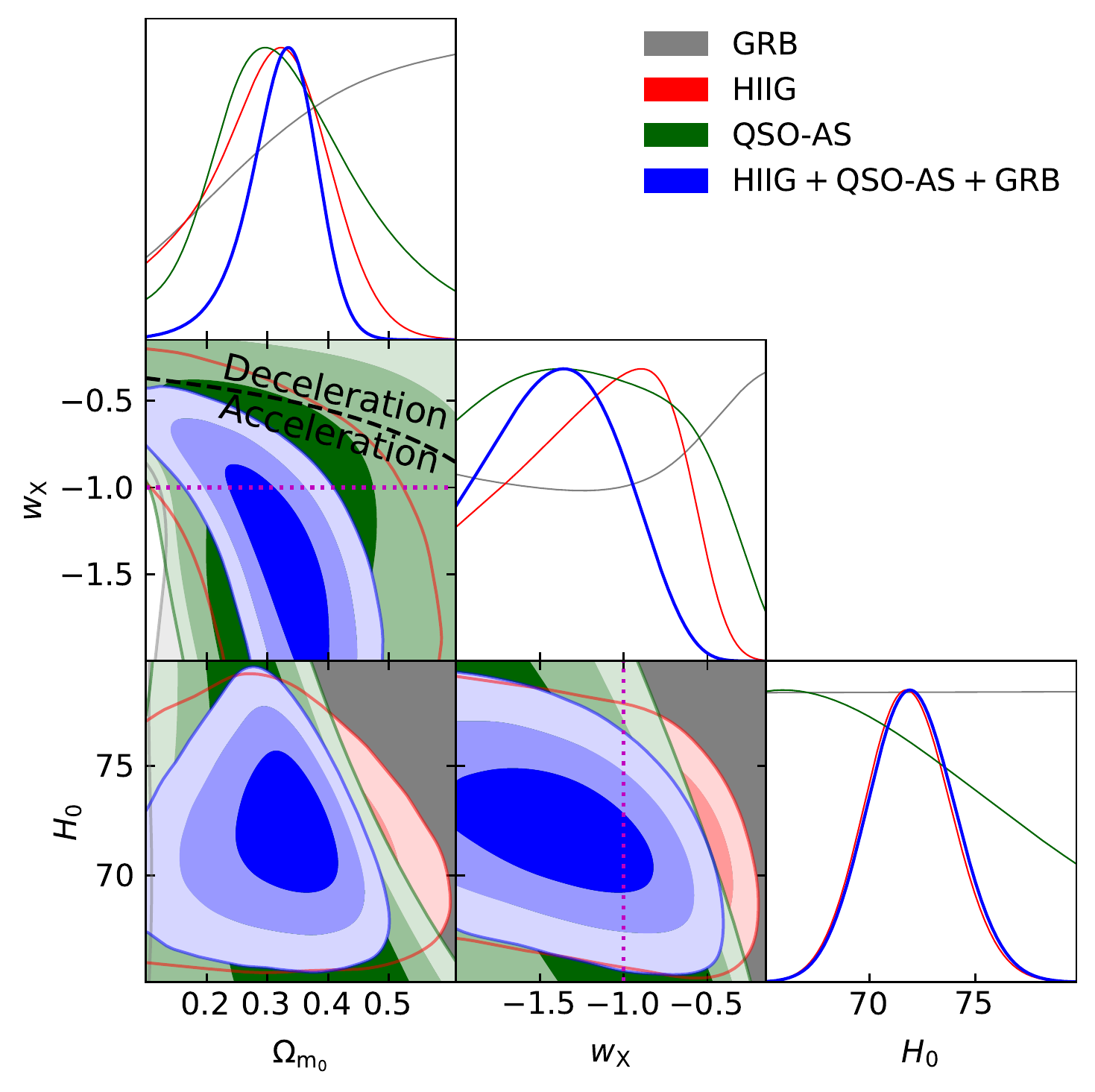}\\
\caption[1$\sigma$, 2$\sigma$, and 3$\sigma$ confidence contours for flat XCDM.]{1$\sigma$, 2$\sigma$, and 3$\sigma$ confidence contours for flat XCDM. The black dotted line is the zero-acceleration line, which divides the parameter space into regions associated with currently-accelerating (below left) and currently-decelerating (above right) cosmological expansion. The magenta lines denote $w_{\rm X}=-1$, i.e. the flat \lcdm\ model.}
\label{fig3}
\end{figure*}

\begin{figure*}
\centering
    \includegraphics[width=3.5in,height=3.5in]{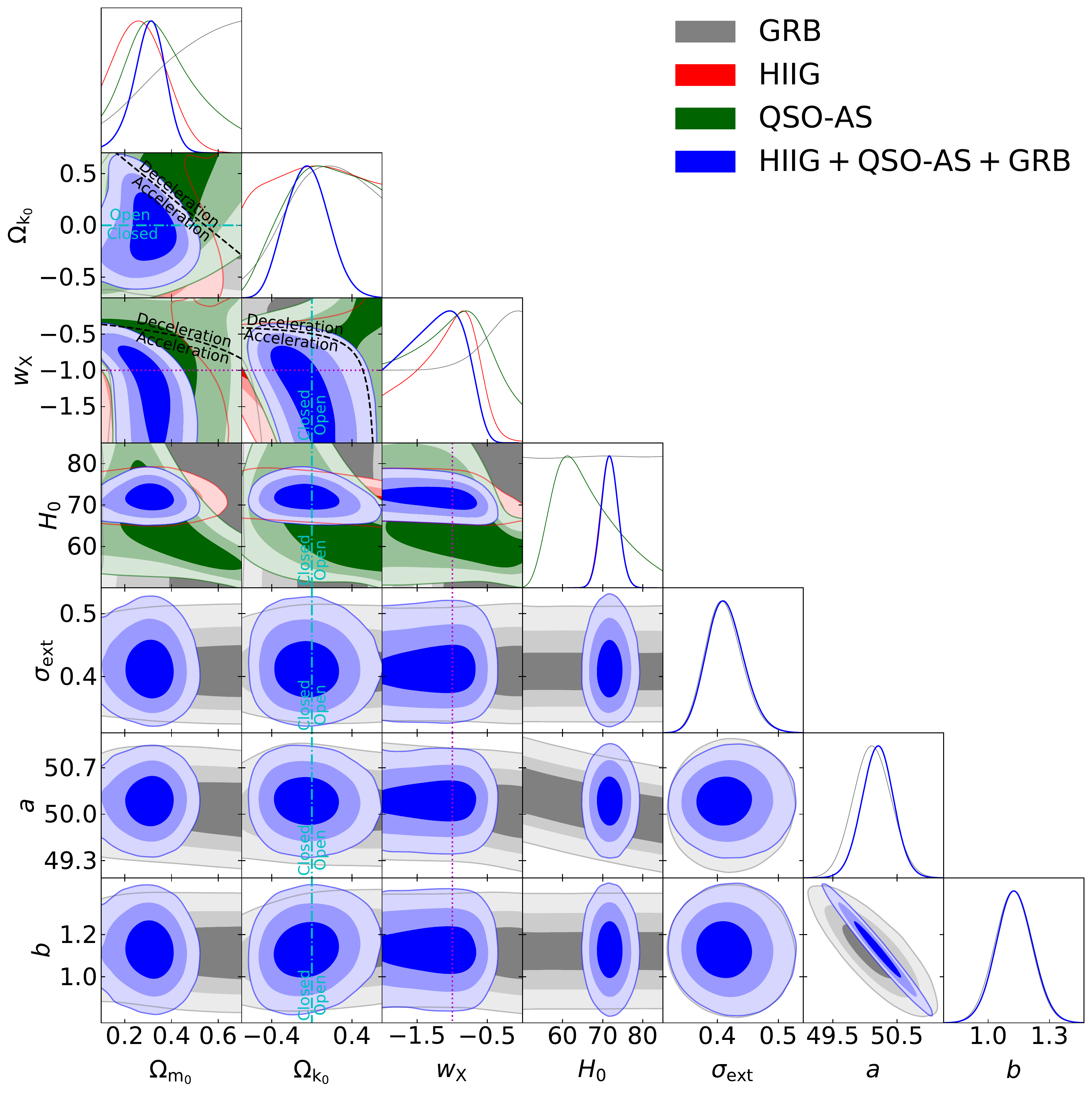}
    \includegraphics[width=3.5in,height=3.5in]{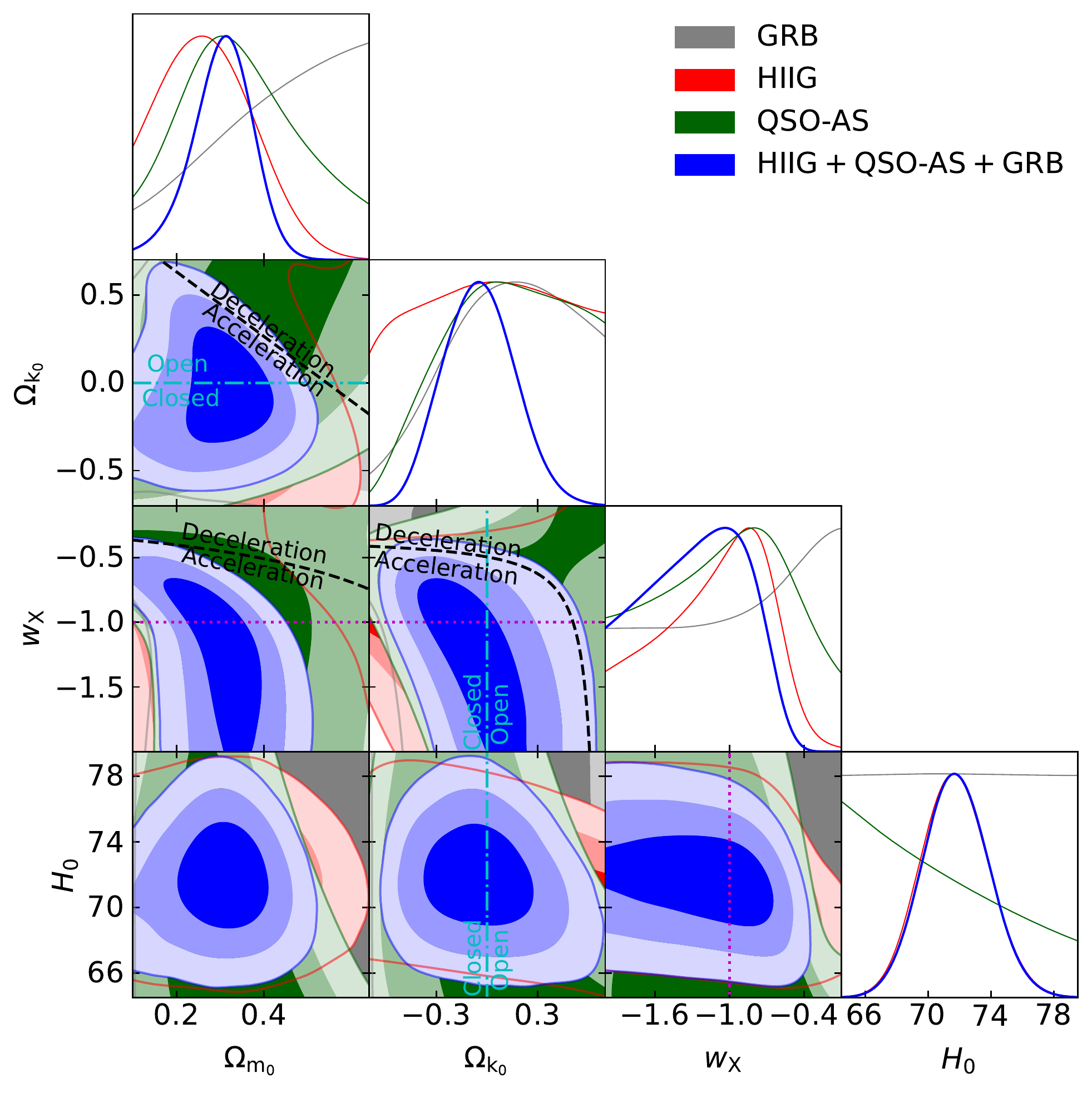}\\
\caption[1$\sigma$, 2$\sigma$, and 3$\sigma$ confidence contours for non-flat XCDM.]{Same as Fig. \ref{fig3} but for non-flat XCDM, where the zero acceleration lines in each of the three subpanels are computed for the third cosmological parameter set to the $H(z)$ + BAO data best-fitting values listed in Table \ref{tab:ch8_BFP}. Currently-accelerating cosmological expansion occurs below these lines. The cyan dash-dot lines represent the flat XCDM case, with closed spatial hypersurfaces either below or to the left. The magenta lines indicate $w_{\rm X} = -1$, i.e. the non-flat \lcdm\ model.}
\label{fig4}
\end{figure*}

We present the posterior one-dimensional (1D) probability distributions and two-dimensional (2D) confidence regions of the cosmological and Amati relation parameters for the six flat and non-flat models in Figs. \ref{fig1}--\ref{fig6}, in gray (GRB), red (HIIG), and green (QSO-AS). The unmarginalized best-fitting parameter values are listed in Table \ref{tab:ch8_BFP}, along with the corresponding $\chi^2$, $-2\ln\mathcal{L}_{\rm max}$, $AIC$, $BIC$, and degrees of freedom $\nu$ (where $\nu \equiv N - n$).\footnote{Note that the $\chi^2$ values listed in Tables \ref{tab:ch8_BFP} and \ref{tab:ch8_BFP2} are computed from the best-fitting parameter values and are not necessarily the minimum (especially when including GRB and QSO-Flux data).} The values of $\Delta\chi^2$, $\Delta AIC$, and $\Delta BIC$ reported in Table \ref{tab:ch8_BFP} are discussed in Section \ref{subsec:comparison}, where we define $\Delta \chi^2$, $\Delta AIC$, and $\Delta BIC$, respectively, as the differences between the values of the $\chi^2$, $AIC$, and $BIC$ associated with a given model and their corresponding minimum values among all models. The marginalized best-fitting parameter values and uncertainties ($\pm 1\sigma$ error bars or $2\sigma$ limits) are given in Table \ref{tab:ch8_1d_BFP}.\footnote{We use the \textsc{python} package \textsc{getdist} \citep{Lewis_2019} to plot these figures and compute the central values (posterior means) and uncertainties of the free parameters listed in Table \ref{tab:ch8_1d_BFP}.}
From Table \ref{tab:ch8_1d_BFP} we find that the QSO-AS constraints on $\Omega_{m0}$ are consistent with other results within a 1$\sigma$ range but with large error bars, ranging from a low of $0.329^{+0.086}_{-0.171}$ (flat \pcdm) to a high of $0.364^{+0.083}_{-0.150}$ (flat \lcdm). 

The QSO-AS constraints on $H_0$ are between $H_0=61.91^{+2.83}_{-4.92}$ km s$^{-1}$ Mpc$^{-1}$ (non-flat \pcdm) and $H_0=68.39^{+6.14}_{-8.98}$ km s$^{-1}$ Mpc$^{-1}$ (flat XCDM), with large error bars and relatively low values for non-flat XCDM and the \pcdm\ models. 

The non-flat models mildly favor open geometry, but are also consistent, given the large error bars, with spatially-flat hypersurfaces (except for non-flat \pcdm, where the open case is favored at $2.76\sigma$). For non-flat \lcdm, non-flat XCDM, and non-flat \pcdm, we find $\Omega_{k0}=0.017^{+0.184}_{-0.277}$, $\Omega_{k0}=0.115^{+0.466}_{-0.293}$, and $\Omega_{k0}=0.254^{+0.304}_{-0.092}$, respectively.\footnote{From Table \ref{tab:ch8_1d_BFP} we see that GRB data are also consistent with flat spatial geometry in the non-flat \lcdm\ and XCDM cases, but also favor, at $2.92\sigma$, open spatial geometry in the case of non-flat \pcdm.}

The fits to the QSO-AS data favor dark energy being a cosmological constant but do not strongly disfavor dark energy dynamics. For flat (non-flat) XCDM, $w_{\rm X}=-1.161^{+0.430}_{-0.679}$ ($w_{\rm X}=-1.030^{+0.593}_{-0.548}$), and for flat (non-flat) \pcdm, $2\sigma$ upper limits of $\alpha$ are $\alpha<2.841$ ($\alpha<4.752$). In the former case, both results are within 1$\sigma$ of $w_{\rm X}=-1$, and in the latter case, both 1D likelihoods peak at $\alpha=0$. 

Constraints on cosmological model parameters derived solely from HIIG data are discussed in Sec. 5.1 of Chapter \ref{Chapter7}, while those derived from GRB data are described in Sec. 5.1 of \cite{Khadka_2020a} (though there are slight differences coming from the different treatments of $H_0$ and the different ranges of flat priors used there and here); both are listed in Table \ref{tab:ch8_1d_BFP} here. In contrast to the HIIG and QSO-AS data sets, the GRB data alone cannot constrain $H_0$ because there is a degeneracy between the intercept parameter ($a$) of the Amati relation and $H_0$; for consistency with the analyses of the HIIG and QSO-AS data, we treat $H_0$ as a free parameter in the GRB data analysis here.

Cosmological constraints obtained using the HIIG, QSO-AS, and GRB data sets are mutually consistent, and are also consistent with those obtained from most other cosmological probes. This is partially a consequence of the larger HIIG, QSO-AS, and GRB data error bars, which lead to relatively weaker constraints on cosmological parameters when each of these data sets is used alone (see Table \ref{tab:ch8_1d_BFP}). However, because the HIIG, QSO-AS, and GRB constraints are mutually consistent, we may jointly analyze these data. Their combined cosmological constraints will therefore be more restrictive than when they are analyzed individually.

We note, from Figs. \ref{fig1}--\ref{fig6}, that a significant part of the likelihood of each of these three data sets lies in the parameter space part with currently-accelerating cosmological expansion.

\subsection{QSO-AS, GRB, and GRB (QGH) joint constraints}
\label{subsec:QGH}

Because the QSO-AS, HIIG, and GRB contours are mutually consistent for all six of the models we study, we jointly analyze these data to obtain QGH constraints.

\begin{figure*}
\centering
    \includegraphics[width=3.5in,height=3.5in]{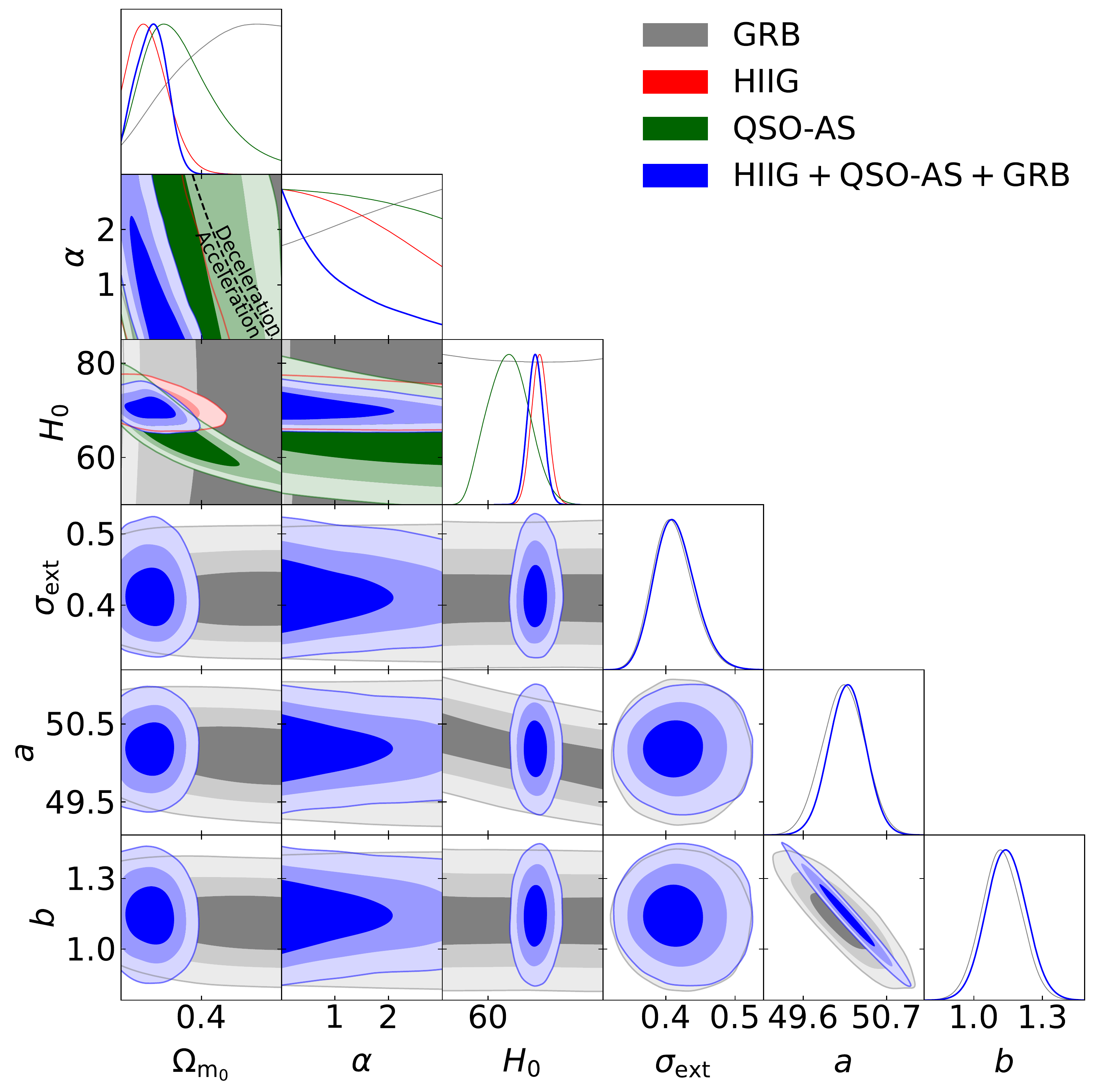}
    \includegraphics[width=3.5in,height=3.5in]{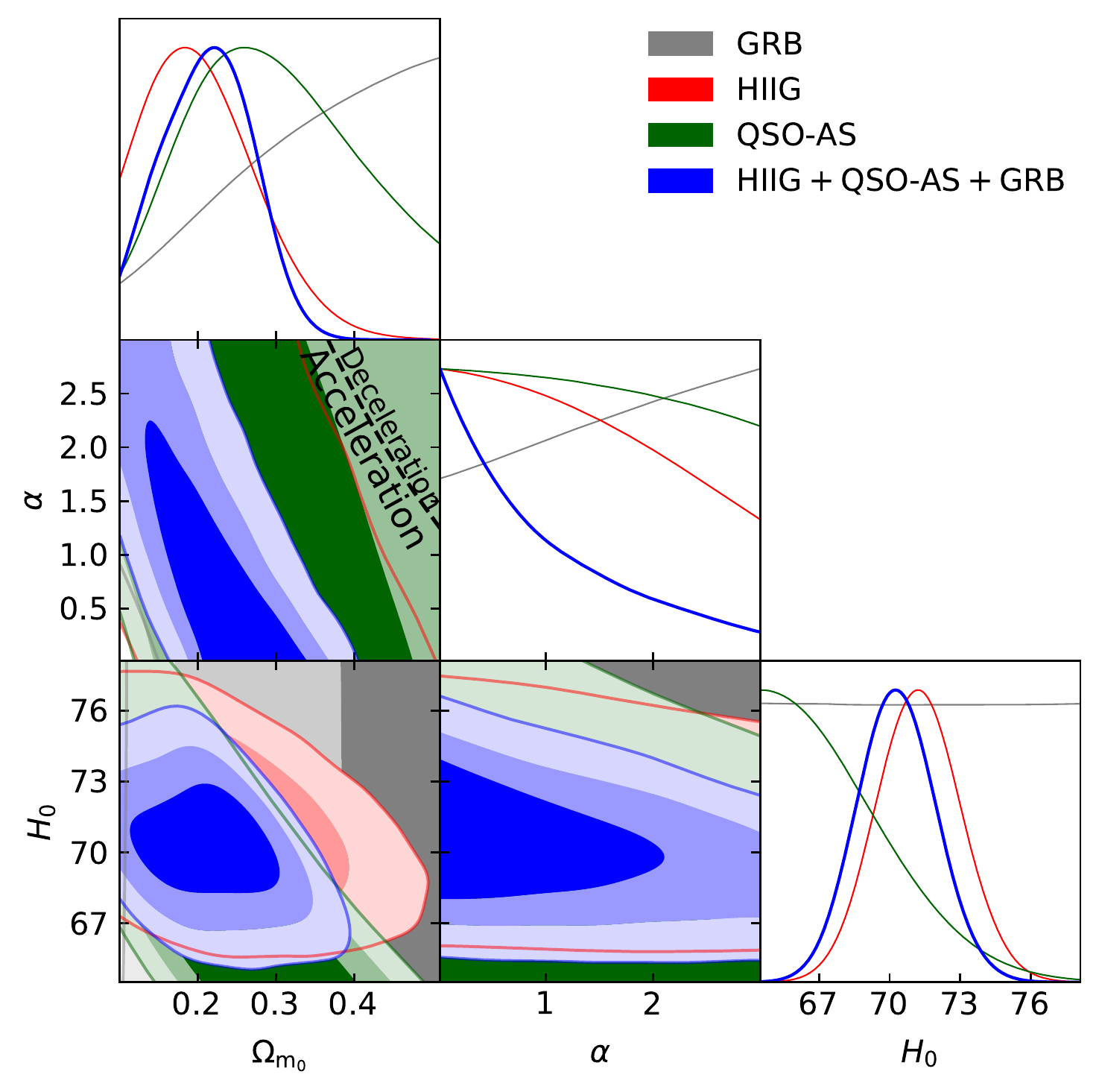}\\
\caption[1$\sigma$, 2$\sigma$, and 3$\sigma$ confidence contours for flat \pcdm.]{1$\sigma$, 2$\sigma$, and 3$\sigma$ confidence contours for flat \pcdm. The black dotted zero-acceleration line splits the parameter space into regions of currently-accelerating (below left) and currently-decelerating (above right) cosmological expansion. The $\alpha = 0$ axis is the flat \lcdm\ model.}
\label{fig5}
\end{figure*}

\begin{figure*}
\centering
    \includegraphics[width=3.5in,height=3.5in]{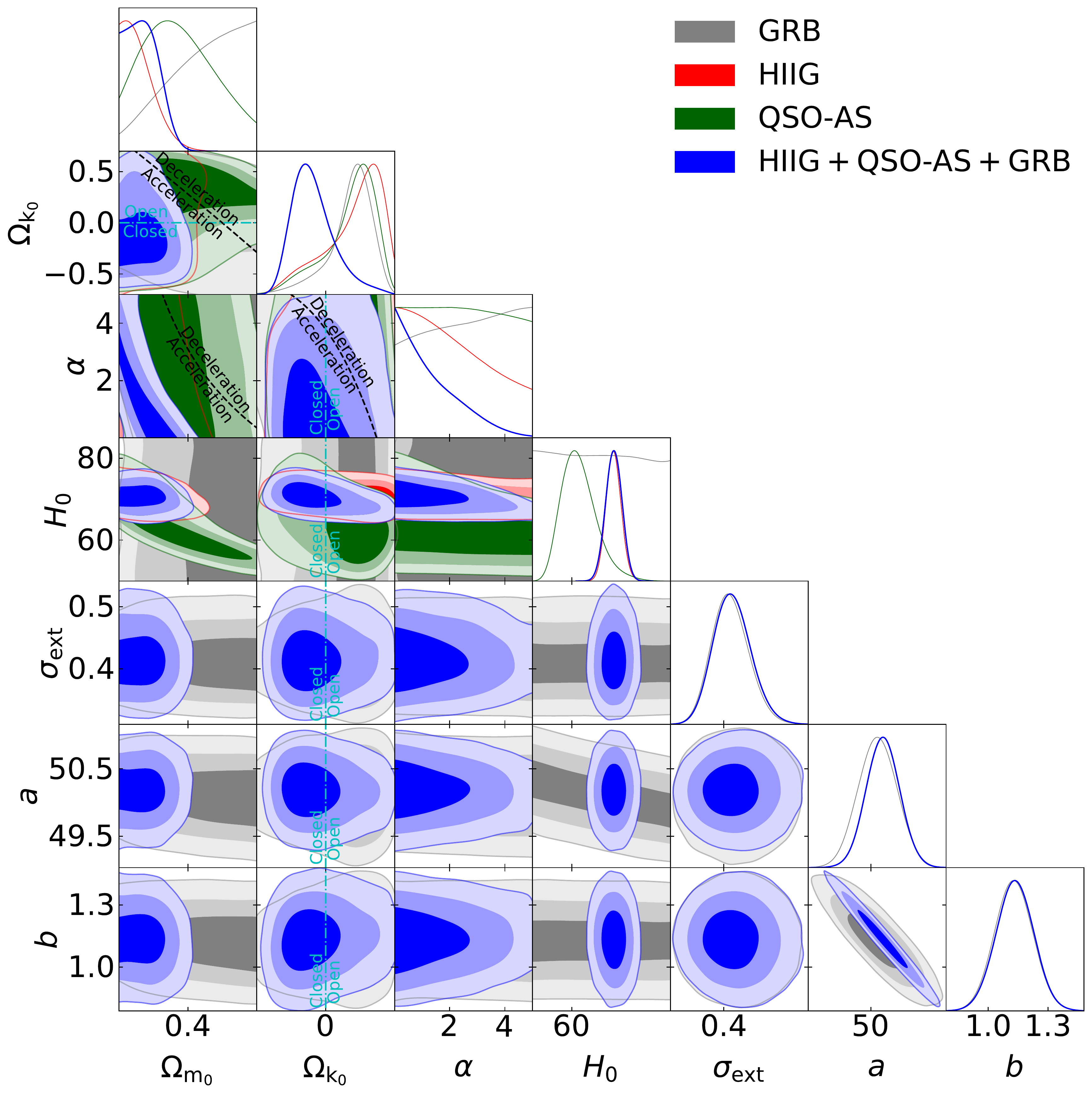}
    \includegraphics[width=3.5in,height=3.5in]{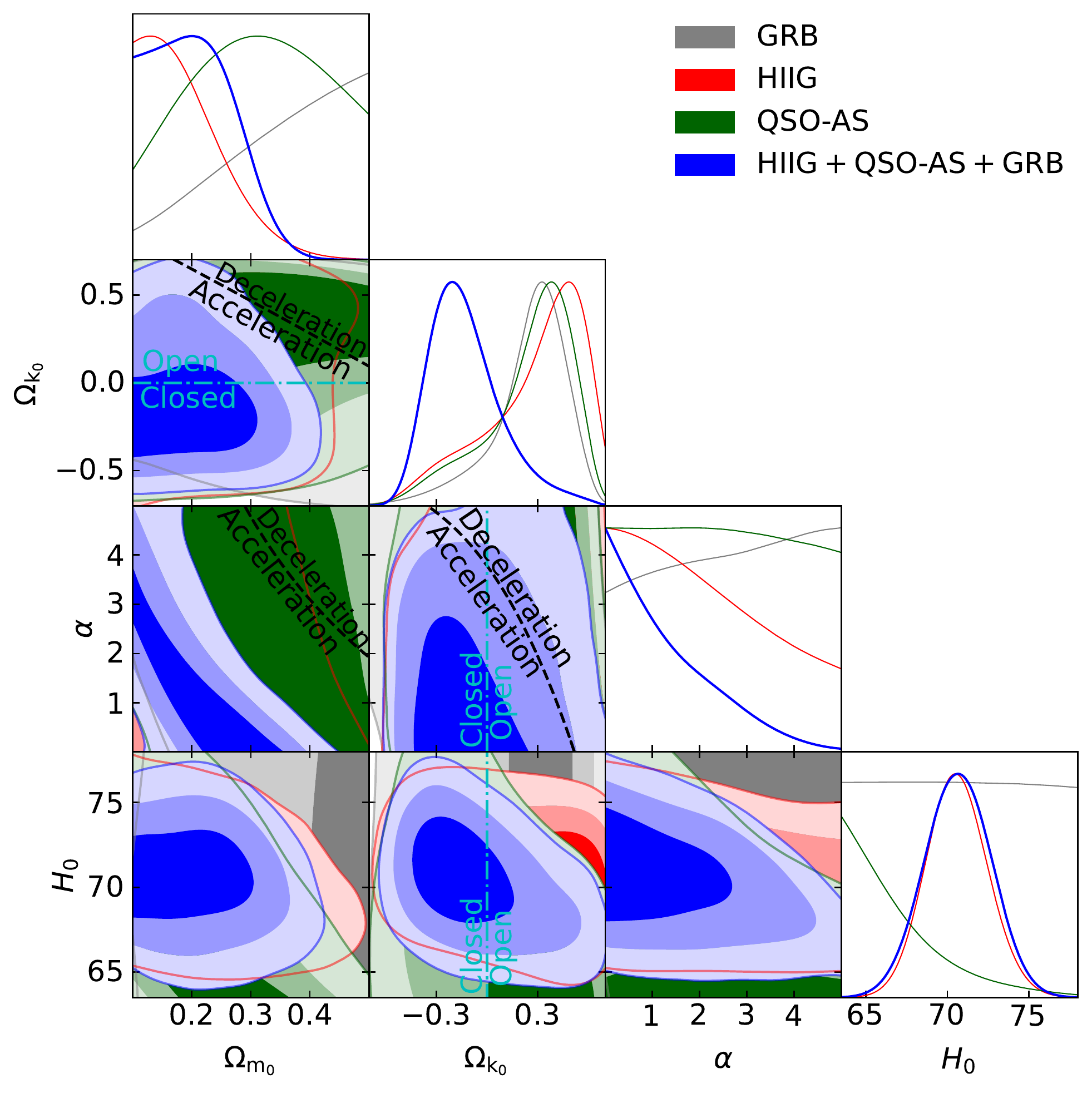}\\
\caption[1$\sigma$, 2$\sigma$, and 3$\sigma$ confidence contours for flat \pcdm.]{Same as Fig. \ref{fig5} but for non-flat \pcdm, where the zero-acceleration lines in each of the sub-panels are computed for the third cosmological parameter set to the $H(z)$ + BAO data best-fitting values listed in Table \ref{tab:ch8_BFP}. Currently-accelerating cosmological expansion occurs below these lines. The cyan dash-dot lines represent the flat \pcdm\ case, with closed spatial geometry either below or to the left. The $\alpha = 0$ axis is the non-flat \lcdm\ model.}
\label{fig6}
\end{figure*}

\begin{figure*}
\centering
    \includegraphics[width=3.5in,height=3.5in]{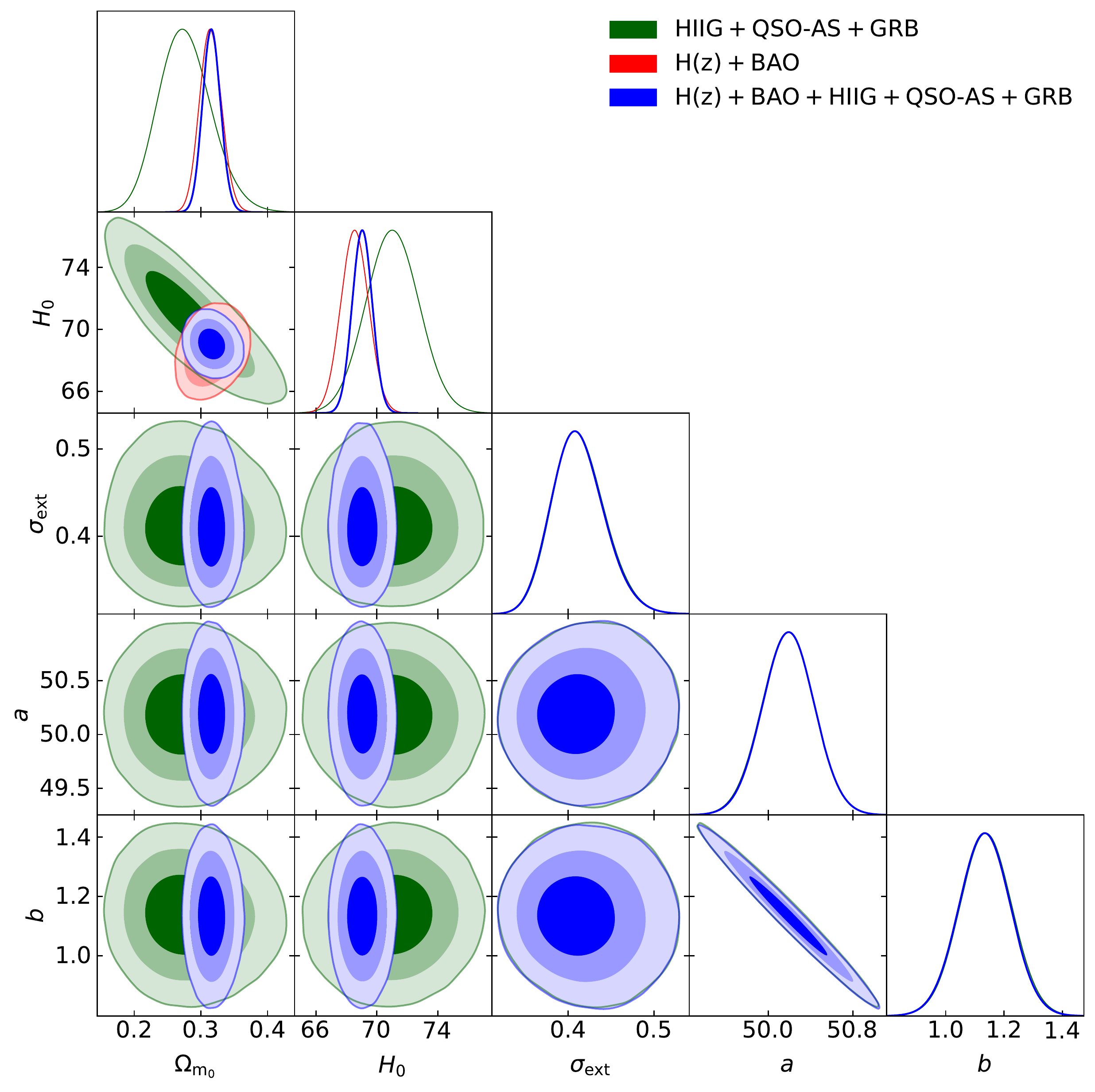}
    \includegraphics[width=3.5in,height=3.5in]{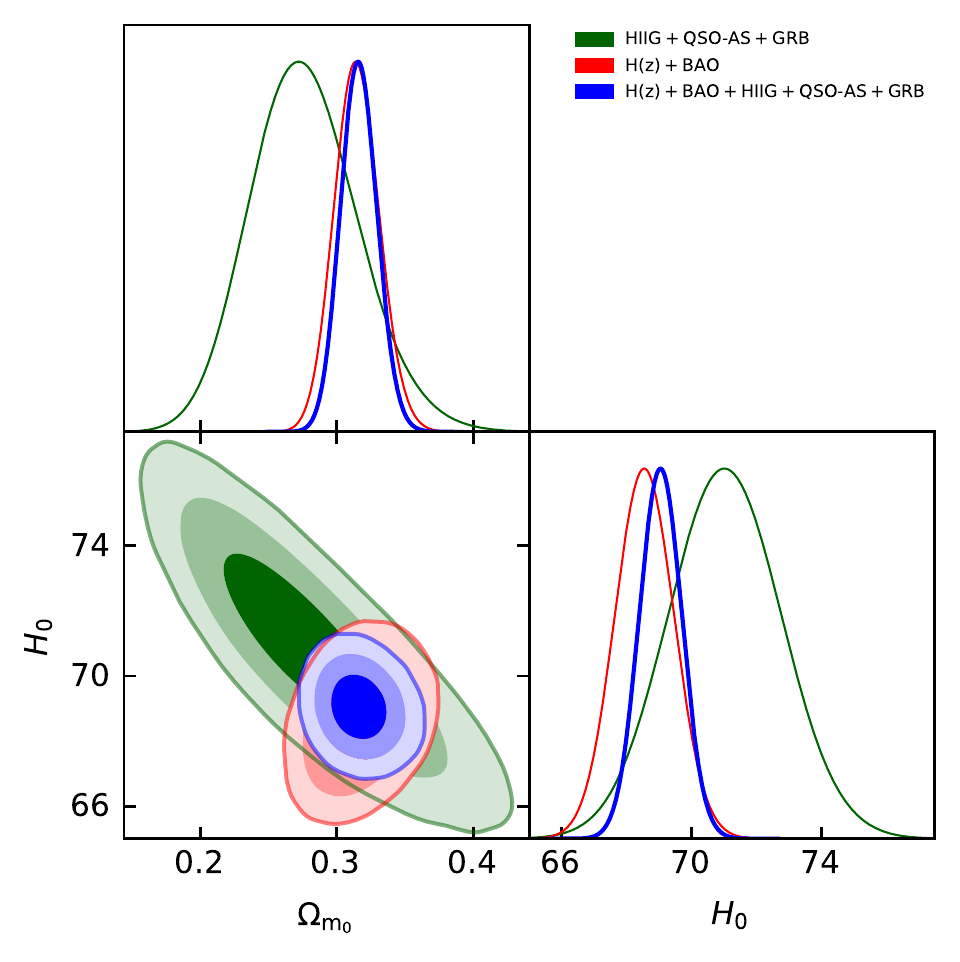}\\
\caption[1$\sigma$, 2$\sigma$, and 3$\sigma$ confidence contours for flat \lcdm\ for different combinations of data.]{Same as Fig. \ref{fig1} (flat \lcdm) but for different combinations of data.}
\label{fig7}
\end{figure*}

\begin{figure*}
\centering
    \includegraphics[width=3.5in,height=3.5in]{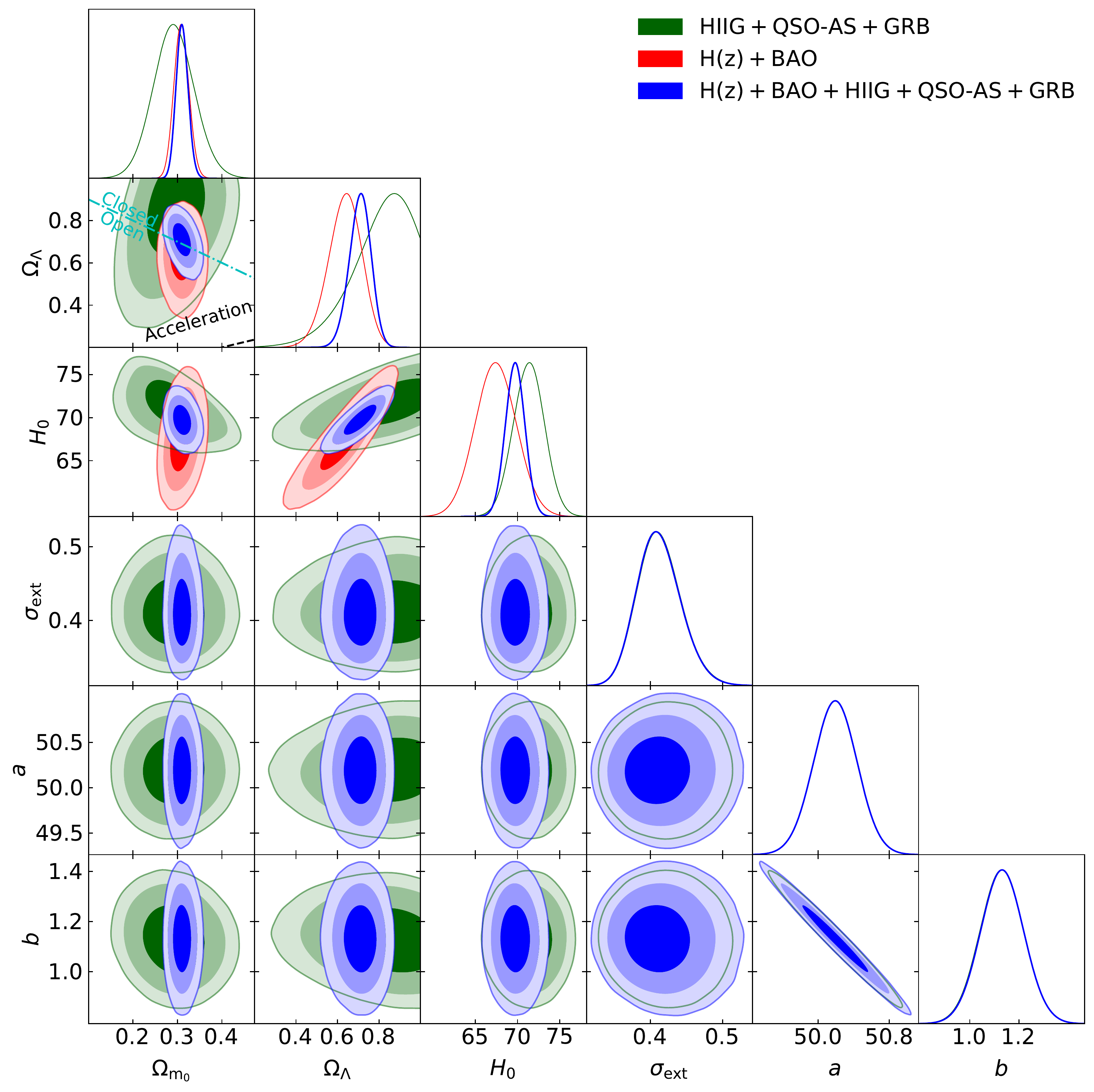}
    \includegraphics[width=3.5in,height=3.5in]{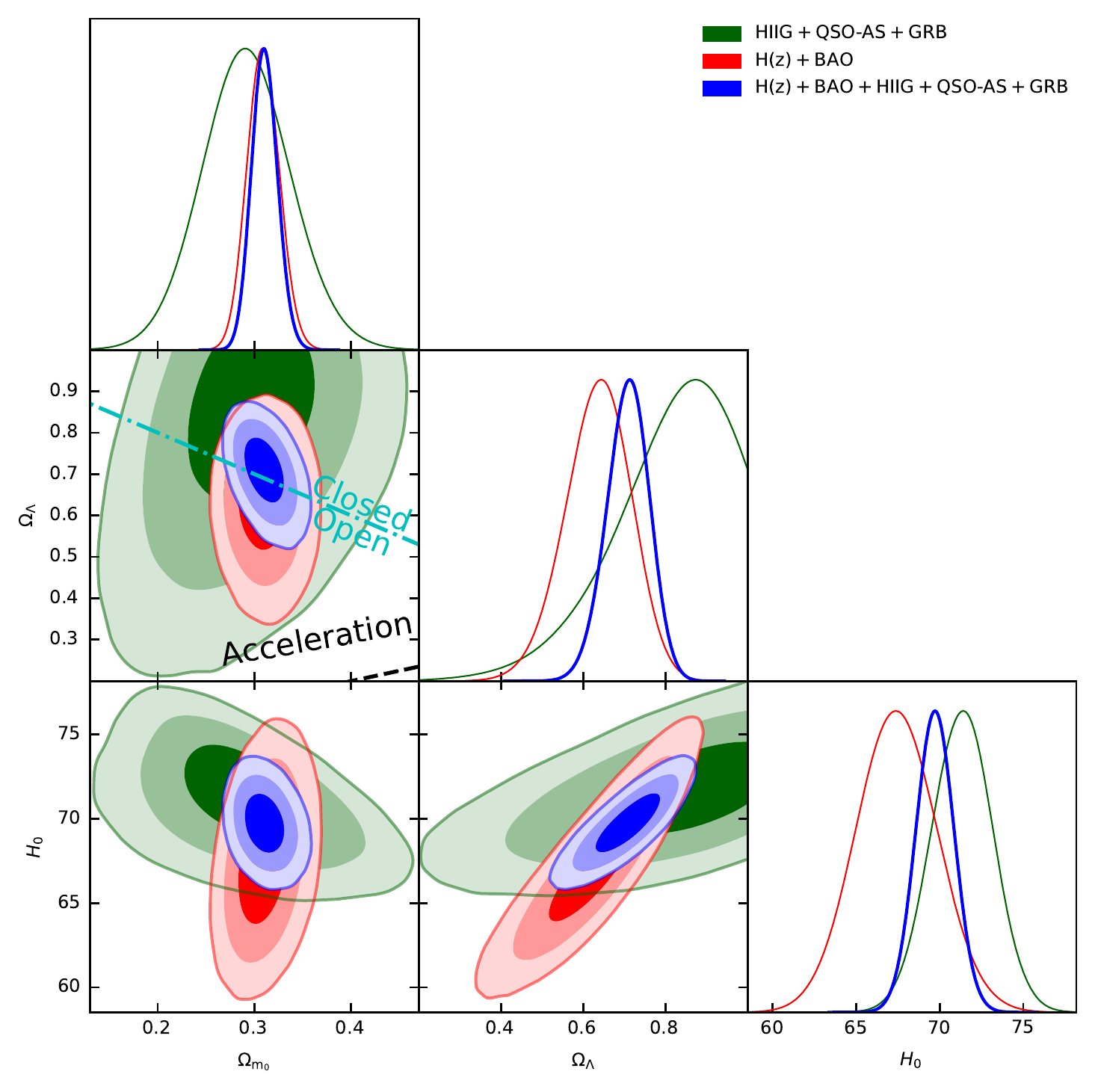}\\
\caption[1$\sigma$, 2$\sigma$, and 3$\sigma$ confidence contours for non-flat \lcdm\ for different combinations of data.]{Same as Fig. \ref{fig2} (non-flat \lcdm) but for different combinations of data.}
\label{fig8}
\end{figure*}

\begin{figure*}
\centering
    \includegraphics[width=3.5in,height=3.5in]{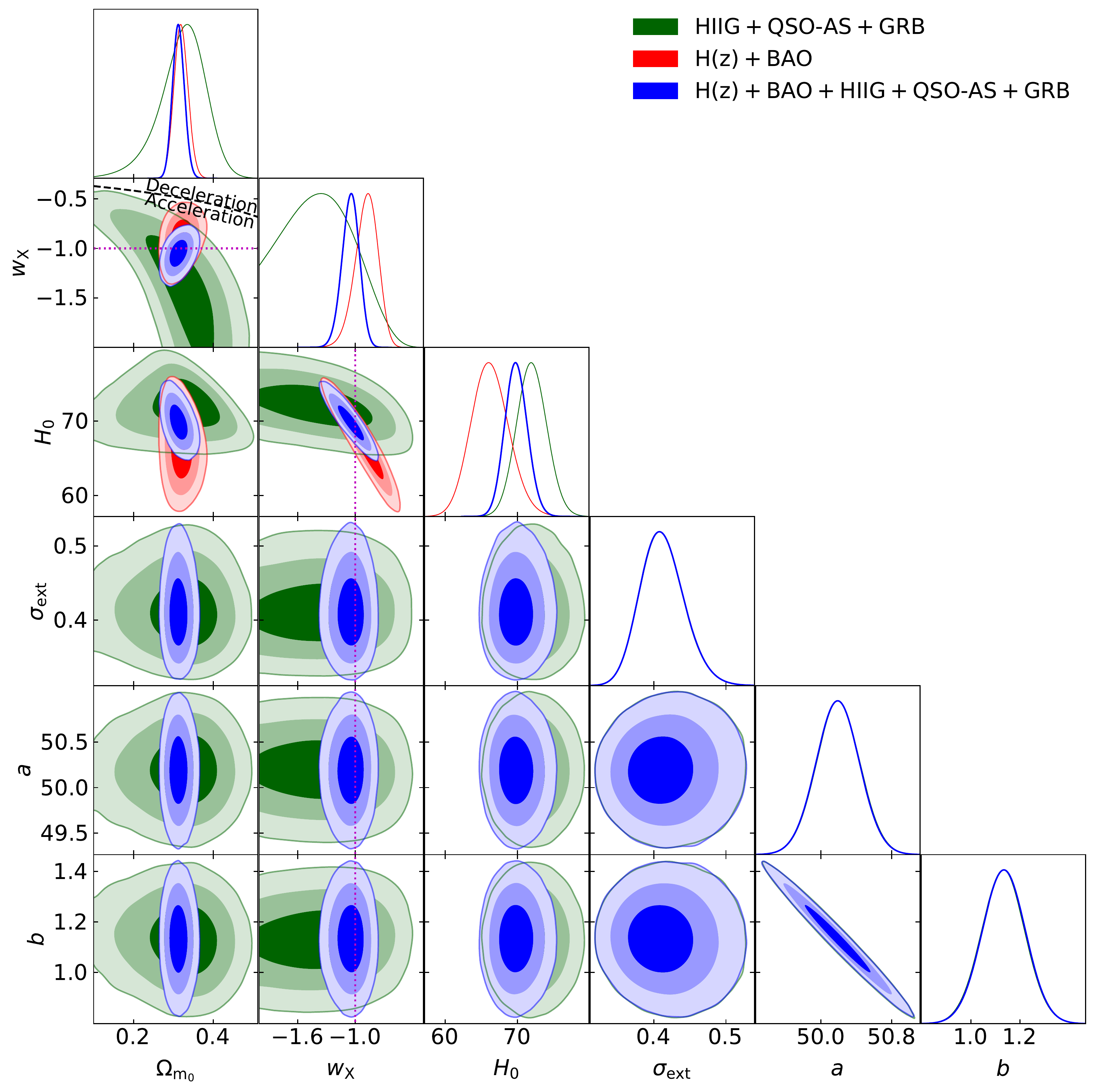}
    \includegraphics[width=3.5in,height=3.5in]{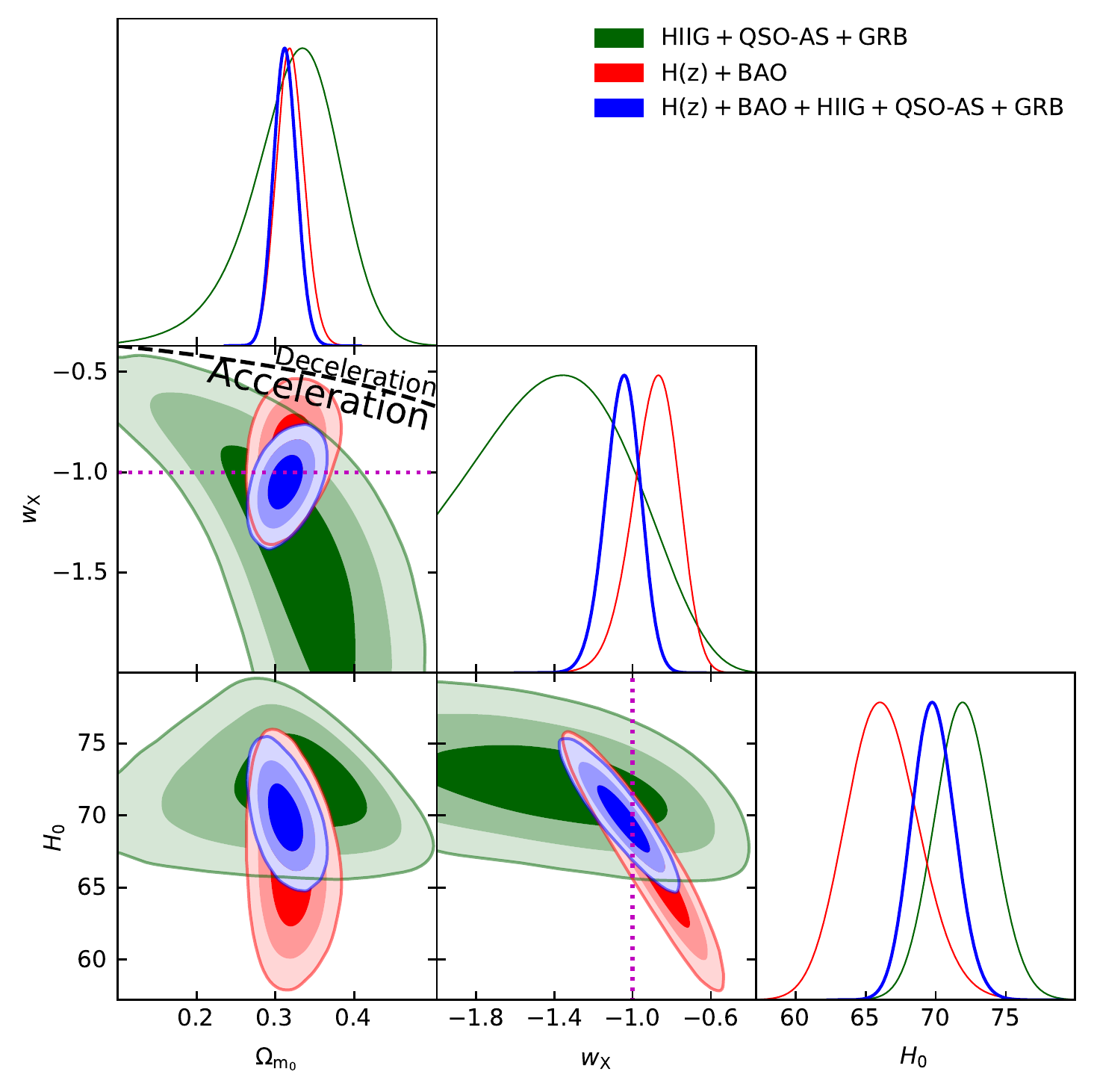}\\
\caption[1$\sigma$, 2$\sigma$, and 3$\sigma$ confidence contours for flat XCDM for different combinations of data.]{Same as Fig. \ref{fig3} (flat XCDM) but for different combinations of data.}
\label{fig9}
\end{figure*}

\begin{figure*}
\centering
    \includegraphics[width=3.5in,height=3.5in]{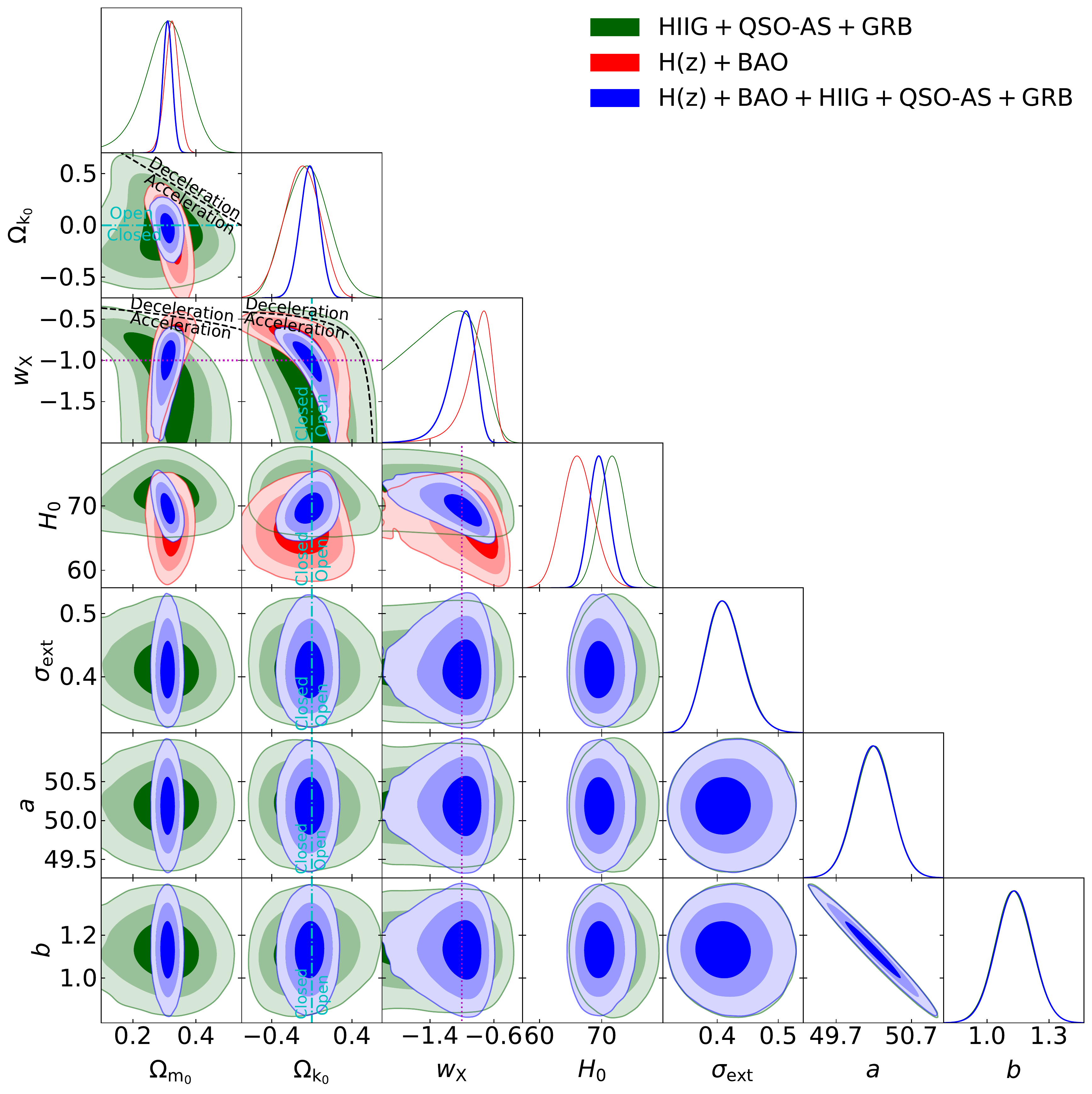}
    \includegraphics[width=3.5in,height=3.5in]{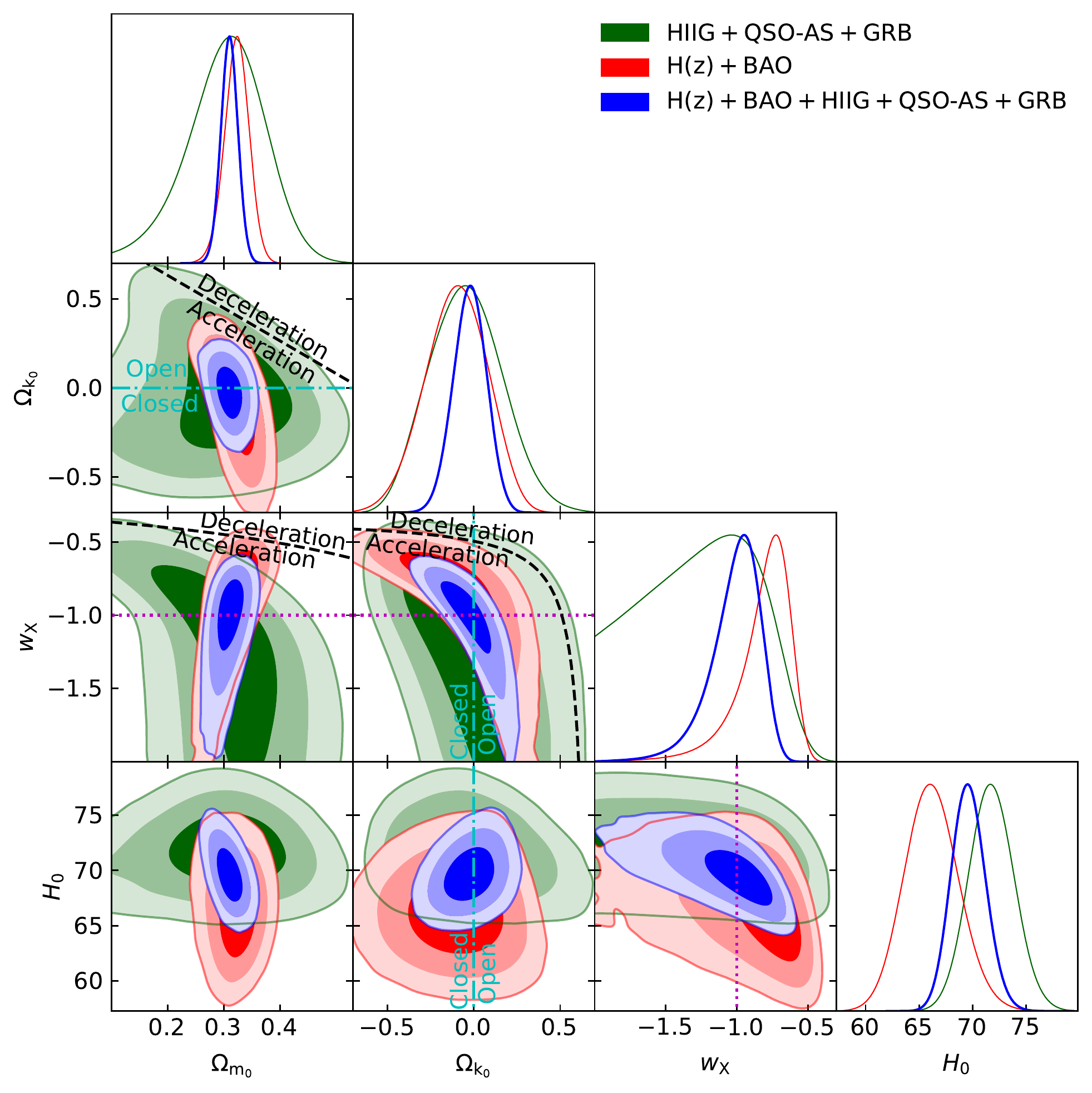}\\
\caption[1$\sigma$, 2$\sigma$, and 3$\sigma$ confidence contours for non-flat XCDM for different combinations of data.]{Same as Fig. \ref{fig4} (non-flat XCDM) but for different combinations of data.}
\label{fig10}
\end{figure*}

\begin{figure*}
\centering
    \includegraphics[width=3.5in,height=3.5in]{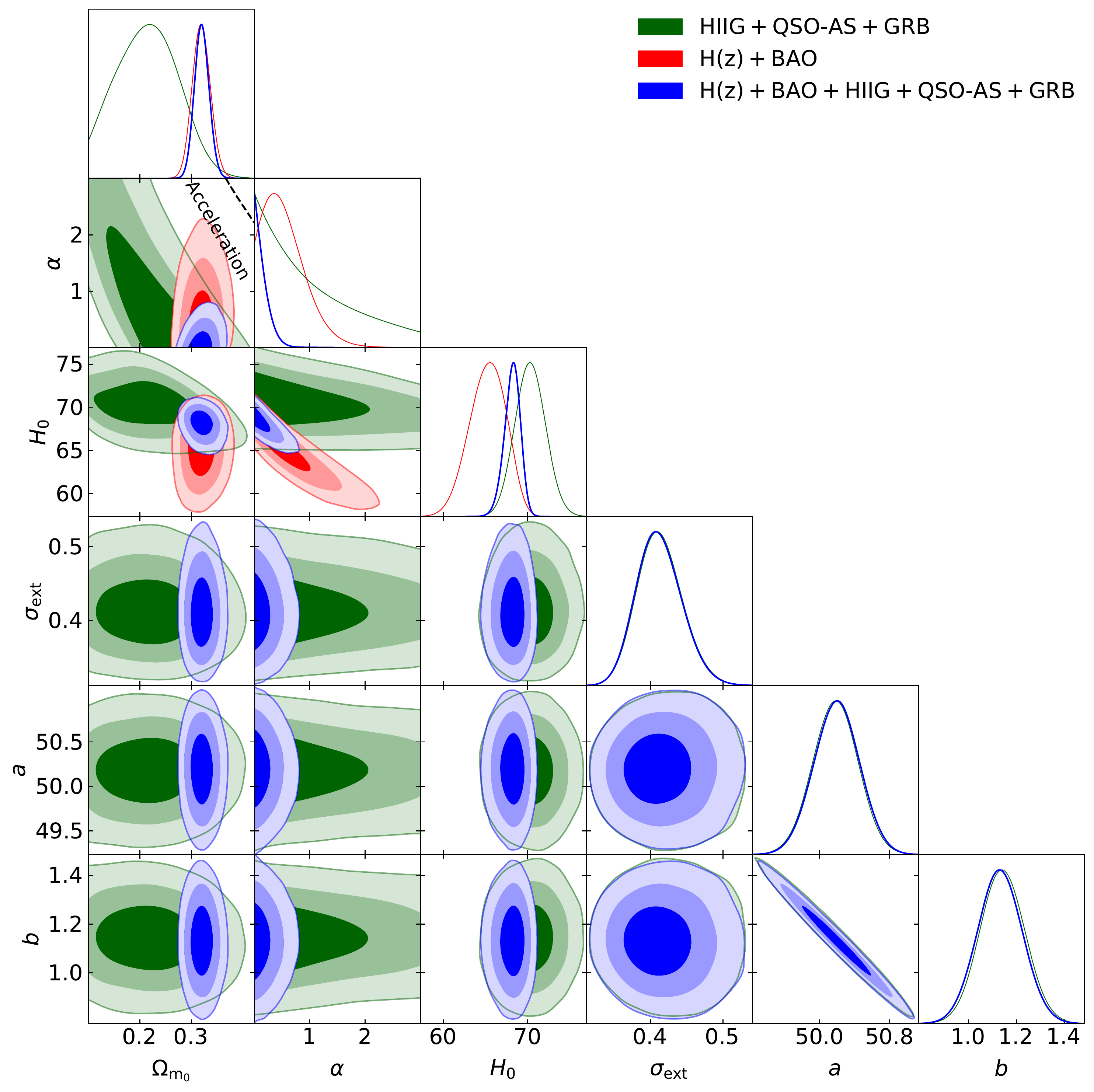}
    \includegraphics[width=3.5in,height=3.5in]{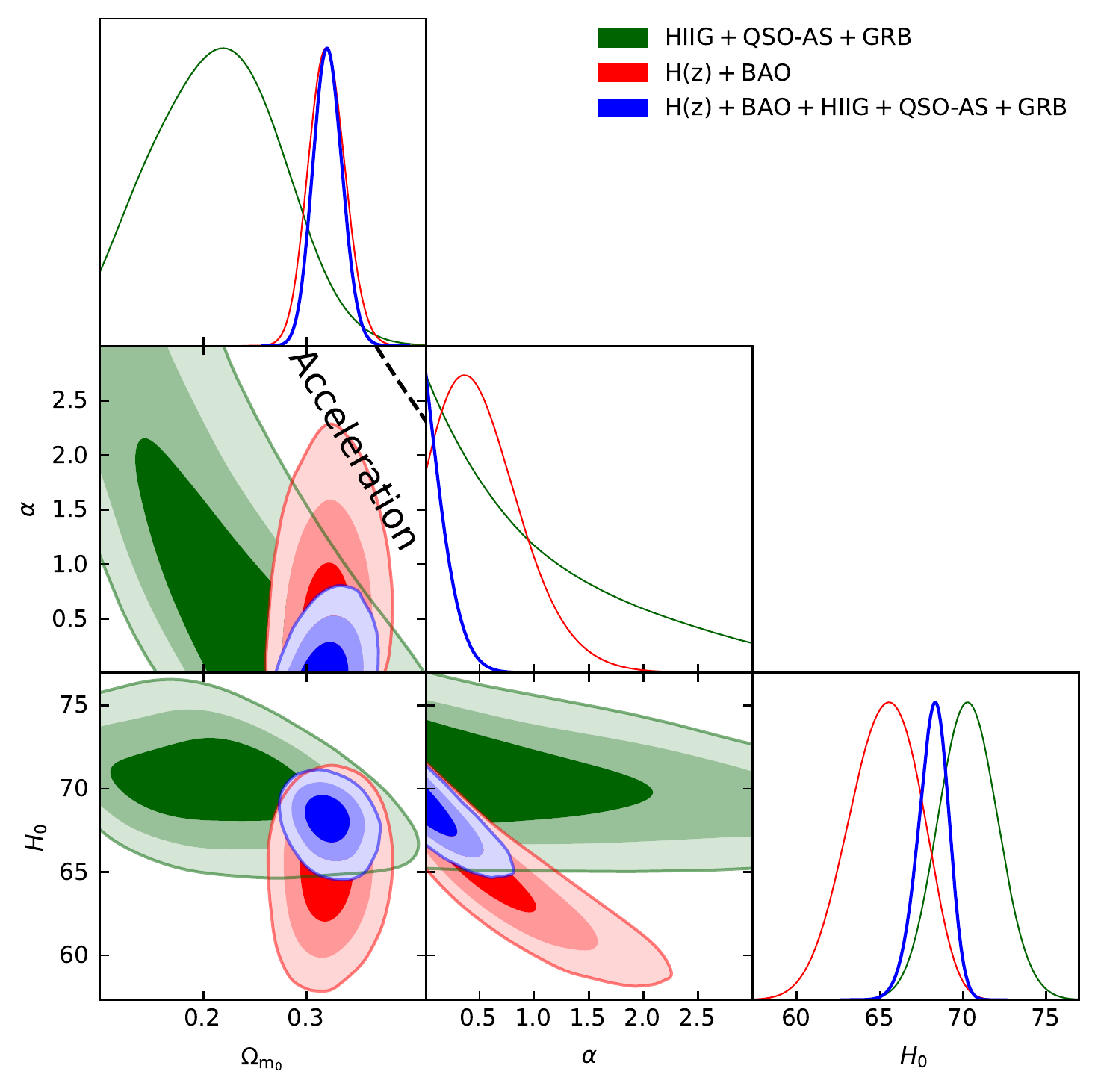}\\
\caption[1$\sigma$, 2$\sigma$, and 3$\sigma$ confidence contours for flat \pcdm\ for different combinations of data.]{Same as Fig. \ref{fig5} (flat \pcdm) but for different combinations of data.}
\label{fig11}
\end{figure*}

\begin{figure*}
\centering
    \includegraphics[width=3.5in,height=3.5in]{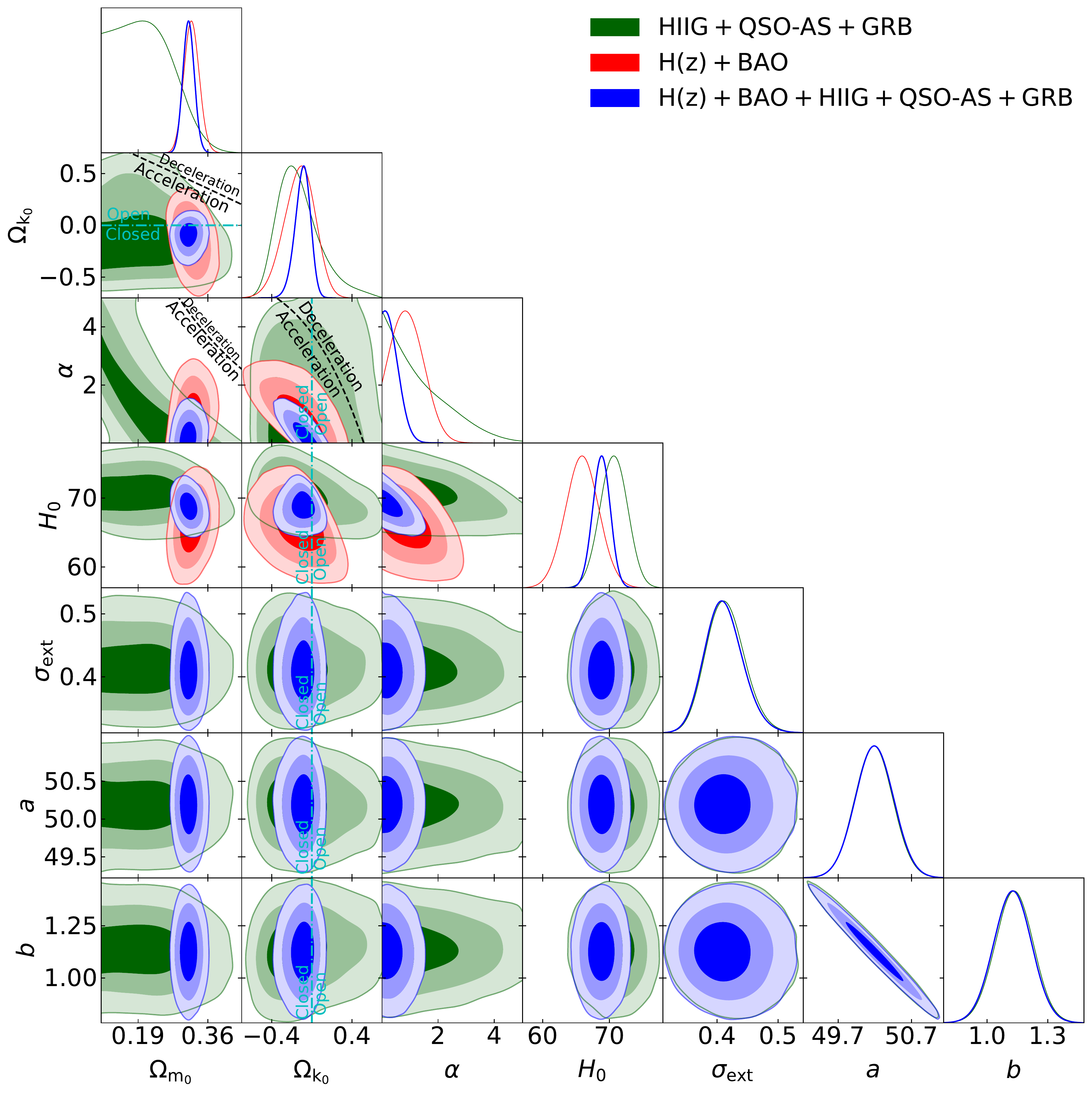}
    \includegraphics[width=3.5in,height=3.5in]{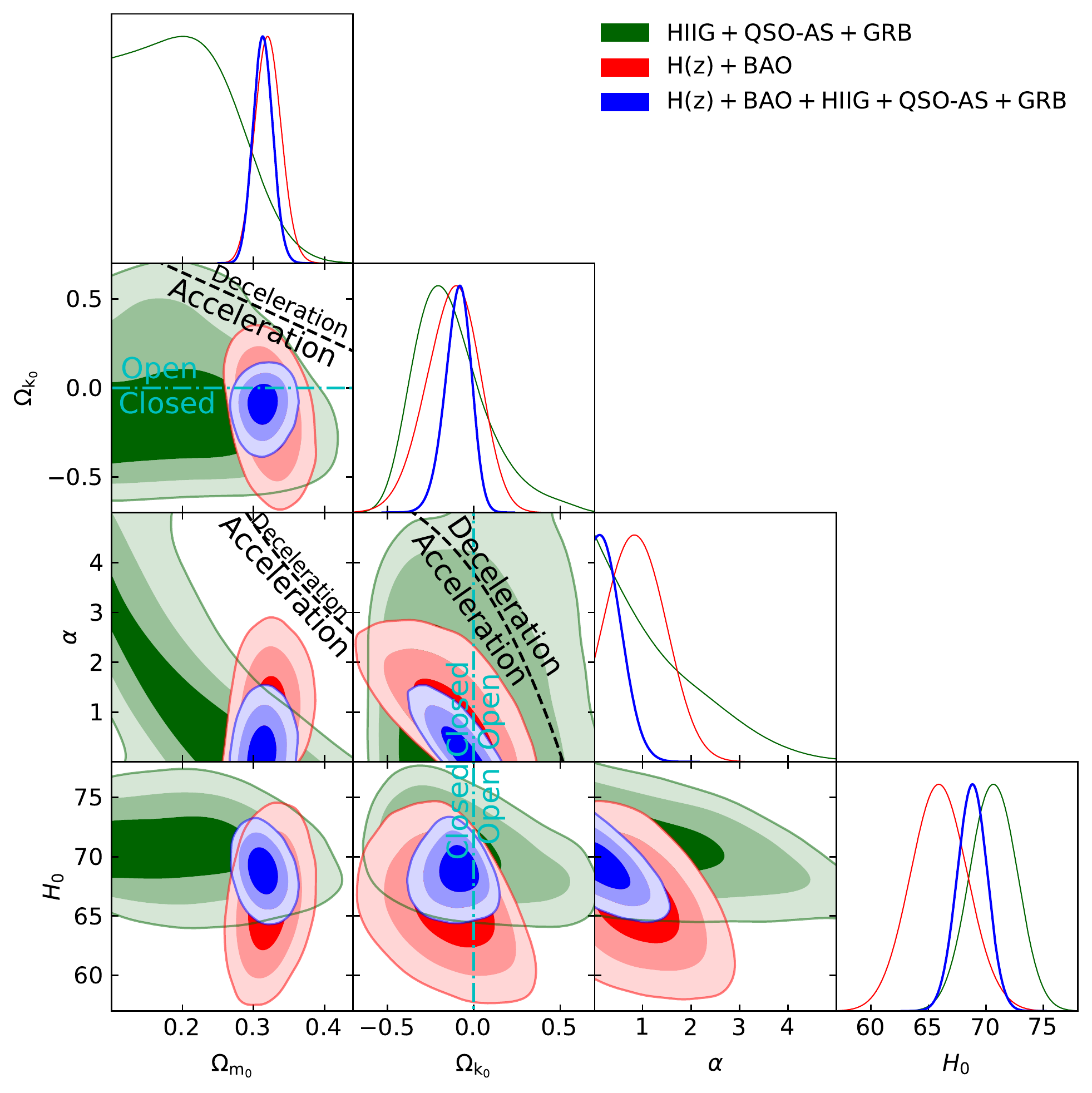}\\
\caption[1$\sigma$, 2$\sigma$, and 3$\sigma$ confidence contours for non-flat \pcdm\ for different combinations of data.]{Same as Fig. \ref{fig6} (non-flat \pcdm) but for different combinations of data.}
\label{fig12}
\end{figure*}

The 1D probability distributions and 2D confidence regions of the cosmological and Amati relation parameters from the QGH data are in Figs. \ref{fig1}--\ref{fig6}, in blue, Figs. \ref{fig7}--\ref{fig12}, in green, and panels (a) of Figs. \ref{ch8_fig01}--\ref{ch8_fig04}, in red. The best-fitting results and uncertainties are in Tables \ref{tab:ch8_BFP} and \ref{tab:ch8_1d_BFP}.

We find that the QGH data combination favors currently-accelerating cosmological expansion.

The fit to the QGH data produces best-fitting values of $\Omega_{m0}$ that lie between $0.205^{+0.044}_{-0.094}$ (non-flat \pcdm) at the low end, and $0.322^{+0.062}_{-0.044}$ (flat XCDM) at the high end. This range is smaller than the ranges within which $\Omega_{m0}$ falls when it is determined from the HIIG, QSO-AS, and GRB data individually, but the low and high ends of the range are still somewhat mutually inconsistent, being 2.66$\sigma$ away from each other. This is a consequence of the low $\Omega_{m0}$ value for non-flat \pcdm; the $\Omega_{m0}$ values for \lcdm\ and XCDM are quite consistent with the recent estimate of \cite{planck2018}. In contrast, the best-fitting values of $H_0$ that we measure from the QGH data are mutually very consistent (within $0.65\sigma$), with $H_0=70.30\pm1.68$ km s$^{-1}$ Mpc$^{-1}$ (flat \pcdm) at the low end of the range and $H_0=72.00^{+1.99}_{-1.98}$ km s$^{-1}$ Mpc$^{-1}$ (flat XCDM) at the high end of the range. These measurements are $0.83\sigma$ (flat XCDM) and $1.70\sigma$ (flat \pcdm) lower than the local Hubble constant measurement of $H_0 = 74.03 \pm 1.42$ km s$^{-1}$ Mpc$^{-1}$ \citep{riess_etal_2019}, and $0.70\sigma$ (flat \pcdm) and $1.16\sigma$ (flat XCDM) higher than the median statistics estimate of $H_0=68 \pm 2.8$ km s$^{-1}$ Mpc$^{-1}$ \citep{chenratmed}.\footnote{Other local expansion rate determinations have slightly lower central values with slightly larger error bars \citep{rigault_etal_2015,86,Dhawan,FernandezArenas,freedman_etal_2019,freedman_etal_2020,rameez_sarkar_2019,Breuvaletal_2020, Efstathiou_2020, Khetan_et_al_2020}. Our $H_0$ measurements are consistent with earlier median statistics estimates \citep{gott_etal_2001,76} and with other recent $H_0$ determinations \citep{chen_etal_2017,DES_2018,Gomez-ValentAmendola2018, planck2018, zhang_2018,dominguez_etal_2019,martinelli_tutusaus_2019,Cuceu_2019,zeng_yan_2019,schoneberg_etal_2019,lin_ishak_2019, Blum_et_al_2020, Lyu_et_al_2020, Philcox_et_al_2020, Zhang_Huang_2020, Birrer_et_al_2020, Denzel_et_al_2020}.}

In contrast to the HIIG, QSO-AS, and GRB only cases, when fitted to the QGH data combination the non-flat models mildly favor closed spatial hypersurfaces. For non-flat \lcdm, non-flat XCDM, and non-flat \pcdm, we find $\Omega_{k0}=-0.093^{+0.092}_{-0.190}$, $\Omega_{k0}=-0.044^{+0.193}_{-0.217}$, and $\Omega_{k0}=-0.124^{+0.127}_{-0.253}$, respectively, with the non-flat \lcdm\ model favoring closed spatial hypersurfaces at 1.01$\sigma$.

The fit to the QGH data combination produces stronger evidence for dark energy dynamics in the flat and non-flat XCDM parametrizations but weaker evidence in the flat and non-flat \pcdm\ models (in comparison to the HIIG and QSO-AS only cases) with tighter error bars on the measured values of $w_{\rm X}$ and $\alpha$. For flat (non-flat) XCDM, $w_{\rm X}=-1.379^{+0.361}_{-0.375}$ ($w_{\rm X}=-1.273^{+0.501}_{-0.321}$), with $w_{\rm X}=-1$ being within the 1$\sigma$ range for non-flat XCDM and being 1.05$\sigma$ larger for flat XCDM. For flat (non-flat) \pcdm, $\alpha<2.584$ ($\alpha<3.414$), where both likelihoods peak at $\alpha=0$.

The constraints on the Amati relation parameters from the QGH data are also model-independent, but with slightly larger central values and smaller error bars for the parameter $a$. A reasonable summary is $\sigma_{\rm ext}=0.413^{+0.026}_{-0.032}$, $a=50.19\pm0.24$, and $b=1.133\pm0.086$.

The QGH cosmological constraints are largely consistent with those from other data, like the constraints from the $H(z)$ + BAO data used in Chapter \ref{Chapter7} and \cite{Khadka_Ratra_2020}, that are shown in red in Figs. \ref{fig7}--\ref{fig12}. We note, however, that there is some mild tension between \pcdm\ $\Omega_{m0}$ values, and between XCDM and \pcdm\ $H_0$ values determined from $H(z)$ + BAO and QGH data, with the $2.46\sigma$ difference between $\Omega_{m0}$ values estimated from the two different data combinations in the non-flat \pcdm\ model being the only somewhat troubling difference (see Table \ref{tab:ch8_1d_BFP}).

\begin{table*}
\centering
\resizebox{\columnwidth}{!}{%
\begin{threeparttable} 
\caption{Unmarginalized best-fitting parameter values for all models from various combinations of data.}\label{tab:ch8_BFP}
\setlength{\tabcolsep}{0.6mm}{
\begin{tabular}{lcccccccccccccccccc}
\toprule
 Model & Data set & $\Omega_{\mathrm{m_0}}$ & $\Omega_{\Lambda}$ & $\Omega_{\mathrm{k_0}}$ & $w_{\mathrm{X}}$ & $\alpha$ & $H_0$\tnote{c} & $\sigma_{\mathrm{ext}}$ & $a$ & $b$ & $\chi^2$ & $\nu$ & $-2\ln\mathcal{L}_{\mathrm{max}}$ & $AIC$ & $BIC$ & $\Delta\chi^2$ & $\Delta AIC$ & $\Delta BIC$ \\
\midrule
Flat \lcdm & GRB & 0.698 & 0.302 & -- & -- & -- & 80.36 & 0.404 & 49.92& 1.113 & 117.98 & 114 & 130.12 & 140.12 & 154.01 & 1.08 & 0.00 & 0.00\\
 & HIIG & 0.276 & 0.724 & -- & -- & -- & 71.81 & -- & -- & -- & 410.75 & 151 & 410.75 & 414.75 & 420.81 & 3.15 & 0.00 & 0.00\\
 & QSO-AS & 0.315 & 0.685 & -- & -- & -- & 68.69 & -- & -- & -- & 352.05 & 118 & 352.05 & 356.05 & 361.62 & 1.76 & 0.00 & 0.00\\
 & QGH\tnote{d} & 0.271 & 0.729 & -- & -- & -- & 71.13 & 0.407 & 50.18 & 1.138 & 879.42 & 387 & 895.05 & 905.05 & 924.91 & 0.12 & 0.00 & 0.00\\
 & $H(z)$ + BAO & 0.314 & 0.686 & -- & -- & -- & 68.53 & -- & -- & -- & 20.82 & 40 & 20.82 & 24.82 & 28.29 & 2.39 & 0.00 & 0.00\\
 & ZBQGH\tnote{e} & 0.317 & 0.683 & -- & -- & -- & 69.06 & 0.404 & 50.19 & 1.134 & 903.61 & 429 & 917.79 & 927.79 & 948.16 & 4.05 & 0.00 & 0.00\\
\\
Non-flat \lcdm & GRB & 0.691 & 0.203 & 0.106 & -- & -- & 77.03 & 0.402 & 49.96 & 1.115 & 117.37 & 113 & 129.96 & 141.96 & 158.64 & 0.47 & 1.84 & 4.63\\
 & HIIG & 0.311 & 1.000 & $-0.311$ & -- & -- & 72.41 & -- & -- & -- & 410.44 & 150 & 410.44 & 416.44 & 425.53 & 2.84 & 1.69 & 4.72\\
 & QSO-AS & 0.266 & 1.000 & $-0.268$ & -- & -- & 74.73 & -- & -- & -- & 351.30 & 117 & 351.30 & 357.30 & 365.66 & 1.01 & 1.25 & 4.04\\
 & QGH\tnote{d} & 0.291 & 0.876 & $-0.167$ & -- & -- & 72.00 & 0.406 & 50.22 & 1.120 & 879.30 & 386 & 894.02 & 906.02 & 929.85 & 0.00 & 0.97 & 4.94\\
 & $H(z)$ + BAO & 0.308 & 0.643 & 0.049 & -- & -- & 67.52 & -- & -- & -- & 20.52 & 39 & 20.52 & 26.52 & 31.73 & 2.09 & 1.70 & 3.44\\
 & ZBQGH\tnote{e} & 0.309 & 0.716 & $-0.025$ & -- & -- & 69.77 & 0.402 & 50.17 & 1.141 & 904.47 & 428 & 917.17 & 929.17 & 953.61 & 4.91 & 1.38 & 5.45\\
\\
Flat XCDM & GRB & 0.102 & -- & -- & $-0.148$ & -- & 55.30 & 0.400 & 50.22 & 1.117 & 118.28 & 113 & 129.79 & 141.79 & 158.47 & 1.38 & 1.67 & 4.46\\
 & HIIG & 0.251 & -- & -- & $-0.899$ & -- & 71.66 & -- & -- & -- & 410.72 & 150 & 410.72 & 416.72 & 425.82 & 3.12 & 1.97 & 5.01\\
 & QSO-AS & 0.267 & -- & -- & $-2.000$ & -- & 81.70 & -- & -- & -- & 351.84 & 117 & 351.84 & 357.84 & 366.20 & 1.55 & 1.79 & 4.58\\
 & QGH\tnote{d} & 0.320 & -- & -- & $-1.306$ & -- & 72.03 & 0.404 & 50.20 & 1.131 & 880.47 & 386 & 894.27 & 906.27 & 930.10 & 1.17 & 1.22 & 5.19\\
 & $H(z)$ + BAO & 0.319 & -- & -- & $-0.865$ & -- & 65.83 & -- & -- & -- & 19.54 & 39 & 19.54 & 25.54 & 30.76 & 1.11 & 0.72 & 2.47\\
 & ZBQGH\tnote{e} & 0.313 & -- & -- & $-1.052$ & -- & 69.90 & 0.407 & 50.19 & 1.132 & 902.09 & 428 & 917.55 & 929.55 & 953.99 & 2.53 & 1.76 & 5.83\\
\\
Non-flat XCDM & GRB & 0.695 & -- & 0.556 & $-1.095$ & -- & 57.64 & 0.399 & 50.13 & 1.133 & 118.43 & 112 & 129.73 & 143.73 & 163.19 & 1.53 & 3.61 & 9.18\\
 & HIIG & 0.100 & -- & $-0.702$ & $-0.655$ & -- & 72.57 & -- & -- & -- & 407.60 & 149 & 407.60 & 415.60 & 427.72 & 0.00 & 0.85 & 6.91\\
 & QSO-AS & 0.100 & -- & $-0.548$ & $-0.670$ & -- & 74.04 & -- & -- & -- & 350.29 & 116 & 350.29 & 358.29 & 369.44 & 0.00 & 2.24 & 7.82\\
 & QGH\tnote{d} & 0.300 & -- & $-0.161$ & $-1.027$ & -- & 80.36 & 0.405 & 50.21 & 1.122 & 879.48 & 385 & 894.01 & 908.01 & 935.81 & 0.18 & 2.96 & 10.90\\
 & $H(z)$ + BAO & 0.327 & -- & $-0.159$ & $-0.730$ & -- & 65.97 & -- & -- & -- & 18.43 & 38 & 18.43 & 26.43 & 33.38 & 0.00 & 1.61 & 5.09\\
 & ZBQGH\tnote{e} & 0.312 & -- & $-0.045$ & $-0.959$ & -- & 69.46 & 0.402 & 50.23 & 1.117 & 904.17 & 427 & 917.07 & 931.07 & 959.58 & 4.61 & 3.28 & 11.42\\
\\
Flat $\phi$CDM & GRB & 0.674 & -- & -- & -- & 2.535 & 84.00 & 0.399 & 49.88 & 1.104 & 119.15 & 113 & 130.14 & 142.14 & 158.82 & 2.25 & 2.02 & 4.81\\
 & HIIG & 0.255 & -- & -- & -- & 0.260 & 71.70 & -- & -- & -- & 410.70 & 150 & 410.70 & 416.70 & 425.80 & 3.10 & 1.95 & 4.99\\
 & QSO-AS & 0.319 & -- & -- & -- & 0.012 & 68.47 & -- & -- & -- & 352.05 & 117 & 352.05 & 358.05 & 366.41 & 1.76 & 2.00 & 4.79\\
 & QGH\tnote{d} & 0.282 & -- & -- & -- & 0.012 & 70.81 & 0.402 & 50.19 & 1.135 & 882.56 & 386 & 895.28 & 907.28 & 931.11 & 3.26 & 2.23 & 6.20\\
 & $H(z)$ + BAO & 0.318 & -- & -- & -- & 0.364 & 66.04 & -- & -- & -- & 19.65 & 39 & 19.65 & 25.65 & 30.86 & 1.22 & 0.83 & 2.57\\
 & ZBQGH\tnote{e} & 0.316 & -- & -- & -- & 0.013 & 69.15 & 0.405 & 50.24 & 1.114 & 903.52 & 428 & 918.12 & 930.12 & 954.56 & 3.96 & 2.33 & 6.40\\
\\
Non-flat $\phi$CDM & GRB & 0.664 & -- & 0.188 & -- & 4.269 & 59.65 & 0.403 & 50.17 & 1.111 & 116.90 & 112 & 129.93 & 143.93 & 163.39 & 0.00 & 3.81 & 9.38\\
 & HIIG & 0.114 & -- & $-0.437$ & -- & 2.680 & 72.14 & -- & -- & -- & 409.91 & 149 & 409.91 & 417.91 & 430.03 & 2.31 & 3.16 & 9.22\\
 & QSO-AS & 0.100 & -- & $-0.433$ & -- & 2.948 & 72.37 & -- & -- & -- & 350.98 & 116 & 350.98 & 358.98 & 370.13 & 0.69 & 2.93 & 8.51\\
 & QGH\tnote{d} & 0.276 & -- & $-0.185$ & -- & $0.145$ & 72.11 & 0.402 & 50.16 & 1.142 & 881.09 & 385 & 894.24 & 908.24 & 936.03 & 1.79 & 3.19 & 11.12\\
 & $H(z)$ + BAO & 0.321 & -- & $-0.137$ & -- & 0.887 & 66.41 & -- & -- & -- & 18.61 & 39 & 18.61 & 26.61 & 33.56 & 0.18 & 1.79 & 5.27\\
 & ZBQGH\tnote{e} & 0.310 & -- & $-0.052$ & -- & 0.193 & 69.06 & 0.411 & 50.21 & 1.126 & 899.56 & 427 & 917.26 & 931.26 & 959.77 & 0.00 & 3.47 & 11.61\\
\bottomrule
\end{tabular}}
\begin{tablenotes}[flushleft]
\item [c] km s$^{-1}$ Mpc$^{-1}$.
\item [d] HIIG + QSO-AS + GRB.
\item [e] $H(z)$ + BAO + HIIG + QSO-AS + GRB.
\end{tablenotes}
\end{threeparttable}%
}
\end{table*}

\subsection{$H(z)$, BAO, QSO-AS, GRB, and HIIG (ZBQGH) constraints}
\label{subsec:ZBQGH}

Given the good mutual consistency between constraints derived from $H(z)$ + BAO data and those derived from QGH data, in this subsection we determine more restrictive joint constraints from the combined $H(z)$, BAO, QSO-AS, GRB, and HIIG (ZBQGH) data on the parameters of our six cosmological models.

The 1D probability distributions and 2D confidence regions of the cosmological and Amati relation parameters for all models from the ZBQGH data are in blue in Figs. \ref{fig7}--\ref{fig12}, and in red in panels (b) of Figs. \ref{ch8_fig01}--\ref{ch8_fig04}. The best-fitting results and uncertainties are in Tables \ref{tab:ch8_BFP} and \ref{tab:ch8_1d_BFP}.

The measured values of $\Omega_{m0}$ here are a little larger, and significantly more restrictively constrained, than the ones in the QGH cases (except for flat XCDM), being between $0.310\pm0.014$ (non-flat XCDM) and $0.320\pm0.013$ (flat \pcdm). The $H_0$ measurements are a little lower, and more tightly constrained, than in the QGH cases, and are in better agreement with the lower median statistics estimate of $H_0$ \citep{chenratmed} than the higher local expansion rate measurement of $H_0$ \citep{riess_etal_2019}, being between $68.16^{+1.01}_{-0.80}$ km s$^{-1}$ Mpc$^{-1}$ (flat \pcdm) and $69.85^{+1.42}_{-1.55}$ km s$^{-1}$ Mpc$^{-1}$ (flat XCDM).

For non-flat \lcdm, non-flat XCDM, and non-flat \pcdm, we measure $\Omega_{k0}=-0.019^{+0.043}_{-0.048}$, $\Omega_{k0}=-0.024^{+0.092}_{-0.093}$, and $\Omega_{k0}=-0.094^{+0.082}_{-0.064}$, respectively, where the central values are a little higher (closer to 0) than what was measured in the QGH cases. The joint constraints are more restrictive, with non-flat \lcdm\ and XCDM within 0.44$\sigma$ and 0.26$\sigma$ of $\Omega_{k0} = 0$, respectively. The non-flat \pcdm\ model, on the other hand, still favors a closed geometry with an $\Omega_{k0}$ that is 1.15$\sigma$ away from zero.

The ZBQGH case has slightly larger measured values and tighter error bars for $w_{\rm X}$ and $\alpha$ than the QGH case, so there is also not much evidence in support of dark energy dynamics. For flat (non-flat) XCDM, $w_{\rm X}=-1.050^{+0.090}_{-0.081}$ ($w_{\rm X}=-1.019^{+0.202}_{-0.099}$). For flat (non-flat) \pcdm, the $2\sigma$ upper limits are $\alpha<0.418$ ($\alpha<0.905$).

The cosmological model-independent constraints from the ZBQGH data combination on the parameters of the Amati relation can be summarized as $\sigma_{\rm ext}=0.412^{+0.026}_{-0.032}$, $a=50.19\pm0.24$, and $b=1.132\pm0.085$.

\subsection{Model comparison}
\label{subsec:comparison}

From Table \ref{tab:ch8_BFP}, we see that the reduced $\chi^2$ values determined from GRB data alone are around unity for all models (being between 1.03 and 1.06) while those values determined from the $H(z)$ + BAO data combination range from 0.48 to 0.53, with the lower reduced $\chi^2$ here being due to the $H(z)$ data (that probably have overestimated error bars). As discussed in Chapters \ref{Chapter5} and \ref{Chapter7}, the cases that involve HIIG and QSO-AS data have a larger reduced $\chi^2$ (between 2.11 and 3.02), which is probably due to underestimated systematic uncertainties in both cases.

Based on the $AIC$ and the $BIC$ (see Table \ref{tab:ch8_BFP}), the flat \lcdm\ model remains the most favored model, across all data combinations, among the six models we study.\footnote{Note that based on the $\Delta \chi^2$ results of Table \ref{tab:ch8_BFP} non-flat \lcdm\ has the minimum $\chi^2$ in the QGH case and non-flat XCDM has the minimum $\chi^2$ in the HIIG, QSO-AS, and $H(z)$ + BAO cases, whereas non-flat \pcdm\ has the minimum $\chi^2$ for the GRB and ZBQGH cases. The $\Delta \chi^2$ values do not, however, penalize a model for having more parameters.} From $\Delta AIC$ and $\Delta BIC$, we find mostly weak or positive evidence against the models we considered, and only in a few cases do we find strong evidence against them. According to $\Delta BIC$, the evidence against non-flat XCDM is strong for the HIIG, QSO-AS, and GRB only cases, and very strong for the QGH and ZBQGH cases. Similarly, the evidence against flat \pcdm\ is strong for the QGH and ZBQGH cases, and the evidence against non-flat \pcdm\ is strong for the HIIG, QSO-AS, and GRB only cases, and very strong for the QGH and ZBQGH cases.

Among these six models, a comparison of the $\Delta BIC$ values from Table \ref{tab:ch8_BFP} shows that the most disfavored model is non-flat \pcdm, and that the second most disfavored model is non-flat XCDM. This is especially true when these models are fitted to the QGH and ZBQGH data combinations, in which cases non-flat \pcdm\ and non-flat XCDM are very strongly disfavored. These models aren't as strongly disfavored by the $AIC$, however; from a comparison of the $\Delta AIC$ values in Table \ref{tab:ch8_BFP}, we see that the evidence against the most disfavored model (non-flat \pcdm) is only positive.

\begin{table*}
\centering
\resizebox{\columnwidth}{!}{%
\begin{threeparttable}
\caption{One-dimensional marginalized best-fitting parameter values and uncertainties ($\pm 1\sigma$ error bars or $2\sigma$ limits) for all models from various combinations of data.}\label{tab:ch8_1d_BFP}
\setlength{\tabcolsep}{0.45mm}{
\begin{tabular}{lcccccccccc}
\toprule
 Model & Data set & $\Omega_{\mathrm{m_0}}$ & $\Omega_{\Lambda}$ & $\Omega_{\mathrm{k_0}}$ & $w_{\mathrm{X}}$ & $\alpha$ & $H_0$\tnote{c} & $\sigma_{\mathrm{ext}}$ & $a$ & $b$ \\
\midrule
Flat \lcdm & GRB & $>0.208$ & -- & -- & -- & -- & -- & $0.411^{+0.026}_{-0.032}$ & $50.16\pm0.27$ & $1.123\pm0.085$\\
 & HIIG & $0.289^{+0.053}_{-0.071}$ & -- & -- & -- & -- & $71.70\pm1.83$ & -- & -- & -- \\
 & QSO-AS & $0.364^{+0.083}_{-0.150}$ & -- & -- & -- & -- & $67.29^{+4.93}_{-5.07}$ & -- & -- & -- \\
 & QGH\tnote{e} & $0.277^{+0.034}_{-0.041}$ & -- & -- & -- & -- & $71.03\pm1.67$ & $0.413^{+0.026}_{-0.032}$ & $50.19\pm0.24$ & $1.138\pm0.085$\\
 & $H(z)$ + BAO & $0.315^{+0.015}_{-0.017}$ & -- & -- & -- & -- & $68.55\pm0.87$ & -- & -- & -- \\
 & ZBQGH\tnote{f} & $0.316\pm0.013$ & -- & -- & -- & -- & $69.05^{+0.62}_{-0.63}$ & $0.412^{+0.026}_{-0.032}$ & $50.19\pm0.23$ & $1.133\pm0.085$\\
\\
Non-flat \lcdm & GRB & $0.463^{+0.226}_{-0.084}$ & $<0.658$\tnote{d} & $-0.007^{+0.251}_{-0.234}$ & -- & -- & -- & $0.412^{+0.026}_{-0.032}$ & $50.17\pm0.28$ & $1.121\pm0.086$\\
 & HIIG & $0.275^{+0.081}_{-0.078}$ & $>0.501$\tnote{d} & $0.094^{+0.237}_{-0.363}$ & -- & -- & $71.50^{+1.80}_{-1.81}$ & -- & -- & -- \\
 & QSO-AS & $0.357^{+0.082}_{-0.135}$ & -- & $0.017^{+0.184}_{-0.277}$ & -- & -- & $67.32^{+4.49}_{-5.44}$ & -- & -- & -- \\
 & QGH\tnote{e} & $0.292\pm0.044$ & $0.801^{+0.191}_{-0.055}$ & $-0.093^{+0.092}_{-0.190}$ & -- & -- & $71.33^{+1.75}_{-1.77}$ & $0.413^{+0.026}_{-0.032}$ & $50.19\pm0.24$ & $1.130\pm0.086$\\
 & $H(z)$ + BAO & $0.309\pm0.016$ & $0.636^{+0.081}_{-0.072}$ & $0.055^{+0.082}_{-0.074}$ & -- & -- & $67.44\pm2.33$ & -- & -- & -- \\
 & ZBQGH\tnote{f} & $0.311^{+0.012}_{-0.014}$ & $0.708^{+0.053}_{-0.046}$ & $-0.019^{+0.043}_{-0.048}$ & -- & -- & $69.72\pm1.10$ & $0.412^{+0.026}_{-0.032}$ & $50.19\pm0.23$ & $1.132\pm0.085$ \\
\\
Flat XCDM & GRB & $>0.366$\tnote{d} & -- & -- & -- & -- & -- & $0.411^{+0.025}_{-0.032}$ & $50.14\pm0.28$ & $1.119\pm0.085$\\
 & HIIG & $0.300^{+0.106}_{-0.083}$ & -- & -- & $-1.180^{+0.560}_{-0.330}$ & -- & $71.85\pm1.96$ & -- & -- & -- \\
 & QSO-AS & $0.349^{+0.090}_{-0.143}$ & -- & -- & $-1.161^{+0.430}_{-0.679}$ & -- & $68.39^{+6.14}_{-8.98}$ & -- & -- & -- \\
 & QGH\tnote{e} & $0.322^{+0.062}_{-0.044}$ & -- & -- & $-1.379^{+0.361}_{-0.375}$ & -- & $72.00^{+1.99}_{-1.98}$ & $0.412^{+0.026}_{-0.032}$ & $50.20\pm0.24$ & $1.130\pm0.085$\\
 & $H(z)$ + BAO & $0.319^{+0.016}_{-0.017}$ & -- & -- & $-0.888^{+0.126}_{-0.098}$ & -- & $66.26^{+2.32}_{-2.63}$ & -- & -- & -- \\
 & ZBQGH\tnote{f} & $0.313^{+0.014}_{-0.015}$ & -- & -- & $-1.050^{+0.090}_{-0.081}$ & -- & $69.85^{+1.42}_{-1.55}$ & $0.412^{+0.026}_{-0.032}$ & $50.19\pm0.24$ & $1.134\pm0.085$ \\
\\
Non-flat XCDM & GRB & $>0.386$\tnote{d} & -- & $0.121^{+0.464}_{-0.275}$ & $>-1.218$\tnote{d} & -- & -- & $0.411^{+0.026}_{-0.032}$ & $50.12\pm0.28$ & $1.122\pm0.087$\\
 & HIIG & $0.275^{+0.084}_{-0.125}$ & -- & $0.011^{+0.457}_{-0.460}$ & $-1.125^{+0.537}_{-0.321}$ & -- & $71.71^{+2.07}_{-2.08}$ & -- & -- & -- \\
 & QSO-AS & $0.359^{+0.111}_{-0.174}$ & -- & $0.115^{+0.466}_{-0.293}$ & $-1.030^{+0.593}_{-0.548}$ & -- & $65.92^{+4.54}_{-9.54}$ & -- & -- & -- \\
 & QGH\tnote{e} & $0.303^{+0.073}_{-0.058}$ & -- & $-0.044^{+0.193}_{-0.217}$ & $-1.273^{+0.501}_{-0.321}$ & -- & $71.77\pm2.02$ & $0.413^{+0.026}_{-0.031}$ & $50.20\pm0.24$ & $1.129\pm0.085$\\
 & $H(z)$ + BAO & $0.323^{+0.021}_{-0.020}$ & -- & $-0.105^{+0.187}_{-0.162}$ & $-0.818^{+0.212}_{-0.071}$ & -- & $66.20^{+2.29}_{-2.55}$ & -- & -- & -- \\
 & ZBQGH\tnote{f} & $0.310\pm0.014$ & -- & $-0.024^{+0.092}_{-0.093}$ & $-1.019^{+0.202}_{-0.099}$ & -- & $69.63^{+1.45}_{-1.62}$ & $0.412^{+0.026}_{-0.031}$ & $50.19\pm0.23$ & $1.132\pm0.085$ \\
\\
Flat $\phi$CDM & GRB & $>0.376$\tnote{d} & -- & -- & -- & -- & -- & $0.411^{+0.025}_{-0.032}$ & $50.13\pm0.28$ & $1.121\pm0.087$\\
 & HIIG & $0.210^{+0.043}_{-0.092}$ & -- & -- & -- & $<2.784$ & $71.23^{+1.79}_{-1.80}$ & -- & -- & -- \\
 & QSO-AS & $0.329^{+0.086}_{-0.171}$ & -- & -- & -- & $<2.841$ & $64.42^{+4.47}_{-4.62}$ & -- & -- & -- \\
 & QGH\tnote{e} & $0.214^{+0.057}_{-0.061}$ & -- & -- & -- & $<2.584$ & $70.30\pm1.68$ & $0.413^{+0.026}_{-0.032}$ & $50.18\pm0.24$ & $1.142\pm0.087$\\
 & $H(z)$ + BAO & $0.319^{+0.016}_{-0.017}$ & -- & -- & -- & $0.550^{+0.169}_{-0.494}$ & $65.25^{+2.25}_{-1.82}$ & -- & -- & -- \\
 & ZBQGH\tnote{f} & $0.320\pm0.013$ & -- & -- & -- & $<0.418$ & $68.16^{+1.01}_{-0.80}$ & $0.412^{+0.027}_{-0.033}$ & $50.20\pm0.24$ & $1.131\pm0.088$ \\
\\
Non-flat $\phi$CDM & GRB & $>0.189$ & -- & $0.251^{+0.247}_{-0.086}$ & -- & -- & -- & $0.411^{+0.026}_{-0.032}$ & $50.11\pm0.28$ & $1.128\pm0.089$\\
 & HIIG & $<0.321$ & -- & $0.291^{+0.348}_{-0.113}$ & -- & $<4.590$ & $70.60^{+1.68}_{-1.84}$ & -- & -- & -- \\
 & QSO-AS & $0.362^{+0.117}_{-0.193}$ & -- & $0.254^{+0.304}_{-0.092}$ & -- & $<4.752$ & $61.91^{+2.83}_{-4.92}$ & -- & -- & -- \\
 & QGH\tnote{e} & $0.205^{+0.044}_{-0.094}$ & -- & $-0.124^{+0.127}_{-0.253}$ & -- & $<3.414$ & $70.66\pm1.90$ & $0.414^{+0.027}_{-0.033}$ & $50.19\pm0.24$ & $1.134\pm0.088$\\
 & $H(z)$ + BAO & $0.321\pm0.017$ & -- & $-0.126^{+0.157}_{-0.130}$ & -- & $0.938^{+0.439}_{-0.644}$ & $65.93\pm2.33$ & -- & -- & -- \\
 & ZBQGH\tnote{f} & $0.313\pm0.013$ & -- & $-0.094^{+0.082}_{-0.064}$ & -- & $<0.905$ & $68.79\pm1.22$ & $0.412^{+0.027}_{-0.033}$ & $50.20\pm0.24$ & $1.126\pm0.087$ \\
\bottomrule
\end{tabular}}
\begin{tablenotes}[flushleft]
\item [c] km s$^{-1}$ Mpc$^{-1}$.
\item [d] This is the 1$\sigma$ limit. The $2\sigma$ limit is set by the prior, and is not shown here.
\item [e] HIIG + QSO-AS + GRB.
\item [f] $H(z)$ + BAO + HIIG + QSO-AS + GRB.
\end{tablenotes}
\end{threeparttable}%
}
\end{table*}

\section{Conclusion}
\label{sec:ch8_conclusion}
We find that cosmological constraints determined from higher-$z$ GRB, \hiig, and QSO-AS data are mutually consistent. It is both reassuring and noteworthy that these higher-$z$ data jointly favor currently-accelerating cosmological expansion, and that their constraints are consistent with the constraints imposed by more widely used and more restrictive $H(z)$ and BAO data. Using a data set consisting of 31 $H(z)$, 11 BAO, 120 QSO-AS, 153 \hiig, and 119 GRB measurements, we jointly constrain the parameters of the GRB Amati relation and of six cosmological models. 

The GRB measurements are of special interest because they reach to $z\sim8.2$ (far beyond the highest $z\sim2.3$ reached by BAO data) and into a much less studied area of redshift space. Current GRB data do not provide very restrictive constraints on cosmological model parameters, but in the near future we expect there to be more GRB observations \citep{Shirokov2020} which should improve the GRB data and provide more restrictive cosmological constraints.

Some of our conclusions do not differ significantly between models and so are model-independent. In particular, for the ZBQGH data (the full data set excluding QSO-Flux data), we find a fairly restrictive summary value of $\Omega_{m0}=0.313 \pm 0.013$ that agrees well with many other recent measurements. From these data we also find a fairly restrictive summary value of $H_0=69.3 \pm 1.2$ \hunit\ that is in better agreement with the results of \cite{chenratmed} and \cite{planck2018} than with the result of \cite{riess_etal_2019}; note that we do not take the $H_0$ tension issue into account (for a review, see \citealp{riess_2019}). The ZBQGH measurements are consistent with flat \lcdm, but do not rule out mild dark energy dynamics or a little spatial curvature energy density. More and better-quality higher-$z$ GRB, \hiig, QSO, and other data will significantly help to test these extensions of flat \lcdm.

\begin{subappendices}
\section{QSO-Flux}
\label{sec:appendix}

QSOs obey a nonlinear relation between their luminosities in the X-ray and UV bands. Using a sample of 808 QSOs in the redshift range $0.061 \leq z \leq 6.280$, \cite{RisalitiandLusso_2015} confirmed that this relation can be written
\begin{equation}
\label{eq:LX-LUV}
  \log L_X=\beta+\gamma\log L_{UV},
\end{equation}
where $L_X$ and $L_{UV}$ are the X-ray and UV luminosities of the QSOs. To make contact with observations, eq. \eqref{eq:LX-LUV} must be expressed in terms of the fluxes $F_X$ and $F_{UV}$ measured at fixed rest-frame wavelengths in the X-ray and UV bands, respectively. With this, eq. (\ref{eq:LX-LUV}) becomes
\begin{equation}
    \log F_X=\beta+(\gamma-1)\log 4\pi+\gamma\log F_{UV} +2(\gamma-1)\log D_{L}.
\end{equation}
Here $D_L$ (defined in eq. \ref{eq:D_L}) is the luminosity distance, which depends on the parameters of our cosmological models. We also treat the slope $\gamma$ and intercept $\beta$ as free parameters in our cosmological model fits.

For QSO-Flux data, the natural log of its likelihood function is
\begin{equation*}
    \ln\mathcal{L_{\rm QF}}=-\frac{1}{2}\sum^{N}_{i=1}\Bigg[\frac{\big[\log(F^{\rm{obs}}_X)_i-\log(F^{\rm{th}}_X)_i\big]^2}{s_i^2}+\ln(2\pi s_i^2)\Bigg], \label{eq:LH_QF} 
\end{equation*}
where $s^2_i = \sigma^2_i + \delta^2$. Here $\sigma_i$ is the uncertainty in $\log\left(F^{\rm obs}_X\right)_i$, and $\delta$ is the global intrinsic dispersion in the data (including the systematic uncertainties), which we treat as a free parameter in our cosmological model fits. We use the \cite{RisalitiandLusso_2019} compilation of 1598 QSO-Flux measurements in the range $0.036 \leq z\leq 5.1003$. The flat priors of cosmological parameters and the Amati relation parameters are in Sec. \ref{sec:ch8_analysis} and, as in \cite{KhadkaandRatra_2020}, the flat priors of the parameters $\delta$, $\gamma$, and $\beta$ are non-zero over $0\leq\delta\leq e^{10}$, $-2\leq\gamma\leq2$, and $0\leq\beta\leq11$, respectively. 

\begin{figure*}
\centering
    \includegraphics[width=3.5in,height=3.5in]{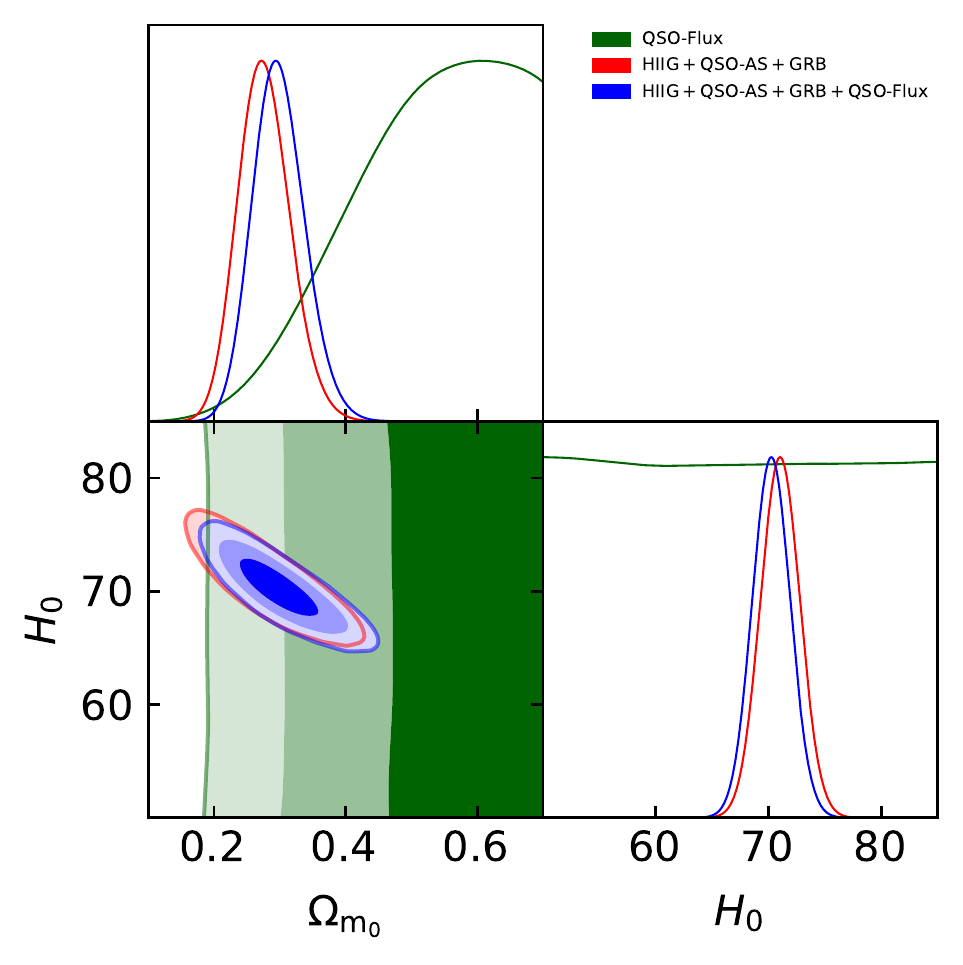}
    \includegraphics[width=3.5in,height=3.5in]{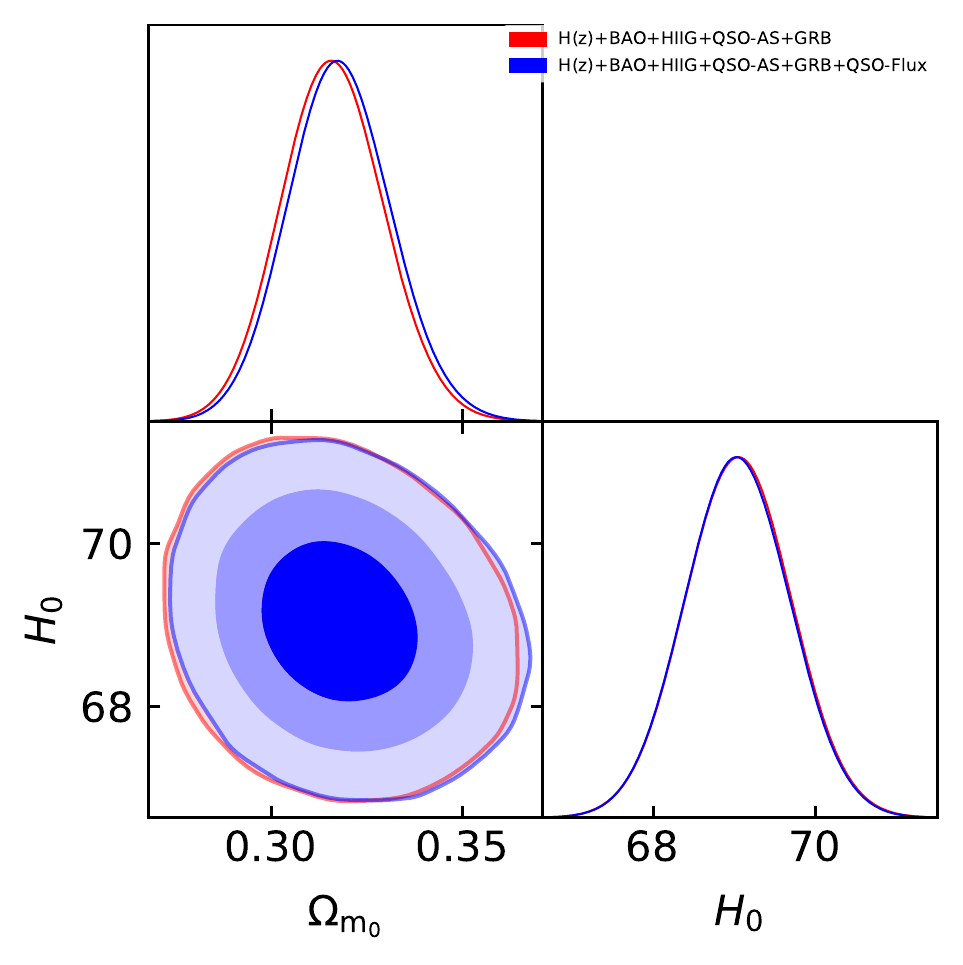}
\caption{Same as Fig. \ref{fig1} (flat \lcdm) but for different combinations of data and showing only cosmological parameters.}
\label{ch8_fig01}
\end{figure*}

\begin{figure*}
\centering
  \subfloat[]{%
    \includegraphics[width=3.5in,height=3.5in]{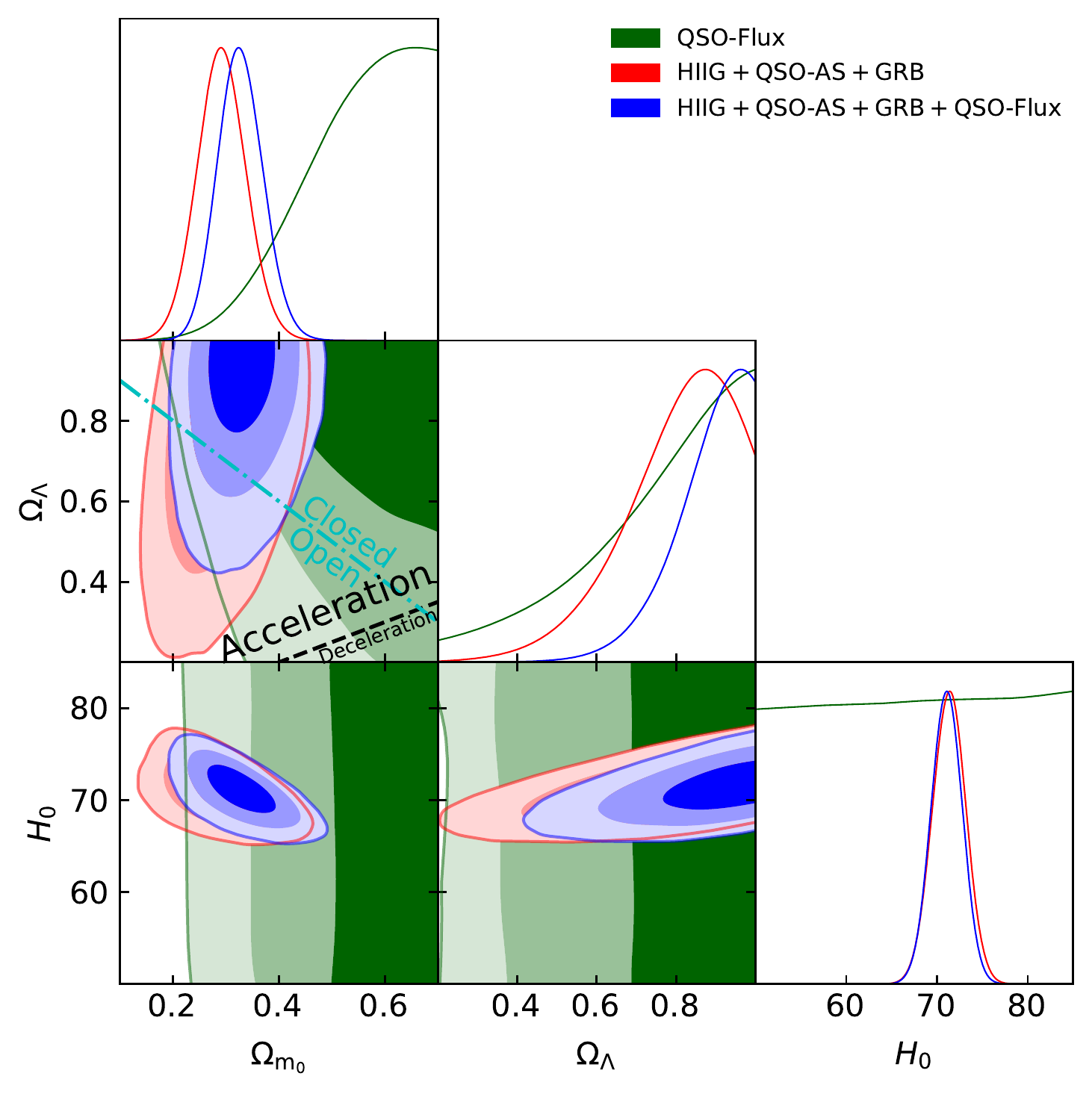}}
  \subfloat[]{%
    \includegraphics[width=3.5in,height=3.5in]{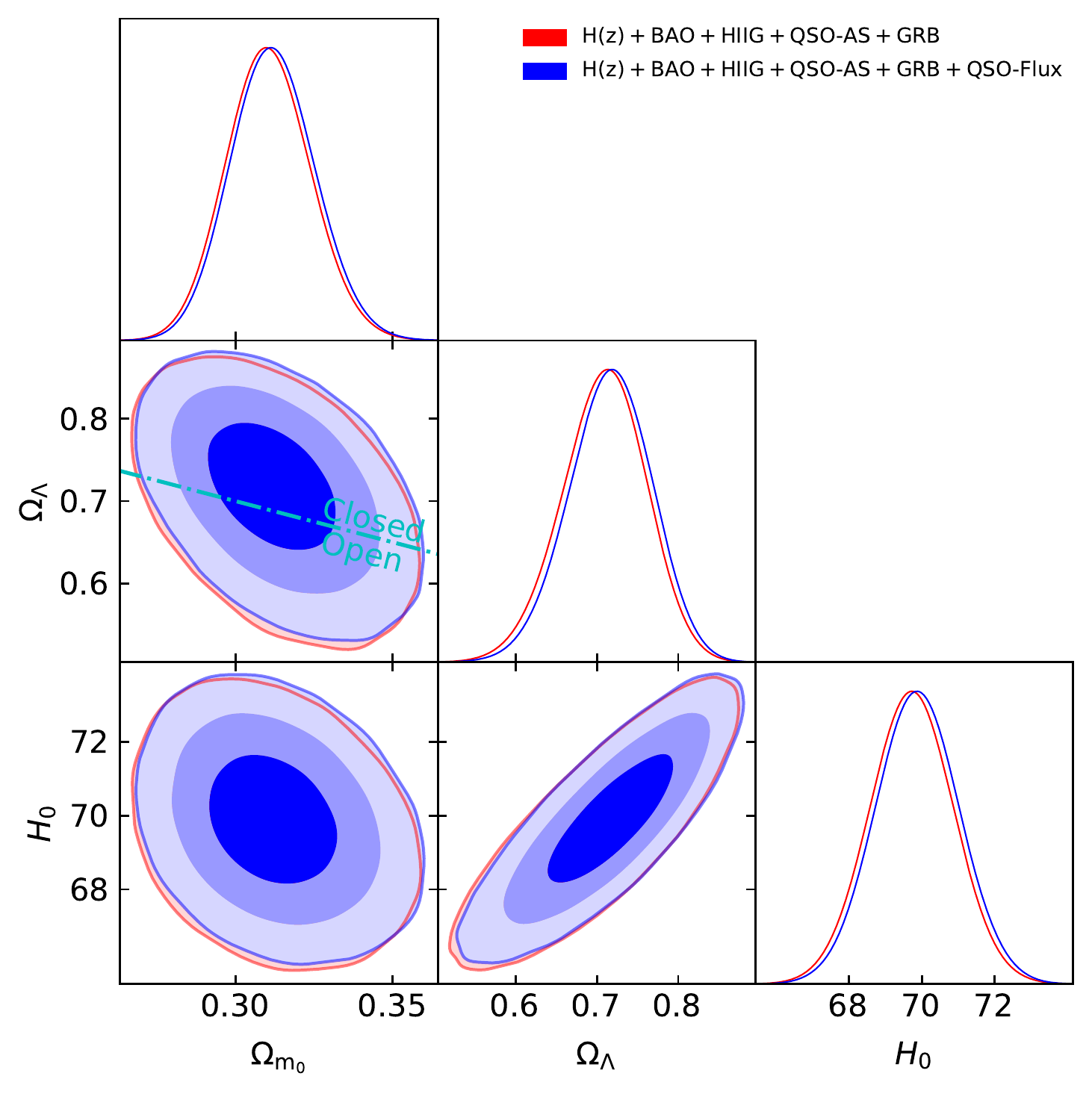}}\\
\caption{Same as Fig. \ref{fig2} (non-flat \lcdm) but for different combinations of data and showing only cosmological parameters.}
\label{ch8_fig02}
\end{figure*}

\begin{figure*}
\centering
  \subfloat[]{%
    \includegraphics[width=3.5in,height=3.5in]{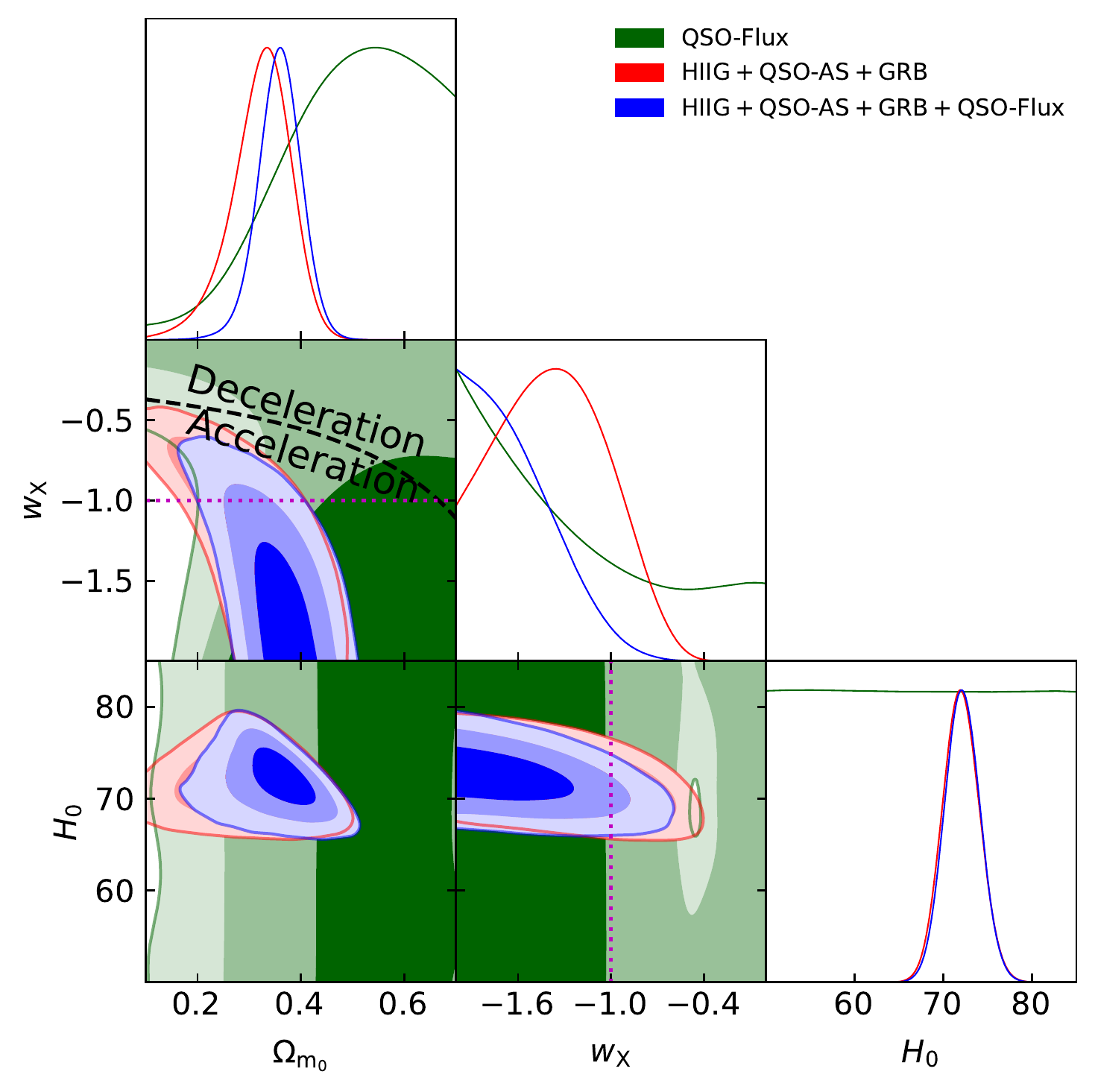}}
  \subfloat[]{%
    \includegraphics[width=3.5in,height=3.5in]{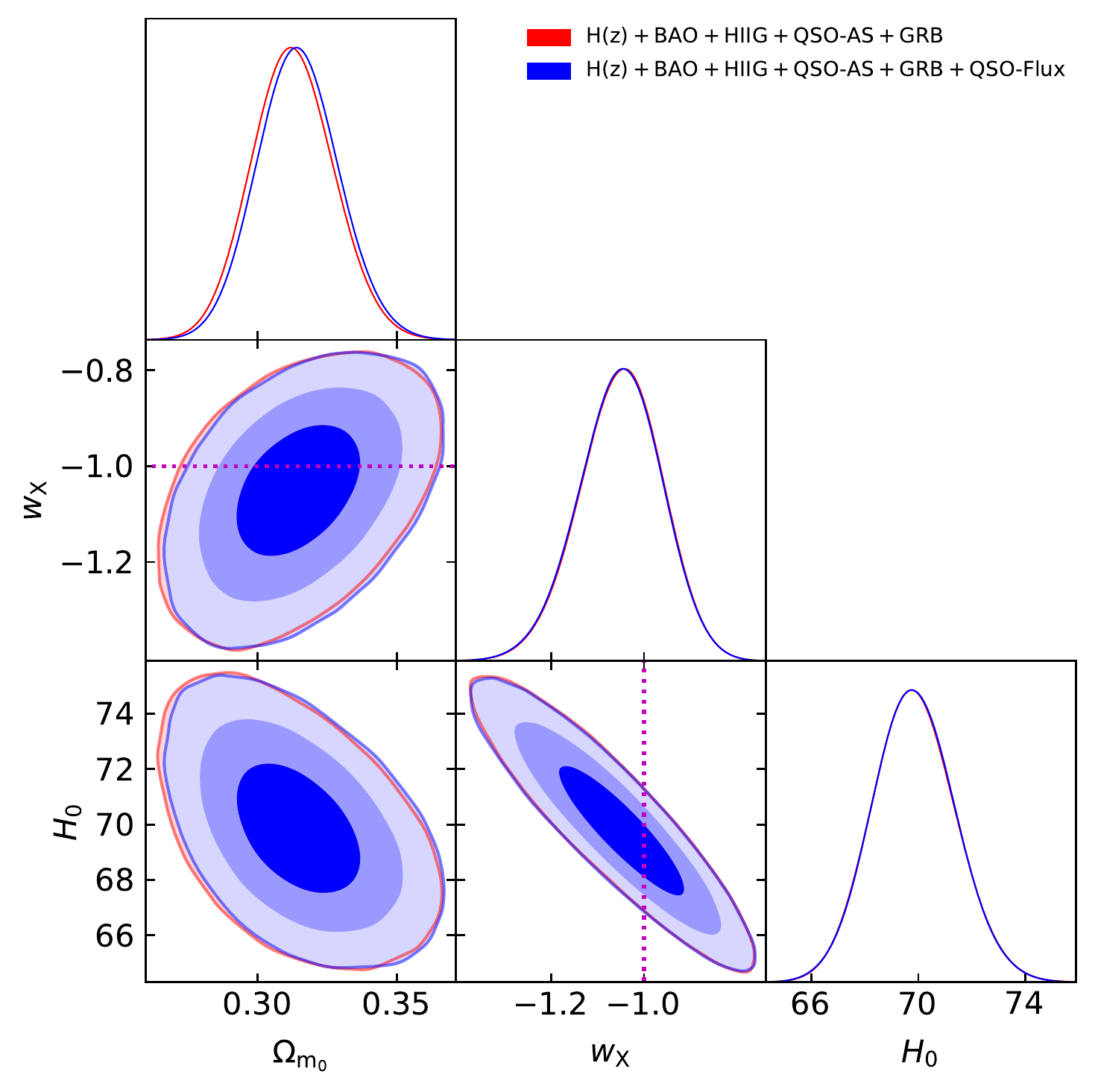}}\\
\caption{Same as Fig. \ref{fig3} (flat XCDM) but for different combinations of data and showing only cosmological parameters.}
\label{ch8_fig03}
\end{figure*}

\begin{figure*}
\centering
    \includegraphics[width=3.5in,height=3.5in]{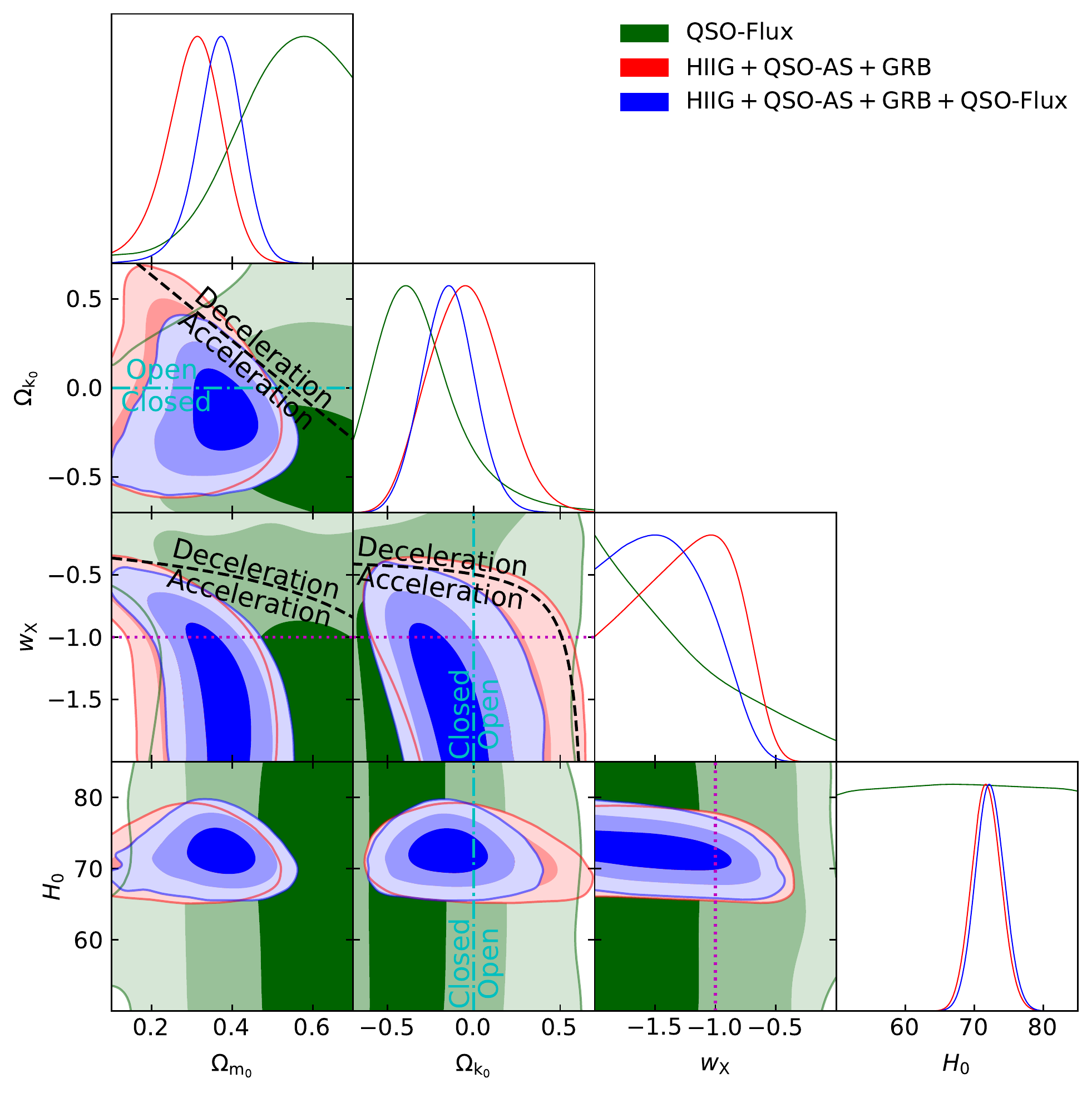}
    \includegraphics[width=3.5in,height=3.5in]{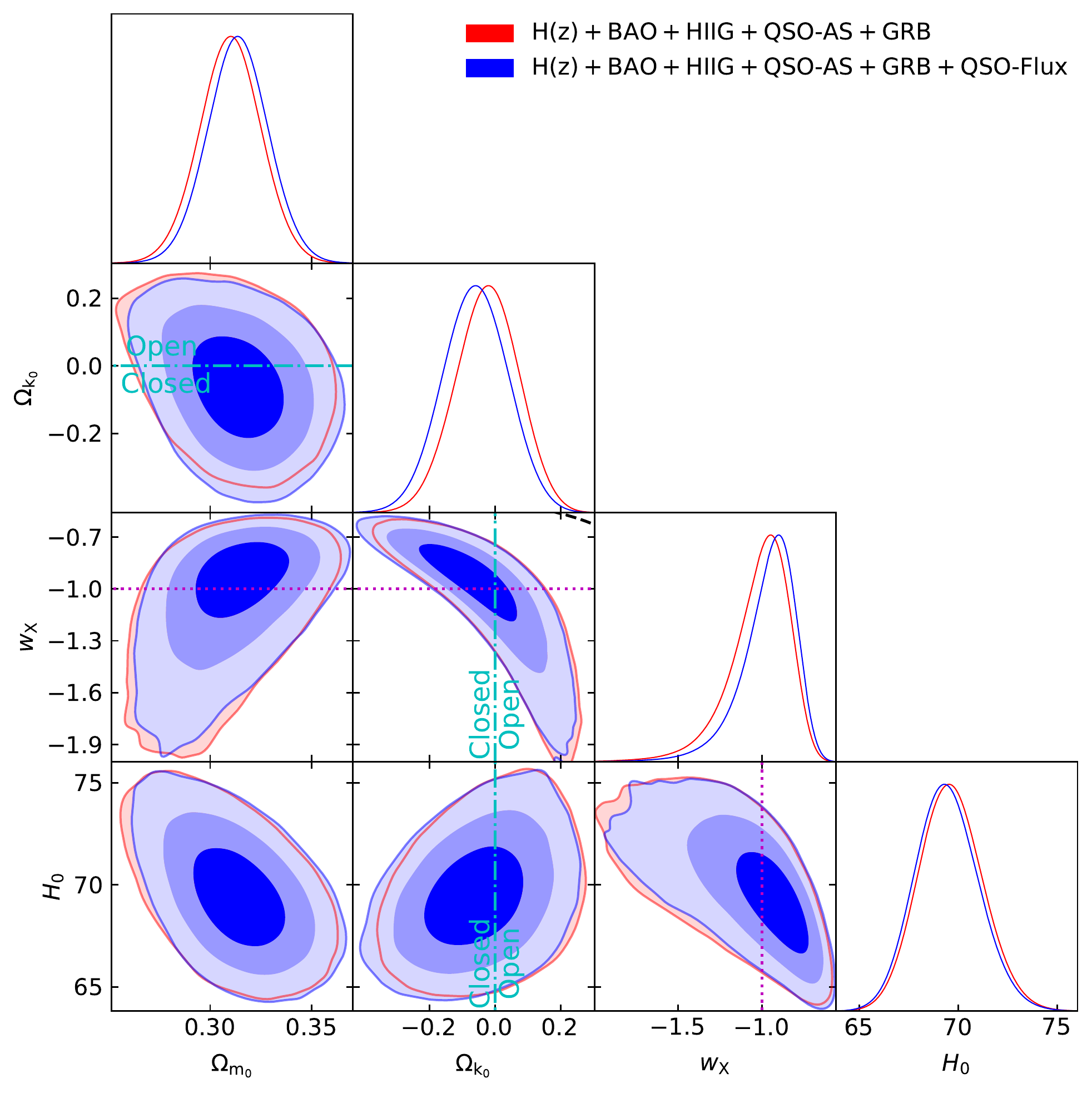}
\caption{Same as Fig. \ref{fig4} (non-flat XCDM) but for different combinations of data and showing only cosmological parameters.}
\label{ch8_fig04}
\end{figure*}

As discussed in \cite{KhadkaandRatra_2020} the QSO-Flux data alone favors large $\Omega_{m0}$ values for the physically-motivated flat and non-flat \lcdm\ and \pcdm\ models. \cite{RisalitiandLusso_2019} and \cite{KhadkaandRatra_2020} note that this is largely a consequence of the $z\sim 2$--5 QSO data. While these large $\Omega_{m0}$ values differ from almost all other measurements of $\Omega_{m0}$, the QSO-Flux data have larger error bars and their cosmological constraint contours are not in conflict with those from other data sets. For these reasons we have used the QSO-Flux data, but in this appendix and not in the main text, and we have not computed QSO-Flux data results for the \pcdm\ cases (these being computationally demanding). We briefly summarize our constraints, listed in Tables \ref{tab:ch8_BFP2} and \ref{tab:ch8_1d_BFP2} and shown in Figs. \ref{ch8_fig01}--\ref{ch8_fig04}, below.

\subsection{QSO-Flux constraints}

Except for flat \lcdm, the constraints on $\Omega_{m0}$ in the QSO-Flux only case are 2$\sigma$ larger than those in the combined ZBQGHF case (see Sec. \ref{sec:ch8_A3}). QSO-Flux data cannot constrain $\alpha$, nor can they constrain $H_0$ (for the same reason that GRB data cannot constrain this parameter; see Section \ref{subsec:GRB}). QSO-Flux data set upper limits on $w_{\rm X}$ for flat and non-flat XCDM, with $w_{\rm X}=-1$ within the 1$\sigma$ range.

\subsection{QSO-AS, GRB, HIIG, and QSO-Flux (QGHF) constraints}

When adding QSO-Flux to QGH data, the joint constraints favor larger $\Omega_{m0}$ and lower $\Omega_{k0}$. In non-flat $\Lambda$CDM closed geometry is favored at 3.24$\sigma$. The $H_0$ constraints are only mildly affected by the addition of the QSO-Flux data. The constraint on $w_{\rm X}$ changes from $-1.379^{+0.361}_{-0.375}$ in the QGH case to $<-1.100$ (2$\sigma$ limit) in the QGHF case for flat XCDM, while for non-flat XCDM, the constraint on $w_{\rm X}$ in the QGHF case is 0.40$\sigma$ lower than that in the QGH case and is 1.80$\sigma$ away from $w_{\rm X}=-1$.

\subsection{$H(z)$, BAO, QSO-AS, GRB, HIIG, and QSO-Flux (ZBQGHF) constraints}
\label{sec:ch8_A3}

When adding QSO-Flux to the ZBQGH combination, the $\Omega_{m0}$ central values are only slightly larger because the $H(z)$ + BAO data dominate this compilation. The joint-constraint central $\Omega_{k0}$ values are lower, and consistent with flat geometry, while the constraints on $H_0$ from this combination are almost unaltered. The constraints on $w_{\rm X}$ are 0.02$\sigma$ lower and 0.23$\sigma$ higher for flat and non-flat XCDM, respectively, both being consistent with $w_{\rm X}=-1$ within 1$\sigma$.

\subsection{Model comparison}

From Table \ref{tab:ch8_BFP2}, we see that the reduced $\chi^2$ of the QSO-Flux case for all models is near unity ($\sim1.01$) and that the reduced $\chi^2$ of cases that include QSO-Flux is brought down to $\sim1.24$--1.26 for all models. Based on the $BIC$ (see Table \ref{tab:ch8_BFP2}), flat \lcdm\ is the most favored model, while based on the $AIC$, non-flat XCDM, flat XCDM, and flat \lcdm\ are the most favored models for the QSO-Flux, QGHF, and ZBQGHF combinations, respectively.\footnote{Note that based on the $\Delta \chi^2$ results of Table \ref{tab:ch8_BFP2} flat \lcdm\ has the minimum $\chi^2$ in the QSO-Flux, QGHF, and ZBQGHF cases.} From $\Delta AIC$ and $\Delta BIC$, we find mostly weak or positive evidence against the models, and only in a few cases do we find strong evidence against our models. According to $\Delta BIC$, the evidence against non-flat XCDM is strong for the QSO-Flux data, and very strong for the QGHF and ZBQGHF data, and the evidence against non-flat \lcdm\ is strong for the ZBQGHF data. According to $\Delta AIC$, the evidence against flat XCDM is strong for the ZBQGHF data.

\begin{sidewaystable*}
\centering
\footnotesize
\begin{threeparttable}
\caption{Unmarginalized best-fitting parameter values for all models from various combinations of data.}\label{tab:ch8_BFP2}
\setlength{\tabcolsep}{0.75mm}{
\begin{tabular}{lccccccccccccccccccccc}
\toprule
 Model & Data set & $\Omega_{\mathrm{m_0}}$ & $\Omega_{\Lambda}$ & $\Omega_{\mathrm{k_0}}$ & $w_{\mathrm{X}}$ & $\alpha$ & $H_0$\tnote{c} & $\sigma_{\mathrm{ext}}$ & $a$ & $b$ & $\delta$ & $\gamma$ & $\beta$ & $\chi^2$ & $\nu$ & $-2\ln\mathcal{L}_{\mathrm{max}}$ & $AIC$ & $BIC$ & $\Delta\chi^2$ & $\Delta AIC$ & $\Delta BIC$ \\
\midrule
Flat \lcdm & QSO-Flux & 0.315 & 0.685 & -- & -- & -- & 68.69 & -- & -- & -- & -- & -- & -- & 1603.28 & 1593 & -50.13 & -40.13 & -13.24 & 0.00 & 1.62 & 0.00\\
 & QGH\tnote{d} & 0.271 & 0.729 & -- & -- & -- & 71.13 & 0.407 & 50.18 & 1.138 & -- & -- & -- & 879.42 & 387 & 895.05 & 905.05 & 924.91 & 0.12 & 0.00 & 0.00\\
 & QGHF\tnote{e} & 0.305 & 0.695 & -- & -- & -- & 70.01 & 0.399 & 50.20 & 1.132 & 0.231 & 0.639 & 7.083 & 2480.01 & 1982 & 848.53 & 864.53 & 909.29 & 0.00 & 2.58 & 0.00\\
 & ZBQGH\tnote{f} & 0.317 & 0.683 & -- & -- & -- & 69.06 & 0.404 & 50.19 & 1.134 & -- & -- & -- & 903.61 & 429 & 917.79 & 927.79 & 948.16 & 1.52 & 0.00 & 0.00\\
 & ZBQGHF\tnote{g} & 0.317 & 0.683 & -- & -- & -- & 69.06 & 0.399 & 50.23 & 1.119 & 0.232 & 0.637 & 7.144 & 2499.87 & 2024 & 870.31 & 886.31 & 931.25 & 0.00 & 0.00 & 0.00\\
\\
Non-flat \lcdm & QSO-Flux & 0.540 & 0.985 & $-0.525$ & -- & -- & 75.75 & -- & -- & -- & 0.230 & 0.611 & 7.888 & 1603.83 & 1592 & -53.25 & -41.25 & -8.99 & 0.55 & 0.50 & 4.25\\
 & QGH\tnote{d} & 0.291 & 0.876 & $-0.167$ & -- & -- & 72.00 & 0.406 & 50.22 & 1.120 & -- & -- & -- & 879.30 & 386 & 894.02 & 906.02 & 929.85 & 0.00 & 0.97 & 4.94\\
 & QGHF\tnote{e} & 0.325 & 0.944 & $-0.269$ & -- & -- & 71.49 & 0.404 & 50.21 & 1.116 & 0.230 & 0.632 & 7.304 & 2486.97 & 1981 & 844.38 & 862.38 & 912.75 & 6.96 & 0.43 & 3.46\\
 & ZBQGH\tnote{f} & 0.309 & 0.716 & $-0.025$ & -- & -- & 69.77 & 0.402 & 50.17 & 1.141 & -- & -- & -- & 904.47 & 428 & 917.17 & 929.17 & 953.61 & 2.38 & 1.38 & 5.45\\
 & ZBQGHF\tnote{g} & 0.309 & 0.709 & $-0.018$ & -- & -- & 69.59 & 0.412 & 50.21 & 1.128 & 0.231 & 0.637 & 7.151 & 2503.43 & 2023 & 869.71 & 887.71 & 938.26 & 3.56 & 1.40 & 7.01 \\
\\
Flat XCDM & QSO-Flux & 0.477 & -- & -- & $-1.988$ & -- & 60.86 & -- & -- & -- & 0.230 & 0.625 & 7.530 & 1604.18 & 1592 & -52.13 & -40.13 & -7.88 & 0.90 & 1.62 & 5.36\\
 & QGH\tnote{d} & 0.320 & -- & -- & $-1.306$ & -- & 72.03 & 0.404 & 50.20 & 1.131 & -- & -- & -- & 880.47 & 386 & 894.27 & 906.27 & 930.10 & 1.17 & 1.22 & 5.19\\
 & QGHF\tnote{e} & 0.370 & -- & -- & $-1.980$ & -- & 73.66 & 0.399 & 50.18 & 1.129 & 0.231 & 0.632 & 7.301 & 2485.59 & 1981 & 843.95 & 861.95 & 912.31 & 5.58 & 0.00 & 3.02\\
 & ZBQGH\tnote{f} & 0.313 & -- & -- & $-1.052$ & -- & 69.90 & 0.407 & 50.19 & 1.132 & -- & -- & -- & 902.09 & 428 & 917.55 & 929.55 & 953.99 & 0.00 & 1.76 & 5.83\\
 & ZBQGHF\tnote{g} & 0.313 & -- & -- & $-1.046$ & -- & 69.84 & 0.401 & 50.18 & 1.134 & 0.231 & 0.635 & 7.215 & 2506.25 & 2023 & 870.16 & 888.16 & 938.71 & 6.38 & 1.85 & 7.46\\
\\
Non-flat XCDM & QSO-Flux & 0.507 & -- & $-0.376$ & $-1.996$ & -- & 75.28 & -- & -- & -- & 0.229 & 0.611 & 7.934 & 1614.59 & 1591 & -55.75 & -41.75 & -4.12 & 11.31 & 0.00 & 9.12\\
 & QGH\tnote{d} & 0.300 & -- & $-0.161$ & $-1.027$ & -- & 80.36 & 0.405 & 50.21 & 1.122 & -- & -- & -- & 879.48 & 385 & 894.01 & 908.01 & 935.81 & 0.18 & 2.96 & 10.90\\
 & QGHF\tnote{e} & 0.395 & -- & $-0.138$ & $-1.639$ & -- & 73.48 & 0.411 & 50.21 & 1.112 & 0.230 & 0.627 & 7.441 & 2486.77 & 1980 & 843.65 & 863.65 & 919.61 & 6.76 & 1.70 & 10.32\\
 & ZBQGH\tnote{f} & 0.312 & -- & $-0.045$ & $-0.959$ & -- & 69.46 & 0.402 & 50.23 & 1.117 & -- & -- & -- & 904.17 & 427 & 917.07 & 931.07 & 959.58 & 2.08 & 3.28 & 11.42\\
 & ZBQGHF\tnote{g} & 0.316 & -- & $-0.089$ & $-0.891$ & -- & 69.05 & 0.410 & 50.23 & 1.111 & 0.230 & 0.633 & 7.247 & 2516.49 & 2022 & 869.25 & 889.25 & 945.41 & 16.62 & 2.94 & 14.16\\
\bottomrule
\end{tabular}}
\begin{tablenotes}[flushleft]
\item [c] km s$^{-1}$ Mpc$^{-1}$.
\item [d] HIIG + QSO-AS + GRB.
\item [e] HIIG + QSO-AS + GRB + QSO-Flux.
\item [f] $H(z)$ + BAO + HIIG + QSO-AS + GRB.
\item [g] $H(z)$ + BAO + HIIG + QSO-AS + GRB + QSO-Flux.
\end{tablenotes}
\end{threeparttable}
\end{sidewaystable*}

\begin{sidewaystable*}
\centering
\scriptsize
\begin{threeparttable}
\caption{One-dimensional marginalized best-fitting parameter values and uncertainties ($\pm 1\sigma$ error bars or $2\sigma$ limits) for all models from various combinations of data.}\label{tab:ch8_1d_BFP2}
\setlength{\tabcolsep}{1.1mm}{
\begin{tabular}{lccccccccccccc}
\toprule
 Model & Data set & $\Omega_{\mathrm{m_0}}$ & $\Omega_{\Lambda}$ & $\Omega_{\mathrm{k_0}}$ & $w_{\mathrm{X}}$ & $\alpha$ & $H_0$\tnote{c} & $\sigma_{\mathrm{ext}}$ & $a$ & $b$ & $\delta$ & $\gamma$ & $\beta$ \\
\midrule
Flat \lcdm & QSO-Flux & $>0.313$ & -- & -- & -- & -- & -- & -- & -- & -- & $0.231\pm0.004$ & $0.626\pm0.011$ & $7.469\pm0.321$ \\
 & QGH\tnote{d} & $0.277^{+0.034}_{-0.041}$ & -- & -- & -- & -- & $71.03\pm1.67$ & $0.413^{+0.026}_{-0.032}$ & $50.19\pm0.24$ & $1.138\pm0.085$ & -- & -- & --\\
 & QGHF\tnote{e} & $0.299^{+0.036}_{-0.043}$ & -- & -- & -- & -- & $70.25^{+1.60}_{-1.61}$ & $0.412^{+0.027}_{-0.032}$ & $50.18\pm0.24$ & $1.136\pm0.085$ & $0.231^{+0.005}_{-0.004}$ & $0.639^{+0.009}_{-0.010}$ & $7.091^{+0.281}_{-0.279}$\\
 & ZBQGH\tnote{f} & $0.316\pm0.013$ & -- & -- & -- & -- & $69.05^{+0.62}_{-0.63}$ & $0.412^{+0.026}_{-0.032}$ & $50.19\pm0.23$ & $1.133\pm0.085$ & -- & -- & --\\
 & ZBQGHF\tnote{g} & $0.318\pm0.013$ & -- & -- & -- & -- & $69.03\pm0.62$ & $0.412^{+0.026}_{-0.032}$ & $50.19\pm0.23$ & $1.133\pm0.084$ & $0.231\pm0.004$ & $0.637\pm0.009$ & $7.146\pm0.268$\\
\\
Non-flat \lcdm & QSO-Flux & $>0.353$ & $>0.357$ & $-0.303^{+0.131}_{-0.252}$ & -- & -- & -- & -- & -- & -- & $0.231^{+0.004}_{-0.005}$ & $0.618\pm0.012$ & $7.709\pm0.366$ \\
 & QGH\tnote{d} & $0.292\pm0.044$ & $0.801^{+0.191}_{-0.055}$ & $-0.093^{+0.092}_{-0.190}$ & -- & -- & $71.33^{+1.75}_{-1.77}$ & $0.413^{+0.026}_{-0.032}$ & $50.19\pm0.24$ & $1.130\pm0.086$ & -- & -- & --\\
 & QGHF\tnote{e} & $0.327^{+0.039}_{-0.043}$ & $>0.691$ & $-0.204^{+0.063}_{-0.125}$ & -- & -- & $71.07\pm1.64$ & $0.413^{+0.027}_{-0.032}$ & $50.20\pm0.24$ & $1.120\pm0.086$ & $0.231\pm0.004$ & $0.632\pm0.010$ & $7.291^{+0.306}_{-0.305}$\\
 & ZBQGH\tnote{f} & $0.311^{+0.012}_{-0.014}$ & $0.708^{+0.053}_{-0.046}$ & $-0.019^{+0.043}_{-0.048}$ & -- & -- & $69.72\pm1.10$ & $0.412^{+0.026}_{-0.032}$ & $50.19\pm0.23$ & $1.132\pm0.085$ & -- & -- & --\\
 & ZBQGHF\tnote{g} & $0.312^{+0.012}_{-0.013}$ & $0.716^{+0.052}_{-0.046}$ & $-0.028\pm0.045$ & -- & -- & $69.88\pm1.10$ & $0.412^{+0.025}_{-0.032}$ & $50.19\pm0.23$ & $1.131\pm0.084$ & $0.231\pm0.004$ & $0.637\pm0.009$ & $7.144\pm0.270$\\
\\
Flat XCDM & QSO-Flux & $0.496^{+0.192}_{-0.069}$ & -- & -- & $<-1.042$\tnote{h} & -- & -- & -- & -- & -- & $0.231\pm0.004$ & $0.624\pm0.011$ & $7.508\pm0.326$ \\
 & QGH\tnote{d} & $0.322^{+0.062}_{-0.044}$ & -- & -- & $-1.379^{+0.361}_{-0.375}$ & -- & $72.00^{+1.99}_{-1.98}$ & $0.412^{+0.026}_{-0.032}$ & $50.20\pm0.24$ & $1.130\pm0.085$ & -- & -- & --\\
 & QGHF\tnote{e} & $0.358^{+0.040}_{-0.038}$ & -- & -- & $<-1.100$ & -- & $72.14\pm1.91$ & $0.411^{+0.026}_{-0.031}$ & $50.20\pm0.23$ & $1.125\pm0.084$ & $0.231\pm0.004$ & $0.633^{+0.009}_{-0.010}$ & $7.268^{+0.287}_{-0.288}$\\
 & ZBQGH\tnote{f} & $0.313^{+0.014}_{-0.015}$ & -- & -- & $-1.050^{+0.090}_{-0.081}$ & -- & $69.85^{+1.42}_{-1.55}$ & $0.412^{+0.026}_{-0.032}$ & $50.19\pm0.24$ & $1.134\pm0.085$ & -- & -- & -- \\
 & ZBQGHF\tnote{g} & $0.315^{+0.013}_{-0.015}$ & -- & -- & $-1.052^{+0.091}_{-0.081}$ & -- & $69.85\pm1.48$ & $0.413^{+0.026}_{-0.032}$ & $50.19\pm0.24$ & $1.133\pm0.086$ & $0.231\pm0.004$ & $0.637\pm0.009$ & $7.135^{+0.270}_{-0.271}$\\
\\
Non-flat XCDM & QSO-Flux & $0.515^{+0.184}_{-0.050}$ & -- & $-0.310^{+0.137}_{-0.289}$ & $<-0.294$ & -- & -- & -- & -- & -- & $0.231^{+0.004}_{-0.005}$ & $0.615\pm0.013$ & $7.817^{+0.398}_{-0.400}$ \\
 & QGH\tnote{d} & $0.303^{+0.073}_{-0.058}$ & -- & $-0.044^{+0.193}_{-0.217}$ & $-1.273^{+0.501}_{-0.321}$ & -- & $71.77\pm2.02$ & $0.413^{+0.026}_{-0.031}$ & $50.20\pm0.24$ & $1.129\pm0.085$ & -- & -- & --\\
 & QGHF\tnote{e} & $0.367^{+0.059}_{-0.048}$ & -- & $-0.146^{+0.143}_{-0.147}$ & $-1.433^{+0.241}_{-0.493}$ & -- & $72.27^{+2.01}_{-1.99}$ & $0.413^{+0.026}_{-0.032}$ & $50.21\pm0.24$ & $1.116\pm0.085$ & $0.231\pm0.004$ & $0.629^{+0.011}_{-0.010}$ & $7.382\pm0.321$\\
 & ZBQGH\tnote{f} & $0.310\pm0.014$ & -- & $-0.024^{+0.092}_{-0.093}$ & $-1.019^{+0.202}_{-0.099}$ & -- & $69.63^{+1.45}_{-1.62}$ & $0.412^{+0.026}_{-0.031}$ & $50.19\pm0.23$ & $1.132\pm0.085$ & -- & -- & -- \\
 & ZBQGHF\tnote{g} & $0.314^{+0.014}_{-0.015}$ & -- & $-0.060^{+0.096}_{-0.095}$ & $-0.968^{+0.184}_{-0.087}$ & -- & $69.43^{+1.43}_{-1.63}$ & $0.412^{+0.026}_{-0.032}$ & $50.19\pm0.24$ & $1.130\pm0.085$ & $0.231\pm0.004$ & $0.636^{+0.009}_{-0.010}$ & $7.182^{+0.278}_{-0.281}$\\
\bottomrule
\end{tabular}}
\begin{tablenotes}[flushleft]
\item [c] km s$^{-1}$ Mpc$^{-1}$.
\item [d] HIIG + QSO-AS + GRB.
\item [e] HIIG + QSO-AS + GRB + QSO-Flux.
\item [f] $H(z)$ + BAO + HIIG + QSO-AS + GRB.
\item [g] $H(z)$ + BAO + HIIG + QSO-AS + GRB + QSO-Flux.
\item [h] This is the 1$\sigma$ limit. The $2\sigma$ limit is set by the prior, and is not shown here.
\end{tablenotes}
\end{threeparttable}
\end{sidewaystable*}

\end{subappendices}

%% file: chapter9.tex
\cleardoublepage

\chapter{Using Pantheon and DES supernova, baryon acoustic oscillation, and Hubble parameter data to constrain the Hubble constant, dark energy dynamics, and spatial curvature}
\chaptermark{SN Ia, BAO, and H(z) constraints}

\newcommand{\Om}{\Omega_{\rm m_0}}
\newcommand{\Ok}{\Omega_{\rm k_0}}
\newcommand{\och}{\Omega_{\rm c_0}\!h^2}
\newcommand{\obhs}{$\Omega_{\rm b_0}\!h^2$}
\newcommand{\ochs}{$\Omega_{\rm c_0}\!h^2$}

\label{Chapter9}

This chapter is based on \cite{Cao_Ryan_Ratra_2021}. Figures and tables by Shulei Cao, from analyses conducted independently by Shulei Cao and Joseph Ryan.

\section{Introduction} \label{sec:intro}

Many observational data sets have been used to place constraints on the parameters of cosmological models, such as the equation of state parameter ($w$) of dark energy.\footnote{For observational constraints on spatial curvature see \cite{Farooq_Mania_Ratra_2015}, \cite{Chen_et_al_2016}, \cite{rana_jain_mahajan_mukherjee_2017}, \cite{Ooba_Ratra_Sugiyama_2017_NFLCDM, Ooba_Ratra_Sugiyama_2017_NFpCDM, Ooba_Ratra_Sugiyama_2017_NFXCDM}, \cite{60}, \cite{Park_Ratra_2018_FpCDM_NFpCDM, Park_Ratra_2018_FXCDM_NFXCDM, Park_Ratra_2018_FLCDM, park_ratra_2020}, \cite{wei_2018}, \cite{DES_2019}, \cite{handley_2019a}, \cite{jesus_etal_2019}, \cite{li_etal_2020}, \cite{geng_etal_2020}, \cite{kumar_etal_2020}, \cite{efstathiou_gratton_2020}, \cite{divalentino_etal_2020}, \cite{divalentino_etal_2020b}, \cite{gao_etal_2020}, \cite{Abbassi_2020}, \cite{Yang_2020}, \cite{Agudelo_Ruiz_2020}, \cite{Velasquez-Toribio_2020}, \cite{Vagnozzi_2020a,Vagnozzi_2020b}, and references therein.} \footnote{For observational constraints on the \pcdm\ model see \cite{yashar_et_al_2009}, \cite{Samushia_2010}, \cite{20}, \cite{Avsajanishvili_2015}, \cite{Sola_etal_2017}, \cite{Sola_perez_gomez_2018,sola_gomez_perez_2019}, \cite{32}, \cite{Ooba_Ratra_Sugiyama_2017_NFpCDM,Ooba_Ratra_Sugiyama_2018_FpCDM}, \cite{sangwan_tripathi_jassal_2018}, \cite{singh_etal_2019}, \cite{Khadka_2020a,KhadkaandRatra_2020,Khadka_Ratra_2020,Khadka_2020d}, \cite{Urena-Lopez_2020}, and references therein.} In Chapter \ref{Chapter8}, we used Hubble parameter ($H(z)$), baryon acoustic oscillation (BAO), quasar angular size (QSO), quasar X-ray and UV flux, \hii\ starburst galaxy (\hiig), and gamma-ray burst (GRB) data to constrain this parameter (among others). The tightest constraints on $w$, we found, come from low-redshift $H(z)$ (cosmic chronometer) and BAO (standard ruler) data, with the standard candle data (\hiig\ and GRB) giving very broad constraints. In this paper we combine measurements of the distances to 1255 Type Ia supernovae (SNe Ia) with our set of $H(z)$ and BAO data (along with QSO and \hiig\ observations) to obtain tight cosmological parameter constraints.

The usefulness of SN Ia data to cosmology is well-known. SN Ia measurements revealed the accelerated expansion of the Universe over twenty years ago, and they are employed today to place constraints on cosmological parameters and to break parameter degeneracies. Over this time period, the sample size of SN Ia distance measurements has grown considerably, and the analysis and mitigation of systematic uncertainties has improved \citep{DES_2019c, DES_2019d}. Supernovae are therefore a reasonably empirically well-understood cosmological probe\footnote{Though the relatively simpler physics underlying cosmic microwave background (CMB) anisotropies and BAO makes those probes better understood than SNe Ia.}, and so can be used to obtain reliable constraints on cosmological model parameters.

In previous chapters we relied on CMB-derived values of the baryon density\footnote{Here $\Omega_{\rm b_0}$ is the baryon density parameter and $h=H_0/(100\ \rm{km \ s^{-1} \ Mpc^{-1}})$.} $\Omega_{b0}h^2$ in order to compute the size of the sound horizon $r_{s}$. The size of the sound horizon is needed to calibrate the BAO scale (see Table \ref{tab:ch9_BAO}), so the constraints we derived from our BAO measurements were indirectly dependent on CMB physics. \cite{Park_Ratra_2018_FpCDM_NFpCDM, Park_Ratra_2018_FXCDM_NFXCDM, Park_Ratra_2018_FLCDM} computed $\Omega_{b0}h^2$ within each of the six models we study (namely flat/non-flat \lcdm, flat/non-flat XCDM, and flat/non-flat \pcdm) from CMB data using primordial energy density fluctuation power spectra $P(k)$ appropriate for flat and curved geometries \citep{Lucchin_1985, ratra_1989,ratra_2017,ratra_peebles_1995}. Other power spectra have been considered in the non-flat case \citep{Lesgourgues_2014,Bonga_2016,handley_2019,Thavanesan_2021}. Since we do not make use of $P(k)$, the controversy associated with $P(k)$ in non-flat models is avoided in our analyses here.

The constraints from $H(z)$ + BAO data and from SN Ia data are not inconsistent, and so these data can be jointly used to constrain cosmological parameters. \cite{park_ratra_2019b} used $H(z)$, BAO, and Pantheon SN Ia apparent magnitude (SN-Pantheon) measurements in such a joint analysis. Here we use a more recent BAO data compilation and new DES-3yr binned SN Ia apparent magnitude (SN-DES) data. We find for all combinations of data we study here that all or almost all of the favored parameter space corresponds to currently accelerating cosmological expansion. The joint analysis of $H(z)$, BAO, and SN Ia data gives fairly model-independent determinations of the Hubble constant, $H_0=68.8\pm1.8\ \rm{km \ s^{-1} \ Mpc^{-1}}$, and the non-relativistic matter density parameter, $\Omega_{\rm m_0}=0.294\pm0.020$. The estimate of $H_0$ is in better agreement with the median statistics $H_0 = 68 \pm 2.8$ \hunit\ estimate of \cite{chenratmed} and the \cite{planck2018} estimate of $H_0 = 67.4 \pm 0.5$ \hunit\ than with the local $H_0 = 74.03 \pm 1.42$ \hunit\ measurement of \cite{riess_etal_2019}. The combined measurements are consistent with the spatially flat \lcdm\ model, but also favor some dark energy dynamics, as well as a little non-zero spatial curvature energy density. More restrictive constraints are derived when these data are combined with QSO and \hiig\ data.
\begin{comment}
This paper is organized as follows. Section \ref{sec:model} summarizes the models we analyze. In Section \ref{sec:data} the data used are introduced and our method of analyzing these data is described in Section \ref{sec:analysis}. We present our results in Section \ref{sec:results}, and our conclusions in Section \ref{sec:conclusion}.
\end{comment}
\section{Cosmological models}
\label{sec:model}

We seek to obtain constraints on the parameters of the flat and non-flat \lcdm, XCDM, and \pcdm\ models and to compare how well these models fit the observations we study (see Chapter \ref{Chapter3} for the details of these models). Our approach here differs from that of earlier chapters in that, instead of varying the non-relativistic matter density parameter $\om$ as a free parameter, we vary the baryonic ($\Omega_{b0}h^2$) and cold dark matter (\ochs) densities as free parameters, treating $\om$ as a derived parameter.\footnote{We do this to eliminate the dependence of the BAO data on CMB physics; see Section \ref{sec:data} for details.} 
\section{Data}
\label{sec:data}

In this paper, we use a combination of $H(z)$, BAO, SN-Pantheon, SN-DES, QSO, and \hiig\ data to constrain the cosmological models we study. 

The $H(z)$ data, compiled in Table \ref{tab:H(z)_data}, consist of 31 measurements spanning the redshift range $0.070 \leq z \leq 1.965$. The BAO data, which have been updated relative to Chapter \ref{Chapter7}, consist of 11 measurements spanning the redshift range $0.38 \leq z \leq 2.334$, listed in Table \ref{tab:ch9_BAO}. 

The SN-Pantheon data, compiled by \cite{scolnic_et_al_2018}, consist of 1048 SN Ia measurements spanning the redshift range $0.01<z<2.3$. The SN-DES data, compiled by \cite{DES_2019d}, consist of 20 binned measurements of 207 SN Ia measurements spanning the redshift range $0.015 \leq z \leq 0.7026$.

The QSO data, listed in Table 1 of \cite{Cao_et_al2017b}, consist of 120 measurements of the angular size
\begin{equation}
    \theta(z) = \frac{l_{\rm m}}{D_{A}(z)},
\end{equation}
spanning the redshift range $0.462 \leq z \leq 2.73$. $l_{\rm m}$ is the characteristic linear size of the quasars in the sample. This quantity is determined by using the Gaussian Process method to reconstruct the expansion history of the Universe from 24 cosmic chronometer measurements over $z < 1.2$. This $H(z)$ function is used to reconstruct the angular size distance $D_{A}(z)$, which can then be used to compute $l_{\rm m}$ given measurements $(\theta_{\rm obs}(z)$) of quasar angular sizes. QSO and $H(z)$ data are therefore somewhat correlated, but the error bars on the constraints derived from QSO data are so large that we do not believe this correlation to be an issue. 

The \hiig\ data consist of 107 low redshift ($0.0088 \leq z \leq 0.16417$) measurements, used in \cite{Chavez_2014} (recalibrated by \citealp{G-M_2019}), and 46 high redshift ($0.636427 \leq z \leq 2.42935$) measurements.

\begin{table}
\centering
\scriptsize
\begin{threeparttable}
\caption{BAO data.}\label{tab:ch9_BAO}
\setlength{\tabcolsep}{0.8mm}{
\begin{tabular}{lccc}
\toprule
$z$ & Measurement\tnote{a} & Value & Ref.\\
\midrule
$0.38$ & $D_M\left(r_{s,{\rm fid}}/r_s\right)$ & 1512.39 & \cite{Alam_et_al_2017}\tnote{b}\\
$0.38$ & $H(z)\left(r_s/r_{s,{\rm fid}}\right)$ & 81.2087 & \cite{Alam_et_al_2017}\tnote{b}\\
$0.51$ & $D_M\left(r_{s,{\rm fid}}/r_s\right)$ & 1975.22 & \cite{Alam_et_al_2017}\tnote{b}\\
$0.51$ & $H(z)\left(r_s/r_{s,{\rm fid}}\right)$ & 90.9029 & \cite{Alam_et_al_2017}\tnote{b}\\
$0.61$ & $D_M\left(r_{s,{\rm fid}}/r_s\right)$ & 2306.68 & \cite{Alam_et_al_2017}\tnote{b}\\
$0.61$ & $H(z)\left(r_s/r_{s,{\rm fid}}\right)$ & 98.9647 & \cite{Alam_et_al_2017}\tnote{b}\\
$0.122$ & $D_V\left(r_{s,{\rm fid}}/r_s\right)$ & $539\pm17$ & \cite{Carter_2018}\\
$0.81$ & $D_A/r_s$ & $10.75\pm0.43$ & \cite{DES_2019b}\\
$1.52$ & $D_V\left(r_{s,{\rm fid}}/r_s\right)$ & $3843\pm147$ & \cite{3}\\
$2.334$ & $D_M/r_s$ & 37.5 & \cite{duMas2020}\tnote{c}\\
$2.334$ & $D_H/r_s$ & 8.99 & \cite{duMas2020}\tnote{c}\\
\bottomrule
\end{tabular}}
\begin{tablenotes}[flushleft]
\item[a] $D_M$, $D_V$, $r_s$, $r_{s, {\rm fid}}$, $D_A$, and $D_M$ have units of Mpc, while $H(z)$ has units of \hunit.
\item[b] The six measurements from \cite{Alam_et_al_2017} are correlated; see equation (20) of \cite{Ryan_Chen_Ratra_2019} for their correlation matrix.
\item[c] The two measurements from \cite{duMas2020} are correlated; see equation \eqref{CovM1} below for their correlation matrix.
\end{tablenotes}
\end{threeparttable}
\end{table}

The covariance matrix $\textbf{C}$ for the BAO data, taken from \cite{Alam_et_al_2017}, is given by eq. (\ref{covmat}). For the BAO data from \cite{duMas2020}, the covariance matrix is
\begin{equation}
\label{CovM1}
    C =
    \begin{bmatrix}
    1.3225 & -0.1009 \\
    -0.1009 & 0.0380
    \end{bmatrix}.
\end{equation}
The scale of BAO measurements is set by the sound horizon ($r_{s}$) during the epoch of radiation drag. To compute this quantity, we use the approximate formula \citep{PhysRevD.92.123516}
\begin{equation}
\label{tab:ch9_sh}
    r_s=\frac{55.154\exp{[-72.3(\Omega_{\nu_0}h^2+0.0006)^2]}}{(\Omega_{b0}\!h^2)^{0.12807}(\Omega_{c0}h^2 + \Omega_{b0}h^2)^{0.25351}} \hspace{1mm}{\rm Mpc}.
\end{equation}
In earlier chapters we did not vary $\Omega_{b0}h^2$ as a free parameter. Instead we used CMB-derived, model-dependent values of $\Omega_{b0}h^2$ to compute $r_{s}$. Because we vary $\Omega_{b0}h^2$ as a free parameter in this chapter, our computations of the sound horizon (and therefore our calibration of the scale of our BAO measurements) are fully independent of CMB physics (at the cost of enlarging the parameter space and so somewhat weakening the constraints).

Following \cite{Conley_et_al_2011} and \cite{Deng_Wei_2018}, we define the theoretical magnitude of a supernova to be
\begin{equation}
\label{eq:m_th}
    m_{\rm th} = 5\log\mathcal{D}_{L}(z) + \mathcal{M},
\end{equation}
where $\mathcal{M}$ is a nuisance parameter to be marginalized over, and $\mathcal{D}_{L}(z)$ is
\begin{equation}
    \mathcal{D}_{L}(z) \equiv \left(1 + z_{\rm hel}\right) \int_0^{z_{\rm cmb}} \frac{d\tilde{z}}{E\left(\tilde{z}\right)}.
\end{equation}
In this equation, $z_{\rm hel}$ is the heliocentric redshift, and $z_{\rm cmb}$ is the CMB-frame redshift. In \cite{Conley_et_al_2011}, equation \ref{eq:D_L} is called the ``Hubble-constant free luminosity distance'', because $E(z)$ does not contain $H_0$. In our case, because we use $h$, $\Omega_{b0}h^2$, and $\Omega_{c0}h^2$ as free parameters, our expansion rate function (and thus our luminosity distance) depends on the Hubble constant. We therefore obtain weak constraints on $H_0$ from the supernova data, unlike \cite{Conley_et_al_2011} and \cite{Deng_Wei_2018} (see Section \ref{sec:results}, below).

\section{Data Analysis Methodology}
\label{sec:analysis}

We use the \textsc{python} module \textsc{emcee} \citep{Foreman-Mackey_Hogg_Lang_Goodman_2013} to maximize the likelihood functions, thereby determining the constraints on the free parameters (see Chapter \ref{Chapter7} for details about this method). In our analyses here the priors on the cosmological parameters are different from zero (and flat) over the ranges $0.005 \leq \Omega_{b0}h^2 \leq 0.1$, $0.001 \leq \Omega_{c0}h^2 \leq 0.99$, $0.2 \leq h \leq 1.0$, $-3 \leq w_{\rm X} \leq 0.2$, $-0.7 \leq \Omega_{k0} \leq 0.7$, and $0 < \alpha \leq 10$. $\Omega_{m0}$ is a derived parameter and depends on $h$.

The likelihood functions of $H(z)$, BAO, \hiig, and QSO data are described in Chapters \ref{Chapter7} and \ref{Chapter8}. For the SN Ia (SN-Pantheon and SN-DES) data, the likelihood function is
\be
\label{eq:LH_SN}
    \mathcal{L}_{\rm SN}= e^{-\chi^2_{\rm SN}/2},
\ee
where, as in \cite{park_ratra_2019b}, $\chi^2_{\rm SN}$ takes the form of equation (C1) in Appendix C of \cite{Conley_et_al_2011} with $\mathcal{M}$ being marginalized. The covariance matrices of the SN Ia data, $\textbf{C}_{\rm SN}$ are the sum of the diagonal statistical uncertainty covariance matrices, $C_{\rm stat} = {\rm diag}(\sigma^2_{\rm SN})$, and the systematic uncertainty covariance matrices, $C_{\rm sys}$: $C_{\rm SN} = C_{\rm stat} + C_{\rm sys}$.\footnote{Note that the covariance matrices for the SN-DES data are the ones described in eq. (18) of \cite{DES_2019d}.} $\sigma_{\rm SN}$ are the SN Ia statistical uncertainties.

As in Chapter \ref{Chapter8}, we use the Akaike Information Criterion ($AIC$) and the Bayesian Information Criterion ($BIC$) to compare the quality of models with different numbers of parameters, where
\be
\label{eq:ch9_AIC}
    AIC=-2\ln \mathcal{L}_{\rm max} + 2n,
\ee
and
\be
\label{eq:ch9_BIC}
    BIC=-2\ln \mathcal{L}_{\rm max} + n\ln N.
\ee
In the preceding equations, $\mathcal{L}_{\rm max}$, $n$, and $N$ are the maximum value of the considered likelihood function, the number of free parameters in the given model, and the number of used data points (e.g., for SN-Pantheon $N=1048$), respectively.

\section{Results}
\label{sec:results}

\begin{figure*}
\centering
    \includegraphics[width=3.5in,height=3.5in]{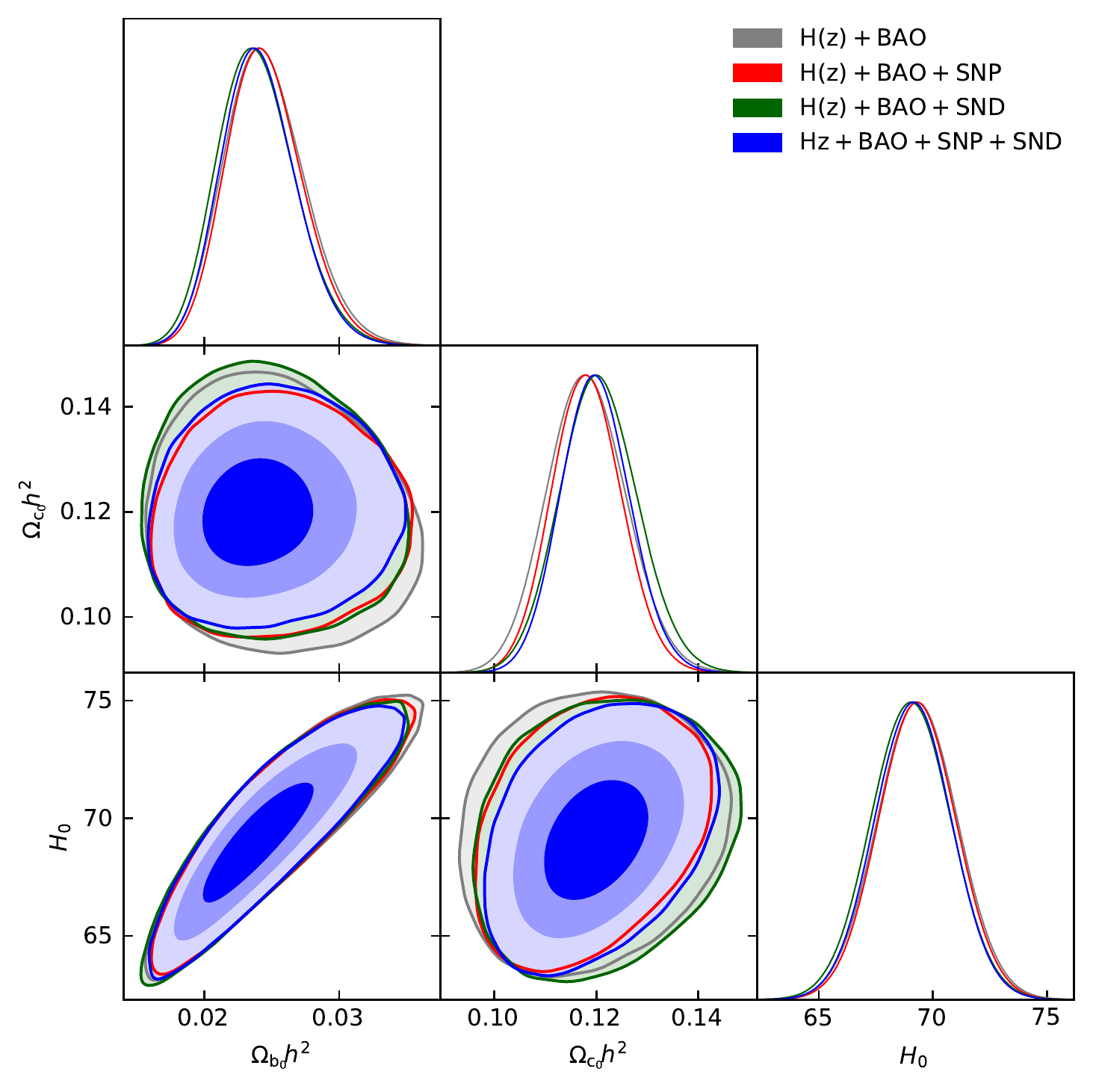}
    \includegraphics[width=3.5in,height=3.5in]{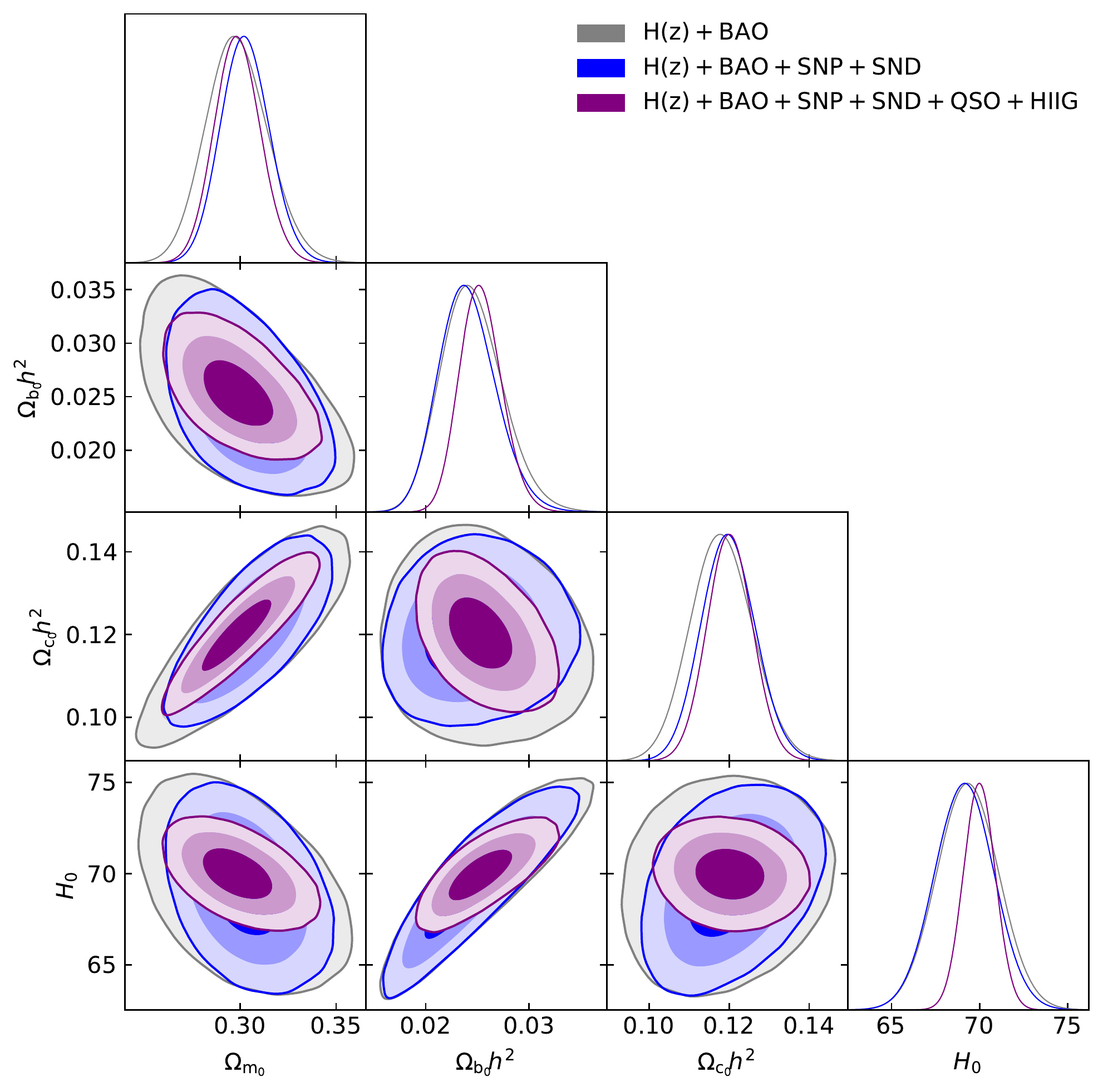}\\
\caption[1$\sigma$, 2$\sigma$, and 3$\sigma$ confidence contours for flat \lcdm.]{1$\sigma$, 2$\sigma$, and 3$\sigma$ confidence contours for flat \lcdm, where the right panel is the comparison including derived cosmological matter density parameter $\Omega_{m0}$. In all cases, the favored parameter space is associated with currently-accelerating cosmological expansion.}
\label{ch9_fig1}
\end{figure*}

\begin{figure*}
\centering
    \includegraphics[width=3.5in,height=3.5in]{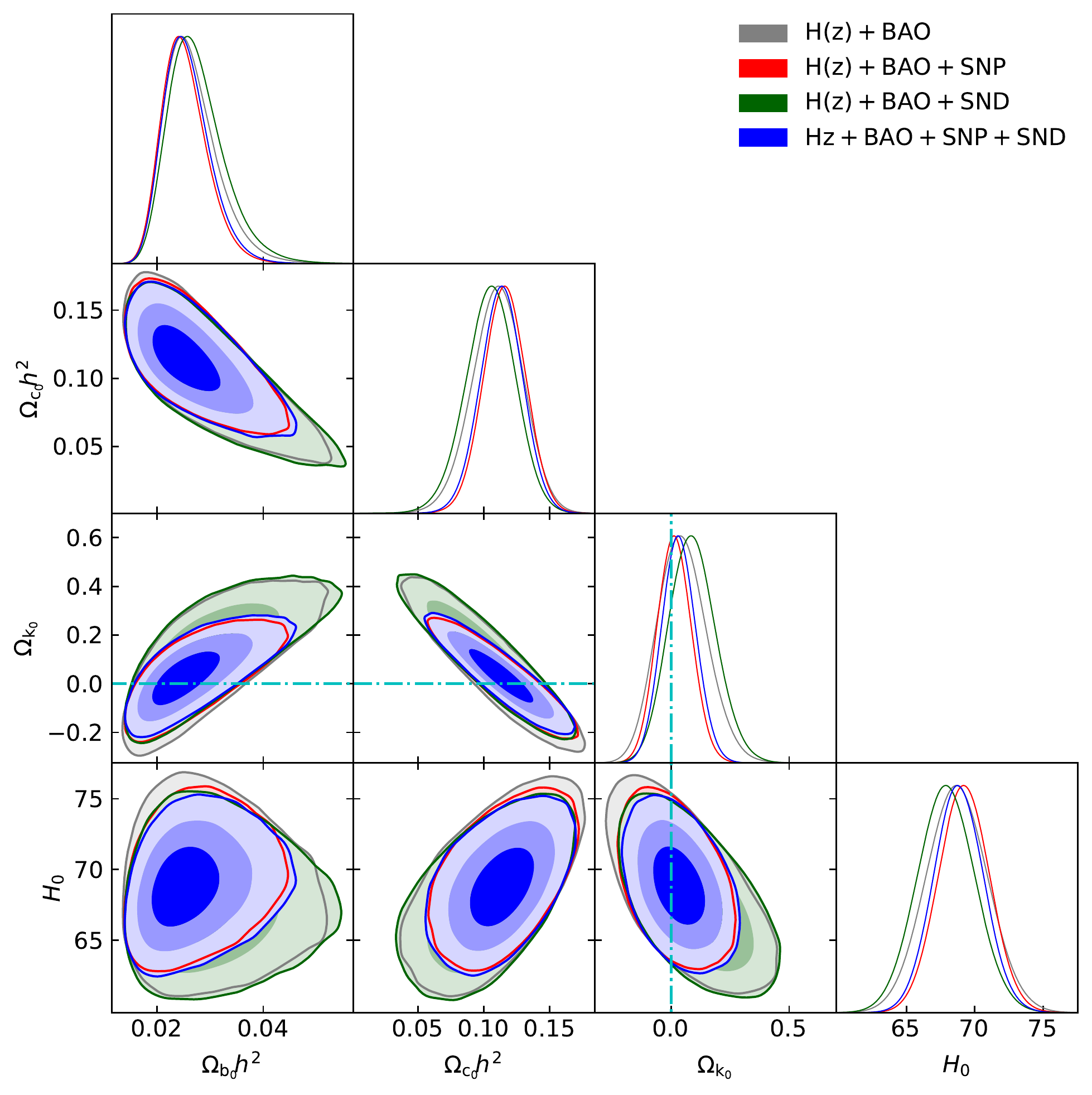}
    \includegraphics[width=3.5in,height=3.5in]{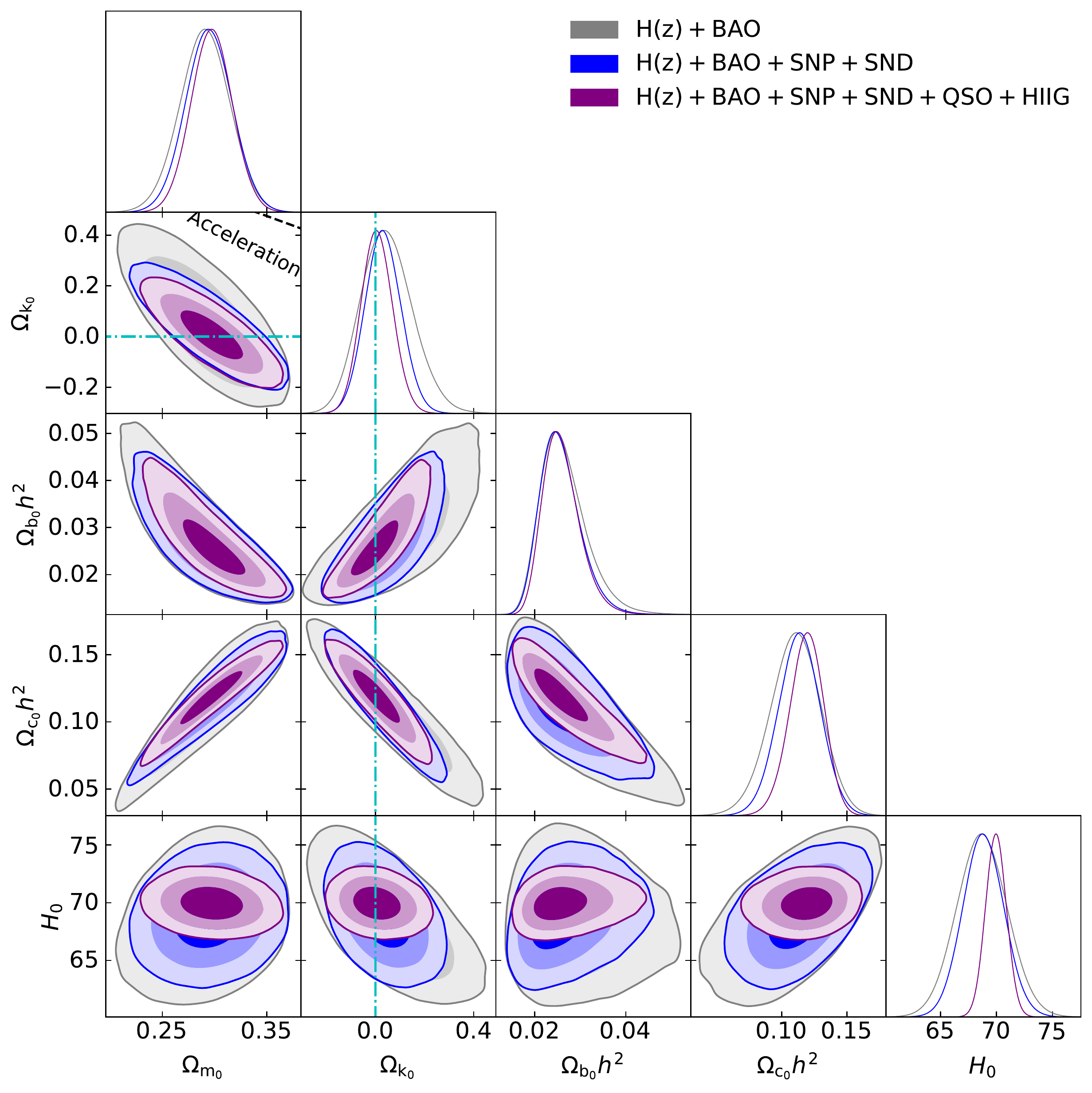}\\
\caption[1$\sigma$, 2$\sigma$, and 3$\sigma$ confidence contours for non-flat \lcdm.]{Same as Fig. \ref{ch9_fig1} but for non-flat \lcdm, where the cyan dash-dot lines represent the flat \lcdm\ case, with closed spatial hypersurfaces either below or to the left. The black dotted line in the right subpanel is the zero-acceleration line, which divides the parameter space into regions associated with currently-accelerating (below left) and currently-decelerating (above right) cosmological expansion. In all cases, the favored parameter space is associated with currently-accelerating cosmological expansion.}
\label{ch9_fig2}
\end{figure*}

\begin{figure*}
\centering
    \includegraphics[width=3.5in,height=3.5in]{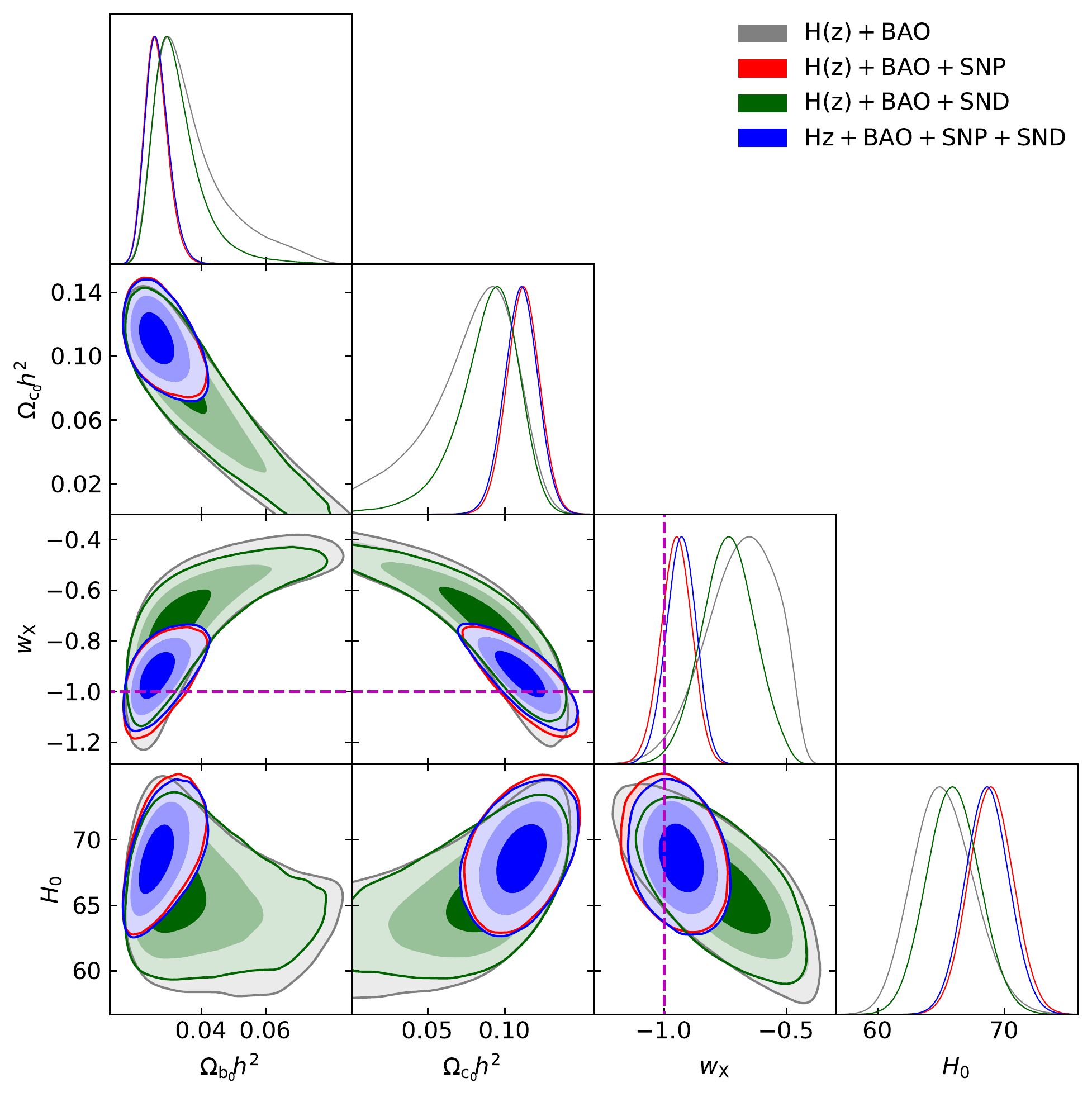}
    \includegraphics[width=3.5in,height=3.5in]{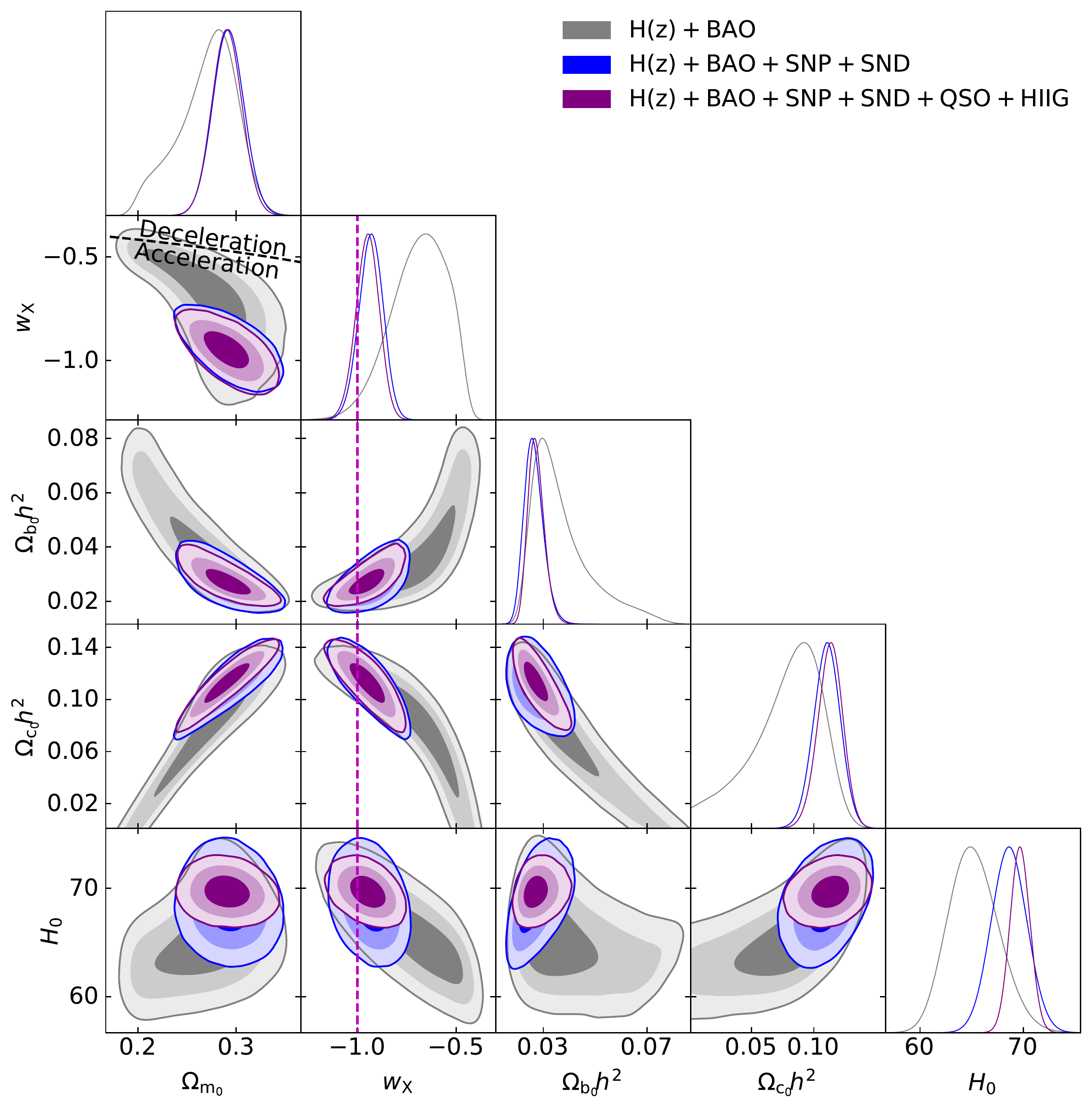}\\
\caption[1$\sigma$, 2$\sigma$, and 3$\sigma$ confidence contours for flat XCDM.]{1$\sigma$, 2$\sigma$, and 3$\sigma$ confidence contours for flat XCDM. The black dotted line in the right panel is the zero-acceleration line, which divides the parameter space into regions associated with currently-accelerating (below) and currently-decelerating (above) cosmological expansion. In all cases, almost all of the favored parameter space is associated with currently-accelerating cosmological expansion. The magenta lines denote $w_{\rm X}=-1$, i.e. the flat \lcdm\ model.}
\label{ch9_fig3}
\end{figure*}

\begin{figure*}
\centering
    \includegraphics[width=3.5in,height=3.5in]{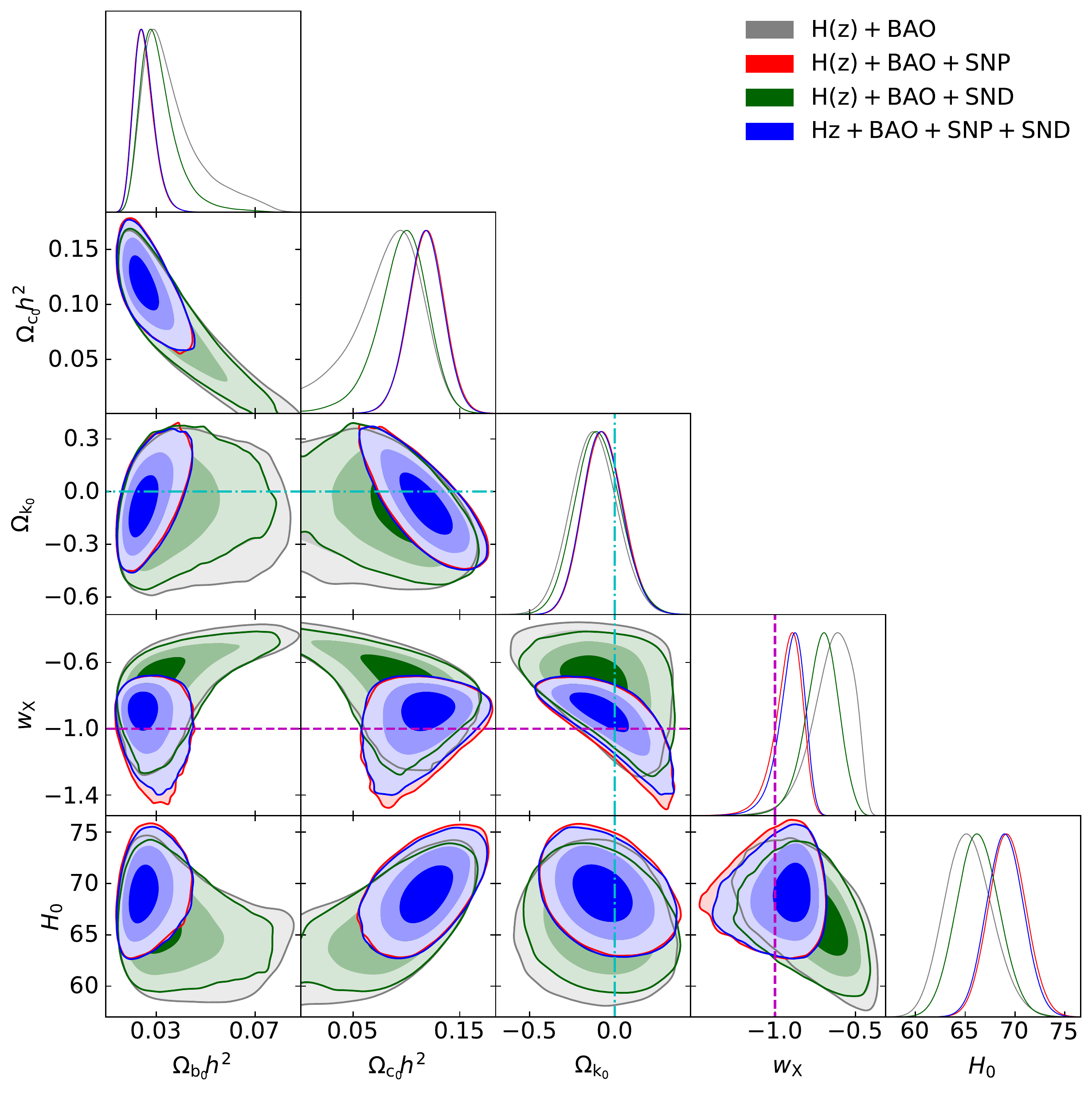}
    \includegraphics[width=3.5in,height=3.5in]{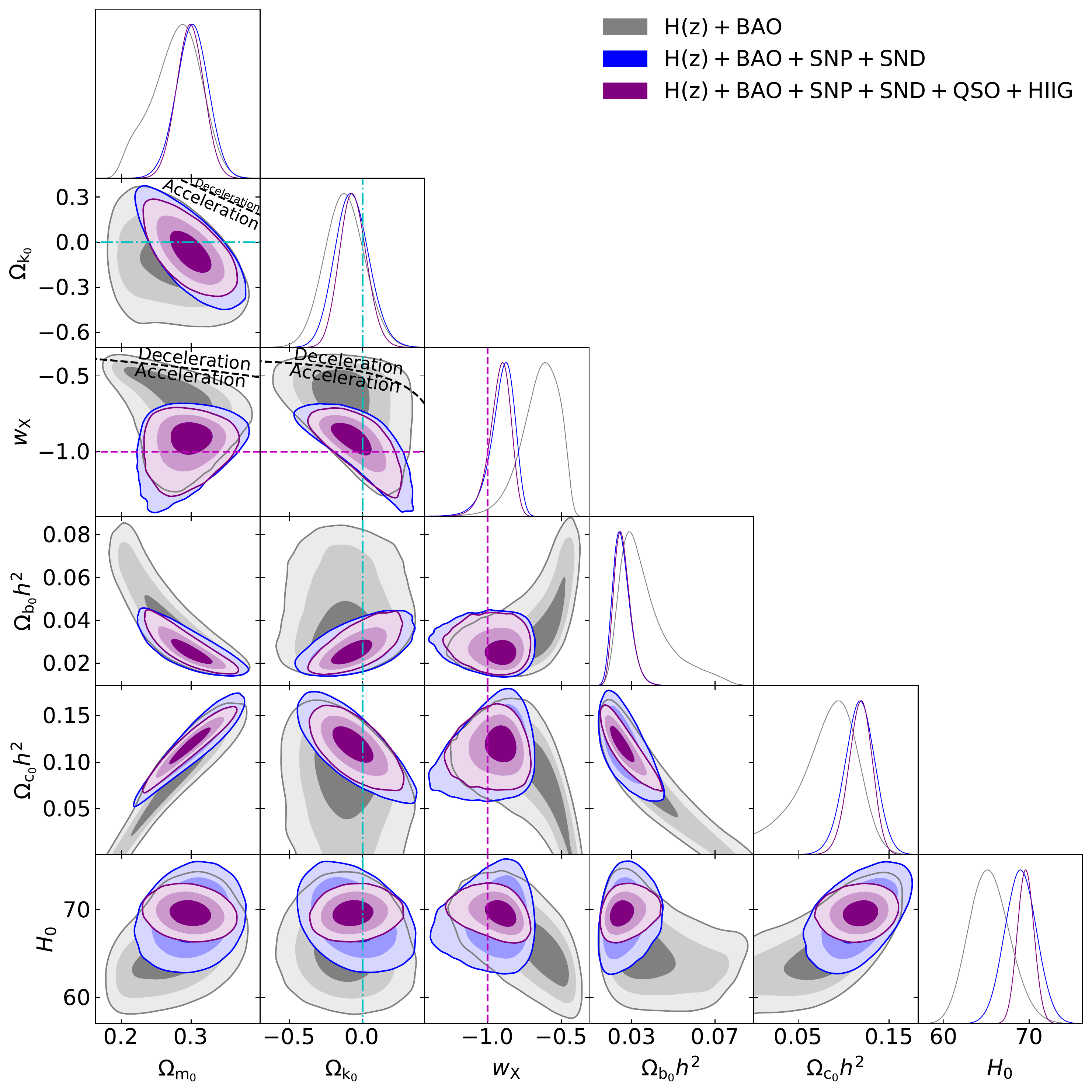}\\
\caption[1$\sigma$, 2$\sigma$, and 3$\sigma$ confidence contours for non-flat XCDM.]{Same as Fig. \ref{ch9_fig3} but for non-flat XCDM, where the zero acceleration lines in each of the three subpanels of the right panel are computed for the third cosmological parameter set to the $H(z)$ + BAO data best-fitting values listed in Table \ref{tab:ch9_BFP}. Currently-accelerating cosmological expansion occurs below these lines. The cyan dash-dot lines represent the flat XCDM case, with closed spatial hypersurfaces either below or to the left. In all cases, almost all of the favored parameter space is associated with currently-accelerating cosmological expansion. The magenta lines indicate $w_{\rm X} = -1$, i.e. the non-flat \lcdm\ model.}
\label{ch9_fig4}
\end{figure*}

\begin{figure*}
\centering
    \includegraphics[width=3.5in,height=3.5in]{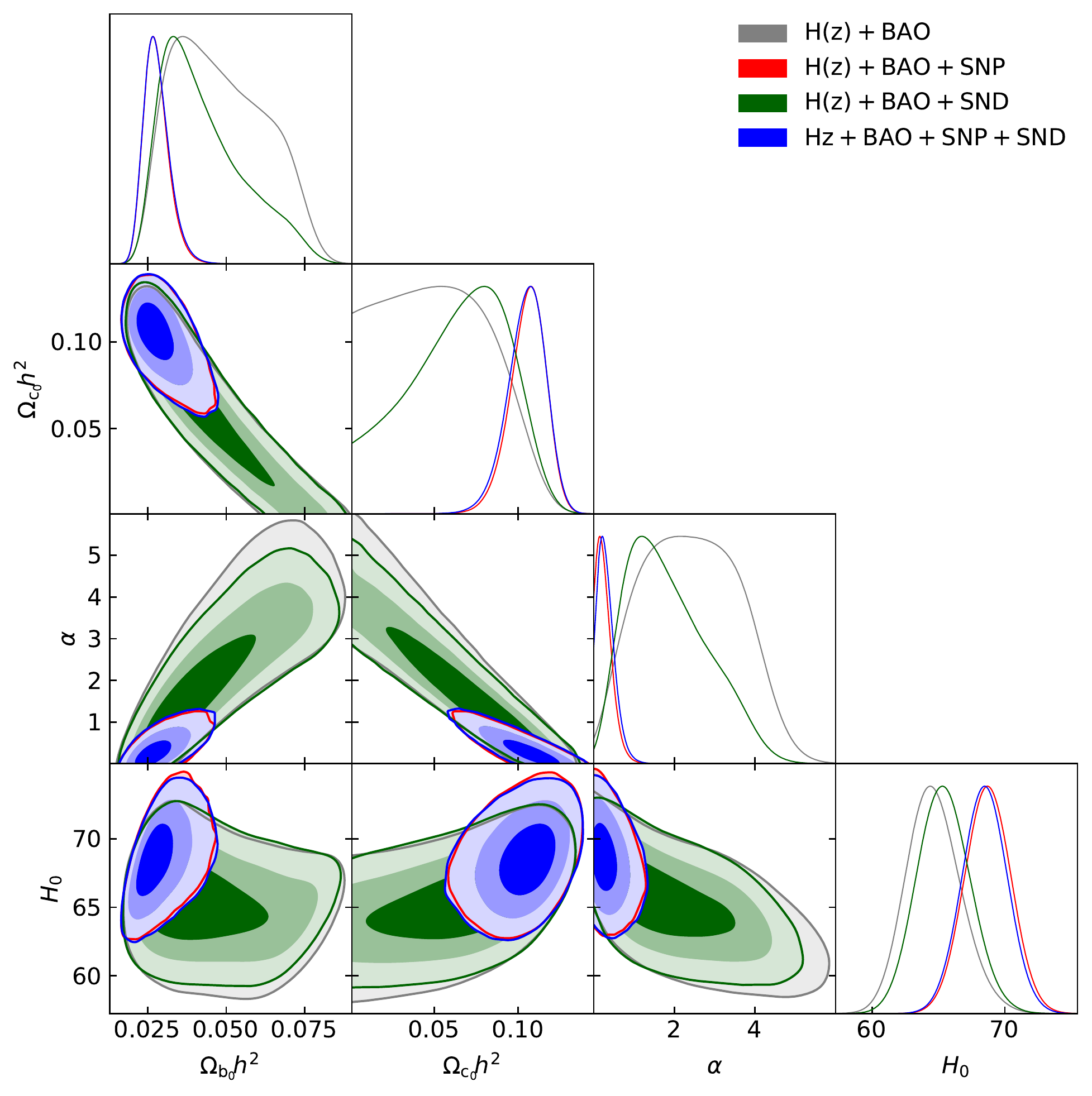}
    \includegraphics[width=3.5in,height=3.5in]{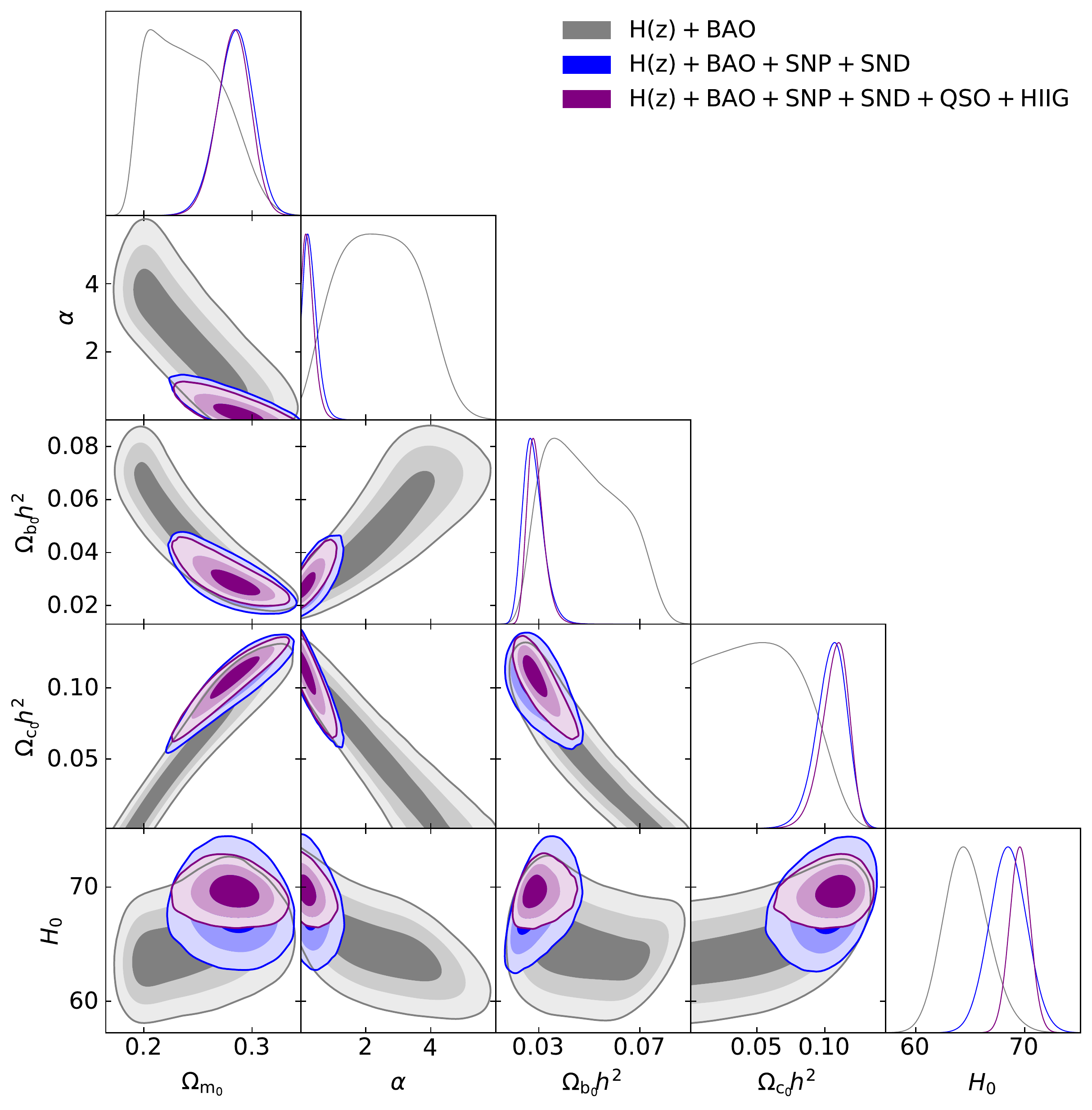}\\
\caption[1$\sigma$, 2$\sigma$, and 3$\sigma$ confidence contours for flat \pcdm.]{1$\sigma$, 2$\sigma$, and 3$\sigma$ confidence contours for flat \pcdm. In all cases, the favored parameter space is associated with currently-accelerating cosmological expansion. The $\alpha = 0$ axis is the flat \lcdm\ model.}
\label{ch9_fig5}
\end{figure*}

\begin{figure*}
\centering
    \includegraphics[width=3.5in,height=3.5in]{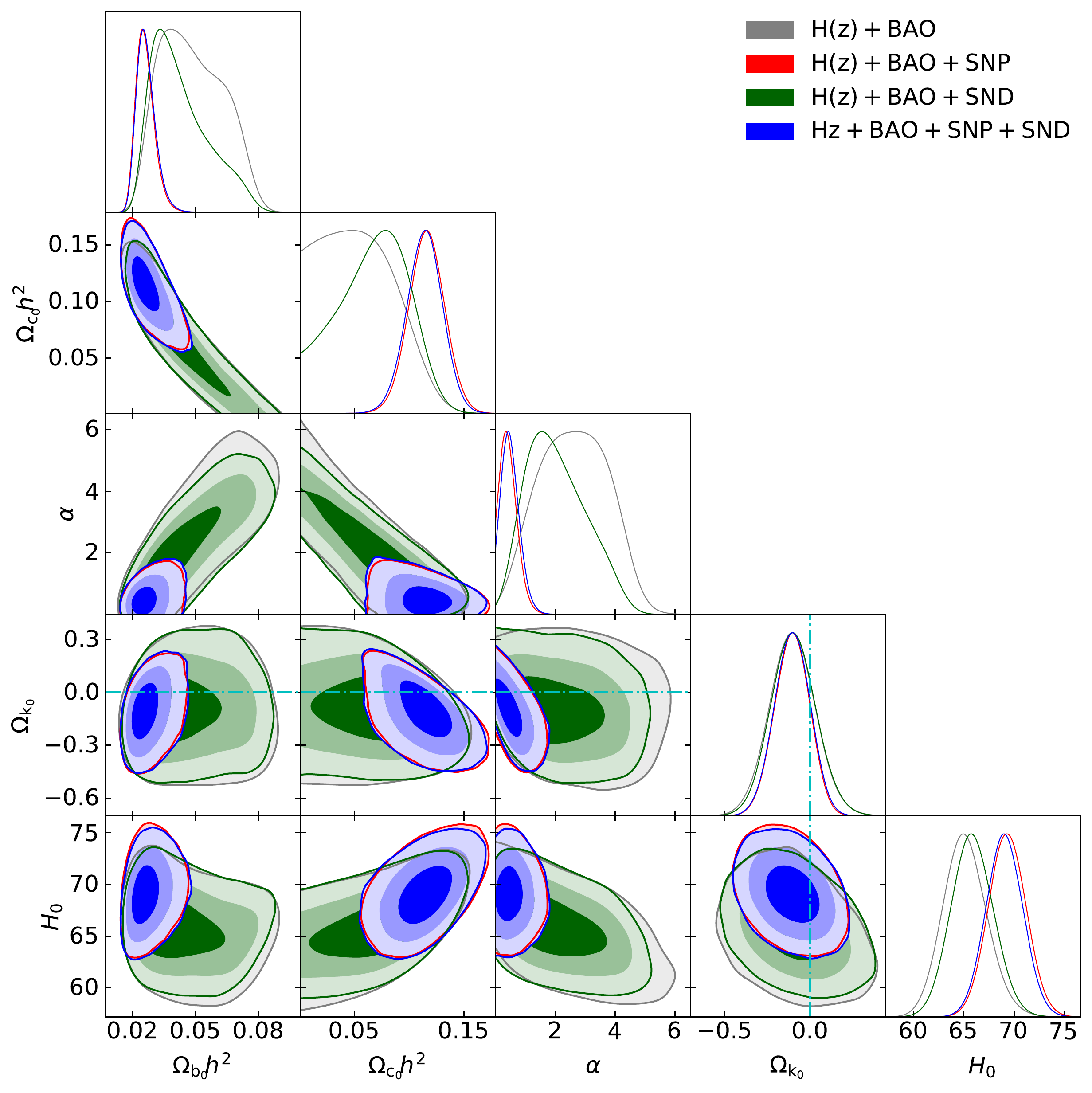}
    \includegraphics[width=3.5in,height=3.5in]{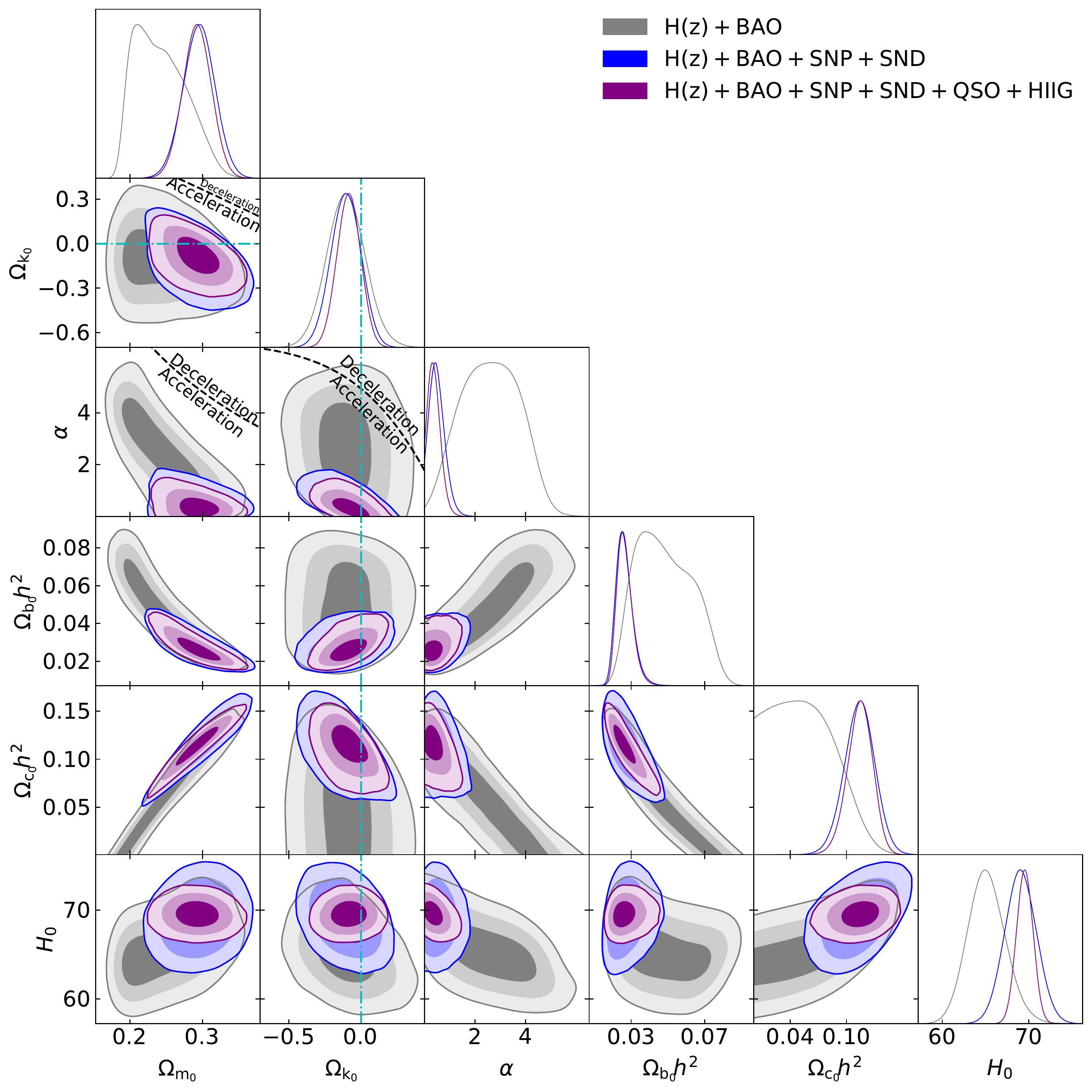}\\
\caption[1$\sigma$, 2$\sigma$, and 3$\sigma$ confidence contours for non-flat \pcdm.]{Same as Fig. \ref{ch9_fig5} but for non-flat \pcdm, where the zero-acceleration lines in each of the sub-panels of the right panel are computed for the third cosmological parameter set to the $H(z)$ + BAO data best-fitting values listed in Table \ref{tab:ch9_BFP}. Currently-accelerating cosmological expansion occurs below these lines. In all cases, almost all of the favored parameter space is associated with currently-accelerating cosmological expansion. The cyan dash-dot lines represent the flat \pcdm\ case, with closed spatial geometry either below or to the left. The $\alpha = 0$ axis is the non-flat \lcdm\ model.}
\label{ch9_fig6}
\end{figure*}

The posterior one-dimensional (1D) probability distributions and two-dimensional (2D) confidence regions of the cosmological parameters for the six flat and non-flat models are shown in Figs. \ref{ch9_fig1}--\ref{ch9_fig6}, in gray ($H(z)$+BAO), red ($H(z)$ + BAO + SN-Pantheon, ZBP), green ($H(z)$ + BAO + SN-DES, ZBD), blue ($H(z)$ + BAO + SN-Pantheon + SN-DES, ZBPD), and purple ($H(z)$ + BAO + SN-Pantheon + SN-DES + QSO + \hiig, ZBPDQH). We list the unmarginalized best-fitting parameter values, as well as the corresponding $\chi^2$, $AIC$, $BIC$, and degrees of freedom $\nu$ ($\nu \equiv N - n$) for all models and data combinations, in Table \ref{tab:ch9_BFP}. The marginalized best-fitting parameter values and uncertainties ($\pm 1\sigma$ error bars or $2\sigma$ limits), for all models and data combinations, are listed in Table \ref{tab:ch9_1d_BFP}.\footnote{The \textsc{python} package \textsc{getdist} \citep{Lewis_2019} is used to analyze the samples.}

\subsection{$H(z)$ + BAO, ZBP, and ZBD constraints}
\label{subsec:HzB}

The 1D marginalized $H(z)$ + BAO constraints on the cosmological parameters are listed in Table \ref{tab:ch9_1d_BFP}. These are (slightly) different from the ones obtained by \cite{Khadka_2020d}, because of the different treatments of both the prior ranges and the coefficient $\kappa$ in the $\phi$CDM models.\footnote{We treated $\kappa$ as a derived constant determined from the parameter $\alpha$ (see eq. \ref{eq:ch3_kappa_def}), while \cite{Khadka_2020d} treated it as a constant derived from the energy budget equation.}

The $H(z)$, BAO, and SN-Pantheon data combinations have previously been studied \citep{park_ratra_2019b}. Relative to that analysis, we use the updated BAO data, shown in Table \ref{tab:ch9_BAO}, in our analysis here. In the ZBP case, we find that the determinations of $\Omega_{k0}$ are more consistent with flat spatial hypersurfaces than what \cite{park_ratra_2019b} found and dark energy dynamics favors less deviation from a cosmological constant in the XCDM cases, while favoring a somewhat stronger deviation from $\alpha=0$ in the non-flat $\phi$CDM case.

Because the $H(z)$, BAO, and SN-DES constraints are consistent across all six of the models we study, we also perform a joint analysis of these data to determine ZBD constraints. Relative to the ZBP constraints, the measured values of $\Omega_{b0}h^2$, $\Omega_{c0}h^2$, and $\Omega_{m0}$ are a little higher, lower, and lower (except for flat $\Lambda$CDM) than those values measured from the ZBP case, respectively. Given the error bars, these differences are not statistically significant. The measured values of $H_0$ are lower than those for the ZBP case. The non-flat XCDM and $\phi$CDM models favor more and less closed geometry than in the ZBP case. The non-flat $\Lambda$CDM model favors more open geometry than in the ZBP case. The constraints for all three non-flat models are consistent with spatially flat hypersurfaces. The fits to the ZBD data produce stronger evidence for dark energy dynamics than the fits to the ZBP data.

\begin{table*}
\centering
\resizebox{\columnwidth}{!}{%
\begin{threeparttable}
\caption{Unmarginalized best-fitting parameter values for all models from various combinations of data.}\label{tab:ch9_BFP}
\setlength{\tabcolsep}{1.5mm}{
\begin{tabular}{lcccccccccccc}
\toprule
Model & Data set & $\Omega_{\mathrm{b_0}}\!h^2$ & $\Omega_{\mathrm{c_0}}\!h^2$ & $\Omega_{\mathrm{m_0}}$ & $\Omega_{\mathrm{k_0}}$ & $w_{\mathrm{X}}$ & $\alpha$ & $H_0$\tnote{a} & $\chi^2$ & $\nu$ & $AIC$ & $BIC$ \\
\midrule
Flat \lcdm & $H(z)$ + BAO & 0.0240 & 0.1179 & 0.299 & -- & -- & -- & 69.11 & 23.64 & 39 & 29.64 & 34.86\\
 & ZBP\tnote{b} & 0.0240 & 0.1180 & 0.299 & -- & -- & -- & 69.10 & 1053.22 & 1087 & 1059.22 & 1074.21\\
 & ZBD\tnote{c} & 0.0234 & 0.1203 & 0.305 & -- & -- & -- & 68.82 & 50.83 & 59 & 56.83 & 63.21\\
 & ZBPD\tnote{d} & 0.0236 & 0.1196 & 0.303 & -- & -- & -- & 68.91 & 1080.46 & 1107 & 1086.46 & 1101.50\\
 & ZBPDQH\tnote{e} & 0.0251 & 0.1203 & 0.299 & -- & -- & -- & 69.92 & 1844.99 & 1380 & 1850.99 & 1866.69\\
\\
Non-flat \lcdm & $H(z)$ + BAO & 0.0248 & 0.1136 & 0.294 & 0.026 & -- & -- & 68.75 & 23.58 & 38 & 31.58 & 38.53\\
 & ZBP\tnote{b} & 0.0241 & 0.1172 & 0.298 & 0.004 & -- & -- & 69.06 & 1053.22 & 1086 & 1061.22 & 1081.20\\
 & ZBD\tnote{c} & 0.0258 & 0.1081 & 0.292 & 0.071 & -- & -- & 67.92 & 50.28 & 58 & 58.28 & 66.79\\
 & ZBPD\tnote{d} & 0.0245 & 0.1150 & 0.297 & 0.023 & -- & -- & 68.68 & 1080.35 & 1106 & 1088.35 & 1108.40\\
 & ZBPDQH\tnote{e} & 0.0249 & 0.1209 & 0.300 & $-0.004$ & -- & -- & 69.93 & 1844.99 & 1379 & 1852.99 & 1873.92\\
\\
Flat XCDM & $H(z)$ + BAO & 0.0323 & 0.0860 & 0.280 & -- & $-0.696$ & -- & 65.12 & 19.65 & 38 & 27.65 & 34.60\\
 & ZBP\tnote{b} & 0.0254 & 0.1120 & 0.292 & -- & $-0.951$ & -- & 68.72 & 1052.63 & 1086 & 1060.63 & 1080.61\\
 & ZBD\tnote{c} & 0.0300 & 0.0934 & 0.286 & -- & $-0.752$ & -- & 65.90 & 45.46 & 58 & 53.46 & 61.97\\
 & ZBPD\tnote{d} & 0.0256 & 0.1107 & 0.293 & -- & $-0.932$ & -- & 68.43 & 1079.23 & 1106 & 1087.23 & 1107.28\\
 & ZBPDQH\tnote{e} & 0.0268 & 0.1136 & 0.291 & -- & $-0.949$ & -- & 69.63 & 1844.27 & 1379 & 1852.27 & 1873.20\\
\\
Non-flat XCDM & $H(z)$ + BAO & 0.0302 & 0.0956 & 0.294 & $-0.155$ & $-0.650$ & -- & 65.55 & 18.31 & 37 & 28.31 & 37.00\\
 & ZBP\tnote{b} & 0.0234 & 0.1231 & 0.307 & $-0.103$ & $-0.895$ & -- & 69.25 & 1051.82 & 1085 & 1061.82 & 1086.79\\
 & ZBD\tnote{c} & 0.0277 & 0.1046 & 0.301 & $-0.136$ & $-0.711$ & -- & 66.45 & 44.34 & 57 & 54.34 & 64.98\\
 & ZBPD\tnote{d} & 0.0236 & 0.1220 & 0.307 & $-0.107$ & $-0.877$ & -- & 68.98 & 1078.36 & 1105 & 1088.36 & 1113.42\\
 & ZBPDQH\tnote{e} & 0.0242 & 0.1217 & 0.303 & $-0.092$ & $-0.900$ & -- & 69.54 & 1843.25 & 1378 & 1853.25 & 1879.41\\
\\
Flat $\phi$CDM & $H(z)$ + BAO & 0.0361 & 0.0758 & 0.264 & -- & -- & 1.484 & 65.30 & 19.48 & 38 & 27.48 & 34.43\\
 & ZBP\tnote{b} & 0.0260 & 0.1145 & 0.292 & -- & -- & 0.101 & 69.51 & 1051.46 & 1086 & 1059.46 & 1079.44\\
 & ZBD\tnote{c} & 0.0328 & 0.0860 & 0.273 & -- & -- & 1.061 & 66.16 & 45.17 & 58 & 53.17 & 61.68\\
 & ZBPD\tnote{d} & 0.0254 & 0.1102 & 0.292 & -- & -- & 0.168 & 68.35 & 1078.18 & 1106 & 1086.18 & 1106.22\\
 & ZBPDQH\tnote{e} & 0.0264 & 0.1135 & 0.290 & -- & -- & 0.132 & 69.57 & 1842.95 & 1379 & 1850.95 & 1871.88\\
\\
Non-flat $\phi$CDM & $H(z)$ + BAO & 0.0354 & 0.0811 & 0.269 & $-0.148$ & -- & 1.819 & 66.06 & 18.16 & 37 & 28.16 & 36.85\\
 & ZBP\tnote{b} & 0.0234 & 0.1225 & 0.305 & $-0.133$ & -- & 0.393 & 69.32 & 1050.31 & 1085 & 1060.31 & 1085.28\\
 & ZBD\tnote{c} & 0.0319 & 0.0933 & 0.282 & $-0.140$ & -- & 1.411 & 66.84 & 44.09 & 57 & 54.09 & 64.72\\
 & ZBPD\tnote{d} & 0.0256 & 0.1159 & 0.298 & $-0.080$ & -- & 0.377 & 69.09 & 1077.13 & 1105 & 1087.13 & 1112.19\\
 & ZBPDQH\tnote{e} & 0.0258 & 0.1155 & 0.293 & $-0.078$ & -- & 0.354 & 69.55 & 1842.00 & 1378 & 1852.00 & 1878.16\\
\bottomrule
\end{tabular}}
\begin{tablenotes}[flushleft]
\item [a] \hunit.
\item [b] $H(z)$ + BAO + SN-Pantheon.
\item [c] $H(z)$ + BAO + SN-DES.
\item [d] $H(z)$ + BAO + SN-Pantheon + SN-DES.
\item [e] $H(z)$ + BAO + SN-Pantheon + SN-DES + QSO + \hiig.
\end{tablenotes}
\end{threeparttable}%
}
\end{table*}

\subsection{$H(z)$, BAO, SN-Pantheon, and SN-DES (ZBPD) constraints}
\label{subsec:ZBPD}

The results of the previous three subsections show that, when combined with $H(z)$ + BAO data, SN-Pantheon data produce tighter constraints on almost all cosmological parameters, than do SN-DES data (with a few exceptions including $\Omega_{b0}h^2$ for non-flat \lcdm, $\Omega_{c0}h^2$ for non-flat \pcdm, and $\Omega_{m0}$ and $H_0$ for flat and non-flat \pcdm). Since the $H(z)$ + BAO, SN-Pantheon, and SN-DES data constraints are not inconsistent, it is useful to derive constraints from an analysis of the combined $H(z)$, BAO, SN-Pantheon, and SN-DES (ZBPD) data. The results of such an analysis are presented in this subsection. We discuss these results in some detail here because, as discussed in Sec. \ref{subsec:ch9_comparison}, we believe that the constraints we obtain from the ZBPD data combination are more reliable than the constraints we obtain from the other data combinations we study.

The measured values of $\Omega_{b0}h^2$ range from a low of $0.0241^{+0.0024}_{-0.0030}$ (flat \lcdm) to a high of $0.0279^{+0.0031}_{-0.0048}$ (flat \pcdm) and those of $\Omega_{c0}h^2$ range from a low of $0.1047^{+0.0125}_{-0.0096}$ (flat \pcdm) to a high of $0.1199\pm0.0067$ (flat \lcdm). The derived constraints on $\Omega_{m0}$ range from a low of $0.284^{+0.017}_{-0.016}$ (flat \pcdm) to a high of $0.303\pm0.013$ (flat \lcdm). These measurements are consistent with what is measured by \cite{planck2018}. In particular, for flat \lcdm, comparing to the TT,TE,EE+lowE+lensing results in Table 2 of \cite{planck2018} the error bars we find here for $\Omega_{b0}h^2$, \ochs\!, and $\Omega_{m0}$ are a factor of 18, 5.6, and 1.8, respectively, larger than the \textit{Planck} error bars, and our estimates here for the quantities differ from the \textit{Planck} estimates by 0.58$\sigma$, 0.015$\sigma$, and 0.82$\sigma$, respectively.

The constraints on $H_0$ are between $H_0=68.48^{+1.71}_{-1.70}$ \hunit\ (flat \pcdm) and $H_0=69.14\pm1.68$ \hunit\ (flat \lcdm), which are $0.35\sigma$ (flat \lcdm) and $0.15\sigma$ (flat \pcdm) higher than the median statistics estimate of $H_0=68 \pm 2.8$ \hunit\ \citep{chenratmed}, and $2.22\sigma$ (flat \lcdm) and $2.50\sigma$ (flat \pcdm) lower than the local Hubble constant measurement of $H_0 = 74.03 \pm 1.42$ \hunit\ \citep{riess_etal_2019}. For flat \lcdm\ our $H_0$ error bar is a factor of 3.1 larger than that from the \textit{Planck} data and our $H_0$ estimate is 1.01$\sigma$ higher than that of \textit{Planck}.

For non-flat \lcdm, non-flat XCDM, and non-flat \pcdm, we find $\Omega_{k0}=0.032\pm0.072$, $\Omega_{k0}=-0.071^{+0.110}_{-0.123}$, and $\Omega_{k0}=-0.105\pm0.104$, respectively, with non-flat \pcdm\ favoring closed geometry at 1.01$\sigma$. The non-flat XCDM and \pcdm\ models favor closed geometry, while the non-flat \lcdm\ model favors open geometry. The constraints for non-flat \lcdm\ and XCDM models are consistent with spatially flat hypersurfaces.

The fits to the ZBPD data favor dark energy dynamics, where for flat (non-flat) XCDM, $w_{\rm X}=-0.932\pm0.061$ ($w_{\rm X}=-0.904^{+0.098}_{-0.058}$), with best-fitting value being 1.11$\sigma$ (1.66$\sigma$) away from $w_{\rm X}=-1$; and for flat (non-flat) \pcdm, $\alpha=0.320^{+0.108}_{-0.277}$ ($\alpha=0.509^{+0.212}_{-0.370}$), with best-fitting value being 1.16$\sigma$ (1.38$\sigma$) away from $\alpha=0$.

\begin{table*}
\centering
\resizebox{\columnwidth}{!}{%
\begin{threeparttable}
\caption{One-dimensional marginalized best-fitting parameter values and uncertainties ($\pm 1\sigma$ error bars or $2\sigma$ limits) for all models from various combinations of data.}\label{tab:ch9_1d_BFP}
\setlength{\tabcolsep}{0.7mm}{
\begin{tabular}{lcccccccc}
\toprule
Model & Data set & $\Omega_{\mathrm{b_0}}\!h^2$ & $\Omega_{\mathrm{c_0}}\!h^2$ & $\Omega_{\mathrm{m_0}}$ & $\Omega_{\mathrm{k_0}}$ & $w_{\mathrm{X}}$ & $\alpha$ & $H_0$\tnote{a}\\
\midrule
Flat \lcdm & $H(z)$ + BAO & $0.0245^{+0.0026}_{-0.0032}$ & $0.1182\pm0.0077$ & $0.298^{+0.015}_{-0.017}$ & -- & -- & -- & $69.33\pm1.75$ \\
 & ZBP\tnote{b} & $0.0245^{+0.0025}_{-0.0031}$ & $0.1182\pm0.0068$ & $0.298\pm0.013$ & -- & -- & -- & $69.32\pm1.70$ \\
 & ZBD\tnote{c} & $0.0239^{+0.0025}_{-0.0032}$ & $0.1206\pm0.0076$ & $0.305^{+0.015}_{-0.017}$ & -- & -- & -- & $69.04\pm1.74$ \\
 & ZBPD\tnote{d} & $0.0241^{+0.0024}_{-0.0030}$ & $0.1199\pm0.0067$ & $0.303\pm0.013$ & -- & -- & -- & $69.14\pm1.68$ \\
 & ZBPDQH\tnote{e} & $0.0253^{+0.0019}_{-0.0022}$ & $0.1202\pm0.0057$ & $0.299\pm0.012$ & -- & -- & -- & $69.98\pm0.91$ \\
\\
Non-flat \lcdm & $H(z)$ + BAO & $0.0265^{+0.0035}_{-0.0059}$ & $0.1104\pm0.0192$ & $0.291\pm0.024$ & $0.047^{+0.095}_{-0.112}$ & -- & -- & $68.71\pm2.24$ \\
 & ZBP\tnote{b} & $0.0253^{+0.0033}_{-0.0049}$ & $0.1158^{+0.0161}_{-0.0160}$ & $0.296\pm0.022$ & $0.013\pm0.073$ & -- & -- & $69.22\pm1.86$ \\
 & ZBD\tnote{c} & $0.0276^{+0.0038}_{-0.0062}$ & $0.1049^{+0.0188}_{-0.0187}$ & $0.288\pm0.024$ & $0.090^{+0.093}_{-0.106}$ & -- & -- & $67.92\pm2.10$ \\
 & ZBPD\tnote{d} & $0.0257^{+0.0033}_{-0.0050}$ & $0.1133\pm0.0160$ & $0.295\pm0.022$ & $0.032\pm0.072$ & -- & -- & $68.83\pm1.82$ \\
 & ZBPDQH\tnote{e} & $0.0260^{+0.0031}_{-0.0046}$ & $0.1188^{+0.0138}_{-0.0123}$ & $0.297\pm0.020$ & $0.007\pm0.063$ & -- & -- & $69.95\pm0.93$ \\
\\
Flat XCDM & $H(z)$ + BAO & $0.0372^{+0.0045}_{-0.0138}$ & $0.0777^{+0.0351}_{-0.0182}$ & $0.270^{+0.036}_{-0.022}$ & -- & $-0.688^{+0.174}_{-0.109}$ & -- & $65.22^{+2.21}_{-2.64}$ \\
 & ZBP\tnote{b} & $0.0261^{+0.0030}_{-0.0041}$ & $0.1118\pm0.0105$ & $0.292\pm0.016$ & -- & $-0.951\pm0.063$ & -- & $68.91\pm1.76$ \\
 & ZBD\tnote{c} & $0.0331^{+0.0038}_{-0.0091}$ & $0.0881^{+0.0235}_{-0.0137}$ & $0.279^{+0.027}_{-0.019}$ & -- & $-0.739^{+0.110}_{-0.108}$ & -- & $65.95\pm2.08$ \\
 & ZBPD\tnote{d} & $0.0264^{+0.0031}_{-0.0042}$ & $0.1105\pm0.0107$ & $0.292\pm0.016$ & -- & $-0.932\pm0.061$ & -- & $68.62\pm1.73$ \\
 & ZBPDQH\tnote{e} & $0.0273^{+0.0026}_{-0.0035}$ & $0.1131^{+0.0104}_{-0.0095}$ & $0.291\pm0.015$ & -- & $-0.949\pm0.059$ & -- & $69.67^{+0.97}_{-0.96}$ \\
\\
Non-flat XCDM & $H(z)$ + BAO & $0.0367^{+0.0049}_{-0.0145}$ & $0.0822^{+0.0376}_{-0.0233}$ & $0.278^{+0.041}_{-0.030}$ & $-0.122^{+0.137}_{-0.136}$ & $-0.647^{+0.159}_{-0.084}$ & -- & $65.39^{+2.18}_{-2.59}$ \\
 & ZBP\tnote{b} & $0.0251^{+0.0031}_{-0.0049}$ & $0.1186\pm0.0167$ & $0.301\pm0.023$ & $-0.066^{+0.111}_{-0.124}$ & $-0.923^{+0.104}_{-0.060}$ & -- & $69.24\pm1.87$ \\
 & ZBD\tnote{c} & $0.0315^{+0.0039}_{-0.0091}$ & $0.0956^{+0.0260}_{-0.0190}$ & $0.290^{+0.031}_{-0.026}$ & $-0.099\pm0.133$ & $-0.714^{+0.116}_{-0.089}$ & -- & $66.30\pm2.14$ \\
 & ZBPD\tnote{d} & $0.0253^{+0.0032}_{-0.0048}$ & $0.1178^{+0.0166}_{-0.0165}$ & $0.301\pm0.023$ & $-0.071^{+0.110}_{-0.123}$ & $-0.904^{+0.098}_{-0.058}$ & -- & $69.00\pm1.85$ \\
 & ZBPDQH\tnote{e} & $0.0256^{+0.0030}_{-0.0046}$ & $0.1182^{+0.0136}_{-0.0121}$ & $0.299\pm0.020$ & $-0.063^{+0.087}_{-0.097}$ & $-0.919^{+0.085}_{-0.056}$ & -- & $69.59\pm0.97$ \\
\\
Flat $\phi$CDM & $H(z)$ + BAO & $0.0480^{+0.0113}_{-0.0195}$ & $0.0524^{+0.0246}_{-0.0427}$ & $0.240^{+0.024}_{-0.044}$ & -- & -- & $2.418^{+1.197}_{-1.331}$ & $64.67^{+1.86}_{-2.22}$ \\
 & ZBP\tnote{b} & $0.0278^{+0.0030}_{-0.0046}$ & $0.1055^{+0.0119}_{-0.0091}$ & $0.284\pm0.016$ & -- & -- & $<0.666$ & $68.71^{+1.73}_{-1.74}$ \\
 & ZBD\tnote{c} & $0.0429^{+0.0071}_{-0.0170}$ & $0.0641^{+0.0371}_{-0.0235}$ & $0.251^{+0.038}_{-0.031}$ & -- & -- & $1.863^{+0.674}_{-1.316}$ & $65.41^{+1.91}_{-2.08}$ \\
 & ZBPD\tnote{d} & $0.0279^{+0.0031}_{-0.0048}$ & $0.1047^{+0.0125}_{-0.0096}$ & $0.284^{+0.017}_{-0.016}$ & -- & -- & $0.320^{+0.108}_{-0.277}$ & $68.48^{+1.71}_{-1.70}$ \\
 & ZBPDQH\tnote{e} & $0.0289^{+0.0025}_{-0.0040}$ & $0.1073^{+0.0116}_{-0.0081}$ & $0.283^{+0.016}_{-0.014}$ & -- & -- & $0.261^{+0.067}_{-0.254}$ & $69.57\pm0.94$ \\
\\
Non-flat $\phi$CDM & $H(z)$ + BAO & $0.0482^{+0.0126}_{-0.0190}$ & $0.0544^{+0.0194}_{-0.0497}$ & $0.242^{+0.024}_{-0.046}$ & $-0.103\pm0.132$ & -- & $2.618^{+1.213}_{-1.226}$ & $65.14^{+2.02}_{-2.29}$\\
 & ZBP\tnote{b} & $0.0260^{+0.0033}_{-0.0051}$ & $0.1159^{+0.0163}_{-0.0161}$ & $0.296\pm0.022$ & $-0.106\pm0.102$ & -- & $0.454^{+0.174}_{-0.372}$ & $69.33\pm1.86$ \\
 & ZBD\tnote{c} & $0.0427^{+0.0076}_{-0.0177}$ & $0.0670^{+0.0379}_{-0.0282}$ & $0.253^{+0.037}_{-0.039}$ & $-0.097\pm0.130$ & -- & $2.058^{+0.779}_{-1.269}$ & $65.86\pm2.09$ \\
 & ZBPD\tnote{d} & $0.0264^{+0.0034}_{-0.0052}$ & $0.1139\pm0.0161$ & $0.295\pm0.022$ & $-0.105\pm0.104$ & -- & $0.509^{+0.212}_{-0.370}$ & $69.06^{+1.84}_{-1.83}$ \\
 & ZBPDQH\tnote{e} & $0.0265^{+0.0031}_{-0.0048}$ & $0.1142^{+0.0141}_{-0.0123}$ & $0.293\pm0.020$ & $-0.085\pm0.081$ & -- & $0.399^{+0.159}_{-0.313}$ & $69.53\pm0.95$ \\
\bottomrule
\end{tabular}}
\begin{tablenotes}[flushleft]
\item [a] \hunit.
\item [b] $H(z)$ + BAO + SN-Pantheon.
\item [c] $H(z)$ + BAO + SN-DES.
\item [d] $H(z)$ + BAO + SN-Pantheon + SN-DES.
\item [e] $H(z)$ + BAO + SN-Pantheon + SN-DES + QSO + \hiig.
\end{tablenotes}
\end{threeparttable}%
}
\end{table*}

\subsection{ZBPDQH constraints}
\label{subsec:ZBPDQH}
Since the constraints derived from $H(z)$, BAO, SN-Pantheon, SN-DES, QSO, and \hiig\ data are not inconsistent, in this subsection we jointly analyze ZBPDQH data to determine more restrictive constraints on the cosmological parameters (though as discussed in Sec. \ref{subsec:ch9_comparison}, we believe these constraints to be less reliable than those that stem from the HzSNPD combination, so we only describe the broad outlines here).

For flat $\Lambda$CDM, the error bars we derive for $\Omega_{b0}h^2$, $\Omega_{c0}h^2$, and $\Omega_{m0}$ are larger than the \textit{Planck} error bars, though our central estimates of these quantities are broadly consistent with those derived from \textit{Planck}.

In a similar fashion, we find larger error bars on $H_0$ in flat $\Lambda$CDM than does \textit{Planck}, though our central estimate is higher than theirs. Generally, the constraints we derive on $H_0$ are more consistent with the median statistics estimate of $H_0=68 \pm 2.8$ km s$^{-1}$ Mpc$^{-1}$ \citep{chenratmed}, than with the local Hubble constant measurement of $H_0 = 74.03 \pm 1.42$ km s$^{-1}$ Mpc$^{-1}$ \citep{riess_etal_2019}.

We find mild evidence for spatial curvature, with non-flat XCDM and $\phi$CDM favoring closed geometry, and non-flat $\Lambda$CDM mildly favoring open geometry. The constraints from non-flat $\Lambda$CDM and XCDM are consistent with spatially flat hypersurfaces to within less than 1$\sigma$.

Additionally, we find mild evidence for dark energy dynamics, with the best-fitting value of $w_{\rm X}$ being 0.86$\sigma$ (1.45$\sigma$) away from $w_{\rm X}=-1$ in flat (non-flat) XCDM, and the best-fitting value of $\alpha$ being 1.03$\sigma$ (1.27$\sigma$) away from $\alpha=0$ in flat (non-flat) \pcdm.

\subsection{Model comparison}
\label{subsec:ch9_comparison}

The values of $\Delta\chi^2$, $\Delta AIC$, $\Delta BIC$, and the reduced $\chi^2$ ($\chi^2/\nu$) are reported in Table \ref{tab:ch9_cab}, where $\Delta \chi^2$, $\Delta AIC$, and $\Delta BIC$, respectively, are defined as the differences between the values of the $\chi^2$, $AIC$, and $BIC$ for a given model and their corresponding minimum values among all models. From Table \ref{tab:ch9_cab}, we see that the reduced $\chi^2$ values determined from the $H(z)$ + BAO data combination range from 0.49 to 0.62, which is probably due to the $H(z)$ data having overestimated error bars (see Chapter \ref{Chapter8} for discussions of the systematic errors of these data). As discussed in Chapter \ref{Chapter5} and Chapter \ref{Chapter7}, the underestimated systematic uncertainties in QSO and \hiig\ data\footnote{Roberto Terlevich and his colleagues are currently investigating the systematic uncertainties of the \hiig\ data, the results of which they plan to publish in a future paper (Roberto Terlevich, private communication, 2021).} result in larger reduced $\chi^2$ ($\sim1.34$) for the models in the ZBPDQH case. The reduced $\chi^2$ values for the ZBP and ZBPD cases are around unity for all models and for the ZBD case range from 0.77 to 0.87. Of the combinations we study here, on the basis of these reduced $\chi^2$ values, the ZBPD constraints should be viewed as the most reliable ones.

We find that based on the $AIC$ and $BIC$, flat \lcdm\ and flat \pcdm\ are the most favored models in different data combination cases. The $\Delta AIC$ results show that the most favored model is flat \lcdm\ in the ZBP case, while the most favored model is flat \pcdm\ in the rest of the data combinations. The $\Delta BIC$ results show that the most favored model is flat \pcdm\ in the $H(z)$ + BAO and ZBD cases, and is flat \lcdm\ in the remaining cases. For both $\Delta AIC$ and $\Delta BIC$ results, the most disfavored model is non-flat \lcdm\ in the $H(z)$ + BAO and ZBD cases, and is non-flat XCDM in all other cases, with positive evidence against non-flat \lcdm\ and either positive or very strong evidence (depending on the data combination) against non-flat XCDM.

Overall, the $\Delta AIC$ results show no strong evidence against any model, and neither do the $\Delta BIC$ results for the $H(z)$ + BAO and ZBD cases. However, in the ZBP and ZBPDQH cases, the $\Delta BIC$ results show strong evidence against the non-flat \lcdm\ and flat XCDM models, and very strong evidence against the non-flat \pcdm\ and XCDM models. In the ZBPD case, the evidence against flat XCDM and flat \pcdm\ is positive, the evidence against non-flat \lcdm\ is strong, and the evidence against non-flat \pcdm\ and non-flat XCDM is very strong. Based on the $\Delta \chi^2$ results, non-flat \pcdm\ has the minimum $\chi^2$ in all cases. 

In summary, the ZBPD data favor flat \pcdm\ ($AIC$) or flat \lcdm\ ($BIC$) among the six models we study here.

\begin{table*}
\centering
\resizebox{\columnwidth}{!}{%
\begin{threeparttable}
\caption{$\Delta \chi^2$, $\Delta AIC$, $\Delta BIC$, and $\chi^2_{\mathrm{min}}/\nu$ values.}\label{tab:ch9_cab}
\setlength{\tabcolsep}{2.0mm}{
\begin{tabular}{lccccccc}
\toprule
 Quantity & Data set & Flat \lcdm & Non-flat \lcdm & Flat XCDM & Non-flat XCDM & Flat \pcdm & Non-flat \pcdm\\
\hline
 & $H(z)$ + BAO & 5.48 & 5.42 & 1.49 & 0.15 & 1.32 & 0.00\\
 & ZBP\tnote{a} & 2.91 & 2.91 & 2.32 & 1.51 & 1.15 & 0.00 \\
$\Delta \chi^2$ & ZBD\tnote{b} & 6.74 & 6.19 & 1.37 & 0.25 & 1.08 & 0.00\\
 & ZBPD\tnote{c} & 3.33 & 3.22 & 2.10 & 1.23 & 1.05 & 0.00\\
 & ZBPDQH\tnote{d} & 2.99 & 2.99 & 2.27 & 1.25 & 0.95 & 0.00\\
 \\
 & $H(z)$ + BAO & 2.16 & 4.10 & 0.17 & 0.83 & 0.00 & 0.68\\
 & ZBP\tnote{a} & 0.00 & 2.00 & 1.41 & 2.60 & 0.24 & 1.09 \\
$\Delta AIC$ & ZBD\tnote{b} & 3.66 & 5.11 & 0.29 & 1.17 & 0.00 & 0.92\\
 & ZBPD\tnote{c} & 0.28 & 2.17 & 1.05 & 2.18 & 0.00 & 0.95\\
 & ZBPDQH\tnote{d} & 0.04 & 2.04 & 1.32 & 2.30 & 0.00 & 1.05\\
 \\
 & $H(z)$ + BAO & 0.43 & 4.10 & 0.17 & 2.57 & 0.00 & 2.42\\
 & ZBP\tnote{a} & 0.00 & 6.99 & 6.40 & 12.58 & 5.23 & 11.07 \\
$\Delta BIC$ & ZBD\tnote{b} & 1.53 & 5.11 & 0.29 & 3.30 & 0.00 & 3.04\\
 & ZBPD\tnote{c} & 0.00 & 6.90 & 5.78 & 11.92 & 4.72 & 10.69\\
 & ZBPDQH\tnote{d} & 0.00 & 7.23 & 6.51 & 12.72 & 5.19 & 11.47\\
 \\
 & $H(z)$ + BAO & 0.61 & 0.62 & 0.52 & 0.49 & 0.51 & 0.49\\
 & ZBP\tnote{a} & 0.97 & 0.97 & 0.97 & 0.97 & 0.97 & 0.97 \\
$\chi^2_{\mathrm{min}}/\nu$ & ZBD\tnote{b} & 0.86 & 0.87 & 0.78 & 0.78 & 0.78 & 0.77\\
 & ZBPD\tnote{c} & 0.98 & 0.98 & 0.98 & 0.98 & 0.97 & 0.97\\
 & ZBPDQH\tnote{d} & 1.34 & 1.34 & 1.34 & 1.34 & 1.34 & 1.34\\
\bottomrule
\end{tabular}}
\begin{tablenotes}[flushleft]
\item [a] $H(z)$ + BAO + SN-Pantheon.
\item [b] $H(z)$ + BAO + SN-DES.
\item [c] $H(z)$ + BAO + SN-Pantheon + SN-DES.
\item [d] $H(z)$ + BAO + SN-Pantheon + SN-DES + QSO + \hiig.
\end{tablenotes}
\end{threeparttable}%
}
\end{table*}

\section{Conclusion}
\label{sec:conclusion}

By analyzing a total of 1383 measurements, consisting of 31 $H(z)$, 11 BAO, 1048 SN-Pantheon, 20 SN-DES, 120 QSO, and 153 HIIG data points, we jointly constrain cosmological parameters in six flat and non-flat cosmological models.

From the constraints derived using the cosmological models, we can identify some relatively model-independent features. As discussed in Sec. \ref{subsec:ch9_comparison}, the $H(z)$ + BAO + SN-Pantheon + SN-DES (ZBPD) data combination produces the most reliable constraints. In particular, for the ZBPD data combination, we find a reasonable and fairly restrictive summary value of $\Omega_{m0}=0.294 \pm 0.020$,\footnote{Here we take the summary central value to be the mean of the two of six central-most values. As for the uncertainty, we call the difference between the two central-most values twice the systematic uncertainty and the average of the two central-most error bars the statistical uncertainty, and compute the summary error bar as the quadrature sum of the two uncertainties.} which is in good agreement with many other recent measurements (e.g. $0.315\pm0.007$ from \citealp{planck2018}). A fairly restrictive summary value of $H_0=68.8\pm1.8$ km s$^{-1}$ Mpc$^{-1}$ is found to be in better agreement with the estimates of \cite{chenratmed} and \cite{planck2018} than with the measurement of \cite{riess_etal_2019}; note that the constraints from BAO data do not depend on physics of the early Universe (with $\Omega_{b0}h^2$ being a free parameter that is fitted to the data used here). There is some room for dark energy dynamics or a little spatial curvature energy density in the ZBPD constraints, but based on $AIC$ and $BIC$ criteria, flat \pcdm\ or flat \lcdm\ are the best candidate models.

%% file: chapter6.tex
\cleardoublepage

\chapter{Evidence against Ryskin's model of cosmic acceleration}\chaptermark{Against Ryskin}

\newcommand{\obh}{\Omega_b h^2}

\label{Chapter6}

This chapter is based on \cite{Ryan_2020}.

\section{Introduction} \label{sec:Intro}
I gave a brief overview of the motivation behind Ryskin's model, and of the prediction it makes for the form that the Hubble parameter takes as a function of $z$, in Sec. \ref{sec:ch3_Ryskin_model}. Here I analyze the model's fit to a set of cosmic chronometer and standard ruler data, and compare the quality of this fit to that of the standard $\Lambda$CDM model.

According to \cite{Ryskin}, his model accurately fits the Hubble diagram built from SNe Ia data, but we will see in this chapter that there are other data sets with which Ryskin's model is much less compatible. In addition to predicting a value of the Hubble constant ($H_0$) that is larger than the values obtained from the CMB and from local measurements (see \cite{planck2018} and \cite{riess2018}, respectively, for these measurements), Ryskin's model fails to predict the trend in high-redshift ($z \gtrsim 1$) Hubble parameter data when its predicted Hubble parameter curve is plotted together with these data.
\begin{comment}
Ryskin's model does not predict the trend in high redshift ($z \gtrsim 1$) $H(z)$ data measurements of the Hubble parameter $H(z)$ for $z \gtrsim 1$, and predicts implausible values of $H_0$ and $\om$ when it is fitted to distance measurements derived from baryon acoustic oscillation (``BAO") data. The fit to angular size measurements from quasars (``QSO") data is inconclusive.
\end{comment}
Recently, another group found that Ryskin's model can not accurately describe structure formation (see \citealp{Revisit_Ryskin}), while leaving open the possibility that other types of observations may be compatible with this model. This analysis presented in this chapter is complementary to, and independent of, the analysis of \cite{Revisit_Ryskin}; I will show that none of the data sets I have collected favor Ryskin's model over \lcdm, making it unlikely that Ryskin's model will be saved by future measurements.

The Hubble parameter derived in \cite{Ryskin} accurately fits the Hubble diagram constructed from SNe Ia data, which Ryskin takes as evidence that his model may be able to explain the origin of cosmic acceleration of the Universe without invoking dark energy. My goal in this paper is to test Eq. \ref{eq:RyskinH} against several sets of observational data (containing measurements at higher redshifts than the SNe Ia measurements used in \citealp{Ryskin}), to determine whether or not Ryskin's model can fit these data sets as well as it fits the currently available SNe Ia data.

\section{Analysis} \label{sec:Analysis}
\subsection{Data} \label{subsec:Data}
In this chapter I use 31 measurements of the Hubble parameter $H(z)$, 11 distance measurements derived from baryon acoustic oscillation (``BAO") data, and 120 quasar (``QSO") angular size measurements. The $H(z)$ data can be found in Table \ref{tab:H(z)_data}; see also Chapter \ref{Chapter4} for a description. The BAO data I use here are listed in Table \ref{tab:ch5_BAO_data}. My method of analyzing these data is slightly different from the method employed in Chapter \ref{Chapter5}; see below for a discussion. The QSO data are listed in \cite{Cao_et_al2017b}; see that paper and Chapter \ref{Chapter5} for a description and discussion.
\begin{comment}
[Maybe just describe the observables; don't say anything about the method until you get to the method section.]
[Rewrite this paragraph. Start with "In brief, the quantities to be compared to measurements are..." and describe those quantities, instead of just providing their mathematical definitions.]
When Ryskin's model is fit to the $H(z)$ data, I directly compare the functional form of $H(z)$ that his model predicts to the measured values of $H(z)$ that are listed in \cite{Ryan_Doshi_Ratra_2018}. For the BAO data, I use eq. \ref{eq:RyskinH} to calculate the distances $D_H(z)$, $D_M(z)$, and $D_V(z)$ (see \cite{Ryan_Chen_Ratra_2019} or \cite{Hogg} for a definition of these). The predicted values of these distances, scaled by the sound horizon $r_{\rm S}$, are the quantities I compared to the measurements listed in \cite{Ryan_Chen_Ratra_2019}. For the QSO data, the predicted quantity to be compared to measurement is $\theta(z) = l_{m}/D_{A}(z)$, where $l_m = 11.03$ pc is a characteristic linear scale and $D_A(z)$ can be computed from $H(z)$. See \cite{Ryan_Chen_Ratra_2019} or \cite{Hogg} for a definition of $D_A(z)$, and \cite{Cao_et_al2017b} for a discussion of the characteristic scale $l_m$.
\end{comment}

For the $H(z)$ data set, the measured quantity is $H(z)$ itself, namely the Hubble parameter as a function of the redshift $z$. For the BAO data, the measured quantities are a set of distances $D_H(z)$, $D_M(z)$, and $D_V(z)$ (see Chapter \ref{Chapter2}) and $H(z)$, scaled by the value that the sound horizon $r_{\rm S}$ takes at the drag epoch. This latter quantity depends on the homogeneous part of the dimensionless matter density parameter ($\Omega_{m0}$), the Hubble constant ($H_0$), the dimensionless baryon density parameter ($\Omega_b h^2$), and the CMB temperature ($T_{\rm CMB}$), thereby making $r_{\rm S}$ a model-dependent quantity.\footnote{$h := H_0/(100 \hspace{1mm} {\rm km}^{-1} \hspace{1mm} {\rm Mpc}^{-1})$} In this chapter I use the same fitting formula that was used in Chapter \ref{Chapter5} to compute $r_{\rm S}$, and I use the same CMB temperature (from \citealp{Fixsen}), but depending on the analysis method (see below) I marginalize over $\Omega_b h^2$ and/or $\Omega_{m0}$. In all other respects my treatment of the BAO data here is the same as the treatment described in Chapter \ref{Chapter5}. Finally, the measured quantity in the QSO data set is $\theta(z) = l_{m}/D_{A}(z)$, where $l_m = 11.03 \pm 0.25$ pc is a characteristic linear scale\footnote{Specifically, $l_m$ is the radius at which the jets of the QSOs tend to become opaque (when observed at frequency $f \sim 2$ GHz; see \citealp{Cao_et_al2017b}).} and $D_A(z)$, the angular size distance, can be computed from $H(z)$. See Chapter \ref{Chapter2} for a definition of $D_A(z)$, and \cite{Cao_et_al2017b} for a discussion of the characteristic linear scale $l_m$.
\begin{comment}
In order to use BAO as standard rulers, it is necessary to compute the sound horizon $r_{\rm S}$. [say more here; describe/cite] This quantity depends on the homogeneous part of the dimensionless matter density parameter ($\Omega_{m0}$), the Hubble parameter ($H_0$), the dimensionless baryon density parameter ($\Omega_b h^2$), and the CMB temperature at recombination ($T_{\rm CMB}$), thereby making $r_{\rm S}$ a model-dependent quantity.\footnote{$h := H_0/(100 \hspace{1mm} {\rm km}^{-1} \hspace{1mm} {\rm Mpc}^{-1})$} In this paper I use the same fitting formula that was used in \cite{Ryan_Chen_Ratra_2019} to calculate $r_{\rm S}$, and I use the same CMB temperature (from \cite{Fixsen}), but depending on the analysis method (see below) I marginalize over $\Omega_{m0}$ and/or $\Omega_b h^2$, because the only adjustable parameter in Ryskin's model is $H_0$, in contrast to the models considered in \cite{Ryan_Chen_Ratra_2019}, in which $\Omega_{m0}$ and $\Omega_b h^2$ were both adjustable parameters (we set $\Omega_b h^2$ to a fixed but model-dependent value, and varied $\Omega_{m0}$ freely). Additionally, for cases in which it is necessary to use 

In all other respects my treatment of the BAO data here is the same as the treatment described in \cite{Ryan_Chen_Ratra_2019}.
\end{comment}
\subsection{Methods} \label{subsec:Methods}
I have chosen to analyze Ryskin's model according to two methods. In the first method, I compute the value that $H_0$ takes when the likelihood function, defined by
\begin{equation}
    \mathcal{L}(H_0) = \int e^{\chi^2(H_0)/2}\uppi(p_n) dp_n
\end{equation}
is maximized, within each data set separately and in full combination. I also compute the minimum $\chi^2$ corresponding to the best-fitting $H_0$, for which
\begin{equation}
\label{eq:chi2min}
    \chi^2_{\rm min} = -2{\rm ln}\left(\mathcal{L}_{\rm max}\right)
\end{equation}
This is very similar to the methods employed in Chapters \ref{Chapter4} and \ref{Chapter5}; see those chapters for details regarding the form that the $\chi^2$ function takes when it is computed within each model and for each data set. The prior function, $\uppi(p_n)$, is necessary to deal with the nuisance parameters $\Omega_{m0}$ and $\Omega_b h^2$ that enter the analysis through the calculation of the sound horizon $r_{\rm S}$ (see above). This prior function has the form $\uppi\left(\om, \obh\right) = \uppi\left(\om\right)\uppi\left(\obh\right)$, where \begin{equation}
\label{eq:om_prior}
  \uppi\left(\om\right) =
    \begin{cases}
      1 & \text{if $0.10 < \om < 0.70$}\\
      0 & \text{otherwise},
    \end{cases}       
\end{equation}
and
\begin{equation}
  \uppi\left(\obh\right) =
    \begin{cases}
      1 & \text{if $0.01000 < \obh < 0.05000$}\\
      0 & \text{otherwise}.
    \end{cases}       
\end{equation}
The main difference between this first analysis method and the analyses of Chapters \ref{Chapter4} and \ref{Chapter5} is that I do not compare the best-fitting value of $H_0$ or the minimum value of $\chi^2$ in Ryskin's model directly to any other models (although the best-fitting $H_0$ can be compared to the measurements of $H_0$ made by \cite{planck2018} and \cite{riess2018}; see below). For the $\chi^2$ function I simply compare the minimum value of $\chi^2$ to the number of degrees of freedom $\nu$ (defined below), and conclude that the fit to the data is poor if $\chi^2_{\rm min}/\nu >> 1$. Additionally, I split the BAO data into two subsets, called ``BAO1" (containing all BAO measurements) and ``BAO2" (which excludes the measurements at $z > 2$), respectively, because the $\chi^2$ function for the $H(z)$ + QSO + BAO1 data combination is so large that the corresponding likelihood function evaluates to zero, and so can't be plotted. This is telling, because it suggests that Ryskin's model can not fit the observational data at high redshift (see also the discussion in Sec. \ref{sec:Results}). The combined fit therefore uses the $H(z)$ + QSO + BAO2 data combination (see Table \ref{tab:ch6_BFP} and Fig. \ref{fig:L(H0)}).

\begin{comment}
The method I use to analyze the data here is the same as the method described in \cite{Ryan_Chen_Ratra_2019} $\Omega_b h^2$ (as mentioned above). This means that when I calculate the likelihood function $\mathcal{L}(H_0)$, I integrate over a prior on $\Omega_b h^2$, like so:

\begin{equation}
\label{eq:Likelihood}
    \mathcal{L}(H_0) = \int e^{-\chi^2/2}\uppi\left(\Omega_b h^2\right) d\left(\Omega_b h^2\right).
\end{equation}

In eq. \ref{eq:Likelihood}, $\chi$ is a function of $H_0$, $\Omega_{m0}$, $\Omega_b h^2$, and the prior function is

\begin{equation}
  \uppi(\obh) =
    \begin{cases}
      1 & \text{if $0.02221 < \obh < 0.02305$}\\
      0 & \text{otherwise}
    \end{cases}       
\end{equation}

I have chosen the upper and lower limits on this prior to coincide with the greatest and least values, respectively, listed in Table [...] of \cite{Ryan_Chen_Ratra_2019}. For the $H(z)$ and QSO data, I only found it necessary to calculate a one-dimensional likelihood function $\mathcal{L}(H_0)$, because Ryskin's model only depends on $H_0$. For the BAO data, I found it necessary to calculate a two-dimensional likelihood function $\mathcal{L}(H_0, \Omega_{m0})$, because a comparison of the BAO measurements to Ryskin's model requires $\Omega_{m0}$ as input (through the $\Omega_{m0}$-dependence of $r_{\rm S}$). I therefore present two-dimensional confidence contours in $H_0$-$\Omega_{m0}$ space, as well as one-dimensional marginalized contours for $H_0$ and $\Omega_{m0}$, for the BAO data.
\end{comment}

In my second analysis method, I directly compare the quality of the fit obtained in Ryskin's model to the quality of the fit obtained with a simple flat \lcdm\ model to each data set, considered separately. For the $H(z)$ and QSO data this is quite simple: all that is necessary is to plot either the function
\begin{equation}
    H(z) = H_0\sqrt{\om(1 + z)^3 + 1 - \om}
\end{equation}
(for the $H(z)$ data), or the function
\begin{equation}
    \theta(z) = \frac{l_m}{D_{\rm A}(z)}
\end{equation}
(for the QSO data) that is predicted by Ryskin's model, together with the same functions as predicted by \lcdm, and see which predicted function better fits the overall trend in the data (see Figs. \ref{fig:QSO_Lcomp}-\ref{fig:Hz_vs_z}). For the BAO data this kind of direct curve fitting is difficult to do, because several of the measurements are correlated (meaning that they do not have independent uncertainties), and the set as a whole consists of measurements of different things. In order to compare Ryskin's model to \lcdm\ using these data, I therefore redid the analysis of Chapter \ref{Chapter5} (in that paper the constraints from BAO alone were not presented), allowing $H_0$ and $\Omega_{m0}$ to vary freely, with $\Omega_b h^2 = 0.02225$ for both Ryskin's model and \lcdm\footnote{This is the value that $\obh$ takes in the \lcdm\ model, previously used in Chapter \ref{Chapter5}, and originally computed from Planck 2015 TT + lowP + lensing CMB anisotropy data in \cite{Park_Ratra_2018_FLCDM}. It is also possible to marginalize over $H_0$ and $\Omega_{m0}$ so as to obtain a best-fitting value of $\Omega_b h^2$ in Ryskin's model, and then use this value instead of $\Omega_b h^2 = 0.02225$ in the two-parameter fits. Doing this turns out to be rather uninformative, however, as the best-fitting values of $\Omega_{m0}$ and $H_0$ that one obtains in this case turn out to be nearly identical to those obtained using $\Omega_b h^2 = 0.02225$.}.

\section{Results} \label{sec:Results}
\begin{figure*}
    \includegraphics[scale=1]{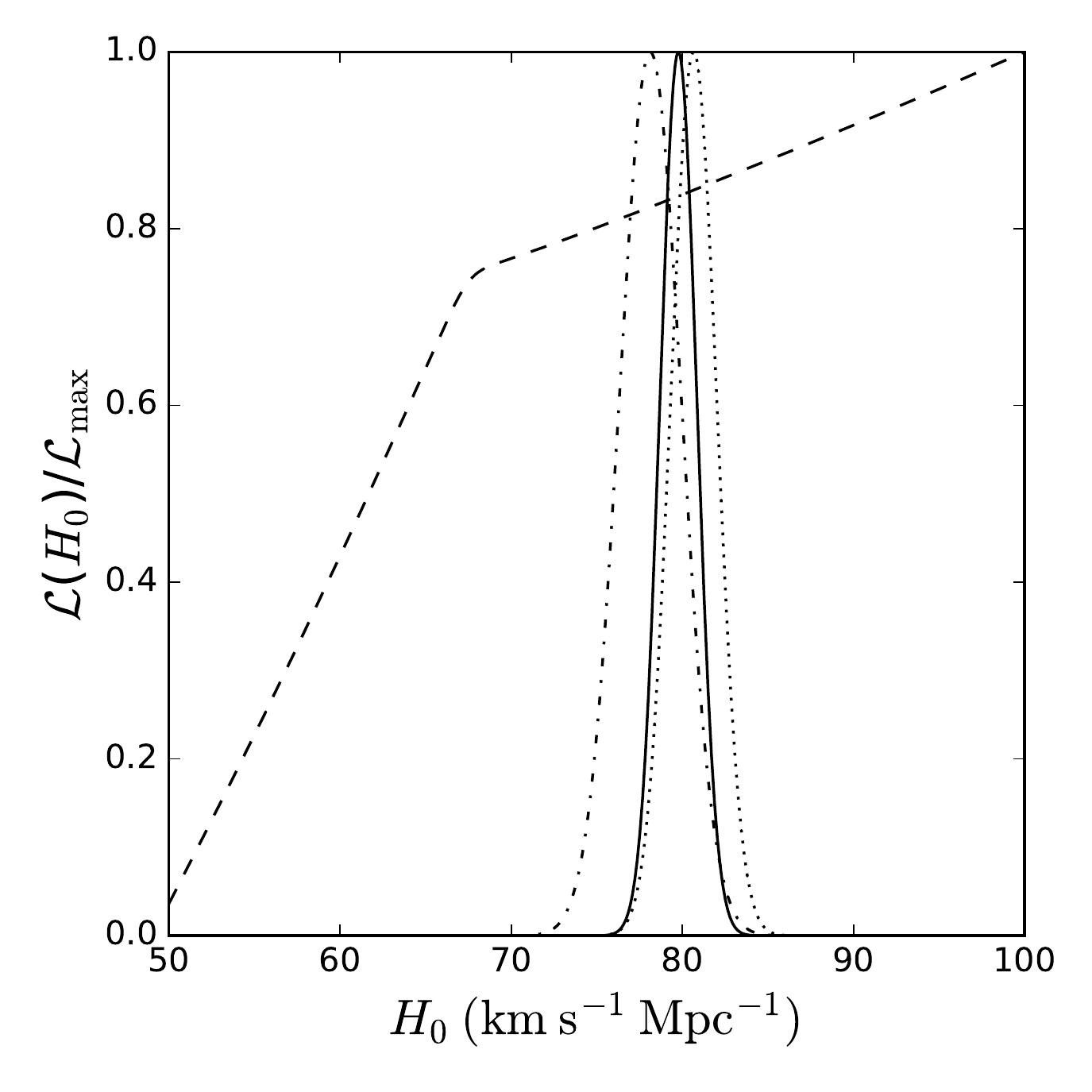}
    \caption[Likelihood functions for $H_0$ according to Ryskin's model.]{Likelihood functions for $H_0$ according to Ryskin's model. The dot-dashed curve represents the fit from the $H(z)$ data, the dotted curved represents the fit from the QSO data, the dashed curve represents the fit from the BAO2 data, and the solid curve represents the fit from the combined $H(z)$ + QSO + BAO2 data. See text for discussion.}
    \label{fig:L(H0)}
\end{figure*}

My results for the fit of Ryskin's model to the data are presented in Table \ref{tab:ch6_BFP} and Fig. \ref{fig:L(H0)}. In the first column of Table \ref{tab:ch6_BFP} I list the data combination, in the second column I list the one-dimensional best-fitting values of $H_0$ with their respective 1$\sigma$ and 2$\sigma$ uncertainties ($\sigma$ here being defined in the same way as the one-sided confidence limits used in Chapter \ref{Chapter5}), and in the third column I list the corresponding value of $\chi^2_{\rm min}/\nu$, where $\chi^2_{\rm min}$ is computed from Eq. \ref{eq:chi2min}, and $\nu$ is the number of degrees of freedom:
\begin{equation}
\label{eq:nu}
    \nu = N - n - 1.
\end{equation}
In the above equation $N$ is the number of data points and $n$ is the number of model parameters.

\begin{table}
    \centering
    \caption[Best-fitting central values of $H_0$ (with 1 and 2$\sigma$ error bars).]{Best-fitting central values of $H_0$ (with 1 and 2$\sigma$ error bars) for the data combinations I considered.}
    \begin{tabular}{c|c|c}
        \hline
        Data set & $H_0$ (km s$^{-1}$ Mpc$^{-1}$) & $\chi^2_{\rm min}/\nu$\\
        \hline
        $H(z)$ & $78.12^{+1.82+3.64}_{-1.82-3.63}$ & 2.20\\
        QSO & 80.66$^{+1.35+2.70}_{-1.35-2.70}$ & 3.14\\
        BAO1 & $100_{-19.27-32.72}$ & 133.12\\
        BAO2 & $100_{-26.78-41.79}$ & 14.48\\
        $H(z)$ + QSO + BAO2 & $79.77^{+1.08+2.17}_{-1.08-2.17}$ & 3.38\\
        \hline
    \end{tabular}
    \label{tab:ch6_BFP}
\end{table}
\begin{comment}
\begin{table}
    \centering
    \caption{Best-fitting mean values of $H_0$ (with 1 and 2$\sigma$ error bars) for the data combinations I considered.}
    \begin{tabular}{c|c|c|c}
        \hline
        Data set & $H_0$ (km s$^{-1}$ Mpc$^{-1}$) & $\chi^2_{\rm min}$ & $\nu$\\
        \hline
        $H(z)$ & $78.12^{+1.82+3.64}_{-1.82-3.63}$ & 63.81 & 29\\
        QSO & 80.66$^{+1.35+2.70}_{-1.35-2.70}$ & 373.76 & 119\\
        BAO1 & $100_{-19.27-32.72}$ & 1198.07 & 9\\
        BAO2 & $100_{-26.78-41.79}$ & 101.35 & 7\\
        $H(z)$ + QSO + BAO2 & $79.77^{+1.08+2.17}_{-1.08-2.17}$ & 540.53 & 160\\
        \hline
    \end{tabular}
    \label{tab:ch6_BFP}
\end{table}
\end{comment}

From Table \ref{tab:ch6_BFP} and Fig. \ref{fig:L(H0)}, one can see that the best-fitting value of $H_0$ from the combination $H(z)$ + QSO + BAO2 (which gives the tightest error bars) is a little over 3$\sigma$ away from the measurement of $H_0 = 74.03 \pm 1.42$ km$^{-1}$ Mpc$^{-1}$ made by \cite{riess2018}, and over 10$\sigma$ away from the measurement of $H_0 = 67.4 \pm 0.5$ km$^{-1}$ Mpc$^{-1}$ made by \cite{planck2018} (here $\sigma$ is equal to $\sqrt{1.08^2 + \sigma_l^2}$ where $\sigma_l$ is the uncertainty of either of the two measurements given above). The agreement, therefore, between the predicted value of $H_0$ under Ryskin's model and the measurements from \cite{planck2018} and \cite{riess2018} is not very good. Further, the value of $\chi^2_{\rm min}/\nu$ ranges from 2.20 to 133.12 for the fit of Ryskin's model to each data set, and is equal to 3.38 for the $H(z)$ + QSO + BAO2 data combination. This suggests, independently of the comparison to the $H_0$ measurements made by \cite{planck2018} and \cite{riess2018}, that Ryskin's model does not provide a good fit to the data listed in Table \ref{tab:ch6_BFP}. In Fig. \ref{fig:L(H0)}, the dashed curve represents the likelihood function computed from BAO2, the dot-dashed curve represents the likelihood function computed from $H(z)$ data, the dotted curve represents the likelihood function computed from QSO data, and the solid curve represents the product of these likelihood functions. Here again one can see how far away the value of $H_0$ predicted by Ryskin's model is from the measurements made by \cite{planck2018} and \cite{riess2018} when $H_0$ is fitted to the $H(z)$ + QSO + BAO2 data combination.

\begin{figure*}
\begin{multicols}{2}
    \includegraphics[width=\linewidth]{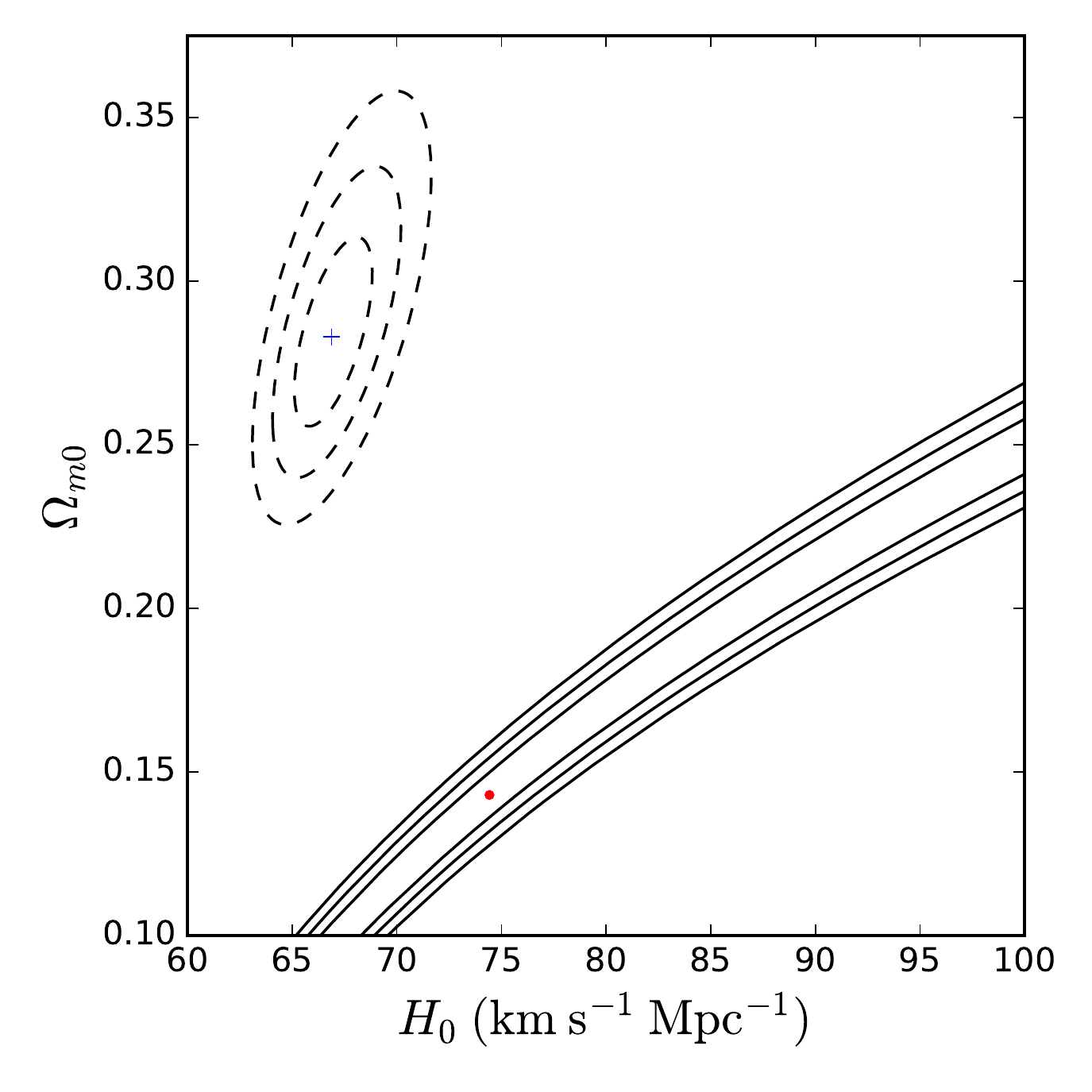}\par
    \includegraphics[width=\linewidth]{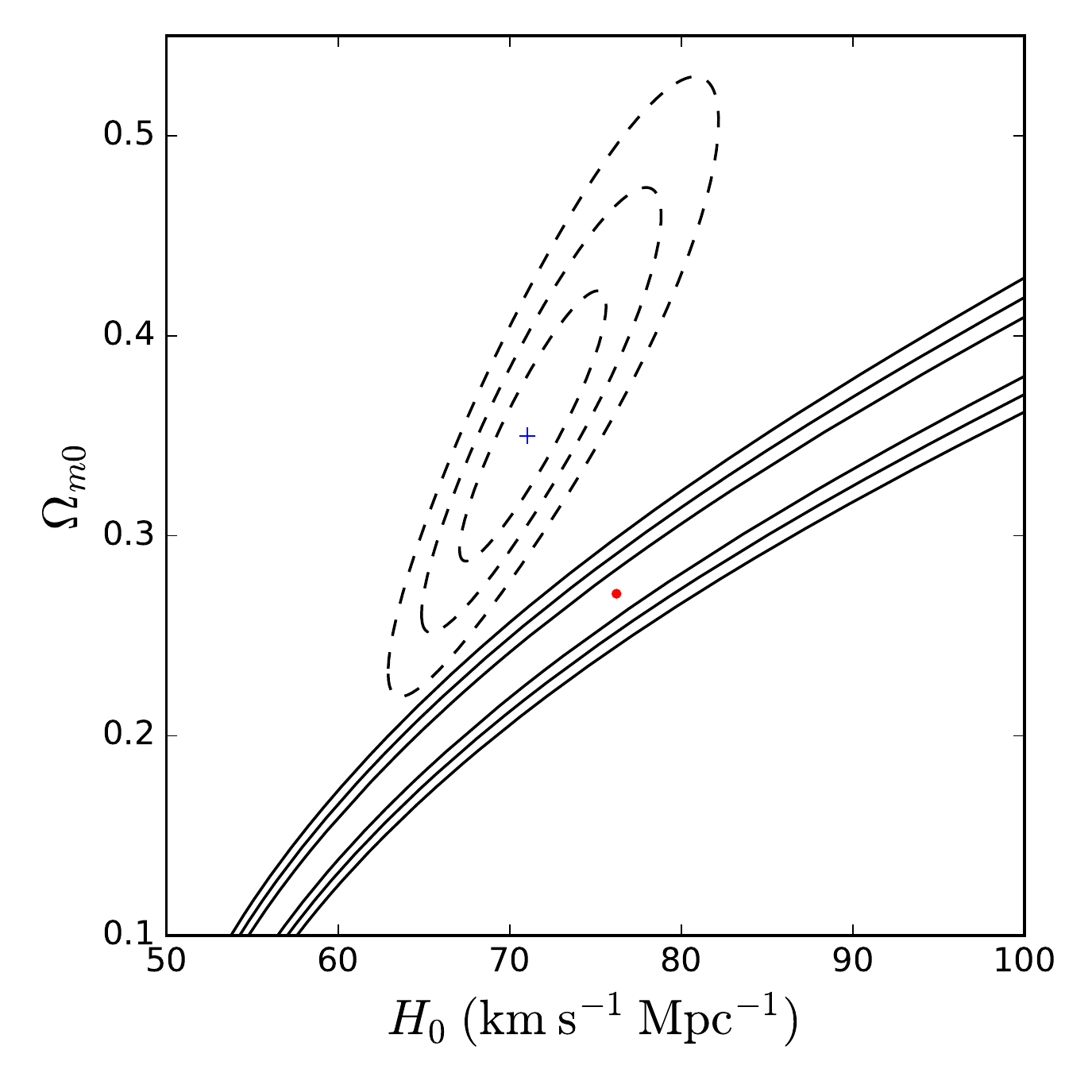}\par
\end{multicols}
\caption[Flat \lcdm\ model versus Ryskin's model with BAO data.]{Flat \lcdm\ model versus Ryskin's model with BAO data. The left panel corresponds to the BAO1 subset of the BAO data, and the right panel corresponds to the BAO2 subset of the same. In both columns I have plotted 1, 2, and 3$\sigma$ confidence contours and best-fitting points in $H_0$-$\Omega_{m0}$ space for both \lcdm\ and Ryskin's model. See text for discussion.}
\label{fig:Lcomp_2D}
\end{figure*}

\begin{table}
    \centering
    \caption[One- and two-dimensional best-fitting values of $H_0$ and $\Omega_{m0}$ for the BAO1 and BAO2 data combinations.]{One- and two-dimensional best-fitting values of $H_0$ and $\Omega_{m0}$ for the BAO1 and BAO2 data combinations. Here $H_0$ has units of km s$^{-1}$ Mpc$^{-1}$ and $\chi^2_{\rm min}/\nu$ pertains to the two-dimensional fit.}
    \begin{tabular}{c|c|c|c|c|c}
        \hline
        Model & Data set & $H_0$ & $\Omega_{m0}$ & $\left(H_0, \Omega_{m0}\right)$ & $\chi^2_{\rm min}/\nu$\\
        \hline
        Ryskin & BAO1 & $100.0_{-21.43-31.06}$ & $0.237_{-0.0794-0.127}^{+0.00963+0.175}$ & (74.43, 0.143) & 147.87\\
         & BAO2 & $100.0_{-28.38-42.02}^{}$ & $0.373_{-0.145-0.249}^{+0.0167+0.0305}$ & (76.22, 0.271) & 14.56\\
        \lcdm & BAO1 & $66.91^{+1.322+2.715}_{-1.152-2.254}$ & $0.284^{+0.0205+0.0424}_{-0.0179-0.0348}$ & (66.88, 0.283) & 0.954\\
         & BAO2 & $71.28_{-2.593-5.003}^{+3.100+6.375}$ & $0.354^{+0.0491+0.102}_{-0.0409-0.0793}$ & (71.03, 0.350) & 0.650\\
        \hline
    \end{tabular}
    \label{tab:ch6_BFP_2D_1D_Lcomp}
\end{table}

\begin{comment}
\begin{table}
    \centering
    \caption{One-dimensional best-fitting values of $H_0$ and $\Omega_{m0}$, respectively, for the BAO1 and BAO2 data combinations.}
    \begin{tabular}{c|c|c|c|c|c}
        \hline
        Model & Data set & $H_0$ (km s$^{-1}$ Mpc$^{-1}$) & $\chi^2_{\rm min}\left(H_0\right)/\nu$ & $\Omega_{m0}$ & $\chi^2_{\rm min}\left(\om\right)/\nu$\\
        \hline
        Ryskin & BAO1 & $100.0_{-21.43-31.06}$ & 132.39 & $0.237_{-0.0794-0.127}^{+0.00963+0.175}$ & 131.14\\
         & BAO2 & $100.0_{-28.38-42.02}^{}$ & 13.54 & $0.373_{-0.145-0.249}^{+0.0167+0.0305}$ & 12.02\\
        \lcdm & BAO1 & $66.91^{+1.322+2.715}_{-1.152-2.254}$ & 1.58 & $0.284^{+0.0205+0.0424}_{-0.0179-0.0348}$ & 0.65\\
         & BAO2 & $71.28_{-2.593-5.003}^{+3.100+6.375}$ & 1.42 & $0.354^{+0.0491+0.102}_{-0.0409-0.0793}$ & 0.24\\
        \hline
    \end{tabular}
    \label{tab:ch6_BFP_1D_Lcomp}
\end{table}
\end{comment}
\begin{figure*}
    \centering
    \includegraphics{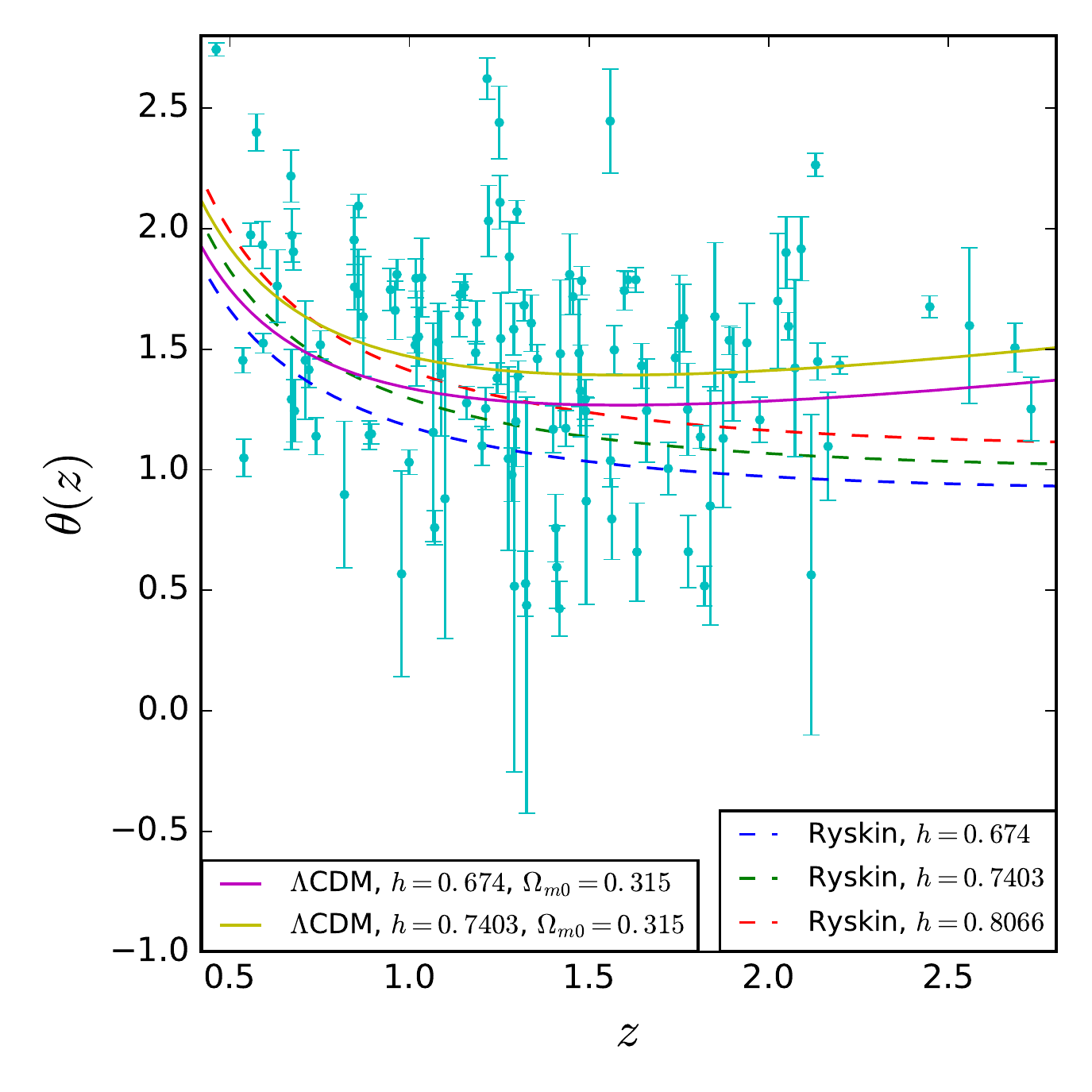}
    \caption[Sound horizon parameter $\theta(z)$ versus redshift $z$ for $\Lambda$CDM and for Ryskin's model.]{Sound horizon parameter $\theta(z)$ versus redshift $z$ for $\Lambda$CDM and for Ryskin's model. The dashed blue, green, and red curves represent $\theta(z)$ as predicted by Ryskin's model, and the solid purple and gold curves represent $\theta(z)$ as predicted by spatially-flat \lcdm. The values $(h, \om) = (0.674, 0.315)$ come from \cite{planck2018}, $h = 0.7403$ comes from \cite{riess2018}, and $h = 0.8066$ comes from the fit of $\theta(z)$ to the QSO data using Ryskin's model. See text for discussion.}
    \label{fig:QSO_Lcomp}
\end{figure*}

\begin{figure*}
    \includegraphics[scale=1]{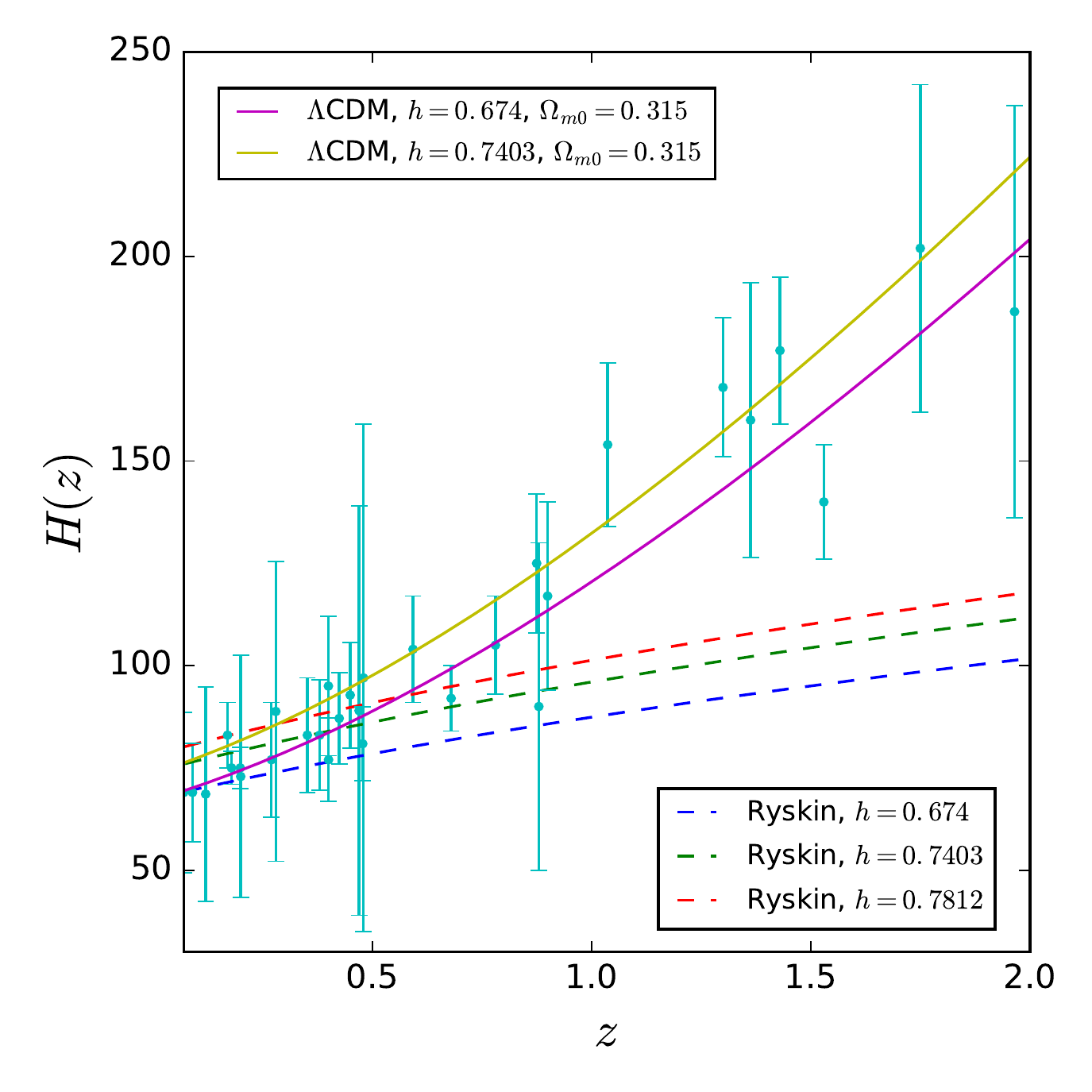}
    \caption[Hubble parameter $H(z)$ versus redshift $z$ for $\Lambda$CDM and for Ryskin's model.]{Hubble parameter $H(z)$ versus redshift $z$ for $\Lambda$CDM and for Ryskin's model. The dashed blue, green, and red curves represent $H(z)$ as predicted by Ryskin's model, and the solid purple and gold curves represent $H(z)$ as predicted by spatially-flat \lcdm. The values $(h, \om) = (0.674, 0.315)$ come from \cite{planck2018}, $h = 0.7403$ comes from \cite{riess2018}, and $h = 0.7812$ comes from the fit of $H(z)$ to the $H(z)$ data using Ryskin's model. See text for discussion.}
    \label{fig:Hz_vs_z}
\end{figure*}

The results of the comparison between Ryskin's model and the spatially-flat \lcdm\ model using BAO data are presented in Table \ref{tab:ch6_BFP_2D_1D_Lcomp} and in Fig. \ref{fig:Lcomp_2D}. In Table \ref{tab:ch6_BFP_2D_1D_Lcomp}, the first column lists the model, the second column lists the data set (here ``BAO1" and ``BAO2" have the same meanings as in the previous analysis method), the third and fourth columns list the one-dimensional best-fitting values of $H_0$ and $\Omega_{m0}$, respectively, the fifth column lists the two-dimensional best-fitting values of $H_0$ and $\Omega_{m0}$, and the sixth column lists the value of $\chi^2_{\rm min}/\nu$ corresponding to each model and data set. In Fig. \ref{fig:Lcomp_2D} the left panel shows the two-dimensional constraints on $H_0$ and $\om$ for the BAO1 data combination, and the right panel shows the two-dimensional constraints on $H_0$ and $\om$ for the BAO2 data combination. In both panels the solid contours correspond to Ryskin's model and the dashed contours correspond to \lcdm. From the figure, one can see that even when \lcdm\ and Ryskin's model are both fitted to the low-redshift BAO data (namely BAO2), the confidence contours for both of these models are disjoint to 3$\sigma$. When \lcdm\ and Ryskin's model are fitted to a data combination that includes high-redshift BAO data (namely BAO1), the confidence contours for both models are even more disjoint than when these models are fitted to the low-redshift data.
\begin{comment}
In Fig. \ref{fig:Lcomp_2D}, one can see that for the BAO2 data combination the confidence contours of both Ryskin's model and \lcdm\ are very stretched in $H_0$-$\Omega_{m0}$ space, both lying very close together when the lower baryon density is used in Ryskin's model (the confidence contours of Ryskin's model when the higher baryon density is used are disjoint from the other two, at least to 3$\sigma$). When the high redshift BAO measurements are included in the analysis, however, it can clearly be seen from Fig. \ref{fig:Lcomp_2D} that the confidence contours of Ryskin's model and \lcdm\ are completely disjoint to 3$\sigma$, no matter which baryon density is used. 
\end{comment}
Fig. \ref{fig:Lcomp_2D} therefore indicates that an investigator who wishes to save Ryskin's model from the BAO measurements I have used must fall on the horns of a dilemma: while it is possible to accommodate a smaller value of $H_0$ within Ryskin's model (i.e. one that is in better agreement with the measurements made by either \cite{planck2018} or \cite{riess2018} and with the \lcdm\ constraints computed here), this is only possible at the cost of predicting an implausibly small value of $\Omega_{m0}$. Similarly, a predicted value of $\Omega_{m0}$ within Ryskin's model that is consistent with the value of $\Omega_{m0}$ predicted by \lcdm\ requires an implausibly large predicted value of $H_0$ within Ryskin's model. Finally, it is clear from Table \ref{tab:ch6_BFP_2D_1D_Lcomp} that Ryskin's model, when it is fitted to either BAO1 or BAO2, has a much larger value of $\chi^2_{\rm min}/\nu$ than the \lcdm\ model, when the \lcdm\ model is fitted to the same data combinations. Ryskin's model therefore provides a much poorer fit to the BAO data (especially high-redshift BAO data) than does the standard \lcdm\ model.

In Fig. \ref{fig:QSO_Lcomp} I have plotted $\theta(z)$ vs $z$ for the values of $\theta(z)$ predicted by both Ryskin's model and spatially-flat \lcdm. The dashed blue, green, and red curves represent Ryskin's predicted $\theta(z)$ with $h$ set to 0.674, 0.7403, and 0.7812, respectively. The solid purple and gold curves represent the theoretical curves of $\theta(z)$ as calculated in the spatially-flat \lcdm\ model with ($h$, $\om$) set to (0.674, 0.315) and (0.7403, 0.315), respectively. This plot is, in my view, rather inconclusive with regard to whether spatially-flat \lcdm\ or Ryskin's model provides a better fit to the data, as the measurements are dispersed widely on the plot, and many of them have very large error bars, so the overall trend in the data is difficult to see. It is clear from the theoretical curves, however, that Ryskin's model predicts a very different angular size than \lcdm\ for $z \gtrsim 1.5$, so a stronger case against Ryskin's model from QSO data could potentially be made with more high-redshift measurements (or more precise low-redshift measurements).

The \lcdm\ model departs even more radically from Ryskin's model, at high redshift, when their respective theoretical $H(z)$ curves are plotted against $H(z)$ data. In Fig. \ref{fig:Hz_vs_z} the dashed curves represent Ryskin's predicted $H(z)$, with the blue, green, and red curves corresponding to $h = 0.674$, $h = 0.7403$, and $h = 0.7812$, respectively. The solid curves represent $H(z)$ as predicted by the \lcdm\ model, where the purple curve corresponds to $(h, \om) = (0.674, 0.315)$, and the gold curve corresponds to $(h, \om) = 0.7403, 0.315)$. From the figure, one can see that at low redshift ($z \lesssim 1$), Ryskin's model appears to fit the data as well as \lcdm, owing to the large error bars on the measurements. At high redshift ($z \gtrsim 1$), the data clearly diverge from the curves predicted by Ryskin's model, and the \lcdm\ curves match the upward trend. 
\begin{comment}
Additionally, it is possible to observe a phase of decelerated expansion at $z \gtrsim 0.75$ in a plot of $H(z)/(1+z)$ versus $z$,\footnote{See \cite{Farooq_Ranjeet_Crandall_Ratra_2017} and references therein.} and in Fig. 5 one can clearly see that Rykin's model fails to capture this decelerated phase, while $\Lambda$CDM captures both it and the present accelerated phase. 
\end{comment}
The \lcdm\ model therefore provides a much better fit to high-redshift $H(z)$ data than does Rykin's model.

\section{Conclusion}

I conclude, based on these results and the earlier findings of \cite{Revisit_Ryskin}, that Ryskin's model of emergent cosmic acceleration does not provide an adequate fit to available cosmological data, and so cannot replace the standard spatially-flat \lcdm\ cosmological model. The fit to the SNe Ia data presented in Ryskin's original paper is primarily a fit to low-redshift ($z \lesssim 1$) measurements; as can be seen from Figs. \ref{fig:Lcomp_2D} and \ref{fig:Hz_vs_z}, as well as Tables \ref{tab:ch6_BFP} and \ref{tab:ch6_BFP_2D_1D_Lcomp}, low-redshift measurements do not clearly distinguish between Ryskin's model and the \lcdm\ model when the predictions of these models are compared to the data. High-redshift measurements, on the other hand (chiefly $H(z)$ and BAO measurements at $z \gtrsim 1$) can distinguish between these two models, and the high-redshift data clearly favor \lcdm\ over Ryskin's model.

%% file: chapter10.tex
\cleardoublepage

\chapter[Power law constraints]{Constraints on power law cosmology from cosmic chronometer, standard ruler, and standard candle data}

\label{Chapter10}

This chapter is based on \cite{Ryan_power_law}.

\section{Introduction}
The power law model was described in Sec. \ref{sec:ch3_power_law_model}; this chapter will present the results of the comparison of that model to the standard $\Lambda$CDM model, when both are confronted with a set of standard candle, standard ruler, and cosmic chronometer data.

Many investigators have found that a power law model with $\beta \approx 1$ is favored by various independent low-redshift probes, such as cosmic chronometers ($H(z)$) \cite{Dev_Jain_Lohiya_2008}, gravitational lensing statistics \cite{Dev_Safonova_Deepak_Lohiya_2002}, Type Ia supernovae (SNe Ia) \cite{Dev_Sethi_Lohiya_2001, Kumar_2012, Rani_et_al_2015, Sethi_Dev_Jain_2005}, baryon acoustic oscillations (BAO) \cite{Shafer_2016, Tutusaus_et_al_2016}, quasar angular sizes (QSO) \cite{Jain_Dev_Alcaniz_2003}, galaxy cluster gas mass fractions \cite{Zhu_Alcaniz_Liu_2008}, and the combination of $H(z)$ + BAO + SNe Ia + gamma-ray burst distance moduli (GRB) \cite{Haridasu_AAP_2017}. Other data sets, however, favor $\beta \approx$ 1.2-1.6 \cite{Dev_Jain_Lohiya_2008, Dolgov_Halenka_Tkachev_2014, Kumar_2012, Shafer_2016, Tutusaus_et_al_2016}; see Table \ref{tab:beta_fits}.

\begin{comment}
Several investigators have claimed that the power law model with $\beta \approx 1$ provides a good fit to several kinds of low redshift data such as cosmic chronometer ($H(z)$) \citep{Dev_Jain_Lohiya_2008}, gravitational lensing statistics \citep{Dev_Safonova_Deepak_Lohiya_2002}, Type Ia supernova (SN Ia) \citep{Dev_Sethi_Lohiya_2001, Kumar_2012, Rani_et_al_2015, Sethi_Dev_Jain_2005}, baryon acoustic oscillation (BAO) \citep{Shafer_2016, Tutusaus_et_al_2016}, quasar angular size (QSO) \citep{Jain_Dev_Alcaniz_2003}, and cluster gas mass fraction data \citep{Zhu_Alcaniz_Liu_2008} (though other data sets favor $\beta \approx$ 1.2-1.6 \citep{Dev_Jain_Lohiya_2008, Dolgov_Halenka_Tkachev_2014, Kumar_2012, Shafer_2016, Tutusaus_et_al_2016}). See Table \ref{tab:beta_fits}.
\end{comment}

\begin{comment}
Some studies have also found that a power law model with $\beta = 1$ can explain the observed abundances of [certain] elements \citep{Lohiya_Batra_Mahajan_Mukherjee_1999, Sethi_Batra_Lohiya_1999}, and can thus account for primordial nucleosynthesis. These conclusions have been challenged by other studies, which find that the observed abundances of [helium, deuterium, lithium...] set a severe constraint on the value of $\beta$ within the power law model, requiring $\beta \approx$ 0.55-0.58 \cite{Kaplinghat_Steigman_Tkachev_1999, Kaplinghat_Steigman_Walker_2000, Kumar_2012}.
\end{comment}
Some studies have also found that a power law model with $\beta = 1$ can produce the right amount of primordial helium to match current observations \citep{Lohiya_Batra_Mahajan_Mukherjee_1999, Sethi_Batra_Lohiya_1999}, and so may be able to account for the synthesis of other light elements. These conclusions are challenged by the results of other studies, which find that $\beta \approx$ 0.55-0.58 is required to produce the right abundances \citep{Kaplinghat_Steigman_Walker_2000, Kaplinghat_Steigman_Tkachev_1999, Kumar_2012}. If these latter studies are correct, then the values of $\beta$ favored by primordial nucleosynthesis are clearly disjoint with those favored by low redshift measurements, and it is difficult to see how they can be reconciled without introducing extra complexity to the power law model (such as the addition of a mechanism that forces $\beta$ to change its value between the two eras; see e.g. \cite{Gumjudpai, Gumjudpai_Thepsuriya_2012, Kaeonikhom_Gumjudpai_Saridakis_2011, Rangdee_Gumjudpai_2014, Wei_2004}).\footnote{For recent efforts to provide an account of primordial nucleosynthesis within the power law model, see \cite{Singh_Lohiya_2015_arXiv, Singh_Lohiya_2015_JCAP}.} Given that the power law model is intended to be a simpler alternative to the \lcdm\ model, such additional complexity seems unjustified, and the power law model appears to be ruled out on these grounds.

\begin{table*}
    \caption{Fits to power law exponent from other low redshift measurements.}
    \label{tab:beta_fits}
    \centering
    \resizebox{\columnwidth}{!}{%
    \begin{tabular}{ccc}
    \hline
    \hline
         Reference & $\beta$ & Data type(s) used \\
         \hline
         \cite{Dev_Jain_Lohiya_2008} & $1.07^{+0.11}_{-0.08}$ & $H(z)$\\
         & $1.42^{+0.08}_{-0.07}$ & SN Ia\\
         \cite{Dev_Safonova_Deepak_Lohiya_2002} & $1.09\pm0.3$ & Gravitational lensing statistics\\
         & $1.13^{+0.4}_{-0.3}$ & \\
         \cite{Dev_Sethi_Lohiya_2001} & $1.004\pm0.043$ & SN Ia\\
         \cite{Dolgov_Halenka_Tkachev_2014} & $1.52 \pm 0.15$ & SN Ia\\
         & $1.55 \pm 0.13$ & \\
         & $1.3$ & BAO\\
         \cite{Jain_Dev_Alcaniz_2003} & $1.0\pm0.3$ & QSO\\
         \cite{Kumar_2012} & $1.22_{-0.16}^{+0.21}$ & $H(z)$\\
         & $1.61_{-0.12}^{+0.14}$ & SN Ia\\
         \cite{Rani_et_al_2015} & $1.05_{-0.066}^{+0.071}$ & $H(z)$\\
         & $1.44^{+0.26}_{-0.18}$ & SN Ia\\
         \cite{Sethi_Dev_Jain_2005} & $1.04^{+0.07}_{-0.06}$ & SN Ia\\
         \cite{Shafer_2016} & $0.93$ & BAO\\
         & $1.44$-$1.56$ & SN Ia\\
         \cite{Tutusaus_et_al_2016} & $0.908\pm0.019$ & BAO\\
         & $1.55\pm0.13$ & SN Ia\\
         \cite{Zhu_Alcaniz_Liu_2008} & $1.14\pm0.05$ & Galaxy cluster gas mass fraction\\
         \cite{Haridasu_AAP_2017} & $1.08 \pm 0.04$ & $H(z)$ + BAO + SNe Ia + GRB\\
    \hline
    \hline
    \end{tabular}%
    }
\end{table*}

\begin{comment}
Table caption: [check these against Table 2 in \cite{Dev_Jain_Lohiya_2008} and other tables in other papers.]
\end{comment}

\begin{comment}
\cite{Gumjudpai} & & $H(z)$, BAO, SNe Ia, QSO ages, cluster gas mass fraction\\
\cite{Rangdee_Gumjudpai_2014} & & BAO, SNe Ia, $H_0$\\
\cite{Kaeonikhom_Gumjudpai_Saridakis_2011} & & BAO, $H_0$\\
\end{comment}

\begin{comment}
[supernova data [add \cite{Dev_Sethi_Lohiya_2001} after you've read it] [lensing statistics; add \cite{Dev_Safonova_Deepak_Lohiya_2002} you've finished reading it] certain combinations of cosmological data [GIVE NUMBERS (make table?)] \cite{Zhu_Alcaniz_Liu_2008}, \cite{Sethi_Dev_Jain_2005}, maybe also \cite{Dolgov_Halenka_Tkachev_2014} [are you sure that this is what Zhu et al claim?][Power law with $\beta \approx 1$ fits older QSO data \cite{Jain_Dev_Alcaniz_2003}[others]
\end{comment}

\begin{comment}
[Regarding Kaplinghat et al 1999, see 'Comment on "Observational constraints on power-law cosmologies"', and also see "Pouring cold water", which I believe rebuts "Comment" (maybe Kumar MNRAS does, too, and maybe also look at Sethi Dev Jain 2005 PLB).]
\end{comment}

A defender of the power law model who does not wish to make the model more complex by introducing a time-variable $\beta$ could attempt to save it by arguing that:

1.) The findings of \cite{Kaplinghat_Steigman_Walker_2000}, \cite{Kaplinghat_Steigman_Tkachev_1999}, and \cite{Kumar_2012} are simply incorrect, and the power law exponent has the value $\beta \approx 1$ during both the nucleosynthesis era and the present era, or

2.) The Universe only undergoes power law expansion at late times, and the power law model with $\beta \approx 1$ adequately describes low redshift observations only.

The latter option is, on its face, plausible. After all, the standard \lcdm\ model holds that the Universe follows power law expansion during both the matter-dominated and radiation-dominated eras, so it might be reasonable to limit the scope of the power law model by suggesting that it only applies after the era of nucleosynthesis.\footnote{In \cite{Kolb_1989}, one of the earliest papers on the subject, the author proposes that, if a hypothetical form of matter called ``K-matter'' were to dominate the energy budget at late times, this would lead to a ``coasting'' cosmic expansion with $\beta = 1$ (with $\beta$ taking on different values in earlier eras).} We must be careful not to push this argument too far, however, because any scale factor $a(t)$ can presumably be approximated by a power law over some arbitrarily short time period. What is at issue is not whether the Universe follows power law expansion during some (relatively) brief portion or portions of its history, but whether it follows power law expansion throughout all (or most) of its history. If it can be shown that the power law model fits low redshift observational data as well as or better than \lcdm\ over some appreciable range of redshifts, then option (2) is validated (and option 1 may be validated as well, if one can marshal a strong argument against the findings of \cite{Kaplinghat_Steigman_Walker_2000}, \cite{Kaplinghat_Steigman_Tkachev_1999}, \cite{Kumar_2012}). If, on the other hand, the power law model fails to provide a good fit to the available low redshift data, then both (1) and (2) are falsified.

\begin{comment}
Both of these options can be falsified by testing the power law model against low redshift data, because if (1.) is correct (and if the power law exponent is assumed not to vary in time), then $\beta$ will have the same value during the era of nucleosynthesis that it has now
\end{comment}

\begin{comment}
[Defenders of the power law model] could argue that the Universe only undergoes power law expansion at late times; by limiting the scope of the power law model, they can plausibly argue that the constraints set by primordial nucleosynthesis on the parameter[s] of this model can be avoided. [This is not a controversial move. After all, it is well-known that the scale factor can be approximated by a power law during the eras of radiation domination and matter domination. The question that I seek to answer in this paper is whether or not the power law model is a good description in the present, $\Lambda$-dominated era. [Be careful with this claim. I think $\Lambda$ only takes over around z of about 3/4 (see H(z) plot.]]
\end{comment}
\begin{comment}
While it is true that the power law model can provide an adequate fit to some data, these data tend to be [either limited in number] or of a single kind ([only standard candle data, for example]). It has been shown in the literature, however, that when the power law model is fitted to several independent data sets consisting of different kinds of measurements, [it performs poorly. \cite{Rani_et_al_2015, Shafer_2016, Tutusaus_et_al_2016} [others].
\end{comment}

A few studies \citep{Rani_et_al_2015, Shafer_2016, Tutusaus_et_al_2016, Haridasu_AAP_2017} have been conducted along these lines. These studies find that, when the power law model is fitted to multiple independent data sets ($H(z)$ alone and $H(z)$ + BAO + SNe Ia + CMB in \citealp{Rani_et_al_2015}, BAO + SNe Ia in \citealp{Shafer_2016}, BAO + SNe Ia + CMB in \citealp{Tutusaus_et_al_2016}, and $H(z)$ + BAO + SNe Ia + GRB in \citealp{Haridasu_AAP_2017}), it performs poorly compared to \lcdm. Here I continue in this vein by fitting the power law and \lcdm\ models to a data set consisting of cosmic chronometer, standard ruler, and standard candle data,  some of which have not yet been used to test the power law model (see Sec. \ref{sec:Data} for a description of the data).
\begin{comment}
Here I test this claim by fitting the power law model, along with a fiducial \lcdm\ model, to several combinations of cosmic chronometer, standard ruler, and standard candle data, some of which have not previously been used to test the power law model (See Sec. \ref{sec:Data} for a description of the measurements I use). [Only a few papers (\cite{Shafer_2016, Tutusaus_et_al_2016}) have performed model comparison tests of the power law cosmology.  Mention this [as a deficiency of earlier studies? That is, several papers that examined PL only fitted beta; they didn't do model comparison so their conclusions aren't very compelling], and connect what they did with what you did].
\end{comment}
I use simple model comparison statistics (the same as those used in \cite{Rani_et_al_2015, Shafer_2016, Tutusaus_et_al_2016, Haridasu_AAP_2017}; see Sec. \ref{sec:Methods}) to compare the quality of the fit in both cases. I discuss my results in Sec. \ref{sec:ch10_Results} and draw my conclusions in Sec. \ref{sec:Conclusion}.

\section{Data}
\label{sec:Data}

\begin{table*}
    \centering
    \caption[Data sets.]{Data sets used in this chapter.}
    \begin{tabular}{ccc}
    \hline
    \hline
         Data type & Number of data points & Redshift range\\
         \hline
         $H(z)$ & 31 & $0.070 \leq z \leq 1.965$ \\
         BAO & 11 & $0.38 \leq z \leq 2.334$\\
         QSO & 120 & $0.462 \leq z \leq 2.73$\\
         GRB & 119 & $0.48 \leq z \leq 8.2$\\
         HIIG & 153 & $0.0088 \leq z \leq 2.42935$\\
         \hline
         \hline
    \end{tabular}
    \label{tab:Data}
\end{table*}

In Table \ref{tab:Data} I list the types of measurements I use, the number of measurements of each type, and the redshift ranges within which the measurements lie. The cosmic chronometer data consist of measurements of the Hubble parameter as a function of the redshift $z$ ($H(z)$), listed in Table \ref{tab:H(z)_data}. To fit the power law and \lcdm\ models to the cosmic chronometer data, I compute $H(z)$ theoretically using eqs. (\ref{eq:ch3_Hz_PL}) and (\ref{eq:E(z)_LCDM}).

I use two sets of standard ruler measurements in this paper. The first set consists of measurements of the quantities $H(z)$, $D_{\rm H}(z)$, $D_{\rm M}(z)$, $D_{\rm A}(z)$, and $D_{\rm V}(z)$, scaled by the value that the sound horizon $r_{\rm s}$ takes at the baryon drag epoch (see Chapter \ref{Chapter2} for definitions of the various distance functions listed above). These measurements are the same as those listed in Table \ref{tab:ch9_BAO}. See Chapters \ref{Chapter4}-\ref{Chapter9} for more details, and for references to the original literature. To compute the size of the sound horizon, I use the approximate formula
\begin{equation}
\label{eq:r_s}
    r_{\rm s}=55.154\frac{{\rm exp}[-72.3(\Omega_{\nu 0}h^2+0.0006)^2]}{(\Omega_{\rm b 0}h^2)^{0.12807}(\Omega_{m0}h^2 - \Omega_{\nu 0}h^2)^{0.25351}} \hspace{1mm}{\rm Mpc},
\end{equation}
where $\Omega_{m0}$, $\Omega_{b0}$, and $\Omega_{\nu 0}$ are the dimensionless energy density parameters of non-relativistic matter, of baryons, and of neutrinos, respectively, and $h := H_0/100$ km s$^{-1}$ Mpc$^{-1}$ \citep{PhysRevD.92.123516}. Following \cite{Carter_2018}, I set $\Omega_{\nu 0} = 0.0014$, which leaves two additional free parameters ($\Omega_{m0}$ and $\Omega_{b0}h^2$) when the power law model is fitted to data combinations containing BAO data, and one additional free parameter ($\Omega_{b0}h^2$) when the \lcdm\ model is fitted to the same data combinations. The second set of standard ruler data consists of measurements of the angular sizes $\theta_{\rm obs}$, in milliarcseconds (mas), of intermediate-luminosity quasars (QSO). The angular size of a quasar can be computed theoretically via
\begin{equation}
    \label{eq:th_th}
    \theta_{\rm th}(z) = \frac{l_{\rm m}}{D_{\rm A}(z)},
\end{equation}
where $l_{\rm m} = 11.03 \pm 0.25$ pc is the characteristic linear size of the quasars in the sample. This quantity can then be compared to $\theta_{\rm obs}$ to determine the quality of the fit of the given model to the QSO data. The QSO angular size measurements are listed, and $l_{\rm m}$ is determined, in \cite{Cao_et_al2017b}; see that paper and Chapter \ref{Chapter5} for discussion and details.

\begin{comment}
[\footnote{Specifically, $l_m$ is the radius at which the jets of the QSOs tend to become opaque (when observed at frequency $f \sim 2$ GHz; see Ref. \cite{Cao_et_al2017b}).}]
\end{comment}

I use two sets of standard candle data in this paper. The first set consists of measurements of the luminosities, fluxes, and velocity dispersions of HII starburst galaxies (HIIG), from which the distance moduli of these galaxies can be computed. The HIIG data consist of a low redshift ($0.0088 \leq z \leq 0.16417$) set of 107 measurements from \cite{Chavez_2014}, plus a high redshift ($0.636427 \leq z \leq 2.42935$) set of 46 measurements from \cite{G-M_2019}. Subsets of these data, which were generously provided to me by Ana Luisa Gonz\'{a}lez-Mor\'{a}n,\footnote{Private communications, 2019 and 2020.} have been used in several studies to constrain cosmological parameters (\citealp{Chavez_2012, Chavez_2016, G-M_2019, Terlevich_2015}; see also Chapters \ref{Chapter7}-\ref{Chapter9}). See Chapter \ref{Chapter7} for a detailed description of how the distance modulus can be computed. Briefly, if one knows the luminosity $L$, flux $f$, and velocity dispersion $\sigma$ of an HII galaxy, one can use these quantities to compute a distance modulus $\mu_{\rm obs}$. This quantity can then be compared to the theoretical distance modulus
\begin{equation}
\label{eq:mu_theo}
    \mu_{\rm th} = 5{\rm log}D_{\rm L}(z) + 25,
\end{equation}
where $D_{\rm L}(z)$ is given by eq. (\ref{eq:D_L}), to determine the quality of the model's fit to the data. The second set of standard candle data consists of measurements of the bolometric fluence $S_{\rm bolo}$ and observed peak energy $E_{\rm p, obs}$ of 119 gamma-ray bursts from \cite{Dirirsa_2019} (GRB). Given a knowledge of the bolometric fluence of a source, one can compute the energy radiated isotropically in the source's rest frame
\begin{equation}
    \label{eq:E_iso}
    E_{\rm iso} = \frac{4\pi D_{\rm L}^2}{1 + z}S_{\rm bolo}.
\end{equation}
GRBs can be standardized through the Amati relation \citep{Amati_2009, Amati2008}
\begin{equation}
\label{eq:ch10_Amati}
    {\rm log} E_{\rm iso} = a + b{\rm log}\left[\left(1 + z\right)E_{\rm p, obs}\right],
\end{equation}
which connects the observed peak energy of a given GRB to its isotropic radiated energy (here $a$ and $b$ are free parameters which I vary when fitting the power law and \lcdm\ models to the GRB data). The GRB likelihood function also contains a parameter which describes the extrinsic scatter of the GRBs in the sample ($\sigma_{\rm ext}$) \citep{D'Agostini_2005}. As in Chapter \ref{Chapter8}, I vary this parameter freely when fitting the power law and \lcdm\ models to the GRB data. By comparing the value of ${\rm log} E_{\rm iso}$ as computed from eq. (\ref{eq:E_iso}) to that computed from eq. (\ref{eq:ch10_Amati}), one can determine the quality of the model fit. For more details about the GRB analysis, see Chapter \ref{Chapter8} and \cite{Khadka_Ratra_2020}.

There is some overlap between the cosmic chronometer data I use in this paper and those that were used in \cite{Dev_Jain_Lohiya_2008}, \cite{Kumar_2012}, and \cite{Rani_et_al_2015} to constrain the parameters of the simple (constant $\beta$) power law model. Many of the measurements these authors used are the same as mine, although I use a larger, more up-to-date set (which is the same as the set of $H(z)$ data used to constrain the simple power law model in \cite{Haridasu_AAP_2017}, though I add one point from \cite{Ratsimbazafy_et_al_2017}). I use a different sample of QSO data than does \cite{Jain_Dev_Alcaniz_2003}, and my BAO measurements have all been updated relative to those of \cite{Dolgov_Halenka_Tkachev_2014}, \cite{Shafer_2016}, \cite{Tutusaus_et_al_2016}, and \cite{Haridasu_AAP_2017}. GRB data were used to constrain the power law model in \cite{Haridasu_AAP_2017}, and many, but not all, of these data are the same as those I use here (additionally, my data set is larger and contains newer measurements). To my knowledge, HIIG data have never been used to constrain the power law model. Because these data are independent of the $H(z)$, BAO, QSO, and GRB data sets, I obtain tight constraints on the parameters of the power law model when I fit it to these data in combination with the $H(z)$, BAO, QSO, and GRB data (see Sec. \ref{sec:ch10_Results}).

\section{Methods}
\label{sec:Methods}
The fiducial model I use in this chapter is the spatially flat \lcdm\ model. The Hubble parameter of this model is given by eq. (\ref{eq:E(z)_LCDM}) with $\Omega_{k0} = 0$.

\begin{comment}
[Only two studies have done model comparison...] [In contrast to \cite{Tutusaus_et_al_2016}, which uses SN, CMB, and BAO data, I rely solely on low redshift ($z \lesssim 8$) measurements, so as to test whether or not the power law model can adequately describe the evolution of the late Universe, independently of early-Universe physics.]
\end{comment}

\begin{comment}
Do I need this?:
[[DON'T MENTION THIS UNTIL SEC. 4]Some of the BAO measurements are correlated, and some are not. See [cite] and Sec. \ref{sec:Methods} for a description of how I analyze these data].
\end{comment}

The methods that I use to compare the power law model to the \lcdm\ model are largely the same as methods that have been used previously in Chapters \ref{Chapter4}-\ref{Chapter9}, as well as \cite{Rani_et_al_2015}, \cite{Shafer_2016}, and \cite{Tutusaus_et_al_2016}, which I briefly summarize here. For each combination of data that I study, I compute the quantity
\begin{equation}
    \chi^2_{\rm min} := -2{\rm ln}\mathcal{L}_{\rm max}
\end{equation}
where the likelihood function $\mathcal{L}$ depends on the parameters of the model under consideration. The likelihood function takes a different form depending on the data combination that is used to compute it; these forms are described in Chapters \ref{Chapter7} and \ref{Chapter8}.\footnote{Some of the BAO data that I use are correlated, so it is necessary to take their covariance matrices into account when computing $\chi^2_{\rm min}$. See Chapters \ref{Chapter5} and \ref{Chapter9} for the covariance matrices of the correlated data.} For models having the same number of parameters, the best-fitting model to the data is that which has a smaller value of $\chi^2_{\rm min}$. As in Chapters \ref{Chapter7} and \ref{Chapter8}, I use the \textsc{Python} module \textsc{emcee} \cite{Foreman-Mackey_Hogg_Lang_Goodman_2013} to sample the likelihood function $\mathcal{L}$, and I use the \textsc{Python} module \textsc{getdist} \cite{Lewis_2019} both to generate the one- and two-dimensional likelihood contours shown in left and right panels of Fig. \ref{fig:ZBQGH_marginalized} and to compute the one-dimensional marginalized best-fitting values (sample means) and 68\% uncertainties (two-sided limits) of the model parameters.

When comparing models with different numbers of parameters, the $\chi^2$ function is not necessarily the most informative statistic to use, because it gives the same weight to simple models that it gives to complex models. For this reason, I also use the corrected Akaike Information Criterion:
\begin{equation}
    {\rm AICc} := {\rm AIC} + \frac{2n(n+1)}{N - n - 1},
\end{equation}
where
\begin{equation}
    {\rm AIC} := \chi^2_{\rm min} + 2n,
\end{equation}
is the Akaike Information Criterion (suitable in the limit that $N >> n$), and the Bayes Information Criterion:
\begin{equation}
    {\rm BIC} := \chi^2_{\rm min} + n {\rm ln} N.
\end{equation}
\cite{Liddle_2007}. In the equations above, $n$ is the number of model parameters and $N$ is the number of data points.\footnote{In previous work (see Chapters \ref{Chapter4}-\ref{Chapter9}), my collaborators and I used the AIC and BIC to compare the quality of cosmological model fits to data. Here I use the AICc in place of the AIC because the AICc is more appropriate for smaller data sets (like the $H(z)$ and BAO sets), because it approaches the AIC in the limit that $N$ is large, and to facilitate the comparison of my results with the results of \cite{Shafer_2016} and \cite{Tutusaus_et_al_2016}, both of which used the AICc in their analyses.} The AICc and BIC punish models that have a greater number of parameters in favor of models with fewer parameters. In this sense, the AICc and BIC provide a quantitative basis for choosing which model, among a set of models, provides the most parsimonious fit to a given set of data.

\section{Results and Discussion}
\label{sec:ch10_Results}

\begin{table*}
    \caption{Best-fitting parameters of the power law model.}
    \label{tab:PL_BFP_Om}
    \centering
    \resizebox{\columnwidth}{!}{%
    \begin{tabular}{cccccccccccc}
    \hline
    \hline
    Data type & $H_0$ (km s$^{-1}$ Mpc$^{-1}$) & $\beta$ & $\om$ & $\Omega_{b0}h^2$ & $a$ & $b$ & $\sigma_{\rm ext}$ & $\nu$ & $\chi^2_{\rm min}/\nu$ & AICc & BIC\\
    \hline
         $H(z)$ & 61.92 & 0.9842 & - & - & - & - & - & 29 & 0.5721 & 21.02 & 23.46 \\
         %\hline
         BAO & 89.78 & 0.9206 & 0.6192 & 0.03819 & - & - & - & 7 & 1.513 & 25.26 & 20.18\\
         QSO & 61.83 & 0.9673 & - & - & - & - & - & 118 & 2.991 & 357.1 & 362.6\\
         %\hline
         BAO+QSO & 60.57 & 0.9213 & 0.2030 & 0.07690 & - & - & - & 127 & 2.864 & 372.0 & 383.2\\
         %\hline
         GRB & 72.34 & 0.7530 & - & - & 49.99 & 1.115 & 0.4010 & 114 & 1.138 & 140.3 & 153.7\\
         HIIG & 70.99 & 1.251 & - & - & - & - & - & 151 & 2.725 & 415.5 & 421.5\\
         %\hline
         GRB+HIIG & 70.31 & 1.158 & - & - & 50.12 & 1.157 & 0.4066 & 267 & 2.039 & 554.7 & 572.5\\      
         %\hline
         All Data & 63.06 & 0.9470 & 0.2234 & 0.06706 & 50.16 & 1.144 & 0.4025 & 427 & 2.229 & 966.2 & 994.4\\
    \hline
    \hline
    \end{tabular}%
    }
\end{table*}

The best-fitting values of the parameters of the power law model (namely, those that minimize the $\chi^2$ function), are recorded in columns 2-8 of Table \ref{tab:PL_BFP_Om}. The number of degrees of freedom, 
\begin{equation}
    \nu := N - n
\end{equation}
is recorded in column 9 of this table. Columns 10-12 record, respectively, the minimum value of the reduced $\chi^2$ function, and the minimum values of the AICc and BIC. Similarly, the best-fitting values of the parameters of the \lcdm\ model are recorded in columns 2-7 of Table \ref{tab:LCDM_BFP_Om}, with the number of degrees of freedom in column 8, the minimum value of the reduced $\chi^2$ function in column 9, and the minimum values of the AICc and BIC in columns 10 and 11, respectively.

In Table \ref{tab:1d_BFP_PL}, in columns 2 and 3, I record the sample means and two-sided uncertainties of the marginalized parameters of the power law model (I exclude the parameters $\Omega_{m0}$, $\Omega_{b0}h^2$, $a$, $b$, $\sigma_{\rm ext}$ from this table because these are nuisance parameters for the power law model). In column 4 I record the sample mean and two-sided uncertainties (computed from the sample mean and two-sided uncertainties of $\beta$) of the current value of the deceleration parameter
\begin{equation}
    \label{eq:q}
    q_0 = \frac{1}{\beta} - 1.
\end{equation}
In column 5 I record $\Delta \chi^2_{\rm min}$, which I define as the difference between the value of $\chi^2_{\rm min}$ as computed within the power law model for a given data combination, and the value of $\chi^2_{\rm min}$ as computed within the \lcdm\ model for the same data combination. The relative probabilities $e^{-\Delta {\rm AICc}/2}$ and $e^{-\Delta {\rm BIC}/2}$ of the power law model I record in columns 6 and 7, where $\Delta$AICc and $\Delta$BIC are defined in the same way as $\Delta \chi^2_{\rm min}$. In columns 2 and 3 of Table \ref{tab:1d_BFP_LCDM}, I record the sample means and two-sided uncertainties of the marginalized parameters of the \lcdm\ model, excluding the nuisance parameters $\Omega_{b0}h^2$, $a$, $b$, and $\sigma_{\rm ext}$. In column 4 of Table \ref{tab:1d_BFP_LCDM} I record the sample mean and two-sided uncertainties (computed from the sample mean and two-sided uncertainties of $\Omega_{m0}$) of the current value of the deceleration parameter
\begin{equation}
    q_0 = \frac{\Omega_{m0}}{2} - \Omega_{\Lambda} = \frac{3}{2}\Omega_{m0} - 1.
\end{equation}
The prior probabilities of all parameters are flat, and non-zero within the ranges $20$ km s$^{-1}$ Mpc$^{-1}$ $\leq H_0 \leq 100$ km s$^{-1}$ Mpc$^{-1}$, $0.25 \leq \beta \leq 4$, $0.1 \leq \Omega_{m0} \leq 0.7$, $0.005 \leq \Omega_{b0}h^2 \leq 0.1$, $40 \leq a \leq 60$, $0 \leq b \leq 5$, and $0 \leq \sigma_{\rm ext} \leq 10$.

\begin{table*}
    \caption{Best-fitting parameters of the \lcdm\ model.}
    \label{tab:LCDM_BFP_Om}
    \centering
    \resizebox{\columnwidth}{!}{%
    \begin{tabular}{ccccccccccc}
    \hline
    \hline
    Data type & $H_0$ (km s$^{-1}$ Mpc$^{-1}$) & $\om$ & $\Omega_{b0}h^2$ & $a$ & $b$ & $\sigma_{\rm ext}$ & $\nu$ & $\chi^2_{\rm min}/\nu$ & AICc & BIC\\
    \hline
         $H(z)$ & 68.15 & 0.3196 & - & - & - & - & 29 & 0.5000 & 18.93 & 21.37\\
         %\hline
         BAO & 74.01 & 0.2967 & 0.03133 & - & - & - & 8 & 1.124 & 18.43 & 16.19\\
         QSO & 68.69 & 0.3154 & - & - & - & - & 118 & 2.983 & 356.1 & 361.6\\
         %\hline
         BAO+QSO & 69.51 & 0.2971 & 0.02459 & - & - & - & 128 & 2.821 & 367.3 & 375.7\\
         %\hline
         GRB & 75.65 & 0.7000 & - & 49.98 & 1.108 & 0.4012 & 114 & 1.141 & 140.6 & 154.0\\
         HIIG & 71.81 & 0.2756 & - & - & - & - & 151 & 2.720 & 414.8 & 420.8\\
         %\hline
         GRB+HIIG & 71.45 & 0.2950 & - & 50.17 & 1.136 & 0.4035 & 267 & 2.031 & 552.5 & 570.3\\
         %\hline
         All Data & 70.07 & 0.2949 & 0.02542 & 50.19 & 1.135 & 0.4040 & 428 & 2.148 & 931.6 & 955.8\\
    \hline
    \hline
    \end{tabular}%
    }
\end{table*}

The two-dimensional confidence contours and one-dimensional likelihoods of the power law model, for several combinations of data, are shown in the left panel of Fig. \ref{fig:ZBQGH_marginalized}. The contours and likelihoods associated with the $H(z)$ data are shown as dotted blue curves, those associated with the BAO + QSO data combination are shown as dash-dotted red curves, those associated with the GRB + HIIG combination are shown as dashed green curves, and those associated with the combination of all the data are shown as solid black curves (I combine the standard ruler and standard candle data in these plots to reduce visual clutter). The right panel of Fig. \ref{fig:ZBQGH_marginalized} shows the two-dimensional confidence contours and one-dimensional likelihoods of the \lcdm\ model, for the same data combinations.

From the marginalized parameter fits in Table \ref{tab:1d_BFP_PL}, I find that the best-fitting value of $\beta$ from the $H(z)$, QSO, GRB, and HIIG data is consistent with $\beta = 1$ to within 1-2$\sigma$, in agreement with many of the studies quoted in Table \ref{tab:beta_fits}. This translates to the best-fitting value of $q_0$ being within 1-2$\sigma$ of $q_0 = 0$ for each of these data sets, consistent with a coasting universe. The BAO data, however, are not consistent with $\beta = 1$, the best-fitting value of $\beta$ for this data set being more than 4$\sigma$ away from unity. This means, as reflected in the best-fitting $q_0$ value, that when the power law model is fitted to the BAO data, these data favor a slowly decelerating universe (rather than a coasting one) to more than 4$\sigma$. The BAO + QSO combination also favors a slowly decelerating universe to more than 4$\sigma$. When these data are combined with the $H(z)$, GRB, and HIIG data, the error bars on $\beta$ and $q_0$ tighten, and the central values of these parameters move slightly closer to $\beta = 1$ and $q_0 = 0$, respectively, though the best-fitting value of $q_0$ is still inconsistent with a coasting universe to more than 3$\sigma$ (see also Fig. \ref{fig:ZBQGH_marginalized}).

\begin{comment}
consistent with a coasting universe \cite{Dev_Sethi_Lohiya_2001, Dev_Safonova_Deepak_Lohiya_2002}
\end{comment}

\begin{comment}
\cite{Dev_Safonova_Deepak_Lohiya_2002, Jain_Dev_Alcaniz_2003, Rani_et_al_2015, Zhu_Alcaniz_Liu_2008} [others?][The studies quoted in Table \ref{tab:beta_fits}
\end{comment}

From Tables \ref{tab:PL_BFP_Om} and \ref{tab:LCDM_BFP_Om}, we can see that the best-fitting power law model has greater $\chi^2/\nu$, AICc, and BIC values than the best-fitting \lcdm\ model across all data combinations, except when these models are fitted to GRB data alone. In this case, the power law model provides a slightly better fit to the data. When we examine the relative probabilities $e^{-\Delta{\rm AICc}/2}$ and $e^{-\Delta{\rm BIC}/2}$ in Table \ref{tab:1d_BFP_PL}, we find that the power law model produces a slightly better fit to the GRB data than does \lcdm. This preference for the power law model over the \lcdm\ model is unique to the GRB data, however, as all other data combinations favor the \lcdm\ model, with the relative probability of the power law model ranging from a high of $0.7047$ (HIIG data) to a low of $4.151 \times 10^{-9}$ (full data set). 

\begin{comment}
[Similar to how Ryskin's model fits supernovae data; compare to other papers that use SN or other standard candle data. Maybe say something about how my collaborators and I are currently looking at SN data; would be interesting to use SN data in conjunction with my other data sets, this will be explored in future work etc. Is Melia's model favored by standard candles?]
\end{comment}

\begin{table*}
    \caption[Marginalized best-fitting parameters and model comparison statistics for the power law model.]{Marginalized best-fitting parameters and model comparison statistics for the power law model. The BAO data alone do not place a tight upper limit on the best-fitting value of $H_0$, and the GRB data do not constrain $H_0$ at all, so these limits are omitted from the table.}
    \label{tab:1d_BFP_PL}
    \centering
    \resizebox{\columnwidth}{!}{%
    \begin{tabular}{ccccccc}
    \hline
    \hline
    Data type & $H_0$ (km s$^{-1}$ Mpc$^{-1}$) & $\beta$ & $q_0$ & $\Delta\chi^2_{\rm min}$ & $e^{-\Delta{\rm AICc}/2}$ & $e^{-\Delta{\rm BIC}/2}$\\
    \hline
         $H(z)$ & $62.46^{+2.693}_{-2.694}$ & $1.013^{+0.06983}_{-0.1038}$ & $-0.01283^{+0.1012}_{-0.06805}$ & 2.090 & 0.3517 & 0.3517\\
         %\hline
         BAO & $72.68_{-8.788}$ & $0.9211^{+0.01653}_{-0.01652}$ & $0.08566_{-0.01948}^{+0.01947}$ & 1.595 & 0.03288 & 0.1360\\
         QSO & $63.22^{+4.088}_{-4.091}$ & $1.045^{+0.1142}_{-0.2054}$ & $-0.04306_{-0.1046}^{+0.1881}$ & 0.9498 & 0.6065 & 0.6065\\
         %\hline
         BAO+QSO & $60.60 \pm 1.108$ & $0.9219^{+0.01645}_{-0.01646}$ & $0.08472_{-0.01946}^{+0.01937}$ & 2.626 & 0.09537 & 0.02352\\
         %\hline
         GRB & - & $0.8707_{-0.2782}^{+0.1197}$ & $0.1485^{+0.3670}_{-0.1579}$ & -0.3534 & 1.162 & 1.162\\
         HIIG & $71.30_{-1.814}^{+1.813}$ & $1.310_{-0.1988}^{+0.1219}$ & $-0.2366^{+0.1158}_{-0.07103}$ & 0.6886 & 0.7047 & 0.7047\\
         %\hline
         GRB+HIIG & $70.58 \pm 1.755$ & $1.199^{+0.1016}_{-0.1535}$ & $-0.1660_{-0.07067}^{+0.1068}$ & 2.209 & 0.3329 & 0.3329\\
         %\hline
         All Data & $63.11^{+0.7886}_{-0.7890}$ & $0.9466^{+0.01593}_{-0.01594}$ & $0.05641_{-0.01778}^{+0.01779}$ & 32.53 & $3.067 \times 10^{-8}$ & $4.151 \times 10^{-9}$\\
    \hline
    \hline
    \end{tabular}%
    }
\end{table*}

\begin{comment}
BAO & $72.68^{+26.87}_{-8.785}$ & $0.9211^{+0.01653}_{-0.01652}$ & $0.08566_{-0.01948}^{+0.01947}$ & 1.595 & 0.03288 & 0.1360\\
GRB & $59.99^{+27.21}_{-27.22}$ & $0.8707_{-0.2782}^{+0.1197}$ & $0.1485^{+0.3670}_{-0.1579}$ & -0.3534 & 1.162 & 1.162\\
\end{comment}

\begin{comment}
In a similar fashion, \cite{Dolgov_Halenka_Tkachev_2014}, \cite{Rani_et_al_2015}, and \cite{Sethi_Dev_Jain_2005} [others; see spreadsheet]
\end{comment}

It is interesting that the best case to be made for the power law model comes from the standard candle data, as the GRB data favor the power law model and the HIIG do not strongly disfavor it. In a similar fashion, \cite{Dolgov_Halenka_Tkachev_2014}, \cite{Rani_et_al_2015}, and \cite{Sethi_Dev_Jain_2005} find that standard candle data (in the form SN Ia measurements) alone do not rule out or do not strongly disfavor the power law model. However, when the GRB and HIIG data are combined, with each other and with the cosmic chronometer and standard ruler data, it is the \lcdm\ model that comes out on top. Cosmic chronometer ($H(z)$) data alone also do not favor the power law model, and neither does the standard ruler (BAO + QSO) combination. Of these three data sets, the QSO set has the least discriminating power, perhaps because of the wide dispersion of the measurements it contains (see the lower left panel of Fig. \ref{fig:Hz/1+z}); as with the HIIG data, the power law model is not strongly ruled out by QSO data alone. The BAO data have the most discriminating power of any solo data set, the fit of the power law model to these data having the smallest relative probabilities compared to \lcdm. This is also true of the standard candle set (BAO + QSO), which gives a smaller relative probability than either the cosmic chronometer or standard candle (GRB + HIIG) set when the power law model is fitted to this data combination. When the power law model is fitted to the full data set, the relative probabilities decrease drastically, to the point that the power law model appears to be very strongly ruled out, at $z \lesssim 8$, in favor of the \lcdm\ model. These results are in broad agreement with the findings of \cite{Rani_et_al_2015}, \cite{Shafer_2016}, \cite{Tutusaus_et_al_2016}, and \cite{Haridasu_AAP_2017}, although they differ somewhat in the details. In particular, using a set of $H(z)$ data that is slightly different from mine, \cite{Rani_et_al_2015} find much stronger evidence against the power law model than I do.\footnote{They quote $\chi^2_{\rm min}/\nu = 1.8131$ for the fit of the power law model to their $H(z)$ data, and $\chi^2_{\rm min}/\nu = 0.7174$ for the fit of the \lcdm\ model to these data, for a difference of $\Delta \chi^2_{\rm min}/\nu = 1.096$. For these same models, I find only $\Delta \chi^2_{\rm min}/\nu = 0.0721$} Both \cite{Shafer_2016} and \cite{Tutusaus_et_al_2016} use BAO data (a smaller set than mine) to evaluate the power law model. Contrary to my results, the BAO measurements they use favor the power law model, although they both find that the power law model is strongly disfavored when BAO data are combined with independent probes (SN Ia in \citealp{Shafer_2016}, and \cite{Tutusaus_et_al_2016} and SN Ia + CMB in \citealp{Tutusaus_et_al_2016}). Using $H(z)$ + BAO + SNe IA + GRB data, \cite{Haridasu_AAP_2017} also find that the power law model is strongly ruled out in favor of \lcdm\ ($\Delta$BIC = 28.02), though their combined data set prefers a slightly larger value of $\beta$ ($1.08 \pm 0.04$) than my combined data set, with larger error bars.

\begin{comment}
[the power law model becomes an indefensible alternative to the \lcdm\ model.]
\end{comment}

\begin{table*}
    \centering
    \caption[Marginalized best-fitting parameters of the \lcdm\ model.]{Marginalized best-fitting parameters of the \lcdm\ model. The BAO data alone do not place a tight upper limit on the value of $H_0$, and the GRB data do not constrain $H_0$ at all, so these limits are excluded from the table.}
    \begin{tabular}{cccc}
    Data type & $H_0$ & $\Omega_{m0}$ & $q_0$\\
    \hline
    \hline
         $H(z)$ & $67.73^{+3.078}_{-3.077}$ & $0.3323^{+0.04983}_{-0.06988}$ & $-0.5016^{+0.07474}_{-0.1048}$\\
         %\hline
         BAO & $83.47_{-4.272}$ & $0.2982^{+0.01547}_{-0.01771}$ & $-0.5527^{+0.2321}_{-0.02657}$\\
         QSO & $67.28^{+4.901}_{-5.039}$ & $0.3642^{+0.08152}_{-0.1503}$ & $-0.4537^{+0.1223}_{-0.2254}$\\
         %\hline
         BAO+QSO & $69.58 \pm 1.379$ & $0.2975^{+0.01529}_{-0.01746}$ & $-0.5538^{+0.02294}_{-0.02619}$\\
         %\hline
         GRB & - & $0.4767_{-0.07217}$ & $-0.2850_{-0.10826}$ \\
         HIIG & $71.70^{+1.819}_{-1.820}$ & $0.2893^{+0.05099}_{-0.07016}$ & $-0.5661^{+0.07649}_{-0.1052}$ \\
         %\hline
         GRB+HIIG & $71.41^{+1.794}_{-1.795}$ & $0.3073^{+0.05140}_{-0.07055}$ & $-0.5391^{+0.07710}_{-0.1058}$ \\
         %\hline
         All Data & $70.13 \pm 0.9590$ & $0.2943^{+0.01368}_{-0.01523}$ & $-0.5586^{+0.2052}_{-0.2284}$ \\
    \hline
    \hline
    \end{tabular}
    \label{tab:1d_BFP_LCDM}
\end{table*}

\begin{comment}
BAO & $83.47^{+16.39}_{-4.272}$ & $0.2982^{+0.01547}_{-0.01771}$ & $-0.5527^{+0.2321}_{-0.02657}$\\
GRB & $60.08 \pm 27.17$ & $0.4767^{+0.2214}_{-0.07217}$ & $-0.2850^{+0.3321}_{-0.10826}$ \\
\end{comment}

That the power law model is ruled out in favor of the \lcdm\ model, from an analysis of $H(z)$, BAO, QSO, GRB, and HIIG data, is a strong statement, and should not be accepted uncritically. Though I believe I have made a good case against the power law model, a few caveats must also be mentioned:

\begin{figure*}
\caption[Constraints on power law and \lcdm\ model parameters.]{The left panel shows one- and two-dimensional constraints on the parameters of the power law model from several combinations of data, and the right panel shows one- and two-dimensional constraints on the \lcdm\ model from the same combinations of data (nuisance parameters excluded).}
\label{fig:ZBQGH_marginalized}
\centering
    \resizebox{\columnwidth}{!}{%
    \includegraphics[scale=1]{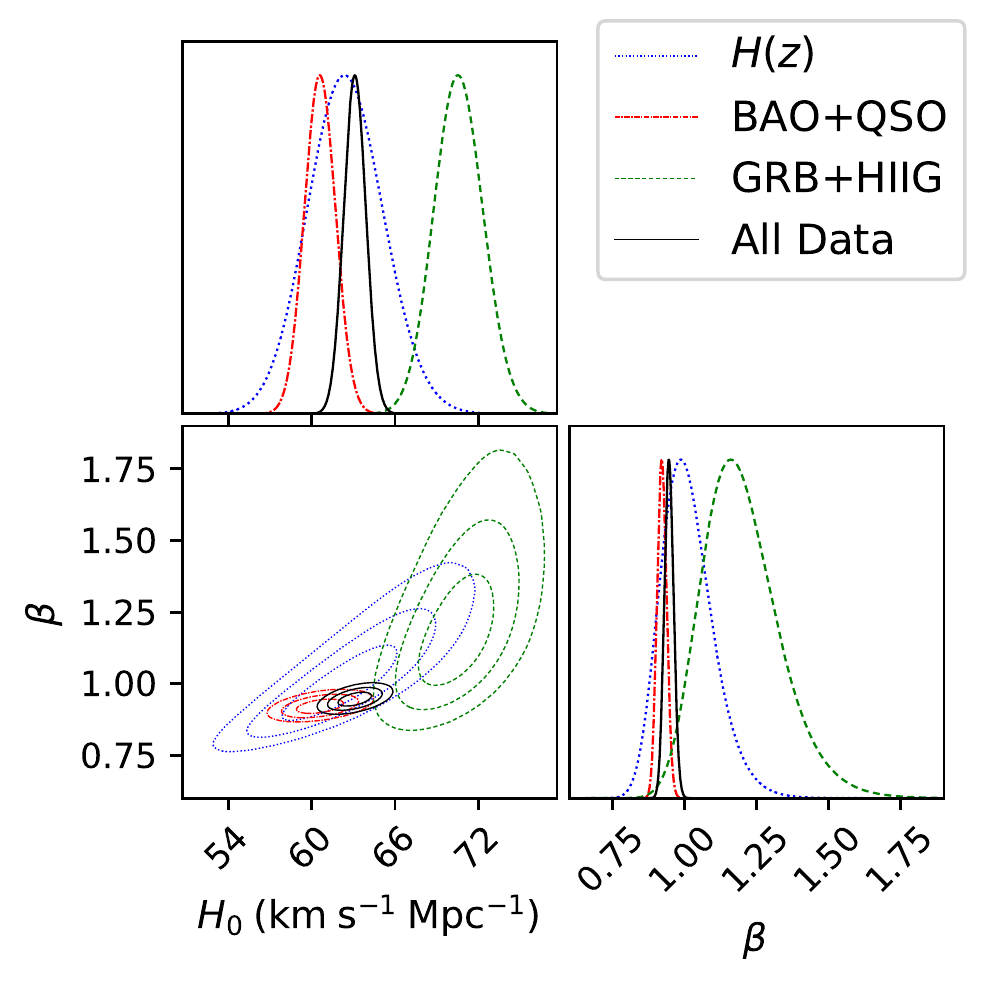}
    \includegraphics[scale=1]{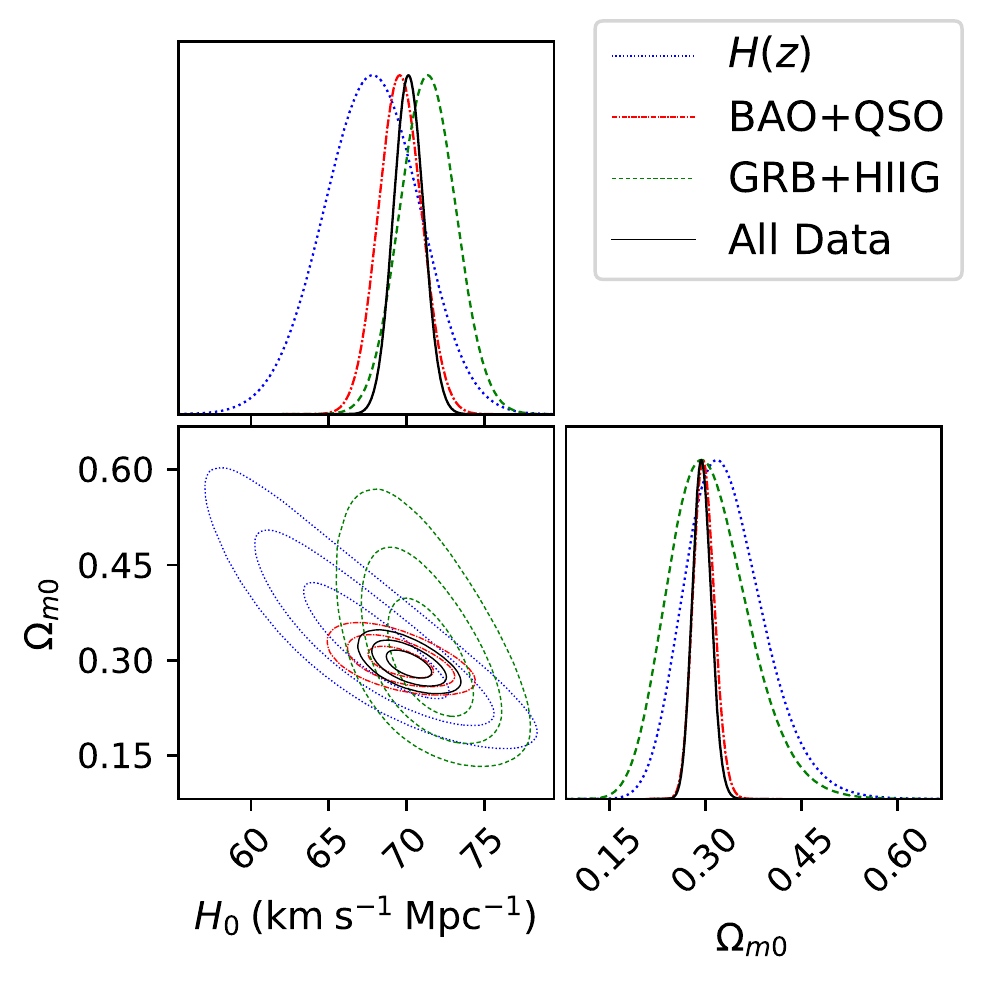}%
    }
\end{figure*}

\begin{comment}
\caption{One- and two-dimensional constraints on the parameters of the power law model. Blue dotted curves correspond to the $H(z)$ data, red dash-dotted curves correspond to the standard ruler (BAO + QSO) data, green dashed curves correspond to the standard candle (GRB + HIIG) data, and the black solid curves correspond to the full data set.}
\end{comment}

1.) The results that are shown in Tables \ref{tab:PL_BFP_Om}-\ref{tab:1d_BFP_LCDM} do not take the finite detection significance of the BAO data into account. As discussed in \cite{Ruiz_et_al_2012}, \cite{Shafer_2016}, and \cite{Tutusaus_et_al_2016}, for a weak BAO signal, one must account for the possibility that the BAO feature in the large-scale matter power spectrum has not actually been detected. To do this, one must replace the standard gaussian $\chi^2$ function $\chi^2_{\rm G} := -2{\rm ln}\mathcal{L}_{\rm G}$ with
\begin{equation}
\label{eq:finite_chi2}
    \chi^2 := \frac{\chi^2_{\rm G}}{\sqrt{1 + \left(\frac{S}{N}\right)^{-4}\chi^4_{\rm G}}},
\end{equation}
where $S/N$, the signal-to-noise ratio, is the detection significance of the BAO feature. As described in Chapter \ref{Chapter9}, three of the BAO measurements I use in this paper are uncorrelated, and the rest are correlated. To test the robustness of my results, I replaced the gaussian likelihoods of the uncorrelated BAO measurements with their counterparts defined by eq. (\ref{eq:finite_chi2}) and performed the BAO analysis again. I found no significant change in the results when I did this, which is perhaps not surprising; in Fig. 11 of \cite{Ruiz_et_al_2012}, the authors show that accounting for the finite detection significance of BAO data only has the effect of widening the confidence contours (primarily the $3\sigma$ contour) a little, and that this widening almost disappears when BAO data are combined with other probes. With that said, I did not investigate the effect of the detection significance of the correlated BAO data on the model fits, and I do not know how large the effect is for these data. However, based on the above considerations as they apply to the uncorrelated BAO data, I do not expect the effect of the detection significance of the uncorrelated BAO data to be a significant factor affecting the validity of my results.

\begin{comment}
2.) The fit to all data combinations ($H(z)$ data excepted), across both models, gives reduced $\chi^2$ values that are all greater than unity (and are, on average, $\geq 2$). The fit to the $H(z)$ data produces, for both the power law and \lcdm\ models, reduced $\chi^2$ values comparable to 0.5. The larger reduced $\chi^2$ values suggest that neither the power law model nor the \lcdm\ model is a particularly good fit to the data (though the reduced $\chi^2$ values of the \lcdm\ model are consistently lower than those of the power law model for all data combinations, GRB alone excepted), or that the uncertainties of these data combinations have been underestimated, or both. The reduced $\chi^2$ values of the $H(z)$ data suggest, [contrariwise][to the contrary] on the other hand, that the uncertainties of these data have been overestimated. This possible overestimation of the $H(z)$ uncertainties has previously been noted by myself and my collaborators (see [cite]), and is [perhaps] apparent in Fig. \ref{fig:Hz/1+z}. [Though I believe that the case against the power law model from my findings is strong ...] [I may be overstating the case with regard to BAO and GRB data, here. Their reduced $\chi^2$ values are mostly around 1.1, so they're probably fine. It's really only the QSO and HIIG data that have too-large reduced chi-squared values; maybe I should focus on these instead].
\end{comment}

2.) The fit to the QSO and HIIG data, across both models, gives reduced $\chi^2$ values that are all $>2$. The fit to the $H(z)$ data gives, for both the power law and \lcdm\ models, reduced $\chi^2$ values comparable to 0.5. The larger reduced $\chi^2$ values suggest that neither the power law model nor the \lcdm\ model is a particularly good fit to the QSO or the HIIG data (though the reduced $\chi^2$ values of the \lcdm\ model are consistently lower for these data than those of the power law model), or that the uncertainties of these data have been underestimated, or both.\footnote{As in Chapters \ref{Chapter7}-\ref{Chapter9}, I only consider the statistical errors of the HIIG data. The systematic uncertainties of the HIIG data are the subject of an ongoing investigation by Roberto Terlevich and his colleagues, the results of which will be published in a forthcoming paper (Roberto Terlevich, private communication, 2021).} The reduced $\chi^2$ values of the $H(z)$ data suggest, on the other hand, that the uncertainties of these data have been overestimated. The possible overestimation of the $H(z)$ uncertainties has previously been noted by myself and my collaborators (see Chapter \ref{Chapter8}), and is perhaps apparent in Fig. \ref{fig:Hz/1+z}. My collaborators and I have also previously noted the possible underestimation of the QSO and HIIG error bars in Chapters \ref{Chapter5}, and \ref{Chapter7}. That the power law model has consistently higher values of $\chi^2_{\rm min}$ for all data sets (GRB excepted) alone and in combination (with the measurements presumably having mostly independent systematics), argues against its validity as a model of cosmic expansion for $z \lesssim 8$, though the argument could be made stronger with a better understanding of the error bars on the measurements.
\begin{comment}
\begin{figure*}
    \centering
    \includegraphics[scale=1.24]{FLCDM_ZBQGH_combined_H0.pdf}
    \caption[One- and two-dimensional constraints on the parameters of the \lcdm\ model.]{One- and two-dimensional constraints on the parameters of the \lcdm\ model. Blue dotted curves correspond to the $H(z)$ data, red dash-dotted curves correspond to the standard ruler (BAO + QSO) data, green dashed curves correspond to the standard candle (GRB + HIIG) data, and the black solid curves correspond to the full data set.}
    \label{fig:FLCDM_ZBQGH_marginalized}
\end{figure*}
\end{comment}

3.) The $H(z)$ data are somewhat correlated with the QSO data. These data are correlated because some cosmic chronometer data were used to obtain the characteristic angular size $l_m$ of the QSO data. As described in \cite{Cao_et_al2017b}, using the Gaussian Process method \citep{Seikel_Clarkson_Smith_2012}, 24 $H(z)$ measurements at $z \leq 1.2$ were interpolated to produce a cosmological model independent Hubble parameter function $H(z)$. This function was then integrated to produce the angular diameter distances used, in conjunction with angular size measurements $\theta_{\rm obs}$, to obtain $l_m = 11.03 \pm 0.25$ pc. This correlation was previously noted in Chapter \ref{Chapter9} and I currently believe that the parameter constraints from QSO data alone are wide enough that the correlation between these data and $H(z)$ data is not significant. With that said, the magnitude of this correlation is not currently known in detail, and a defender of the power law model could point to this as a weakness of my study. One could solve this problem by treating $l_m$ as a free parameter in the cosmological model fits, although this tends to produce parameter constraints that are so wide as to be nearly uninformative.\footnote{Shulei Cao, private communication, 2021.} My collaborators and I are currently working to understand this issue better.

\section{Direct comparison of models to data}

\begin{figure*}
    \caption[Power law and \lcdm\ model predictions compared to data.]{In all panels, the abbreviation ``PL'' denotes the power law model and $h := H_0/(100\hspace{1mm}{\rm km}\hspace{1mm}{\rm s}^{-1}\hspace{1mm}{\rm Mpc}^{-1})$. In the lower right panel the HIIG data are represented by blue dots, and each GRB datum is represented by a purple ``x''.}
    \label{fig:Hz/1+z}
    \centering
    \resizebox{\columnwidth}{!}{%
    \includegraphics[scale=1]{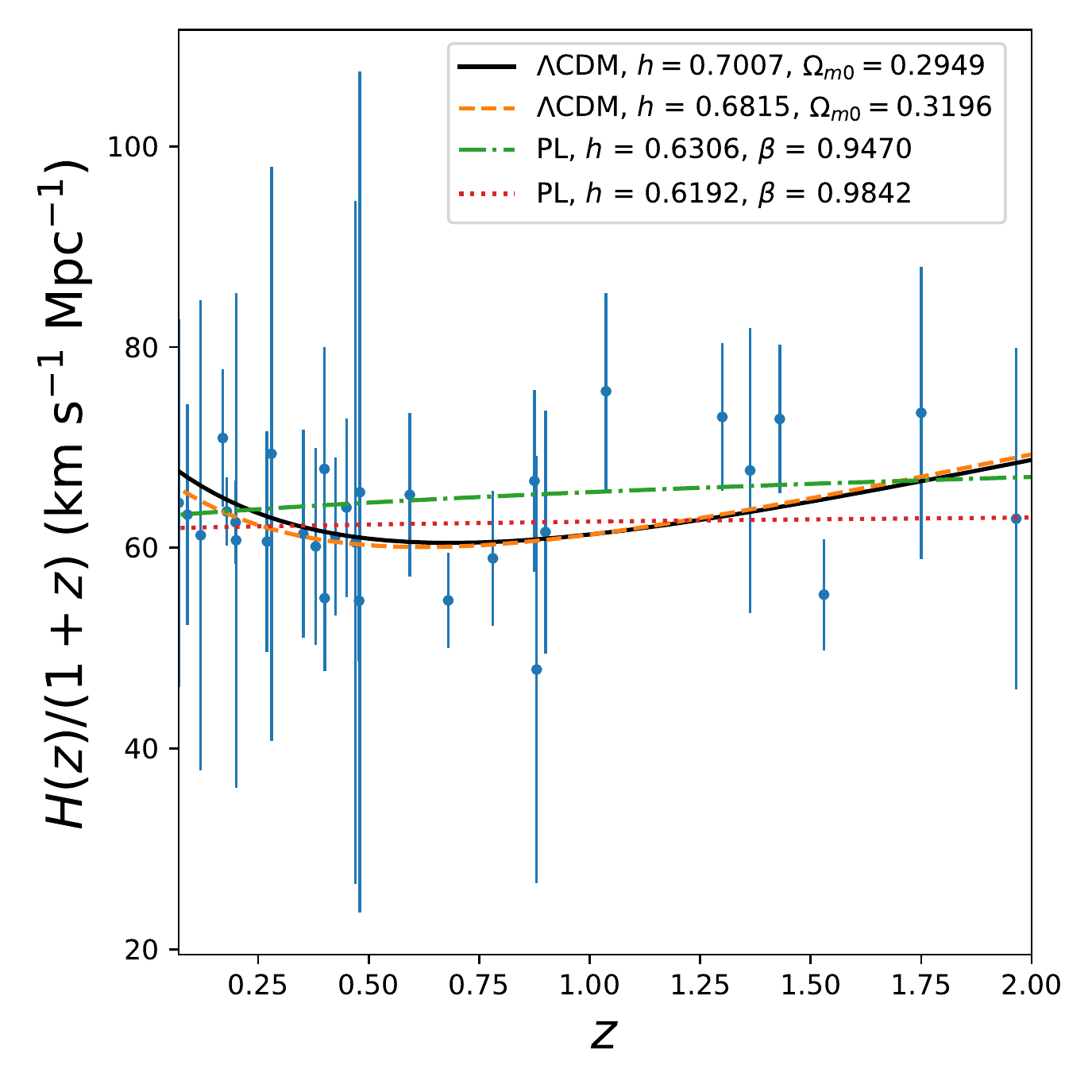}
    \includegraphics[scale=1]{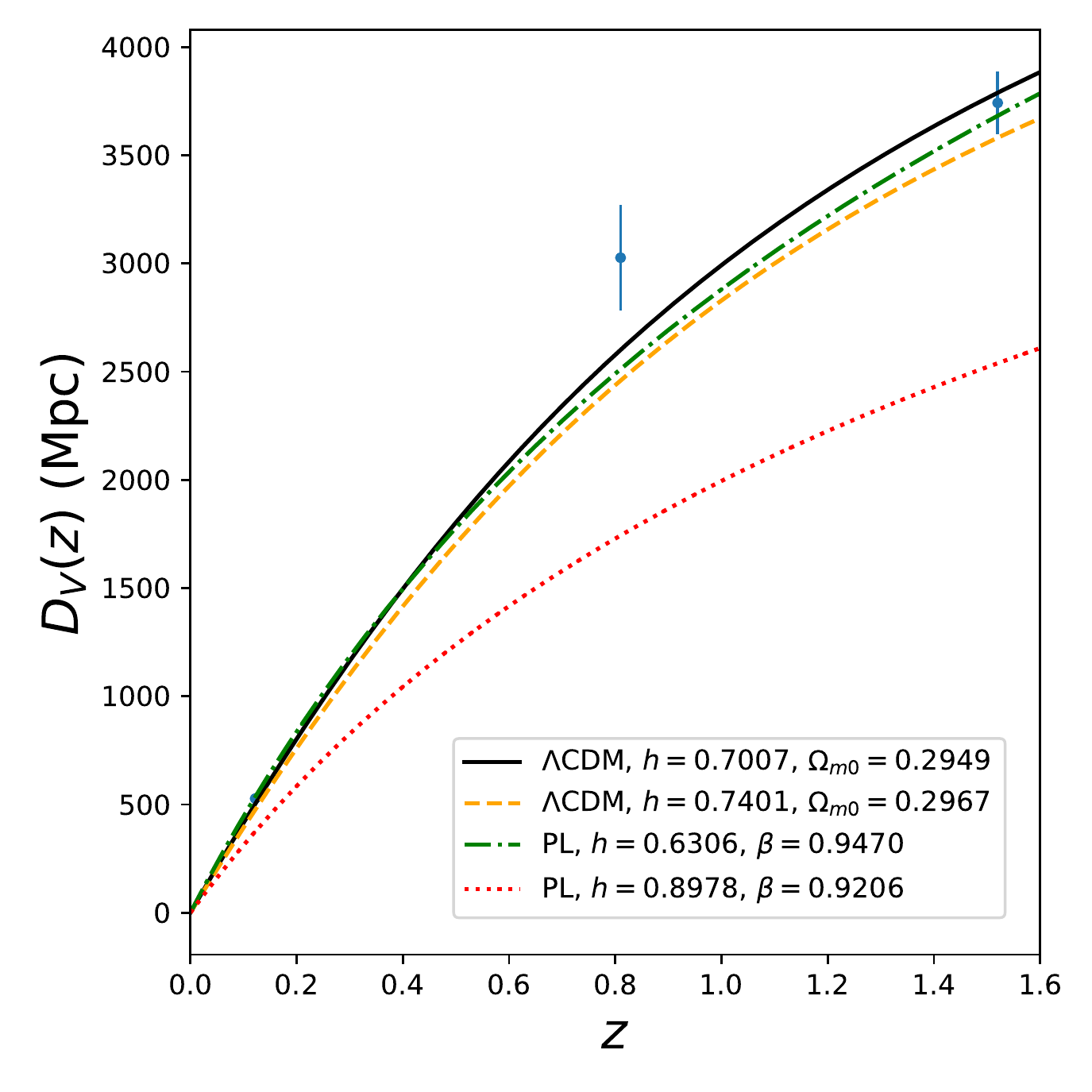}
    }
    \centering
    \resizebox{\columnwidth}{!}{%
    \includegraphics[scale=1]{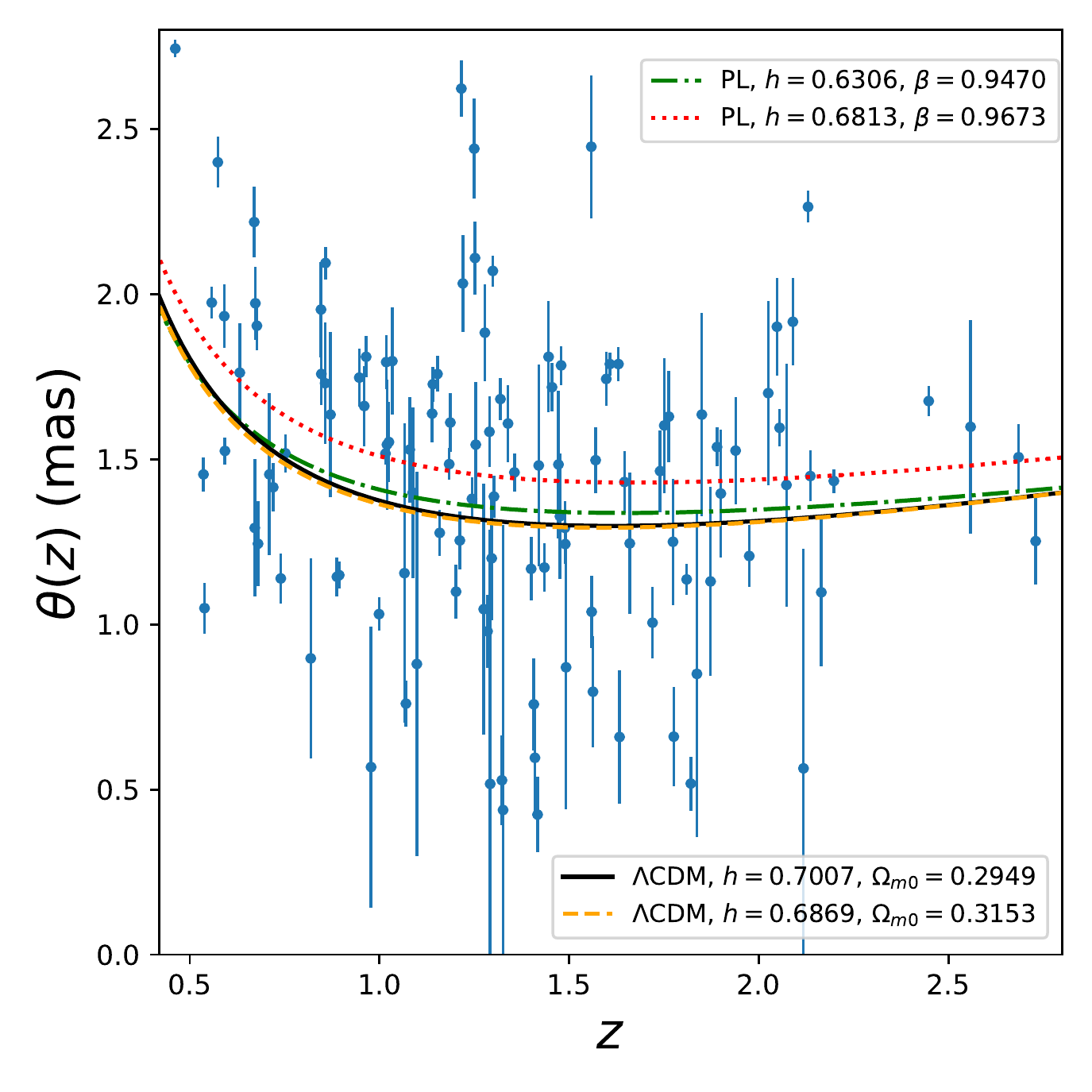}
    \includegraphics[scale=1]{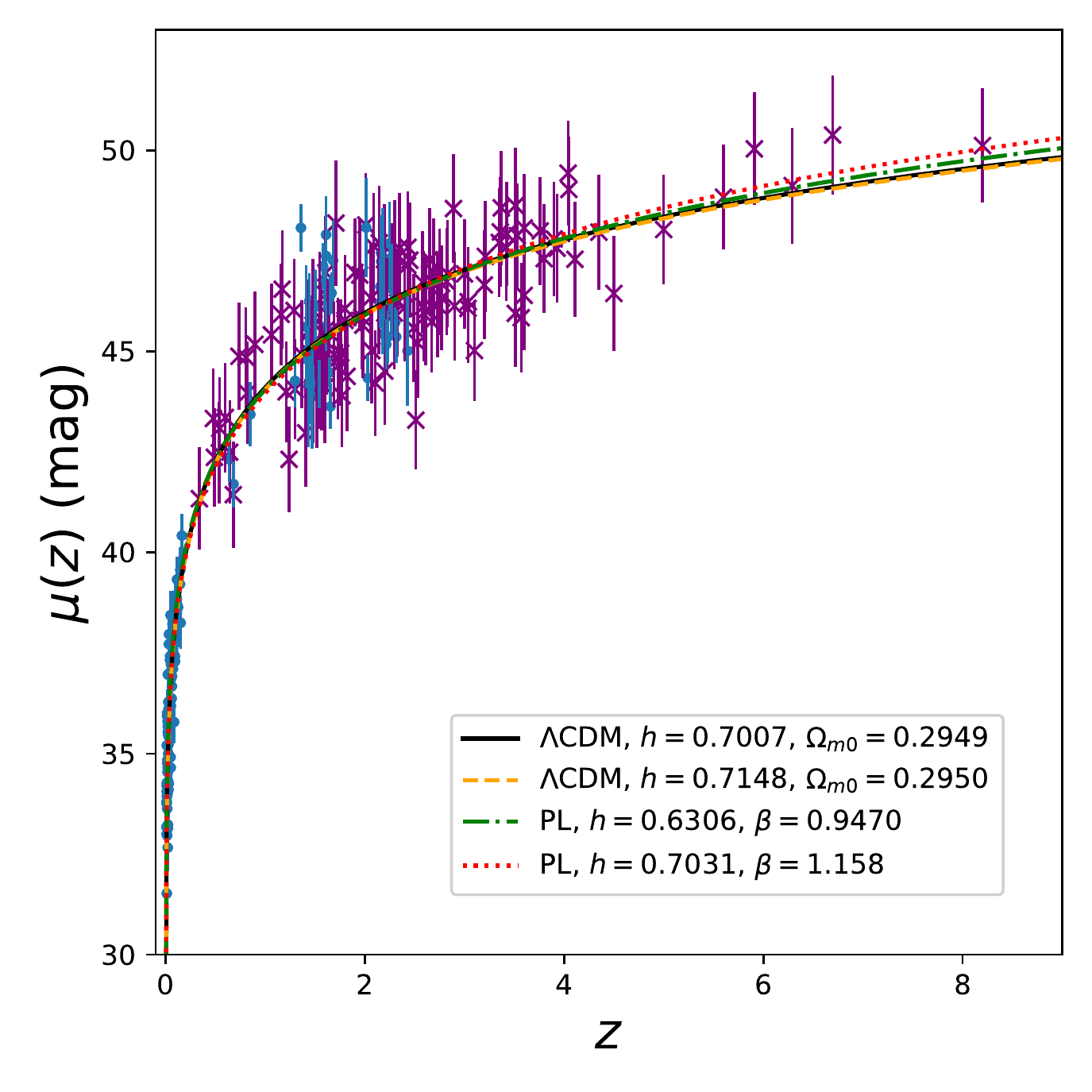}%
    }
\end{figure*}

\begin{comment}
\caption{The black solid curve denotes the \lcdm\ model fitted to the full data set, the orange dashed curve denotes the \lcdm\ model fitted to the $H(z)$ data, the green dash-dotted curve denotes the power law model fitted to the full data set, and the red dotted curve denotes the power law model fitted to the $H(z)$ data.}
\end{comment}

Here I plot the predictions of the power law and \lcdm\ models together with the various data sets I use. In the upper left panel of Fig. \ref{fig:Hz/1+z} I plot $\frac{H(z)}{1 + z}$ versus $z$, where the blue dots represent the $H(z)$ measurements and the curves represent the predicted value of $\frac{H(z)}{1 + z}$, as a function of redshift, for the power law and \lcdm\ models when these models are fitted either to the full data set or to the $H(z)$ data alone. From the figure, we can see that the power law model fails to account for deceleration-acceleration transition which occurs around $z \sim 0.75$.\footnote{For a discussion of the deceleration-acceleration transition, see e.g. \cite{Farooq_Ranjeet_Crandall_Ratra_2017}.} This is backed up by the analyses of \cite{Kumar_2012} and \cite{Rani_et_al_2015} (although a stronger case for this could be made using $H(z)$ data with smaller error bars). 

The upper right panel of Fig. \ref{fig:Hz/1+z} shows a plot of the measured value of the volume-averaged angular diameter distance $D_{\rm V}(z)$, at three different redshifts, from the uncorrelated BAO measurements shown in Table \ref{tab:ch9_BAO}.\footnote{I did not use the correlated measurements because these do not have independent error bars.} To obtain the central value and error bars of $D_{\rm V}(z)$ at $z = 0.81$, I computed
\begin{equation}
    D_{\rm V}(z) = \left[cz(1 + z)^2 \frac{D_{\rm A}(z)^2}{H(z)}\right]^{1/3}
\end{equation}
from the central value and uncertainty of the $D_{\rm A}(z)$ measurement at $z = 0.81$, along with the median central value and median uncertainty of the two $H(z)$ measurements at $z = 0.70$ and $z = 0.90$ from Table \ref{tab:H(z)_data}.\footnote{Although the sound horizon $r_s$ (which sets the scale of the BAO measurements) is a function of the model parameters, I found that both the \lcdm\ and power law models predict $r_s = 144.23$ Mpc when the best-fitting parameters of each model (from the full data set) are used to compute it. This means that the $D_{\rm V}(z)$ data are effectively model-independent.}  The curves shown in the upper right panel of Fig. \ref{fig:Hz/1+z} represent the predicted values of $D_{\rm V}(z)$ for the \lcdm\ and power law models. Although the power law model appears to be ruled out when fitted to the BAO data alone (as it severely under-predicts the values of $D_{\rm V}(z)$ at all redshifts), when it is fitted to the full data set its predictions are nearly indistinguishable (within the error bars of the measurements) from those of the \lcdm\ model. A stronger (or perhaps weaker) case against the power law model, however, could presumably be made with more independent measurements, as the dispersion of the values of $D_{\rm V}(z)$ can not be readily inferred from such a small data set.

\begin{comment}
From this plot it is clear that the predictions of the \lcdm\ and power law models, with regard to $D_{\rm V}(z)$, do not differ as much as the predictions these models make for $H(z)$.
\end{comment}

\begin{comment}
[[REWRITE]From this plot it is clear that the predictions of the power law and \lcdm\ models are very different, and can easily be distinguished with only three measurements. This accounts, in part, for the low relative probability of the power law model seen in Table \ref{tab:1d_BFP_PL} (I say ``in part'' because the uncorrelated measurements are not included here). A stronger (or perhaps weaker) case, however, could be made with more independent measurements, as the dispersion of the values of $D_{\rm V}(z)$ can not be readily inferred from such a small data set.]
\end{comment}
\begin{comment}
\begin{figure*}
    \centering
    \includegraphics[scale=0.6341]{BAO_LCDM_vs_PL.pdf}
    \caption{Volume-averaged angular diameter distance $D_{\rm V}(z)$ versus redshift $z$. The black solid curve denotes the \lcdm\ model fitted to the full data set, the orange dashed curve denotes the \lcdm\ model fitted to the BAO data, the green dash-dotted curve denotes the power law model fitted to the full data set, and the red dotted curve denotes the power law model fitted to the BAO data.}
    \label{fig:BAO}
\end{figure*}
\end{comment}
Large dispersion is a particular problem for the QSO data, as the angular size measurements $\theta(z)$ do not show a clear trend with increasing redshift. This, coupled with the fact that the \lcdm\ and power law model predictions of the angular size are very similar over the range of the QSO data, means that these data do not clearly favor one model over the other (see the lower left panel of Fig. \ref{fig:Hz/1+z}). Similarly, the \lcdm\ and power law model predictions of the distance modulus $\mu(z)$ are almost identical over the redshift range containing the HIIG and GRB data. Although these data show a clear trend with increasing redshift, the predictions of the \lcdm\ and power law models only begin to diverge around $z \approx 4$, a redshift beyond which most of the data lie. These data therefore, like the QSO data, do not strongly favor one model over the other (see the lower right panel of Fig. \ref{fig:Hz/1+z}).\footnote{To plot the GRB data model-independently, I used values of $a$ and $b$ computed from the average of the (nearly model-independent) best-fitting values of those parameters that are listed in the bottom rows of Tables \ref{tab:PL_BFP_Om} and \ref{tab:LCDM_BFP_Om}.}
\begin{comment}
\begin{figure*}
    \centering
    \includegraphics[scale=0.6341]{th_LCDM_vs_PL.pdf}
    \caption{Angular size $\theta(z)$ versus redshift $z$. The black solid curve denotes the \lcdm\ model fitted to the full data set, the orange dashed curve denotes the \lcdm\ model fitted to the QSO data, the green dash-dotted curve denotes the power law model fitted to the full data set, and the red dotted curve denotes the power law model fitted to the QSO data.}
    \label{fig:th}
\end{figure*}

\begin{figure*}
    \centering
    \includegraphics[scale=0.6341]{mu_LCDM_vs_PL.pdf}
    \caption{Distance modulus $\mu(z)$ versus redshift $z$. The black solid curve denotes the \lcdm\ model fitted to the full data set, the orange dashed curve denotes the \lcdm\ model fitted to the HIIG data, the green dash-dotted curve denotes the power law model fitted to the full data set, and the red dotted curve denotes the power law model fitted to the HIIG data. The HIIG data are represented here by blue dots, and the GRB data are each represented by a purple ``x''.}
    \label{fig:mu}
\end{figure*}
\end{comment}
\begin{comment}
\begin{figure}
    \centering
    \includegraphics{DA_LCDM_vs_PL.pdf}
    \caption{$D_{\rm A}(z)$ has units of Mpc.}
    \label{fig:DA}
\end{figure}
\end{comment}

\section{Conclusion}
\label{sec:Conclusion}

In this chapter, I analyzed a set of cosmic chronometer ($H(z)$), standard ruler (BAO and QSO), and standard candle (GRB and HIIG) data to find out whether or not the power law model fits these data as well as or better than the standard \lcdm\ model. Using simple model comparison statistics, similar to what I and many others have used to test alternatives to \lcdm, I found that the power law model does not provide a good fit to the data, compared to \lcdm. The power law model is therefore not a viable candidate to replace the \lcdm\ model at $z \lesssim 8$.

My results are consistent with, and complementary to, other recent studies which have investigated the fit of the power law model to low redshift data. These results, along with the constraints set by primordial nucleosynthesis, show that the simple power law model with a constant exponent $\beta$ does not adequately describe the evolution of the Universe over the course of its history. 

%% file: conclusion.tex
\chapter{Conclusion}

\label{Conclusion}
In this work we examined several models of cosmic acceleration from multiple combinations of standard candle, standard ruler, and cosmic chronometer data. An interesting result is that, across all models and most data combinations, we find somewhat model-independent best-fitting values on the matter density parameter $\Omega_{m0} \approx 0.3 \pm 0.015$, and on the Hubble constant $H_0 \approx 69 \pm 1.4$ km s$^{-1}$ Mpc$^{-1}$ (to compute these, I simply took the average of the summary statistics given in Chapters \ref{Chapter7}-\ref{Chapter9}).

The $H(z)$ + BAO data combination provides relatively tight constraints on the parameters of all the models we studied. QSO-AS additionally tighten these parameter constraints, but not by much (owing to the large error bars, and the wide dispersion, of the measurements). HIIG data, in contrast, can give tighter constraints in certain directions of the parameter space. In particular, HIIG data tighten constraints on dark energy parameters and on the Hubble constant (these effects being most pronounced in the flat $\phi$CDM model). SNe Ia data favor flat spatial hypersurfaces, but are consistent with some dark energy dynamics in the form of $\phi$CDM (these data also favor $\Lambda$CDM). QSO + HIIG more strongly constrain $H_0$ than SNe Ia, but SNe Ia more strongly constrain the matter density parameters and dynamical dark energy parameters. Finally, while GRB measurements do not provide very tight constraints on cosmological model parameters, they do reach farther in redshift space (out to $z \sim 8.2$) than the other probes we used in this work. GRB measurements are consistent with those of the other probes we considered, and so can be combined with them, extending the range of cosmological model parameter constraints into a little-studied region of redshift space. We hope that, with the coming of more GRB observations in the future, these higher-redshift constraints can be used to pin down the dynamics of dark energy accurately and precisely.

Overall, this work shows that as more observational data continue to be developed, the late-time picture of the Universe established by the $\Lambda$CDM model (of an accelerated expansion powered by a cosmological constant $\Lambda$) continues to hold up to scrutiny. While we find some indications of nonflat spatial hypersurfaces and of dynamical dark energy, the current best model of the background expansion remains the $\Lambda$CDM model. Additionally, although more exotic alternatives can be proposed (such as Ryskin's model and the power law model), and while these models may be consistent with a limited set of data occupying a narrow range in redshift, these models do a worse job (in some cases a much worse job) of fitting the larger data sets, covering a broader redshift range, that we have considered here.

That we find, using a large set of observational data, results consistent with large-scale spatial flatness, is significant. Many other studies continue to find broadly similar results, and this lends ever more support to the current inflationary paradigm. There remain, however, some indications in the literature of observational data favoring non-flat spatial hypersurfaces. Continuing development of late-time observational probes, especially of those (like HIIG, QSO-AS, and GRB) currently having wide error bars, will help to settle this question in the future.

%% file: appendixA.tex
% +--------------------------------------------------------------------+
% | Appendix A Page (Optional)                                         
% +--------------------------------------------------------------------+

\cleardoublepage
\chapter{Data}
\label{AppendixA}

In this appendix I list the $H(z)$, QSO, GRB, and SNe Ia data used elsewhere in this work. To obtain the HIIG data, the reader can contact Ana Luisa Gonz\'{a}lez-Mor\'{a}n (see, for example, \citealp{G-M_2019}).

%%
%Section: H(z) data
%%
\section{$H(z)$ data}
\label{sec:A1}

See Chapter \ref{Chapter4} for description and discussion.

\begin{table*}
\centering
\caption[$H(z)$ data.]{$H(z)$ data. $H(z)$ and $\sigma_H$ have units of ${\rm km}\hspace{1mm}{\rm s}^{-1}{\rm Mpc}^{-1}$.}
\label{tab:H(z)_data}
\begin{tabular}{cccc}
\hline
$z$ & $H(z)$ & $\sigma_{H}$ & Ref.\\
\hline
0.07 & 69 & 19.6 & \cite{73}\\
\hline
0.09 & 69 & 12 & \cite{69}\\
\hline
0.12	& 68.6 & 26.2 & \cite{73}\\
\hline
0.17	& 83 & 8 & \cite{69}\\
\hline
0.179	& 75 & 4 & \cite{70}\\
\hline
0.199	& 75 & 5 & \cite{70}\\
\hline
0.20	& 72.9 & 29.6 & \cite{73}\\
\hline
0.27	& 77 & 14 & \cite{69}\\
\hline
0.28	& 88.8 & 36.6 & \cite{73}\\
\hline
0.352	& 83 & 14 & \cite{70}\\
\hline
0.3802 & 83 & 13.5 & \cite{68}\\
\hline
0.4	& 95 & 17 & \cite{69}\\
\hline
0.4004 & 77 & 10.2 & \cite{68}\\
\hline
0.4247 & 87.1 & 11.2 & \cite{68}\\
\hline
0.4497 & 92.8 & 12.9 & \cite{68}\\
\hline
0.47 & 89 & 50 & \cite{Ratsimbazafy_et_al_2017}\\
\hline
0.4783 & 80.9 & 9 & \cite{68}\\
\hline
0.48	& 97 & 62 & \cite{71}\\
\hline
0.593	& 104	& 13 & \cite{70}\\
\hline
0.68	& 92 & 8 & \cite{70}\\
\hline
0.781	& 105	& 12 & \cite{70}\\
\hline
0.875	& 125	& 17 & \cite{70}\\
\hline
0.88	& 90 & 40 & \cite{71}\\
\hline
0.90	& 117 & 23 & \cite{69}\\
\hline
1.037	& 154	& 20 & \cite{70}\\
\hline
1.3	& 168	& 17 & \cite{69}\\
\hline
1.363	& 160	& 33.6 & \cite{72}\\
\hline
1.43	& 177	& 18 & \cite{69}\\
\hline
1.53	& 140	& 14 & \cite{69}\\
\hline
1.75	& 202	& 40 & \cite{69}\\
\hline
1.965	& 186.5 & 50.4 & \cite{72}\\
\hline
\end{tabular}
\end{table*}

\newpage
\begin{comment}
%%
%Section: BAO data
%%
\section{BAO data}
\label{sec:A2}

\begin{table}[h]
\centering
\caption{BAO data. $D_M \left(r_{s,{\rm fid}}/r_s\right)$ and $D_V \left(r_{s,{\rm fid}}/r_s\right)$ have units of Mpc, while $H(z)\left(r_s/r_{s,{\rm fid}}\right)$ has units of ${\rm km}\hspace{1mm}{\rm s}^{-1}{\rm Mpc}^{-1}$ and $r_s$ has units of Mpc.}
\label{tab:BAO_data_old}
\begin{tabular}{ccccc}
\hline
$z$ & Measurement & Value & $\sigma$ & Ref.\\
\hline
$0.38$ & $D_M\left(r_{s,{\rm fid}}/r_s\right)$ & 1518 & 22 & \cite{Alam_et_al_2017}\\
\hline
$0.51$ & $D_M\left(r_{s,{\rm fid}}/r_s\right)$ & 1977 & 27 & \cite{Alam_et_al_2017}\\
\hline
$0.61$ & $D_M\left(r_{s,{\rm fid}}/r_s\right)$ & 2283 & 32 & \cite{Alam_et_al_2017}\\
\hline
$0.38$ & $H(z)\left(r_s/r_{s,{\rm fid}}\right)$ & 81.5 & 1.9 & \cite{Alam_et_al_2017}\\
\hline
$0.51$ & $H(z)\left(r_s/r_{s,{\rm fid}}\right)$ & 90.4 & 1.9 & \cite{Alam_et_al_2017}\\
\hline
$0.61$ & $H(z)\left(r_s/r_{s,{\rm fid}}\right)$ & 97.3 & 2.1 & \cite{Alam_et_al_2017}\\
\hline
$0.106$ & $r_s/D_V$ & 0.336 & 0.015 & \cite{10}\\
\hline
$0.15$ & $D_V\left(r_{s,{\rm fid}}/r_s\right)$ & $664$ & $25$ & \cite{2}\\
\hline
$1.52$ & $D_V\left(r_{s,{\rm fid}}/r_s\right)$ & $3855$ & $170$ & \cite{3}\\
\hline
$2.33$ & $\frac{\left(D_H\right)^{0.7} \left(D_{M}\right)^{0.3}}{r_s}$ & 13.94 & 0.35 & \cite{9}\\
\hline
$2.36$ & $c/\left(r_s H(z)\right)$ & 9.0 & 0.3 & \cite{11}\\
\hline
\end{tabular}
\end{table}
\end{comment}
%%
%Section: QSO data
%%
\section{QSO data}
\label{sec:A3}

See Chapter \ref{Chapter5} for a discussion of the QSO data, which are listed below. The redshift $z$ of each measurement is recorded in the first column, the angular size $\theta$ in milliarcseconds (mas) is recorded in the second column, and the uncertainty $\sigma$ on each measurement is recorded in the third column.

\begin{verbatim}
0.462	2.743	0.027
0.5362	1.454	0.052
0.539	1.049	0.077
0.558	1.974	0.048
0.574	2.399	0.077
0.591	1.933	0.097
0.5928	1.525	0.041
0.632	1.762	0.15
0.67	2.218	0.108
0.6715	1.292	0.208
0.673	1.972	0.11
0.677	1.904	0.075
0.68	1.244	0.129
0.71	1.454	0.245
0.72	1.415	0.074
0.74	1.139	0.076
0.752	1.518	0.058
0.819	0.897	0.304
0.846	1.953	0.144
0.847	1.758	0.094
0.857	1.73	0.183
0.858	2.094	0.049
0.871	1.635	0.249
0.888	1.144	0.059
0.894	1.149	0.041
0.947	1.747	0.087
0.96	1.661	0.121
0.965	1.81	0.063
0.978	0.568	0.426
0.999	1.031	0.051
1.016	1.517	0.033
1.018	1.794	0.081
1.02	1.544	0.197
1.025	1.552	0.122
1.034	1.797	0.163
1.066	1.155	0.453
1.07	0.76	0.07
1.08	1.529	0.16
1.088	1.399	0.258
1.0987	0.88	0.581
1.139	1.638	0.086
1.141	1.727	0.053
1.153	1.758	0.054
1.159	1.277	0.069
1.184	1.485	0.048
1.187	1.611	0.089
1.202	1.099	0.081
1.212	1.254	0.088
1.216	2.622	0.085
1.22	2.032	0.147
1.244	1.38	0.064
1.25	2.44	0.151
1.252	2.109	0.111
1.254	1.544	0.19
1.275	1.046	0.381
1.278	1.883	0.146
1.285	0.979	0.111
1.29	1.583	0.107
1.292	0.517	0.77
1.296	1.2	0.187
1.299	2.07	0.047
1.302	1.387	0.064
1.319	1.682	0.064
1.323	0.528	0.135
1.326	0.438	0.863
1.339	1.608	0.117
1.356	1.46	0.058
1.4	1.168	0.096
1.407	0.758	0.14
1.41	0.596	0.171
1.417	0.424	0.114
1.42	1.481	0.306
1.435	1.172	0.074
1.446	1.81	0.168
1.455	1.718	0.073
1.472	1.484	0.223
1.476	1.327	0.19
1.479	1.784	0.059
1.489	1.292	0.082
1.49	1.243	0.059
1.492	0.87	0.429
1.5586	2.446	0.216
1.5595	1.038	0.109
1.563	0.796	0.168
1.57	1.497	0.1
1.598	1.743	0.081
1.6077	1.788	0.034
1.6301	1.788	0.051
1.633	0.659	0.203
1.646	1.431	0.093
1.66	1.245	0.214
1.72	1.005	0.108
1.74	1.464	0.124
1.751	1.602	0.205
1.763	1.629	0.139
1.774	1.25	0.191
1.776	0.66	0.15
1.81	1.136	0.047
1.821	0.518	0.082
1.837	0.85	0.495
1.85	1.635	0.307
1.873	1.13	0.286
1.89	1.537	0.059
1.9	1.396	0.194
1.939	1.526	0.163
1.975	1.207	0.094
2.025	1.7	0.279
2.048	1.901	0.148
2.055	1.595	0.057
2.073	1.422	0.368
2.09	1.916	0.132
2.118	0.564	0.664
2.13	2.264	0.048
2.136	1.449	0.077
2.165	1.097	0.223
2.198	1.434	0.036
2.448	1.676	0.045
2.5584	1.598	0.323
2.685	1.506	0.101
2.73	1.252	0.132
\end{verbatim}

\section{GRB data}
\label{sec:GRB_data}

In this section I list the GRB data used in Chapters \ref{Chapter8}, \ref{Chapter9}, and \ref{Chapter10}. The first group consists of 25 measurements from Table 2 of \cite{Dirirsa_2019}, and the second group consists of 94 measurements from Table 5 of \cite{Dirirsa_2019}.

In both groups, the first column records the redshift $z$ of each measurement, the second column records the error on the redshift, the third column records the peak energy $E_{\rm p}$ emitted by the GRB, the fourth column is the error on the peak energy, the fourth column records the bolometric fluence divided by $10^5$ ($S_{\rm bolo}/10^{5}$), and the fifth column is the error on $S_{\rm bolo}/10^5$.

\begin{verbatim}
3.51	0     1424.42  35.24    9.24  0.09  
2.53	0     2119.788  119.06  22.40  0.29 
1.406	0     1546.86  37.25    83.54  1.16    
1.17	0     19334.10  652.25  49.91  1.36 
0.807	0     137.84  14.93     0.71  0.03     
2.06	0     2428.51  160.80   8.10  0.17  
1.758	0     985.66  13.20     9.20  0.12     
2.33	0     1320.18  50.90    4.89  0.06  
0.6439	0     370.15  4.97      17.42  0.12    
2.40	0     1163.20  28.54    4.85  0.05  
2.49	0     1601.40  32.19    11.40  0.11 
0.3399	0     294.25  5.86      31.72  0.20    
2.2		0     1214.47  26.24    20.49  0.25 
3.512	0     8675.78  852.66   6.14  0.09     
1.567	0     797.62  18.05     11.74  0.17    
1.368	0     1370.82  27.68    11.88  0.16    
1.063	0     202.63  20.10     0.75  0.04     
0.49	0     60.32  1.93       2.25  0.04  
0.8969	0     857.81  33.08     4.43  0.08     
2.1062	0     868.63  13.85     17.90  0.13    
1.822	0     2146.57   21.71   39.05  0.22    
0.544	0     236.91  4.55      5.72  0.09     
0.736	0     1221.71  81.87    7.99  0.20     
3.57	0     2060.09  138.07   15.76  0.39 
4.35	0.15  6953.87  1188.77  10.40  0.24 
\end{verbatim}

\begin{verbatim}
3.9    0 1783.60  374.85  2.36  0.77 
0.54   0 146.49  23.9     5.75  0.64 
3.6    0 809.60  135.70   0.70  0.07 
1.44   0 312.32  48.8     1.39  0.23 
1.73   0 387.23  244.07   3.56  0.55 
2.22   0 740.60  322.0    3.32  0.68 
1.46   0 223.86  70.11    1.55  0.23 
1.77   0 421.04  13.85    9.32  0.02 
1.61   0 572.25  50.95    2.76  0.21 
0.82   0 218.40  20.93    2.73  0.24 
2.83   0 1164.32  49.79   2.51  0.01 
3.36   0 1117.47  241.11  1.05  0.08 
5.0    0 894.00  240.0    1.06  0.11 
2.89   0 420.44  124.58   0.18  0.03 
0.68   0 519.87  88.88    69.47  8.72
1.73   0 417.38  54.56    4.62  0.59 
1.8    0 129.97  10.27    0.44  0.02 
1.48   0 68.45  18.60     0.15  0.02 
3.8    0 274.33  93.04    0.43  0.07 
2.67   0 157.49  20.92    0.74  0.07 
3.93   0 1651.55  123.25  2.69  0.23 
3.1    0 156.62  0.04     1.59  0.18 
2.2    0 243.20  12.8     0.87  0.07 
0.6    0 247.54  100.61   4.84  0.12 
3.76   0 1003.94  137.98  0.99  0.17 
1.3    0 128.63  6.89     1.73 0.06  
2.27   0 2063.37  101.37  4.56  0.09 
3.6    0 496.80 151.8     1.88  0.25 
5.91   0 2031.54  483.7   0.49  0.09 
2.09   0 911.83  132.65   0.82  0.05 
2.01   0 186.07  31.56    0.08  0.01 
1.16   0 191.80  8.62     0.46  0.04 
0.48   0 81.35  5.92      1.29  0.07 
1.24   0 881.77  24.62    75.21  4.76
1.29   0 405.86  22.93    1.05  0.10 
1.69   0 547.68  83.53    4.75  0.16 
4.04   0 221.85  37.31    0.05  0.01 
2.73   0 447.60  22.38    1.69  0.03 
1.21   0 176.61  4.42     2.53  0.04 
1.44   0 115.00  25.0     0.42  0.03 
1.48   0 766.00  30.0     14.6  1.50 
1.489  0 572.00  143.0    2.30  0.50 
1.52   0 328.00  55.0     6.40  0.60 
1.547  0 934.00  148.0    13.30  1.30
1.547  0 1149.00  166.0   3.48  0.63 
1.563  0 44.00  33.0      0.21  0.06 
1.6    0 1724.0  466.0    35.80  5.80
1.604  0 2077  286        5.32  0.590
1.608  0 1567  384        2.35  0.59 
1.619  0 423.0  42.0      2.60  0.40 
1.6398 0 650.0  55.0      3.40  0.28 
1.71   0 280.0  190.0     0.11  0.034
1.8    0 627.0  65.0      2.027  0.48
1.9    0 290.0  100.0     0.38  0.01 
1.95   0 906.0  272.0     1.50  0.30 
1.9685 0 261.0  52.0      0.96  0.09 
1.98   0 289.0  66.0      1.30  0.10 
2.07   0 310.0  20.0      2.60  0.60
2.14   0 186.0  24.0      0.50  0.06 
2.145  0 1013.0  160.0    0.87  0.40 
2.198  0 415.0  111.0     0.47  0.16 
2.296  0 784.0  285.0     3.40  0.50 
2.3    0 266.0  117.0     0.27  0.04 
2.346  0 539.0  200.0     0.51  0.05 
2.43   0 514.0  102.0     0.73  0.07 
2.433  0 584.0  180.0     0.56  0.14 
2.452  0 2000.0  700.0    3.08  0.53 
2.512  0 47.23  1.08      1.71  0.33 
2.58   0 147.0  14.0      0.27  0.057
2.591  0 1741.0  227.0    7.86  1.37 
2.612  0 1325.0  277.0    6.40  0.50 
2.65   0 128.0  26.0      0.14  0.02 
2.69   0 376.0  100.0     0.64  0.058
2.752  0 230.0  66.0      0.47  0.044
2.77   0 505.0  34.0      1.67  0.17 
2.821  0 1333.0  107.0    3.50  0.20 
2.9    0 467.0  110.0     1.90  0.40 
3.0    0 536.0  172.0     1.09  0.17 
3.036  0 1691.0  226.0    8.96  0.48 
3.038  0 234.0  93.0      0.81  0.095
3.2    0 448.0  148.0     1.20  0.10 
3.21   0 105.0  21.0      0.12  0.06 
3.35   0 1470.0  180.0    1.82  0.20 
3.37   0 270.0  113.0     0.12  0.04 
3.42   0 685.0  133.0     0.87  0.11 
3.425  0 279.0  28.0      0.23  0.04 
3.53   0 285.0  34.0      0.25  0.04 
4.048  0 394.0  46.0      0.14  0.03 
4.109  0 971.0  390.0     1.96  0.38 
4.5    0 987.0  416.0     4.70  0.80 
5.6    0 475.0  47.0      0.27  0.04 
6.29   0 3178  1094.0     2.00  0.20 
6.695  0 710.0  350.0     0.12  0.035
8.2    0 491.0  200.0     0.12  0.032
\end{verbatim}

\section{SNe Ia data}
\label{sec:SNe_Ia_data}

The Pantheon data and covariance matrix files are too large to reproduce here. They can be downloaded from: \url{https://github.com/dscolnic/Pantheon}. For the analysis described in Chapter \ref{Chapter9}, we used ``lcparam\_full\_long\_zhel.txt'' and ``sys\_full\_long.txt''. In ``lcparam\_full\_long\_zhel.txt'', column 1 records the CMB-frame redshift $z_{\rm CMB}$, column 2 records the heliocentric redshift $z_{\rm hel}$, column 4 records the magnitude $m$, and column 5 records the uncertainty on the magnitude $\delta m$. ``sys\_full\_long.txt'' contains the systematic errors on the measurements. For an example of a \textsc{Python} code that can be used to extract the data from these files, see below.

\begin{verbatim}
Pdata = genfromtxt('/homes/jwryan/emcee_stable/SNe_paper/Pantheon/
    lcparam_full_long_zhel.txt', encoding=None, dtype=None)

zcmb = []
zhel = []
m_ob = []
DM_ob = []
for i in range(0, 1048, 1):
    zcmb.append(Pdata[i][1])
    zhel.append(Pdata[i][2])
    m_ob.append(Pdata[i][4])
    DM_ob.append(Pdata[i][5])

sys_err = loadtxt('/homes/jwryan/emcee_stable/SNe_paper/Pantheon/
    sys_full_long.txt', unpack=True)

sys_P = []
V = []
sys_err_count = 0

for i in range(1, 1098305, 1):
    V.append(sys_err[i])
    sys_err_count += 1
    if sys_err_count == 1048:
        sys_P.append(V)
        V = []
        sys_err_count = 0

dm_ob = array(DM_ob)        
CovP = diag(dm_ob**2) + sys_P
CinvP = inv(matrix(CovP)) #This is the inverse of the covariance matrix.
\end{verbatim}

The supernova data from the Dark Energy Survey can be downloaded from: \url{http://desdr-server.ncsa.illinois.edu/despublic/sn_files/y3/tar_files/05-COSMOLOGY.tar.gz}. These files have the same structure as the Pantheon data files. For an example of a \textsc{Python} code that can be used to extract the data from these files, see below.

\begin{verbatim}
Ddata = genfromtxt('/homes/jwryan/emcee_stable/SNe_paper/DES/
    05-COSMOLOGY/COSMOLOGY_INPUTS/lcparam_DESonly.txt', encoding=None,
    dtype=None)

zcmb_D = []
zhel_D = []
m_ob_D = []
DM_ob_D = []
for i in range(0, 20, 1):
    zcmb_D.append(Ddata[i][1])
    zhel_D.append(Ddata[i][2])
    m_ob_D.append(Ddata[i][4])
    DM_ob_D.append(Ddata[i][5])

sys_err = loadtxt('/homes/jwryan/emcee_stable/SNe_paper/DES/
    05-COSMOLOGY/COSMOLOGY_INPUTS/sys_DESonly_ALLSYS.txt',
    unpack=True)

sys_D = []
V = []
sys_err_count = 0

for i in range(1, 401, 1):
    V.append(sys_err[i])
    sys_err_count += 1
    if sys_err_count == 20:
        sys_D.append(V)
        V = []
        sys_err_count = 0

dm_ob_D = array(DM_ob_D)        
CovD = diag(dm_ob_D**2) + sys_D
CinvD = inv(matrix(CovD)) #This is the inverse covariance matrix.
\end{verbatim}

%% file: appendixC.tex
% +--------------------------------------------------------------------+
% | Appendix C Page (Optional)                                         
% +--------------------------------------------------------------------+

\cleardoublepage
\chapter{Codes}
\label{Appendix C}

%%
%Nonflat phiCDM code
%%
\section{Non-flat $\phi$CDM model code}
\label{sec:phiCDM_code}

In this section I provide representative samples of the codes I used to compute constraints on the parameters of the non-flat $\phi$CDM model, in Chapter \ref{Chapter5}, using $H(z)$ and BAO data. Constraints on the parameters of the flat $\phi$CDM model can be obtained from this code by setting $K = 1$. I have not included the codes I used to compute constraints on the parameters of the flat/non-flat $\Lambda$CDM model and the flat/non-flat XCDM parametrization, partly to keep this appendix to a reasonable length, and partly because these codes are fairly simple to write and to run. 

The codes in this section were computed on the Beocat Research Cluster at Kansas State University, and written in the \textsc{Python} language. Below I provide the \textsc{Python} codes, condensed and edited for readability, along with their associated .bash scripts (.bash scripts are used to submit jobs to the cluster via a linux terminal; for more information, see: \url{https://support.beocat.ksu.edu/BeocatDocs/index.php?title=Main_Page}). Each of the .bash scripts begins with ``\#!/bin/sh'', and each of the Python codes begins with ``\# -*- coding: utf-8 -*-''.

Directly below is the .bash script I used to submit my non-flat $\phi$CDM code. It submits the code as an array job in 2500 steps, meaning in this case that the code gets broken into 2500 distinct pieces that run independently. This is a form of parallelization, intended to ensure that the code finishes within a reasonable amount of time, as the non-flat $\phi$CDM likelihood function is very time-expensive to compute.
\begin{verbatim}

#!/bin/sh

#SBATCH --job-name=Nonflat_phiCDM2_array

#SBATCH --array=1-2500:1

#SBATCH --mem-per-cpu=4G   # Memory per core, use --mem= for memory per node
#SBATCH --time=145:00:00   # Use the form DD-HH:MM:SS
#SBATCH --nodes=1
#SBATCH --ntasks-per-node=1

#SBATCH --mail-user=jwryan@phys.ksu.edu
#SBATCH --mail-type=ALL    # same as =BEGIN,FAIL,END

module load Anaconda3

python Nonflat_phiCDM2_fixed.py $SLURM_ARRAY_TASK_ID
\end{verbatim}

Below is the code I used to compute the four-dimensional likelihood function for the non-flat $\phi$CDM model. There is a nested loop at the beginning of the loop that runs over the variables N and Kr. N corresponds to a single step along the $\alpha$ axis of the parameter space, and Kr sets ranges for the curvature scale $K$. Because I run this code as an array job, there are 2500 separate instances of it running simultaneously (in principle). Each instance occupies a single point along the $\alpha$ axis, while stepping through 50 points along the $K$ axis, 61 points along the $\Omega_{m0}$ axis, and 351 points along the $H_0$ axis. Therefore each instance computes the likelihood function 1,070,550 times, independently. This represents a vast improvement in the overall computation time of the code; if the code were not run in parallel, the likelihood function would need to be computed 2,676,375,000 times, serially.

\begin{verbatim}

# -*- coding: utf-8 -*-

import sys

n = sys.argv[1]

count = 0
for N in range(1, 501, 1):
    for Kr in range(0, 5, 1):
        count += 1
        if Kr == 0:
            Ka = -2.10
            Kb = -1.60
        if Kr == 1:
            Ka = -1.60
            Kb = -1.10
        if Kr == 2:
            Ka = -1.10
            Kb = -0.61
        if Kr == 3:
            Ka = -0.60
            Kb = -0.10
        if Kr == 4:
            Ka = -0.10
            Kb = 0.40
        if str(count) == n:
            break
    if str(count) == n:
        break

from scipy.integrate import odeint #ODE solver
import numpy
import math
from math import sin, sqrt
from numpy import savetxt, loadtxt, matrix
from numpy.linalg import inv

z0 = [2.36, 2.33, 1.52, 0.61, 0.51, 0.38, 0.15, 0.106]

DM_obs = [1512.39, 1975.22, 2306.68]
H_obs = [81.2087, 90.9029, 98.9647]
##Data points from DR12 website. File "BAO_consensus_results_dM_Hz.txt"

cHobs = 9.0 #Font-Ribera
dobs = 0.336 #Beutler (Farooq)
B_obs = 13.94 #Bautista

rfid = 147.60 #Planck, Table 4 of 1502.01589.

z_obs, Hz_obs, sigHobs = loadtxt('Hzdata.dat',unpack = True)
##From Table 1 of 1607.03537v2, refs. 4,6,7,10 excluded. H(z) data.

##Covariance matrix from DR12 website. "BAO_consensus_covtot_dM_Hz.txt".
Cov = matrix([[624.707,23.729,325.332,8.34963,157.386,3.57778],
              [23.729,5.60873,11.6429,2.33996,6.39263,0.968056],
              [325.332,11.6429,905.777,29.3392,515.271,14.1013],
              [8.34963,2.33996,29.3392,5.42327,16.1422,2.85334],
              [157.386,6.39263,515.271,16.1422,1375.12,40.4327],
              [3.57778,0.968056,14.1013,2.85334,40.4327,6.25936]])
Cinv = inv(Cov)
del Cov

Ok1 = numpy.array([])
chi2Hzonly = numpy.array([])
chi22B = numpy.array([])
chi2AB = numpy.array([]) 

c = 299792458./1000. #Speed of light in km/s
Th = 2.7255/2.7 #T_CMB/2.7, from Eisenstein and Hu 1998
pi = numpy.pi
m = 1.
n = 5.
t0 = 10.**(-5.)
tf = 150.
dt = t0
t = numpy.arange(0., tf, dt)
a0 = t0**(2./3.) #I assumed a ~ t^(2/3) in the early universe.

def phiCDM(w, t, zz):
    
    p, v, a = w #I used p for phi, v for d(phi)/dt, and a for the
    ##scale factor.
    al, k, m, K = zz #These are the parameters of the model. See below.

    f = [v,
         -3.*v*(((4./(9.*a**3.))) + (1./12.)*(v**2. + (k*m)/(p**(al)))
            - K/(a**2.))**(1./2.)
            + ((k*al*m)/2.)/(p**(al + 1.)), 
        (((4./9.)/a) + (((a**2.)/12.))*(v**2. 
        + (k*m)*(p**(-al))) - K)**(1./2.)]
    return f

def rs(H0, O, Obh2): #Sound horizon. Formula from Eisenstein and Hu, 1998.
    h = H0/100.
    Ob = Obh2/(h**2.) #Baryon density. Fiducial value: 0.02303
    b2 = 0.238*((O*(h**2.))**(0.223))
    b1 = 0.313*((O*(h**2.))**(-0.419))*(1. + (0.607*(O*(h**2))**(0.674)))
    zd = 1291.*(((O*(h**2.))**(0.251))/(1. 
               + (0.659*(O*(h**2.))**(0.828))))*(1. 
               + (b1*((Ob*(h**2.)))**(b2)))
    Rd = 31.5*(Ob*(h**2.))*(Th**(-4.))*(1000./zd)
    zeq = 25000.*(O*(h**2.))*(Th**(-4.))
    Req = 31.5*(Ob*(h**2.))*(Th**(-4.))*(1000./zeq)
    keq = 0.0746*(O*(h**2.))*(Th**(-2.))
    A = (1. + Rd)**(1./2.)
    B = (Rd + Req)**(1./2.)
    C = 1. + (Req)**(1./2.)
    return (2./(3.*keq))*((6./Req)**(1./2.))*(numpy.log((A + B)/C))

def E(O, red, Ok_0, Ophiz):
    return (O*((1. + red)**3.) + (Ok_0)*((1. + red)**2.) + Ophiz)**(1./2.)
    
def D_M(H0, q, O, Ok_0):
    if Ok_0 == 0.:
        return (c/H0)*h0*afin*rr[q]
    if Ok_0 < 0.:
        return (c/H0)*(1/(sqrt(-Ok_0)))*(sin((sqrt(-Ok_0))*(h0*afin*rr[q])))
    if Ok_0 > 0.:
        return (c/H0)*(1/(sqrt(Ok_0)))*(sinh((sqrt(Ok_0))*(h0*afin*rr[q])))

def chi_sq(H0, O, Ok_0):
    rsc = rfid/(rs(67.81, 0.308, 0.02226))
    DM_th = []
    H_th = []
    for q in range(7, -1, -1):
        z1 = z0[q]
        H1 = H0*E(O, z1, Ok_0, O_phi_z[q])
        DM = D_M(H0, q, O, Ok_0)
        y = (H0/c)*DM
        if 3 <= q <= 5:
            DM_th.append(D_M(H0, q, O, Ok_0))
            H_th.append(H1)
        if z1 == 0.15:
            r = rs(H0, O, 0.02303)/rs(67, 0.31, 0.048*0.67*0.67)
            DV = (c/H0)*(((y**2.)*z1)/(E(O, z1, Ok_0, O_phi_z[q])))**(1./3.)
            DVobs1 = 664.*r #Ross
            unc1 = 25.*r
            chi2DV1 = ((DV - DVobs1)**2.)/(unc1**2.)
        if z1 == 0.106:
            DV = (c/H0)*(((y**2.)*z1)/(E(O, z1, Ok_0, O_phi_z[q])))**(1./3.)
            dth = (rs(H0, O, 0.02303)*rsc)/DV #Distilled parameter
            chi2d = ((dth - dobs)**2.)/(0.015**2.)
        if z1 == 1.52:
            r = rs(H0, O, 0.02303)/rs(67.6, 0.31, 0.022)
            DV = (c/H0)*(((y**2.)*z1)/(E(O, z1, Ok_0, O_phi_z[q])))**(1./3.)
            DVobs2 = 3843.*r #Ata
            unc2 = 147.*r
            chi2DV2 = ((DV - DVobs2)**2.)/(unc2**2.)
        if z1 == 2.33:
            DH = c/H1
            F = DH**(0.7)
            G = DM**(0.3)
            B_th = (F*G)/(rs(H0, O, 0.02303)*rsc)
            chi2B = ((B_th - B_obs)**2.)/(0.35**2.)
        if z1 == 2.36:
            cHth = (c/(rs(H0, O, 0.02303)*rsc))*(1./H1)
            chi2H = ((cHth - cHobs)**2.)/(0.3**2.)
    r = rs(H0, O, 0.02303)/rs(67.6, 0.31, 0.022)
    Delta = numpy.array([(DM_th[0]/r - DM_obs[0]), 
                         (r*H_th[0] - H_obs[0]),
                         (DM_th[1]/r - DM_obs[1]),
                         (r*H_th[1] - H_obs[1]),
                         (DM_th[2]/r - DM_obs[2]),
                         (r*H_th[2] - H_obs[2])]) 
    prod = (Cinv.dot(Delta)).T
    chi_sq1 = Delta.dot(prod)
    chi_sq_11 = chi_sq1.item((0,0))
    return ((chi2H + chi2B + chi2d + chi_sq_11 + chi2DV1 + chi2DV2),
        (chi2H + chi2B))

def chi2h(H0, O, Ok_0, Ophiz):
    return ((1/sigHobs)*(H0*E(O, z_obs, Ok_0, Ophiz) - Hz_obs))**2   

def Ofunc(d):
    Omegam1 = (4./9.)*(1./(sol[d,2])**3.)
    Omegaphi1 = (1./12.)*(((sol[d,1])**2.) + k/((sol[d,0])**al))
    Omegak1 = -K/((sol[d,2])**2.)
    Omegam = Omegam1/(Omegam1 + Omegaphi1 + Omegak1)
    return Omegam, Omegam1, Omegaphi1, Omegak1

KK = numpy.arange(Ka, Kb, 0.01)
al = N/100.
Om = numpy.arange(0.10, 0.71, 0.01)

k = (8./3.)*((al + 4.)/(al + 2.))*(((2./3.)*(al*(al + 2.))))**(al/2.) #This 
##is kappa, from eq. (2) of arXiv:1307.7399v1.

##Initial conditions on phi, d(phi)/dt, a.
p0 = (((2./3.)*(al*(al + 2.)))**(1./2.))*(t0)**(2./(al + 2.)) #Initial value
##of phi.
v0 = (((8./3.)*al*(1./(al + 2.)))**(1./2.))/(t0)**(al/(al + 2.)) 
##Initial value of d(phi)/dt.
w0 = [p0, v0, a0]    

for K in KK:
    zz = [al, k, m, K]
        
    #solution array
    sol = odeint(phiCDM, w0, t, args=(zz,))
          
    for O in Om:
        
        for b in range(0, int(tf/dt), 1):
            if (O >= Ofunc(b)[0]):
                break
        afin = sol[b,2]
        h0 = (sol[b+1, 2] - afin)/afin
        Ok_0 = Ofunc(b)[3]/(Ofunc(b)[1] + Ofunc(b)[2] + Ofunc(b)[3])

        Omegap = ((4./9.)*(1./(sol[b,2])**3.) 
                + (1./12.)*(((sol[b,1])**2.) + k/((sol[b,0])**al)) 
                + -K/((sol[b,2])**2.))
        rr = []
        O_phi_z = []
        O_phi_z1 = []
        
        for z1 in z0:
            qt = 0
            t1 = []
            r1 = []
            for d in range(0, b+1, 1):
                if ((sol[d,2])/afin >= 1./(1. + z1)):
                    t1.append(d)
                    r1.append(1./(sol[d,2]))
                    qt += 1
                if qt == 1:
                    Omegaphi_z = (1./12.)*(((sol[d,1])**2.) 
                    + k/((sol[d,0])**al))
            if qt == 1:
                rr.append(sum(r1))
            else:
                rr.append(numpy.trapz(r1, t1))
            O_phi_z.append(Omegaphi_z/((4./9.)*(1./(sol[b,2])**3.) 
            + (1./12.)*(((sol[b,1])**2.) + k/((sol[b,0])**al)) 
            + -K/((sol[b,2])**2.)))                
        
        for qs in range(0, len(z_obs), 1):
            for d in range(0, b+1, 1):
                if ((sol[d,2])/afin >= 1./(1. + z_obs[qs])):
                    Omegaphi2 = (1./12.)*(((sol[d,1])**2.) 
                    + k/((sol[d,0])**al))
                    O_phi_z1.append(Omegaphi2/((4./9.)*(1./(sol[b,2])**3.) 
                    + (1./12.)*(((sol[b,1])**2.) + k/((sol[b,0])**al)) 
                    + -K/((sol[b,2])**2.)))
                    break
        
        for H0 in numpy.arange(50., 85.1, 0.1):
            chi2hh = sum(chi2h(H0, O, Ok_0, O_phi_z1))   
            chi2Hzonly = numpy.append(chi2Hzonly, chi2hh)
            chi22B = numpy.append(chi22B, chi_sq(H0, O, Ok_0)[1])
            chi2AB = numpy.append(chi2AB, chi_sq(H0, O, Ok_0)[0])
            Ok1 = numpy.append(Ok1, Ok_0)
            
#savetxt('phiCDM_curved_alpha_' + str(NN) + '.dat', a1)
#savetxt('phiCDM_curved_Omega_m0_' + str(NN) + '.dat', O1)
savetxt('phiCDM_curved_Omega_k0_' + str(count) + '.dat', Ok1)
#savetxt('phiCDM_curved_chi^2_H(z)only' + str(count) + '.dat', chi2Hzonly)
#savetxt('phiCDM_curved_chi^2_H(z)_2B' + str(count) + '.dat', chi22B)
savetxt('phiCDM_curved_chi^2_AB_' + str(count) + '.dat', chi2AB)
\end{verbatim}

The above code produces 2500 output files, so it is necessary to combine them before they can be analyzed. Due to time and memory limitations, I found it necessary to do this in a somewhat complicated fashion. In the first step of this process, I used the code ``data\_compiler\_3.py'' to condense the original set of 2500 files to a set of 500 files by adding the files from the set of 2500 together in groups of 5. Basically, this means that the reduced set of 500 files corresponds to a single step along the $\alpha$ axis, with 250 steps along the $K$ axis, 61 steps along the $\Omega_{m0}$ axis, and 351 steps along the $H_0$ axis. The .bash script associated with this code is below, followed by the code itself.

\begin{verbatim}

#!/bin/sh

#SBATCH --job-name=data_compiler_nonflat_phiCDM

#SBATCH --array=1-2:1

#SBATCH --mem-per-cpu=4G   # Memory per core, use --mem= for memory per node
#SBATCH --time=145:00:00   # Use the form DD-HH:MM:SS
#SBATCH --nodes=1
#SBATCH --ntasks-per-node=1

#SBATCH --mail-user=jwryan@phys.ksu.edu
#SBATCH --mail-type=ALL    # same as =BEGIN,FAIL,END

module load Anaconda3

python data_compiler_3.py $SLURM_ARRAY_TASK_ID
\end{verbatim}

\begin{verbatim}

# -*- coding: utf-8 -*-

from numpy import savetxt, loadtxt
import numpy
import sys

n = sys.argv[1]
syscount = 0

for M in range(0, 2, 1):
    syscount += 1
    if str(syscount) == n:
        break

l = ['phiCDM_curved_chi^2_AB_',
     'phiCDM_curved_Omega_k0_']
 
l2 = ['Nonflat_phiCDM_chi2_AB_',
      'Nonflat_phiCDM_Omega_k0_']

N = 2500 #number of files
x = numpy.array([])
c = 0
d = 0
    
for N in range(1, N+1, 1):
    c += 1
    a = loadtxt(l[M] + str(N) + '.dat', unpack=True)
    x = numpy.append(x, a)
    del a
    
    if c == 5:
        d += 1
        #print(d)
        savetxt(l2[M] + str(d) + '.dat', x)
        x = numpy.array([])
        c = 0
\end{verbatim}

Next, the output files are marginalized, over one parameter at a time. The .bash script for this is below.

\begin{verbatim}

#!/bin/sh

#SBATCH --job-name=4D_to_3D

#SBATCH --array=1-1500:1

#SBATCH --mem-per-cpu=1G   # Memory per core, use --mem= for memory per node
#SBATCH --time=26:00:00   # Use the form DD-HH:MM:SS
#SBATCH --nodes=1
#SBATCH --ntasks-per-node=1

#SBATCH --mail-user=jwryan@phys.ksu.edu
#SBATCH --mail-type=ALL    # same as =BEGIN,FAIL,END

module load Anaconda3/5.0.1

python 4D_to_3D.py $SLURM_ARRAY_TASK_ID
\end{verbatim}

What follows is the text of ``4D\_to\_3D.py''. This code is run as an array job in 1500 steps because at this point in the process the likelihood files have not yet been compiled along the $\alpha$ axis, and because the files have also been interpolated along the $\Omega_{k0}$ axis. Essentially, I ran this code twice. In the first instance, I ran those portions of the code contained within the quoted comments (containing the Nearest Neighbor interpolating function) to interpolate the likelihood function along the $\Omega_{k0}$ axis. I found it necessary to do this because a uniformly-spaced set of $K$ inputs does not necessarily correspond to a uniformly-spaced set of $\Omega_{k0}$ outputs. I chose to interpolate along the $\Omega_{k0}$ axis in 300 steps because this seemed to strike a good balance between computation time and fineness of resolution. I then ran the code a second time, feeding into it the interpolated likelihood function files (those having `Rescale' in their names), to obtain a set of files correspond to a marginalized likelihood function (in three dimensions rather than four).

\begin{verbatim}

# -*- coding: utf-8 -*-

from numpy import exp, split, dstack, array, trapz, loadtxt, savetxt, 
from numpy import arange, linspace
from scipy.interpolate import NearestNDInterpolator as ND
import sys

Oml = 61
b = 500 #alpha length
Okl = 300 #Omega_k0 length
H0l = 351

n = sys.argv[1]
syscount = 0

for N in range(1, 501, 1):
    for Hcase in range(1, 2, 1):
        for mcase in range(1, 4, 1):
            syscount += 1
            if str(syscount) == n:
                break
        if str(syscount) == n:
            break
    if str(syscount) == n:
        break

#alpha = []
O_m = []
H_0 = []
for N2 in range(1, 301, 1):
    for N3 in arange(0.10, 0.71, 0.01):
        for N4 in arange(50, 85.001, 0.1):
            #alpha.append(N)
            O_m.append(N3)
            H_0.append(N4)
            
O_m0 = array(O_m)
H0 = array(H_0)
del O_m

if Hcase == 1:
    HC = 'AB'
if Hcase == 2:
    HC = 'Hz_2B'
if Hcase == 3:
    HC = 'Hz_AB'

L = loadtxt('Rescale_Nonflat_phiCDM_L_' + HC + '_' + str(N) + '.dat',
unpack=True)
O_k0 = loadtxt('Rescale_Nonflat_phiCDM_Omega_k0_' + HC + '_' + str(N)
    + '.dat',
    unpack=True)
#L = exp(-chi2/2)
#del chi2
    
if mcase == 1: #Marginalize over loop 2
    a = split(H0, Okl)
    t = split(O_m0, Okl)
    v = split(L, Okl)
    
    del O_m0, L, H0
    
    c = []
    d = []
    e = []
    
    for p in range(0, Okl, 1):
        c.append((dstack(split(a[p], Oml))).flatten())
        d.append((dstack(split(t[p], Oml))).flatten())
        e.append((dstack(split(v[p], Oml))).flatten())
    
    del a, t, v
    H02 = (array(c)).flatten()
    Om2 = (array(d)).flatten()
    L2 = (array(e)).flatten()
    del c, d, e
    """
    Ok_int = []
    Om_int = []
    L_int = []
    H0_int = []
    
    f = ND((O_k0, H02, Om2), L2, rescale=True)
    for p in linspace(min(O_k0), max(O_k0), 300):
        for q in arange(50, 85.001, 0.1):
            for r in arange(0.10, 0.71, 0.01):
                Ok_int.append(p)
                Om_int.append(r)
                H0_int.append(q)
                L_int.append(f(p, q, r))
                
    del H02, Om2, L2
    """        
    Omk = []
    Om = []
    L = []
    L3 = []
    alp = []
    H0 = []
    
    c = 0
    for q in range(0, len(L2), 1):
        c += 1
        Om.append(Om2[q])
        L3.append(L2[q])
        if c == Oml:
            L.append(abs(trapz(L3, Om)))
            Omk.append(O_k0[q])
            H0.append(H02[q])
            alp.append(N/100)
            c = 0
            Om = []
            L3 = []
            
    savetxt('3D_' + HC + '_Nonflat_phiCDM_Omega_k0_mcase1_' + str(N)
    + '.dat', Omk)
    savetxt('3D_' + HC + '_Nonflat_phiCDM_L_mcase1_' + str(N) + '.dat', L)
    savetxt('3D_' + HC + '_Nonflat_phiCDM_alpha_mcase1_' + str(N) + '.dat',
    alp)
    savetxt('3D_' + HC + '_Nonflat_phiCDM_H0_mcase1_' + str(N) + '.dat', H0)
        
if mcase == 2: #Marginalize over loop 3
    """
    Ok_int = []
    Om_int = []
    L_int = []
    alp_int = []
    H0_int = []
    
    f = ND((O_k0, O_m0, H_0), L, rescale=True)
    for p in linspace(min(O_k0), max(O_k0), 300):
        for r in arange(0.10, 0.71, 0.01):
            for q in arange(50, 85.001, 0.1):
                Ok_int.append(p)
                Om_int.append(r)
                L_int.append(f(p, r, q))
                H0_int.append(q)
            
    Ok = []
    Om = []
    chi_2 = []

    del O_m0, O_k0, f
    """  
    
    c = 0
    Ok1 = []
    Om1 = []
    L2 = []
    H02 = []
    al1 = []
    L3 = []

    for q in range(0, len(L), 1):
        c += 1
        H02.append(H0[q])
        L2.append(L[q])
        
        if c == H0l:
            L3.append(abs(trapz(L2, H02)))
            Ok1.append(O_k0[q])
            Om1.append(O_m0[q])
            al1.append(N/100)
            c = 0
            H02 = []
            L2 = []
            
    savetxt('3D_' + HC + '_Nonflat_phiCDM_alpha_mcase2_' + str(N) + '.dat',
        al1)
    savetxt('3D_' + HC + '_Nonflat_phiCDM_Omega_k0_mcase2_' 
        + str(N) + '.dat', Ok1)
    savetxt('3D_' + HC + '_Nonflat_phiCDM_L_mcase2_' + str(N) 
        + '.dat', L3)
    savetxt('3D_' + HC + '_Nonflat_phiCDM_Omega_m0_mcase2_' 
        + str(N) + '.dat', Om1)
    
if mcase == 3: #Marginalizing over loop 1
    a = split(O_m0, Okl)
    t = split(O_k0, Okl)
    v = split(L, Okl)
    w = split(H0, Okl)
    
    del O_k0, O_m0, L, H0        
    
    c = []
    d = []
    e = []
    g = []
    for p in range(0, Okl, 1):
        c.append((dstack(split(a[p], Oml))).flatten())
        d.append((dstack(split(t[p], Oml))).flatten())
        e.append((dstack(split(v[p], Oml))).flatten())
        g.append((dstack(split(w[p], Oml))).flatten())
        
    del a, t, v, w

    Om2 = dstack(split((array(c)).flatten(), Okl)).flatten()
    Ok2 = dstack(split((array(d)).flatten(), Okl)).flatten()
    L2 = dstack(split((array(e)).flatten(), Okl)).flatten()
    H2 = dstack(split((array(g)).flatten(), Okl)).flatten()
    
    del c, d, e, g
    """
    Ok_int = []
    Om_int = []
    L_int = []
    H0_int = []
    
    f = ND((Ok2, Om2, H2), L2, rescale=True)
    for q in arange(50, 85.001, 0.1):
        for r in arange(0.10, 0.71, 0.01):
            for p in linspace(min(Ok2), max(Ok2), 300):
                Ok_int.append(p)
                Om_int.append(r)
                L_int.append(f(q,r,p))
                H0_int.append(q)
    """
    Ok3 = []
    Om3 = []
    H3 = []
    L3 = []
    L4 = []
    al = []
    
    c = 0
    for q in range(0, len(L2), 1):
        c += 1
        if abs(Ok2[q]) <= 0.50:
            Ok3.append(Ok2[q])
            L3.append(L2[q])
        
        if c == 300:
            H3.append(H2[q])
            Om3.append(Om2[q])
            L4.append(abs(trapz(L3, Ok3)))
            al.append(N/100)
            c = 0
            Ok3 = []
            L3 = []   
            
    savetxt('3D_' + HC + '_Nonflat_phiCDM_alpha_mcase3_' 
        + str(N) + '.dat', al)
    savetxt('3D_' + HC + '_Nonflat_phiCDM_H0_mcase3_' + str(N) + '.dat', H3)
    savetxt('3D_' + HC + '_Nonflat_phiCDM_L_mcase3_' + str(N) + '.dat', L4)
    savetxt('3D_' + HC + '_Nonflat_phiCDM_Omega_m0_mcase3_'
        + str(N) + '.dat', Om3)
\end{verbatim}

At this point in the process, the likelihood files are small enough that they can be compiled along the $\alpha$ axis. ``chi2\_combiner.py'', below, does this.

\begin{verbatim}

#!/bin/sh

#SBATCH --job-name=chi2_combiner

#SBATCH --array=1-9:1

#SBATCH --mem-per-cpu=1G   # Memory per core, use --mem= for memory per node
#SBATCH --time=23:00:00   # Use the form DD-HH:MM:SS
#SBATCH --nodes=1
#SBATCH --ntasks-per-node=1

#SBATCH --mail-user=jwryan@phys.ksu.edu
#SBATCH --mail-type=ALL    # same as =BEGIN,FAIL,END

module load Anaconda3

python chi2_combiner.py $SLURM_ARRAY_TASK_ID
\end{verbatim}

\begin{verbatim}

# -*- coding: utf-8 -*-

from numpy import array, loadtxt, savetxt, append
import sys, os
    
for mcase in range(1, 4, 1):
    for Hcase in range(1, 4, 1):
        syscount += 1
        if str(syscount) == n:
            break
    if str(syscount) == n:
        break
        
if Hcase == 1:
    HC = 'H(z)only'
if Hcase == 2:
    HC = 'Hz_2B'
if Hcase == 3:
    HC = 'Hz_AB'

aa = []
bb = []
cc = []
dd = []

if mcase == 1:
    for N in range(1, 501, 1):
        a = loadtxt('3D_' + HC + '_Nonflat_phiCDM_Omega_k0_mcase1_'
            + str(N) + '.dat', unpack=True)
        b = loadtxt('3D_' + HC + '_Nonflat_phiCDM_L_mcase1_'
            + str(N) + '.dat', unpack=True)
        c = loadtxt('3D_' + HC + '_Nonflat_phiCDM_alpha_mcase1_'
            + str(N) + '.dat', unpack=True)
        d = loadtxt('3D_' + HC + '_Nonflat_phiCDM_H0_mcase1_'
            + str(N) + '.dat', unpack=True)
    
        aa = append(aa, a)
        bb = append(bb, b)
        cc = append(cc, c)
        dd = append(dd, d)
        
    savetxt('3D_' + HC + '_Nonflat_phiCDM_Omega_k0_mcase1_compiled.dat', aa)
    savetxt('3D_' + HC + '_Nonflat_phiCDM_L_mcase1_compiled.dat', bb)
    savetxt('3D_' + HC + '_Nonflat_phiCDM_alpha_mcase1_compiled.dat', cc)
    savetxt('3D_' + HC + '_Nonflat_phiCDM_H0_mcase1_compiled.dat', dd)
    
if mcase == 2:
    for N in range(1, 501, 1):
        a = loadtxt('3D_' + HC + '_Nonflat_phiCDM_Omega_k0_mcase2_'
            + str(N) + '.dat', unpack=True)
        b = loadtxt('3D_' + HC + '_Nonflat_phiCDM_L_mcase2_'
            + str(N) + '.dat', unpack=True)
        c = loadtxt('3D_' + HC + '_Nonflat_phiCDM_alpha_mcase2_'
            + str(N) + '.dat', unpack=True)
        d = loadtxt('3D_' + HC + '_Nonflat_phiCDM_Omega_m0_mcase2_'
            + str(N) + '.dat', unpack=True)

        aa = append(aa, a)
        bb = append(bb, b)
        cc = append(cc, c)
        dd = append(dd, d)    
        
    savetxt('3D_' + HC + '_Nonflat_phiCDM_Omega_k0_mcase2_compiled.dat', aa)
    savetxt('3D_' + HC + '_Nonflat_phiCDM_L_mcase2_compiled.dat', bb)
    savetxt('3D_' + HC + '_Nonflat_phiCDM_alpha_mcase2_compiled.dat', cc)
    savetxt('3D_' + HC + '_Nonflat_phiCDM_Omega_m0_mcase2_compiled.dat', dd)
    
if mcase == 3:
    for N in range(1, 501, 1):
        a = loadtxt('3D_' + HC + '_Nonflat_phiCDM_Omega_m0_mcase3_'
            + str(N) + '.dat', unpack=True)
        b = loadtxt('3D_' + HC + '_Nonflat_phiCDM_L_mcase3_'
            + str(N) + '.dat', unpack=True)
        c = loadtxt('3D_' + HC + '_Nonflat_phiCDM_alpha_mcase3_'
            + str(N) + '.dat', unpack=True)
        d = loadtxt('3D_' + HC + '_Nonflat_phiCDM_H0_mcase3_'
            + str(N) + '.dat', unpack=True)
    
        aa = append(aa, a)
        bb = append(bb, b)
        cc = append(cc, c)
        dd = append(dd, d)
        
    savetxt('3D_' + HC + '_Nonflat_phiCDM_Omega_m0_mcase3_compiled.dat', aa)
    savetxt('3D_' + HC + '_Nonflat_phiCDM_L_mcase3_compiled.dat', bb)
    savetxt('3D_' + HC + '_Nonflat_phiCDM_alpha_mcase3_compiled.dat', cc)
    savetxt('3D_' + HC + '_Nonflat_phiCDM_H0_mcase3_compiled.dat', dd)

"""
#This portion of the code can be used to combine,
#for example, the H(z) likelihood function with 
#the BAO likelihood function to obtain the 
#H(z) + BAO likelihood function. This must be 
#done before running ``data_compiler_3.py''. 
#When doing this, it is also necessary to 
#change the line #SBATCH --array=1-9:1 
#in the ``chi2_combiner'' .bash script to #SBATCH --array=1-500:1.

n = sys.argv[1]
#syscount = 0

for N in range(1, 501, 1):
    #syscount += 1
    if str(N) == n:
        break
        
#for N in range(1, 501, 1):
#a = loadtxt('Nonflat_phiCDM_chi2_AB_' + str(N) + '.dat', unpack=True)
c = loadtxt('Nonflat_phiCDM_chi2_H(z)only_' + str(N) + '.dat', unpack=True)

#d = a + c
#del a

#savetxt('Nonflat_phiCDM_chi2_Hz_AB_' + str(N) + '.dat', d)
#del d

b = loadtxt('Nonflat_phiCDM_chi2_AB_' + str(N) + '.dat', unpack=True)

e = b + c
del b, c

savetxt('Nonflat_phiCDM_chi2_Hz_AB_' + str(N) + '.dat', e)
del e

#os.remove('Nonflat_phiCDM_chi2_AB_' + str(N) + '.dat')
#os.remove('Nonflat_phiCDM_chi2_2B_' + str(N) + '.dat')
"""
\end{verbatim}

This code marginalizes the likelihood function further, reducing it from a three-dimensional function to a two-dimensional function.

\begin{verbatim}

#!/bin/sh

#SBATCH --job-name=3D_to_2D

#SBATCH --array=1-9:1

#SBATCH --mem-per-cpu=6G   # Memory per core, use --mem= for memory per node
#SBATCH --time=26:00:00   # Use the form DD-HH:MM:SS
#SBATCH --nodes=1
#SBATCH --ntasks-per-node=1

#SBATCH --mail-user=jwryan@phys.ksu.edu
#SBATCH --mail-type=ALL    # same as =BEGIN,FAIL,END

module load Anaconda3/5.0.1

python 3D_to_2D.py $SLURM_ARRAY_TASK_ID
\end{verbatim}

\begin{verbatim}

# -*- coding: utf-8 -*-

from numpy import exp, split, dstack, array, trapz, loadtxt, 
from numpy import savetxt, arange, linspace
from scipy.interpolate import NearestNDInterpolator as ND
from scipy.integrate import quad
import sys

Oml = 61
alphal = 500
Okl = 300
H0l = 351

n = sys.argv[1]
syscount = 0

for Hcase in range(1, 2, 1):
    for mcase in range(1, 4, 1):
        for mc in range(1, 4, 1):
            syscount += 1
            if str(syscount) == n:
                break
        if str(syscount) == n:
            break
    if str(syscount) == n:
        break

if Hcase == 1:
    HC = 'AB'
if Hcase == 2:
    HC = 'Hz_2B'
if Hcase == 3:
    HC = 'Hz_AB'
    
if mcase == 1:
    O_k = loadtxt(HC + '_Nonflat_phiCDM_Omega_k0_mcase1_compiled_3D.dat', 
        unpack=True)
    L = loadtxt(HC + '_Nonflat_phiCDM_L_mcase1_compiled_3D.dat', unpack=True)
    alpha = loadtxt(HC + '_Nonflat_phiCDM_alpha_mcase1_compiled_3D.dat', 
        unpack=True)
    H0 = loadtxt(HC + '_Nonflat_phiCDM_H0_mcase1_compiled_3D.dat',
        unpack=True)

    if mc == 1:                      
        c = 0
        L2 = []
        Omk2 = []
        al2 = []
        H02 = []
        L3 = []
        
        for q in range(0, len(L), 1):
            c += 1
            H02.append(H0[q])
            L2.append(L[q])
        
            if c == H0l:
                L3.append(abs(trapz(L2, H02)))
                al2.append(alpha[q])
                Omk2.append(O_k[q])
                c = 0
                H02 = []
                L2 = []
                
        savetxt('2D_' + HC + '_Nonflat_phiCDM_L(alOk)_mcase1_' + str(mc)
            + '.dat', L3)
        savetxt('2D_' + HC + '_Nonflat_phiCDM_alpha_mcase1_' + str(mc)
            + '.dat', al2)
        savetxt('2D_' + HC + '_Nonflat_phiCDM_Omega_k0_mcase1_' + str(mc)
            + '.dat', Omk2)
        
    if mc == 2:
        t = split(O_k, alphal)
        del O_k
        v = split(L, alphal)
        del  L
        w = split(alpha, alphal)
        del alpha
        u = split(H0, alphal)
        del H0
        
        d = []
        e = []
        g = []
        h = []
        
        for p in range(0, alphal, 1):
            d.append((dstack(split(t[p], Okl))).flatten())
            e.append((dstack(split(v[p], Okl))).flatten())
            g.append((dstack(split(w[p], Okl))).flatten())
            h.append((dstack(split(u[p], Okl))).flatten())
        
        del t, v, w, u
        Omk = (array(d)).flatten()
        del d
        L2 = (array(e)).flatten()
        del e
        alpha = (array(g)).flatten()
        del g
        H0 = (array(h)).flatten()
          
        Omk2 = []
        L = []
        L3 = []
        al2 = []
        H02 = []
        c = 0
        for q in range(0, len(L2), 1):
            c += 1
            if abs(Omk[q]) <= 0.5:
                Omk2.append(Omk[q])
                L3.append(L2[q])
            
            if c == Okl:
                L.append(abs(trapz(L3, Omk2)))
                al2.append(alpha[q])
                H02.append(H0[q])
                c = 0
                Omk2 = []
                L3 = []
        
        del L2, Omk       
        savetxt('2D_' + HC + '_Nonflat_phiCDM_L(alH)_mcase1_' + str(mc)
            + '.dat', L)
        del L
        savetxt('2D_' + HC + '_Nonflat_phiCDM_alpha_mcase1_' + str(mc)
            + '.dat', al2)
        del al2
        savetxt('2D_' + HC + '_Nonflat_phiCDM_H0_mcase1_'
            + str(mc) + '.dat', H02)
        
    if mc == 3:
        a = split(alpha, alphal)
        del alpha
        v = split(L, alphal)
        del L
        t = split(O_k, alphal)
        del O_k
        w = split(H0, alphal)
        del H0
        
        c = []
        e = []
        g = []
        h = []
        for p in range(0, alphal, 1):
            c.append((dstack(split(a[p], Okl))).flatten())
            e.append((dstack(split(v[p], Okl))).flatten())
            g.append((dstack(split(t[p], Okl))).flatten())
            h.append((dstack(split(w[p], Okl))).flatten())
            
        del a, v, t, w
    
        a2 = dstack(split((array(c)).flatten(), alphal)).flatten()
        del c
        L2 = dstack(split((array(e)).flatten(), alphal)).flatten()
        del e
        Omk = dstack(split((array(g)).flatten(), alphal)).flatten()
        H0 = dstack(split((array(h)).flatten(), alphal)).flatten()
        del g, h
        
        f = ND((H0, Omk, a2), L2, rescale=True)
        mnO = min(Omk)
        mxO = max(Omk)
        del H0, a2, L2, Omk
            
        a3 = []
        L3 = []
        Omk2 = []
        H02 = []
        
        for N in arange(50, 85.001, 0.1):
            for N2 in linspace(mnO, mxO, 300):
                for N3 in arange(0.01, 5.01, 0.01):
                    H02.append(N)
                    Omk2.append(N2)
                    a3.append(N3)
                    L3.append(f(N, N2, N3))

        a4 = []
        L4 = []
        L5 = []
        Omk3 = []
        H03 = []
        
        c = 0
        for q in range(0, len(L3), 1):
            c += 1
            a4.append(a3[q])
            L4.append(L3[q])
            
            if c == alphal:
                L5.append(abs(trapz(L4, a4)))
                Omk3.append(Omk2[q])
                H03.append(H02[q])
                c = 0
                a4 = []
                L4 = [] 
        
        del a3, L3, Omk2, H02
        
        savetxt('2D_' + HC + '_Nonflat_phiCDM_L(HOk)_mcase1_'
            + str(mc) + '.dat', L5)
        savetxt('2D_' + HC + '_Nonflat_phiCDM_Omega_k0_mcase1_' 
            + str(mc) + '.dat', Omk3)
        savetxt('2D_' + HC + '_Nonflat_phiCDM_H0_mcase1_' + 
            str(mc) + '.dat', H03)
        
if mcase == 2:
    O_m = loadtxt(HC + '_Nonflat_phiCDM_Omega_m0_mcase2_compiled_3D.dat', 
        unpack=True)
    L = loadtxt(HC + '_Nonflat_phiCDM_L_mcase2_compiled_3D.dat', unpack=True)
    alpha = loadtxt(HC + '_Nonflat_phiCDM_alpha_mcase2_compiled_3D.dat', 
        unpack=True)
    O_k = loadtxt(HC + '_Nonflat_phiCDM_Omega_k0_mcase2_compiled_3D.dat', 
        unpack=True)

    if mc == 1:
        c = 0
        L2 = []
        Om2 = []
        L3 = []
        Ok2 = []
        alpha2 = []
        
        for q in range(0, len(L), 1):
            c += 1
            Om2.append(O_m[q])
            L2.append(L[q])
        
            if c == Oml:
                L3.append(abs(trapz(L2, Om2)))
                Ok2.append(O_k[q])
                alpha2.append(alpha[q])
                c = 0
                Om2 = []
                L2 = []
        
        del O_k, alpha, L      
        savetxt('2D_' + HC + '_Nonflat_phiCDM_L(alOk)_mcase2_' + str(mc)
            + '.dat', L3)
        savetxt('2D_' + HC + '_Nonflat_phiCDM_alpha_mcase2_' + str(mc)
            + '.dat', alpha2)
        savetxt('2D_' + HC + '_Nonflat_phiCDM_Omega_k0_mcase2_' + str(mc)
            + '.dat', Ok2)
        
    if mc == 2:
        a = split(alpha, alphal)
        del alpha
        t = split(O_m, alphal)
        del O_m
        v = split(L, alphal)
        del  L
        w = split(O_k, alphal)
        del O_k
        
        d = []
        e = []
        f = []
        g = []
        
        for p in range(0, alphal, 1):
            d.append((dstack(split(a[p], Okl))).flatten())
            e.append((dstack(split(t[p], Okl))).flatten())
            f.append((dstack(split(v[p], Okl))).flatten())
            g.append((dstack(split(w[p], Okl))).flatten())
        
        del t, v, a, w
        alp = (array(d)).flatten()
        del d
        Om = (array(e)).flatten()
        del e
        L2 = (array(f)).flatten()
        del f
        Omk = (array(g)).flatten()
        del g
          
        Omk3 = []
        L = []
        L3 = []
        Omk2 = []
        alp2 = []
        Om2 = []
        
        c = 0
        for q in range(0, len(L2), 1):
            c += 1
            if abs(Omk[q]) <= 0.5:
                Omk2.append(Omk[q])
                L3.append(L2[q])
            
            if c == Okl:
                L.append(abs(trapz(L3, Omk2)))
                Om2.append(Om[q])
                alp2.append(alp[q])
                c = 0
                Omk2 = []
                L3 = []
        
        del L2, Om, alp, Omk         
        savetxt('2D_' + HC + '_Nonflat_phiCDM_L(alOm)_mcase2_' + str(mc)
            + '.dat', L)
        savetxt('2D_' + HC + '_Nonflat_phiCDM_alpha_mcase2_' + str(mc)
            + '.dat', alp2)
        savetxt('2D_' + HC + '_Nonflat_phiCDM_Omega_m0_mcase2_' 
            + str(mc) + '.dat', Om2)
        
    if mc == 3:
        a = split(alpha, alphal)
        del alpha
        t = split(L, alphal)
        del L
        v = split(O_m, alphal)
        del O_m
        w = split(O_k, alphal)
        del O_k        
        
        c = []
        e = []
        f = []
        g = []
        for p in range(0, alphal, 1):
            c.append((dstack(split(a[p], Okl))).flatten())
            e.append((dstack(split(t[p], Okl))).flatten())
            f.append((dstack(split(v[p], Okl))).flatten())
            g.append((dstack(split(w[p], Okl))).flatten())
            
        del a, t, v, w
    
        a2 = dstack(split((array(c)).flatten(), alphal)).flatten()
        del c
        L2 = dstack(split((array(e)).flatten(), alphal)).flatten()
        del e
        Om2 = dstack(split((array(f)).flatten(), alphal)).flatten()
        del f
        Ok2 = dstack(split((array(g)).flatten(), alphal)).flatten()
        del g
        
        f = ND((Om2, Ok2, a2), L2, rescale=True)
        del Om2, a2, L2
            
        a3 = []
        L3 = []
        Om3 = []
        Ok3 = []
        for N in arange(0.10, 0.71, 0.01):
            for N2 in linspace(min(Ok2), max(Ok2), 300):
                for N3 in arange(0.01, 5.01, 0.01):
                    Om3.append(N)
                    Ok3.append(N2)
                    a3.append(N3)
                    L3.append(f(N, N2, N3))
        del Ok2
        a4 = []
        L4 = []
        L5 = []
        Om4 = []
        Ok4 = []
        
        c = 0
        for q in range(0, len(L3), 1):
            c += 1
            a4.append(a3[q])
            L4.append(L3[q])
            
            if c == alphal:
                L5.append(abs(trapz(L4, a4)))
                Om4.append(Om3[q])
                Ok4.append(Ok3[q])
                c = 0
                a4 = []
                L4 = []
        del Om3, Ok3, a3, L3
        
        savetxt('2D_' + HC + '_Nonflat_phiCDM_L(OkOm)_mcase2_' 
            + str(mc) + '.dat', L5)
        savetxt('2D_' + HC + '_Nonflat_phiCDM_Omega_k0_mcase2_' 
            + str(mc) + '.dat', Ok4)
        savetxt('2D_' + HC + '_Nonflat_phiCDM_Omega_m0_mcase2_' 
            + str(mc) + '.dat', Om4)
    
if mcase == 3:
    O_m = loadtxt(HC + '_Nonflat_phiCDM_Omega_m0_mcase3_compiled_3D.dat', 
        unpack=True)
    L = loadtxt(HC + '_Nonflat_phiCDM_L_mcase3_compiled_3D.dat', unpack=True)
    alpha = loadtxt(HC + '_Nonflat_phiCDM_alpha_mcase3_compiled_3D.dat', 
        unpack=True)
    H0 = loadtxt(HC + '_Nonflat_phiCDM_H0_mcase3_compiled_3D.dat',
        unpack=True)
    
    if mc == 1:     
        c = 0
        L2 = []
        Om = []
        L3 = []
        al2 = []
        H02 = []
    
        for q in range(0, len(L), 1):
            c += 1
            Om.append(O_m[q])
            L2.append(L[q])
        
            if c == Oml:
                L3.append(abs(trapz(L2, Om)))
                al2.append(alpha[q])
                H02.append(H0[q])
                c = 0
                Om = []
                L2 = []
        
        del L2, O_m, alpha, H0
        savetxt('2D_' + HC + '_Nonflat_phiCDM_L(alH)_mcase3_' + str(mc)
            + '.dat', L3)
        savetxt('2D_' + HC + '_Nonflat_phiCDM_alpha_mcase3_' + str(mc)
            + '.dat', al2)
        savetxt('2D_' + HC + '_Nonflat_phiCDM_H0_mcase3_' + str(mc)
            + '.dat', H02)
        del L3
        
    if mc == 2:
        a = split(alpha, alphal)
        del alpha
        v = split(H0, alphal)
        del H0
        t = split(L, alphal)
        del  L
        w = split(O_m, alphal)
        del O_m
        
        d = []
        e = []
        f = []
        g = []
        
        for p in range(0, alphal, 1):
            d.append((dstack(split(a[p], H0l))).flatten())
            e.append((dstack(split(t[p], H0l))).flatten())
            f.append((dstack(split(v[p], H0l))).flatten())
            g.append((dstack(split(w[p], H0l))).flatten())
        
        del t, v, a, w
        a2 = (array(d)).flatten()
        del d
        L2 = (array(e)).flatten()
        del e
        H2 = (array(f)).flatten()
        del f
        Om2 = (array(g)).flatten()
          
        H3 = []
        L = []
        L3 = []
        Om3 = []
        a3 = []
        
        c = 0
        for q in range(0, len(L2), 1):
            c += 1
            H3.append(H2[q])
            L3.append(L2[q])
            
            if c == H0l:
                L.append(abs(trapz(L3, H3)))
                a3.append(a2[q])
                Om3.append(Om2[q])
                c = 0
                H3 = []
                L3 = []
        
        del L2, H2, Om2, a2
        savetxt('2D_' + HC + '_Nonflat_phiCDM_L(alOm)_mcase3_' + str(mc)
            + '.dat', L)
        savetxt('2D_' + HC + '_Nonflat_phiCDM_alpha_mcase3_' + str(mc)
            + '.dat', a3)
        savetxt('2D_' + HC + '_Nonflat_phiCDM_Omega_m0_mcase3_' + str(mc)
            + '.dat', Om3)
        del L
        
    if mc == 3:
        a = split(alpha, alphal)
        del alpha
        t = split(L, alphal)
        del L
        v = split(H0, alphal)
        del H0
        w = split(O_m, alphal)
        del O_m
        
        c = []
        e = []
        f = []
        g = []
        for p in range(0, alphal, 1):
            c.append((dstack(split(a[p], H0l))).flatten())
            e.append((dstack(split(t[p], H0l))).flatten())
            f.append((dstack(split(v[p], H0l))).flatten())
            g.append((dstack(split(w[p], H0l))).flatten())
            
        del a, v, t, w
    
        a2 = dstack(split((array(c)).flatten(), alphal)).flatten()
        del c
        L2 = dstack(split((array(e)).flatten(), alphal)).flatten()
        del e
        H2 = dstack(split((array(f)).flatten(), alphal)).flatten()
        del f
        Om2 = dstack(split((array(g)).flatten(), alphal)).flatten()
        del g
    
        a3 = []
        L3 = []
        L4 = []
        Om3 = []
        H3 = []
        
        c = 0
        for q in range(0, len(L2), 1):
            c += 1
            a3.append(a2[q])
            L3.append(L2[q])
            
            if c == alphal:
                L4.append(abs(trapz(L3, a3)))
                Om3.append(Om2[q])
                H3.append(H2[q])
                c = 0
                a3 = []
                L3 = []
        
        del L2, a2, H2, Om2
        savetxt('2D_' + HC + '_Nonflat_phiCDM_L(OmH)_mcase3_' + str(mc)
            + '.dat', L4)
        savetxt('2D_' + HC + '_Nonflat_phiCDM_Omega_m0_mcase3_' + str(mc)
            + '.dat', Om3)
        savetxt('2D_' + HC + '_Nonflat_phiCDM_H0_mcase3_' + str(mc)
            + '.dat', H3)
\end{verbatim}

This code carries out the final step of the marginalization process, producing one-dimensional best-fitting values of the cosmological model parameters along with 1- and 2$\sigma$ confidence intervals of these best-fitting values.

\begin{verbatim}

#!/bin/sh

#SBATCH --job-name=1d_Nonflat_phiCDM

#SBATCH --array=1-48:1

#SBATCH --mem-per-cpu=1G   # Memory per core, use --mem= for memory per node
#SBATCH --time=26:00:00   # Use the form DD-HH:MM:SS
#SBATCH --nodes=1
#SBATCH --ntasks-per-node=1

#SBATCH --mail-user=jwryan@phys.ksu.edu
#SBATCH --mail-type=ALL    # same as =BEGIN,FAIL,END

module load Anaconda3/5.0.1

python 1d_Nonflat_phiCDM.py $SLURM_ARRAY_TASK_ID
\end{verbatim}

\begin{verbatim}

# -*- coding: utf-8 -*-

from numpy import linspace, dstack, split, savetxt, loadtxt, 
from numpy import exp, log, trapz, arange, 
from numpy import pi, sqrt, array
from scipy.interpolate import NearestNDInterpolator as ND
from scipy.interpolate import interp1d
from scipy.integrate import quad
import matplotlib.pyplot as plt
import sys

Oml = 61
alphal = 500
H0l = 351
Okl = 300

b1 = 0.6827
b2 = 0.9545
xinc = 1/100000
Linc = 1/100

n = sys.argv[1]
syscount = 0

for data in range(1, 5, 1):
    if data == 1:
        d = 'Hz_AB'
    if data == 2:
        d = 'Hz_2B'
    if data == 3:
        d = 'H(z)only'
    if data == 4:
        d = 'AB'
            
    for mcase in range(1, 7, 1):
        for case in range(0, 2, 1):
            syscount += 1
            if str(syscount) == n:
                break
        if str(syscount) == n:
            break
    if str(syscount) == n:
        break
           
if mcase == 1:
    Ml = Oml*Okl
    x0 = loadtxt('2D_' + d + '_Nonflat_phiCDM_Omega_m0_mcase2_3.dat',
        unpack=True)
    y0 = loadtxt('2D_' + d + '_Nonflat_phiCDM_Omega_k0_mcase2_3.dat',
        unpack=True)
    z0 = loadtxt('2D_' + d + '_Nonflat_phiCDM_L(OkOm)_mcase2_3.dat',
        unpack=True)
    
    if case == 0:
        """
        Omega_k0 marginalized
        """
        
        xlab = '$\Omega_{m0}$'
        xlab2 = 'Om'
        Il = Okl
        x1 = x0
        y1 = y0
        z1 = z0
        
    if case == 1:
        """
        Omega_m0 marginalized
        """
        
        xlab = '$\Omega_{k0}$'
        xlab2 = 'Ok'
        Il = Oml
        y1 = dstack(split(x0, Oml)).flatten()
        x1 = dstack(split(y0, Oml)).flatten()
        z1 = dstack(split(z0, Oml)).flatten()        
        
    z2 = []
    y2 = []
    x = []
    y = []
    
    c = 0
    for q in range(0, Ml, 1):
        c += 1
        if case == 0:
            if abs(y1[q]) <= 0.50:
                z2.append(z1[q])
                y2.append(y1[q])
        if case == 1:
            z2.append(z1[q])
            y2.append(y1[q])
        
        if c == Il:
            y.append(abs(trapz(z2, y2)))
            x.append(x1[q])
            
            z2 = []
            y2 = []
            c = 0
    
    del y1, x1, z1
                
if mcase == 2:
    Ml = alphal*Oml
    y0 = loadtxt('2D_' + d + '_Nonflat_phiCDM_Omega_m0_mcase2_2.dat',
        unpack=True)
    x0 = loadtxt('2D_' + d + '_Nonflat_phiCDM_alpha_mcase2_2.dat',
        unpack=True)
    z0 = loadtxt('2D_' + d + '_Nonflat_phiCDM_L(alOm)_mcase2_2.dat',
        unpack=True)
    
    if case == 0:
        """
        Omega_m0 marginalized
        """
        
        xlab = '$\\alpha$'
        xlab2 = 'al'
        Il = Oml
        x1 = x0
        y1 = y0
        z1 = z0
        del x0, y0, z0
        
    if case == 1:
        """
        alpha marginalized
        """
        
        xlab = '$\Omega_{m0}$'
        xlab2 = 'Om'
        Il = alphal
        y1 = dstack(split(x0, alphal)).flatten()
        del x0
        x1 = dstack(split(y0, alphal)).flatten()
        del y0
        z1 = dstack(split(z0, alphal)).flatten()
        del z0

    z2 = []
    y2 = []
    x = []
    y = []
    
    c = 0
    for q in range(0, Ml, 1):
        c += 1
        z2.append(z1[q])
        y2.append(y1[q])
        
        if c == Il:
            y.append(abs(trapz(z2, y2)))
            x.append(x1[q])
            
            z2 = []
            y2 = []
            c = 0
    
    del y1, x1, z1
                    
if mcase == 3:
    Ml = H0l*Oml
    y0 = loadtxt('2D_' + d + '_Nonflat_phiCDM_H0_mcase3_3.dat', 
        unpack=True)
    x0 = loadtxt('2D_' + d + '_Nonflat_phiCDM_Omega_m0_mcase3_3.dat',
        unpack=True)
    z0 = loadtxt('2D_' + d + '_Nonflat_phiCDM_L(OmH)_mcase3_3.dat',
        unpack=True)
    
    if case == 0:
        """
        H0 marginalized
        """
        
        xlab = '$\Omega_{m0}$'
        xlab2 = 'Om'
        Il = H0l
        x1 = x0
        y1 = y0
        z1 = z0
        del x0, y0, z0
        
    if case == 1:
        """
        Omega_m0 marginalized
        """
        
        xlab = '$H_0$'
        xlab2 = 'H0'
        Il = Oml
        y1 = dstack(split(x0, Oml)).flatten()
        del x0
        x1 = dstack(split(y0, Oml)).flatten()
        del y0
        z1 = dstack(split(z0, Oml)).flatten()
        del z0
           
    z2 = []
    y2 = []
    x = []
    y = []
    
    c = 0
    for q in range(0, Ml, 1):
        c += 1
        z2.append(z1[q])
        y2.append(y1[q])
        
        if c == Il:
            y.append(abs(trapz(z2, y2)))
            x.append(x1[q])
            
            z2 = []
            y2 = []
            c = 0
    
    del y1, x1, z1
    
if mcase == 4:
    Ml = alphal*Okl
    x0 = loadtxt('2D_' + d + '_Nonflat_phiCDM_alpha_mcase1_1.dat',
        unpack=True)
    y0 = loadtxt('2D_' + d + '_Nonflat_phiCDM_Omega_k0_mcase1_1.dat',
        unpack=True)
    z0 = loadtxt('2D_' + d + '_Nonflat_phiCDM_L(alOk)_mcase1_1.dat',
        unpack=True)
    
    if case == 0:
        """
        Omega_k0 marginalized
        """
        
        xlab = '$\\alpha$'
        xlab2 = 'al'
        Il = Okl
        x1 = x0
        y1 = y0
        z1 = z0
        del x0, y0, z0
        
    if case == 1:
        """
        alpha marginalized
        """
        
        z11 = []
        y11 = []
        x11 = []
        q = 0
        
        f = ND((x0, y0), z0, rescale=True)
        
        for N2 in arange(0.01, 5.01, 0.01):
            for N in linspace(min(y0), max(y0), 300):
                z11.append(f(N2, N))
                y11.append(N)
                x11.append(N2)
                q += 1
        
        xlab = '$\Omega_{k0}$'
        xlab2 = 'Ok'
        Il = alphal
        y1 = dstack(split(array(x11), alphal)).flatten()
        del x0
        x1 = dstack(split(array(y11), alphal)).flatten()
        del y0
        z1 = dstack(split(array(z11), alphal)).flatten()
        del z0
                   
    z2 = []
    y2 = []
    x = []
    y = []
    
    c = 0
    for q in range(0, Ml, 1):
        c += 1
        if case == 0:
            if abs(y1[q]) <= 0.50:
                z2.append(z1[q])
                y2.append(y1[q])
        if case == 1:
            z2.append(z1[q])
            y2.append(y1[q])
        
        if c == Il:
            y.append(abs(trapz(z2, y2)))
            x.append(x1[q])
            
            z2 = []
            y2 = []
            c = 0
    
    del y1, x1, z1
    
if mcase == 5:
    Ml = H0l*Okl
    y0 = loadtxt('2D_' + d + '_Nonflat_phiCDM_Omega_k0_mcase1_3.dat',
        unpack=True)
    x0 = loadtxt('2D_' + d + '_Nonflat_phiCDM_H0_mcase1_3.dat', 
        unpack=True)
    z0 = loadtxt('2D_' + d + '_Nonflat_phiCDM_L(HOk)_mcase1_3.dat',
        unpack=True)
    
    if case == 0:
        """
        Omega_k0 marginalized
        """
        
        xlab = '$H_0$'
        xlab2 = 'H0'
        Il = Okl
        x1 = x0
        y1 = y0
        z1 = z0
        del x0, y0, z0
        
    if case == 1:
        """
        H0 marginalized
        """
        
        xlab = '$\Omega_{k0}$'
        xlab2 = 'Ok'
        Il = H0l
        y1 = dstack(split(x0, H0l)).flatten()
        del x0
        x1 = dstack(split(y0, H0l)).flatten()
        del y0
        z1 = dstack(split(z0, H0l)).flatten()
        del z0
           
    z2 = []
    y2 = []
    x = []
    y = []
    
    c = 0
    for q in range(0, Ml, 1):
        c += 1
        if case == 0:
            if abs(y1[q]) <= 0.50:
                z2.append(z1[q])
                y2.append(y1[q])
        if case == 1:
            z2.append(z1[q])
            y2.append(y1[q])
        
        if c == Il:
            y.append(abs(trapz(z2, y2)))
            x.append(x1[q])
            
            z2 = []
            y2 = []
            c = 0
    
    del y1, x1, z1
    
if mcase == 6:
    Ml = H0l*alphal
    x0 = loadtxt('2D_' + d + '_Nonflat_phiCDM_alpha_mcase1_2.dat',
        unpack=True)
    y0 = loadtxt('2D_' + d + '_Nonflat_phiCDM_H0_mcase1_2.dat', 
        unpack=True)
    z0 = loadtxt('2D_' + d + '_Nonflat_phiCDM_L(alH)_mcase1_2.dat',
        unpack=True)
    
    if case == 0:
        """
        H0 marginalized
        """
        
        xlab = '$\\alpha$'
        xlab2 = 'al'
        Il = H0l
        x1 = x0
        y1 = y0
        z1 = z0
        del x0, y0, z0
        
    if case == 1:
        """
        alpha marginalized
        """
        
        xlab = '$H_0$'
        xlab2 = 'H0'
        Il = alphal
        y1 = dstack(split(x0, alphal)).flatten()
        del x0
        x1 = dstack(split(y0, alphal)).flatten()
        del y0
        z1 = dstack(split(z0, alphal)).flatten()
        del z0
           
    z2 = []
    y2 = []
    x = []
    y = []
    
    c = 0
    for q in range(0, Ml, 1):
        c += 1
        z2.append(z1[q])
        y2.append(y1[q])
        
        if c == Il:
            y.append(abs(trapz(z2, y2)))
            x.append(x1[q])
            
            z2 = []
            y2 = []
            c = 0
    
    del y1, x1, z1

savetxt('Nonflat_phiCDM_L(' + xlab2 + ')_' + d + '_mcase' + str(mcase)
    + '.dat', y)
savetxt('Nonflat_phiCDM_' + xlab2 + '_' + d + '_mcase' + str(mcase)
    + '_1D.dat', x)
  
a = []
for upp in range(1, 2, 1):
    print(mcase, case)
    if upp == 1:
        upper = True
        limlab = 'upper'
    if upp == 2:
        upper = False
        limlab = 'lower'
                           
    L = interp1d(x, y, kind='cubic')
    
    candL1 = []
    candr1 = []
    candp1 = []
    
    candL2 = []
    candr2 = []
    candp2 = []
    
    xmax = x[y.index(max(y))]
    
    if upper == True:
        k = 1
        if xlab2 == 'Ok':
            xlim = 0.5
            xmax = -0.5
        else:
            xlim = max(x)
    if upper == False:
        k = -1
        if xlab2 == 'Ok':
            xlim = -0.5
        else:
            xlim = min(x)
            
    z = arange(xmax, xlim, k*xinc)
    
    for r in z:
        LCO1tot, error = quad(lambda m: L(m), xmax, xlim)
        LCO1, error = quad(lambda m: L(m), xmax, r)
        LCO = abs(LCO1/LCO1tot)
        #print(LCO, r)
        if abs((LCO - b1)/b1) < Linc:
            candL1.append(LCO)
            candr1.append(r)
            candp1.append(abs((LCO - b1)/b1))
        if abs((LCO - b2)/b2) < Linc:
            candL2.append(LCO)
            candr2.append(r)
            candp2.append(abs((LCO - b2)/b2))
        if LCO >= b2 + 0.01:
            break
    
    pmin = candp1.index(min(candp1))   
    Lmin = candL1[pmin]
    rmin = candr1[pmin]
    
    a.append(mcase)
    a.append(Lmin)
    a.append(rmin)
    a.append(pmin)
    
    pmin = candp2.index(min(candp2))   
    Lmin = candL2[pmin]
    rmin = candr2[pmin]
    
    a.append(Lmin)
    a.append(rmin)
    a.append(pmin)
    a.append(xmax)

savetxt('Nonflat_phiCDM_L(' + xlab2 + ')_' + d + '_mcase' + 
    str(mcase) + '.dat', y)
savetxt('Nonflat_phiCDM_' + xlab2 + '_' + d + '_mcase' + 
    str(mcase) + '_1D.dat', x)
savetxt('Nonflat_phiCDM_data_' + xlab2 + '_' + d + '_mcase' + 
    str(mcase) + '.dat', a)
\end{verbatim}

This code, in two parts, can be run at any point. It simply computes the minimum value of the four-dimensional (that is, unmarginalized) $\chi^2$ function, and the best-fitting values of the cosmological parameters corresponding to this minimum $\chi^2$. The first part (``phiCDM\_3D\_BFP.py'') locates the minimum $\chi^2$ within each of the set of 500 files corresponding to steps along the $\alpha$ axis, and the second part (``phiCDM\_3D\_BFP\_2.py'') finds the minimum of these minima.

\begin{verbatim}

#!/bin/sh

#SBATCH --job-name=Nonflat_phiCDM_BFP

#SBATCH --array=1-500:1

#SBATCH --mem-per-cpu=1G   # Memory per core, use --mem= for memory per node
#SBATCH --time=26:00:00   # Use the form DD-HH:MM:SS
#SBATCH --nodes=1
#SBATCH --ntasks-per-node=1

#SBATCH --mail-user=jwryan@phys.ksu.edu
#SBATCH --mail-type=ALL    # same as =BEGIN,FAIL,END

module load Anaconda3/5.0.1

python phiCDM_3D_BFP.py
\end{verbatim}

\begin{verbatim}

# -*- coding: utf-8 -*-

import sys
from numpy import loadtxt, savetxt

n = sys.argv[1]
syscount = 0
for x in range(1, 501, 1):
    for p in range(1, 5, 1):
        syscount += 1
        if str(syscount) == n:
            break
    if str(syscount) == n:
        break
        
if p == 4:
    l = 'AB_'
    arr = []
    g = loadtxt('Nonflat_phiCDM_chi2_Hz_' + l + str(x) + '.dat',
        unpack=True)
    f = loadtxt('Nonflat_phiCDM_chi2_' + 'H(z)only_' + str(x) + '.dat',
        unpack=True)
    d = g - f
    del g, f
    dd = list(d)
    del d

if p == 2:
    l = 'Hz_2B_'
    arr = []
    d = loadtxt('Nonflat_phiCDM_chi2_' + l + str(x) + '.dat', unpack=True)
    dd = list(d)
    del d

if p == 3:
    l = 'H(z)only_'
    arr = []
    d = loadtxt('Nonflat_phiCDM_chi2_' + l + str(x) + '.dat', unpack=True)
    dd = list(d)
    del d

if p == 1:
    l = 'Hz_AB_'
    arr = []
    d = loadtxt('Nonflat_phiCDM_chi2_' + l + str(x) + '.dat', unpack=True)
    dd = list(d)
    del d

qmin = dd.index(min(dd))

arr.append(x)
arr.append(min(dd))
arr.append(qmin)
    
savetxt('Nonflat_phiCDM_3D_chi2min_' + l + str(x) + '.dat', arr)
\end{verbatim}

\begin{verbatim}

#!/bin/sh

#SBATCH --job-name=Nonflat_phiCDM_BFP

#SBATCH --mem-per-cpu=1G   # Memory per core, use --mem= for memory per node
#SBATCH --time=26:00:00   # Use the form DD-HH:MM:SS
#SBATCH --nodes=1
#SBATCH --ntasks-per-node=1

#SBATCH --mail-user=jwryan@phys.ksu.edu
#SBATCH --mail-type=ALL    # same as =BEGIN,FAIL,END

module load Anaconda3/5.0.1

python phiCDM_3D_BFP_2.py
\end{verbatim}

\begin{verbatim}

# -*- coding: utf-8 -*-

from numpy import loadtxt, arange

for Hc in range(3, 4, 1):
    if Hc == 1:
        l = 'H(z)only_'
    if Hc == 2:
        l = 'Hz_2B_'
    if Hc == 3:
        l = 'AB_'

    x = loadtxt('Nonflat_phiCDM_3D_chi2min_' + l + '1' + '.dat',
        unpack=True)
    x0 = x[0]
    x1 = x[1]
    x2 = x[2]
    
    for n in range(2, 501, 1):
        y = loadtxt('Nonflat_phiCDM_3D_chi2min_' + l + str(n) + '.dat',
            unpack=True)
        y0 = y[0]
        y1 = y[1]
        y2 = y[2]
        if y1 < x1:
            x0 = y0
            x1 = y1
            x2 = y2
            
    print(l)
    print('100*alpha = ', x0)
    print('chi^2 = ', x1)
    print(x2)
    
    Ok = loadtxt('Nonflat_phiCDM_Omega_k0_' + str(int(x0)) + '.dat',
        unpack=True)
    print('Omega_k = ', Ok[int(x2)])
    
    count = 0
    for K in arange(-2.10, 0.40, 0.01):
        for O in arange(0.10, 0.71, 0.01):
            for H in arange(50, 85.1, 0.1):
                if count == x2:
                    break
                count += 1
            if count == x2:
                break
        if count == x2:
            break
            
    print('Omega_m = ', O)
    print('H0 = ', H)
\end{verbatim}

\section{\textsc{emcee} codes}

Here I wish to thank Shulei Cao for technical advice.

%%
%Power law code
%%
\subsection{Power law model codes}
\label{sec:power_law_code}

Here I provide a version of the code I used to compute the constraints on the power law model using $H(z)$, BAO, QSO-AS, GRB, and HIIG data (see Chapter \ref{Chapter10}). The text of the code has been condensed slightly and edited for readability. I have included this code in this appendix mainly to serve as an illustration of how to apply the \textsc{emcee} module to the analysis of a simple cosmological model. As in the previous section, I provide the .bash script followed by the \textsc{Python} code itself.

\begin{verbatim}

#!/bin/sh

#SBATCH --job-name=PL_Om

#SBATCH --mem-per-cpu=1G   # Memory per core, use --mem= for memory per node
#SBATCH --time=23:00:00   # Use the form DD-HH:MM:SS
#SBATCH --nodes=1
#SBATCH --ntasks-per-node=20
#SBATCH --constraint="mages|elves|heroes|dwarves|moles|wizards"
#SBATCH --gres=killable:1

#SBATCH --mail-user=jwryan@phys.ksu.edu

#SBATCH --mail-type=ALL    # same as =BEGIN,FAIL,END

module purge
module load Python/3.7.0-iomkl-2018b
source ~/virtualenvs/emcee_2/bin/activate

export PYTHONDONTWRITEBYTECODE=1
PYTHON_BINARY=$(which python)
export OMP_NUM_THREADS=1
host=`hostname`

time mpirun ${PYTHON_BINARY} Power_law_Om_2.py
echo "Finished run on 32 cores on $host"
\end{verbatim}

\begin{verbatim}

# -*- coding: utf-8 -*-

from numpy import genfromtxt, log, matrix, loadtxt, exp, arange, savetxt
from numpy import array, random, inf, isfinite, pi, log10
from numpy import sin, sinh, sqrt, cos, cosh, identity, diag, isnan
import emcee
import sys
from scipy.integrate import quad
from numpy.linalg import inv
from emcee.utils import MPIPool
    
lab = 'ZBQGH'

z_obsQ, th_obs, sig_th_obs = loadtxt('QSO_120.txt', unpack=True) #QSO data.

z_obs, Hz_obs, sigHobs = loadtxt('H(z)data.dat',unpack = True)
##From Table 1 of 1607.03537v2, refs. 4,6,7,10 excluded. H(z) data.

##DM_obs, H_obs, A_obs, s, and Cov are from Table 8 of 1607.03155v1.
DM_obs = array([1512.39, 1975.22, 2306.68])
H_obs = array([81.2087, 90.9029, 98.9647])
##Data points from DR12 website. File "BAO_consensus_results_dM_Hz.txt"

##Covariance matrix from DR12 website. "BAO_consensus_covtot_dM_Hz.txt".
Cov = matrix([[624.707,23.729,325.332,8.34963,157.386,3.57778],
              [23.729,5.60873,11.6429,2.33996,6.39263,0.968056],
              [325.332,11.6429,905.777,29.3392,515.271,14.1013],
              [8.34963,2.33996,29.3392,5.42327,16.1422,2.85334],
              [157.386,6.39263,515.271,16.1422,1375.12,40.4327],
              [3.57778,0.968056,14.1013,2.85334,40.4327,6.25936]])
Cinv = inv(Cov)
del Cov

Cov2 = matrix([[1.3225, -0.1009],[-0.1009, 0.0380]])
Cinv2 = inv(Cov2)
del Cov2
##See arXiv:2101.08817

c = 299792458/1000 #Speed of light in km/s
Th = 2.7255/2.7 #T_CMB/2.7, from Eisenstein and Hu 1998
                #and Fixsen 0911.1955
lm = 11.03 #Standard rod in units of pc
On = 0.0014 #Neutrino density

ls, lf, zH, els, elf, ezH = loadtxt('HIIG_minus_GEHR.txt', unpack=True)
f = 10**(lf)
fu = elf*f*(log(10))
#From an email sent to me by Ana Luisa Gonzalez-Moran. 153 HIIG measurements

G25 = loadtxt('25_GRB.txt', unpack=True) #GRB data.
G94 = loadtxt('94_GRB.txt', unpack=True) #GRB data.

zG = []
zGe = []
Ep_obs = []
Ep_obs_err = []
Sb_obs = []
Sb_obs_err = []

for q in range(0, len(G25[0]), 1):
    zG.append(G25[0][q])
    zGe.append(G25[1][q])
    Ep_obs.append(G25[2][q])
    Ep_obs_err.append(G25[3][q])
    Sb_obs.append((10**(-5))*G25[4][q])
    Sb_obs_err.append((10**(-5))*G25[5][q])
for q in range(0, len(G94[0]), 1):
    zG.append(G94[0][q])
    zGe.append(G94[1][q])
    Ep_obs.append(G94[2][q])
    Ep_obs_err.append(G94[3][q])
    Sb_obs.append((10**(-5))*G94[4][q])
    Sb_obs_err.append((10**(-5))*G94[5][q])

def logL(paras): #natural logarithm of the likelihood function.
    h, beta, Om, Obh2 = paras[0], paras[1], paras[2], paras[3]
    aG, bG, s_ext = paras[4], paras[5], paras[6]
    
    a = 33.268
    asig = 0.083
    b = 5.022
    bsig = 0.058
    ##Above: HIIG slope and intercept.

    def H(h, beta, z): #Hubble parameter
        return (100*h)*((1 + z)**(1/beta))
        
    def E(h, beta, z): #Expansion parameter
        return H(h, beta, z)/(100*h)
    
    def D_M(h, beta, z): #Transverse comoving distance
    ##(see astro-ph/9905116v4).
        dH = c/(100*h)
        I, error = quad(lambda m: 1/(E(h, beta, m)), 0, z)
        return I*dH
    
    def rs(h, Om, Obh2): #Sound horizon.
        Onh2 = On*(h**2)
        Omh2 = Om*(h**2)
        A = 55.154*(exp(-72.3*((Onh2 + 0.0006)**2)))
        B = ((Omh2 - Onh2)**0.25351)*((Obh2)**0.12807)
        return A/B
    
    def chi_sq(h, beta, Om, Obh2): #Chi^2 for correlated BAO data.
        r = rs(h, Om, Obh2)/147.78
        DM_th = []
        H_th = []
        for z in array([0.38, 0.51, 0.61]):
            DM_th.append(D_M(h, beta, z))
            H_th.append(H(h, beta, z))
        Delta = array([(DM_th[0]/r - DM_obs[0]), 
                             (r*H_th[0] - H_obs[0]),
                             (DM_th[1]/r - DM_obs[1]),
                             (r*H_th[1] - H_obs[1]),
                             (DM_th[2]/r - DM_obs[2]),
                             (r*H_th[2] - H_obs[2])])        
        prod = (Cinv.dot(Delta)).T
        chi_sq1 = Delta.dot(prod)
        return chi_sq1.item((0,0))
        
    def chi2_2(h, beta, Om, Obh2): #Chi^2 for BAO data.
        for z1 in array([0.122, 0.81, 1.52, 2.334]):
            dH = c/(100*h)
            if z1 == 0.122:
                r = rs(h, Om, Obh2)/147.5
                y = (D_M(h, beta, z1))/dH
                DV = dH*(((y**2)*z1)/(E(h, beta, z1)))**(1/3) 
                ##Distance parameter
                DVobs3 = 539*r #Carter et al., doi:10.1093/mnras/sty2405
                unc3 = 17*r
                chi2DV3 = ((DV - DVobs3)**2)/(unc3**2)
            if z1 == 0.81:
                DA = D_M(h, beta, z1)/(1 + z1)
                DA_obs = 10.75
                DA_unc = 0.43
                chi2DA = ((DA/(rs(h, Om, Obh2)) - DA_obs)**2)/(DA_unc**2)
            if z1 == 1.52:
                r = rs(h, Om, Obh2)/147.78
                y = (D_M(h, beta, z1))/dH
                DV = dH*(((y**2)*z1)/(E(h, beta, z1)))**(1/3) 
                ##Distance parameter
                DVobs2 = 3843*r #Ata
                unc2 = 147*r
                chi2DV2 = ((DV - DVobs2)**2)/(unc2**2)
            if z1 == 2.334:
                DH = c/(H(h, beta, z1))
                DH_obs = 8.99
                DM_obs = 37.5
                Delta2 = array([(D_M(h, beta, z1)/(rs(h, Om, Obh2))
                    - DM_obs),
                    (DH/(rs(h, Om, Obh2)) - DH_obs)])
                prod2 = (Cinv2.dot(Delta2)).T
                chi2hm = Delta2.dot(prod2)
                chi2HM = chi2hm.item((0,0))
                
        return ((chi2DA + chi2HM + chi_sq(h, beta, Om, Obh2)
        + chi2DV3 + chi2DV2), 
            (chi2HM))
    
    def chi2h(h, beta): #Chi^2 for H(z).
        return sum(((1/sigHobs)*(H(h, beta, z_obs) - Hz_obs))**2)
        
    def th(h, beta, z): #Eq. (2) of 1708.08635v1
        #D_A = D_M(1 + z) (see astro-ph/9905116v4) so l_m/D_A = 
        #l_m(1 + z)/D_M.
        #D_M has units of Mpc, so the 10^6 factor converts it to pc.
        #1 rad = 2.06265*(10^8) milliarcsec.
        return ((180/pi)*(3600*1000))*(1 + z)*(lm/((10**6)*D_M(h, beta, z)))
    
    def chi2_Q(h, beta): #Eq. (6) of 1708.08635v1
        N = 0
        for i in range(0, 120, 1):
            A = (th(h, beta, z_obsQ[i]) - th_obs[i])**2)
            B = (sig_th_obs[i] + 0.1*th_obs[i])**2
            N += A/B 
        return N
        
    def mu_obs(v, f, a, b): #Observed distance modulus.
        return 2.5*(b*v + a) - 2.5*log10(f*(10**(0))) - 100.2
    
    def mu_th(h, beta, z): #Theoretical distance modulus.
        return 5*log10((1 + z)*D_M(h, beta, z)) + 25
        
    def unc1(v, vu, f, fu, a, asig, b, bsig): #uncertainty in \mu_obs
        return 2.5*sqrt((b*vu)**2 + (fu/(f*log(10)))**2 + 
        (bsig*v)**2 + (asig**2))
        
    def unc3(h, beta, z, zu): 
    ##uncertainty in \mu_th due to redshift uncertainty
        I, error = quad(lambda m: 1/(E(h, beta, m)), 0, z)
        y = 1
        A = (c/(100*h))*((1 + z)/D_M(h, beta, z))*(y/E(h, beta, z))
        B = 5/((1 + z)*log(10))
        return B*(A + 1)*zu    
       
    def chi2_4(h, beta, a, b, asig, bsig): 
    ##Takes HIIG_minus_GEHR.txt as input.
        NN = 0
        for i in range(0, len(zH), 1):
            A = mu_obs(ls[i], f[i], a, b) - mu_th(h, beta, zH[i]))**2
            B = unc1(ls[i], els[i], f[i], fu[i], a, asig, b, bsig)**2
            C = unc3(h, beta, zH[i], ezH[i])**2
            D = B + C
            NN += A/D
        return NN
        
    def D_L(h, beta, z): #Luminosity distance.
        return (1 + z)*D_M(h, beta, z)
    
    def E_iso(h, beta, z, Sb_obs): #Isotropic energy radiated by a GRB.
        return (1/(1 + z))*(4*pi*
        ((D_L(h, beta, z)*(3.086*(10**24)))**2)*Sb_obs)

    def frac_err_z(h, beta, z): 
    ##uncertainty in E_{iso} due to redshift uncertainty
        I, error = quad(lambda m: 1/(E(h, beta, m)), 0, z)
        y = 1
        A = (2*(c/(100*h))*(y/E(h, beta, z)))/(D_M(h, beta, z))
        return (1/(1 + z)) + A

    def chi2_G(h, beta, aG, bG, s_ext): #Chi^2 for GRB.
        gsum = 0
        for i in range(0, 119, 1):
            A = log10(E_iso(h, beta, zG[i], Sb_obs[i])) - 
                bG*log10(Ep_obs[i]) - aG
            s2 = (((Sb_obs_err[i]/Sb_obs[i])/(log(10)))**2 
                + (bG**2)*((Ep_obs_err[i]/Ep_obs[i])/(log(10)))**2 
                + s_ext**2 + 
                (((frac_err_z(h, beta, zG[i])/log(10))**2)*(zGe[i]**2)))
            gsum += (A**2)/s2 + log(2*pi*s2)
        return gsum
    
    Larg = (-1/2)*(chi2h(h, beta) + chi2_2(h, beta, Om, Obh2)[0] 
        + chi2_Q(h, beta) + chi2_G(h, beta, aG, bG, s_ext)
        + chi2_4(h, beta, a, b, asig, bsig))
        
    if isnan(Larg)==True:
        Larg=-inf
    else:
        Larg=Larg
    
    return Larg

def lnprior(paras):
    h, beta, Om, Obh2 = paras[0], paras[1], paras[2], paras[3]
    aG, bG, s_ext = paras[4], paras[5], paras[6]
        if ((0.2 <= h <= 1) and (0.25 <= beta <= 4) and (0.1 <= Om <= 0.7) 
        and (0.005 <= Obh2 <= 0.1) 
        and (40 <= aG <= 60) and (0 <= bG <= 5) and (0 <= s_ext <= 10)):
            return 0.0
    return -inf

def lnprob(paras):
    lp = lnprior(paras)
    if not isfinite(lp):
        return -inf
    return lp + logL(paras)
    
with MPIPool() as pool:
    if not pool.is_master():
        pool.wait()
        sys.exit(0)
        
    guess = array([0.70, 1, 0.3, 0.0225, 50, 2.5, 5])
    ##Initial values of parameters.
    ndim, nwalkers = 7, 100
    ##Number of parameters, number of walkers.
        
    p0 = random.rand(ndim*nwalkers).reshape((nwalkers, ndim))*1e-4 + guess
    ##Randomly perturbs positions of walkers around initial guess.
    sampler = emcee.EnsembleSampler(nwalkers, ndim, lnprob, pool=pool)
    pos, prob, state = sampler.run_mcmc(p0, 2000)
    sampler.reset()
    ##Sampler resets after 2000 steps.
    
    sampler.run_mcmc(pos, 20000)
    ##Sampler runs for a total of 20000 steps.
    chain = sampler.flatchain.copy()
    chi2 = -2*(sampler.flatlnprobability)
    savetxt('/homes/jwryan/emcee_stable/Results/
        Power_law/PL_Om_' + lab + '_chi2.txt', chi2)
    savetxt('/homes/jwryan/emcee_stable/Results/
        Power_law/PL_Om_' + lab + '.txt', chain)
\end{verbatim}

The following code analyzes the output chains produced by the code shown above, and makes two- and one-dimensional parameter constraint plots.

\begin{verbatim}

#!/bin/sh

#SBATCH --job-name=PL_Om_plotter

#SBATCH --mem-per-cpu=30G   # Memory per core, use --mem= for memory per node
#SBATCH --time=23:00:00   # Use the form DD-HH:MM:SS
#SBATCH --nodes=1
#SBATCH --ntasks-per-node=1
#SBATCH --constraint="mages|elves|heroes|dwarves|moles|wizards"
#SBATCH --gres=killable:1

#SBATCH --mail-user=jwryan@phys.ksu.edu

#SBATCH --mail-type=ALL    # same as =BEGIN,FAIL,END

module purge
module load Python/3.8.2-GCCcore-9.3.0
module load matplotlib
source ~/virtualenvs/emcee_2/bin/activate

export PYTHONDONTWRITEBYTECODE=1
PYTHON_BINARY=$(which python)
export OMP_NUM_THREADS=1
host=`hostname`

time ${PYTHON_BINARY} Plotter.py $SLURM_ARRAY_TASK_ID
echo "Finished run on 1 core on $host"
\end{verbatim}

\begin{verbatim}

# -*- coding: utf-8 -*-

from __future__ import print_function
from getdist import plots, MCSamples
import matplotlib.pyplot as plt
import sys
import emcee
from numpy import loadtxt, log, array, reshape, arange
from scipy.interpolate import interp1d
    
lab = 'ZBQGH'

chain = loadtxt(lab + '.txt', unpack=True)
print('chain loaded')

h = chain[0,:]
beta = chain[1,:]
Om = chain[2,:]
Obh2 = chain[3,:]
aG = chain[4,:]
bG = chain[5,:]
s_ext = chain[6,:]
del chain

chi2 = loadtxt(lab + '_chi2.txt', unpack=True)
print('chi2 loaded')

print(min(chi2), "chi2 minimum")
print(h[list(chi2).index(min(chi2))], "<< h best-fitting value")
print(beta[list(chi2).index(min(chi2))], "<< beta best-fitting value")
print(Om[list(chi2).index(min(chi2))], "<< Om best-fitting value")
print(Obh2[list(chi2).index(min(chi2))], "<< Obh2 best-fitting value")
print(aG[list(chi2).index(min(chi2))], "<< aG best-fitting value")
print(bG[list(chi2).index(min(chi2))], "<< bG best-fitting value")
print(s_ext[list(chi2).index(min(chi2))], "<< s_ext best-fitting value")

samps = array([h, beta, Om, Obh2, aG, bG, s_ext])
del h, beta, Om, Obh2, aG, bG, s_ext, chi2
samps = samps.T
names = ["h","beta","Om","Obh2","aG","bG","s_ext"]
labels = [r"H_0", r"\beta", r"\Omega_{m0}", 
    r"\Omega_{b0}h^2", r"a", r"b", r"\sigma_{ext}"]
samples = MCSamples(samples=samps, names = names, labels = labels)

g = plots.getSubplotPlotter()
samples.updateSettings({'contours': [0.6827, 0.9545, 0.9973]})
g.settings.alpha_filled_add=0.4
g.settings.num_plot_contours = 3
#g.settings.axes_fontsize = 0.1
g.settings.axis_tick_step = 0.06

g.triangle_plot([samples], names, filled_compare=False, 
    line_args={'ls':'solid', 'color':'black'})

g.export(lab + '.pdf') 
##Saves two-dimensional constraint contours to a .pdf file.

a=samples3.getMeans()
stats = samples3.getMargeStats()
lims0 = stats.parWithName('h').limits
lims1 = stats.parWithName('beta').limits
lims2 = stats.parWithName('Om').limits
lims3 = stats.parWithName('Obh2').limits
lims4 = stats.parWithName('aG').limits
lims5 = stats.parWithName('bG').limits
lims6 = stats.parWithName('s_ext').limits

print("means = ", a) 
##Prints one-dimensional sample means (best-fitting values).

for (conf, lim0, lim1, lim2, lim3, lim4, lim5, lim6
    in zip(samples3.contours,lims0, 
    lims1, lims2, lims3, lims4, lims5, lims6)):
    print('h %s%% lower: %.5f upper: %.5f (%s)'%(conf, lim0.lower - a[0], 
        lim0.upper - a[0], lim0.limitType()))
    print('beta %s%% lower: %.5f upper: %.5f (%s)'%(conf, lim1.lower - a[1],
        lim1.upper - a[1], lim1.limitType()))
    """
    print('Omega_{m0} %s%% lower: %.5f upper: %.5f (%s)'%(conf, lim2.lower, 
        lim2.upper, lim2.limitType()))
    print('Omega_{b0}h^2 %s%% lower: %.5f upper: %.5f (%s)'%(conf, 
        lim3.lower, lim3.upper, lim3.limitType()))
    print('aG %s%% lower: %.5f upper: %.5f (%s)'%(conf, lim4.lower,
        lim4.upper, lim4.limitType()))
    print('bG %s%% lower: %.5f upper: %.5f (%s)'%(conf, lim5.lower,
        lim5.upper, lim5.limitType()))
    print('s_ext %s%% lower: %.5f upper: %.5f (%s)'%(conf, lim6.lower,
        lim6.upper, lim6.limitType()))
    """
##Above loop prints two-sided confidence intervals on marginalized parameters.
\end{verbatim}

\subsection{Non-flat $\phi$CDM model codes}
\label{sec:phiCDM_emcee}

Here I provide a representative version of a \textsc{Python} code that can be used to compute constraints on the parameters of the non-flat $\phi$CDM model, from $H(z)$ and BAO data, using \textsc{emcee}. It is much simpler to obtain cosmological model constraints using \textsc{emcee} compared to the method shown in Sec. \ref{sec:phiCDM_code}. All that is required to produce a likelihood function is a code similar what I have copied below. Two-dimensional confidence contours, as well as one-dimensional sample means and confidence intervals, can be obtained from the output of this code using the same kind of code as what I used to obtain confidence contours and sample means for the power law model, above.

\begin{verbatim}
#!/bin/sh

#SBATCH --job-name=NFpCDM

#SBATCH --mem-per-cpu=1G   # Memory per core, use --mem= for memory per node
#SBATCH --time=10-00:00:00   # Use the form DD-HH:MM:SS
#SBATCH --nodes=1
#SBATCH --ntasks-per-node=48
#SBATCH --constraint="elves|heroes|dwarves|moles|wizards"
#SBATCH --gres=killable:1

#SBATCH --mail-user=jwryan@phys.ksu.edu
#SBATCH --mail-type=ALL    # same as =BEGIN,FAIL,END

module purge
module load Python/3.7.0-iomkl-2018b
source ~/virtualenvs/emcee_2/bin/activate

export PYTHONDONTWRITEBYTECODE=1
PYTHON_BINARY=$(which python)
export OMP_NUM_THREADS=1
host=`hostname`

time mpirun ${PYTHON_BINARY} Nonflat_phiCDM.py
\end{verbatim}

\begin{verbatim}

# -*- coding: utf-8 -*-

from scipy.integrate import odeint #ODE solver
from numpy import genfromtxt, log, matrix, loadtxt, exp, arange 
from numpy import savetxt, array, random, inf, isfinite, pi 
from numpy import log10, sin, sinh, sqrt, cos, cosh
from numpy import identity, diag, isnan, reshape, column_stack
from numpy.linalg import inv
from getdist import plots, MCSamples
import emcee
import sys
from scipy.integrate import quad
from math import ceil
from emcee.utils import MPIPool

lab = 'ZB'

K = 0
alx = arange(0.01, 3.01, 0.01)
Om = arange(0.10, 0.71, 0.01)
Hrange = arange(50.0, 85.001, 0.01)

z0 = [2.334, 1.52, 0.81, 0.61, 0.51, 0.38, 0.122]

DM_obs = [1512.39, 1975.22, 2306.68]
H_obs = [81.2087, 90.9029, 98.9647]
##Data points from DR12 website. File "BAO_consensus_results_dM_Hz.txt"

z_obsQ, th_obs, sig_th_obs = loadtxt('QSO_120.txt', unpack=True)

z_obs, Hz_obs, sigHobs = loadtxt('H(z)data.dat',unpack = True)
##From Table 1 of 1607.03537v2, refs. 4,6,7,10 excluded.

rfid = 147.60 #Planck, Table 4 of 1502.01589.

##Covariance matrix from DR12 website. "BAO_consensus_covtot_dM_Hz.txt".
Cov = matrix([[624.707,23.729,325.332,8.34963,157.386,3.57778],
              [23.729,5.60873,11.6429,2.33996,6.39263,0.968056],
              [325.332,11.6429,905.777,29.3392,515.271,14.1013],
              [8.34963,2.33996,29.3392,5.42327,16.1422,2.85334],
              [157.386,6.39263,515.271,16.1422,1375.12,40.4327],
              [3.57778,0.968056,14.1013,2.85334,40.4327,6.25936]])
Cinv = inv(Cov)
del Cov

Cov2 = matrix([[1.3225, -0.1009],[-0.1009, 0.0380]])
Cinv2 = inv(Cov2)
del Cov2
##Cov2 calculated from correlation matrix and uncertainties in 
##arXiv:2101.08817

c = 299792458./1000. #Speed of light in km/s
Th = 2.7255/2.7 #T_CMB/2.7, from Eisenstein and Hu 1998
                #and Fixsen 0911.1955
On = 0.0014
m = 1.
t0 = 10.**(-4.)
tf = 150.
dt = 10.**(-4.)
t = arange(t0, tf + t0, dt)
a0 = t0**(2./3.) #I assumed a ~ t^(2/3) in the early universe.

def logL(paras):
    h, Obh2, Och2, al, K = paras[0], paras[1], paras[2], paras[3], paras[4]
        
    H0 = 100*h
    O = (Obh2 + Och2)/(h**2) + On
     
    def phiCDM(w, t, zz):
        
        p, v, a = w #I used p for phi, v for d(phi)/dt, and a for the 
        ##scale factor.
        al, k, m, K = zz #These are the parameters of the model. See below.
    
        f = [v,
             -3.*v*(((4./(9.*a**3.))) + (1./12.)*(v**2. + (k*m)/(p**(al))) 
             - K/(a**2.))**(1./2.)
             + ((k*al*m)/2.)/(p**(al + 1.)), 
            (((4./9.)/a) + (((a**2.)/12.))*(v**2. + 
            (k*m)*(p**(-al))) - K)**(1./2.)]
        return f
    
    def chi2_2(H0, O, al, K):
        k = (8./3.)*((al + 4.)/(al + 2.))*(((2./3.)*(al*(al + 2.))))**(al/2.) 
        #This is kappa,
        #from eq. (2) of arXiv:1307.7399v1.
        
        #initial conditions on phi, d(phi)/dt, a.
        p0 = (((2./3.)*(al*(al + 2.)))**(1./2.))*(t0)**(2./(al + 2.)) 
            #Initial value of phi.
        v0 = (((8./3.)*al*(1./(al + 2.)))**(1./2.))/(t0)**(al/(al + 2.)) 
            #Initial value of d(phi)/dt.
        w0 = [p0, v0, a0]    
        
        zz = [al, k, m, K]
        
        #solution array
        sol = odeint(phiCDM, w0, t, args=(zz,))
        
        def Ome(b):
            Omegam1 = (4./9.)*(1./(sol[b,2])**3.)
            Omegaphi1 = (1./12.)*(((sol[b,1])**2.) + k/((sol[b,0])**al))
            Omegak1 = -K/((sol[b,2])**2.)
            Omegam = Omegam1/(Omegam1 + Omegaphi1 + Omegak1)
            Omegak = Omegak1/(Omegam1 + Omegaphi1 + Omegak1)
            return Omegam, Omegak
        
        def rs(h, Obh2, Och2):
            Onh2 = On*(h**2)
            A = 55.154*(exp(-72.3*((Onh2 + 0.0006)**2)))
            B = ((Obh2 + Och2)**0.25351)*((Obh2)**0.12807)
            return A/B
        
        def E(O, red, Ok_0, Ophiz):
            return (O*((1 + red)**3) + (Ok_0)*((1 + red)**2.) + 
            Ophiz)**(1/2)
            
        def D_M(H0, q, O, Ok_0):
            if Ok_0 == 0:
                return (c/H0)*h0*afin*rr[q]
            if Ok_0 < 0:
                return (c/H0)*(1/(sqrt(-Ok_0)))*(sin((sqrt(-Ok_0))*
                (h0*afin*rr[q])))
            if Ok_0 > 0:
                return (c/H0)*(1/(sqrt(Ok_0)))*(sinh((sqrt(Ok_0))*
                (h0*afin*rr[q])))
        
        def chi_sq(H0, O, Ok_0):
            DM_th = []
            H_th = []
            for q in range(6, -1, -1):
                z1 = z0[q]
                H1 = H0*E(O, z1, Ok_0, O_phi_z[q])
                DM = D_M(H0, q, O, Ok_0)
                y = (H0/c)*DM
                if 3 <= q <= 5:
                    DM_th.append(D_M(H0, q, O, Ok_0))
                    H_th.append(H1)
                if z1 == 0.122:
                    r = rs(h, Obh2, Och2)/147.5
                    DV = ((c/H0)
                        *(((y**2.)*z1)/(E(O, z1, Ok_0, O_phi_z[q])))**(1./3.))
                    DVobs3 = 539*r #Carter et al., doi:10.1093/mnras/sty2405
                    unc3 = 17*r
                    chi2DV3 = ((DV - DVobs3)**2)/(unc3**2)
                if z1 == 0.81:
                    DA = DM/(1 + z1)
                    DA_obs = 10.75
                    DA_unc = 0.43
                    chi2DA = (((DA/(rs(h, Obh2, Och2))
                        - DA_obs)**2)/(DA_unc**2))
                if z1 == 1.52:
                    r = rs(h, Obh2, Och2)/147.78
                    DV = ((c/H0)
                        *(((y**2.)*z1)/(E(O, z1, Ok_0, O_phi_z[q])))**(1./3.))
                    DVobs2 = 3843.*r #Ata
                    unc2 = 147.*r
                    chi2DV2 = ((DV - DVobs2)**2.)/(unc2**2.)
                if z1 == 2.334:
                    DH_obs = 8.99
                    Dm_obs = 37.5
                    DH = c/H1
                    Delta2 = array([(DM/(rs(h, Obh2, Och2)) - Dm_obs),
                        (DH/(rs(h, Obh2, Och2)) - DH_obs)])        
                    prod2 = (Cinv2.dot(Delta2)).T
                    chi2hm = Delta2.dot(prod2)
                    chi2HM = chi2hm.item((0,0))
                    
            r = rs(h, Obh2, Och2)/147.78
            Delta = array([(DM_th[0]/r - DM_obs[0]), 
                                 (r*H_th[0] - H_obs[0]),
                                 (DM_th[1]/r - DM_obs[1]),
                                 (r*H_th[1] - H_obs[1]),
                                 (DM_th[2]/r - DM_obs[2]),
                                 (r*H_th[2] - H_obs[2])]) 
            prod = (Cinv.dot(Delta)).T
            chi_sq1 = Delta.dot(prod)
            chi_sq_11 = chi_sq1.item((0,0))
            return ((chi2DA + chi2HM + chi_sq_11
                + chi2DV3 + chi2DV2), (chi2HM))
           
        def Ofunc(d):
            Omegam1 = (4./9.)*(1./(sol[d,2])**3.)
            Omegaphi1 = (1./12.)*(((sol[d,1])**2.) + k/((sol[d,0])**al))
            Omegak1 = -K/((sol[d,2])**2.)
            Omegam = Omegam1/(Omegam1 + Omegaphi1 + Omegak1)
            return Omegam, Omegam1, Omegaphi1, Omegak1
        
        def chi2h(H0, O, Ok_0):
            return (sum(((1/sigHobs)*((H0*E(O, z_obs, Ok_0, O_phi_z1))
                - Hz_obs))**2))
        
        def D_L(H0, z, q, O, Ok_0):
            return (1 + z)*D_M(H0, q, O, Ok_0)
          
        for b in range(0, ceil(tf/dt), 1):
            if (O >= Ofunc(b)[0]):
                break
        afin = sol[b,2]
        h0 = (sol[b+1, 2] - afin)/afin
        global Ok_0
        Ok_0 = Ofunc(b)[3]/(Ofunc(b)[1] + Ofunc(b)[2] + Ofunc(b)[3])

        Omegap = ((4./9.)*(1./(sol[b,2])**3.) 
                + (1./12.)*(((sol[b,1])**2.) + k/((sol[b,0])**al)) 
                + -K/((sol[b,2])**2.))
                
        rr = []
        O_phi_z = []
        for z1 in z0:
            r = 0
            qt = 0
            t1 = []
            r1 = []
            
            for d in range(0, b+1, 1):
                if ((sol[d,2])/afin >= 1./(1. + z1)):
                    r += 1./(sol[d,2])
                    qt += 1
                if qt == 1:
                    Omegaphi_z = ((1./12.)*(((sol[d,1])**2.)
                        + k/((sol[d,0])**al)))
            rr.append(r)
            O_phi_z.append(Omegaphi_z/((4./9.)*(1./(sol[b,2])**3.) 
            + (1./12.)*(((sol[b,1])**2.) + k/((sol[b,0])**al)) 
            + -K/((sol[b,2])**2.)))        
        cBAO = chi_sq(H0, O, Ok_0)[0]
        
        O_phi_z1 = []
        for qs in range(0, len(z_obs), 1):
            for d in range(0, b+1, 1):
                if ((sol[d,2])/afin >= 1./(1. + z_obs[qs])):
                    Omegaphi2 = ((1./12.)*(((sol[d,1])**2.)
                        + k/((sol[d,0])**al)))
                    O_phi_z1.append(Omegaphi2/((4./9.)*(1./(sol[b,2])**3.) 
                    + (1./12.)*(((sol[b,1])**2.) + k/((sol[b,0])**al)) 
                    + -K/((sol[b,2])**2.)))
                    break
        chi2Hz = chi2h(H0, O, Ok_0)
            
        return chi2Hz + cBAO
    
    Larg = (-1/2)*(chi2_2(H0, O, al, K))
    if isnan(Larg)==True:
        Larg=-inf
    else:
        Larg=Larg
        
    return Larg
    
def lnprior(paras):
    h, Obh2, Och2, al, K = paras[0], paras[1], paras[2], paras[3], paras[4]
    if ((0.2 <= h <= 1) and (0.005 <= Obh2 <= 0.1)
        and (0.001 <= Och2 <= 0.99) 
        and (0.01 <= al <= 3) and (-2.1 <= K <= 0.4)):
        return 0.0        
    return -inf

def lnprob(paras):
    lp = lnprior(paras)
    if not isfinite(lp):
        return -inf, -inf
    lp2 = logL(paras)
    if not isfinite(lp2):
        return lp, -inf
    return lp + lp2, Ok_0

with MPIPool() as pool:
    if not pool.is_master():
        pool.wait()
        sys.exit(0)
    guess = array([0.70, 0.0225, 0.4955, 0.01, 0])
    ndim, nwalkers = 5, 100
    
    p0 = random.rand(ndim*nwalkers).reshape((nwalkers, ndim))*1e-4 + guess
    sampler = emcee.EnsembleSampler(nwalkers, ndim, lnprob, pool=pool)
    pos, prob, state, blobs = sampler.run_mcmc(p0, 1000)
    sampler.reset()
    
    sampler.run_mcmc(pos, 5000)
    OK = reshape(sampler.blobs, 500000, order='F')
    Ch = sampler.flatchain.copy()
    chain = column_stack((Ch, OK))
    #chain = sampler.flatchain.copy()
    chi2 = -2*(sampler.flatlnprobability)
    savetxt('/homes/jwryan/emcee_stable/Results/NFpCDM_' + lab 
    + '_chi2.txt', chi2)
    savetxt('/homes/jwryan/emcee_stable/Results/NFpCDM_' + lab 
    + '.txt', chain)
\end{verbatim}

%% file: etdrtemplate.bbl
\begin{thebibliography}{317}
\expandafter\ifx\csname natexlab\endcsname\relax\def\natexlab#1{#1}\fi
\expandafter\ifx\csname bibnamefont\endcsname\relax
  \def\bibnamefont#1{#1}\fi
\expandafter\ifx\csname bibfnamefont\endcsname\relax
  \def\bibfnamefont#1{#1}\fi
\expandafter\ifx\csname citenamefont\endcsname\relax
  \def\citenamefont#1{#1}\fi
\expandafter\ifx\csname url\endcsname\relax
  \def\url#1{\texttt{#1}}\fi
\expandafter\ifx\csname urlprefix\endcsname\relax\def\urlprefix{URL }\fi
\providecommand{\bibinfo}[2]{#2}
\providecommand{\eprint}[2][]{\url{#2}}

\bibitem[{\citenamefont{Peebles}(1993)}]{Peebles_1993}
\bibinfo{author}{\bibfnamefont{P.}~\bibnamefont{Peebles}},
  \emph{\bibinfo{title}{{Principles of Physical Cosmology}}}
  (\bibinfo{publisher}{Princeton University Press},
  \bibinfo{address}{Princeton, NJ}, \bibinfo{year}{1993}).

\bibitem[{\citenamefont{{Dodelson}}(2003)}]{Dodelson}
\bibinfo{author}{\bibfnamefont{S.}~\bibnamefont{{Dodelson}}},
  \emph{\bibinfo{title}{{Modern Cosmology}}} (\bibinfo{publisher}{Academic
  Press}, \bibinfo{address}{San Diego, CA}, \bibinfo{year}{2003}),
  \urlprefix\url{https://cds.cern.ch/record/1282338?ln=en}.

\bibitem[{\citenamefont{Mukhanov}(2005)}]{Mukhanov}
\bibinfo{author}{\bibfnamefont{V.}~\bibnamefont{Mukhanov}},
  \emph{\bibinfo{title}{{Physical Foundations of Cosmology}}}
  (\bibinfo{publisher}{Cambridge University Press},
  \bibinfo{address}{Cambridge, UK}, \bibinfo{year}{2005}),
  \urlprefix\url{https://cds.cern.ch/record/991646?ln=en}.

\bibitem[{\citenamefont{Weinberg}(2008)}]{weinberg}
\bibinfo{author}{\bibfnamefont{S.}~\bibnamefont{Weinberg}},
  \emph{\bibinfo{title}{{Cosmology}}} (\bibinfo{publisher}{Oxford University
  Press}, \bibinfo{address}{New York, NY}, \bibinfo{year}{2008}),
  \urlprefix\url{https://cds.cern.ch/record/1102255?ln=en}.

\bibitem[{\citenamefont{Zee}(2013)}]{zee}
\bibinfo{author}{\bibfnamefont{A.}~\bibnamefont{Zee}},
  \emph{\bibinfo{title}{{Einstein Gravity in a Nutshell}}}
  (\bibinfo{publisher}{Princeton University Press},
  \bibinfo{address}{Princeton, NJ}, \bibinfo{year}{2013}),
  \urlprefix\url{https://cds.cern.ch/record/1529077?ln=en}.

\bibitem[{\citenamefont{{Thorne} and {Blandford}}(2017)}]{MCP}
\bibinfo{author}{\bibfnamefont{K.~S.} \bibnamefont{{Thorne}}} \bibnamefont{and}
  \bibinfo{author}{\bibfnamefont{R.~D.} \bibnamefont{{Blandford}}},
  \emph{\bibinfo{title}{{Modern Classical Physics}}}
  (\bibinfo{publisher}{Princeton University Press},
  \bibinfo{address}{Princeton, NJ}, \bibinfo{year}{2017}).

\bibitem[{\citenamefont{{Thornton} and {Rex}}(2006)}]{Thornton_Rex}
\bibinfo{author}{\bibfnamefont{S.~T.} \bibnamefont{{Thornton}}}
  \bibnamefont{and} \bibinfo{author}{\bibfnamefont{A.}~\bibnamefont{{Rex}}},
  \emph{\bibinfo{title}{{Modern Physics for Scientists and Engineers}}}
  (\bibinfo{publisher}{Brooks/Cole, Cengage Learning},
  \bibinfo{address}{Belmont, CA}, \bibinfo{year}{2006}), \bibinfo{edition}{3rd}
  ed.

\bibitem[{\citenamefont{{Gray} and {Taylor}}(2007)}]{Gray_Taylor_2007}
\bibinfo{author}{\bibfnamefont{C.~G.} \bibnamefont{{Gray}}} \bibnamefont{and}
  \bibinfo{author}{\bibfnamefont{E.~F.} \bibnamefont{{Taylor}}},
  \bibinfo{journal}{Am. J. Phys.} \textbf{\bibinfo{volume}{75}},
  \bibinfo{pages}{434} (\bibinfo{year}{2007}).

\bibitem[{\citenamefont{{Price}}(2016)}]{Price_2016}
\bibinfo{author}{\bibfnamefont{R.~H.} \bibnamefont{{Price}}},
  \bibinfo{journal}{Am. J. Phys.} \textbf{\bibinfo{volume}{84}},
  \bibinfo{pages}{588} (\bibinfo{year}{2016}).

\bibitem[{\citenamefont{{Lemons}}(2009)}]{Lemons_Thermo}
\bibinfo{author}{\bibfnamefont{D.~S.} \bibnamefont{{Lemons}}},
  \emph{\bibinfo{title}{{Mere Thermodynamics}}} (\bibinfo{publisher}{The Johns
  Hopkins University Press}, \bibinfo{address}{Baltimore, MD},
  \bibinfo{year}{2009}).

\bibitem[{\citenamefont{{Park} et~al.}(2017)\citenamefont{{Park}, {Hyun},
  {Noh}, and {Hwang}}}]{Park_et_al_2017}
\bibinfo{author}{\bibfnamefont{C.-G.} \bibnamefont{{Park}}},
  \bibinfo{author}{\bibfnamefont{H.}~\bibnamefont{{Hyun}}},
  \bibinfo{author}{\bibfnamefont{H.}~\bibnamefont{{Noh}}}, \bibnamefont{and}
  \bibinfo{author}{\bibfnamefont{J.-c.} \bibnamefont{{Hwang}}},
  \bibinfo{journal}{Mon. Not. R. Astron. Soc.} \textbf{\bibinfo{volume}{469}},
  \bibinfo{pages}{1924} (\bibinfo{year}{2017}), \eprint{1611.02139}.

\bibitem[{\citenamefont{{M{\'e}sz{\'a}ros}}(2019)}]{Meszaros_2019}
\bibinfo{author}{\bibfnamefont{A.}~\bibnamefont{{M{\'e}sz{\'a}ros}}},
  \bibinfo{journal}{Astron. Nachr.} \textbf{\bibinfo{volume}{340}},
  \bibinfo{pages}{564} (\bibinfo{year}{2019}), \eprint{1912.07523}.

\bibitem[{\citenamefont{{Secrest} et~al.}(2021)\citenamefont{{Secrest}, {von
  Hausegger}, {Rameez}, {Mohayaee}, {Sarkar}, and
  {Colin}}}]{Secrest_et_al_2021}
\bibinfo{author}{\bibfnamefont{N.~J.} \bibnamefont{{Secrest}}},
  \bibinfo{author}{\bibfnamefont{S.}~\bibnamefont{{von Hausegger}}},
  \bibinfo{author}{\bibfnamefont{M.}~\bibnamefont{{Rameez}}},
  \bibinfo{author}{\bibfnamefont{R.}~\bibnamefont{{Mohayaee}}},
  \bibinfo{author}{\bibfnamefont{S.}~\bibnamefont{{Sarkar}}}, \bibnamefont{and}
  \bibinfo{author}{\bibfnamefont{J.}~\bibnamefont{{Colin}}},
  \bibinfo{journal}{Astrophys. J. Lett.} \textbf{\bibinfo{volume}{908}},
  \bibinfo{eid}{L51} (\bibinfo{year}{2021}), \eprint{2009.14826}.

\bibitem[{\citenamefont{{Gon{\c{c}}alves}
  et~al.}(2018)\citenamefont{{Gon{\c{c}}alves}, {Carvalho}, {Bengaly},
  {Carvalho}, {Bernui}, {Alcaniz}, and {Maartens}}}]{Goncalves_et_al_2018}
\bibinfo{author}{\bibfnamefont{R.~S.} \bibnamefont{{Gon{\c{c}}alves}}},
  \bibinfo{author}{\bibfnamefont{G.~C.} \bibnamefont{{Carvalho}}},
  \bibinfo{author}{\bibfnamefont{J.}~\bibnamefont{{Bengaly}},
  \bibfnamefont{C.~A.~P.}}, \bibinfo{author}{\bibfnamefont{J.~C.}
  \bibnamefont{{Carvalho}}},
  \bibinfo{author}{\bibfnamefont{A.}~\bibnamefont{{Bernui}}},
  \bibinfo{author}{\bibfnamefont{J.~S.} \bibnamefont{{Alcaniz}}},
  \bibnamefont{and}
  \bibinfo{author}{\bibfnamefont{R.}~\bibnamefont{{Maartens}}},
  \bibinfo{journal}{Mon. Not. R. Astron. Soc.} \textbf{\bibinfo{volume}{475}},
  \bibinfo{pages}{L20} (\bibinfo{year}{2018}), \eprint{1710.02496}.

\bibitem[{\citenamefont{{Gon{\c{c}}alves}
  et~al.}(2020)\citenamefont{{Gon{\c{c}}alves}, {Carvalho}, {Andrade},
  {Bengaly}, {Carvalho}, and {Alcaniz}}}]{Goncalves_et_al_2020}
\bibinfo{author}{\bibfnamefont{R.~S.} \bibnamefont{{Gon{\c{c}}alves}}},
  \bibinfo{author}{\bibfnamefont{G.~C.} \bibnamefont{{Carvalho}}},
  \bibinfo{author}{\bibfnamefont{U.}~\bibnamefont{{Andrade}}},
  \bibinfo{author}{\bibfnamefont{C.~A.~P.} \bibnamefont{{Bengaly}}},
  \bibinfo{author}{\bibfnamefont{J.~C.} \bibnamefont{{Carvalho}}},
  \bibnamefont{and}
  \bibinfo{author}{\bibfnamefont{J.}~\bibnamefont{{Alcaniz}}},
  \bibinfo{journal}{ArXiv e-prints}  (\bibinfo{year}{2020}),
  \eprint{2010.06635}.

\bibitem[{\citenamefont{{Yadav} et~al.}(2005)\citenamefont{{Yadav},
  {Bharadwaj}, {Pandey}, and {Seshadri}}}]{Yadav_et_al_2005}
\bibinfo{author}{\bibfnamefont{J.}~\bibnamefont{{Yadav}}},
  \bibinfo{author}{\bibfnamefont{S.}~\bibnamefont{{Bharadwaj}}},
  \bibinfo{author}{\bibfnamefont{B.}~\bibnamefont{{Pandey}}}, \bibnamefont{and}
  \bibinfo{author}{\bibfnamefont{T.~R.} \bibnamefont{{Seshadri}}},
  \bibinfo{journal}{Mon. Not. R. Astron. Soc.} \textbf{\bibinfo{volume}{364}},
  \bibinfo{pages}{601} (\bibinfo{year}{2005}), \eprint{astro-ph/0504315}.

\bibitem[{\citenamefont{{Planck Collaboration}
  et~al.}(2020)\citenamefont{{Planck Collaboration}, {Akrami}, {Ashdown},
  {Aumont}, {Baccigalupi}, {Ballardini}, {Banday}, {Barreiro}, {Bartolo},
  {Basak} et~al.}}]{Planck_2018_Isotropy}
\bibinfo{author}{\bibnamefont{{Planck Collaboration}}},
  \bibinfo{author}{\bibfnamefont{Y.}~\bibnamefont{{Akrami}}},
  \bibinfo{author}{\bibfnamefont{M.}~\bibnamefont{{Ashdown}}},
  \bibinfo{author}{\bibfnamefont{J.}~\bibnamefont{{Aumont}}},
  \bibinfo{author}{\bibfnamefont{C.}~\bibnamefont{{Baccigalupi}}},
  \bibinfo{author}{\bibfnamefont{M.}~\bibnamefont{{Ballardini}}},
  \bibinfo{author}{\bibfnamefont{A.~J.} \bibnamefont{{Banday}}},
  \bibinfo{author}{\bibfnamefont{R.~B.} \bibnamefont{{Barreiro}}},
  \bibinfo{author}{\bibfnamefont{N.}~\bibnamefont{{Bartolo}}},
  \bibinfo{author}{\bibfnamefont{S.}~\bibnamefont{{Basak}}},
  \bibnamefont{et~al.}, \bibinfo{journal}{Astron. Astrophys.}
  \textbf{\bibinfo{volume}{641}}, \bibinfo{eid}{A7} (\bibinfo{year}{2020}),
  \eprint{1906.02552}.

\bibitem[{\citenamefont{{Wu} et~al.}(1999)\citenamefont{{Wu}, {Lahav}, and
  {Rees}}}]{Wu_Lahav_Rees_1999}
\bibinfo{author}{\bibfnamefont{K.~K.~S.} \bibnamefont{{Wu}}},
  \bibinfo{author}{\bibfnamefont{O.}~\bibnamefont{{Lahav}}}, \bibnamefont{and}
  \bibinfo{author}{\bibfnamefont{M.~J.} \bibnamefont{{Rees}}},
  \bibinfo{journal}{Nature} \textbf{\bibinfo{volume}{397}},
  \bibinfo{pages}{225} (\bibinfo{year}{1999}), \eprint{astro-ph/9804062}.

\bibitem[{\citenamefont{{Deng} and {Wei}}(2018)}]{Deng_Wei_2018}
\bibinfo{author}{\bibfnamefont{H.-K.} \bibnamefont{{Deng}}} \bibnamefont{and}
  \bibinfo{author}{\bibfnamefont{H.}~\bibnamefont{{Wei}}},
  \bibinfo{journal}{Eur. Phys. J. C} \textbf{\bibinfo{volume}{78}},
  \bibinfo{eid}{755} (\bibinfo{year}{2018}), \eprint{1806.02773}.

\bibitem[{\citenamefont{{Smoot} et~al.}(1992)\citenamefont{{Smoot}, {Bennett},
  {Kogut}, {Wright}, {Aymon}, {Boggess}, {Cheng}, {de Amici}, {Gulkis},
  {Hauser} et~al.}}]{Smoot_et_al_1992}
\bibinfo{author}{\bibfnamefont{G.~F.} \bibnamefont{{Smoot}}},
  \bibinfo{author}{\bibfnamefont{C.~L.} \bibnamefont{{Bennett}}},
  \bibinfo{author}{\bibfnamefont{A.}~\bibnamefont{{Kogut}}},
  \bibinfo{author}{\bibfnamefont{E.~L.} \bibnamefont{{Wright}}},
  \bibinfo{author}{\bibfnamefont{J.}~\bibnamefont{{Aymon}}},
  \bibinfo{author}{\bibfnamefont{N.~W.} \bibnamefont{{Boggess}}},
  \bibinfo{author}{\bibfnamefont{E.~S.} \bibnamefont{{Cheng}}},
  \bibinfo{author}{\bibfnamefont{G.}~\bibnamefont{{de Amici}}},
  \bibinfo{author}{\bibfnamefont{S.}~\bibnamefont{{Gulkis}}},
  \bibinfo{author}{\bibfnamefont{M.~G.} \bibnamefont{{Hauser}}},
  \bibnamefont{et~al.}, \bibinfo{journal}{Astrophys. J. Lett.}
  \textbf{\bibinfo{volume}{396}}, \bibinfo{pages}{L1} (\bibinfo{year}{1992}).

\bibitem[{\citenamefont{{Ryden} and {Peterson}}(2010)}]{Ryden_Peterson}
\bibinfo{author}{\bibfnamefont{B.}~\bibnamefont{{Ryden}}} \bibnamefont{and}
  \bibinfo{author}{\bibfnamefont{B.~M.} \bibnamefont{{Peterson}}},
  \emph{\bibinfo{title}{{Foundations of Astrophysics}}}
  (\bibinfo{publisher}{Addison-Wesley}, \bibinfo{address}{San Francisco, CA},
  \bibinfo{year}{2010}), \urlprefix\url{https://cds.cern.ch/record/1186260}.

\bibitem[{\citenamefont{{Gradshteyn} and {Ryzhik}}(2015)}]{G&R}
\bibinfo{author}{\bibfnamefont{I.~S.} \bibnamefont{{Gradshteyn}}}
  \bibnamefont{and} \bibinfo{author}{\bibfnamefont{I.~M.}
  \bibnamefont{{Ryzhik}}}, \emph{\bibinfo{title}{{Table of Integrals, Series,
  and Products}}} (\bibinfo{publisher}{Elsevier}, \bibinfo{address}{Waltham,
  MA}, \bibinfo{year}{2015}), \bibinfo{edition}{8th} ed.

\bibitem[{\citenamefont{{Hogg}}(1999)}]{Hogg}
\bibinfo{author}{\bibfnamefont{D.~W.} \bibnamefont{{Hogg}}},
  \bibinfo{journal}{arXiv e-prints}  (\bibinfo{year}{1999}),
  \eprint{astro-ph/9905116}.

\bibitem[{\citenamefont{Farooq}(2013)}]{5}
\bibinfo{author}{\bibfnamefont{M.~O.} \bibnamefont{Farooq}}, Ph.D. thesis,
  \bibinfo{school}{Kansas State U.} (\bibinfo{year}{2013}), \eprint{1309.3710}.

\bibitem[{\citenamefont{{Riess} et~al.}(1998)\citenamefont{{Riess},
  {Filippenko}, {Challis}, {Clocchiatti}, {Diercks}, {Garnavich}, {Gilliland},
  {Hogan}, {Jha}, {Kirshner} et~al.}}]{Riess_et_al_1998}
\bibinfo{author}{\bibfnamefont{A.~G.} \bibnamefont{{Riess}}},
  \bibinfo{author}{\bibfnamefont{A.~V.} \bibnamefont{{Filippenko}}},
  \bibinfo{author}{\bibfnamefont{P.}~\bibnamefont{{Challis}}},
  \bibinfo{author}{\bibfnamefont{A.}~\bibnamefont{{Clocchiatti}}},
  \bibinfo{author}{\bibfnamefont{A.}~\bibnamefont{{Diercks}}},
  \bibinfo{author}{\bibfnamefont{P.~M.} \bibnamefont{{Garnavich}}},
  \bibinfo{author}{\bibfnamefont{R.~L.} \bibnamefont{{Gilliland}}},
  \bibinfo{author}{\bibfnamefont{C.~J.} \bibnamefont{{Hogan}}},
  \bibinfo{author}{\bibfnamefont{S.}~\bibnamefont{{Jha}}},
  \bibinfo{author}{\bibfnamefont{R.~P.} \bibnamefont{{Kirshner}}},
  \bibnamefont{et~al.}, \bibinfo{journal}{Astron. J.}
  \textbf{\bibinfo{volume}{116}}, \bibinfo{pages}{1009} (\bibinfo{year}{1998}),
  \eprint{astro-ph/9805201}.

\bibitem[{\citenamefont{{Perlmutter} et~al.}(1999)\citenamefont{{Perlmutter},
  {Aldering}, {Goldhaber}, {Knop}, {Nugent}, {Castro}, {Deustua}, {Fabbro},
  {Goobar}, {Groom} et~al.}}]{Perlmutter_et_al_1999}
\bibinfo{author}{\bibfnamefont{S.}~\bibnamefont{{Perlmutter}}},
  \bibinfo{author}{\bibfnamefont{G.}~\bibnamefont{{Aldering}}},
  \bibinfo{author}{\bibfnamefont{G.}~\bibnamefont{{Goldhaber}}},
  \bibinfo{author}{\bibfnamefont{R.~A.} \bibnamefont{{Knop}}},
  \bibinfo{author}{\bibfnamefont{P.}~\bibnamefont{{Nugent}}},
  \bibinfo{author}{\bibfnamefont{P.~G.} \bibnamefont{{Castro}}},
  \bibinfo{author}{\bibfnamefont{S.}~\bibnamefont{{Deustua}}},
  \bibinfo{author}{\bibfnamefont{S.}~\bibnamefont{{Fabbro}}},
  \bibinfo{author}{\bibfnamefont{A.}~\bibnamefont{{Goobar}}},
  \bibinfo{author}{\bibfnamefont{D.~E.} \bibnamefont{{Groom}}},
  \bibnamefont{et~al.}, \bibinfo{journal}{Astrophys. J.}
  \textbf{\bibinfo{volume}{517}}, \bibinfo{pages}{565} (\bibinfo{year}{1999}),
  \eprint{astro-ph/9812133}.

\bibitem[{\citenamefont{{Planck Collaboration}}(2018)}]{planck2018_overview}
\bibinfo{author}{\bibnamefont{{Planck Collaboration}}}, \bibinfo{journal}{arXiv
  e-prints}  (\bibinfo{year}{2018}), \bibinfo{note}{arXiv:1807.06205}.

\bibitem[{\citenamefont{{Peebles} and {Ratra}}(2003)}]{Peebles_Ratra}
\bibinfo{author}{\bibfnamefont{P.~J.~E.} \bibnamefont{{Peebles}}}
  \bibnamefont{and} \bibinfo{author}{\bibfnamefont{B.}~\bibnamefont{{Ratra}}},
  \bibinfo{journal}{Rev. Mod. Phys.} \textbf{\bibinfo{volume}{75}},
  \bibinfo{pages}{559} (\bibinfo{year}{2003}), \eprint{astro-ph/0207347}.

\bibitem[{\citenamefont{{Ratra} and {Vogeley}}(2008)}]{Ratra_Vogeley}
\bibinfo{author}{\bibfnamefont{B.}~\bibnamefont{{Ratra}}} \bibnamefont{and}
  \bibinfo{author}{\bibfnamefont{M.~S.} \bibnamefont{{Vogeley}}},
  \bibinfo{journal}{Publ. Astron. Soc. Pac} \textbf{\bibinfo{volume}{120}},
  \bibinfo{pages}{235} (\bibinfo{year}{2008}), \eprint{0706.1565}.

\bibitem[{\citenamefont{{Planck Collaboration}}(2020)}]{planck2018}
\bibinfo{author}{\bibnamefont{{Planck Collaboration}}},
  \bibinfo{journal}{Astron. Astrophys.} \textbf{\bibinfo{volume}{641}},
  \bibinfo{eid}{A6} (\bibinfo{year}{2020}), \eprint{1807.06209}.

\bibitem[{\citenamefont{{Chevallier} and
  {Polarski}}(2001)}]{Chevallier_Polarski_2001}
\bibinfo{author}{\bibfnamefont{M.}~\bibnamefont{{Chevallier}}}
  \bibnamefont{and}
  \bibinfo{author}{\bibfnamefont{D.}~\bibnamefont{{Polarski}}},
  \bibinfo{journal}{Int. J. Mod. Phys. D} \textbf{\bibinfo{volume}{10}},
  \bibinfo{pages}{213} (\bibinfo{year}{2001}), \eprint{gr-qc/0009008}.

\bibitem[{\citenamefont{{Linder}}(2003)}]{Linder_2003}
\bibinfo{author}{\bibfnamefont{E.~V.} \bibnamefont{{Linder}}},
  \bibinfo{journal}{Phys. Rev. Lett.} \textbf{\bibinfo{volume}{90}},
  \bibinfo{eid}{091301} (\bibinfo{year}{2003}), \eprint{astro-ph/0208512}.

\bibitem[{\citenamefont{{Wang} et~al.}(2017{\natexlab{a}})\citenamefont{{Wang},
  {Wang}, and {Li}}}]{Wang_Wang_Li_2017}
\bibinfo{author}{\bibfnamefont{S.}~\bibnamefont{{Wang}}},
  \bibinfo{author}{\bibfnamefont{Y.}~\bibnamefont{{Wang}}}, \bibnamefont{and}
  \bibinfo{author}{\bibfnamefont{M.}~\bibnamefont{{Li}}},
  \bibinfo{journal}{Phys. Rep.} \textbf{\bibinfo{volume}{696}},
  \bibinfo{pages}{1} (\bibinfo{year}{2017}{\natexlab{a}}), \eprint{1612.00345}.

\bibitem[{\citenamefont{{Peebles} and {Ratra}}(1988)}]{6}
\bibinfo{author}{\bibfnamefont{P.~J.~E.} \bibnamefont{{Peebles}}}
  \bibnamefont{and} \bibinfo{author}{\bibfnamefont{B.}~\bibnamefont{{Ratra}}},
  \bibinfo{journal}{Astrophys. J. Lett.} \textbf{\bibinfo{volume}{325}},
  \bibinfo{pages}{L17} (\bibinfo{year}{1988}).

\bibitem[{\citenamefont{{Ratra} and {Peebles}}(1988)}]{ratpeeb88}
\bibinfo{author}{\bibfnamefont{B.}~\bibnamefont{{Ratra}}} \bibnamefont{and}
  \bibinfo{author}{\bibfnamefont{P.~J.~E.} \bibnamefont{{Peebles}}},
  \bibinfo{journal}{Phys. Rev. D} \textbf{\bibinfo{volume}{37}},
  \bibinfo{pages}{3406} (\bibinfo{year}{1988}).

\bibitem[{\citenamefont{{Pavlov} et~al.}(2013)\citenamefont{{Pavlov},
  {Westmoreland}, {Saaidi}, and {Ratra}}}]{pavlov13}
\bibinfo{author}{\bibfnamefont{A.}~\bibnamefont{{Pavlov}}},
  \bibinfo{author}{\bibfnamefont{S.}~\bibnamefont{{Westmoreland}}},
  \bibinfo{author}{\bibfnamefont{K.}~\bibnamefont{{Saaidi}}}, \bibnamefont{and}
  \bibinfo{author}{\bibfnamefont{B.}~\bibnamefont{{Ratra}}},
  \bibinfo{journal}{Phys. Rev. D} \textbf{\bibinfo{volume}{88}},
  \bibinfo{eid}{123513} (\bibinfo{year}{2013}), \eprint{1307.7399}.

\bibitem[{\citenamefont{{Armendariz-Picon}
  et~al.}(2000)\citenamefont{{Armendariz-Picon}, {Mukhanov}, and
  {Steinhardt}}}]{Armendariz-Picon_Mukhanov_Steinhardt_2000}
\bibinfo{author}{\bibfnamefont{C.}~\bibnamefont{{Armendariz-Picon}}},
  \bibinfo{author}{\bibfnamefont{V.}~\bibnamefont{{Mukhanov}}},
  \bibnamefont{and} \bibinfo{author}{\bibfnamefont{P.~J.}
  \bibnamefont{{Steinhardt}}}, \bibinfo{journal}{Phys. Rev. Lett.}
  \textbf{\bibinfo{volume}{85}}, \bibinfo{pages}{4438} (\bibinfo{year}{2000}),
  \eprint{astro-ph/0004134}.

\bibitem[{\citenamefont{{Armendariz-Picon}
  et~al.}(2001)\citenamefont{{Armendariz-Picon}, {Mukhanov}, and
  {Steinhardt}}}]{Aremendariz-Picon_Mukhanov_Steinhardt_2001}
\bibinfo{author}{\bibfnamefont{C.}~\bibnamefont{{Armendariz-Picon}}},
  \bibinfo{author}{\bibfnamefont{V.}~\bibnamefont{{Mukhanov}}},
  \bibnamefont{and} \bibinfo{author}{\bibfnamefont{P.~J.}
  \bibnamefont{{Steinhardt}}}, \bibinfo{journal}{Phys. Rev. D}
  \textbf{\bibinfo{volume}{63}}, \bibinfo{eid}{103510} (\bibinfo{year}{2001}),
  \eprint{astro-ph/0006373}.

\bibitem[{\citenamefont{{Kamenshchik} et~al.}(2001)\citenamefont{{Kamenshchik},
  {Moschella}, and {Pasquier}}}]{Kamenshchik_Moschella_Pasquier_2001}
\bibinfo{author}{\bibfnamefont{A.}~\bibnamefont{{Kamenshchik}}},
  \bibinfo{author}{\bibfnamefont{U.}~\bibnamefont{{Moschella}}},
  \bibnamefont{and}
  \bibinfo{author}{\bibfnamefont{V.}~\bibnamefont{{Pasquier}}},
  \bibinfo{journal}{Phys. Lett. B} \textbf{\bibinfo{volume}{511}},
  \bibinfo{pages}{265} (\bibinfo{year}{2001}), \eprint{gr-qc/0103004}.

\bibitem[{\citenamefont{{Bento} et~al.}(2002)\citenamefont{{Bento},
  {Bertolami}, and {Sen}}}]{Bento_Bertolami_Sen_2002}
\bibinfo{author}{\bibfnamefont{M.~C.} \bibnamefont{{Bento}}},
  \bibinfo{author}{\bibfnamefont{O.}~\bibnamefont{{Bertolami}}},
  \bibnamefont{and} \bibinfo{author}{\bibfnamefont{A.~A.} \bibnamefont{{Sen}}},
  \bibinfo{journal}{Phys. Rev. D} \textbf{\bibinfo{volume}{66}},
  \bibinfo{eid}{043507} (\bibinfo{year}{2002}), \eprint{gr-qc/0202064}.

\bibitem[{\citenamefont{{Copeland} et~al.}(2006)\citenamefont{{Copeland},
  {Sami}, and {Tsujikawa}}}]{Copeland_Sami_Tsujikawa_2006}
\bibinfo{author}{\bibfnamefont{E.~J.} \bibnamefont{{Copeland}}},
  \bibinfo{author}{\bibfnamefont{M.}~\bibnamefont{{Sami}}}, \bibnamefont{and}
  \bibinfo{author}{\bibfnamefont{S.}~\bibnamefont{{Tsujikawa}}},
  \bibinfo{journal}{Int. J. Mod. Phys. D} \textbf{\bibinfo{volume}{15}},
  \bibinfo{pages}{1753} (\bibinfo{year}{2006}), \eprint{hep-th/0603057}.

\bibitem[{\citenamefont{{Yoo} and {Watanabe}}(2012)}]{Yoo_Watanabe_2012}
\bibinfo{author}{\bibfnamefont{J.}~\bibnamefont{{Yoo}}} \bibnamefont{and}
  \bibinfo{author}{\bibfnamefont{Y.}~\bibnamefont{{Watanabe}}},
  \bibinfo{journal}{Int. J. Mod. Phys. D} \textbf{\bibinfo{volume}{21}},
  \bibinfo{eid}{1230002} (\bibinfo{year}{2012}), \eprint{1212.4726}.

\bibitem[{\citenamefont{{Ryskin}}(2015)}]{Ryskin}
\bibinfo{author}{\bibfnamefont{G.}~\bibnamefont{{Ryskin}}},
  \bibinfo{journal}{Astropart. Phys.} \textbf{\bibinfo{volume}{62}},
  \bibinfo{pages}{258} (\bibinfo{year}{2015}), \eprint{1810.07516}.

\bibitem[{\citenamefont{{Ryan}}(2020)}]{Ryan_2020}
\bibinfo{author}{\bibfnamefont{J.}~\bibnamefont{{Ryan}}},
  \bibinfo{journal}{Astropart. Phys.} \textbf{\bibinfo{volume}{118}},
  \bibinfo{eid}{102428} (\bibinfo{year}{2020}), \eprint{2001.00695}.

\bibitem[{\citenamefont{{Ryan}}(2021)}]{Ryan_power_law}
\bibinfo{author}{\bibfnamefont{J.}~\bibnamefont{{Ryan}}},
  \bibinfo{journal}{ArXiv e-prints}  (\bibinfo{year}{2021}),
  \eprint{2102.08457}.

\bibitem[{\citenamefont{{Weinberg}}(1989)}]{Weinberg_1}
\bibinfo{author}{\bibfnamefont{S.}~\bibnamefont{{Weinberg}}},
  \bibinfo{journal}{Rev. Mod. Phys.} \textbf{\bibinfo{volume}{61}},
  \bibinfo{pages}{1} (\bibinfo{year}{1989}).

\bibitem[{\citenamefont{{Weinberg}}(2000)}]{Weinberg_2}
\bibinfo{author}{\bibfnamefont{S.}~\bibnamefont{{Weinberg}}},
  \bibinfo{journal}{ArXiv e-prints}  (\bibinfo{year}{2000}),
  \bibinfo{note}{arXiv:astro-ph/0005265}.

\bibitem[{\citenamefont{{Straumann}}(2002)}]{Straumann}
\bibinfo{author}{\bibfnamefont{N.}~\bibnamefont{{Straumann}}},
  \bibinfo{journal}{arXiv e-prints}  (\bibinfo{year}{2002}),
  \bibinfo{note}{arXiv:gr-qc/0208027}.

\bibitem[{\citenamefont{{Martin}}(2012)}]{Martin}
\bibinfo{author}{\bibfnamefont{J.}~\bibnamefont{{Martin}}},
  \bibinfo{journal}{C. R. Phys.} \textbf{\bibinfo{volume}{13}},
  \bibinfo{pages}{566} (\bibinfo{year}{2012}), \eprint{1205.3365}.

\bibitem[{\citenamefont{{Clifton} et~al.}(2012)\citenamefont{{Clifton},
  {Ferreira}, {Padilla}, and {Skordis}}}]{modgrav}
\bibinfo{author}{\bibfnamefont{T.}~\bibnamefont{{Clifton}}},
  \bibinfo{author}{\bibfnamefont{P.~G.} \bibnamefont{{Ferreira}}},
  \bibinfo{author}{\bibfnamefont{A.}~\bibnamefont{{Padilla}}},
  \bibnamefont{and}
  \bibinfo{author}{\bibfnamefont{C.}~\bibnamefont{{Skordis}}},
  \bibinfo{journal}{Phys. Rep.} \textbf{\bibinfo{volume}{513}},
  \bibinfo{pages}{1} (\bibinfo{year}{2012}), \eprint{1106.2476}.

\bibitem[{\citenamefont{{Dev} et~al.}(2008)\citenamefont{{Dev}, {Jain}, and
  {Lohiya}}}]{Dev_Jain_Lohiya_2008}
\bibinfo{author}{\bibfnamefont{A.}~\bibnamefont{{Dev}}},
  \bibinfo{author}{\bibfnamefont{D.}~\bibnamefont{{Jain}}}, \bibnamefont{and}
  \bibinfo{author}{\bibfnamefont{D.}~\bibnamefont{{Lohiya}}},
  \bibinfo{journal}{arXiv e-prints}  (\bibinfo{year}{2008}),
  \eprint{0804.3491}.

\bibitem[{\citenamefont{{Sethi} et~al.}(2005)\citenamefont{{Sethi}, {Dev}, and
  {Jain}}}]{Sethi_Dev_Jain_2005}
\bibinfo{author}{\bibfnamefont{G.}~\bibnamefont{{Sethi}}},
  \bibinfo{author}{\bibfnamefont{A.}~\bibnamefont{{Dev}}}, \bibnamefont{and}
  \bibinfo{author}{\bibfnamefont{D.}~\bibnamefont{{Jain}}},
  \bibinfo{journal}{Phys. Lett. B} \textbf{\bibinfo{volume}{624}},
  \bibinfo{pages}{135} (\bibinfo{year}{2005}), \eprint{astro-ph/0506255}.

\bibitem[{\citenamefont{{Ryan} et~al.}(2018)\citenamefont{{Ryan}, {Doshi}, and
  {Ratra}}}]{Ryan_Doshi_Ratra_2018}
\bibinfo{author}{\bibfnamefont{J.}~\bibnamefont{{Ryan}}},
  \bibinfo{author}{\bibfnamefont{S.}~\bibnamefont{{Doshi}}}, \bibnamefont{and}
  \bibinfo{author}{\bibfnamefont{B.}~\bibnamefont{{Ratra}}},
  \bibinfo{journal}{Mon. Not. R. Astron. Soc.} \textbf{\bibinfo{volume}{480}},
  \bibinfo{pages}{759} (\bibinfo{year}{2018}), \eprint{1805.06408}.

\bibitem[{\citenamefont{{Samushia} et~al.}(2007)\citenamefont{{Samushia},
  {Chen}, and {Ratra}}}]{24}
\bibinfo{author}{\bibfnamefont{L.}~\bibnamefont{{Samushia}}},
  \bibinfo{author}{\bibfnamefont{G.}~\bibnamefont{{Chen}}}, \bibnamefont{and}
  \bibinfo{author}{\bibfnamefont{B.}~\bibnamefont{{Ratra}}},
  \bibinfo{journal}{ArXiv e-prints}  (\bibinfo{year}{2007}),
  \eprint{0706.1963}.

\bibitem[{\citenamefont{{Yashar} et~al.}(2009)\citenamefont{{Yashar}, {Bozek},
  {Abrahamse}, {Albrecht}, and {Barnard}}}]{yashar_et_al_2009}
\bibinfo{author}{\bibfnamefont{M.}~\bibnamefont{{Yashar}}},
  \bibinfo{author}{\bibfnamefont{B.}~\bibnamefont{{Bozek}}},
  \bibinfo{author}{\bibfnamefont{A.}~\bibnamefont{{Abrahamse}}},
  \bibinfo{author}{\bibfnamefont{A.}~\bibnamefont{{Albrecht}}},
  \bibnamefont{and}
  \bibinfo{author}{\bibfnamefont{M.}~\bibnamefont{{Barnard}}},
  \bibinfo{journal}{Phys. Rev. D} \textbf{\bibinfo{volume}{79}},
  \bibinfo{eid}{103004} (\bibinfo{year}{2009}), \eprint{0811.2253}.

\bibitem[{\citenamefont{{Samushia} and {Ratra}}(2010)}]{samushia_ratra_2010}
\bibinfo{author}{\bibfnamefont{L.}~\bibnamefont{{Samushia}}} \bibnamefont{and}
  \bibinfo{author}{\bibfnamefont{B.}~\bibnamefont{{Ratra}}},
  \bibinfo{journal}{Astrophys. J.} \textbf{\bibinfo{volume}{714}},
  \bibinfo{pages}{1347} (\bibinfo{year}{2010}), \eprint{0905.3836}.

\bibitem[{\citenamefont{{Chen} and
  {Ratra}}(2011{\natexlab{a}})}]{chen_ratra_2011b}
\bibinfo{author}{\bibfnamefont{Y.}~\bibnamefont{{Chen}}} \bibnamefont{and}
  \bibinfo{author}{\bibfnamefont{B.}~\bibnamefont{{Ratra}}},
  \bibinfo{journal}{Phys. Lett. B} \textbf{\bibinfo{volume}{703}},
  \bibinfo{pages}{406} (\bibinfo{year}{2011}{\natexlab{a}}),
  \eprint{1106.4294}.

\bibitem[{\citenamefont{{Campanelli} et~al.}(2012)\citenamefont{{Campanelli},
  {Fogli}, {Kahniashvili}, {Marrone}, and {Ratra}}}]{20}
\bibinfo{author}{\bibfnamefont{L.}~\bibnamefont{{Campanelli}}},
  \bibinfo{author}{\bibfnamefont{G.~L.} \bibnamefont{{Fogli}}},
  \bibinfo{author}{\bibfnamefont{T.}~\bibnamefont{{Kahniashvili}}},
  \bibinfo{author}{\bibfnamefont{A.}~\bibnamefont{{Marrone}}},
  \bibnamefont{and} \bibinfo{author}{\bibfnamefont{B.}~\bibnamefont{{Ratra}}},
  \bibinfo{journal}{European Physical Journal C} \textbf{\bibinfo{volume}{72}},
  \bibinfo{eid}{2218} (\bibinfo{year}{2012}), \eprint{1110.2310}.

\bibitem[{\citenamefont{{Pavlov} et~al.}(2014)\citenamefont{{Pavlov}, {Farooq},
  and {Ratra}}}]{23}
\bibinfo{author}{\bibfnamefont{A.}~\bibnamefont{{Pavlov}}},
  \bibinfo{author}{\bibfnamefont{O.}~\bibnamefont{{Farooq}}}, \bibnamefont{and}
  \bibinfo{author}{\bibfnamefont{B.}~\bibnamefont{{Ratra}}},
  \bibinfo{journal}{Phys. Rev. D} \textbf{\bibinfo{volume}{90}},
  \bibinfo{eid}{023006} (\bibinfo{year}{2014}), \eprint{1312.5285}.

\bibitem[{\citenamefont{{Avsajanishvili}
  et~al.}(2015)\citenamefont{{Avsajanishvili}, {Samushia}, {Arkhipova}, and
  {Kahniashvili}}}]{Avsajanishvili_2015}
\bibinfo{author}{\bibfnamefont{O.}~\bibnamefont{{Avsajanishvili}}},
  \bibinfo{author}{\bibfnamefont{L.}~\bibnamefont{{Samushia}}},
  \bibinfo{author}{\bibfnamefont{N.~A.} \bibnamefont{{Arkhipova}}},
  \bibnamefont{and}
  \bibinfo{author}{\bibfnamefont{T.}~\bibnamefont{{Kahniashvili}}},
  \bibinfo{journal}{ArXiv e-prints}  (\bibinfo{year}{2015}),
  \eprint{1511.09317}.

\bibitem[{\citenamefont{{Sola Peracaula} et~al.}(2016)\citenamefont{{Sola
  Peracaula}, {de Cruz Perez}, and {Gomez-Valent}}}]{26}
\bibinfo{author}{\bibfnamefont{J.}~\bibnamefont{{Sola Peracaula}}},
  \bibinfo{author}{\bibfnamefont{J.}~\bibnamefont{{de Cruz Perez}}},
  \bibnamefont{and}
  \bibinfo{author}{\bibfnamefont{A.}~\bibnamefont{{Gomez-Valent}}},
  \bibinfo{journal}{ArXiv e-prints}  (\bibinfo{year}{2016}),
  \eprint{1606.00450}.

\bibitem[{\citenamefont{{Sol{\`a} Peracaula}
  et~al.}(2018)\citenamefont{{Sol{\`a} Peracaula}, {de Cruz P{\'e}rez}, and
  {G{\'o}mez-Valent}}}]{Sola_perez_gomez_2018}
\bibinfo{author}{\bibfnamefont{J.}~\bibnamefont{{Sol{\`a} Peracaula}}},
  \bibinfo{author}{\bibfnamefont{J.}~\bibnamefont{{de Cruz P{\'e}rez}}},
  \bibnamefont{and}
  \bibinfo{author}{\bibfnamefont{A.}~\bibnamefont{{G{\'o}mez-Valent}}},
  \bibinfo{journal}{Mon. Not. R. Astron. Soc.} \textbf{\bibinfo{volume}{478}},
  \bibinfo{pages}{4357} (\bibinfo{year}{2018}), \eprint{1703.08218}.

\bibitem[{\citenamefont{{Sol{\`a}}
  et~al.}(2017{\natexlab{a}})\citenamefont{{Sol{\`a}}, {G{\'o}mez-Valent}, and
  {de Cruz P{\'e}rez}}}]{Sola_etal_2017}
\bibinfo{author}{\bibfnamefont{J.}~\bibnamefont{{Sol{\`a}}}},
  \bibinfo{author}{\bibfnamefont{A.}~\bibnamefont{{G{\'o}mez-Valent}}},
  \bibnamefont{and} \bibinfo{author}{\bibfnamefont{J.}~\bibnamefont{{de Cruz
  P{\'e}rez}}}, \bibinfo{journal}{Mod. Phys. Lett. A}
  \textbf{\bibinfo{volume}{32}}, \bibinfo{eid}{1750054-144}
  (\bibinfo{year}{2017}{\natexlab{a}}), \eprint{1610.08965}.

\bibitem[{\citenamefont{{Sol{\`a}}
  et~al.}(2017{\natexlab{b}})\citenamefont{{Sol{\`a}}, {G{\'o}mez-Valent}, and
  {de Cruz P{\'e}rez}}}]{30}
\bibinfo{author}{\bibfnamefont{J.}~\bibnamefont{{Sol{\`a}}}},
  \bibinfo{author}{\bibfnamefont{A.}~\bibnamefont{{G{\'o}mez-Valent}}},
  \bibnamefont{and} \bibinfo{author}{\bibfnamefont{J.}~\bibnamefont{{de Cruz
  P{\'e}rez}}}, \bibinfo{journal}{Phys. Lett. B}
  \textbf{\bibinfo{volume}{774}}, \bibinfo{pages}{317}
  (\bibinfo{year}{2017}{\natexlab{b}}), \eprint{1705.06723}.

\bibitem[{\citenamefont{{Sol{\`a}}
  et~al.}(2017{\natexlab{c}})\citenamefont{{Sol{\`a}}, {G{\'o}mez-Valent}, and
  {de Cruz P{\'e}rez}}}]{28}
\bibinfo{author}{\bibfnamefont{J.}~\bibnamefont{{Sol{\`a}}}},
  \bibinfo{author}{\bibfnamefont{A.}~\bibnamefont{{G{\'o}mez-Valent}}},
  \bibnamefont{and} \bibinfo{author}{\bibfnamefont{J.}~\bibnamefont{{de Cruz
  P{\'e}rez}}}, \bibinfo{journal}{Astrophys. J.}
  \textbf{\bibinfo{volume}{836}}, \bibinfo{eid}{43}
  (\bibinfo{year}{2017}{\natexlab{c}}), \eprint{1602.02103}.

\bibitem[{\citenamefont{{Avsajanishvili}
  et~al.}(2017)\citenamefont{{Avsajanishvili}, {Huang}, {Samushia}, and
  {Kahniashvili}}}]{18}
\bibinfo{author}{\bibfnamefont{O.}~\bibnamefont{{Avsajanishvili}}},
  \bibinfo{author}{\bibfnamefont{Y.}~\bibnamefont{{Huang}}},
  \bibinfo{author}{\bibfnamefont{L.}~\bibnamefont{{Samushia}}},
  \bibnamefont{and}
  \bibinfo{author}{\bibfnamefont{T.}~\bibnamefont{{Kahniashvili}}},
  \bibinfo{journal}{ArXiv e-prints}  (\bibinfo{year}{2017}),
  \eprint{1711.11465}.

\bibitem[{\citenamefont{{G{\'o}mez-Valent} and {Sol{\`a}}}(2017)}]{22}
\bibinfo{author}{\bibfnamefont{A.}~\bibnamefont{{G{\'o}mez-Valent}}}
  \bibnamefont{and}
  \bibinfo{author}{\bibfnamefont{J.}~\bibnamefont{{Sol{\`a}}}},
  \bibinfo{journal}{EPL} \textbf{\bibinfo{volume}{120}}, \bibinfo{pages}{39001}
  (\bibinfo{year}{2017}), \eprint{1711.00692}.

\bibitem[{\citenamefont{{Zhai} et~al.}(2017)\citenamefont{{Zhai}, {Blanton},
  {Slosar}, and {Tinker}}}]{32}
\bibinfo{author}{\bibfnamefont{Z.}~\bibnamefont{{Zhai}}},
  \bibinfo{author}{\bibfnamefont{M.}~\bibnamefont{{Blanton}}},
  \bibinfo{author}{\bibfnamefont{A.}~\bibnamefont{{Slosar}}}, \bibnamefont{and}
  \bibinfo{author}{\bibfnamefont{J.}~\bibnamefont{{Tinker}}},
  \bibinfo{journal}{Astrophys. J.} \textbf{\bibinfo{volume}{850}},
  \bibinfo{eid}{183} (\bibinfo{year}{2017}), \eprint{1705.10031}.

\bibitem[{\citenamefont{{Mehrabi} and {Basilakos}}(2018)}]{91}
\bibinfo{author}{\bibfnamefont{A.}~\bibnamefont{{Mehrabi}}} \bibnamefont{and}
  \bibinfo{author}{\bibfnamefont{S.}~\bibnamefont{{Basilakos}}},
  \bibinfo{journal}{ArXiv e-prints}  (\bibinfo{year}{2018}),
  \eprint{1804.10794}.

\bibitem[{\citenamefont{{Sangwan} et~al.}(2018)\citenamefont{{Sangwan},
  {Tripathi}, and {Jassal}}}]{sangwan_tripathi_jassal_2018}
\bibinfo{author}{\bibfnamefont{A.}~\bibnamefont{{Sangwan}}},
  \bibinfo{author}{\bibfnamefont{A.}~\bibnamefont{{Tripathi}}},
  \bibnamefont{and} \bibinfo{author}{\bibfnamefont{H.~K.}
  \bibnamefont{{Jassal}}}, \bibinfo{journal}{ArXiv e-prints}
  (\bibinfo{year}{2018}), \eprint{1804.09350}.

\bibitem[{\citenamefont{{Ooba} et~al.}(2019)\citenamefont{{Ooba}, {Ratra}, and
  {Sugiyama}}}]{Ooba_Ratra_Sugiyama_2018_FpCDM}
\bibinfo{author}{\bibfnamefont{J.}~\bibnamefont{{Ooba}}},
  \bibinfo{author}{\bibfnamefont{B.}~\bibnamefont{{Ratra}}}, \bibnamefont{and}
  \bibinfo{author}{\bibfnamefont{N.}~\bibnamefont{{Sugiyama}}},
  \bibinfo{journal}{Astrophys. Space Sci.} \textbf{\bibinfo{volume}{364}},
  \bibinfo{eid}{176} (\bibinfo{year}{2019}), \eprint{1802.05571}.

\bibitem[{\citenamefont{{Planck Collaboration}}(2016)}]{planck_2016}
\bibinfo{author}{\bibnamefont{{Planck Collaboration}}},
  \bibinfo{journal}{Astron. Astrophys.} \textbf{\bibinfo{volume}{594}},
  \bibinfo{eid}{A13} (\bibinfo{year}{2016}), \eprint{1502.01589}.

\bibitem[{\citenamefont{{Park} and
  {Ratra}}(2019{\natexlab{a}})}]{Park_Ratra_2018_FXCDM_NFXCDM}
\bibinfo{author}{\bibfnamefont{C.-G.} \bibnamefont{{Park}}} \bibnamefont{and}
  \bibinfo{author}{\bibfnamefont{B.}~\bibnamefont{{Ratra}}},
  \bibinfo{journal}{Astrophys. Space Sci.} \textbf{\bibinfo{volume}{364}},
  \bibinfo{eid}{82} (\bibinfo{year}{2019}{\natexlab{a}}), \eprint{1803.05522}.

\bibitem[{\citenamefont{{Sahni} et~al.}(2014)\citenamefont{{Sahni},
  {Shafieloo}, and {Starobinsky}}}]{39}
\bibinfo{author}{\bibfnamefont{V.}~\bibnamefont{{Sahni}}},
  \bibinfo{author}{\bibfnamefont{A.}~\bibnamefont{{Shafieloo}}},
  \bibnamefont{and} \bibinfo{author}{\bibfnamefont{A.~A.}
  \bibnamefont{{Starobinsky}}}, \bibinfo{journal}{Astrophys. J. Lett.}
  \textbf{\bibinfo{volume}{793}}, \bibinfo{eid}{L40} (\bibinfo{year}{2014}),
  \eprint{1406.2209}.

\bibitem[{\citenamefont{{Ding} et~al.}(2015)\citenamefont{{Ding}, {Biesiada},
  {Cao}, {Li}, and {Zhu}}}]{34}
\bibinfo{author}{\bibfnamefont{X.}~\bibnamefont{{Ding}}},
  \bibinfo{author}{\bibfnamefont{M.}~\bibnamefont{{Biesiada}}},
  \bibinfo{author}{\bibfnamefont{S.}~\bibnamefont{{Cao}}},
  \bibinfo{author}{\bibfnamefont{Z.}~\bibnamefont{{Li}}}, \bibnamefont{and}
  \bibinfo{author}{\bibfnamefont{Z.-H.} \bibnamefont{{Zhu}}},
  \bibinfo{journal}{Astrophys. J. Lett.} \textbf{\bibinfo{volume}{803}},
  \bibinfo{eid}{L22} (\bibinfo{year}{2015}), \eprint{1503.04923}.

\bibitem[{\citenamefont{{Sol{\`a}} et~al.}(2015)\citenamefont{{Sol{\`a}},
  {G{\'o}mez-Valent}, and {de Cruz P{\'e}rez}}}]{40}
\bibinfo{author}{\bibfnamefont{J.}~\bibnamefont{{Sol{\`a}}}},
  \bibinfo{author}{\bibfnamefont{A.}~\bibnamefont{{G{\'o}mez-Valent}}},
  \bibnamefont{and} \bibinfo{author}{\bibfnamefont{J.}~\bibnamefont{{de Cruz
  P{\'e}rez}}}, \bibinfo{journal}{Astrophys. J. Lett.}
  \textbf{\bibinfo{volume}{811}}, \bibinfo{eid}{L14} (\bibinfo{year}{2015}),
  \eprint{1506.05793}.

\bibitem[{\citenamefont{{Zheng} et~al.}(2016)\citenamefont{{Zheng}, {Ding},
  {Biesiada}, {Cao}, and {Zhu}}}]{43}
\bibinfo{author}{\bibfnamefont{X.}~\bibnamefont{{Zheng}}},
  \bibinfo{author}{\bibfnamefont{X.}~\bibnamefont{{Ding}}},
  \bibinfo{author}{\bibfnamefont{M.}~\bibnamefont{{Biesiada}}},
  \bibinfo{author}{\bibfnamefont{S.}~\bibnamefont{{Cao}}}, \bibnamefont{and}
  \bibinfo{author}{\bibfnamefont{Z.-H.} \bibnamefont{{Zhu}}},
  \bibinfo{journal}{Astrophys. J.} \textbf{\bibinfo{volume}{825}},
  \bibinfo{eid}{17} (\bibinfo{year}{2016}), \eprint{1604.07910}.

\bibitem[{\citenamefont{{Zhao} et~al.}(2017)\citenamefont{{Zhao}, {Raveri},
  {Pogosian}, {Wang}, {Crittenden}, {Handley}, {Percival}, {Beutler},
  {Brinkmann}, {Chuang} et~al.}}]{42}
\bibinfo{author}{\bibfnamefont{G.-B.} \bibnamefont{{Zhao}}},
  \bibinfo{author}{\bibfnamefont{M.}~\bibnamefont{{Raveri}}},
  \bibinfo{author}{\bibfnamefont{L.}~\bibnamefont{{Pogosian}}},
  \bibinfo{author}{\bibfnamefont{Y.}~\bibnamefont{{Wang}}},
  \bibinfo{author}{\bibfnamefont{R.~G.} \bibnamefont{{Crittenden}}},
  \bibinfo{author}{\bibfnamefont{W.~J.} \bibnamefont{{Handley}}},
  \bibinfo{author}{\bibfnamefont{W.~J.} \bibnamefont{{Percival}}},
  \bibinfo{author}{\bibfnamefont{F.}~\bibnamefont{{Beutler}}},
  \bibinfo{author}{\bibfnamefont{J.}~\bibnamefont{{Brinkmann}}},
  \bibinfo{author}{\bibfnamefont{C.-H.} \bibnamefont{{Chuang}}},
  \bibnamefont{et~al.}, \bibinfo{journal}{Nat. Astron.}
  \textbf{\bibinfo{volume}{1}}, \bibinfo{pages}{627} (\bibinfo{year}{2017}),
  \eprint{1701.08165}.

\bibitem[{\citenamefont{{Zhang}
  et~al.}(2017{\natexlab{a}})\citenamefont{{Zhang}, {Zhang}, {Wang}, {Qi},
  {Wang}, and {Zhao}}}]{41}
\bibinfo{author}{\bibfnamefont{Y.-C.} \bibnamefont{{Zhang}}},
  \bibinfo{author}{\bibfnamefont{H.-Y.} \bibnamefont{{Zhang}}},
  \bibinfo{author}{\bibfnamefont{D.-D.} \bibnamefont{{Wang}}},
  \bibinfo{author}{\bibfnamefont{Y.-H.} \bibnamefont{{Qi}}},
  \bibinfo{author}{\bibfnamefont{Y.-T.} \bibnamefont{{Wang}}},
  \bibnamefont{and} \bibinfo{author}{\bibfnamefont{G.-B.}
  \bibnamefont{{Zhao}}}, \bibinfo{journal}{Res. Astron. Astrophys.}
  \textbf{\bibinfo{volume}{17}}, \bibinfo{eid}{050}
  (\bibinfo{year}{2017}{\natexlab{a}}), \eprint{1703.08293}.

\bibitem[{\citenamefont{{Cao} et~al.}(2017{\natexlab{a}})\citenamefont{{Cao},
  {Duan}, {Meng}, and {Zhang}}}]{33}
\bibinfo{author}{\bibfnamefont{S.-L.} \bibnamefont{{Cao}}},
  \bibinfo{author}{\bibfnamefont{X.-W.} \bibnamefont{{Duan}}},
  \bibinfo{author}{\bibfnamefont{X.-L.} \bibnamefont{{Meng}}},
  \bibnamefont{and} \bibinfo{author}{\bibfnamefont{T.-J.}
  \bibnamefont{{Zhang}}}, \bibinfo{journal}{ArXiv e-prints}
  (\bibinfo{year}{2017}{\natexlab{a}}), \eprint{1712.01703}.

\bibitem[{\citenamefont{{G{\'o}mez-Valent} and {Sol{\`a}}}(2018)}]{35}
\bibinfo{author}{\bibfnamefont{A.}~\bibnamefont{{G{\'o}mez-Valent}}}
  \bibnamefont{and}
  \bibinfo{author}{\bibfnamefont{J.}~\bibnamefont{{Sol{\`a}}}},
  \bibinfo{journal}{ArXiv e-prints}  (\bibinfo{year}{2018}),
  \eprint{1801.08501}.

\bibitem[{\citenamefont{{Gott}}(1982)}]{gott_1982}
\bibinfo{author}{\bibfnamefont{J.~R.} \bibnamefont{{Gott}},
  \bibfnamefont{III}}, \bibinfo{journal}{Nature}
  \textbf{\bibinfo{volume}{295}}, \bibinfo{pages}{304} (\bibinfo{year}{1982}).

\bibitem[{\citenamefont{{Hawking}}(1984)}]{hawking_1984}
\bibinfo{author}{\bibfnamefont{S.~W.} \bibnamefont{{Hawking}}},
  \bibinfo{journal}{Nucl. Phys. B} \textbf{\bibinfo{volume}{239}},
  \bibinfo{pages}{257} (\bibinfo{year}{1984}).

\bibitem[{\citenamefont{{Ratra}}(1985)}]{ratra_1985}
\bibinfo{author}{\bibfnamefont{B.}~\bibnamefont{{Ratra}}},
  \bibinfo{journal}{Phys. Rev. D} \textbf{\bibinfo{volume}{31}},
  \bibinfo{pages}{1931} (\bibinfo{year}{1985}).

\bibitem[{\citenamefont{{Ratra} and {Peebles}}(1994)}]{57}
\bibinfo{author}{\bibfnamefont{B.}~\bibnamefont{{Ratra}}} \bibnamefont{and}
  \bibinfo{author}{\bibfnamefont{P.~J.~E.} \bibnamefont{{Peebles}}},
  \bibinfo{journal}{Astrophys. J. Lett.} \textbf{\bibinfo{volume}{432}},
  \bibinfo{pages}{L5} (\bibinfo{year}{1994}).

\bibitem[{\citenamefont{{Ratra} and {Peebles}}(1995)}]{ratra_peebles_1995}
\bibinfo{author}{\bibfnamefont{B.}~\bibnamefont{{Ratra}}} \bibnamefont{and}
  \bibinfo{author}{\bibfnamefont{P.~J.~E.} \bibnamefont{{Peebles}}},
  \bibinfo{journal}{Phys. Rev. D} \textbf{\bibinfo{volume}{52}},
  \bibinfo{pages}{1837} (\bibinfo{year}{1995}).

\bibitem[{\citenamefont{{Ratra}}(2017)}]{ratra_2017}
\bibinfo{author}{\bibfnamefont{B.}~\bibnamefont{{Ratra}}},
  \bibinfo{journal}{Phys. Rev. D} \textbf{\bibinfo{volume}{96}},
  \bibinfo{eid}{103534} (\bibinfo{year}{2017}), \eprint{1707.03439}.

\bibitem[{\citenamefont{{Ooba} et~al.}(2018{\natexlab{a}})\citenamefont{{Ooba},
  {Ratra}, and {Sugiyama}}}]{Ooba_Ratra_Sugiyama_2017_NFLCDM}
\bibinfo{author}{\bibfnamefont{J.}~\bibnamefont{{Ooba}}},
  \bibinfo{author}{\bibfnamefont{B.}~\bibnamefont{{Ratra}}}, \bibnamefont{and}
  \bibinfo{author}{\bibfnamefont{N.}~\bibnamefont{{Sugiyama}}},
  \bibinfo{journal}{Astrophys. J.} \textbf{\bibinfo{volume}{864}},
  \bibinfo{eid}{80} (\bibinfo{year}{2018}{\natexlab{a}}), \eprint{1707.03452}.

\bibitem[{\citenamefont{{Farooq} et~al.}(2015)\citenamefont{{Farooq}, {Mania},
  and {Ratra}}}]{Farooq_Mania_Ratra_2015}
\bibinfo{author}{\bibfnamefont{O.}~\bibnamefont{{Farooq}}},
  \bibinfo{author}{\bibfnamefont{D.}~\bibnamefont{{Mania}}}, \bibnamefont{and}
  \bibinfo{author}{\bibfnamefont{B.}~\bibnamefont{{Ratra}}},
  \bibinfo{journal}{Astrophys. Space Sci.} \textbf{\bibinfo{volume}{357}},
  \bibinfo{eid}{11} (\bibinfo{year}{2015}), \eprint{1308.0834}.

\bibitem[{\citenamefont{{Chen} et~al.}(2016)\citenamefont{{Chen}, {Ratra},
  {Biesiada}, {Li}, and {Zhu}}}]{Chen_et_al_2016}
\bibinfo{author}{\bibfnamefont{Y.}~\bibnamefont{{Chen}}},
  \bibinfo{author}{\bibfnamefont{B.}~\bibnamefont{{Ratra}}},
  \bibinfo{author}{\bibfnamefont{M.}~\bibnamefont{{Biesiada}}},
  \bibinfo{author}{\bibfnamefont{S.}~\bibnamefont{{Li}}}, \bibnamefont{and}
  \bibinfo{author}{\bibfnamefont{Z.-H.} \bibnamefont{{Zhu}}},
  \bibinfo{journal}{Astrophys. J.} \textbf{\bibinfo{volume}{829}},
  \bibinfo{eid}{61} (\bibinfo{year}{2016}), \eprint{1603.07115}.

\bibitem[{\citenamefont{{Yu} and {Wang}}(2016)}]{yu_wang_2016}
\bibinfo{author}{\bibfnamefont{H.}~\bibnamefont{{Yu}}} \bibnamefont{and}
  \bibinfo{author}{\bibfnamefont{F.~Y.} \bibnamefont{{Wang}}},
  \bibinfo{journal}{Astrophys. J.} \textbf{\bibinfo{volume}{828}},
  \bibinfo{eid}{85} (\bibinfo{year}{2016}), \eprint{1605.02483}.

\bibitem[{\citenamefont{{L'Huillier} and {Shafieloo}}(2017)}]{48}
\bibinfo{author}{\bibfnamefont{B.}~\bibnamefont{{L'Huillier}}}
  \bibnamefont{and}
  \bibinfo{author}{\bibfnamefont{A.}~\bibnamefont{{Shafieloo}}},
  \bibinfo{journal}{J. Cosmol. Astropart. Phys.} \textbf{\bibinfo{volume}{1}},
  \bibinfo{eid}{015} (\bibinfo{year}{2017}), \eprint{1606.06832}.

\bibitem[{\citenamefont{{Farooq} et~al.}(2017)\citenamefont{{Farooq}, {Ranjeet
  Madiyar}, {Crandall}, and {Ratra}}}]{Farooq_Ranjeet_Crandall_Ratra_2017}
\bibinfo{author}{\bibfnamefont{O.}~\bibnamefont{{Farooq}}},
  \bibinfo{author}{\bibfnamefont{F.}~\bibnamefont{{Ranjeet Madiyar}}},
  \bibinfo{author}{\bibfnamefont{S.}~\bibnamefont{{Crandall}}},
  \bibnamefont{and} \bibinfo{author}{\bibfnamefont{B.}~\bibnamefont{{Ratra}}},
  \bibinfo{journal}{Astrophys. J.} \textbf{\bibinfo{volume}{835}},
  \bibinfo{eid}{26} (\bibinfo{year}{2017}), \eprint{1607.03537}.

\bibitem[{\citenamefont{{Wei} and {Wu}}(2017)}]{wei_wu_2017}
\bibinfo{author}{\bibfnamefont{J.-J.} \bibnamefont{{Wei}}} \bibnamefont{and}
  \bibinfo{author}{\bibfnamefont{X.-F.} \bibnamefont{{Wu}}},
  \bibinfo{journal}{Astrophys. J.} \textbf{\bibinfo{volume}{838}},
  \bibinfo{eid}{160} (\bibinfo{year}{2017}), \eprint{1611.00904}.

\bibitem[{\citenamefont{{Rana} et~al.}(2017)\citenamefont{{Rana}, {Jain},
  {Mahajan}, and {Mukherjee}}}]{rana_jain_mahajan_mukherjee_2017}
\bibinfo{author}{\bibfnamefont{A.}~\bibnamefont{{Rana}}},
  \bibinfo{author}{\bibfnamefont{D.}~\bibnamefont{{Jain}}},
  \bibinfo{author}{\bibfnamefont{S.}~\bibnamefont{{Mahajan}}},
  \bibnamefont{and}
  \bibinfo{author}{\bibfnamefont{A.}~\bibnamefont{{Mukherjee}}},
  \bibinfo{journal}{J. Cosmol. Astropart. Phys.} \textbf{\bibinfo{volume}{3}},
  \bibinfo{eid}{028} (\bibinfo{year}{2017}), \eprint{1611.07196}.

\bibitem[{\citenamefont{{Yu} et~al.}(2018)\citenamefont{{Yu}, {Ratra}, and
  {Wang}}}]{60}
\bibinfo{author}{\bibfnamefont{H.}~\bibnamefont{{Yu}}},
  \bibinfo{author}{\bibfnamefont{B.}~\bibnamefont{{Ratra}}}, \bibnamefont{and}
  \bibinfo{author}{\bibfnamefont{F.-Y.} \bibnamefont{{Wang}}},
  \bibinfo{journal}{Astrophys. J.} \textbf{\bibinfo{volume}{856}},
  \bibinfo{eid}{3} (\bibinfo{year}{2018}), \eprint{1711.03437}.

\bibitem[{\citenamefont{{Mitra} et~al.}(2018)\citenamefont{{Mitra},
  {Choudhury}, and {Ratra}}}]{mitra_choudhury_ratra_2018}
\bibinfo{author}{\bibfnamefont{S.}~\bibnamefont{{Mitra}}},
  \bibinfo{author}{\bibfnamefont{T.~R.} \bibnamefont{{Choudhury}}},
  \bibnamefont{and} \bibinfo{author}{\bibfnamefont{B.}~\bibnamefont{{Ratra}}},
  \bibinfo{journal}{Mon. Not. R. Astron. Soc.} \textbf{\bibinfo{volume}{479}},
  \bibinfo{pages}{4566} (\bibinfo{year}{2018}), \eprint{1712.00018}.

\bibitem[{\citenamefont{{Park} and
  {Ratra}}(2019{\natexlab{b}})}]{Park_Ratra_2018_FLCDM}
\bibinfo{author}{\bibfnamefont{C.-G.} \bibnamefont{{Park}}} \bibnamefont{and}
  \bibinfo{author}{\bibfnamefont{B.}~\bibnamefont{{Ratra}}},
  \bibinfo{journal}{Astrophys. J.} \textbf{\bibinfo{volume}{882}},
  \bibinfo{eid}{158} (\bibinfo{year}{2019}{\natexlab{b}}), \eprint{1801.00213}.

\bibitem[{\citenamefont{{Ooba} et~al.}(2018{\natexlab{b}})\citenamefont{{Ooba},
  {Ratra}, and {Sugiyama}}}]{Ooba_Ratra_Sugiyama_2017_NFXCDM}
\bibinfo{author}{\bibfnamefont{J.}~\bibnamefont{{Ooba}}},
  \bibinfo{author}{\bibfnamefont{B.}~\bibnamefont{{Ratra}}}, \bibnamefont{and}
  \bibinfo{author}{\bibfnamefont{N.}~\bibnamefont{{Sugiyama}}},
  \bibinfo{journal}{Astrophys. J.} \textbf{\bibinfo{volume}{869}},
  \bibinfo{eid}{34} (\bibinfo{year}{2018}{\natexlab{b}}), \eprint{1710.03271}.

\bibitem[{\citenamefont{{Ooba} et~al.}(2018{\natexlab{c}})\citenamefont{{Ooba},
  {Ratra}, and {Sugiyama}}}]{Ooba_Ratra_Sugiyama_2017_NFpCDM}
\bibinfo{author}{\bibfnamefont{J.}~\bibnamefont{{Ooba}}},
  \bibinfo{author}{\bibfnamefont{B.}~\bibnamefont{{Ratra}}}, \bibnamefont{and}
  \bibinfo{author}{\bibfnamefont{N.}~\bibnamefont{{Sugiyama}}},
  \bibinfo{journal}{Astrophys. J.} \textbf{\bibinfo{volume}{866}},
  \bibinfo{eid}{68} (\bibinfo{year}{2018}{\natexlab{c}}), \eprint{1712.08617}.

\bibitem[{\citenamefont{{Farooq} and {Ratra}}(2013)}]{Farooq_Ratra_2013}
\bibinfo{author}{\bibfnamefont{O.}~\bibnamefont{{Farooq}}} \bibnamefont{and}
  \bibinfo{author}{\bibfnamefont{B.}~\bibnamefont{{Ratra}}},
  \bibinfo{journal}{Astrophys. J. Lett.} \textbf{\bibinfo{volume}{766}},
  \bibinfo{eid}{L7} (\bibinfo{year}{2013}), \eprint{1301.5243}.

\bibitem[{\citenamefont{{Farooq} et~al.}(2013)\citenamefont{{Farooq},
  {Crandall}, and {Ratra}}}]{farooq_crandall_ratra_2013}
\bibinfo{author}{\bibfnamefont{O.}~\bibnamefont{{Farooq}}},
  \bibinfo{author}{\bibfnamefont{S.}~\bibnamefont{{Crandall}}},
  \bibnamefont{and} \bibinfo{author}{\bibfnamefont{B.}~\bibnamefont{{Ratra}}},
  \bibinfo{journal}{Phys. Lett. B} \textbf{\bibinfo{volume}{726}},
  \bibinfo{pages}{72} (\bibinfo{year}{2013}), \eprint{1305.1957}.

\bibitem[{\citenamefont{{Capozziello} et~al.}(2014)\citenamefont{{Capozziello},
  {Farooq}, {Luongo}, and {Ratra}}}]{88}
\bibinfo{author}{\bibfnamefont{S.}~\bibnamefont{{Capozziello}}},
  \bibinfo{author}{\bibfnamefont{O.}~\bibnamefont{{Farooq}}},
  \bibinfo{author}{\bibfnamefont{O.}~\bibnamefont{{Luongo}}}, \bibnamefont{and}
  \bibinfo{author}{\bibfnamefont{B.}~\bibnamefont{{Ratra}}},
  \bibinfo{journal}{Phys. Rev. D} \textbf{\bibinfo{volume}{90}},
  \bibinfo{eid}{044016} (\bibinfo{year}{2014}), \eprint{1403.1421}.

\bibitem[{\citenamefont{{Moresco}
  et~al.}(2016{\natexlab{a}})\citenamefont{{Moresco}, {Pozzetti}, {Cimatti},
  {Jimenez}, {Maraston}, {Verde}, {Thomas}, {Citro}, {Tojeiro}, and
  {Wilkinson}}}]{68}
\bibinfo{author}{\bibfnamefont{M.}~\bibnamefont{{Moresco}}},
  \bibinfo{author}{\bibfnamefont{L.}~\bibnamefont{{Pozzetti}}},
  \bibinfo{author}{\bibfnamefont{A.}~\bibnamefont{{Cimatti}}},
  \bibinfo{author}{\bibfnamefont{R.}~\bibnamefont{{Jimenez}}},
  \bibinfo{author}{\bibfnamefont{C.}~\bibnamefont{{Maraston}}},
  \bibinfo{author}{\bibfnamefont{L.}~\bibnamefont{{Verde}}},
  \bibinfo{author}{\bibfnamefont{D.}~\bibnamefont{{Thomas}}},
  \bibinfo{author}{\bibfnamefont{A.}~\bibnamefont{{Citro}}},
  \bibinfo{author}{\bibfnamefont{R.}~\bibnamefont{{Tojeiro}}},
  \bibnamefont{and}
  \bibinfo{author}{\bibfnamefont{D.}~\bibnamefont{{Wilkinson}}},
  \bibinfo{journal}{J. Cosmol. Astropart. Phys.}
  \textbf{\bibinfo{volume}{2016}}, \bibinfo{eid}{014}
  (\bibinfo{year}{2016}{\natexlab{a}}), \eprint{1601.01701}.

\bibitem[{\citenamefont{{Jesus} et~al.}(2018)\citenamefont{{Jesus}, {Holanda},
  and {Pereira}}}]{jesus_holanda_pereira_2018}
\bibinfo{author}{\bibfnamefont{J.~F.} \bibnamefont{{Jesus}}},
  \bibinfo{author}{\bibfnamefont{R.~F.~L.} \bibnamefont{{Holanda}}},
  \bibnamefont{and} \bibinfo{author}{\bibfnamefont{S.~H.}
  \bibnamefont{{Pereira}}}, \bibinfo{journal}{J. Cosmol. Astropart. Phys.}
  \textbf{\bibinfo{volume}{5}}, \bibinfo{eid}{073} (\bibinfo{year}{2018}),
  \eprint{1712.01075}.

\bibitem[{\citenamefont{{Haridasu} et~al.}(2018)\citenamefont{{Haridasu},
  {Lukovi{\'c}}, {Moresco}, and {Vittorio}}}]{haridasu_etal_2018}
\bibinfo{author}{\bibfnamefont{B.~S.} \bibnamefont{{Haridasu}}},
  \bibinfo{author}{\bibfnamefont{V.~V.} \bibnamefont{{Lukovi{\'c}}}},
  \bibinfo{author}{\bibfnamefont{M.}~\bibnamefont{{Moresco}}},
  \bibnamefont{and}
  \bibinfo{author}{\bibfnamefont{N.}~\bibnamefont{{Vittorio}}},
  \bibinfo{journal}{J. Cosmol. Astropart. Phys.}
  \textbf{\bibinfo{volume}{2018}}, \bibinfo{eid}{015} (\bibinfo{year}{2018}),
  \eprint{1805.03595}.

\bibitem[{\citenamefont{{Bassett} and {Hlozek}}(2010)}]{75}
\bibinfo{author}{\bibfnamefont{B.}~\bibnamefont{{Bassett}}} \bibnamefont{and}
  \bibinfo{author}{\bibfnamefont{R.}~\bibnamefont{{Hlozek}}},
  \emph{\bibinfo{title}{{Baryon acoustic oscillations}}}
  (\bibinfo{year}{2010}), p. \bibinfo{pages}{246}.

\bibitem[{\citenamefont{{Alam} et~al.}(2017)\citenamefont{{Alam}, {Ata},
  {Bailey}, {Beutler}, {Bizyaev}, {Blazek}, {Bolton}, {Brownstein}, {Burden},
  {Chuang} et~al.}}]{Alam_et_al_2017}
\bibinfo{author}{\bibfnamefont{S.}~\bibnamefont{{Alam}}},
  \bibinfo{author}{\bibfnamefont{M.}~\bibnamefont{{Ata}}},
  \bibinfo{author}{\bibfnamefont{S.}~\bibnamefont{{Bailey}}},
  \bibinfo{author}{\bibfnamefont{F.}~\bibnamefont{{Beutler}}},
  \bibinfo{author}{\bibfnamefont{D.}~\bibnamefont{{Bizyaev}}},
  \bibinfo{author}{\bibfnamefont{J.~A.} \bibnamefont{{Blazek}}},
  \bibinfo{author}{\bibfnamefont{A.~S.} \bibnamefont{{Bolton}}},
  \bibinfo{author}{\bibfnamefont{J.~R.} \bibnamefont{{Brownstein}}},
  \bibinfo{author}{\bibfnamefont{A.}~\bibnamefont{{Burden}}},
  \bibinfo{author}{\bibfnamefont{C.-H.} \bibnamefont{{Chuang}}},
  \bibnamefont{et~al.}, \bibinfo{journal}{Mon. Not. R. Astron. Soc.}
  \textbf{\bibinfo{volume}{470}}, \bibinfo{pages}{2617} (\bibinfo{year}{2017}),
  \eprint{1607.03155}.

\bibitem[{\citenamefont{{Beutler} et~al.}(2011)\citenamefont{{Beutler},
  {Blake}, {Colless}, {Jones}, {Staveley-Smith}, {Campbell}, {Parker},
  {Saunders}, and {Watson}}}]{10}
\bibinfo{author}{\bibfnamefont{F.}~\bibnamefont{{Beutler}}},
  \bibinfo{author}{\bibfnamefont{C.}~\bibnamefont{{Blake}}},
  \bibinfo{author}{\bibfnamefont{M.}~\bibnamefont{{Colless}}},
  \bibinfo{author}{\bibfnamefont{D.~H.} \bibnamefont{{Jones}}},
  \bibinfo{author}{\bibfnamefont{L.}~\bibnamefont{{Staveley-Smith}}},
  \bibinfo{author}{\bibfnamefont{L.}~\bibnamefont{{Campbell}}},
  \bibinfo{author}{\bibfnamefont{Q.}~\bibnamefont{{Parker}}},
  \bibinfo{author}{\bibfnamefont{W.}~\bibnamefont{{Saunders}}},
  \bibnamefont{and} \bibinfo{author}{\bibfnamefont{F.}~\bibnamefont{{Watson}}},
  \bibinfo{journal}{Mon. Not. R. Astron. Soc.} \textbf{\bibinfo{volume}{416}},
  \bibinfo{pages}{3017} (\bibinfo{year}{2011}), \eprint{1106.3366}.

\bibitem[{\citenamefont{{Ross} et~al.}(2015)\citenamefont{{Ross}, {Samushia},
  {Howlett}, {Percival}, {Burden}, and {Manera}}}]{2}
\bibinfo{author}{\bibfnamefont{A.~J.} \bibnamefont{{Ross}}},
  \bibinfo{author}{\bibfnamefont{L.}~\bibnamefont{{Samushia}}},
  \bibinfo{author}{\bibfnamefont{C.}~\bibnamefont{{Howlett}}},
  \bibinfo{author}{\bibfnamefont{W.~J.} \bibnamefont{{Percival}}},
  \bibinfo{author}{\bibfnamefont{A.}~\bibnamefont{{Burden}}}, \bibnamefont{and}
  \bibinfo{author}{\bibfnamefont{M.}~\bibnamefont{{Manera}}},
  \bibinfo{journal}{Mon. Not. R. Astron. Soc.} \textbf{\bibinfo{volume}{449}},
  \bibinfo{pages}{835} (\bibinfo{year}{2015}), \eprint{1409.3242}.

\bibitem[{\citenamefont{{Ata} et~al.}(2018)\citenamefont{{Ata}, {Baumgarten},
  {Bautista}, {Beutler}, {Bizyaev}, {Blanton}, {Blazek}, {Bolton}, {Brinkmann},
  {Brownstein} et~al.}}]{3}
\bibinfo{author}{\bibfnamefont{M.}~\bibnamefont{{Ata}}},
  \bibinfo{author}{\bibfnamefont{F.}~\bibnamefont{{Baumgarten}}},
  \bibinfo{author}{\bibfnamefont{J.}~\bibnamefont{{Bautista}}},
  \bibinfo{author}{\bibfnamefont{F.}~\bibnamefont{{Beutler}}},
  \bibinfo{author}{\bibfnamefont{D.}~\bibnamefont{{Bizyaev}}},
  \bibinfo{author}{\bibfnamefont{M.~R.} \bibnamefont{{Blanton}}},
  \bibinfo{author}{\bibfnamefont{J.~A.} \bibnamefont{{Blazek}}},
  \bibinfo{author}{\bibfnamefont{A.~S.} \bibnamefont{{Bolton}}},
  \bibinfo{author}{\bibfnamefont{J.}~\bibnamefont{{Brinkmann}}},
  \bibinfo{author}{\bibfnamefont{J.~R.} \bibnamefont{{Brownstein}}},
  \bibnamefont{et~al.}, \bibinfo{journal}{Mon. Not. R. Astron. Soc.}
  \textbf{\bibinfo{volume}{473}}, \bibinfo{pages}{4773} (\bibinfo{year}{2018}),
  \eprint{1705.06373}.

\bibitem[{\citenamefont{{Bautista} et~al.}(2017)\citenamefont{{Bautista},
  {Busca}, {Guy}, {Rich}, {Blomqvist}, {du Mas des Bourboux}, {Pieri},
  {Font-Ribera}, {Bailey}, {Delubac} et~al.}}]{9}
\bibinfo{author}{\bibfnamefont{J.~E.} \bibnamefont{{Bautista}}},
  \bibinfo{author}{\bibfnamefont{N.~G.} \bibnamefont{{Busca}}},
  \bibinfo{author}{\bibfnamefont{J.}~\bibnamefont{{Guy}}},
  \bibinfo{author}{\bibfnamefont{J.}~\bibnamefont{{Rich}}},
  \bibinfo{author}{\bibfnamefont{M.}~\bibnamefont{{Blomqvist}}},
  \bibinfo{author}{\bibfnamefont{H.}~\bibnamefont{{du Mas des Bourboux}}},
  \bibinfo{author}{\bibfnamefont{M.~M.} \bibnamefont{{Pieri}}},
  \bibinfo{author}{\bibfnamefont{A.}~\bibnamefont{{Font-Ribera}}},
  \bibinfo{author}{\bibfnamefont{S.}~\bibnamefont{{Bailey}}},
  \bibinfo{author}{\bibfnamefont{T.}~\bibnamefont{{Delubac}}},
  \bibnamefont{et~al.}, \bibinfo{journal}{Astron. Astrophys.}
  \textbf{\bibinfo{volume}{603}}, \bibinfo{eid}{A12} (\bibinfo{year}{2017}),
  \eprint{1702.00176}.

\bibitem[{\citenamefont{{Font-Ribera} et~al.}(2014)\citenamefont{{Font-Ribera},
  {Kirkby}, {Busca}, {Miralda-Escud{\'e}}, {Ross}, {Slosar}, {Rich}, {Aubourg},
  {Bailey}, {Bhardwaj} et~al.}}]{11}
\bibinfo{author}{\bibfnamefont{A.}~\bibnamefont{{Font-Ribera}}},
  \bibinfo{author}{\bibfnamefont{D.}~\bibnamefont{{Kirkby}}},
  \bibinfo{author}{\bibfnamefont{N.}~\bibnamefont{{Busca}}},
  \bibinfo{author}{\bibfnamefont{J.}~\bibnamefont{{Miralda-Escud{\'e}}}},
  \bibinfo{author}{\bibfnamefont{N.~P.} \bibnamefont{{Ross}}},
  \bibinfo{author}{\bibfnamefont{A.}~\bibnamefont{{Slosar}}},
  \bibinfo{author}{\bibfnamefont{J.}~\bibnamefont{{Rich}}},
  \bibinfo{author}{\bibfnamefont{{\'E}.}~\bibnamefont{{Aubourg}}},
  \bibinfo{author}{\bibfnamefont{S.}~\bibnamefont{{Bailey}}},
  \bibinfo{author}{\bibfnamefont{V.}~\bibnamefont{{Bhardwaj}}},
  \bibnamefont{et~al.}, \bibinfo{journal}{J. Cosmol. Astropart. Phys.}
  \textbf{\bibinfo{volume}{5}}, \bibinfo{eid}{027} (\bibinfo{year}{2014}),
  \eprint{1311.1767}.

\bibitem[{\citenamefont{{Eisenstein} and {Hu}}(1998)}]{8}
\bibinfo{author}{\bibfnamefont{D.~J.} \bibnamefont{{Eisenstein}}}
  \bibnamefont{and} \bibinfo{author}{\bibfnamefont{W.}~\bibnamefont{{Hu}}},
  \bibinfo{journal}{Astrophys. J.} \textbf{\bibinfo{volume}{496}},
  \bibinfo{pages}{605} (\bibinfo{year}{1998}), \eprint{astro-ph/9709112}.

\bibitem[{\citenamefont{{Moresco} et~al.}(2012)\citenamefont{{Moresco},
  {Cimatti}, {Jimenez}, {Pozzetti}, {Zamorani}, {Bolzonella}, {Dunlop},
  {Lamareille}, {Mignoli}, {Pearce} et~al.}}]{70}
\bibinfo{author}{\bibfnamefont{M.}~\bibnamefont{{Moresco}}},
  \bibinfo{author}{\bibfnamefont{A.}~\bibnamefont{{Cimatti}}},
  \bibinfo{author}{\bibfnamefont{R.}~\bibnamefont{{Jimenez}}},
  \bibinfo{author}{\bibfnamefont{L.}~\bibnamefont{{Pozzetti}}},
  \bibinfo{author}{\bibfnamefont{G.}~\bibnamefont{{Zamorani}}},
  \bibinfo{author}{\bibfnamefont{M.}~\bibnamefont{{Bolzonella}}},
  \bibinfo{author}{\bibfnamefont{J.}~\bibnamefont{{Dunlop}}},
  \bibinfo{author}{\bibfnamefont{F.}~\bibnamefont{{Lamareille}}},
  \bibinfo{author}{\bibfnamefont{M.}~\bibnamefont{{Mignoli}}},
  \bibinfo{author}{\bibfnamefont{H.}~\bibnamefont{{Pearce}}},
  \bibnamefont{et~al.}, \bibinfo{journal}{J. Cosmol. Astropart. Phys.}
  \textbf{\bibinfo{volume}{8}}, \bibinfo{eid}{006} (\bibinfo{year}{2012}),
  \eprint{1201.3609}.

\bibitem[{\citenamefont{{Moresco}}(2015)}]{72}
\bibinfo{author}{\bibfnamefont{M.}~\bibnamefont{{Moresco}}},
  \bibinfo{journal}{Mon. Not. R. Astron. Soc.} \textbf{\bibinfo{volume}{450}},
  \bibinfo{pages}{L16} (\bibinfo{year}{2015}), \eprint{1503.01116}.

\bibitem[{\citenamefont{{Chen} and {Ratra}}(2011{\natexlab{b}})}]{chenratmed}
\bibinfo{author}{\bibfnamefont{G.}~\bibnamefont{{Chen}}} \bibnamefont{and}
  \bibinfo{author}{\bibfnamefont{B.}~\bibnamefont{{Ratra}}},
  \bibinfo{journal}{Publ. Astron. Soc. Pac.} \textbf{\bibinfo{volume}{123}},
  \bibinfo{pages}{1127} (\bibinfo{year}{2011}{\natexlab{b}}),
  \eprint{1105.5206}.

\bibitem[{\citenamefont{{Gott} et~al.}(2001)\citenamefont{{Gott}, {Vogeley},
  {Podariu}, and {Ratra}}}]{gott_etal_2001}
\bibinfo{author}{\bibfnamefont{J.~R.} \bibnamefont{{Gott}},
  \bibfnamefont{III}}, \bibinfo{author}{\bibfnamefont{M.~S.}
  \bibnamefont{{Vogeley}}},
  \bibinfo{author}{\bibfnamefont{S.}~\bibnamefont{{Podariu}}},
  \bibnamefont{and} \bibinfo{author}{\bibfnamefont{B.}~\bibnamefont{{Ratra}}},
  \bibinfo{journal}{Astrophys. J.} \textbf{\bibinfo{volume}{549}},
  \bibinfo{pages}{1} (\bibinfo{year}{2001}), \eprint{astro-ph/0006103}.

\bibitem[{\citenamefont{{Chen} et~al.}(2003)\citenamefont{{Chen}, {Gott}, and
  {Ratra}}}]{76}
\bibinfo{author}{\bibfnamefont{G.}~\bibnamefont{{Chen}}},
  \bibinfo{author}{\bibfnamefont{J.~R.} \bibnamefont{{Gott}},
  \bibfnamefont{III}}, \bibnamefont{and}
  \bibinfo{author}{\bibfnamefont{B.}~\bibnamefont{{Ratra}}},
  \bibinfo{journal}{Publ. Astron. Soc. Pac.} \textbf{\bibinfo{volume}{115}},
  \bibinfo{pages}{1269} (\bibinfo{year}{2003}), \eprint{astro-ph/0308099}.

\bibitem[{\citenamefont{{Chen} et~al.}(2017)\citenamefont{{Chen}, {Kumar}, and
  {Ratra}}}]{chen_etal_2017}
\bibinfo{author}{\bibfnamefont{Y.}~\bibnamefont{{Chen}}},
  \bibinfo{author}{\bibfnamefont{S.}~\bibnamefont{{Kumar}}}, \bibnamefont{and}
  \bibinfo{author}{\bibfnamefont{B.}~\bibnamefont{{Ratra}}},
  \bibinfo{journal}{Astrophys. J.} \textbf{\bibinfo{volume}{835}},
  \bibinfo{eid}{86} (\bibinfo{year}{2017}), \eprint{1606.07316}.

\bibitem[{\citenamefont{{Wang} et~al.}(2017{\natexlab{b}})\citenamefont{{Wang},
  {Xu}, and {Zhao}}}]{wang_xu_zhao_2017}
\bibinfo{author}{\bibfnamefont{Y.}~\bibnamefont{{Wang}}},
  \bibinfo{author}{\bibfnamefont{L.}~\bibnamefont{{Xu}}}, \bibnamefont{and}
  \bibinfo{author}{\bibfnamefont{G.-B.} \bibnamefont{{Zhao}}},
  \bibinfo{journal}{Astrophys. J.} \textbf{\bibinfo{volume}{849}},
  \bibinfo{eid}{84} (\bibinfo{year}{2017}{\natexlab{b}}), \eprint{1706.09149}.

\bibitem[{\citenamefont{{Lin} and {Ishak}}(2017)}]{Linishak}
\bibinfo{author}{\bibfnamefont{W.}~\bibnamefont{{Lin}}} \bibnamefont{and}
  \bibinfo{author}{\bibfnamefont{M.}~\bibnamefont{{Ishak}}},
  \bibinfo{journal}{Phys. Rev. D} \textbf{\bibinfo{volume}{96}},
  \bibinfo{eid}{083532} (\bibinfo{year}{2017}), \eprint{1708.09813}.

\bibitem[{\citenamefont{{Haridasu}
  et~al.}(2017{\natexlab{a}})\citenamefont{{Haridasu}, {Lukovi{\'c}}, and
  {Vittorio}}}]{81}
\bibinfo{author}{\bibfnamefont{B.~S.} \bibnamefont{{Haridasu}}},
  \bibinfo{author}{\bibfnamefont{V.~V.} \bibnamefont{{Lukovi{\'c}}}},
  \bibnamefont{and}
  \bibinfo{author}{\bibfnamefont{N.}~\bibnamefont{{Vittorio}}},
  \bibinfo{journal}{ArXiv e-prints}  (\bibinfo{year}{2017}{\natexlab{a}}),
  \eprint{1711.03929}.

\bibitem[{\citenamefont{{G{\'o}mez-Valent} and
  {Amendola}}(2018)}]{Gomez-ValentAmendola2018}
\bibinfo{author}{\bibfnamefont{A.}~\bibnamefont{{G{\'o}mez-Valent}}}
  \bibnamefont{and}
  \bibinfo{author}{\bibfnamefont{L.}~\bibnamefont{{Amendola}}},
  \bibinfo{journal}{J. Cosmol. Astropart. Phys.} \textbf{\bibinfo{volume}{4}},
  \bibinfo{eid}{051} (\bibinfo{year}{2018}), \eprint{1802.01505}.

\bibitem[{\citenamefont{{Riess} et~al.}(2016)\citenamefont{{Riess}, {Macri},
  {Hoffmann}, {Scolnic}, {Casertano}, {Filippenko}, {Tucker}, {Reid}, {Jones},
  {Silverman} et~al.}}]{riess2016}
\bibinfo{author}{\bibfnamefont{A.~G.} \bibnamefont{{Riess}}},
  \bibinfo{author}{\bibfnamefont{L.~M.} \bibnamefont{{Macri}}},
  \bibinfo{author}{\bibfnamefont{S.~L.} \bibnamefont{{Hoffmann}}},
  \bibinfo{author}{\bibfnamefont{D.}~\bibnamefont{{Scolnic}}},
  \bibinfo{author}{\bibfnamefont{S.}~\bibnamefont{{Casertano}}},
  \bibinfo{author}{\bibfnamefont{A.~V.} \bibnamefont{{Filippenko}}},
  \bibinfo{author}{\bibfnamefont{B.~E.} \bibnamefont{{Tucker}}},
  \bibinfo{author}{\bibfnamefont{M.~J.} \bibnamefont{{Reid}}},
  \bibinfo{author}{\bibfnamefont{D.~O.} \bibnamefont{{Jones}}},
  \bibinfo{author}{\bibfnamefont{J.~M.} \bibnamefont{{Silverman}}},
  \bibnamefont{et~al.}, \bibinfo{journal}{Astrophys. J.}
  \textbf{\bibinfo{volume}{826}}, \bibinfo{eid}{56} (\bibinfo{year}{2016}),
  \eprint{1604.01424}.

\bibitem[{\citenamefont{{Rigault} et~al.}(2015)\citenamefont{{Rigault},
  {Aldering}, {Kowalski}, {Copin}, {Antilogus}, {Aragon}, {Bailey}, {Baltay},
  {Baugh}, {Bongard} et~al.}}]{rigault_etal_2015}
\bibinfo{author}{\bibfnamefont{M.}~\bibnamefont{{Rigault}}},
  \bibinfo{author}{\bibfnamefont{G.}~\bibnamefont{{Aldering}}},
  \bibinfo{author}{\bibfnamefont{M.}~\bibnamefont{{Kowalski}}},
  \bibinfo{author}{\bibfnamefont{Y.}~\bibnamefont{{Copin}}},
  \bibinfo{author}{\bibfnamefont{P.}~\bibnamefont{{Antilogus}}},
  \bibinfo{author}{\bibfnamefont{C.}~\bibnamefont{{Aragon}}},
  \bibinfo{author}{\bibfnamefont{S.}~\bibnamefont{{Bailey}}},
  \bibinfo{author}{\bibfnamefont{C.}~\bibnamefont{{Baltay}}},
  \bibinfo{author}{\bibfnamefont{D.}~\bibnamefont{{Baugh}}},
  \bibinfo{author}{\bibfnamefont{S.}~\bibnamefont{{Bongard}}},
  \bibnamefont{et~al.}, \bibinfo{journal}{Astrophys. J.}
  \textbf{\bibinfo{volume}{802}}, \bibinfo{eid}{20} (\bibinfo{year}{2015}),
  \eprint{1412.6501}.

\bibitem[{\citenamefont{{Zhang}
  et~al.}(2017{\natexlab{b}})\citenamefont{{Zhang}, {Childress}, {Davis},
  {Karpenka}, {Lidman}, {Schmidt}, and {Smith}}}]{86}
\bibinfo{author}{\bibfnamefont{B.~R.} \bibnamefont{{Zhang}}},
  \bibinfo{author}{\bibfnamefont{M.~J.} \bibnamefont{{Childress}}},
  \bibinfo{author}{\bibfnamefont{T.~M.} \bibnamefont{{Davis}}},
  \bibinfo{author}{\bibfnamefont{N.~V.} \bibnamefont{{Karpenka}}},
  \bibinfo{author}{\bibfnamefont{C.}~\bibnamefont{{Lidman}}},
  \bibinfo{author}{\bibfnamefont{B.~P.} \bibnamefont{{Schmidt}}},
  \bibnamefont{and} \bibinfo{author}{\bibfnamefont{M.}~\bibnamefont{{Smith}}},
  \bibinfo{journal}{Mon. Not. R. Astron. Soc.} \textbf{\bibinfo{volume}{471}},
  \bibinfo{pages}{2254} (\bibinfo{year}{2017}{\natexlab{b}}),
  \eprint{1706.07573}.

\bibitem[{\citenamefont{{Dhawan} et~al.}(2018)\citenamefont{{Dhawan}, {Jha},
  and {Leibundgut}}}]{Dhawan}
\bibinfo{author}{\bibfnamefont{S.}~\bibnamefont{{Dhawan}}},
  \bibinfo{author}{\bibfnamefont{S.~W.} \bibnamefont{{Jha}}}, \bibnamefont{and}
  \bibinfo{author}{\bibfnamefont{B.}~\bibnamefont{{Leibundgut}}},
  \bibinfo{journal}{Astron. Astrophys.} \textbf{\bibinfo{volume}{609}},
  \bibinfo{eid}{A72} (\bibinfo{year}{2018}), \eprint{1707.00715}.

\bibitem[{\citenamefont{{Fern{\'a}ndez Arenas}
  et~al.}(2018)\citenamefont{{Fern{\'a}ndez Arenas}, {Terlevich}, {Terlevich},
  {Melnick}, {Ch{\'a}vez}, {Bresolin}, {Telles}, {Plionis}, and
  {Basilakos}}}]{FernandezArenas}
\bibinfo{author}{\bibfnamefont{D.}~\bibnamefont{{Fern{\'a}ndez Arenas}}},
  \bibinfo{author}{\bibfnamefont{E.}~\bibnamefont{{Terlevich}}},
  \bibinfo{author}{\bibfnamefont{R.}~\bibnamefont{{Terlevich}}},
  \bibinfo{author}{\bibfnamefont{J.}~\bibnamefont{{Melnick}}},
  \bibinfo{author}{\bibfnamefont{R.}~\bibnamefont{{Ch{\'a}vez}}},
  \bibinfo{author}{\bibfnamefont{F.}~\bibnamefont{{Bresolin}}},
  \bibinfo{author}{\bibfnamefont{E.}~\bibnamefont{{Telles}}},
  \bibinfo{author}{\bibfnamefont{M.}~\bibnamefont{{Plionis}}},
  \bibnamefont{and}
  \bibinfo{author}{\bibfnamefont{S.}~\bibnamefont{{Basilakos}}},
  \bibinfo{journal}{Mon. Not. R. Astron. Soc.} \textbf{\bibinfo{volume}{474}},
  \bibinfo{pages}{1250} (\bibinfo{year}{2018}), \eprint{1710.05951}.

\bibitem[{\citenamefont{{Liddle}}(2007)}]{Liddle_2007}
\bibinfo{author}{\bibfnamefont{A.~R.} \bibnamefont{{Liddle}}},
  \bibinfo{journal}{Mon. Not. R. Astron. Soc.} \textbf{\bibinfo{volume}{377}},
  \bibinfo{pages}{L74} (\bibinfo{year}{2007}), \eprint{astro-ph/0701113}.

\bibitem[{\citenamefont{{Ryan} et~al.}(2019)\citenamefont{{Ryan}, {Chen}, and
  {Ratra}}}]{Ryan_Chen_Ratra_2019}
\bibinfo{author}{\bibfnamefont{J.}~\bibnamefont{{Ryan}}},
  \bibinfo{author}{\bibfnamefont{Y.}~\bibnamefont{{Chen}}}, \bibnamefont{and}
  \bibinfo{author}{\bibfnamefont{B.}~\bibnamefont{{Ratra}}},
  \bibinfo{journal}{Mon. Not. R. Astron. Soc.} \textbf{\bibinfo{volume}{488}},
  \bibinfo{pages}{3844} (\bibinfo{year}{2019}), \eprint{1902.03196}.

\bibitem[{\citenamefont{{Park} and
  {Ratra}}(2018)}]{Park_Ratra_2018_FpCDM_NFpCDM}
\bibinfo{author}{\bibfnamefont{C.-G.} \bibnamefont{{Park}}} \bibnamefont{and}
  \bibinfo{author}{\bibfnamefont{B.}~\bibnamefont{{Ratra}}},
  \bibinfo{journal}{Astrophys. J.} \textbf{\bibinfo{volume}{868}},
  \bibinfo{eid}{83} (\bibinfo{year}{2018}), \eprint{1807.07421}.

\bibitem[{\citenamefont{{Mitra} et~al.}(2019)\citenamefont{{Mitra}, {Park},
  {Choudhury}, and {Ratra}}}]{mitra_park_choudhury_ratra_2019}
\bibinfo{author}{\bibfnamefont{S.}~\bibnamefont{{Mitra}}},
  \bibinfo{author}{\bibfnamefont{C.-G.} \bibnamefont{{Park}}},
  \bibinfo{author}{\bibfnamefont{T.~R.} \bibnamefont{{Choudhury}}},
  \bibnamefont{and} \bibinfo{author}{\bibfnamefont{B.}~\bibnamefont{{Ratra}}},
  \bibinfo{journal}{ArXiv e-prints}  (\bibinfo{year}{2019}),
  \eprint{1901.09927}.

\bibitem[{\citenamefont{{Penton} et~al.}(2018)\citenamefont{{Penton}, {Peyton},
  {Zahoor}, and {Ratra}}}]{penton_peyton_zahoor_ratra_2018}
\bibinfo{author}{\bibfnamefont{J.}~\bibnamefont{{Penton}}},
  \bibinfo{author}{\bibfnamefont{J.}~\bibnamefont{{Peyton}}},
  \bibinfo{author}{\bibfnamefont{A.}~\bibnamefont{{Zahoor}}}, \bibnamefont{and}
  \bibinfo{author}{\bibfnamefont{B.}~\bibnamefont{{Ratra}}},
  \bibinfo{journal}{Publ. Astron. Soc. Pac.} \textbf{\bibinfo{volume}{130}},
  \bibinfo{pages}{114001} (\bibinfo{year}{2018}), \eprint{1808.01490}.

\bibitem[{\citenamefont{{Sol{\`a} Peracaula}
  et~al.}(2019)\citenamefont{{Sol{\`a} Peracaula}, {G{\'o}mez-Valent}, and {de
  Cruz P{\'e}rez}}}]{sola_gomez_perez_2019}
\bibinfo{author}{\bibfnamefont{J.}~\bibnamefont{{Sol{\`a} Peracaula}}},
  \bibinfo{author}{\bibfnamefont{A.}~\bibnamefont{{G{\'o}mez-Valent}}},
  \bibnamefont{and} \bibinfo{author}{\bibfnamefont{J.}~\bibnamefont{{de Cruz
  P{\'e}rez}}}, \bibinfo{journal}{Phys. Dark Universe}
  \textbf{\bibinfo{volume}{25}}, \bibinfo{eid}{100311} (\bibinfo{year}{2019}),
  \eprint{1811.03505}.

\bibitem[{\citenamefont{{Wang} et~al.}(2018)\citenamefont{{Wang}, {Pogosian},
  {Zhao}, and {Zucca}}}]{wang_pogosian_zhao_zucca_2018}
\bibinfo{author}{\bibfnamefont{Y.}~\bibnamefont{{Wang}}},
  \bibinfo{author}{\bibfnamefont{L.}~\bibnamefont{{Pogosian}}},
  \bibinfo{author}{\bibfnamefont{G.-B.} \bibnamefont{{Zhao}}},
  \bibnamefont{and} \bibinfo{author}{\bibfnamefont{A.}~\bibnamefont{{Zucca}}},
  \bibinfo{journal}{Astrophys. J. Lett.} \textbf{\bibinfo{volume}{869}},
  \bibinfo{eid}{L8} (\bibinfo{year}{2018}), \eprint{1807.03772}.

\bibitem[{\citenamefont{{Zhang} et~al.}(2018)\citenamefont{{Zhang}, {Lee}, and
  {Geng}}}]{zhang_lee_geng_2018}
\bibinfo{author}{\bibfnamefont{J.-J.} \bibnamefont{{Zhang}}},
  \bibinfo{author}{\bibfnamefont{C.-C.} \bibnamefont{{Lee}}}, \bibnamefont{and}
  \bibinfo{author}{\bibfnamefont{C.-Q.} \bibnamefont{{Geng}}},
  \bibinfo{journal}{ArXiv e-prints}  (\bibinfo{year}{2018}),
  \eprint{1812.06710}.

\bibitem[{\citenamefont{{Qi} et~al.}(2018)\citenamefont{{Qi}, {Cao}, {Zhang},
  {Biesiada}, {Wu}, and {Zhu}}}]{Qi_et_al2018}
\bibinfo{author}{\bibfnamefont{J.-Z.} \bibnamefont{{Qi}}},
  \bibinfo{author}{\bibfnamefont{S.}~\bibnamefont{{Cao}}},
  \bibinfo{author}{\bibfnamefont{S.}~\bibnamefont{{Zhang}}},
  \bibinfo{author}{\bibfnamefont{M.}~\bibnamefont{{Biesiada}}},
  \bibinfo{author}{\bibfnamefont{Y.}~\bibnamefont{{Wu}}}, \bibnamefont{and}
  \bibinfo{author}{\bibfnamefont{Z.-H.} \bibnamefont{{Zhu}}},
  \bibinfo{journal}{ArXiv e-prints}  (\bibinfo{year}{2018}),
  \eprint{1803.01990}.

\bibitem[{\citenamefont{{Park} and
  {Ratra}}(2019{\natexlab{c}})}]{park_ratra_2019b}
\bibinfo{author}{\bibfnamefont{C.-G.} \bibnamefont{{Park}}} \bibnamefont{and}
  \bibinfo{author}{\bibfnamefont{B.}~\bibnamefont{{Ratra}}},
  \bibinfo{journal}{Astrophys. Space Sci.} \textbf{\bibinfo{volume}{364}},
  \bibinfo{eid}{134} (\bibinfo{year}{2019}{\natexlab{c}}), \eprint{1809.03598}.

\bibitem[{\citenamefont{{Mukherjee} et~al.}(2019)\citenamefont{{Mukherjee},
  {Paul}, and {Jassal}}}]{mukherjee_paul_jassal_2019}
\bibinfo{author}{\bibfnamefont{A.}~\bibnamefont{{Mukherjee}}},
  \bibinfo{author}{\bibfnamefont{N.}~\bibnamefont{{Paul}}}, \bibnamefont{and}
  \bibinfo{author}{\bibfnamefont{H.~K.} \bibnamefont{{Jassal}}},
  \bibinfo{journal}{J. Cosmol. Astropart. Phys.} \textbf{\bibinfo{volume}{1}},
  \bibinfo{eid}{005} (\bibinfo{year}{2019}), \eprint{1809.08849}.

\bibitem[{\citenamefont{{DES Collaboration}}(2019{\natexlab{a}})}]{DES_2019}
\bibinfo{author}{\bibnamefont{{DES Collaboration}}}, \bibinfo{journal}{Phys.
  Rev. D} \textbf{\bibinfo{volume}{99}}, \bibinfo{eid}{123505}
  (\bibinfo{year}{2019}{\natexlab{a}}), \eprint{1810.02499}.

\bibitem[{\citenamefont{{Zheng} et~al.}(2019)\citenamefont{{Zheng}, {Melia},
  and {Zhang}}}]{Zheng_2019}
\bibinfo{author}{\bibfnamefont{J.}~\bibnamefont{{Zheng}}},
  \bibinfo{author}{\bibfnamefont{F.}~\bibnamefont{{Melia}}}, \bibnamefont{and}
  \bibinfo{author}{\bibfnamefont{T.-J.} \bibnamefont{{Zhang}}},
  \bibinfo{journal}{ArXiv e-prints}  (\bibinfo{year}{2019}),
  \eprint{1901.05705}.

\bibitem[{\citenamefont{{Ruan} et~al.}(2019)\citenamefont{{Ruan}, {Melia},
  {Chen}, and {Zhang}}}]{ruan_etal_2019}
\bibinfo{author}{\bibfnamefont{C.-Z.} \bibnamefont{{Ruan}}},
  \bibinfo{author}{\bibfnamefont{F.}~\bibnamefont{{Melia}}},
  \bibinfo{author}{\bibfnamefont{Y.}~\bibnamefont{{Chen}}}, \bibnamefont{and}
  \bibinfo{author}{\bibfnamefont{T.-J.} \bibnamefont{{Zhang}}},
  \bibinfo{journal}{Astrophys. J.} \textbf{\bibinfo{volume}{881}},
  \bibinfo{eid}{137} (\bibinfo{year}{2019}), \eprint{1901.06626}.

\bibitem[{\citenamefont{{Witzemann} et~al.}(2018)\citenamefont{{Witzemann},
  {Bull}, {Clarkson}, {Santos}, {Spinelli}, and
  {Weltman}}}]{witzemann_et_al_2018}
\bibinfo{author}{\bibfnamefont{A.}~\bibnamefont{{Witzemann}}},
  \bibinfo{author}{\bibfnamefont{P.}~\bibnamefont{{Bull}}},
  \bibinfo{author}{\bibfnamefont{C.}~\bibnamefont{{Clarkson}}},
  \bibinfo{author}{\bibfnamefont{M.~G.} \bibnamefont{{Santos}}},
  \bibinfo{author}{\bibfnamefont{M.}~\bibnamefont{{Spinelli}}},
  \bibnamefont{and}
  \bibinfo{author}{\bibfnamefont{A.}~\bibnamefont{{Weltman}}},
  \bibinfo{journal}{Mon. Not. R. Astron. Soc.} \textbf{\bibinfo{volume}{477}},
  \bibinfo{pages}{L122} (\bibinfo{year}{2018}), \eprint{1711.02179}.

\bibitem[{\citenamefont{{Wei}}(2018)}]{wei_2018}
\bibinfo{author}{\bibfnamefont{J.-J.} \bibnamefont{{Wei}}},
  \bibinfo{journal}{Astrophys. J.} \textbf{\bibinfo{volume}{868}},
  \bibinfo{eid}{29} (\bibinfo{year}{2018}), \eprint{1806.09781}.

\bibitem[{\citenamefont{{G{\'o}mez-Valent}}(2018)}]{gomez_valent_2018}
\bibinfo{author}{\bibfnamefont{A.}~\bibnamefont{{G{\'o}mez-Valent}}},
  \bibinfo{journal}{ArXiv e-prints}  (\bibinfo{year}{2018}),
  \eprint{1810.02278}.

\bibitem[{\citenamefont{{Chen} et~al.}(2015)\citenamefont{{Chen}, {Geng},
  {Cao}, {Huang}, and {Zhu}}}]{chen_geng_cao_huang_zhu_2015}
\bibinfo{author}{\bibfnamefont{Y.}~\bibnamefont{{Chen}}},
  \bibinfo{author}{\bibfnamefont{C.-Q.} \bibnamefont{{Geng}}},
  \bibinfo{author}{\bibfnamefont{S.}~\bibnamefont{{Cao}}},
  \bibinfo{author}{\bibfnamefont{Y.-M.} \bibnamefont{{Huang}}},
  \bibnamefont{and} \bibinfo{author}{\bibfnamefont{Z.-H.} \bibnamefont{{Zhu}}},
  \bibinfo{journal}{J. Cosmol. Astropart. Phys.} \textbf{\bibinfo{volume}{2}},
  \bibinfo{eid}{010} (\bibinfo{year}{2015}), \eprint{1312.1443}.

\bibitem[{\citenamefont{{Anagnostopoulos} and
  {Basilakos}}(2018)}]{anagnostopoulos_2018}
\bibinfo{author}{\bibfnamefont{F.~K.} \bibnamefont{{Anagnostopoulos}}}
  \bibnamefont{and}
  \bibinfo{author}{\bibfnamefont{S.}~\bibnamefont{{Basilakos}}},
  \bibinfo{journal}{Phys. Rev. D} \textbf{\bibinfo{volume}{97}},
  \bibinfo{eid}{063503} (\bibinfo{year}{2018}), \eprint{1709.02356}.

\bibitem[{\citenamefont{{Mamon} and {Bamba}}(2018)}]{mamon_bamba_2018}
\bibinfo{author}{\bibfnamefont{A.~A.} \bibnamefont{{Mamon}}} \bibnamefont{and}
  \bibinfo{author}{\bibfnamefont{K.}~\bibnamefont{{Bamba}}},
  \bibinfo{journal}{ArXiv e-prints}  (\bibinfo{year}{2018}),
  \eprint{1805.02854}.

\bibitem[{\citenamefont{{Geng} et~al.}(2018)\citenamefont{{Geng}, {Guo},
  {Wang}, {Zhang}, and {Zhang}}}]{geng_et_al_2018}
\bibinfo{author}{\bibfnamefont{J.-J.} \bibnamefont{{Geng}}},
  \bibinfo{author}{\bibfnamefont{R.-Y.} \bibnamefont{{Guo}}},
  \bibinfo{author}{\bibfnamefont{A.-Z.} \bibnamefont{{Wang}}},
  \bibinfo{author}{\bibfnamefont{J.-F.} \bibnamefont{{Zhang}}},
  \bibnamefont{and} \bibinfo{author}{\bibfnamefont{X.}~\bibnamefont{{Zhang}}},
  \bibinfo{journal}{Commun. Theor. Phys.} \textbf{\bibinfo{volume}{70}},
  \bibinfo{pages}{445} (\bibinfo{year}{2018}), \eprint{1806.10735}.

\bibitem[{\citenamefont{{Liu} et~al.}(2018)\citenamefont{{Liu}, {Guo}, {Zhang},
  and {Zhang}}}]{liu_et_al_2018}
\bibinfo{author}{\bibfnamefont{Y.}~\bibnamefont{{Liu}}},
  \bibinfo{author}{\bibfnamefont{R.-Y.} \bibnamefont{{Guo}}},
  \bibinfo{author}{\bibfnamefont{J.-F.} \bibnamefont{{Zhang}}},
  \bibnamefont{and} \bibinfo{author}{\bibfnamefont{X.}~\bibnamefont{{Zhang}}},
  \bibinfo{journal}{ArXiv e-prints}  (\bibinfo{year}{2018}),
  \eprint{1811.12131}.

\bibitem[{\citenamefont{{Cao} et~al.}(2017{\natexlab{b}})\citenamefont{{Cao},
  {Zheng}, {Biesiada}, {Qi}, {Chen}, and {Zhu}}}]{Cao_et_al2017b}
\bibinfo{author}{\bibfnamefont{S.}~\bibnamefont{{Cao}}},
  \bibinfo{author}{\bibfnamefont{X.}~\bibnamefont{{Zheng}}},
  \bibinfo{author}{\bibfnamefont{M.}~\bibnamefont{{Biesiada}}},
  \bibinfo{author}{\bibfnamefont{J.}~\bibnamefont{{Qi}}},
  \bibinfo{author}{\bibfnamefont{Y.}~\bibnamefont{{Chen}}}, \bibnamefont{and}
  \bibinfo{author}{\bibfnamefont{Z.-H.} \bibnamefont{{Zhu}}},
  \bibinfo{journal}{Astron. Astrophys.} \textbf{\bibinfo{volume}{606}},
  \bibinfo{eid}{A15} (\bibinfo{year}{2017}{\natexlab{b}}), \eprint{1708.08635}.

\bibitem[{\citenamefont{{Daly} and {Guerra}}(2002)}]{daly_guerra_2002}
\bibinfo{author}{\bibfnamefont{R.~A.} \bibnamefont{{Daly}}} \bibnamefont{and}
  \bibinfo{author}{\bibfnamefont{E.~J.} \bibnamefont{{Guerra}}},
  \bibinfo{journal}{Astron. J.} \textbf{\bibinfo{volume}{124}},
  \bibinfo{pages}{1831} (\bibinfo{year}{2002}), \eprint{astro-ph/0209503}.

\bibitem[{\citenamefont{{Podariu} et~al.}(2003)\citenamefont{{Podariu}, {Daly},
  {Mory}, and {Ratra}}}]{podariu_daly_mory_ratra_2003}
\bibinfo{author}{\bibfnamefont{S.}~\bibnamefont{{Podariu}}},
  \bibinfo{author}{\bibfnamefont{R.~A.} \bibnamefont{{Daly}}},
  \bibinfo{author}{\bibfnamefont{M.~P.} \bibnamefont{{Mory}}},
  \bibnamefont{and} \bibinfo{author}{\bibfnamefont{B.}~\bibnamefont{{Ratra}}},
  \bibinfo{journal}{Astrophys. J.} \textbf{\bibinfo{volume}{584}},
  \bibinfo{pages}{577} (\bibinfo{year}{2003}), \eprint{astro-ph/0207096}.

\bibitem[{\citenamefont{{Bonamente} et~al.}(2006)\citenamefont{{Bonamente},
  {Joy}, {LaRoque}, {Carlstrom}, {Reese}, and {Dawson}}}]{bonamente_2006}
\bibinfo{author}{\bibfnamefont{M.}~\bibnamefont{{Bonamente}}},
  \bibinfo{author}{\bibfnamefont{M.~K.} \bibnamefont{{Joy}}},
  \bibinfo{author}{\bibfnamefont{S.~J.} \bibnamefont{{LaRoque}}},
  \bibinfo{author}{\bibfnamefont{J.~E.} \bibnamefont{{Carlstrom}}},
  \bibinfo{author}{\bibfnamefont{E.~D.} \bibnamefont{{Reese}}},
  \bibnamefont{and} \bibinfo{author}{\bibfnamefont{K.~S.}
  \bibnamefont{{Dawson}}}, \bibinfo{journal}{Astrophys. J.}
  \textbf{\bibinfo{volume}{647}}, \bibinfo{pages}{25} (\bibinfo{year}{2006}),
  \eprint{astro-ph/0512349}.

\bibitem[{\citenamefont{{Chen} and {Ratra}}(2012)}]{Chen_Ratra_2012}
\bibinfo{author}{\bibfnamefont{Y.}~\bibnamefont{{Chen}}} \bibnamefont{and}
  \bibinfo{author}{\bibfnamefont{B.}~\bibnamefont{{Ratra}}},
  \bibinfo{journal}{Astron. Astrophys.} \textbf{\bibinfo{volume}{543}},
  \bibinfo{eid}{A104} (\bibinfo{year}{2012}), \eprint{1105.5660}.

\bibitem[{\citenamefont{{Gurvits} et~al.}(1999)\citenamefont{{Gurvits},
  {Kellermann}, and {Frey}}}]{gurvits_kellermann_frey_1999}
\bibinfo{author}{\bibfnamefont{L.~I.} \bibnamefont{{Gurvits}}},
  \bibinfo{author}{\bibfnamefont{K.~I.} \bibnamefont{{Kellermann}}},
  \bibnamefont{and} \bibinfo{author}{\bibfnamefont{S.}~\bibnamefont{{Frey}}},
  \bibinfo{journal}{Astron. Astrophys.} \textbf{\bibinfo{volume}{342}},
  \bibinfo{pages}{378} (\bibinfo{year}{1999}), \eprint{astro-ph/9812018}.

\bibitem[{\citenamefont{{Vishwakarma}}(2001)}]{vishwakarma_2001}
\bibinfo{author}{\bibfnamefont{R.~G.} \bibnamefont{{Vishwakarma}}},
  \bibinfo{journal}{Class. Quantum Gravity} \textbf{\bibinfo{volume}{18}},
  \bibinfo{pages}{1159} (\bibinfo{year}{2001}), \eprint{astro-ph/0012492}.

\bibitem[{\citenamefont{{Lima} and {Alcaniz}}(2002)}]{lima_alcaniz_2002}
\bibinfo{author}{\bibfnamefont{J.~A.~S.} \bibnamefont{{Lima}}}
  \bibnamefont{and} \bibinfo{author}{\bibfnamefont{J.~S.}
  \bibnamefont{{Alcaniz}}}, \bibinfo{journal}{Astrophys. J.}
  \textbf{\bibinfo{volume}{566}}, \bibinfo{pages}{15} (\bibinfo{year}{2002}),
  \eprint{astro-ph/0109047}.

\bibitem[{\citenamefont{{Zhu} and {Fujimoto}}(2002)}]{zhu_fujimoto_2002}
\bibinfo{author}{\bibfnamefont{Z.-H.} \bibnamefont{{Zhu}}} \bibnamefont{and}
  \bibinfo{author}{\bibfnamefont{M.-K.} \bibnamefont{{Fujimoto}}},
  \bibinfo{journal}{Astrophys. J.} \textbf{\bibinfo{volume}{581}},
  \bibinfo{pages}{1} (\bibinfo{year}{2002}), \eprint{astro-ph/0212192}.

\bibitem[{\citenamefont{{Chen} and {Ratra}}(2003)}]{Chen_Ratra_2003}
\bibinfo{author}{\bibfnamefont{G.}~\bibnamefont{{Chen}}} \bibnamefont{and}
  \bibinfo{author}{\bibfnamefont{B.}~\bibnamefont{{Ratra}}},
  \bibinfo{journal}{Astrophys. J.} \textbf{\bibinfo{volume}{582}},
  \bibinfo{pages}{586} (\bibinfo{year}{2003}), \eprint{astro-ph/0207051}.

\bibitem[{\citenamefont{{Cao} et~al.}(2017{\natexlab{c}})\citenamefont{{Cao},
  {Biesiada}, {Jackson}, {Zheng}, {Zhao}, and {Zhu}}}]{Cao_et_al2017a}
\bibinfo{author}{\bibfnamefont{S.}~\bibnamefont{{Cao}}},
  \bibinfo{author}{\bibfnamefont{M.}~\bibnamefont{{Biesiada}}},
  \bibinfo{author}{\bibfnamefont{J.}~\bibnamefont{{Jackson}}},
  \bibinfo{author}{\bibfnamefont{X.}~\bibnamefont{{Zheng}}},
  \bibinfo{author}{\bibfnamefont{Y.}~\bibnamefont{{Zhao}}}, \bibnamefont{and}
  \bibinfo{author}{\bibfnamefont{Z.-H.} \bibnamefont{{Zhu}}},
  \bibinfo{journal}{J. Cosmol. Astropart. Phys.} \textbf{\bibinfo{volume}{2}},
  \bibinfo{eid}{012} (\bibinfo{year}{2017}{\natexlab{c}}), \eprint{1609.08748}.

\bibitem[{\citenamefont{{Li} et~al.}(2017)\citenamefont{{Li}, {Cao}, {Zheng},
  {Qi}, {Biesiada}, and {Zhu}}}]{Li_et_al2017}
\bibinfo{author}{\bibfnamefont{X.}~\bibnamefont{{Li}}},
  \bibinfo{author}{\bibfnamefont{S.}~\bibnamefont{{Cao}}},
  \bibinfo{author}{\bibfnamefont{X.}~\bibnamefont{{Zheng}}},
  \bibinfo{author}{\bibfnamefont{J.}~\bibnamefont{{Qi}}},
  \bibinfo{author}{\bibfnamefont{M.}~\bibnamefont{{Biesiada}}},
  \bibnamefont{and} \bibinfo{author}{\bibfnamefont{Z.-H.} \bibnamefont{{Zhu}}},
  \bibinfo{journal}{ArXiv e-prints}  (\bibinfo{year}{2017}),
  \eprint{1708.08867}.

\bibitem[{\citenamefont{{Qi} et~al.}(2017)\citenamefont{{Qi}, {Cao},
  {Biesiada}, {Zheng}, and {Zhu}}}]{Qi_et_al2017}
\bibinfo{author}{\bibfnamefont{J.-Z.} \bibnamefont{{Qi}}},
  \bibinfo{author}{\bibfnamefont{S.}~\bibnamefont{{Cao}}},
  \bibinfo{author}{\bibfnamefont{M.}~\bibnamefont{{Biesiada}}},
  \bibinfo{author}{\bibfnamefont{X.}~\bibnamefont{{Zheng}}}, \bibnamefont{and}
  \bibinfo{author}{\bibfnamefont{Z.-H.} \bibnamefont{{Zhu}}},
  \bibinfo{journal}{Eur. Phys. J. C} \textbf{\bibinfo{volume}{77}},
  \bibinfo{eid}{502} (\bibinfo{year}{2017}), \eprint{1708.08603}.

\bibitem[{\citenamefont{{Xu} et~al.}(2018)\citenamefont{{Xu}, {Cao}, {Qi},
  {Biesiada}, {Zheng}, and {Zhu}}}]{Xu_et_al2018}
\bibinfo{author}{\bibfnamefont{T.}~\bibnamefont{{Xu}}},
  \bibinfo{author}{\bibfnamefont{S.}~\bibnamefont{{Cao}}},
  \bibinfo{author}{\bibfnamefont{J.}~\bibnamefont{{Qi}}},
  \bibinfo{author}{\bibfnamefont{M.}~\bibnamefont{{Biesiada}}},
  \bibinfo{author}{\bibfnamefont{X.}~\bibnamefont{{Zheng}}}, \bibnamefont{and}
  \bibinfo{author}{\bibfnamefont{Z.-H.} \bibnamefont{{Zhu}}},
  \bibinfo{journal}{J. Cosmol. Astropart. Phys.} \textbf{\bibinfo{volume}{6}},
  \bibinfo{eid}{042} (\bibinfo{year}{2018}), \eprint{1708.08631}.

\bibitem[{\citenamefont{{Riess} et~al.}(2018)\citenamefont{{Riess},
  {Casertano}, {Yuan}, {Macri}, {Anderson}, {MacKenty}, {Bowers}, {Clubb},
  {Filippenko}, {Jones} et~al.}}]{riess2018}
\bibinfo{author}{\bibfnamefont{A.~G.} \bibnamefont{{Riess}}},
  \bibinfo{author}{\bibfnamefont{S.}~\bibnamefont{{Casertano}}},
  \bibinfo{author}{\bibfnamefont{W.}~\bibnamefont{{Yuan}}},
  \bibinfo{author}{\bibfnamefont{L.}~\bibnamefont{{Macri}}},
  \bibinfo{author}{\bibfnamefont{J.}~\bibnamefont{{Anderson}}},
  \bibinfo{author}{\bibfnamefont{J.~W.} \bibnamefont{{MacKenty}}},
  \bibinfo{author}{\bibfnamefont{J.~B.} \bibnamefont{{Bowers}}},
  \bibinfo{author}{\bibfnamefont{K.~I.} \bibnamefont{{Clubb}}},
  \bibinfo{author}{\bibfnamefont{A.~V.} \bibnamefont{{Filippenko}}},
  \bibinfo{author}{\bibfnamefont{D.~O.} \bibnamefont{{Jones}}},
  \bibnamefont{et~al.}, \bibinfo{journal}{Astrophys. J.}
  \textbf{\bibinfo{volume}{855}}, \bibinfo{eid}{136} (\bibinfo{year}{2018}),
  \eprint{1801.01120}.

\bibitem[{\citenamefont{{DES Collaboration}}(2018)}]{DES_2018}
\bibinfo{author}{\bibnamefont{{DES Collaboration}}}, \bibinfo{journal}{Mon.
  Not. R. Astron. Soc.} \textbf{\bibinfo{volume}{480}}, \bibinfo{pages}{3879}
  (\bibinfo{year}{2018}), \eprint{1711.00403}.

\bibitem[{\citenamefont{{\lowercase{Da} Silva} and
  {Cavalcanti}}(2018)}]{daSilva}
\bibinfo{author}{\bibfnamefont{G.~P.} \bibnamefont{{\lowercase{Da} Silva}}}
  \bibnamefont{and} \bibinfo{author}{\bibfnamefont{A.~G.}
  \bibnamefont{{Cavalcanti}}}, \bibinfo{journal}{Braz. J. Phys.}
  \textbf{\bibinfo{volume}{48}}, \bibinfo{pages}{521} (\bibinfo{year}{2018}),
  \eprint{1805.06849}.

\bibitem[{\citenamefont{{Zhang}}(2018)}]{zhang_2018}
\bibinfo{author}{\bibfnamefont{J.}~\bibnamefont{{Zhang}}},
  \bibinfo{journal}{Publ. Astron. Soc. Pac.} \textbf{\bibinfo{volume}{130}},
  \bibinfo{pages}{084502} (\bibinfo{year}{2018}).

\bibitem[{\citenamefont{{Zhang} and {Huang}}(2018)}]{zhanghuang}
\bibinfo{author}{\bibfnamefont{X.}~\bibnamefont{{Zhang}}} \bibnamefont{and}
  \bibinfo{author}{\bibfnamefont{Q.-G.} \bibnamefont{{Huang}}},
  \bibinfo{journal}{ArXiv e-prints}  (\bibinfo{year}{2018}),
  \eprint{1812.01877}.

\bibitem[{\citenamefont{{Zhang} et~al.}(2019)\citenamefont{{Zhang}, {Huang},
  and {Li}}}]{zhanghuangli}
\bibinfo{author}{\bibfnamefont{X.}~\bibnamefont{{Zhang}}},
  \bibinfo{author}{\bibfnamefont{Q.-G.} \bibnamefont{{Huang}}},
  \bibnamefont{and} \bibinfo{author}{\bibfnamefont{X.-D.} \bibnamefont{{Li}}},
  \bibinfo{journal}{Mon. Not. R. Astron. Soc.} \textbf{\bibinfo{volume}{483}},
  \bibinfo{pages}{1655} (\bibinfo{year}{2019}), \eprint{1801.07403}.

\bibitem[{\citenamefont{{Moresco}
  et~al.}(2016{\natexlab{b}})\citenamefont{{Moresco}, {Jimenez}, {Verde},
  {Cimatti}, {Pozzetti}, {Maraston}, and {Thomas}}}]{morescoOK}
\bibinfo{author}{\bibfnamefont{M.}~\bibnamefont{{Moresco}}},
  \bibinfo{author}{\bibfnamefont{R.}~\bibnamefont{{Jimenez}}},
  \bibinfo{author}{\bibfnamefont{L.}~\bibnamefont{{Verde}}},
  \bibinfo{author}{\bibfnamefont{A.}~\bibnamefont{{Cimatti}}},
  \bibinfo{author}{\bibfnamefont{L.}~\bibnamefont{{Pozzetti}}},
  \bibinfo{author}{\bibfnamefont{C.}~\bibnamefont{{Maraston}}},
  \bibnamefont{and} \bibinfo{author}{\bibfnamefont{D.}~\bibnamefont{{Thomas}}},
  \bibinfo{journal}{J. Cosmol. Astropart. Phys.} \textbf{\bibinfo{volume}{12}},
  \bibinfo{eid}{039} (\bibinfo{year}{2016}{\natexlab{b}}), \eprint{1604.00183}.

\bibitem[{\citenamefont{{Zheng} et~al.}(2017)\citenamefont{{Zheng}, {Biesiada},
  {Cao}, {Qi}, and {Zhu}}}]{zheng_biesiada_cao_qi_zhu_2017}
\bibinfo{author}{\bibfnamefont{X.}~\bibnamefont{{Zheng}}},
  \bibinfo{author}{\bibfnamefont{M.}~\bibnamefont{{Biesiada}}},
  \bibinfo{author}{\bibfnamefont{S.}~\bibnamefont{{Cao}}},
  \bibinfo{author}{\bibfnamefont{J.}~\bibnamefont{{Qi}}}, \bibnamefont{and}
  \bibinfo{author}{\bibfnamefont{Z.-H.} \bibnamefont{{Zhu}}},
  \bibinfo{journal}{J. Cosmol. Astropart. Phys.} \textbf{\bibinfo{volume}{10}},
  \bibinfo{eid}{030} (\bibinfo{year}{2017}), \eprint{1705.06204}.

\bibitem[{\citenamefont{{Cao} et~al.}(2020)\citenamefont{{Cao}, {Ryan}, and
  {Ratra}}}]{Cao_Ryan_Ratra_2020}
\bibinfo{author}{\bibfnamefont{S.}~\bibnamefont{{Cao}}},
  \bibinfo{author}{\bibfnamefont{J.}~\bibnamefont{{Ryan}}}, \bibnamefont{and}
  \bibinfo{author}{\bibfnamefont{B.}~\bibnamefont{{Ratra}}},
  \bibinfo{journal}{Mon. Not. R. Astron. Soc.} \textbf{\bibinfo{volume}{497}},
  \bibinfo{pages}{3191} (\bibinfo{year}{2020}), \eprint{2005.12617}.

\bibitem[{\citenamefont{{Siegel} et~al.}(2005)\citenamefont{{Siegel},
  {Guzm{\'a}n}, {Gallego}, {Ordu{\~n}a L{\'o}pez}, and {Rodr{\'\i}guez
  Hidalgo}}}]{Siegel_2005}
\bibinfo{author}{\bibfnamefont{E.~R.} \bibnamefont{{Siegel}}},
  \bibinfo{author}{\bibfnamefont{R.}~\bibnamefont{{Guzm{\'a}n}}},
  \bibinfo{author}{\bibfnamefont{J.~P.} \bibnamefont{{Gallego}}},
  \bibinfo{author}{\bibfnamefont{M.}~\bibnamefont{{Ordu{\~n}a L{\'o}pez}}},
  \bibnamefont{and}
  \bibinfo{author}{\bibfnamefont{P.}~\bibnamefont{{Rodr{\'\i}guez Hidalgo}}},
  \bibinfo{journal}{Mon. Not. R. Astron. Soc.} \textbf{\bibinfo{volume}{356}},
  \bibinfo{pages}{1117} (\bibinfo{year}{2005}), \eprint{astro-ph/0410612}.

\bibitem[{\citenamefont{{Plionis} et~al.}(2009)\citenamefont{{Plionis},
  {Terlevich}, {Basilakos}, {Bresolin}, {Terlevich}, {Melnick}, and
  {Georgantopoulos}}}]{Plionis_2009}
\bibinfo{author}{\bibfnamefont{M.}~\bibnamefont{{Plionis}}},
  \bibinfo{author}{\bibfnamefont{R.}~\bibnamefont{{Terlevich}}},
  \bibinfo{author}{\bibfnamefont{S.}~\bibnamefont{{Basilakos}}},
  \bibinfo{author}{\bibfnamefont{F.}~\bibnamefont{{Bresolin}}},
  \bibinfo{author}{\bibfnamefont{E.}~\bibnamefont{{Terlevich}}},
  \bibinfo{author}{\bibfnamefont{J.}~\bibnamefont{{Melnick}}},
  \bibnamefont{and}
  \bibinfo{author}{\bibfnamefont{I.}~\bibnamefont{{Georgantopoulos}}}, in
  \emph{\bibinfo{booktitle}{Journal of Physics Conference Series}}
  (\bibinfo{year}{2009}), vol. \bibinfo{volume}{189} of
  \emph{\bibinfo{series}{J. Phys. Conf. Ser.}}, p. \bibinfo{pages}{012032},
  \eprint{0903.0131}.

\bibitem[{\citenamefont{{Plionis} et~al.}(2010)\citenamefont{{Plionis},
  {Terlevich}, {Basilakos}, {Bresolin}, {Terlevich}, {Melnick}, and
  {Chavez}}}]{Plionis_2010}
\bibinfo{author}{\bibfnamefont{M.}~\bibnamefont{{Plionis}}},
  \bibinfo{author}{\bibfnamefont{R.}~\bibnamefont{{Terlevich}}},
  \bibinfo{author}{\bibfnamefont{S.}~\bibnamefont{{Basilakos}}},
  \bibinfo{author}{\bibfnamefont{F.}~\bibnamefont{{Bresolin}}},
  \bibinfo{author}{\bibfnamefont{E.}~\bibnamefont{{Terlevich}}},
  \bibinfo{author}{\bibfnamefont{J.}~\bibnamefont{{Melnick}}},
  \bibnamefont{and} \bibinfo{author}{\bibfnamefont{R.}~\bibnamefont{{Chavez}}},
  in \emph{\bibinfo{booktitle}{American Institute of Physics Conference
  Series}}, edited by \bibinfo{editor}{\bibfnamefont{J.-M.}
  \bibnamefont{{Alimi}}} \bibnamefont{and}
  \bibinfo{editor}{\bibfnamefont{A.}~\bibnamefont{{Fu{\"o}zfa}}}
  (\bibinfo{year}{2010}), vol. \bibinfo{volume}{1241} of
  \emph{\bibinfo{series}{AIP Conf. Proc.}}, pp. \bibinfo{pages}{267--276},
  \eprint{0911.3198}.

\bibitem[{\citenamefont{{Plionis} et~al.}(2011)\citenamefont{{Plionis},
  {Terlevich}, {Basilakos}, {Bresolin}, {Terlevich}, {Melnick}, and
  {Chavez}}}]{Plionis_2011}
\bibinfo{author}{\bibfnamefont{M.}~\bibnamefont{{Plionis}}},
  \bibinfo{author}{\bibfnamefont{R.}~\bibnamefont{{Terlevich}}},
  \bibinfo{author}{\bibfnamefont{S.}~\bibnamefont{{Basilakos}}},
  \bibinfo{author}{\bibfnamefont{F.}~\bibnamefont{{Bresolin}}},
  \bibinfo{author}{\bibfnamefont{E.}~\bibnamefont{{Terlevich}}},
  \bibinfo{author}{\bibfnamefont{J.}~\bibnamefont{{Melnick}}},
  \bibnamefont{and} \bibinfo{author}{\bibfnamefont{R.}~\bibnamefont{{Chavez}}},
  \bibinfo{journal}{Mon. Not. R. Astron. Soc.} \textbf{\bibinfo{volume}{416}},
  \bibinfo{pages}{2981} (\bibinfo{year}{2011}), \eprint{1106.4558}.

\bibitem[{\citenamefont{{Mania} and {Ratra}}(2012)}]{Mania_2012}
\bibinfo{author}{\bibfnamefont{D.}~\bibnamefont{{Mania}}} \bibnamefont{and}
  \bibinfo{author}{\bibfnamefont{B.}~\bibnamefont{{Ratra}}},
  \bibinfo{journal}{Phys. Lett. B} \textbf{\bibinfo{volume}{715}},
  \bibinfo{pages}{9} (\bibinfo{year}{2012}), \eprint{1110.5626}.

\bibitem[{\citenamefont{{Ch{\'a}vez} et~al.}(2016)\citenamefont{{Ch{\'a}vez},
  {Plionis}, {Basilakos}, {Terlevich}, {Terlevich}, {Melnick}, {Bresolin}, and
  {Gonz{\'a}lez-Mor{\'a}n}}}]{Chavez_2016}
\bibinfo{author}{\bibfnamefont{R.}~\bibnamefont{{Ch{\'a}vez}}},
  \bibinfo{author}{\bibfnamefont{M.}~\bibnamefont{{Plionis}}},
  \bibinfo{author}{\bibfnamefont{S.}~\bibnamefont{{Basilakos}}},
  \bibinfo{author}{\bibfnamefont{R.}~\bibnamefont{{Terlevich}}},
  \bibinfo{author}{\bibfnamefont{E.}~\bibnamefont{{Terlevich}}},
  \bibinfo{author}{\bibfnamefont{J.}~\bibnamefont{{Melnick}}},
  \bibinfo{author}{\bibfnamefont{F.}~\bibnamefont{{Bresolin}}},
  \bibnamefont{and} \bibinfo{author}{\bibfnamefont{A.~L.}
  \bibnamefont{{Gonz{\'a}lez-Mor{\'a}n}}}, \bibinfo{journal}{Mon. Not. R.
  Astron. Soc.} \textbf{\bibinfo{volume}{462}}, \bibinfo{pages}{2431}
  (\bibinfo{year}{2016}), \eprint{1607.06458}.

\bibitem[{\citenamefont{{Wei} et~al.}(2016)\citenamefont{{Wei}, {Wu}, and
  {Melia}}}]{Wei_2016}
\bibinfo{author}{\bibfnamefont{J.-J.} \bibnamefont{{Wei}}},
  \bibinfo{author}{\bibfnamefont{X.-F.} \bibnamefont{{Wu}}}, \bibnamefont{and}
  \bibinfo{author}{\bibfnamefont{F.}~\bibnamefont{{Melia}}},
  \bibinfo{journal}{Mon. Not. R. Astron. Soc.} \textbf{\bibinfo{volume}{463}},
  \bibinfo{pages}{1144} (\bibinfo{year}{2016}), \eprint{1608.02070}.

\bibitem[{\citenamefont{{Yennapureddy} and {Melia}}(2017)}]{Yennapureddy_2017}
\bibinfo{author}{\bibfnamefont{M.~K.} \bibnamefont{{Yennapureddy}}}
  \bibnamefont{and} \bibinfo{author}{\bibfnamefont{F.}~\bibnamefont{{Melia}}},
  \bibinfo{journal}{J. Cosmol. Astropart. Phys.}
  \textbf{\bibinfo{volume}{2017}}, \bibinfo{eid}{029} (\bibinfo{year}{2017}),
  \eprint{1711.03454}.

\bibitem[{\citenamefont{{Gonz{\'a}lez-Mor{\'a}n}
  et~al.}(2019)\citenamefont{{Gonz{\'a}lez-Mor{\'a}n}, {Ch{\'a}vez},
  {Terlevich}, {Terlevich}, {Bresolin}, {Fern{\'a}ndez-Arenas}, {Plionis},
  {Basilakos}, {Melnick}, and {Telles}}}]{G-M_2019}
\bibinfo{author}{\bibfnamefont{A.~L.} \bibnamefont{{Gonz{\'a}lez-Mor{\'a}n}}},
  \bibinfo{author}{\bibfnamefont{R.}~\bibnamefont{{Ch{\'a}vez}}},
  \bibinfo{author}{\bibfnamefont{R.}~\bibnamefont{{Terlevich}}},
  \bibinfo{author}{\bibfnamefont{E.}~\bibnamefont{{Terlevich}}},
  \bibinfo{author}{\bibfnamefont{F.}~\bibnamefont{{Bresolin}}},
  \bibinfo{author}{\bibfnamefont{D.}~\bibnamefont{{Fern{\'a}ndez-Arenas}}},
  \bibinfo{author}{\bibfnamefont{M.}~\bibnamefont{{Plionis}}},
  \bibinfo{author}{\bibfnamefont{S.}~\bibnamefont{{Basilakos}}},
  \bibinfo{author}{\bibfnamefont{J.}~\bibnamefont{{Melnick}}},
  \bibnamefont{and} \bibinfo{author}{\bibfnamefont{E.}~\bibnamefont{{Telles}}},
  \bibinfo{journal}{Mon. Not. R. Astron. Soc.} \textbf{\bibinfo{volume}{487}},
  \bibinfo{pages}{4669} (\bibinfo{year}{2019}), \eprint{1906.02195}.

\bibitem[{\citenamefont{{Wan} et~al.}(2019)\citenamefont{{Wan}, {Cao}, {Melia},
  and {Zhang}}}]{Wan_2019}
\bibinfo{author}{\bibfnamefont{H.-Y.} \bibnamefont{{Wan}}},
  \bibinfo{author}{\bibfnamefont{S.-L.} \bibnamefont{{Cao}}},
  \bibinfo{author}{\bibfnamefont{F.}~\bibnamefont{{Melia}}}, \bibnamefont{and}
  \bibinfo{author}{\bibfnamefont{T.-J.} \bibnamefont{{Zhang}}},
  \bibinfo{journal}{Phys. Dark Universe} \textbf{\bibinfo{volume}{26}},
  \bibinfo{eid}{100405} (\bibinfo{year}{2019}), \eprint{1910.14024}.

\bibitem[{\citenamefont{{Wu} et~al.}(2020)\citenamefont{{Wu}, {Cao}, {Zhang},
  {Liu}, {Liu}, {Geng}, and {Lian}}}]{Wu_2020}
\bibinfo{author}{\bibfnamefont{Y.}~\bibnamefont{{Wu}}},
  \bibinfo{author}{\bibfnamefont{S.}~\bibnamefont{{Cao}}},
  \bibinfo{author}{\bibfnamefont{J.}~\bibnamefont{{Zhang}}},
  \bibinfo{author}{\bibfnamefont{T.}~\bibnamefont{{Liu}}},
  \bibinfo{author}{\bibfnamefont{Y.}~\bibnamefont{{Liu}}},
  \bibinfo{author}{\bibfnamefont{S.}~\bibnamefont{{Geng}}}, \bibnamefont{and}
  \bibinfo{author}{\bibfnamefont{Y.}~\bibnamefont{{Lian}}},
  \bibinfo{journal}{Astrophys. J.} \textbf{\bibinfo{volume}{888}},
  \bibinfo{eid}{113} (\bibinfo{year}{2020}), \eprint{1911.10959}.

\bibitem[{\citenamefont{{Melnick} et~al.}(2000)\citenamefont{{Melnick},
  {Terlevich}, and {Terlevich}}}]{Melnick_2000}
\bibinfo{author}{\bibfnamefont{J.}~\bibnamefont{{Melnick}}},
  \bibinfo{author}{\bibfnamefont{R.}~\bibnamefont{{Terlevich}}},
  \bibnamefont{and}
  \bibinfo{author}{\bibfnamefont{E.}~\bibnamefont{{Terlevich}}},
  \bibinfo{journal}{Mon. Not. R. Astron. Soc.} \textbf{\bibinfo{volume}{311}},
  \bibinfo{pages}{629} (\bibinfo{year}{2000}), \eprint{astro-ph/9908346}.

\bibitem[{\citenamefont{{Ch{\'a}vez} et~al.}(2012)\citenamefont{{Ch{\'a}vez},
  {Terlevich}, {Terlevich}, {Plionis}, {Bresolin}, {Basilakos}, and
  {Melnick}}}]{Chavez_2012}
\bibinfo{author}{\bibfnamefont{R.}~\bibnamefont{{Ch{\'a}vez}}},
  \bibinfo{author}{\bibfnamefont{E.}~\bibnamefont{{Terlevich}}},
  \bibinfo{author}{\bibfnamefont{R.}~\bibnamefont{{Terlevich}}},
  \bibinfo{author}{\bibfnamefont{M.}~\bibnamefont{{Plionis}}},
  \bibinfo{author}{\bibfnamefont{F.}~\bibnamefont{{Bresolin}}},
  \bibinfo{author}{\bibfnamefont{S.}~\bibnamefont{{Basilakos}}},
  \bibnamefont{and}
  \bibinfo{author}{\bibfnamefont{J.}~\bibnamefont{{Melnick}}},
  \bibinfo{journal}{Mon. Not. R. Astron. Soc.} \textbf{\bibinfo{volume}{425}},
  \bibinfo{pages}{L56} (\bibinfo{year}{2012}), \eprint{1203.6222}.

\bibitem[{\citenamefont{{Ch{\'a}vez} et~al.}(2014)\citenamefont{{Ch{\'a}vez},
  {Terlevich}, {Terlevich}, {Bresolin}, {Melnick}, {Plionis}, and
  {Basilakos}}}]{Chavez_2014}
\bibinfo{author}{\bibfnamefont{R.}~\bibnamefont{{Ch{\'a}vez}}},
  \bibinfo{author}{\bibfnamefont{R.}~\bibnamefont{{Terlevich}}},
  \bibinfo{author}{\bibfnamefont{E.}~\bibnamefont{{Terlevich}}},
  \bibinfo{author}{\bibfnamefont{F.}~\bibnamefont{{Bresolin}}},
  \bibinfo{author}{\bibfnamefont{J.}~\bibnamefont{{Melnick}}},
  \bibinfo{author}{\bibfnamefont{M.}~\bibnamefont{{Plionis}}},
  \bibnamefont{and}
  \bibinfo{author}{\bibfnamefont{S.}~\bibnamefont{{Basilakos}}},
  \bibinfo{journal}{Mon. Not. R. Astron. Soc.} \textbf{\bibinfo{volume}{442}},
  \bibinfo{pages}{3565} (\bibinfo{year}{2014}), \eprint{1405.4010}.

\bibitem[{\citenamefont{{Terlevich} et~al.}(2015)\citenamefont{{Terlevich},
  {Terlevich}, {Melnick}, {Ch{\'a}vez}, {Plionis}, {Bresolin}, and
  {Basilakos}}}]{Terlevich_2015}
\bibinfo{author}{\bibfnamefont{R.}~\bibnamefont{{Terlevich}}},
  \bibinfo{author}{\bibfnamefont{E.}~\bibnamefont{{Terlevich}}},
  \bibinfo{author}{\bibfnamefont{J.}~\bibnamefont{{Melnick}}},
  \bibinfo{author}{\bibfnamefont{R.}~\bibnamefont{{Ch{\'a}vez}}},
  \bibinfo{author}{\bibfnamefont{M.}~\bibnamefont{{Plionis}}},
  \bibinfo{author}{\bibfnamefont{F.}~\bibnamefont{{Bresolin}}},
  \bibnamefont{and}
  \bibinfo{author}{\bibfnamefont{S.}~\bibnamefont{{Basilakos}}},
  \bibinfo{journal}{Mon. Not. R. Astron. Soc.} \textbf{\bibinfo{volume}{451}},
  \bibinfo{pages}{3001} (\bibinfo{year}{2015}), \eprint{1505.04376}.

\bibitem[{\citenamefont{{Risaliti} and {Lusso}}(2015)}]{RisalitiandLusso_2015}
\bibinfo{author}{\bibfnamefont{G.}~\bibnamefont{{Risaliti}}} \bibnamefont{and}
  \bibinfo{author}{\bibfnamefont{E.}~\bibnamefont{{Lusso}}},
  \bibinfo{journal}{Astrophys. J.} \textbf{\bibinfo{volume}{815}},
  \bibinfo{eid}{33} (\bibinfo{year}{2015}), \eprint{1505.07118}.

\bibitem[{\citenamefont{{Risaliti} and {Lusso}}(2019)}]{RisalitiandLusso_2019}
\bibinfo{author}{\bibfnamefont{G.}~\bibnamefont{{Risaliti}}} \bibnamefont{and}
  \bibinfo{author}{\bibfnamefont{E.}~\bibnamefont{{Lusso}}},
  \bibinfo{journal}{Nat. Astron.} \textbf{\bibinfo{volume}{3}},
  \bibinfo{pages}{272} (\bibinfo{year}{2019}), \eprint{1811.02590}.

\bibitem[{\citenamefont{{Yang} et~al.}(2019)\citenamefont{{Yang}, {Banerjee},
  and {Colg{\'a}in}}}]{Yang_2019}
\bibinfo{author}{\bibfnamefont{T.}~\bibnamefont{{Yang}}},
  \bibinfo{author}{\bibfnamefont{A.}~\bibnamefont{{Banerjee}}},
  \bibnamefont{and} \bibinfo{author}{\bibfnamefont{E.~{\'O}.}
  \bibnamefont{{Colg{\'a}in}}}, \bibinfo{journal}{ArXiv e-prints}
  (\bibinfo{year}{2019}), \eprint{1911.01681}.

\bibitem[{\citenamefont{{Khadka} and
  {Ratra}}(2020{\natexlab{a}})}]{KhadkaandRatra_2020}
\bibinfo{author}{\bibfnamefont{N.}~\bibnamefont{{Khadka}}} \bibnamefont{and}
  \bibinfo{author}{\bibfnamefont{B.}~\bibnamefont{{Ratra}}},
  \bibinfo{journal}{ArXiv e-prints}  (\bibinfo{year}{2020}{\natexlab{a}}),
  \eprint{2004.09979}.

\bibitem[{\citenamefont{{Khadka} and
  {Ratra}}(2020{\natexlab{b}})}]{Khadka_2020a}
\bibinfo{author}{\bibfnamefont{N.}~\bibnamefont{{Khadka}}} \bibnamefont{and}
  \bibinfo{author}{\bibfnamefont{B.}~\bibnamefont{{Ratra}}},
  \bibinfo{journal}{Mon. Not. R. Astron. Soc.} \textbf{\bibinfo{volume}{492}},
  \bibinfo{pages}{4456} (\bibinfo{year}{2020}{\natexlab{b}}),
  \eprint{1909.01400}.

\bibitem[{\citenamefont{{Zheng} et~al.}(2020)\citenamefont{{Zheng}, {Liao},
  {Biesiada}, {Cao}, {Liu}, and {Zhu}}}]{Zheng_2020}
\bibinfo{author}{\bibfnamefont{X.}~\bibnamefont{{Zheng}}},
  \bibinfo{author}{\bibfnamefont{K.}~\bibnamefont{{Liao}}},
  \bibinfo{author}{\bibfnamefont{M.}~\bibnamefont{{Biesiada}}},
  \bibinfo{author}{\bibfnamefont{S.}~\bibnamefont{{Cao}}},
  \bibinfo{author}{\bibfnamefont{T.-H.} \bibnamefont{{Liu}}}, \bibnamefont{and}
  \bibinfo{author}{\bibfnamefont{Z.-H.} \bibnamefont{{Zhu}}},
  \bibinfo{journal}{Astrophys. J.} \textbf{\bibinfo{volume}{892}},
  \bibinfo{eid}{103} (\bibinfo{year}{2020}), \eprint{2002.09909}.

\bibitem[{\citenamefont{{Lamb} and {Reichart}}(2000)}]{Lamb_2000}
\bibinfo{author}{\bibfnamefont{D.~Q.} \bibnamefont{{Lamb}}} \bibnamefont{and}
  \bibinfo{author}{\bibfnamefont{D.~E.} \bibnamefont{{Reichart}}},
  \bibinfo{journal}{Astrophys. J.} \textbf{\bibinfo{volume}{536}},
  \bibinfo{pages}{1} (\bibinfo{year}{2000}), \eprint{astro-ph/9909002}.

\bibitem[{\citenamefont{{Demianski} et~al.}(2019)\citenamefont{{Demianski},
  {Piedipalumbo}, {Sawant}, and {Amati}}}]{Demianski_2019}
\bibinfo{author}{\bibfnamefont{M.}~\bibnamefont{{Demianski}}},
  \bibinfo{author}{\bibfnamefont{E.}~\bibnamefont{{Piedipalumbo}}},
  \bibinfo{author}{\bibfnamefont{D.}~\bibnamefont{{Sawant}}}, \bibnamefont{and}
  \bibinfo{author}{\bibfnamefont{L.}~\bibnamefont{{Amati}}},
  \bibinfo{journal}{ArXiv e-prints}  (\bibinfo{year}{2019}),
  \eprint{1911.08228}.

\bibitem[{\citenamefont{{Riess} et~al.}(2019)\citenamefont{{Riess},
  {Casertano}, {Yuan}, {Macri}, and {Scolnic}}}]{riess_etal_2019}
\bibinfo{author}{\bibfnamefont{A.~G.} \bibnamefont{{Riess}}},
  \bibinfo{author}{\bibfnamefont{S.}~\bibnamefont{{Casertano}}},
  \bibinfo{author}{\bibfnamefont{W.}~\bibnamefont{{Yuan}}},
  \bibinfo{author}{\bibfnamefont{L.~M.} \bibnamefont{{Macri}}},
  \bibnamefont{and}
  \bibinfo{author}{\bibfnamefont{D.}~\bibnamefont{{Scolnic}}},
  \bibinfo{journal}{Astrophys. J.} \textbf{\bibinfo{volume}{876}},
  \bibinfo{eid}{85} (\bibinfo{year}{2019}), \eprint{1903.07603}.

\bibitem[{\citenamefont{{Carter} et~al.}(2018)\citenamefont{{Carter},
  {Beutler}, {Percival}, {Blake}, {Koda}, and {Ross}}}]{Carter_2018}
\bibinfo{author}{\bibfnamefont{P.}~\bibnamefont{{Carter}}},
  \bibinfo{author}{\bibfnamefont{F.}~\bibnamefont{{Beutler}}},
  \bibinfo{author}{\bibfnamefont{W.~J.} \bibnamefont{{Percival}}},
  \bibinfo{author}{\bibfnamefont{C.}~\bibnamefont{{Blake}}},
  \bibinfo{author}{\bibfnamefont{J.}~\bibnamefont{{Koda}}}, \bibnamefont{and}
  \bibinfo{author}{\bibfnamefont{A.~J.} \bibnamefont{{Ross}}},
  \bibinfo{journal}{Mon. Not. R. Astron. Soc.} \textbf{\bibinfo{volume}{481}}
  (\bibinfo{year}{2018}), \eprint{1803.01746}.

\bibitem[{\citenamefont{{DES Collaboration}}(2019{\natexlab{b}})}]{DES_2019b}
\bibinfo{author}{\bibnamefont{{DES Collaboration}}}, \bibinfo{journal}{Mon.
  Not. R. Astron. Soc.} \textbf{\bibinfo{volume}{483}}, \bibinfo{pages}{4866}
  (\bibinfo{year}{2019}{\natexlab{b}}), \eprint{1712.06209}.

\bibitem[{\citenamefont{{\MakeLowercase{D}e Sainte Agathe}
  et~al.}(2019)\citenamefont{{\MakeLowercase{D}e Sainte Agathe}, {Balland}, {du
  Mas des Bourboux}, {Busca}, {Blomqvist}, {Guy}, {Rich}, {Font-Ribera},
  {Pieri}, {Bautista} et~al.}}]{Agathe}
\bibinfo{author}{\bibfnamefont{V.}~\bibnamefont{{\MakeLowercase{D}e Sainte
  Agathe}}}, \bibinfo{author}{\bibfnamefont{C.}~\bibnamefont{{Balland}}},
  \bibinfo{author}{\bibfnamefont{H.}~\bibnamefont{{du Mas des Bourboux}}},
  \bibinfo{author}{\bibfnamefont{N.~G.} \bibnamefont{{Busca}}},
  \bibinfo{author}{\bibfnamefont{M.}~\bibnamefont{{Blomqvist}}},
  \bibinfo{author}{\bibfnamefont{J.}~\bibnamefont{{Guy}}},
  \bibinfo{author}{\bibfnamefont{J.}~\bibnamefont{{Rich}}},
  \bibinfo{author}{\bibfnamefont{A.}~\bibnamefont{{Font-Ribera}}},
  \bibinfo{author}{\bibfnamefont{M.~M.} \bibnamefont{{Pieri}}},
  \bibinfo{author}{\bibfnamefont{J.~E.} \bibnamefont{{Bautista}}},
  \bibnamefont{et~al.}, \bibinfo{journal}{Astron. Astrophys.}
  \textbf{\bibinfo{volume}{629}}, \bibinfo{eid}{A85} (\bibinfo{year}{2019}),
  \eprint{1904.03400}.

\bibitem[{\citenamefont{{Erb} et~al.}(2006)\citenamefont{{Erb}, {Steidel},
  {Shapley}, {Pettini}, {Reddy}, and {Adelberger}}}]{Erb_2006}
\bibinfo{author}{\bibfnamefont{D.~K.} \bibnamefont{{Erb}}},
  \bibinfo{author}{\bibfnamefont{C.~C.} \bibnamefont{{Steidel}}},
  \bibinfo{author}{\bibfnamefont{A.~E.} \bibnamefont{{Shapley}}},
  \bibinfo{author}{\bibfnamefont{M.}~\bibnamefont{{Pettini}}},
  \bibinfo{author}{\bibfnamefont{N.~A.} \bibnamefont{{Reddy}}},
  \bibnamefont{and} \bibinfo{author}{\bibfnamefont{K.~L.}
  \bibnamefont{{Adelberger}}}, \bibinfo{journal}{Astrophys. J.}
  \textbf{\bibinfo{volume}{646}}, \bibinfo{pages}{107} (\bibinfo{year}{2006}),
  \eprint{astro-ph/0604041}.

\bibitem[{\citenamefont{{Masters} et~al.}(2014)\citenamefont{{Masters},
  {McCarthy}, {Siana}, {Malkan}, {Mobasher}, {Atek}, {Henry}, {Martin},
  {Rafelski}, {Hathi} et~al.}}]{Masters_2014}
\bibinfo{author}{\bibfnamefont{D.}~\bibnamefont{{Masters}}},
  \bibinfo{author}{\bibfnamefont{P.}~\bibnamefont{{McCarthy}}},
  \bibinfo{author}{\bibfnamefont{B.}~\bibnamefont{{Siana}}},
  \bibinfo{author}{\bibfnamefont{M.}~\bibnamefont{{Malkan}}},
  \bibinfo{author}{\bibfnamefont{B.}~\bibnamefont{{Mobasher}}},
  \bibinfo{author}{\bibfnamefont{H.}~\bibnamefont{{Atek}}},
  \bibinfo{author}{\bibfnamefont{A.}~\bibnamefont{{Henry}}},
  \bibinfo{author}{\bibfnamefont{C.~L.} \bibnamefont{{Martin}}},
  \bibinfo{author}{\bibfnamefont{M.}~\bibnamefont{{Rafelski}}},
  \bibinfo{author}{\bibfnamefont{N.~P.} \bibnamefont{{Hathi}}},
  \bibnamefont{et~al.}, \bibinfo{journal}{Astrophys. J.}
  \textbf{\bibinfo{volume}{785}}, \bibinfo{eid}{153} (\bibinfo{year}{2014}),
  \eprint{1402.0510}.

\bibitem[{\citenamefont{{Maseda} et~al.}(2014)\citenamefont{{Maseda}, {van der
  Wel}, {Rix}, {da Cunha}, {Pacifici}, {Momcheva}, {Brammer}, {Meidt}, {Franx},
  {van Dokkum} et~al.}}]{Maseda_2014}
\bibinfo{author}{\bibfnamefont{M.~V.} \bibnamefont{{Maseda}}},
  \bibinfo{author}{\bibfnamefont{A.}~\bibnamefont{{van der Wel}}},
  \bibinfo{author}{\bibfnamefont{H.-W.} \bibnamefont{{Rix}}},
  \bibinfo{author}{\bibfnamefont{E.}~\bibnamefont{{da Cunha}}},
  \bibinfo{author}{\bibfnamefont{C.}~\bibnamefont{{Pacifici}}},
  \bibinfo{author}{\bibfnamefont{I.}~\bibnamefont{{Momcheva}}},
  \bibinfo{author}{\bibfnamefont{G.~B.} \bibnamefont{{Brammer}}},
  \bibinfo{author}{\bibfnamefont{S.~E.} \bibnamefont{{Meidt}}},
  \bibinfo{author}{\bibfnamefont{M.}~\bibnamefont{{Franx}}},
  \bibinfo{author}{\bibfnamefont{P.}~\bibnamefont{{van Dokkum}}},
  \bibnamefont{et~al.}, \bibinfo{journal}{Astrophys. J.}
  \textbf{\bibinfo{volume}{791}}, \bibinfo{eid}{17} (\bibinfo{year}{2014}),
  \eprint{1406.3351}.

\bibitem[{\citenamefont{{Aubourg} et~al.}(2015)\citenamefont{{Aubourg},
  {Bailey}, {Bautista}, {Beutler}, {Bhardwaj}, {Bizyaev}, {Blanton},
  {Blomqvist}, {Bolton}, {Bovy} et~al.}}]{PhysRevD.92.123516}
\bibinfo{author}{\bibfnamefont{{\'E}.}~\bibnamefont{{Aubourg}}},
  \bibinfo{author}{\bibfnamefont{S.}~\bibnamefont{{Bailey}}},
  \bibinfo{author}{\bibfnamefont{J.~E.} \bibnamefont{{Bautista}}},
  \bibinfo{author}{\bibfnamefont{F.}~\bibnamefont{{Beutler}}},
  \bibinfo{author}{\bibfnamefont{V.}~\bibnamefont{{Bhardwaj}}},
  \bibinfo{author}{\bibfnamefont{D.}~\bibnamefont{{Bizyaev}}},
  \bibinfo{author}{\bibfnamefont{M.}~\bibnamefont{{Blanton}}},
  \bibinfo{author}{\bibfnamefont{M.}~\bibnamefont{{Blomqvist}}},
  \bibinfo{author}{\bibfnamefont{A.~S.} \bibnamefont{{Bolton}}},
  \bibinfo{author}{\bibfnamefont{J.}~\bibnamefont{{Bovy}}},
  \bibnamefont{et~al.}, \bibinfo{journal}{Phys. Rev. D}
  \textbf{\bibinfo{volume}{92}}, \bibinfo{eid}{123516} (\bibinfo{year}{2015}),
  \eprint{1411.1074}.

\bibitem[{\citenamefont{{Gordon} et~al.}(2003)\citenamefont{{Gordon},
  {Clayton}, {Misselt}, {Landolt}, and {Wolff}}}]{Gordon_2003}
\bibinfo{author}{\bibfnamefont{K.~D.} \bibnamefont{{Gordon}}},
  \bibinfo{author}{\bibfnamefont{G.~C.} \bibnamefont{{Clayton}}},
  \bibinfo{author}{\bibfnamefont{K.~A.} \bibnamefont{{Misselt}}},
  \bibinfo{author}{\bibfnamefont{A.~U.} \bibnamefont{{Landolt}}},
  \bibnamefont{and} \bibinfo{author}{\bibfnamefont{M.~J.}
  \bibnamefont{{Wolff}}}, \bibinfo{journal}{Astrophys. J.}
  \textbf{\bibinfo{volume}{594}}, \bibinfo{pages}{279} (\bibinfo{year}{2003}),
  \eprint{astro-ph/0305257}.

\bibitem[{\citenamefont{{Riess}}(2019)}]{riess_2019}
\bibinfo{author}{\bibfnamefont{A.~G.} \bibnamefont{{Riess}}},
  \bibinfo{journal}{Nat. Rev. Phys.} \textbf{\bibinfo{volume}{2}},
  \bibinfo{pages}{10} (\bibinfo{year}{2019}), \eprint{2001.03624}.

\bibitem[{\citenamefont{{Moresco} et~al.}(2018)\citenamefont{{Moresco},
  {Jimenez}, {Verde}, {Pozzetti}, {Cimatti}, and {Citro}}}]{Setting_stage_1}
\bibinfo{author}{\bibfnamefont{M.}~\bibnamefont{{Moresco}}},
  \bibinfo{author}{\bibfnamefont{R.}~\bibnamefont{{Jimenez}}},
  \bibinfo{author}{\bibfnamefont{L.}~\bibnamefont{{Verde}}},
  \bibinfo{author}{\bibfnamefont{L.}~\bibnamefont{{Pozzetti}}},
  \bibinfo{author}{\bibfnamefont{A.}~\bibnamefont{{Cimatti}}},
  \bibnamefont{and} \bibinfo{author}{\bibfnamefont{A.}~\bibnamefont{{Citro}}},
  \bibinfo{journal}{Astrophys. J.} \textbf{\bibinfo{volume}{868}},
  \bibinfo{eid}{84} (\bibinfo{year}{2018}), \eprint{1804.05864}.

\bibitem[{\citenamefont{{Moresco} et~al.}(2020)\citenamefont{{Moresco},
  {Jimenez}, {Verde}, {Cimatti}, and {Pozzetti}}}]{Setting_stage_2}
\bibinfo{author}{\bibfnamefont{M.}~\bibnamefont{{Moresco}}},
  \bibinfo{author}{\bibfnamefont{R.}~\bibnamefont{{Jimenez}}},
  \bibinfo{author}{\bibfnamefont{L.}~\bibnamefont{{Verde}}},
  \bibinfo{author}{\bibfnamefont{A.}~\bibnamefont{{Cimatti}}},
  \bibnamefont{and}
  \bibinfo{author}{\bibfnamefont{L.}~\bibnamefont{{Pozzetti}}},
  \bibinfo{journal}{arXiv e-prints}  (\bibinfo{year}{2020}),
  \eprint{2003.07362}.

\bibitem[{\citenamefont{{Foreman-Mackey}
  et~al.}(2013)\citenamefont{{Foreman-Mackey}, {Hogg}, {Lang}, and
  {Goodman}}}]{Foreman-Mackey_Hogg_Lang_Goodman_2013}
\bibinfo{author}{\bibfnamefont{D.}~\bibnamefont{{Foreman-Mackey}}},
  \bibinfo{author}{\bibfnamefont{D.~W.} \bibnamefont{{Hogg}}},
  \bibinfo{author}{\bibfnamefont{D.}~\bibnamefont{{Lang}}}, \bibnamefont{and}
  \bibinfo{author}{\bibfnamefont{J.}~\bibnamefont{{Goodman}}},
  \bibinfo{journal}{Publ. Astron. Soc. Pac.} \textbf{\bibinfo{volume}{125}},
  \bibinfo{pages}{306} (\bibinfo{year}{2013}), \eprint{1202.3665}.

\bibitem[{\citenamefont{Cahill}(2013)}]{Cahill_2013}
\bibinfo{author}{\bibfnamefont{K.}~\bibnamefont{Cahill}},
  \emph{\bibinfo{title}{{Physical Mathematics}}} (\bibinfo{publisher}{Cambridge
  University Press}, \bibinfo{address}{New York, NY}, \bibinfo{year}{2013}).

\bibitem[{\citenamefont{{von Toussaint}}(2011)}]{von_Toussaint_2011}
\bibinfo{author}{\bibfnamefont{U.}~\bibnamefont{{von Toussaint}}},
  \bibinfo{journal}{Rev. Mod. Phys.} \textbf{\bibinfo{volume}{83}},
  \bibinfo{pages}{943} (\bibinfo{year}{2011}),
  \urlprefix\url{https://link.aps.org/doi/10.1103/RevModPhys.83.943}.

\bibitem[{\citenamefont{{Goodman} and {Weare}}(2010)}]{Goodman_Weare_2010}
\bibinfo{author}{\bibfnamefont{J.}~\bibnamefont{{Goodman}}} \bibnamefont{and}
  \bibinfo{author}{\bibfnamefont{J.}~\bibnamefont{{Weare}}},
  \bibinfo{journal}{Comm. App. Math. Comp. Sci.} \textbf{\bibinfo{volume}{5}},
  \bibinfo{pages}{65} (\bibinfo{year}{2010}).

\bibitem[{\citenamefont{{Melnick} et~al.}(2017)\citenamefont{{Melnick},
  {Telles}, {Bordalo}, {Ch{\'a}vez}, {Fern{\'a}ndez-Arenas}, {Terlevich},
  {Terlevich}, {Bresolin}, {Plionis}, and {Basilakos}}}]{Melnick_2017}
\bibinfo{author}{\bibfnamefont{J.}~\bibnamefont{{Melnick}}},
  \bibinfo{author}{\bibfnamefont{E.}~\bibnamefont{{Telles}}},
  \bibinfo{author}{\bibfnamefont{V.}~\bibnamefont{{Bordalo}}},
  \bibinfo{author}{\bibfnamefont{R.}~\bibnamefont{{Ch{\'a}vez}}},
  \bibinfo{author}{\bibfnamefont{D.}~\bibnamefont{{Fern{\'a}ndez-Arenas}}},
  \bibinfo{author}{\bibfnamefont{E.}~\bibnamefont{{Terlevich}}},
  \bibinfo{author}{\bibfnamefont{R.}~\bibnamefont{{Terlevich}}},
  \bibinfo{author}{\bibfnamefont{F.}~\bibnamefont{{Bresolin}}},
  \bibinfo{author}{\bibfnamefont{M.}~\bibnamefont{{Plionis}}},
  \bibnamefont{and}
  \bibinfo{author}{\bibfnamefont{S.}~\bibnamefont{{Basilakos}}},
  \bibinfo{journal}{Astron. Astrophys.} \textbf{\bibinfo{volume}{599}},
  \bibinfo{eid}{A76} (\bibinfo{year}{2017}), \eprint{1612.01974}.

\bibitem[{\citenamefont{{Lewis}}(2019)}]{Lewis_2019}
\bibinfo{author}{\bibfnamefont{A.}~\bibnamefont{{Lewis}}},
  \bibinfo{journal}{ArXiv e-prints}  (\bibinfo{year}{2019}),
  \eprint{1910.13970}.

\bibitem[{\citenamefont{{Freedman} et~al.}(2019)\citenamefont{{Freedman},
  {Madore}, {Hatt}, {Hoyt}, {Jang}, {Beaton}, {Burns}, {Lee}, {Monson},
  {Neeley} et~al.}}]{freedman_etal_2019}
\bibinfo{author}{\bibfnamefont{W.~L.} \bibnamefont{{Freedman}}},
  \bibinfo{author}{\bibfnamefont{B.~F.} \bibnamefont{{Madore}}},
  \bibinfo{author}{\bibfnamefont{D.}~\bibnamefont{{Hatt}}},
  \bibinfo{author}{\bibfnamefont{T.~J.} \bibnamefont{{Hoyt}}},
  \bibinfo{author}{\bibfnamefont{I.~S.} \bibnamefont{{Jang}}},
  \bibinfo{author}{\bibfnamefont{R.~L.} \bibnamefont{{Beaton}}},
  \bibinfo{author}{\bibfnamefont{C.~R.} \bibnamefont{{Burns}}},
  \bibinfo{author}{\bibfnamefont{M.~G.} \bibnamefont{{Lee}}},
  \bibinfo{author}{\bibfnamefont{A.~J.} \bibnamefont{{Monson}}},
  \bibinfo{author}{\bibfnamefont{J.~R.} \bibnamefont{{Neeley}}},
  \bibnamefont{et~al.}, \bibinfo{journal}{Astrophys. J.}
  \textbf{\bibinfo{volume}{882}}, \bibinfo{eid}{34} (\bibinfo{year}{2019}),
  \eprint{1907.05922}.

\bibitem[{\citenamefont{{Freedman} et~al.}(2020)\citenamefont{{Freedman},
  {Madore}, {Hoyt}, {Jang}, {Beaton}, {Lee}, {Monson}, {Neeley}, and
  {Rich}}}]{freedman_etal_2020}
\bibinfo{author}{\bibfnamefont{W.~L.} \bibnamefont{{Freedman}}},
  \bibinfo{author}{\bibfnamefont{B.~F.} \bibnamefont{{Madore}}},
  \bibinfo{author}{\bibfnamefont{T.}~\bibnamefont{{Hoyt}}},
  \bibinfo{author}{\bibfnamefont{I.~S.} \bibnamefont{{Jang}}},
  \bibinfo{author}{\bibfnamefont{R.}~\bibnamefont{{Beaton}}},
  \bibinfo{author}{\bibfnamefont{M.~G.} \bibnamefont{{Lee}}},
  \bibinfo{author}{\bibfnamefont{A.}~\bibnamefont{{Monson}}},
  \bibinfo{author}{\bibfnamefont{J.}~\bibnamefont{{Neeley}}}, \bibnamefont{and}
  \bibinfo{author}{\bibfnamefont{J.}~\bibnamefont{{Rich}}},
  \bibinfo{journal}{Astrophys. J.} \textbf{\bibinfo{volume}{891}},
  \bibinfo{eid}{57} (\bibinfo{year}{2020}), \eprint{2002.01550}.

\bibitem[{\citenamefont{{Rameez} and {Sarkar}}(2019)}]{rameez_sarkar_2019}
\bibinfo{author}{\bibfnamefont{M.}~\bibnamefont{{Rameez}}} \bibnamefont{and}
  \bibinfo{author}{\bibfnamefont{S.}~\bibnamefont{{Sarkar}}},
  \bibinfo{journal}{ArXiv e-prints}  (\bibinfo{year}{2019}),
  \eprint{1911.06456}.

\bibitem[{\citenamefont{{Dom{\'\i}nguez}
  et~al.}(2019)\citenamefont{{Dom{\'\i}nguez}, {Wojtak}, {Finke}, {Ajello},
  {Helgason}, {Prada}, {Desai}, {Paliya}, {Marcotulli}, and
  {Hartmann}}}]{dominguez_etal_2019}
\bibinfo{author}{\bibfnamefont{A.}~\bibnamefont{{Dom{\'\i}nguez}}},
  \bibinfo{author}{\bibfnamefont{R.}~\bibnamefont{{Wojtak}}},
  \bibinfo{author}{\bibfnamefont{J.}~\bibnamefont{{Finke}}},
  \bibinfo{author}{\bibfnamefont{M.}~\bibnamefont{{Ajello}}},
  \bibinfo{author}{\bibfnamefont{K.}~\bibnamefont{{Helgason}}},
  \bibinfo{author}{\bibfnamefont{F.}~\bibnamefont{{Prada}}},
  \bibinfo{author}{\bibfnamefont{A.}~\bibnamefont{{Desai}}},
  \bibinfo{author}{\bibfnamefont{V.}~\bibnamefont{{Paliya}}},
  \bibinfo{author}{\bibfnamefont{L.}~\bibnamefont{{Marcotulli}}},
  \bibnamefont{and} \bibinfo{author}{\bibfnamefont{D.~H.}
  \bibnamefont{{Hartmann}}}, \bibinfo{journal}{Astrophys. J.}
  \textbf{\bibinfo{volume}{885}}, \bibinfo{eid}{137} (\bibinfo{year}{2019}),
  \eprint{1903.12097}.

\bibitem[{\citenamefont{{Martinelli} and
  {Tutusaus}}(2019)}]{martinelli_tutusaus_2019}
\bibinfo{author}{\bibfnamefont{M.}~\bibnamefont{{Martinelli}}}
  \bibnamefont{and}
  \bibinfo{author}{\bibfnamefont{I.}~\bibnamefont{{Tutusaus}}},
  \bibinfo{journal}{Symmetry} \textbf{\bibinfo{volume}{11}},
  \bibinfo{pages}{986} (\bibinfo{year}{2019}), \eprint{1906.09189}.

\bibitem[{\citenamefont{{Cuceu} et~al.}(2019)\citenamefont{{Cuceu}, {Farr},
  {Lemos}, and {Font-Ribera}}}]{Cuceu_2019}
\bibinfo{author}{\bibfnamefont{A.}~\bibnamefont{{Cuceu}}},
  \bibinfo{author}{\bibfnamefont{J.}~\bibnamefont{{Farr}}},
  \bibinfo{author}{\bibfnamefont{P.}~\bibnamefont{{Lemos}}}, \bibnamefont{and}
  \bibinfo{author}{\bibfnamefont{A.}~\bibnamefont{{Font-Ribera}}},
  \bibinfo{journal}{J. Cosmol. Astropart. Phys.}
  \textbf{\bibinfo{volume}{2019}}, \bibinfo{eid}{044} (\bibinfo{year}{2019}),
  \eprint{1906.11628}.

\bibitem[{\citenamefont{{Zeng} and {Yan}}(2019)}]{zeng_yan_2019}
\bibinfo{author}{\bibfnamefont{H.}~\bibnamefont{{Zeng}}} \bibnamefont{and}
  \bibinfo{author}{\bibfnamefont{D.}~\bibnamefont{{Yan}}},
  \bibinfo{journal}{Astrophys. J.} \textbf{\bibinfo{volume}{882}},
  \bibinfo{eid}{87} (\bibinfo{year}{2019}), \eprint{1907.10965}.

\bibitem[{\citenamefont{{Sch{\"o}neberg}
  et~al.}(2019)\citenamefont{{Sch{\"o}neberg}, {Lesgourgues}, and
  {Hooper}}}]{schoneberg_etal_2019}
\bibinfo{author}{\bibfnamefont{N.}~\bibnamefont{{Sch{\"o}neberg}}},
  \bibinfo{author}{\bibfnamefont{J.}~\bibnamefont{{Lesgourgues}}},
  \bibnamefont{and} \bibinfo{author}{\bibfnamefont{D.~C.}
  \bibnamefont{{Hooper}}}, \bibinfo{journal}{J. Cosmol. Astropart. Phys.}
  \textbf{\bibinfo{volume}{2019}}, \bibinfo{eid}{029} (\bibinfo{year}{2019}),
  \eprint{1907.11594}.

\bibitem[{\citenamefont{{Lin} and {Ishak}}(2019)}]{lin_ishak_2019}
\bibinfo{author}{\bibfnamefont{W.}~\bibnamefont{{Lin}}} \bibnamefont{and}
  \bibinfo{author}{\bibfnamefont{M.}~\bibnamefont{{Ishak}}},
  \bibinfo{journal}{ArXiv e-prints}  (\bibinfo{year}{2019}),
  \eprint{1909.10991}.

\bibitem[{\citenamefont{{Zhang} and {Huang}}(2019)}]{zhang_huang_2019}
\bibinfo{author}{\bibfnamefont{X.}~\bibnamefont{{Zhang}}} \bibnamefont{and}
  \bibinfo{author}{\bibfnamefont{Q.-G.} \bibnamefont{{Huang}}},
  \bibinfo{journal}{ArXiv e-prints}  (\bibinfo{year}{2019}),
  \eprint{1911.09439}.

\bibitem[{\citenamefont{{Cao} et~al.}(2021{\natexlab{a}})\citenamefont{{Cao},
  {Ryan}, {Khadka}, and {Ratra}}}]{Cao_Ryan_Khadka_Ratra}
\bibinfo{author}{\bibfnamefont{S.}~\bibnamefont{{Cao}}},
  \bibinfo{author}{\bibfnamefont{J.}~\bibnamefont{{Ryan}}},
  \bibinfo{author}{\bibfnamefont{N.}~\bibnamefont{{Khadka}}}, \bibnamefont{and}
  \bibinfo{author}{\bibfnamefont{B.}~\bibnamefont{{Ratra}}},
  \bibinfo{journal}{Mon. Not. R. Astron. Soc.} \textbf{\bibinfo{volume}{501}},
  \bibinfo{pages}{1520} (\bibinfo{year}{2021}{\natexlab{a}}),
  \eprint{2009.12953}.

\bibitem[{\citenamefont{{Scolnic} et~al.}(2018)\citenamefont{{Scolnic},
  {Jones}, {Rest}, {Pan}, {Chornock}, {Foley}, {Huber}, {Kessler}, {Narayan},
  {Riess} et~al.}}]{scolnic_et_al_2018}
\bibinfo{author}{\bibfnamefont{D.~M.} \bibnamefont{{Scolnic}}},
  \bibinfo{author}{\bibfnamefont{D.~O.} \bibnamefont{{Jones}}},
  \bibinfo{author}{\bibfnamefont{A.}~\bibnamefont{{Rest}}},
  \bibinfo{author}{\bibfnamefont{Y.~C.} \bibnamefont{{Pan}}},
  \bibinfo{author}{\bibfnamefont{R.}~\bibnamefont{{Chornock}}},
  \bibinfo{author}{\bibfnamefont{R.~J.} \bibnamefont{{Foley}}},
  \bibinfo{author}{\bibfnamefont{M.~E.} \bibnamefont{{Huber}}},
  \bibinfo{author}{\bibfnamefont{R.}~\bibnamefont{{Kessler}}},
  \bibinfo{author}{\bibfnamefont{G.}~\bibnamefont{{Narayan}}},
  \bibinfo{author}{\bibfnamefont{A.~G.} \bibnamefont{{Riess}}},
  \bibnamefont{et~al.}, \bibinfo{journal}{Astrophys. J.}
  \textbf{\bibinfo{volume}{859}}, \bibinfo{eid}{101} (\bibinfo{year}{2018}),
  \eprint{1710.00845}.

\bibitem[{\citenamefont{{\MakeLowercase{E}BOSS
  Collaboration}}(2020)}]{eBOSS_2020}
\bibinfo{author}{\bibnamefont{{\MakeLowercase{E}BOSS Collaboration}}},
  \bibinfo{journal}{ArXiv e-prints} \bibinfo{eid}{arXiv:2007.08991}
  (\bibinfo{year}{2020}).

\bibitem[{\citenamefont{{Lamb} and {Reichart}}(2001)}]{Lamb2001}
\bibinfo{author}{\bibfnamefont{D.~Q.} \bibnamefont{{Lamb}}} \bibnamefont{and}
  \bibinfo{author}{\bibfnamefont{D.~E.} \bibnamefont{{Reichart}}}, in
  \emph{\bibinfo{booktitle}{Gamma-ray Bursts in the Afterglow Era}}, edited by
  \bibinfo{editor}{\bibfnamefont{E.}~\bibnamefont{{Costa}}},
  \bibinfo{editor}{\bibfnamefont{F.}~\bibnamefont{{Frontera}}},
  \bibnamefont{and} \bibinfo{editor}{\bibfnamefont{J.}~\bibnamefont{{Hjorth}}}
  (\bibinfo{year}{2001}), p. \bibinfo{pages}{226}, \eprint{astro-ph/0108099}.

\bibitem[{\citenamefont{{Amati} et~al.}(2002)\citenamefont{{Amati}, {Frontera},
  {Tavani}, {in't Zand }, {Antonelli}, {Costa}, {Feroci}, {Guidorzi}, {Heise},
  {Masetti} et~al.}}]{Amati2002}
\bibinfo{author}{\bibfnamefont{L.}~\bibnamefont{{Amati}}},
  \bibinfo{author}{\bibfnamefont{F.}~\bibnamefont{{Frontera}}},
  \bibinfo{author}{\bibfnamefont{M.}~\bibnamefont{{Tavani}}},
  \bibinfo{author}{\bibfnamefont{J.~J.~M.} \bibnamefont{{in't Zand }}},
  \bibinfo{author}{\bibfnamefont{A.}~\bibnamefont{{Antonelli}}},
  \bibinfo{author}{\bibfnamefont{E.}~\bibnamefont{{Costa}}},
  \bibinfo{author}{\bibfnamefont{M.}~\bibnamefont{{Feroci}}},
  \bibinfo{author}{\bibfnamefont{C.}~\bibnamefont{{Guidorzi}}},
  \bibinfo{author}{\bibfnamefont{J.}~\bibnamefont{{Heise}}},
  \bibinfo{author}{\bibfnamefont{N.}~\bibnamefont{{Masetti}}},
  \bibnamefont{et~al.}, \bibinfo{journal}{Astron. Astrophys.}
  \textbf{\bibinfo{volume}{390}}, \bibinfo{pages}{81} (\bibinfo{year}{2002}),
  \eprint{astro-ph/0205230}.

\bibitem[{\citenamefont{{Amati} et~al.}(2008)\citenamefont{{Amati}, {Guidorzi},
  {Frontera}, {Della Valle}, {Finelli}, {Landi}, and {Montanari}}}]{Amati2008}
\bibinfo{author}{\bibfnamefont{L.}~\bibnamefont{{Amati}}},
  \bibinfo{author}{\bibfnamefont{C.}~\bibnamefont{{Guidorzi}}},
  \bibinfo{author}{\bibfnamefont{F.}~\bibnamefont{{Frontera}}},
  \bibinfo{author}{\bibfnamefont{M.}~\bibnamefont{{Della Valle}}},
  \bibinfo{author}{\bibfnamefont{F.}~\bibnamefont{{Finelli}}},
  \bibinfo{author}{\bibfnamefont{R.}~\bibnamefont{{Landi}}}, \bibnamefont{and}
  \bibinfo{author}{\bibfnamefont{E.}~\bibnamefont{{Montanari}}},
  \bibinfo{journal}{Mon. Not. R. Astron. Soc.} \textbf{\bibinfo{volume}{391}},
  \bibinfo{pages}{577} (\bibinfo{year}{2008}), \eprint{0805.0377}.

\bibitem[{\citenamefont{{Amati} et~al.}(2009)\citenamefont{{Amati}, {Frontera},
  and {Guidorzi}}}]{Amati_2009}
\bibinfo{author}{\bibfnamefont{L.}~\bibnamefont{{Amati}}},
  \bibinfo{author}{\bibfnamefont{F.}~\bibnamefont{{Frontera}}},
  \bibnamefont{and}
  \bibinfo{author}{\bibfnamefont{C.}~\bibnamefont{{Guidorzi}}},
  \bibinfo{journal}{Astron. Astrophys.} \textbf{\bibinfo{volume}{508}},
  \bibinfo{pages}{173} (\bibinfo{year}{2009}), \eprint{0907.0384}.

\bibitem[{\citenamefont{{Ghirlanda} et~al.}(2004)\citenamefont{{Ghirlanda},
  {Ghisellini}, and {Lazzati}}}]{Ghirlanda2004}
\bibinfo{author}{\bibfnamefont{G.}~\bibnamefont{{Ghirlanda}}},
  \bibinfo{author}{\bibfnamefont{G.}~\bibnamefont{{Ghisellini}}},
  \bibnamefont{and}
  \bibinfo{author}{\bibfnamefont{D.}~\bibnamefont{{Lazzati}}},
  \bibinfo{journal}{Astrophys. J.} \textbf{\bibinfo{volume}{616}},
  \bibinfo{pages}{331} (\bibinfo{year}{2004}), \eprint{astro-ph/0405602}.

\bibitem[{\citenamefont{{Demianski} and {Piedipalumbo}}(2011)}]{Demianski2011}
\bibinfo{author}{\bibfnamefont{M.}~\bibnamefont{{Demianski}}} \bibnamefont{and}
  \bibinfo{author}{\bibfnamefont{E.}~\bibnamefont{{Piedipalumbo}}},
  \bibinfo{journal}{Mon. Not. R. Astron. Soc.} \textbf{\bibinfo{volume}{415}},
  \bibinfo{pages}{3580} (\bibinfo{year}{2011}), \eprint{1104.5614}.

\bibitem[{\citenamefont{{Wang} et~al.}(2015)\citenamefont{{Wang}, {Dai}, and
  {Liang}}}]{Fyan2015}
\bibinfo{author}{\bibfnamefont{F.~Y.} \bibnamefont{{Wang}}},
  \bibinfo{author}{\bibfnamefont{Z.~G.} \bibnamefont{{Dai}}}, \bibnamefont{and}
  \bibinfo{author}{\bibfnamefont{E.~W.} \bibnamefont{{Liang}}},
  \bibinfo{journal}{New Astron. Rev.} \textbf{\bibinfo{volume}{67}},
  \bibinfo{pages}{1} (\bibinfo{year}{2015}), \eprint{1504.00735}.

\bibitem[{\citenamefont{{Amati} et~al.}(2019)\citenamefont{{Amati},
  {D'Agostino}, {Luongo}, {Muccino}, and {Tantalo}}}]{Amati2019}
\bibinfo{author}{\bibfnamefont{L.}~\bibnamefont{{Amati}}},
  \bibinfo{author}{\bibfnamefont{R.}~\bibnamefont{{D'Agostino}}},
  \bibinfo{author}{\bibfnamefont{O.}~\bibnamefont{{Luongo}}},
  \bibinfo{author}{\bibfnamefont{M.}~\bibnamefont{{Muccino}}},
  \bibnamefont{and}
  \bibinfo{author}{\bibfnamefont{M.}~\bibnamefont{{Tantalo}}},
  \bibinfo{journal}{Mon. Not. R. Astron. Soc.} \textbf{\bibinfo{volume}{486}},
  \bibinfo{pages}{L46} (\bibinfo{year}{2019}), \eprint{1811.08934}.

\bibitem[{\citenamefont{{Wang} et~al.}(2016)\citenamefont{{Wang}, {Wang},
  {Cheng}, and {Dai}}}]{Wang_2016}
\bibinfo{author}{\bibfnamefont{J.~S.} \bibnamefont{{Wang}}},
  \bibinfo{author}{\bibfnamefont{F.~Y.} \bibnamefont{{Wang}}},
  \bibinfo{author}{\bibfnamefont{K.~S.} \bibnamefont{{Cheng}}},
  \bibnamefont{and} \bibinfo{author}{\bibfnamefont{Z.~G.} \bibnamefont{{Dai}}},
  \bibinfo{journal}{Astron. Astrophys.} \textbf{\bibinfo{volume}{585}},
  \bibinfo{eid}{A68} (\bibinfo{year}{2016}), \eprint{1509.08558}.

\bibitem[{\citenamefont{{Demianski} et~al.}(2017)\citenamefont{{Demianski},
  {Piedipalumbo}, {Sawant}, and {Amati}}}]{Demianski_2017a}
\bibinfo{author}{\bibfnamefont{M.}~\bibnamefont{{Demianski}}},
  \bibinfo{author}{\bibfnamefont{E.}~\bibnamefont{{Piedipalumbo}}},
  \bibinfo{author}{\bibfnamefont{D.}~\bibnamefont{{Sawant}}}, \bibnamefont{and}
  \bibinfo{author}{\bibfnamefont{L.}~\bibnamefont{{Amati}}},
  \bibinfo{journal}{Astron. Astrophys.} \textbf{\bibinfo{volume}{598}},
  \bibinfo{eid}{A112} (\bibinfo{year}{2017}), \eprint{1610.00854}.

\bibitem[{\citenamefont{{Fana Dirirsa} et~al.}(2019)\citenamefont{{Fana
  Dirirsa}, {Razzaque}, {Piron}, {Arimoto}, {Axelsson}, {Kocevski}, {Longo},
  {Ohno}, and {Zhu}}}]{Dirirsa_2019}
\bibinfo{author}{\bibfnamefont{F.}~\bibnamefont{{Fana Dirirsa}}},
  \bibinfo{author}{\bibfnamefont{S.}~\bibnamefont{{Razzaque}}},
  \bibinfo{author}{\bibfnamefont{F.}~\bibnamefont{{Piron}}},
  \bibinfo{author}{\bibfnamefont{M.}~\bibnamefont{{Arimoto}}},
  \bibinfo{author}{\bibfnamefont{M.}~\bibnamefont{{Axelsson}}},
  \bibinfo{author}{\bibfnamefont{D.}~\bibnamefont{{Kocevski}}},
  \bibinfo{author}{\bibfnamefont{F.}~\bibnamefont{{Longo}}},
  \bibinfo{author}{\bibfnamefont{M.}~\bibnamefont{{Ohno}}}, \bibnamefont{and}
  \bibinfo{author}{\bibfnamefont{S.}~\bibnamefont{{Zhu}}},
  \bibinfo{journal}{Astrophys. J.} \textbf{\bibinfo{volume}{887}},
  \bibinfo{eid}{13} (\bibinfo{year}{2019}), \eprint{1910.07009}.

\bibitem[{\citenamefont{{Khadka} and
  {Ratra}}(2020{\natexlab{c}})}]{Khadka_Ratra_2020}
\bibinfo{author}{\bibfnamefont{N.}~\bibnamefont{{Khadka}}} \bibnamefont{and}
  \bibinfo{author}{\bibfnamefont{B.}~\bibnamefont{{Ratra}}},
  \bibinfo{journal}{Mon. Not. R. Astron. Soc.} \textbf{\bibinfo{volume}{499}},
  \bibinfo{pages}{391} (\bibinfo{year}{2020}{\natexlab{c}}),
  \eprint{2007.13907}.

\bibitem[{\citenamefont{{Wei} and {Melia}}(2020)}]{Wei_Melia_2020}
\bibinfo{author}{\bibfnamefont{J.-J.} \bibnamefont{{Wei}}} \bibnamefont{and}
  \bibinfo{author}{\bibfnamefont{F.}~\bibnamefont{{Melia}}},
  \bibinfo{journal}{Astrophys. J.} \textbf{\bibinfo{volume}{888}},
  \bibinfo{eid}{99} (\bibinfo{year}{2020}), \eprint{1912.00668}.

\bibitem[{\citenamefont{{Khadka} and
  {Ratra}}(2020{\natexlab{d}})}]{Khadka_2020b}
\bibinfo{author}{\bibfnamefont{N.}~\bibnamefont{{Khadka}}} \bibnamefont{and}
  \bibinfo{author}{\bibfnamefont{B.}~\bibnamefont{{Ratra}}},
  \bibinfo{journal}{Mon. Not. R. Astron. Soc.} \textbf{\bibinfo{volume}{497}},
  \bibinfo{pages}{263} (\bibinfo{year}{2020}{\natexlab{d}}),
  \eprint{2004.09979}.

\bibitem[{\citenamefont{{Liang} and {Zhang}}(2005)}]{Liang2005}
\bibinfo{author}{\bibfnamefont{E.}~\bibnamefont{{Liang}}} \bibnamefont{and}
  \bibinfo{author}{\bibfnamefont{B.}~\bibnamefont{{Zhang}}},
  \bibinfo{journal}{Astrophys. J.} \textbf{\bibinfo{volume}{633}},
  \bibinfo{pages}{611} (\bibinfo{year}{2005}), \eprint{astro-ph/0504404}.

\bibitem[{\citenamefont{Muccino}(2020)}]{Muccino_2020}
\bibinfo{author}{\bibfnamefont{M.}~\bibnamefont{Muccino}},
  \bibinfo{journal}{Symmetry} \textbf{\bibinfo{volume}{12}},
  \bibinfo{pages}{1118} (\bibinfo{year}{2020}).

\bibitem[{\citenamefont{{Liu} and {Wei}}(2015)}]{Liu_Wei_2015}
\bibinfo{author}{\bibfnamefont{J.}~\bibnamefont{{Liu}}} \bibnamefont{and}
  \bibinfo{author}{\bibfnamefont{H.}~\bibnamefont{{Wei}}},
  \bibinfo{journal}{Gen. Relativ. Gravit.} \textbf{\bibinfo{volume}{47}},
  \bibinfo{eid}{141} (\bibinfo{year}{2015}), \eprint{1410.3960}.

\bibitem[{\citenamefont{{D'Agostini}}(2005)}]{D'Agostini_2005}
\bibinfo{author}{\bibfnamefont{G.}~\bibnamefont{{D'Agostini}}},
  \bibinfo{journal}{ArXiv e-prints} \bibinfo{eid}{physics/0511182}
  (\bibinfo{year}{2005}).

\bibitem[{\citenamefont{{Breuval} et~al.}(2020)\citenamefont{{Breuval},
  {Kervella}, {Anderson}, {Riess}, {Arenou}, {Trahin}, {M{\'e}rand},
  {Gallenne}, {Gieren}, {Storm} et~al.}}]{Breuvaletal_2020}
\bibinfo{author}{\bibfnamefont{L.}~\bibnamefont{{Breuval}}},
  \bibinfo{author}{\bibfnamefont{P.}~\bibnamefont{{Kervella}}},
  \bibinfo{author}{\bibfnamefont{R.~I.} \bibnamefont{{Anderson}}},
  \bibinfo{author}{\bibfnamefont{A.~G.} \bibnamefont{{Riess}}},
  \bibinfo{author}{\bibfnamefont{F.}~\bibnamefont{{Arenou}}},
  \bibinfo{author}{\bibfnamefont{B.}~\bibnamefont{{Trahin}}},
  \bibinfo{author}{\bibfnamefont{A.}~\bibnamefont{{M{\'e}rand}}},
  \bibinfo{author}{\bibfnamefont{A.}~\bibnamefont{{Gallenne}}},
  \bibinfo{author}{\bibfnamefont{W.}~\bibnamefont{{Gieren}}},
  \bibinfo{author}{\bibfnamefont{J.}~\bibnamefont{{Storm}}},
  \bibnamefont{et~al.}, \bibinfo{journal}{Astron. Astrophys.}
  \textbf{\bibinfo{volume}{643}}, \bibinfo{eid}{A115} (\bibinfo{year}{2020}),
  \eprint{2006.08763}.

\bibitem[{\citenamefont{{Efstathiou}}(2020)}]{Efstathiou_2020}
\bibinfo{author}{\bibfnamefont{G.}~\bibnamefont{{Efstathiou}}},
  \bibinfo{journal}{ArXiv e-prints} \bibinfo{eid}{arXiv:2007.10716}
  (\bibinfo{year}{2020}).

\bibitem[{\citenamefont{{Khetan} et~al.}(2020)\citenamefont{{Khetan}, {Izzo},
  {Branchesi}, {Wojtak}, {Cantiello}, {Murugeshan}, {Cappellaro}, {Della
  Valle}, {Gall}, {Hjorth} et~al.}}]{Khetan_et_al_2020}
\bibinfo{author}{\bibfnamefont{N.}~\bibnamefont{{Khetan}}},
  \bibinfo{author}{\bibfnamefont{L.}~\bibnamefont{{Izzo}}},
  \bibinfo{author}{\bibfnamefont{M.}~\bibnamefont{{Branchesi}}},
  \bibinfo{author}{\bibfnamefont{R.}~\bibnamefont{{Wojtak}}},
  \bibinfo{author}{\bibfnamefont{M.}~\bibnamefont{{Cantiello}}},
  \bibinfo{author}{\bibfnamefont{C.}~\bibnamefont{{Murugeshan}}},
  \bibinfo{author}{\bibfnamefont{A.~A.~E.} \bibnamefont{{Cappellaro}}},
  \bibinfo{author}{\bibfnamefont{M.}~\bibnamefont{{Della Valle}}},
  \bibinfo{author}{\bibfnamefont{C.}~\bibnamefont{{Gall}}},
  \bibinfo{author}{\bibfnamefont{J.}~\bibnamefont{{Hjorth}}},
  \bibnamefont{et~al.}, \bibinfo{journal}{ArXiv e-prints}
  \bibinfo{eid}{arXiv:2008.07754} (\bibinfo{year}{2020}), \eprint{2008.07754}.

\bibitem[{\citenamefont{{Blum} et~al.}(2020)\citenamefont{{Blum}, {Castorina},
  and {Simonovi{\'c}}}}]{Blum_et_al_2020}
\bibinfo{author}{\bibfnamefont{K.}~\bibnamefont{{Blum}}},
  \bibinfo{author}{\bibfnamefont{E.}~\bibnamefont{{Castorina}}},
  \bibnamefont{and}
  \bibinfo{author}{\bibfnamefont{M.}~\bibnamefont{{Simonovi{\'c}}}},
  \bibinfo{journal}{Astrophys. J. Lett.} \textbf{\bibinfo{volume}{892}},
  \bibinfo{eid}{L27} (\bibinfo{year}{2020}), \eprint{2001.07182}.

\bibitem[{\citenamefont{{Lyu} et~al.}(2020)\citenamefont{{Lyu}, {Haridasu},
  {Viel}, and {Xia}}}]{Lyu_et_al_2020}
\bibinfo{author}{\bibfnamefont{M.-Z.} \bibnamefont{{Lyu}}},
  \bibinfo{author}{\bibfnamefont{B.~S.} \bibnamefont{{Haridasu}}},
  \bibinfo{author}{\bibfnamefont{M.}~\bibnamefont{{Viel}}}, \bibnamefont{and}
  \bibinfo{author}{\bibfnamefont{J.-Q.} \bibnamefont{{Xia}}},
  \bibinfo{journal}{Astrophys. J.} \textbf{\bibinfo{volume}{900}},
  \bibinfo{eid}{160} (\bibinfo{year}{2020}), \eprint{2001.08713}.

\bibitem[{\citenamefont{{Philcox} et~al.}(2020)\citenamefont{{Philcox},
  {Ivanov}, {Simonovi{\'c}}, and {Zaldarriaga}}}]{Philcox_et_al_2020}
\bibinfo{author}{\bibfnamefont{O.~H.~E.} \bibnamefont{{Philcox}}},
  \bibinfo{author}{\bibfnamefont{M.~M.} \bibnamefont{{Ivanov}}},
  \bibinfo{author}{\bibfnamefont{M.}~\bibnamefont{{Simonovi{\'c}}}},
  \bibnamefont{and}
  \bibinfo{author}{\bibfnamefont{M.}~\bibnamefont{{Zaldarriaga}}},
  \bibinfo{journal}{J. Cosmol. Astropart. Phys.}
  \textbf{\bibinfo{volume}{2020}}, \bibinfo{eid}{032} (\bibinfo{year}{2020}),
  \eprint{2002.04035}.

\bibitem[{\citenamefont{{Zhang} and {Huang}}(2020)}]{Zhang_Huang_2020}
\bibinfo{author}{\bibfnamefont{X.}~\bibnamefont{{Zhang}}} \bibnamefont{and}
  \bibinfo{author}{\bibfnamefont{Q.-G.} \bibnamefont{{Huang}}},
  \bibinfo{journal}{ArXiv e-prints} \bibinfo{eid}{arXiv:2006.16692}
  (\bibinfo{year}{2020}), \eprint{2006.16692}.

\bibitem[{\citenamefont{{Birrer} et~al.}(2020)\citenamefont{{Birrer}, {Shajib},
  {Galan}, {Millon}, {Treu}, {Agnello}, {Auger}, {Chen}, {Christensen},
  {Collett} et~al.}}]{Birrer_et_al_2020}
\bibinfo{author}{\bibfnamefont{S.}~\bibnamefont{{Birrer}}},
  \bibinfo{author}{\bibfnamefont{A.~J.} \bibnamefont{{Shajib}}},
  \bibinfo{author}{\bibfnamefont{A.}~\bibnamefont{{Galan}}},
  \bibinfo{author}{\bibfnamefont{M.}~\bibnamefont{{Millon}}},
  \bibinfo{author}{\bibfnamefont{T.}~\bibnamefont{{Treu}}},
  \bibinfo{author}{\bibfnamefont{A.}~\bibnamefont{{Agnello}}},
  \bibinfo{author}{\bibfnamefont{M.}~\bibnamefont{{Auger}}},
  \bibinfo{author}{\bibfnamefont{G.~C.~F.} \bibnamefont{{Chen}}},
  \bibinfo{author}{\bibfnamefont{L.}~\bibnamefont{{Christensen}}},
  \bibinfo{author}{\bibfnamefont{T.}~\bibnamefont{{Collett}}},
  \bibnamefont{et~al.}, \bibinfo{journal}{ArXiv e-prints}
  \bibinfo{eid}{arXiv:2007.02941} (\bibinfo{year}{2020}), \eprint{2007.02941}.

\bibitem[{\citenamefont{{Denzel} et~al.}(2021)\citenamefont{{Denzel}, {Coles},
  {Saha}, and {Williams}}}]{Denzel_et_al_2020}
\bibinfo{author}{\bibfnamefont{P.}~\bibnamefont{{Denzel}}},
  \bibinfo{author}{\bibfnamefont{J.~P.} \bibnamefont{{Coles}}},
  \bibinfo{author}{\bibfnamefont{P.}~\bibnamefont{{Saha}}}, \bibnamefont{and}
  \bibinfo{author}{\bibfnamefont{L.~L.~R.} \bibnamefont{{Williams}}},
  \bibinfo{journal}{Mon. Not. R. Astron. Soc.} \textbf{\bibinfo{volume}{501}},
  \bibinfo{pages}{784} (\bibinfo{year}{2021}), \eprint{2007.14398}.

\bibitem[{\citenamefont{{Shirokov} et~al.}(2020)\citenamefont{{Shirokov},
  {Sokolov}, {Lovyagin}, {Amati}, {Baryshev}, {Sokolov}, and
  {Gorokhov}}}]{Shirokov2020}
\bibinfo{author}{\bibfnamefont{S.~I.} \bibnamefont{{Shirokov}}},
  \bibinfo{author}{\bibfnamefont{I.~V.} \bibnamefont{{Sokolov}}},
  \bibinfo{author}{\bibfnamefont{N.~Y.} \bibnamefont{{Lovyagin}}},
  \bibinfo{author}{\bibfnamefont{L.}~\bibnamefont{{Amati}}},
  \bibinfo{author}{\bibfnamefont{Y.~V.} \bibnamefont{{Baryshev}}},
  \bibinfo{author}{\bibfnamefont{V.~V.} \bibnamefont{{Sokolov}}},
  \bibnamefont{and} \bibinfo{author}{\bibfnamefont{V.~L.}
  \bibnamefont{{Gorokhov}}}, \bibinfo{journal}{Mon. Not. R. Astron. Soc.}
  \textbf{\bibinfo{volume}{496}}, \bibinfo{pages}{1530} (\bibinfo{year}{2020}),
  \eprint{2006.00981}.

\bibitem[{\citenamefont{{Cao} et~al.}(2021{\natexlab{b}})\citenamefont{{Cao},
  {Ryan}, and {Ratra}}}]{Cao_Ryan_Ratra_2021}
\bibinfo{author}{\bibfnamefont{S.}~\bibnamefont{{Cao}}},
  \bibinfo{author}{\bibfnamefont{J.}~\bibnamefont{{Ryan}}}, \bibnamefont{and}
  \bibinfo{author}{\bibfnamefont{B.}~\bibnamefont{{Ratra}}},
  \bibinfo{journal}{ArXiv e-prints}  (\bibinfo{year}{2021}{\natexlab{b}}),
  \eprint{2101.08817}.

\bibitem[{\citenamefont{{Park} and {Ratra}}(2020)}]{park_ratra_2020}
\bibinfo{author}{\bibfnamefont{C.-G.} \bibnamefont{{Park}}} \bibnamefont{and}
  \bibinfo{author}{\bibfnamefont{B.}~\bibnamefont{{Ratra}}},
  \bibinfo{journal}{Phys. Rev. D} \textbf{\bibinfo{volume}{101}},
  \bibinfo{eid}{083508} (\bibinfo{year}{2020}), \eprint{1908.08477}.

\bibitem[{\citenamefont{{Handley}}(2019{\natexlab{a}})}]{handley_2019a}
\bibinfo{author}{\bibfnamefont{W.}~\bibnamefont{{Handley}}},
  \bibinfo{journal}{arXiv e-prints}  (\bibinfo{year}{2019}{\natexlab{a}}),
  \eprint{1908.09139}.

\bibitem[{\citenamefont{{Jesus} et~al.}(2020)\citenamefont{{Jesus}, {Valentim},
  {Moraes}, and {Malheiro}}}]{jesus_etal_2019}
\bibinfo{author}{\bibfnamefont{J.~F.} \bibnamefont{{Jesus}}},
  \bibinfo{author}{\bibfnamefont{R.}~\bibnamefont{{Valentim}}},
  \bibinfo{author}{\bibfnamefont{P.~H.~R.~S.} \bibnamefont{{Moraes}}},
  \bibnamefont{and}
  \bibinfo{author}{\bibfnamefont{M.}~\bibnamefont{{Malheiro}}},
  \bibinfo{journal}{Mon. Not. R. Astron. Soc.}  (\bibinfo{year}{2020}),
  \eprint{1907.01033}.

\bibitem[{\citenamefont{{Li} et~al.}(2020)\citenamefont{{Li}, {Du}, and
  {Xu}}}]{li_etal_2020}
\bibinfo{author}{\bibfnamefont{E.-K.} \bibnamefont{{Li}}},
  \bibinfo{author}{\bibfnamefont{M.}~\bibnamefont{{Du}}}, \bibnamefont{and}
  \bibinfo{author}{\bibfnamefont{L.}~\bibnamefont{{Xu}}},
  \bibinfo{journal}{Mon. Not. R. Astron. Soc.} \textbf{\bibinfo{volume}{491}},
  \bibinfo{pages}{4960} (\bibinfo{year}{2020}), \eprint{1903.11433}.

\bibitem[{\citenamefont{{Geng} et~al.}(2020)\citenamefont{{Geng}, {Hsu}, {Yin},
  and {Zhang}}}]{geng_etal_2020}
\bibinfo{author}{\bibfnamefont{C.-Q.} \bibnamefont{{Geng}}},
  \bibinfo{author}{\bibfnamefont{Y.-T.} \bibnamefont{{Hsu}}},
  \bibinfo{author}{\bibfnamefont{L.}~\bibnamefont{{Yin}}}, \bibnamefont{and}
  \bibinfo{author}{\bibfnamefont{K.}~\bibnamefont{{Zhang}}},
  \bibinfo{journal}{Chin. Phys. C} \textbf{\bibinfo{volume}{44}},
  \bibinfo{eid}{105104} (\bibinfo{year}{2020}), \eprint{2002.05290}.

\bibitem[{\citenamefont{{Kumar} et~al.}(2020)\citenamefont{{Kumar}, {Jain},
  {Mahajan}, {Mukherjee}, and {Rani}}}]{kumar_etal_2020}
\bibinfo{author}{\bibfnamefont{D.}~\bibnamefont{{Kumar}}},
  \bibinfo{author}{\bibfnamefont{D.}~\bibnamefont{{Jain}}},
  \bibinfo{author}{\bibfnamefont{S.}~\bibnamefont{{Mahajan}}},
  \bibinfo{author}{\bibfnamefont{A.}~\bibnamefont{{Mukherjee}}},
  \bibnamefont{and} \bibinfo{author}{\bibfnamefont{N.}~\bibnamefont{{Rani}}},
  \bibinfo{journal}{ArXiv e-prints}  (\bibinfo{year}{2020}),
  \eprint{2002.06354}.

\bibitem[{\citenamefont{{Efstathiou} and
  {Gratton}}(2020)}]{efstathiou_gratton_2020}
\bibinfo{author}{\bibfnamefont{G.}~\bibnamefont{{Efstathiou}}}
  \bibnamefont{and}
  \bibinfo{author}{\bibfnamefont{S.}~\bibnamefont{{Gratton}}},
  \bibinfo{journal}{Mon. Not. R. Astron. Soc.} \textbf{\bibinfo{volume}{496}},
  \bibinfo{pages}{L91} (\bibinfo{year}{2020}), \eprint{2002.06892}.

\bibitem[{\citenamefont{{Di Valentino}
  et~al.}(2020{\natexlab{a}})\citenamefont{{Di Valentino}, {Melchiorri}, and
  {Silk}}}]{divalentino_etal_2020}
\bibinfo{author}{\bibfnamefont{E.}~\bibnamefont{{Di Valentino}}},
  \bibinfo{author}{\bibfnamefont{A.}~\bibnamefont{{Melchiorri}}},
  \bibnamefont{and} \bibinfo{author}{\bibfnamefont{J.}~\bibnamefont{{Silk}}},
  \bibinfo{journal}{ArXiv e-prints} \bibinfo{eid}{arXiv:2003.04935}
  (\bibinfo{year}{2020}{\natexlab{a}}).

\bibitem[{\citenamefont{{Di Valentino}
  et~al.}(2020{\natexlab{b}})\citenamefont{{Di Valentino}, {Melchiorri}, and
  {Silk}}}]{divalentino_etal_2020b}
\bibinfo{author}{\bibfnamefont{E.}~\bibnamefont{{Di Valentino}}},
  \bibinfo{author}{\bibfnamefont{A.}~\bibnamefont{{Melchiorri}}},
  \bibnamefont{and} \bibinfo{author}{\bibfnamefont{J.}~\bibnamefont{{Silk}}},
  \bibinfo{journal}{Nature Astronomy} \textbf{\bibinfo{volume}{4}},
  \bibinfo{pages}{196} (\bibinfo{year}{2020}{\natexlab{b}}),
  \eprint{1911.02087}.

\bibitem[{\citenamefont{{Gao} et~al.}(2020)\citenamefont{{Gao}, {Chen}, and
  {Zheng}}}]{gao_etal_2020}
\bibinfo{author}{\bibfnamefont{C.}~\bibnamefont{{Gao}}},
  \bibinfo{author}{\bibfnamefont{Y.}~\bibnamefont{{Chen}}}, \bibnamefont{and}
  \bibinfo{author}{\bibfnamefont{J.}~\bibnamefont{{Zheng}}},
  \bibinfo{journal}{Res. Astron. Astrophys.} \textbf{\bibinfo{volume}{20}},
  \bibinfo{eid}{151} (\bibinfo{year}{2020}), \eprint{2004.09291}.

\bibitem[{\citenamefont{{Abbassi} and {Abbassi}}(2020)}]{Abbassi_2020}
\bibinfo{author}{\bibfnamefont{M.~H.} \bibnamefont{{Abbassi}}}
  \bibnamefont{and} \bibinfo{author}{\bibfnamefont{A.~H.}
  \bibnamefont{{Abbassi}}}, \bibinfo{journal}{J. Cosmol. Astropart. Phys.}
  \textbf{\bibinfo{volume}{2020}}, \bibinfo{eid}{042} (\bibinfo{year}{2020}),
  \eprint{2006.06347}.

\bibitem[{\citenamefont{{Yang} and {Gong}}(2020)}]{Yang_2020}
\bibinfo{author}{\bibfnamefont{Y.}~\bibnamefont{{Yang}}} \bibnamefont{and}
  \bibinfo{author}{\bibfnamefont{Y.}~\bibnamefont{{Gong}}},
  \bibinfo{journal}{ArXiv e-prints}  (\bibinfo{year}{2020}),
  \eprint{2007.05714}.

\bibitem[{\citenamefont{{Agudelo Ruiz} et~al.}(2020)\citenamefont{{Agudelo
  Ruiz}, {Fabris}, {Velasquez-Toribio}, and {Shapiro}}}]{Agudelo_Ruiz_2020}
\bibinfo{author}{\bibfnamefont{J.~A.} \bibnamefont{{Agudelo Ruiz}}},
  \bibinfo{author}{\bibfnamefont{J.~C.} \bibnamefont{{Fabris}}},
  \bibinfo{author}{\bibfnamefont{A.~M.} \bibnamefont{{Velasquez-Toribio}}},
  \bibnamefont{and} \bibinfo{author}{\bibfnamefont{I.~L.}
  \bibnamefont{{Shapiro}}}, \bibinfo{journal}{Gravit. Cosmol.}
  \textbf{\bibinfo{volume}{26}}, \bibinfo{pages}{316} (\bibinfo{year}{2020}),
  \eprint{2007.12636}.

\bibitem[{\citenamefont{{Vel{\'a}squez-Toribio} and
  {Fabris}}(2020)}]{Velasquez-Toribio_2020}
\bibinfo{author}{\bibfnamefont{A.~M.} \bibnamefont{{Vel{\'a}squez-Toribio}}}
  \bibnamefont{and} \bibinfo{author}{\bibfnamefont{J.~C.}
  \bibnamefont{{Fabris}}}, \bibinfo{journal}{ArXiv e-prints}
  (\bibinfo{year}{2020}), \eprint{2008.12741}.

\bibitem[{\citenamefont{{Vagnozzi}
  et~al.}(2020{\natexlab{a}})\citenamefont{{Vagnozzi}, {Di Valentino},
  {Gariazzo}, {Melchiorri}, {Mena}, and {Silk}}}]{Vagnozzi_2020a}
\bibinfo{author}{\bibfnamefont{S.}~\bibnamefont{{Vagnozzi}}},
  \bibinfo{author}{\bibfnamefont{E.}~\bibnamefont{{Di Valentino}}},
  \bibinfo{author}{\bibfnamefont{S.}~\bibnamefont{{Gariazzo}}},
  \bibinfo{author}{\bibfnamefont{A.}~\bibnamefont{{Melchiorri}}},
  \bibinfo{author}{\bibfnamefont{O.}~\bibnamefont{{Mena}}}, \bibnamefont{and}
  \bibinfo{author}{\bibfnamefont{J.}~\bibnamefont{{Silk}}},
  \bibinfo{journal}{ArXiv e-prints}  (\bibinfo{year}{2020}{\natexlab{a}}),
  \eprint{2010.02230}.

\bibitem[{\citenamefont{{Vagnozzi}
  et~al.}(2020{\natexlab{b}})\citenamefont{{Vagnozzi}, {Loeb}, and
  {Moresco}}}]{Vagnozzi_2020b}
\bibinfo{author}{\bibfnamefont{S.}~\bibnamefont{{Vagnozzi}}},
  \bibinfo{author}{\bibfnamefont{A.}~\bibnamefont{{Loeb}}}, \bibnamefont{and}
  \bibinfo{author}{\bibfnamefont{M.}~\bibnamefont{{Moresco}}},
  \bibinfo{journal}{ArXiv e-prints}  (\bibinfo{year}{2020}{\natexlab{b}}),
  \eprint{2011.11645}.

\bibitem[{\citenamefont{{Samushia} et~al.}(2010)\citenamefont{{Samushia},
  {Dev}, {Jain}, and {Ratra}}}]{Samushia_2010}
\bibinfo{author}{\bibfnamefont{L.}~\bibnamefont{{Samushia}}},
  \bibinfo{author}{\bibfnamefont{A.}~\bibnamefont{{Dev}}},
  \bibinfo{author}{\bibfnamefont{D.}~\bibnamefont{{Jain}}}, \bibnamefont{and}
  \bibinfo{author}{\bibfnamefont{B.}~\bibnamefont{{Ratra}}},
  \bibinfo{journal}{Phys. Lett. B} \textbf{\bibinfo{volume}{693}},
  \bibinfo{pages}{509} (\bibinfo{year}{2010}), \eprint{0906.2734}.

\bibitem[{\citenamefont{{Singh} et~al.}(2019)\citenamefont{{Singh}, {Sangwan},
  and {Jassal}}}]{singh_etal_2019}
\bibinfo{author}{\bibfnamefont{A.}~\bibnamefont{{Singh}}},
  \bibinfo{author}{\bibfnamefont{A.}~\bibnamefont{{Sangwan}}},
  \bibnamefont{and} \bibinfo{author}{\bibfnamefont{H.~K.}
  \bibnamefont{{Jassal}}}, \bibinfo{journal}{J. Cosmol. Astropart. Phys.}
  \textbf{\bibinfo{volume}{2019}}, \bibinfo{eid}{047} (\bibinfo{year}{2019}),
  \eprint{1811.07513}.

\bibitem[{\citenamefont{{Khadka} and
  {Ratra}}(2020{\natexlab{e}})}]{Khadka_2020d}
\bibinfo{author}{\bibfnamefont{N.}~\bibnamefont{{Khadka}}} \bibnamefont{and}
  \bibinfo{author}{\bibfnamefont{B.}~\bibnamefont{{Ratra}}},
  \bibinfo{journal}{ArXiv e-prints}  (\bibinfo{year}{2020}{\natexlab{e}}),
  \eprint{2012.09291}.

\bibitem[{\citenamefont{{Ure{\~n}a-L{\'o}pez} and
  {Roy}}(2020)}]{Urena-Lopez_2020}
\bibinfo{author}{\bibfnamefont{L.~A.} \bibnamefont{{Ure{\~n}a-L{\'o}pez}}}
  \bibnamefont{and} \bibinfo{author}{\bibfnamefont{N.}~\bibnamefont{{Roy}}},
  \bibinfo{journal}{Phys. Rev. D} \textbf{\bibinfo{volume}{102}},
  \bibinfo{eid}{063510} (\bibinfo{year}{2020}), \eprint{2007.08873}.

\bibitem[{\citenamefont{{DES Collaboration}}(2019{\natexlab{c}})}]{DES_2019c}
\bibinfo{author}{\bibnamefont{{DES Collaboration}}},
  \bibinfo{journal}{Astrophys. J. Lett.} \textbf{\bibinfo{volume}{872}},
  \bibinfo{eid}{L30} (\bibinfo{year}{2019}{\natexlab{c}}), \eprint{1811.02374}.

\bibitem[{\citenamefont{{DES Collaboration}}(2019{\natexlab{d}})}]{DES_2019d}
\bibinfo{author}{\bibnamefont{{DES Collaboration}}},
  \bibinfo{journal}{Astrophys. J.} \textbf{\bibinfo{volume}{874}},
  \bibinfo{eid}{150} (\bibinfo{year}{2019}{\natexlab{d}}), \eprint{1811.02377}.

\bibitem[{\citenamefont{{Lucchin} and {Matarrese}}(1985)}]{Lucchin_1985}
\bibinfo{author}{\bibfnamefont{F.}~\bibnamefont{{Lucchin}}} \bibnamefont{and}
  \bibinfo{author}{\bibfnamefont{S.}~\bibnamefont{{Matarrese}}},
  \bibinfo{journal}{Phys. Rev. D} \textbf{\bibinfo{volume}{32}},
  \bibinfo{pages}{1316} (\bibinfo{year}{1985}).

\bibitem[{\citenamefont{{Ratra}}(1989)}]{ratra_1989}
\bibinfo{author}{\bibfnamefont{B.}~\bibnamefont{{Ratra}}},
  \bibinfo{journal}{Phys. Rev. D} \textbf{\bibinfo{volume}{40}},
  \bibinfo{pages}{3939} (\bibinfo{year}{1989}).

\bibitem[{\citenamefont{{Lesgourgues} and {Tram}}(2014)}]{Lesgourgues_2014}
\bibinfo{author}{\bibfnamefont{J.}~\bibnamefont{{Lesgourgues}}}
  \bibnamefont{and} \bibinfo{author}{\bibfnamefont{T.}~\bibnamefont{{Tram}}},
  \bibinfo{journal}{J. Cosmol. Astropart. Phys.}
  \textbf{\bibinfo{volume}{2014}}, \bibinfo{eid}{032} (\bibinfo{year}{2014}),
  \eprint{1312.2697}.

\bibitem[{\citenamefont{{Bonga} et~al.}(2016)\citenamefont{{Bonga}, {Gupt}, and
  {Yokomizo}}}]{Bonga_2016}
\bibinfo{author}{\bibfnamefont{B.}~\bibnamefont{{Bonga}}},
  \bibinfo{author}{\bibfnamefont{B.}~\bibnamefont{{Gupt}}}, \bibnamefont{and}
  \bibinfo{author}{\bibfnamefont{N.}~\bibnamefont{{Yokomizo}}},
  \bibinfo{journal}{J. Cosmol. Astropart. Phys.}
  \textbf{\bibinfo{volume}{2016}}, \bibinfo{eid}{031} (\bibinfo{year}{2016}),
  \eprint{1605.07556}.

\bibitem[{\citenamefont{{Handley}}(2019{\natexlab{b}})}]{handley_2019}
\bibinfo{author}{\bibfnamefont{W.}~\bibnamefont{{Handley}}},
  \bibinfo{journal}{Phys. Rev. D} \textbf{\bibinfo{volume}{100}},
  \bibinfo{eid}{123517} (\bibinfo{year}{2019}{\natexlab{b}}),
  \eprint{1907.08524}.

\bibitem[{\citenamefont{{Thavanesan} et~al.}(2021)\citenamefont{{Thavanesan},
  {Werth}, and {Handley}}}]{Thavanesan_2021}
\bibinfo{author}{\bibfnamefont{A.}~\bibnamefont{{Thavanesan}}},
  \bibinfo{author}{\bibfnamefont{D.}~\bibnamefont{{Werth}}}, \bibnamefont{and}
  \bibinfo{author}{\bibfnamefont{W.}~\bibnamefont{{Handley}}},
  \bibinfo{journal}{Phys. Rev. D} \textbf{\bibinfo{volume}{103}},
  \bibinfo{eid}{023519} (\bibinfo{year}{2021}), \eprint{2009.05573}.

\bibitem[{\citenamefont{{\MakeLowercase{D}u Mas des Bourboux}
  et~al.}(2020)\citenamefont{{\MakeLowercase{D}u Mas des Bourboux}, {Rich},
  {Font-Ribera}, {de Sainte Agathe}, {Farr}, {Etourneau}, {Le Goff}, {Cuceu},
  {Balland}, {Bautista} et~al.}}]{duMas2020}
\bibinfo{author}{\bibfnamefont{H.}~\bibnamefont{{\MakeLowercase{D}u Mas des
  Bourboux}}}, \bibinfo{author}{\bibfnamefont{J.}~\bibnamefont{{Rich}}},
  \bibinfo{author}{\bibfnamefont{A.}~\bibnamefont{{Font-Ribera}}},
  \bibinfo{author}{\bibfnamefont{V.}~\bibnamefont{{de Sainte Agathe}}},
  \bibinfo{author}{\bibfnamefont{J.}~\bibnamefont{{Farr}}},
  \bibinfo{author}{\bibfnamefont{T.}~\bibnamefont{{Etourneau}}},
  \bibinfo{author}{\bibfnamefont{J.-M.} \bibnamefont{{Le Goff}}},
  \bibinfo{author}{\bibfnamefont{A.}~\bibnamefont{{Cuceu}}},
  \bibinfo{author}{\bibfnamefont{C.}~\bibnamefont{{Balland}}},
  \bibinfo{author}{\bibfnamefont{J.~E.} \bibnamefont{{Bautista}}},
  \bibnamefont{et~al.}, \bibinfo{journal}{Astrophys. J.}
  \textbf{\bibinfo{volume}{901}}, \bibinfo{eid}{153} (\bibinfo{year}{2020}),
  \eprint{2007.08995}.

\bibitem[{\citenamefont{{Conley} et~al.}(2011)\citenamefont{{Conley}, {Guy},
  {Sullivan}, {Regnault}, {Astier}, {Balland}, {Basa}, {Carlberg}, {Fouchez},
  {Hardin} et~al.}}]{Conley_et_al_2011}
\bibinfo{author}{\bibfnamefont{A.}~\bibnamefont{{Conley}}},
  \bibinfo{author}{\bibfnamefont{J.}~\bibnamefont{{Guy}}},
  \bibinfo{author}{\bibfnamefont{M.}~\bibnamefont{{Sullivan}}},
  \bibinfo{author}{\bibfnamefont{N.}~\bibnamefont{{Regnault}}},
  \bibinfo{author}{\bibfnamefont{P.}~\bibnamefont{{Astier}}},
  \bibinfo{author}{\bibfnamefont{C.}~\bibnamefont{{Balland}}},
  \bibinfo{author}{\bibfnamefont{S.}~\bibnamefont{{Basa}}},
  \bibinfo{author}{\bibfnamefont{R.~G.} \bibnamefont{{Carlberg}}},
  \bibinfo{author}{\bibfnamefont{D.}~\bibnamefont{{Fouchez}}},
  \bibinfo{author}{\bibfnamefont{D.}~\bibnamefont{{Hardin}}},
  \bibnamefont{et~al.}, \bibinfo{journal}{Astrophys. J., Suppl. Ser.}
  \textbf{\bibinfo{volume}{192}}, \bibinfo{eid}{1} (\bibinfo{year}{2011}),
  \eprint{1104.1443}.

\bibitem[{\citenamefont{{Huang} et~al.}(2020)\citenamefont{{Huang}, {Gao}, and
  {Xu}}}]{Revisit_Ryskin}
\bibinfo{author}{\bibfnamefont{Z.}~\bibnamefont{{Huang}}},
  \bibinfo{author}{\bibfnamefont{H.}~\bibnamefont{{Gao}}}, \bibnamefont{and}
  \bibinfo{author}{\bibfnamefont{H.}~\bibnamefont{{Xu}}},
  \bibinfo{journal}{Astropart. Phys.} \textbf{\bibinfo{volume}{114}},
  \bibinfo{pages}{77} (\bibinfo{year}{2020}), \eprint{1905.02441}.

\bibitem[{\citenamefont{{Fixsen}}(2009)}]{Fixsen}
\bibinfo{author}{\bibfnamefont{D.~J.} \bibnamefont{{Fixsen}}},
  \bibinfo{journal}{Astrophys. J.} \textbf{\bibinfo{volume}{707}},
  \bibinfo{pages}{916} (\bibinfo{year}{2009}), \eprint{0911.1955}.

\bibitem[{\citenamefont{{Dev \textit{et
  al}.}}(2002)}]{Dev_Safonova_Deepak_Lohiya_2002}
\bibinfo{author}{\bibfnamefont{A.}~\bibnamefont{{Dev \textit{et al}.}}},
  \bibinfo{journal}{Phys. Lett. B} \textbf{\bibinfo{volume}{548}},
  \bibinfo{pages}{12} (\bibinfo{year}{2002}), \eprint{astro-ph/0204150}.

\bibitem[{\citenamefont{{Dev} et~al.}(2001)\citenamefont{{Dev}, {Sethi}, and
  {Lohiya}}}]{Dev_Sethi_Lohiya_2001}
\bibinfo{author}{\bibfnamefont{A.}~\bibnamefont{{Dev}}},
  \bibinfo{author}{\bibfnamefont{M.}~\bibnamefont{{Sethi}}}, \bibnamefont{and}
  \bibinfo{author}{\bibfnamefont{D.}~\bibnamefont{{Lohiya}}},
  \bibinfo{journal}{Phys. Lett. B} \textbf{\bibinfo{volume}{504}},
  \bibinfo{pages}{207} (\bibinfo{year}{2001}), \eprint{astro-ph/0008193}.

\bibitem[{\citenamefont{{Kumar}}(2012)}]{Kumar_2012}
\bibinfo{author}{\bibfnamefont{S.}~\bibnamefont{{Kumar}}},
  \bibinfo{journal}{Mon. Not. R. Astron. Soc.} \textbf{\bibinfo{volume}{422}},
  \bibinfo{pages}{2532} (\bibinfo{year}{2012}), \eprint{1109.6924}.

\bibitem[{\citenamefont{{Rani} et~al.}(2015)\citenamefont{{Rani},
  {Altaibayeva}, {Shahalam}, {Singh}, and {Myrzakulov}}}]{Rani_et_al_2015}
\bibinfo{author}{\bibfnamefont{S.}~\bibnamefont{{Rani}}},
  \bibinfo{author}{\bibfnamefont{A.}~\bibnamefont{{Altaibayeva}}},
  \bibinfo{author}{\bibfnamefont{M.}~\bibnamefont{{Shahalam}}},
  \bibinfo{author}{\bibfnamefont{J.~K.} \bibnamefont{{Singh}}},
  \bibnamefont{and}
  \bibinfo{author}{\bibfnamefont{R.}~\bibnamefont{{Myrzakulov}}},
  \bibinfo{journal}{J. Cosmol. Astropart. Phys.}
  \textbf{\bibinfo{volume}{2015}}, \bibinfo{eid}{031} (\bibinfo{year}{2015}),
  \eprint{1404.6522}.

\bibitem[{\citenamefont{{Shafer}}(2015)}]{Shafer_2016}
\bibinfo{author}{\bibfnamefont{D.~L.} \bibnamefont{{Shafer}}},
  \bibinfo{journal}{Phys. Rev. D} \textbf{\bibinfo{volume}{91}},
  \bibinfo{eid}{103516} (\bibinfo{year}{2015}), \eprint{1502.05416}.

\bibitem[{\citenamefont{{Tutusaus} et~al.}(2016)\citenamefont{{Tutusaus},
  {Lamine}, {Blanchard}, {Dupays}, {Zolnierowski}, {Cohen-Tanugi}, {Ealet},
  {Escoffier}, {Le F{\`e}vre}, {Ili{\'c}} et~al.}}]{Tutusaus_et_al_2016}
\bibinfo{author}{\bibfnamefont{I.}~\bibnamefont{{Tutusaus}}},
  \bibinfo{author}{\bibfnamefont{B.}~\bibnamefont{{Lamine}}},
  \bibinfo{author}{\bibfnamefont{A.}~\bibnamefont{{Blanchard}}},
  \bibinfo{author}{\bibfnamefont{A.}~\bibnamefont{{Dupays}}},
  \bibinfo{author}{\bibfnamefont{Y.}~\bibnamefont{{Zolnierowski}}},
  \bibinfo{author}{\bibfnamefont{J.}~\bibnamefont{{Cohen-Tanugi}}},
  \bibinfo{author}{\bibfnamefont{A.}~\bibnamefont{{Ealet}}},
  \bibinfo{author}{\bibfnamefont{S.}~\bibnamefont{{Escoffier}}},
  \bibinfo{author}{\bibfnamefont{O.}~\bibnamefont{{Le F{\`e}vre}}},
  \bibinfo{author}{\bibfnamefont{S.}~\bibnamefont{{Ili{\'c}}}},
  \bibnamefont{et~al.}, \bibinfo{journal}{Phys. Rev. D}
  \textbf{\bibinfo{volume}{94}}, \bibinfo{eid}{103511} (\bibinfo{year}{2016}),
  \eprint{1610.03371}.

\bibitem[{\citenamefont{{Jain} et~al.}(2003)\citenamefont{{Jain}, {Dev}, and
  {Alcaniz}}}]{Jain_Dev_Alcaniz_2003}
\bibinfo{author}{\bibfnamefont{D.}~\bibnamefont{{Jain}}},
  \bibinfo{author}{\bibfnamefont{A.}~\bibnamefont{{Dev}}}, \bibnamefont{and}
  \bibinfo{author}{\bibfnamefont{J.~S.} \bibnamefont{{Alcaniz}}},
  \bibinfo{journal}{Class. Quantum Gravity} \textbf{\bibinfo{volume}{20}},
  \bibinfo{pages}{4485} (\bibinfo{year}{2003}), \eprint{astro-ph/0302025}.

\bibitem[{\citenamefont{{Zhu} et~al.}(2008)\citenamefont{{Zhu}, {Hu},
  {Alcaniz}, and {Liu}}}]{Zhu_Alcaniz_Liu_2008}
\bibinfo{author}{\bibfnamefont{Z.-H.} \bibnamefont{{Zhu}}},
  \bibinfo{author}{\bibfnamefont{M.}~\bibnamefont{{Hu}}},
  \bibinfo{author}{\bibfnamefont{J.~S.} \bibnamefont{{Alcaniz}}},
  \bibnamefont{and} \bibinfo{author}{\bibfnamefont{Y.~X.} \bibnamefont{{Liu}}},
  \bibinfo{journal}{Astron. Astrophys.} \textbf{\bibinfo{volume}{483}},
  \bibinfo{pages}{15} (\bibinfo{year}{2008}), \eprint{0712.3602}.

\bibitem[{\citenamefont{{Haridasu}
  et~al.}(2017{\natexlab{b}})\citenamefont{{Haridasu}, {Lukovi{\'c}},
  {D'Agostino}, and {Vittorio}}}]{Haridasu_AAP_2017}
\bibinfo{author}{\bibfnamefont{B.~S.} \bibnamefont{{Haridasu}}},
  \bibinfo{author}{\bibfnamefont{V.~V.} \bibnamefont{{Lukovi{\'c}}}},
  \bibinfo{author}{\bibfnamefont{R.}~\bibnamefont{{D'Agostino}}},
  \bibnamefont{and}
  \bibinfo{author}{\bibfnamefont{N.}~\bibnamefont{{Vittorio}}},
  \bibinfo{journal}{Astron. Astrophys.} \textbf{\bibinfo{volume}{600}},
  \bibinfo{eid}{L1} (\bibinfo{year}{2017}{\natexlab{b}}), \eprint{1702.08244}.

\bibitem[{\citenamefont{{Dolgov} et~al.}(2014)\citenamefont{{Dolgov},
  {Halenka}, and {Tkachev}}}]{Dolgov_Halenka_Tkachev_2014}
\bibinfo{author}{\bibfnamefont{A.}~\bibnamefont{{Dolgov}}},
  \bibinfo{author}{\bibfnamefont{V.}~\bibnamefont{{Halenka}}},
  \bibnamefont{and}
  \bibinfo{author}{\bibfnamefont{I.}~\bibnamefont{{Tkachev}}},
  \bibinfo{journal}{J. Cosmol. Astropart. Phys.}
  \textbf{\bibinfo{volume}{2014}}, \bibinfo{eid}{047} (\bibinfo{year}{2014}),
  \eprint{1406.2445}.

\bibitem[{\citenamefont{{Lohiya} et~al.}(1999)\citenamefont{{Lohiya}, {Batra},
  {Mahajan}, and {Mukherjee}}}]{Lohiya_Batra_Mahajan_Mukherjee_1999}
\bibinfo{author}{\bibfnamefont{D.}~\bibnamefont{{Lohiya}}},
  \bibinfo{author}{\bibfnamefont{A.}~\bibnamefont{{Batra}}},
  \bibinfo{author}{\bibfnamefont{S.}~\bibnamefont{{Mahajan}}},
  \bibnamefont{and}
  \bibinfo{author}{\bibfnamefont{A.}~\bibnamefont{{Mukherjee}}},
  \bibinfo{journal}{arXiv e-prints}  (\bibinfo{year}{1999}),
  \eprint{nucl-th/9902022}.

\bibitem[{\citenamefont{{Sethi} et~al.}(1999)\citenamefont{{Sethi}, {Batra},
  and {Lohiya}}}]{Sethi_Batra_Lohiya_1999}
\bibinfo{author}{\bibfnamefont{M.}~\bibnamefont{{Sethi}}},
  \bibinfo{author}{\bibfnamefont{A.}~\bibnamefont{{Batra}}}, \bibnamefont{and}
  \bibinfo{author}{\bibfnamefont{D.}~\bibnamefont{{Lohiya}}},
  \bibinfo{journal}{Phys. Rev. D} \textbf{\bibinfo{volume}{60}},
  \bibinfo{eid}{108301} (\bibinfo{year}{1999}).

\bibitem[{\citenamefont{{Kaplinghat} et~al.}(2000)\citenamefont{{Kaplinghat},
  {Steigman}, and {Walker}}}]{Kaplinghat_Steigman_Walker_2000}
\bibinfo{author}{\bibfnamefont{M.}~\bibnamefont{{Kaplinghat}}},
  \bibinfo{author}{\bibfnamefont{G.}~\bibnamefont{{Steigman}}},
  \bibnamefont{and} \bibinfo{author}{\bibfnamefont{T.~P.}
  \bibnamefont{{Walker}}}, \bibinfo{journal}{Phys. Rev. D}
  \textbf{\bibinfo{volume}{61}}, \bibinfo{eid}{103507} (\bibinfo{year}{2000}),
  \eprint{astro-ph/9911066}.

\bibitem[{\citenamefont{{Kaplinghat} et~al.}(1999)\citenamefont{{Kaplinghat},
  {Steigman}, {Tkachev}, and {Walker}}}]{Kaplinghat_Steigman_Tkachev_1999}
\bibinfo{author}{\bibfnamefont{M.}~\bibnamefont{{Kaplinghat}}},
  \bibinfo{author}{\bibfnamefont{G.}~\bibnamefont{{Steigman}}},
  \bibinfo{author}{\bibfnamefont{I.}~\bibnamefont{{Tkachev}}},
  \bibnamefont{and} \bibinfo{author}{\bibfnamefont{T.~P.}
  \bibnamefont{{Walker}}}, \bibinfo{journal}{Phys. Rev. D}
  \textbf{\bibinfo{volume}{59}}, \bibinfo{eid}{043514} (\bibinfo{year}{1999}),
  \eprint{astro-ph/9805114}.

\bibitem[{\citenamefont{{Gumjudpai}}(2013)}]{Gumjudpai}
\bibinfo{author}{\bibfnamefont{B.}~\bibnamefont{{Gumjudpai}}},
  \bibinfo{journal}{Mod. Phys. Lett. A} \textbf{\bibinfo{volume}{28}},
  \bibinfo{eid}{1350122} (\bibinfo{year}{2013}), \eprint{1307.4552}.

\bibitem[{\citenamefont{{Gumjudpai} and
  {Thepsuriya}}(2012)}]{Gumjudpai_Thepsuriya_2012}
\bibinfo{author}{\bibfnamefont{B.}~\bibnamefont{{Gumjudpai}}} \bibnamefont{and}
  \bibinfo{author}{\bibfnamefont{K.}~\bibnamefont{{Thepsuriya}}},
  \bibinfo{journal}{Astrophys. Space Sci.} \textbf{\bibinfo{volume}{342}},
  \bibinfo{pages}{537} (\bibinfo{year}{2012}), \eprint{1207.2920}.

\bibitem[{\citenamefont{{Kaeonikhom} et~al.}(2011)\citenamefont{{Kaeonikhom},
  {Gumjudpai}, and {Saridakis}}}]{Kaeonikhom_Gumjudpai_Saridakis_2011}
\bibinfo{author}{\bibfnamefont{C.}~\bibnamefont{{Kaeonikhom}}},
  \bibinfo{author}{\bibfnamefont{B.}~\bibnamefont{{Gumjudpai}}},
  \bibnamefont{and} \bibinfo{author}{\bibfnamefont{E.~N.}
  \bibnamefont{{Saridakis}}}, \bibinfo{journal}{Phys. Lett. B}
  \textbf{\bibinfo{volume}{695}}, \bibinfo{pages}{45} (\bibinfo{year}{2011}),
  \eprint{1008.2182}.

\bibitem[{\citenamefont{{Rangdee} and
  {Gumjudpai}}(2014)}]{Rangdee_Gumjudpai_2014}
\bibinfo{author}{\bibfnamefont{R.}~\bibnamefont{{Rangdee}}} \bibnamefont{and}
  \bibinfo{author}{\bibfnamefont{B.}~\bibnamefont{{Gumjudpai}}},
  \bibinfo{journal}{Astrophys. Space Sci.} \textbf{\bibinfo{volume}{349}},
  \bibinfo{pages}{975} (\bibinfo{year}{2014}), \eprint{1210.5550}.

\bibitem[{\citenamefont{{Wei}}(2004)}]{Wei_2004}
\bibinfo{author}{\bibfnamefont{Y.-H.} \bibnamefont{{Wei}}},
  \bibinfo{journal}{ArXiv e-prints}  (\bibinfo{year}{2004}),
  \eprint{astro-ph/0405368}.

\bibitem[{\citenamefont{{Singh} and
  {Lohiya}}(2015{\natexlab{a}})}]{Singh_Lohiya_2015_arXiv}
\bibinfo{author}{\bibfnamefont{P.}~\bibnamefont{{Singh}}} \bibnamefont{and}
  \bibinfo{author}{\bibfnamefont{D.}~\bibnamefont{{Lohiya}}},
  \bibinfo{journal}{arXiv e-prints}  (\bibinfo{year}{2015}{\natexlab{a}}),
  \eprint{1510.04961}.

\bibitem[{\citenamefont{{Singh} and
  {Lohiya}}(2015{\natexlab{b}})}]{Singh_Lohiya_2015_JCAP}
\bibinfo{author}{\bibfnamefont{P.}~\bibnamefont{{Singh}}} \bibnamefont{and}
  \bibinfo{author}{\bibfnamefont{D.}~\bibnamefont{{Lohiya}}},
  \bibinfo{journal}{J. Cosmol. Astropart. Phys.}
  \textbf{\bibinfo{volume}{2015}}, \bibinfo{eid}{061}
  (\bibinfo{year}{2015}{\natexlab{b}}), \eprint{1312.7706}.

\bibitem[{\citenamefont{{Kolb}}(1989)}]{Kolb_1989}
\bibinfo{author}{\bibfnamefont{E.~W.} \bibnamefont{{Kolb}}},
  \bibinfo{journal}{Astrophys. J.} \textbf{\bibinfo{volume}{344}},
  \bibinfo{pages}{543} (\bibinfo{year}{1989}).

\bibitem[{\citenamefont{{Ratsimbazafy}
  et~al.}(2017)\citenamefont{{Ratsimbazafy}, {Loubser}, {Crawford}, {Cress},
  {Bassett}, {Nichol}, and {V{\"a}is{\"a}nen}}}]{Ratsimbazafy_et_al_2017}
\bibinfo{author}{\bibfnamefont{A.~L.} \bibnamefont{{Ratsimbazafy}}},
  \bibinfo{author}{\bibfnamefont{S.~I.} \bibnamefont{{Loubser}}},
  \bibinfo{author}{\bibfnamefont{S.~M.} \bibnamefont{{Crawford}}},
  \bibinfo{author}{\bibfnamefont{C.~M.} \bibnamefont{{Cress}}},
  \bibinfo{author}{\bibfnamefont{B.~A.} \bibnamefont{{Bassett}}},
  \bibinfo{author}{\bibfnamefont{R.~C.} \bibnamefont{{Nichol}}},
  \bibnamefont{and}
  \bibinfo{author}{\bibfnamefont{P.}~\bibnamefont{{V{\"a}is{\"a}nen}}},
  \bibinfo{journal}{Mon. Not. R. Astron. Soc.} \textbf{\bibinfo{volume}{467}},
  \bibinfo{pages}{3239} (\bibinfo{year}{2017}), \eprint{1702.00418}.

\bibitem[{\citenamefont{{Ruiz} et~al.}(2012)\citenamefont{{Ruiz}, {Shafer},
  {Huterer}, and {Conley}}}]{Ruiz_et_al_2012}
\bibinfo{author}{\bibfnamefont{E.~J.} \bibnamefont{{Ruiz}}},
  \bibinfo{author}{\bibfnamefont{D.~L.} \bibnamefont{{Shafer}}},
  \bibinfo{author}{\bibfnamefont{D.}~\bibnamefont{{Huterer}}},
  \bibnamefont{and} \bibinfo{author}{\bibfnamefont{A.}~\bibnamefont{{Conley}}},
  \bibinfo{journal}{Phys. Rev. D} \textbf{\bibinfo{volume}{86}},
  \bibinfo{eid}{103004} (\bibinfo{year}{2012}), \eprint{1207.4781}.

\bibitem[{\citenamefont{{Seikel} et~al.}(2012)\citenamefont{{Seikel},
  {Clarkson}, and {Smith}}}]{Seikel_Clarkson_Smith_2012}
\bibinfo{author}{\bibfnamefont{M.}~\bibnamefont{{Seikel}}},
  \bibinfo{author}{\bibfnamefont{C.}~\bibnamefont{{Clarkson}}},
  \bibnamefont{and} \bibinfo{author}{\bibfnamefont{M.}~\bibnamefont{{Smith}}},
  \bibinfo{journal}{J. Cosmol. Astropart. Phys.}
  \textbf{\bibinfo{volume}{2012}}, \bibinfo{eid}{036} (\bibinfo{year}{2012}),
  \eprint{1204.2832}.

\bibitem[{\citenamefont{{Zhang} et~al.}(2014)\citenamefont{{Zhang}, {Zhang},
  {Yuan}, {Liu}, {Zhang}, and {Sun}}}]{73}
\bibinfo{author}{\bibfnamefont{C.}~\bibnamefont{{Zhang}}},
  \bibinfo{author}{\bibfnamefont{H.}~\bibnamefont{{Zhang}}},
  \bibinfo{author}{\bibfnamefont{S.}~\bibnamefont{{Yuan}}},
  \bibinfo{author}{\bibfnamefont{S.}~\bibnamefont{{Liu}}},
  \bibinfo{author}{\bibfnamefont{T.-J.} \bibnamefont{{Zhang}}},
  \bibnamefont{and} \bibinfo{author}{\bibfnamefont{Y.-C.} \bibnamefont{{Sun}}},
  \bibinfo{journal}{Res. Astron. Astrophys.} \textbf{\bibinfo{volume}{14}},
  \bibinfo{eid}{1221-1233} (\bibinfo{year}{2014}), \eprint{1207.4541}.

\bibitem[{\citenamefont{{Simon} et~al.}(2005)\citenamefont{{Simon}, {Verde},
  and {Jimenez}}}]{69}
\bibinfo{author}{\bibfnamefont{J.}~\bibnamefont{{Simon}}},
  \bibinfo{author}{\bibfnamefont{L.}~\bibnamefont{{Verde}}}, \bibnamefont{and}
  \bibinfo{author}{\bibfnamefont{R.}~\bibnamefont{{Jimenez}}},
  \bibinfo{journal}{Phys. Rev. D} \textbf{\bibinfo{volume}{71}},
  \bibinfo{eid}{123001} (\bibinfo{year}{2005}), \eprint{astro-ph/0412269}.

\bibitem[{\citenamefont{{Stern} et~al.}(2010)\citenamefont{{Stern}, {Jimenez},
  {Verde}, {Kamionkowski}, and {Stanford}}}]{71}
\bibinfo{author}{\bibfnamefont{D.}~\bibnamefont{{Stern}}},
  \bibinfo{author}{\bibfnamefont{R.}~\bibnamefont{{Jimenez}}},
  \bibinfo{author}{\bibfnamefont{L.}~\bibnamefont{{Verde}}},
  \bibinfo{author}{\bibfnamefont{M.}~\bibnamefont{{Kamionkowski}}},
  \bibnamefont{and} \bibinfo{author}{\bibfnamefont{S.~A.}
  \bibnamefont{{Stanford}}}, \bibinfo{journal}{J. Cosmol. Astropart. Phys.}
  \textbf{\bibinfo{volume}{2}}, \bibinfo{eid}{008} (\bibinfo{year}{2010}),
  \eprint{0907.3149}.

\end{thebibliography}
